\documentclass[useAMS,usenatbib]{mn2e}

\usepackage{natbib}
\usepackage{amsmath}
\usepackage{epsfig}
\usepackage{afterpage}
\usepackage{subfigure}
\title[The radio source population at high frequency.]{The radio source population at high frequency: follow-up of the 15-GHz 9C survey.}
\author[R.C. Bolton et al.]
{R.C.~Bolton$^1$,\thanks{E-mail:r.bolton@mrao.cam.ac.uk (RCB)} 
G.~Cotter$^2$,
G.G.~Pooley$^1$, 
J.M.~Riley$^1$,
E.M.~Waldram$^1$,
\newauthor
 C.J.~Chandler$^3$,
 B.S.~Mason$^4$, 
 T.J.~Pearson$^5$,
 A.C.S.~Readhead$^5$\\
$^1$Cavendish Astrophysics, Dep. of Physics, Madingley Road, Cambridge, CB3 0HE\\
$^2$Oxford Astrophysics, Denys Wilkinson Building, Keble Road, Oxford, OX1 3RH\\
$^3$NRAO VLA, Array Operations Center, P.O. Box O, 1003 Lopezville Road, Socorro, NM 87801-0387, USA\\
$^4$NRAO Greenbank Telescope, P.O. Box 2, Rt. 28/92, Green Bank, WV 24944-0002, USA\\
$^5$California Institute of Technology, 1201 East California Blvd, Pasadena CA 91125, USA}

\newcounter{con}
\begin{document}

\date{\today}

\maketitle

\label{firstpage}

\begin{abstract}
We have carried out extensive radio and optical follow-up of 176 sources from the 15\,GHz 9th Cambridge survey. Optical identifications have been found for 155 of the radio sources; optical images are given with radio maps overlaid. The continuum radio spectrum of each source spanning the frequency range 1.4 - 43\,GHz is also given.

Two flux-limited samples are defined, one containing 124 sources complete to 25\,mJy and one of 70 sources complete to 60\,mJy. Between one fifth and one quarter of sources from these flux-limited samples display convex radio spectra, rising between 1.4 and 4.8\,GHz. These rising-spectrum sources make up a much larger fraction of the radio source population at this high selection frequency than in lower frequency surveys. 

We find that by using non-simultaneous survey flux density measurements at 1.4 and 15\,GHz to remove steep spectrum objects, the efficiency of selecting objects with spectra rising between 1.4 and 4.8\,GHz (as seen in simultaneous measurements) can be raised to 49\,percent without compromising the completeness of the rising spectrum sample.
\end{abstract}


\section{Introduction}
 In this paper we present the results of our follow-up work on the 9th Cambridge survey (9C hereafter). The 9C survey was carried out at 15\,GHz with the Ryle Telescope \citep[RT see][]{J1}, primarily motivated by the need to identify foreground sources in the fields surveyed by the Cambridge/Jodrell/Tenerife cosmic microwave background experiment, the Very Small Array \citep[VSA, e.g.][]{W2}. A rastering technique was used to scan the fields with each possible detection being followed up with pointed observations to confirm its existence and measure its flux density or to rule it out: see \citet{W1} for a full description of 9C. The survey fields were chosen to contain few very bright radio sources, but otherwise should be representative of the radio sky in general. 15\,GHz is the highest radio frequency at which an extensive survey has been carried out, so 9C provides new insights into the properties of the radio source population.

Current models of radio source growth that consider the effects of self absorption on the synchrotron emission from young sources indicate that very young radio sources (tens to hundreds of years old) should have radio spectra which peak between about one gigahertz and a few tens of gigahertz \citep{odea,snellen}, with younger sources having spectra peaking at higher frequencies than older sources. Any radio survey is biased toward the selection of sources with spectra peaking close to the selection frequency, hence 9C should provide a means of generating samples rich in sources peaking at close to 15\,GHz and thereby testing the models of source growth in very young sources.

We have selected 176 sources from the 9C survey (155 of which are from complete flux-limited samples) and carried out multi-frequency simultaneous radio observations to obtain the continuum radio spectra and maps. \it{R}\normalfont-band optical imaging was performed in order to make optical identifications (IDs). 

The layout of this paper is as follows. In \S \ref{sec:data} we discuss sample selection, data acquisition and data reduction. In \S \ref{sec:results} we present the results -- the radio flux densities, radio maps and optical counterpart data. In \S \ref{sec:stats} we discuss the sample statistics with regard to the radio spectra, radio morphology, optical colour and optical morphology.  In \S \ref{sec:compare} we compare these results with previous work and in \S \ref{GPS} we consider a means of increasing the efficiency of selecting rising spectrum sources. We summarise our findings in \S \ref{sec:discussion}.

\section{data acquisition}
\label{sec:data}
\subsection{Selecting the sample}
 
Two complete samples of sources were selected from the first three regions of the 9C survey, coincident with the VSA fields at $00^{\mbox{\small{h}}} 20^{\mbox{\small{m}}} +30^{\circ}$, $09^{\mbox{\small{h}}} 40^{\mbox{\small{m}}} +32^{\circ}$ and $15^{\mbox{\small{h}}} 40^{\mbox{\small{m}}} +43^{\circ}$(J2000.0). Sample A is complete to 25 mJy and contains 124 sources selected from regions in all three fields, a total area of 176 square degrees. Sample B is complete to 60 mJy, with 70 sources in a total area of 246 square degrees; it consists of all sources in sample A above 60 mJy (39 sources), plus 31 additional sources from a further region of the 9h field. 
                                                                                
Additionally 21 9C sources (sample C) were observed which were either outside the sample areas or had flux densities lower than the selection limit -- these do not form a complete sample.

\subsection {Radio observations and data reduction} Simultaneous continuum snapshot observations were made for each source at frequencies of 1.4, 4.8, 22 and 43\,GHz  with the VLA of the National Radio Astronomy Observatory\footnote{The National Radio Astronomy Observatory is a facility of the National Science
Foundation operated under cooperative agreement by Associated Universities, Inc.} (table \ref{radio_obs}) and at 15\,GHz with the RT. In addition, 51 sources from the $ 00^{\mbox{\small{h}}}$ field were observed within a few months at 31\,GHz with the Owens Valley Radio Observatory (OVRO) 40m telescope.

The \it{uv}\normalfont-plane coverage of the VLA differs significantly for the different sets of observations. The data from Jan 2002 were taken in the $\vec{\mbox{DA}}$ move configuration and only a few antennas had been moved into their A-array positions; Although at 4.8\,GHz, for example, there are baselines out to 500\,kilo-$\lambda$, the majority are less than 15\,kilo-$\lambda$ and the resulting beam is messy. In order to obtain good flux density estimates and a smooth beam, the central portion of the \it{uv}\normalfont-plane (corresponding to the D-configuration baselines) was used; after this, the full \it{uv}\normalfont-coverage was used to look for structure on smaller angular scales

The VLA data were reduced using the NRAO \texttt{AIPS} package. For each dataset maps were made and cleaned with the \texttt{AIPS} task \texttt{IMAGR}. Self-calibration was applied to those maps with sufficiently high signal to noise ratio -- typically sources with point-like components having flux densities of around 40\,mJy or greater. In each case one or more rounds of self-calibration, in phase and in amplitude and phase, were carried out to maximise the signal to noise ratio of the final map. Time spent on source was typically about 40 s at 1.4\,GHz, 60 s at 4.8\,GHz, 100 s at 22\,GHz and 120 s at 43\,GHz, giving typical rms noise values of 0.5, 0.4, 0.8 and 2\,mJy respectively. The VLA flux calibration is assumed to be good to about 1\,percent, 2\,percent, 5\,percent and 10\,percent at 1.4, 4.8, 22 and 43\,GHz respectively.

Each source was observed for about 20 minutes with the RT; the rms noise is about 0.9\,mJy and the calibration uncertainties are approximately 3\,percent.

The OVRO 40-m observations were carried out between 2002 Jan and 2002 July. The rms noise on the flux density measurements is typically 1 mJy, but is often  higher for the brighter sources. Flux calibration uncertainties are about 5\,percent. 

\begin{table*}
 \centering 
 \caption{Summary of VLA observations. Numbers in brackets are of sources in sample B that are also in sample A. The $09^{\mbox{\small{h}}}$ sample was originally only complete to 60\,mJy but subsequently one quarter of the area was filled in (in 2001 Dec) by studying the 10 sources in the central region with 9C flux densities between 25\,mJy and 60\,mJy. The remainder of the $09^{\mbox{\small{h}}}$ region is complete to only 60\,mJy, hence the 31 sources that are only part of Sample B observed in 2001 Jan. \label{radio_obs}}
 \begin{tabular}{@{}l@{\hspace{0.2cm}}l@{\hspace{0.2cm}}l@{\hspace{0.2cm}}l@{\hspace{0.2cm}}c@{\hspace{0.2cm}}c@{\hspace{0.2cm}}c@{\hspace{0.2cm}}c@{\hspace{0.2cm}}c@{\hspace{0.2cm}}c@{\hspace{0.2cm}}c@{}}
 \hline
VLA run  & Field & Config. & Obs.date & \multicolumn{4}l{Synthesised beam size (arcsec $\times$ arcsec)} & \multicolumn{3}c{Number in each sample}\\ 
 number &      &         &          & 1.4\,GHz & 4.8\,GHz & 22\,GHz & 43\,GHz                                     &   A & B   &  C  \\
\hline
1 & $00^{\mbox{\small{h}}}$ & D & 2001 Nov & $51\times49$ & $15\times14$ & $3.6\times3.1$ & $2.2\times1.6$ & 37 & (11) & 12  \\
2 & $00^{\mbox{\small{h}}}$ & BnA & 2002 May & $4.2\times1.9$ & $1.3\times0.5$ & $0.29\times0.11$ & $0.13\times0.08$ & 12 & (3) & - \\ 	
\hline 
3 & $09^{\mbox{\small{h}}}$ & $\vec{\mbox{AB}}$ & 2001 Jan & $1.8\times1.4$ & $0.52\times0.39$ & $0.15\times0.09$ & $0.09\times0.03$ & 4 & (4 +) 31 & 7 \\
4 & $09^{\mbox{\small{h}}}$ & D & 2001 Dec & $55\times46$ & $16\times14$ & $4.2\times3.1$ & $2.5\times1.5$ & 10 & - & - \\ 
\hline
5 & $15^{\mbox{\small{h}}}$ & $\vec{\mbox{DA}}$ & 2002 Jan & $77\times52$ & $12\times9.2$ & $2.4\times1.5$ & $1.9\times1.9$ & 61 & (21) & 2 \\ 
\hline
\end{tabular}
\end{table*}

\subsection{Optical observations}
Optical imaging using the Kron-Cousins \it{R}\normalfont-band filter (650\,nm) was carried out with the 60-inch telescope at the Palomar Observatory on 2002 March 5, 6, July 12, 13, 14 December 3, 4 and 2003 January 28, 31. Typically one 900-s exposure was made for each object, with more time given to fainter objects when possible. The limiting magnitude is $R\approx21$ (3 sigma within an aperture the same angular size as the seeing) and the seeing is generally around 1-2\,arcsec. The rising spectrum, flat spectrum and compact steep spectrum (CSS) radio sources were selected for imaging, priority being given to those with no optical counterpart in the APM catalogue of the first Palomar Observatory sky survey \citep[POSS-I: see][for a description of the APM project]{M1}. 

The red POSS-I \it{E }\normalfont plates correspond approximately to \it{R}\normalfont-band and hence we refer to the APM catalogue red magnitudes as \it{R}\normalfont-band in this work. The red images from the second Palomar Observatory sky survey \citep[POSS-II:][]{R1} that appear in digitised form in the second digitised sky survey \citep[DSS2; see, e.g.][]{M2} are also approximately in the optical \it{R}\normalfont-band.

The weather conditions were not photometric so standards were not used to calibrate photometry. Instead, the zero-point for each image was found by using the magnitudes given in the APM catalogue for several point-like sources in each 60-inch image. For each object the \texttt{IRAF} task \texttt{phot} was used to calculate magnitudes in a number of interactively chosen apertures after subtraction of the sky background (measured from a chosen annulus). The 60-inch images reach saturation for objects brighter than $R \approx 16$ and the APM photometry is accurate to 0.25 magnitudes down to about $R=18.0$. Thus only those (point-like) objects with magnitudes between 17 and 18 (sometimes 19 for sparse fields) were used to calculate the zero-point. The measured zero-points within each image have a typical standard deviation of 0.15 magnitudes. The total error given for each object is the quadratic sum of the errors in the zero-point and the APM catalogue.

For bright objects ($R<17.0$) the APM magnitude is given. For all objects not observed with the Palomar 60-inch telescope we present DSS2 \it{R}\normalfont-band images, complete to  $R\approx 20.8$. We quote magnitudes from the APM catalogue for these objects, and bootstrap DSS2 to APM (in the same manner as for the 60-inch data) if the object is too faint to be in the APM catalogue. Where no \it{R}\normalfont-band counterpart was seen in either the 60-inch or the DSS2 \it{R}\normalfont-band image, the DSS2 \it{O}\normalfont-band (blue) and \it{I}\normalfont-band (infrared) images were studied. Only one such object was found in another band: 9CJ1531+4048 has a blue APM magnitude $O=22.14$. Where possible the optical colour of each source is calculated by comparing the \it{R}\normalfont-band measurement with the \it{O}\normalfont-band magnitude in APM.

\section{Results}
\label{sec:results}
\begin{figure*}
\includegraphics*[bb=50 60 510 760, width=15cm]{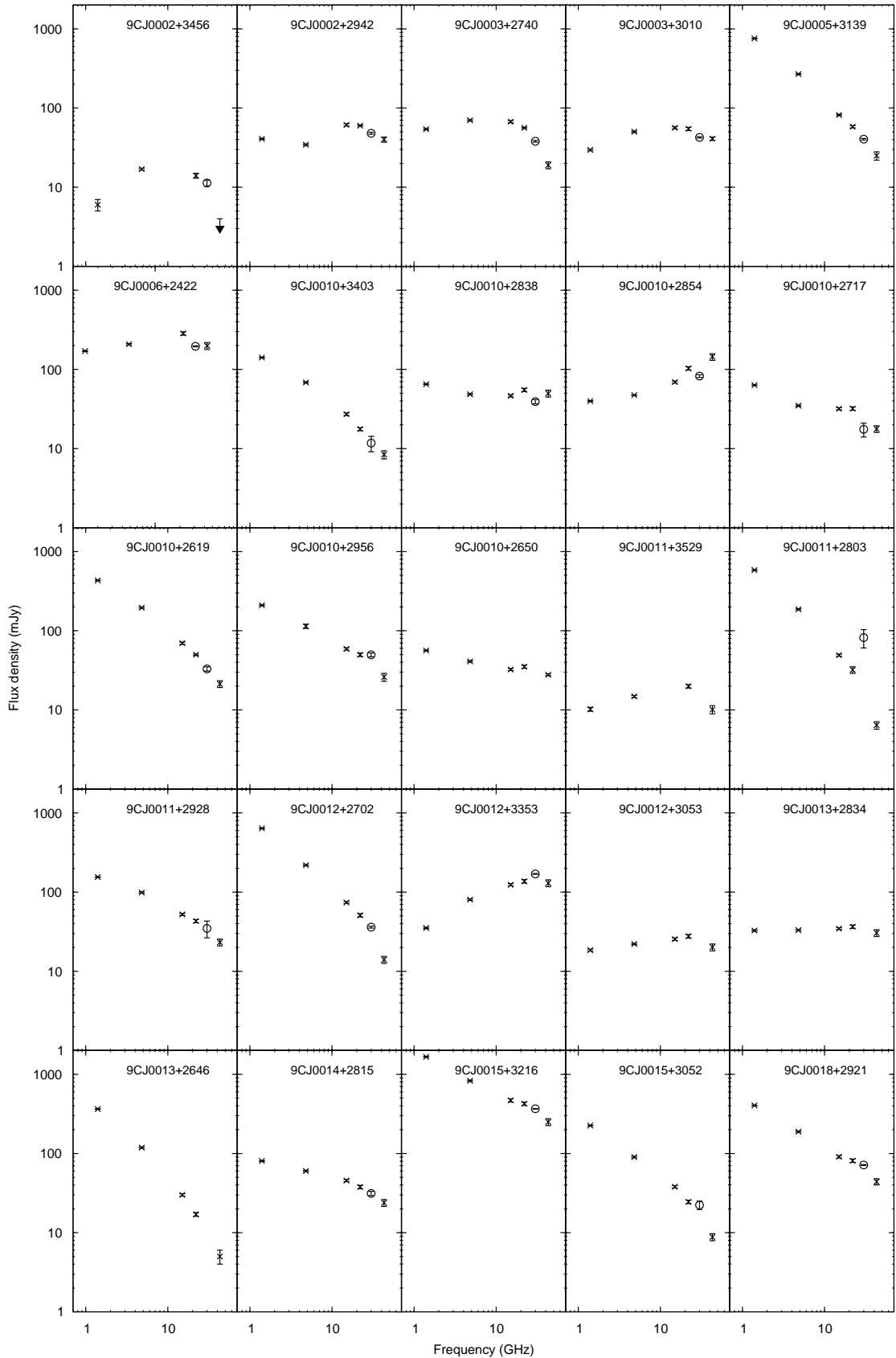}\caption{Radio spectra for sources 9CJ0002+3456 to 9CJ0018+2921.\label{spectra}}
\end{figure*}

\begin{figure*}
\includegraphics*[bb=50 592 176 760, width=4cm]{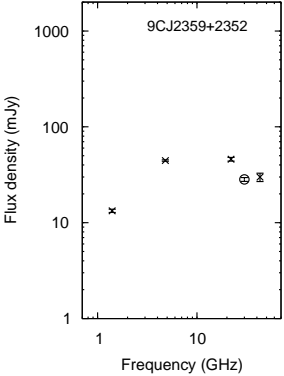}\caption{Radio spectrum for source 9CJ2359+2352. \label{spectra_end}}
\end{figure*}

\begin{figure*}
 \begin{center}
 \mbox{
\subfigure[9CJ0002+3456 (P60 \it{R}\normalfont)]{\epsfig{figure=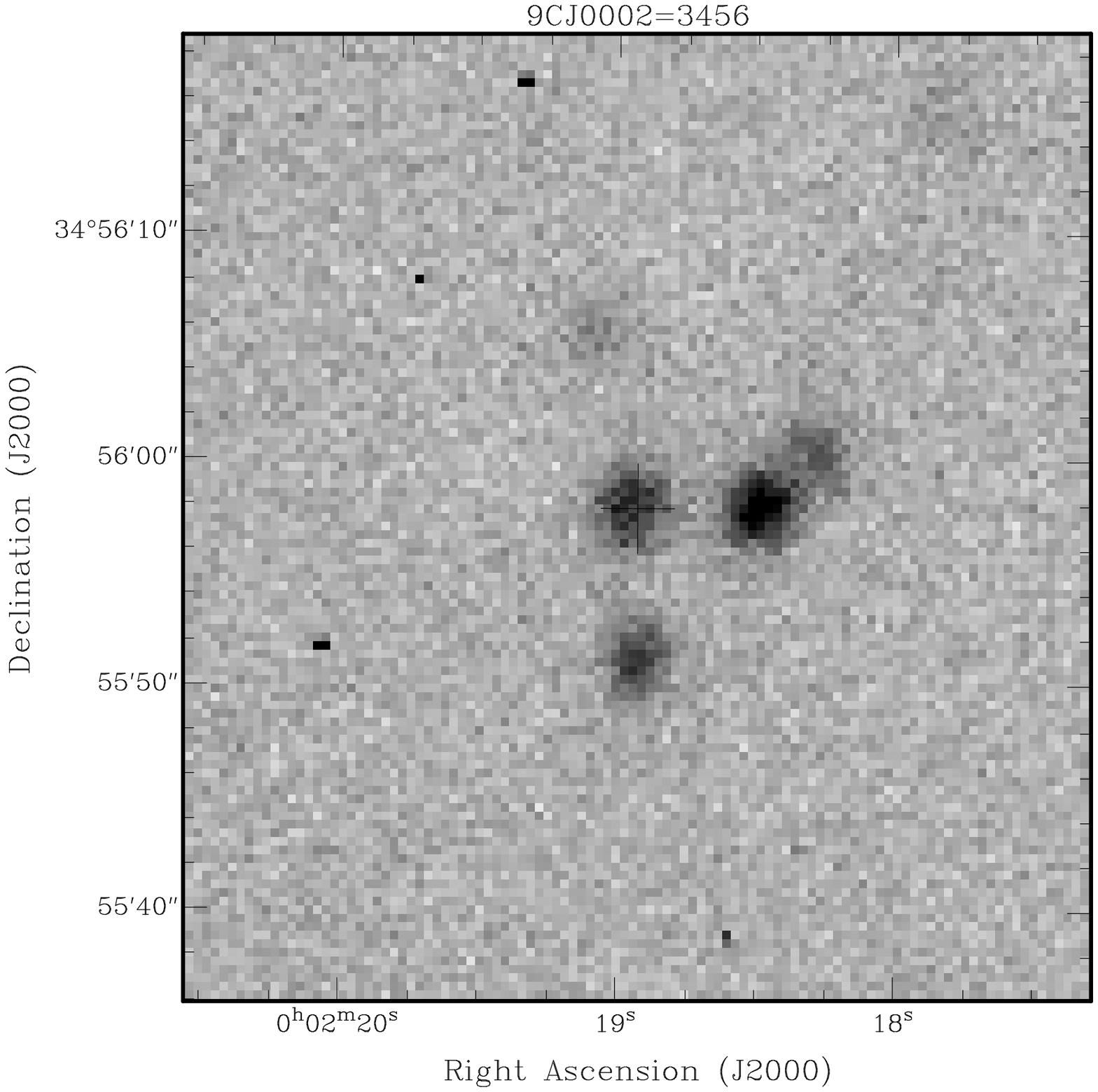,width=4.0cm,clip=}} \quad
\subfigure[9CJ0002+2942 (DSS2 \it{R}\normalfont)]{\epsfig{figure=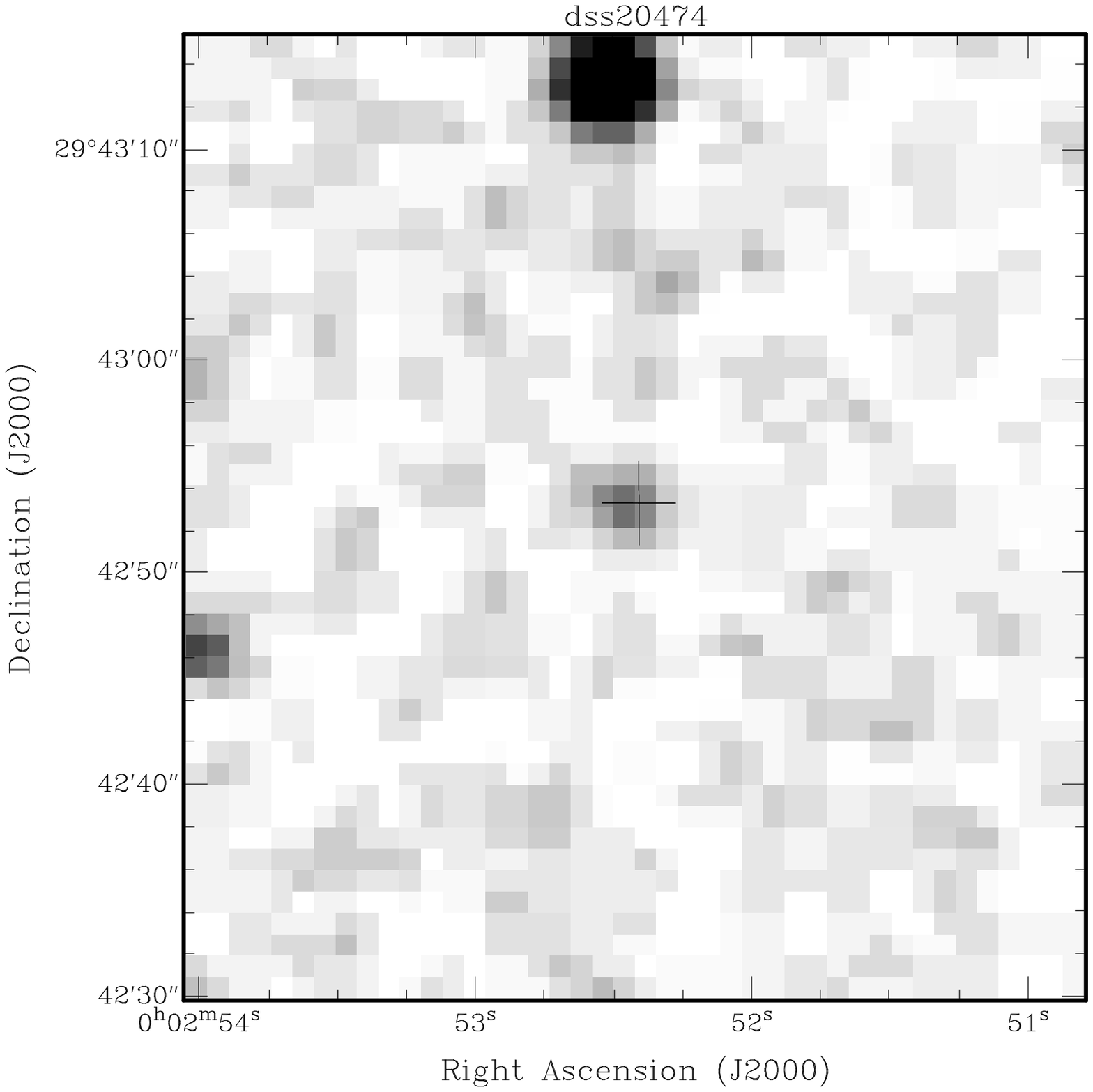,width=4.0cm,clip=}} \quad
\subfigure[9CJ0003+2740 (P60 \it{R}\normalfont)]{\epsfig{figure=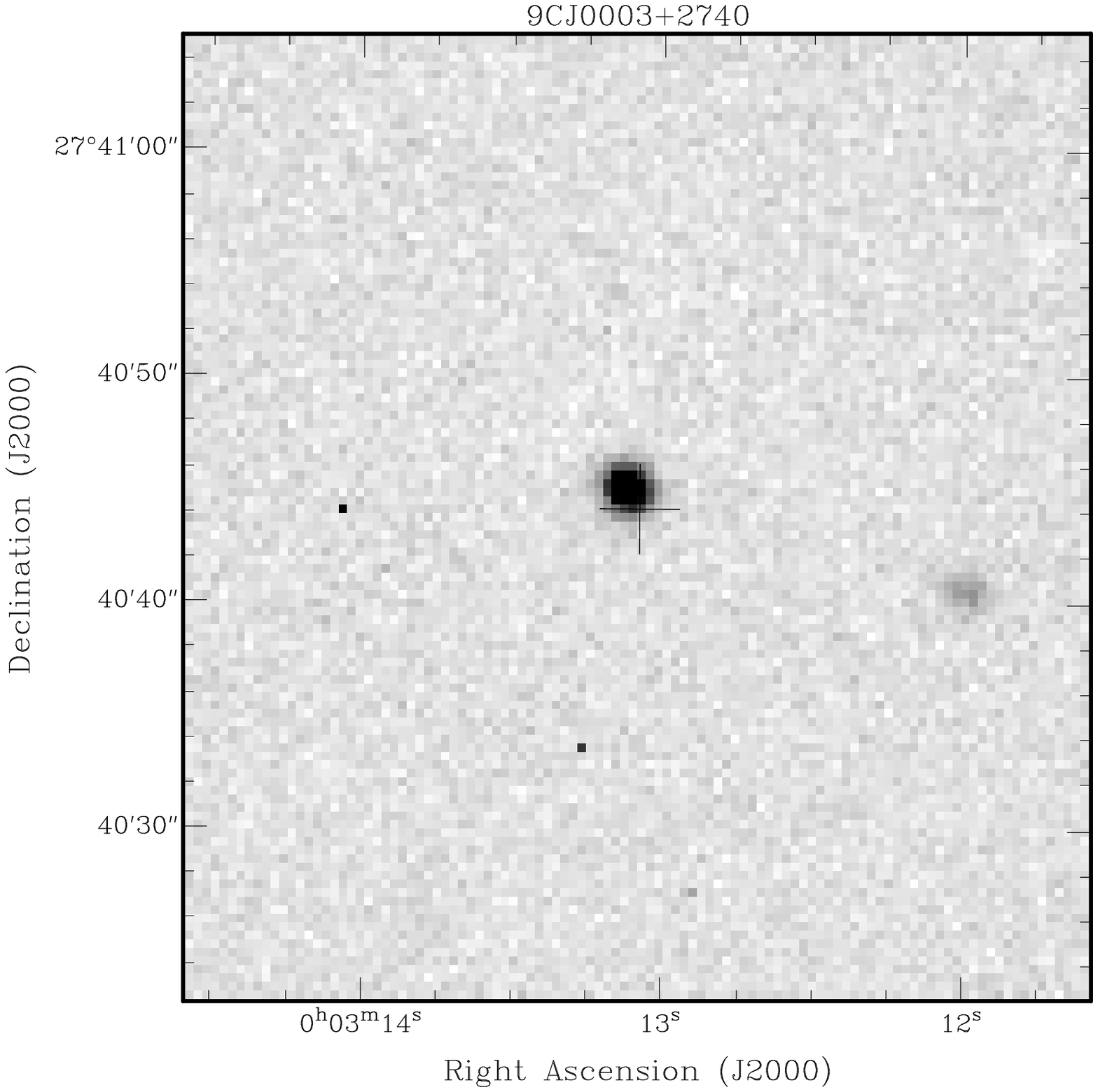,width=4.0cm,clip=}}
 }
 \mbox{
\subfigure[9CJ0003+3010 (P60 \it{R}\normalfont)]{\epsfig{figure=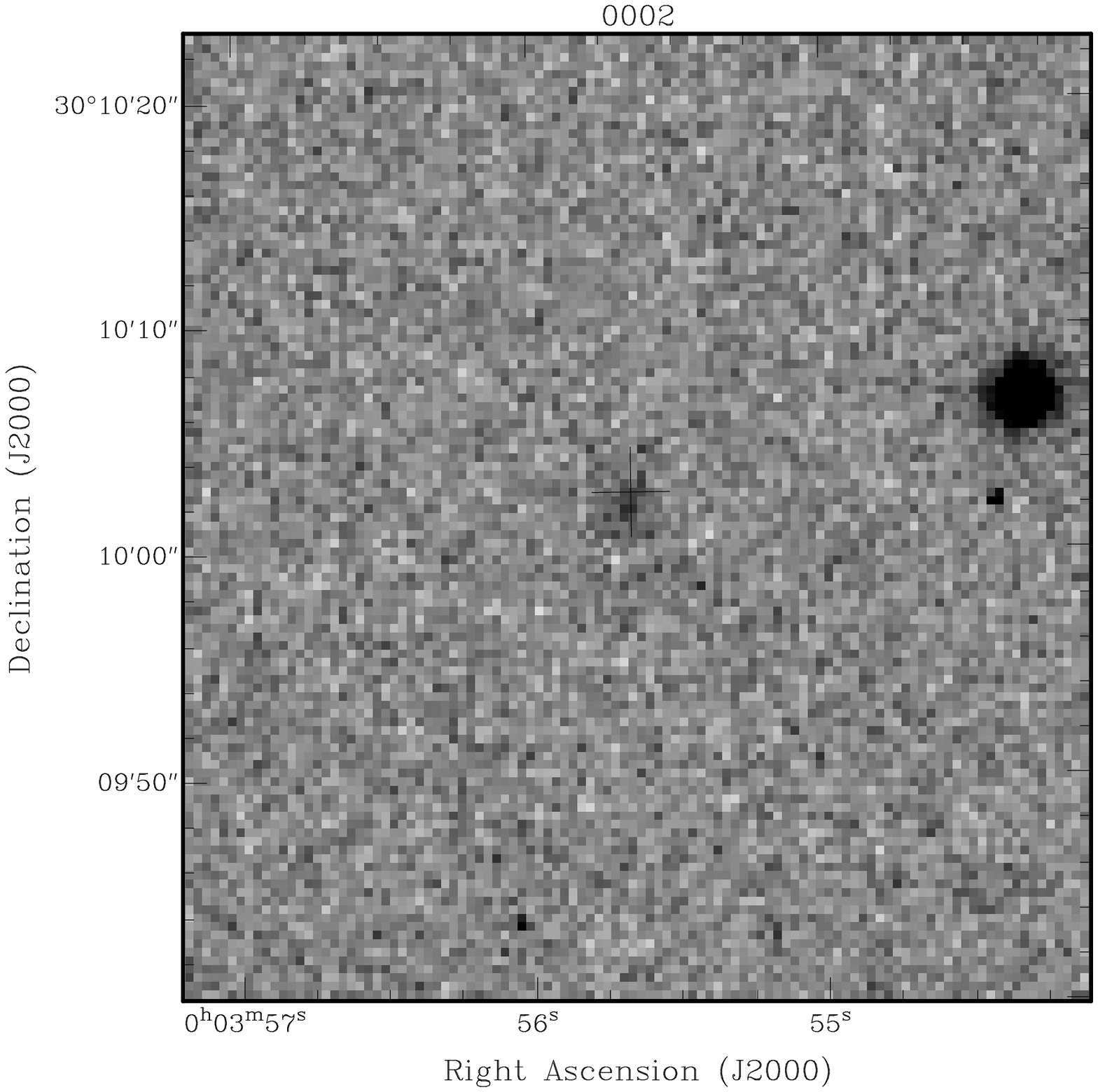,width=4.0cm,clip=}}\quad 
\subfigure[9CJ0005+3139 (DSS2 \it{R}\normalfont)]{\epsfig{figure=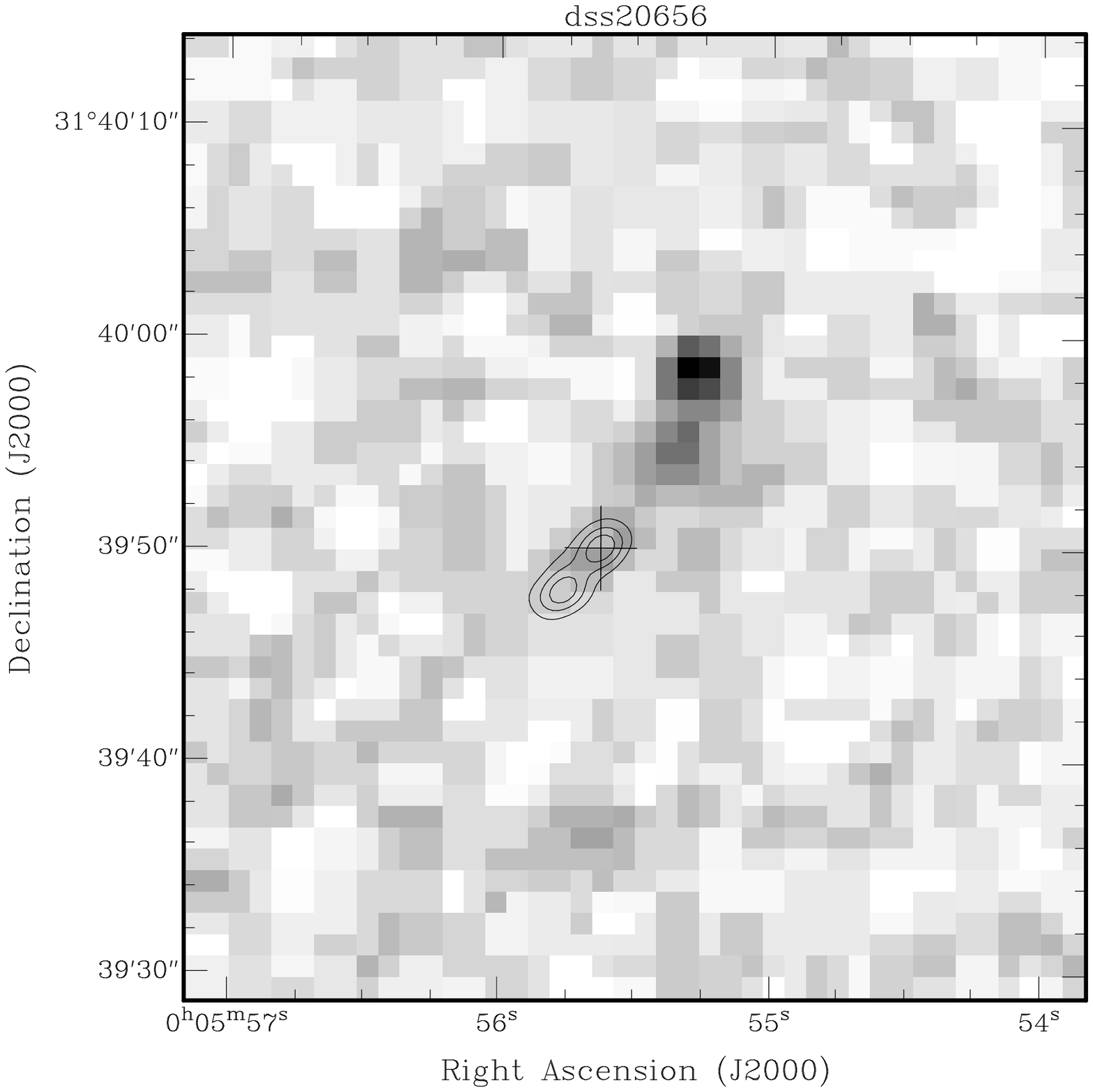,width=4.0cm,clip=}\label{a}}\quad
\subfigure[9CJ0006+2422 (DSS2 \it{R}\normalfont)]{\epsfig{figure=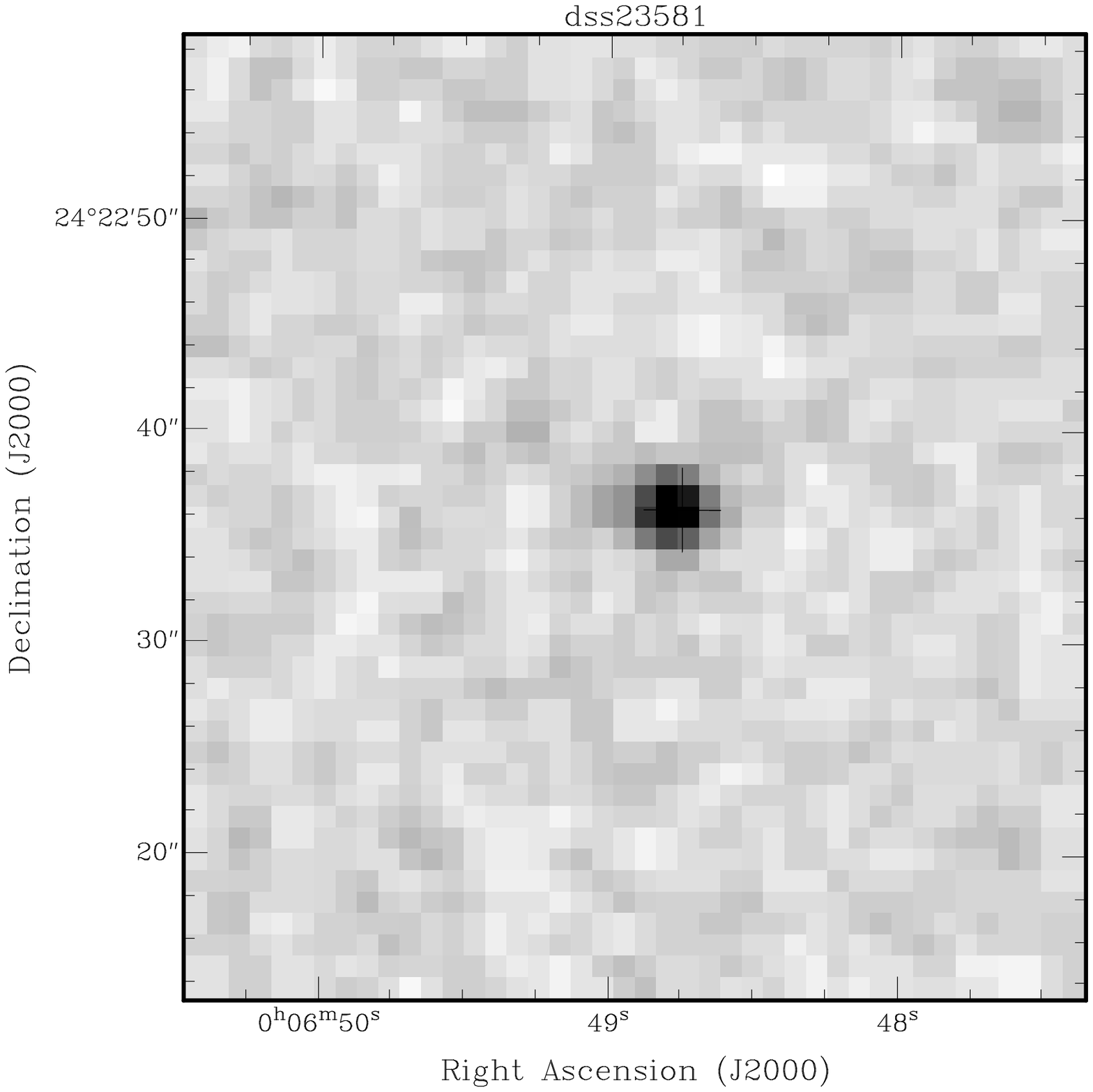,width=4.0cm,clip=}}
 }
 \mbox{
\subfigure[9CJ0010+3403 (DSS2 \it{R}\normalfont)]{\epsfig{figure=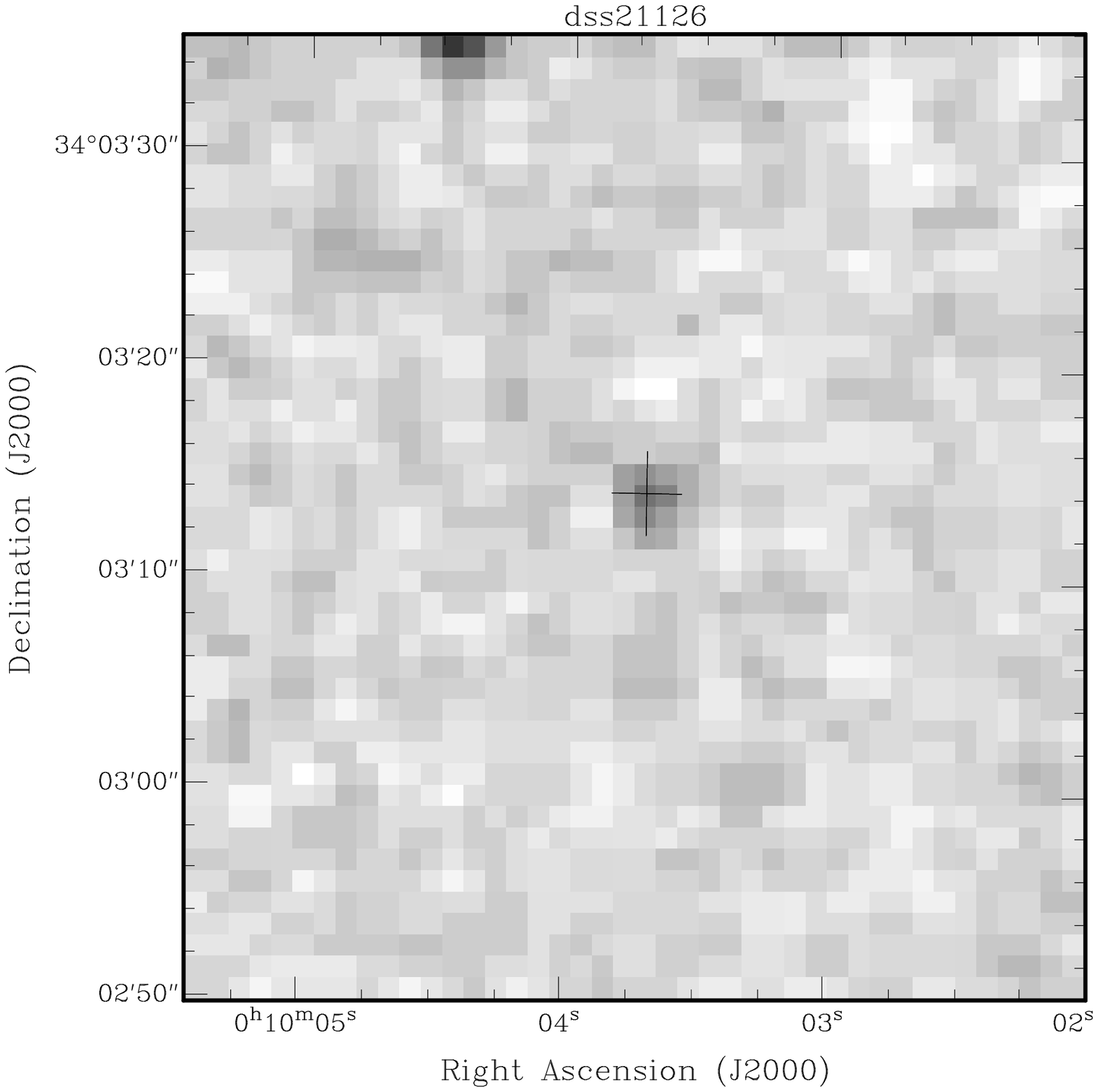,width=4.0cm,clip=}}\quad
\subfigure[9CJ0010+2838 (P60 \it{R}\normalfont)]{\epsfig{figure=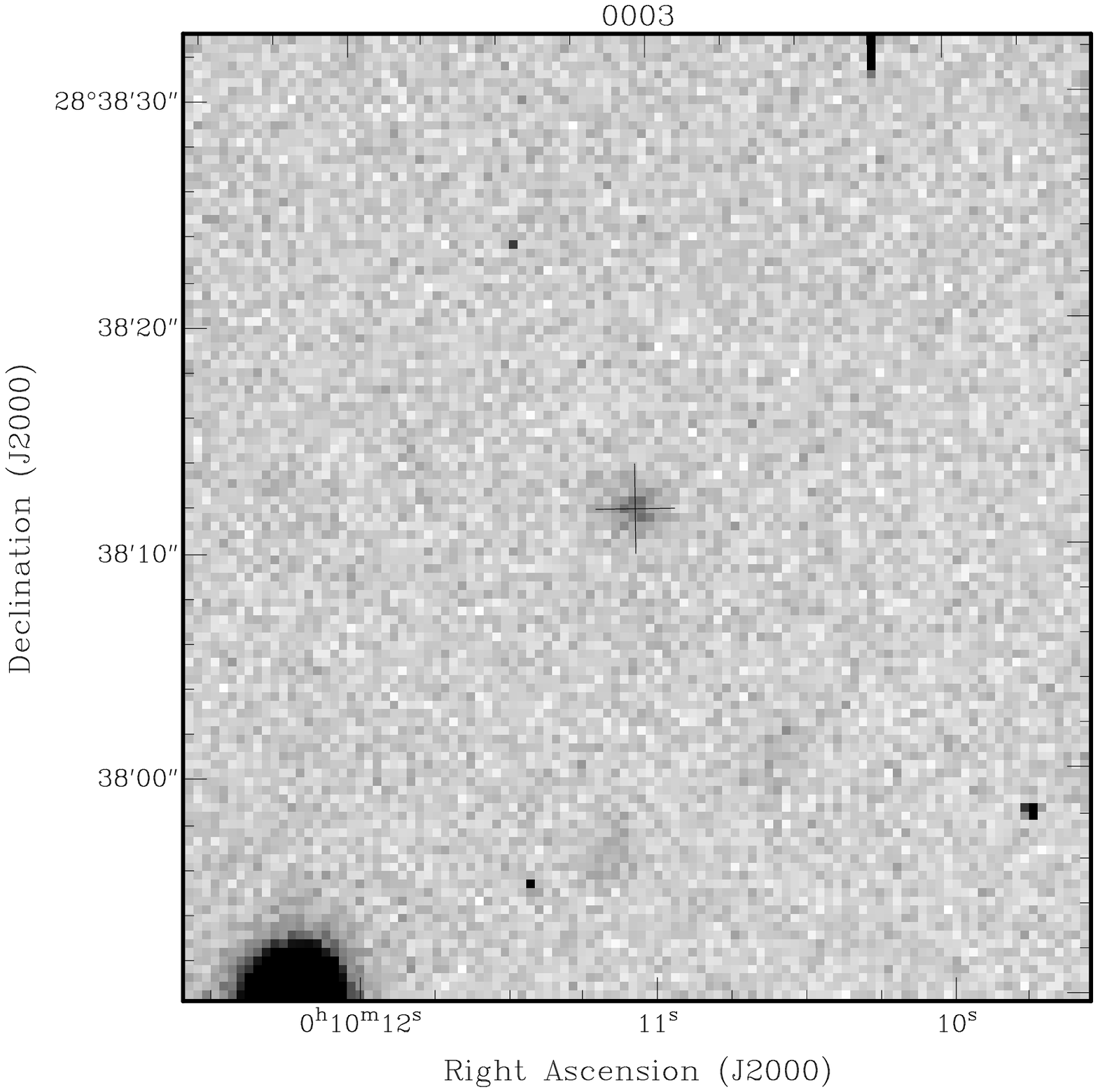,width=4.0cm,clip=}}\quad
\subfigure[9CJ0010+2854 (P60 \it{R}\normalfont)]{\epsfig{figure=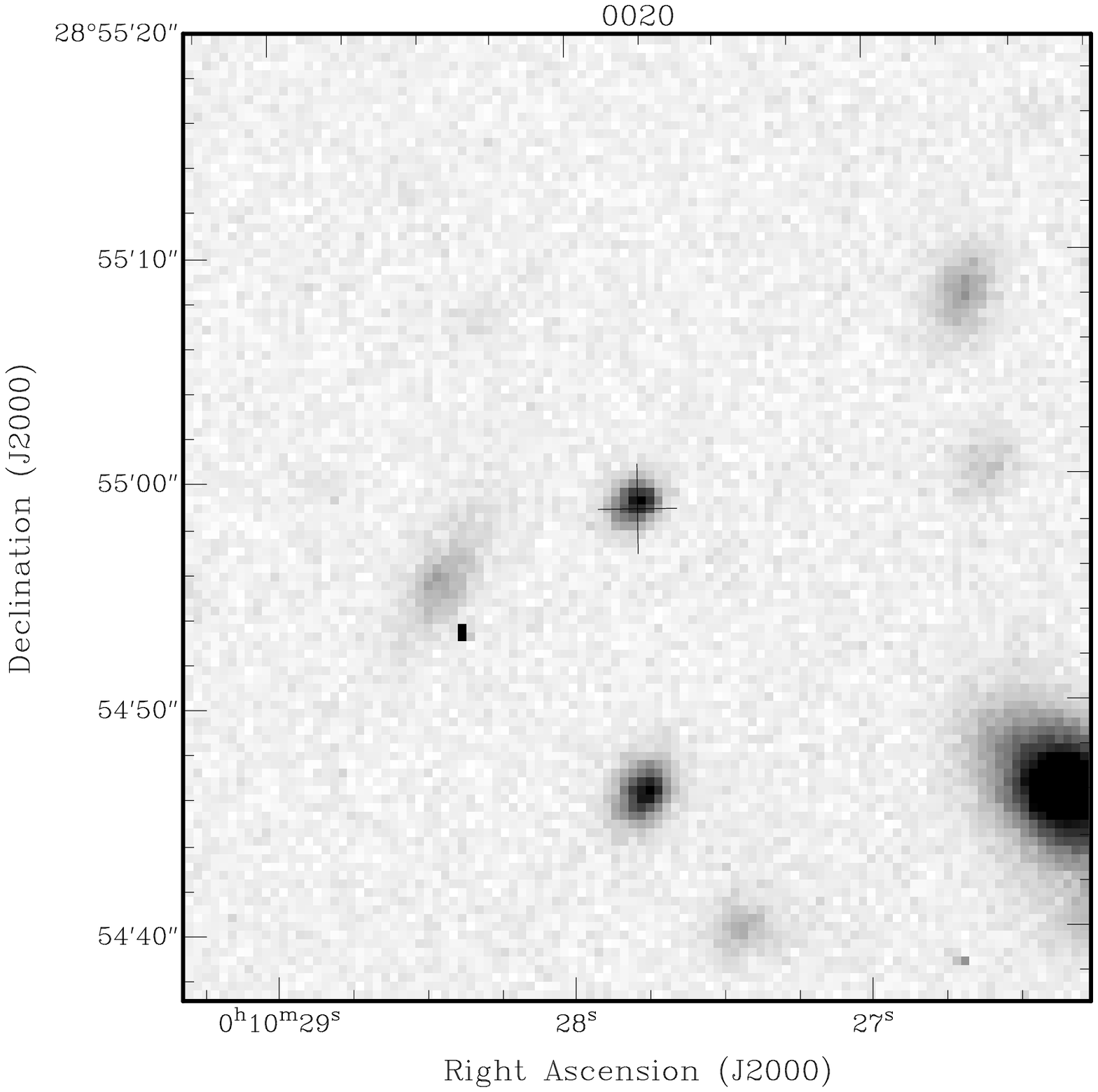,width=4.0cm,clip=}}
 }
 \mbox{
\subfigure[9CJ0010+2717 (P60 \it{R}\normalfont)]{\epsfig{figure=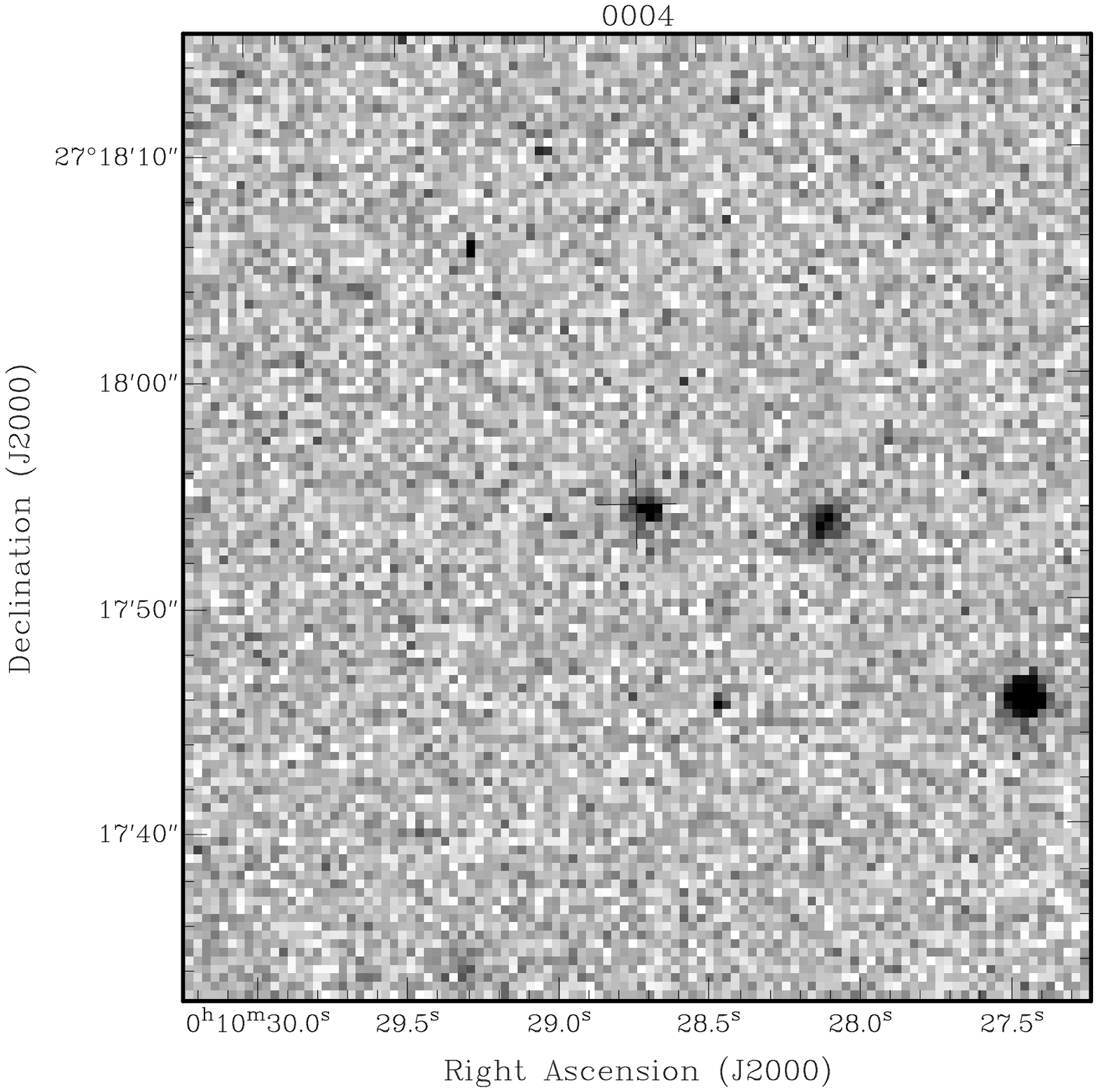,width=4.0cm,clip=}}\quad	 
\subfigure[9CJ0010+2619 (P60 \it{R}\normalfont)]{\epsfig{figure=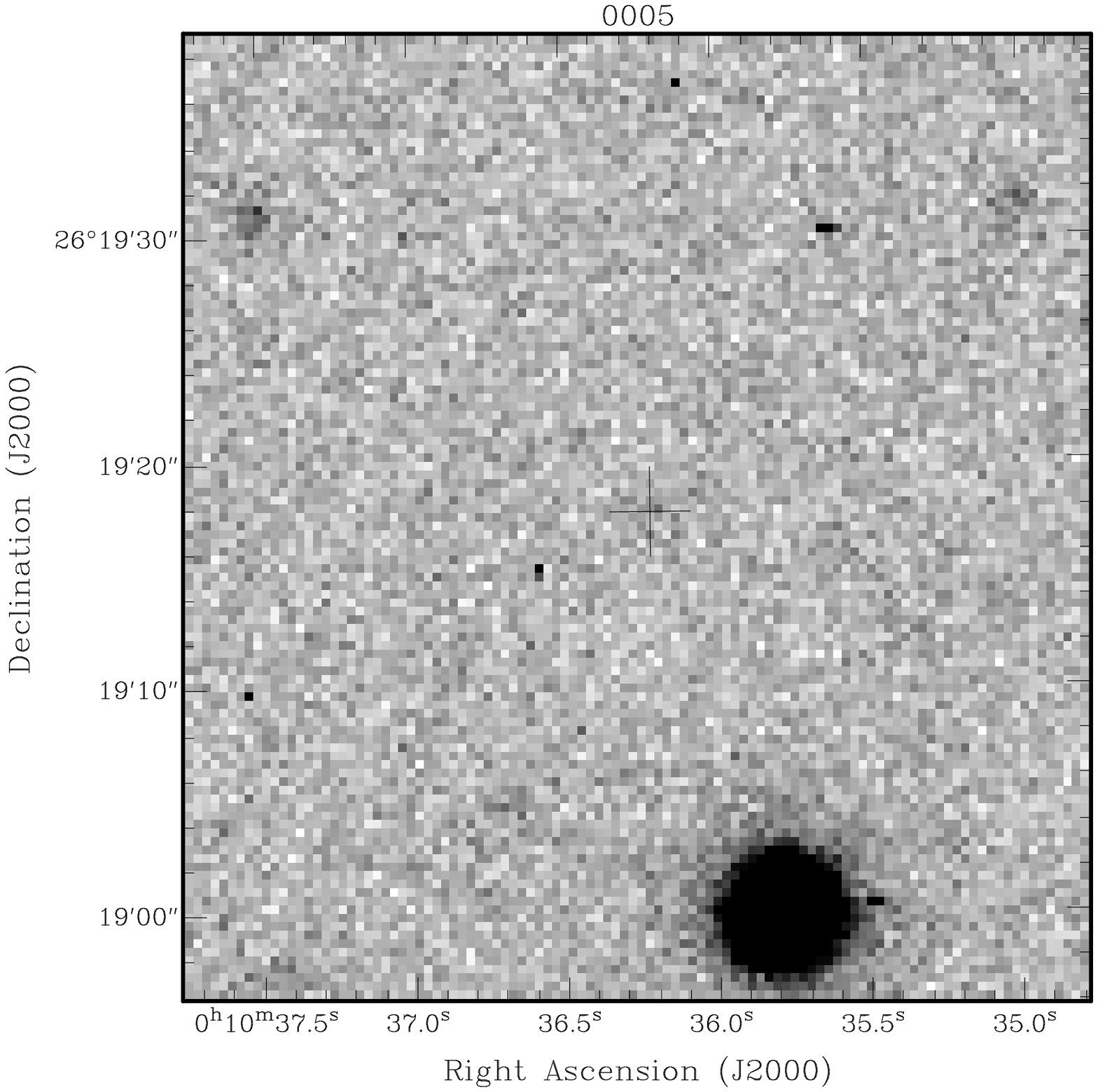,width=4.0cm,clip=}}\quad	 
\subfigure[9CJ0010+2956 (DSS2 \it{R}\normalfont)]{\epsfig{figure=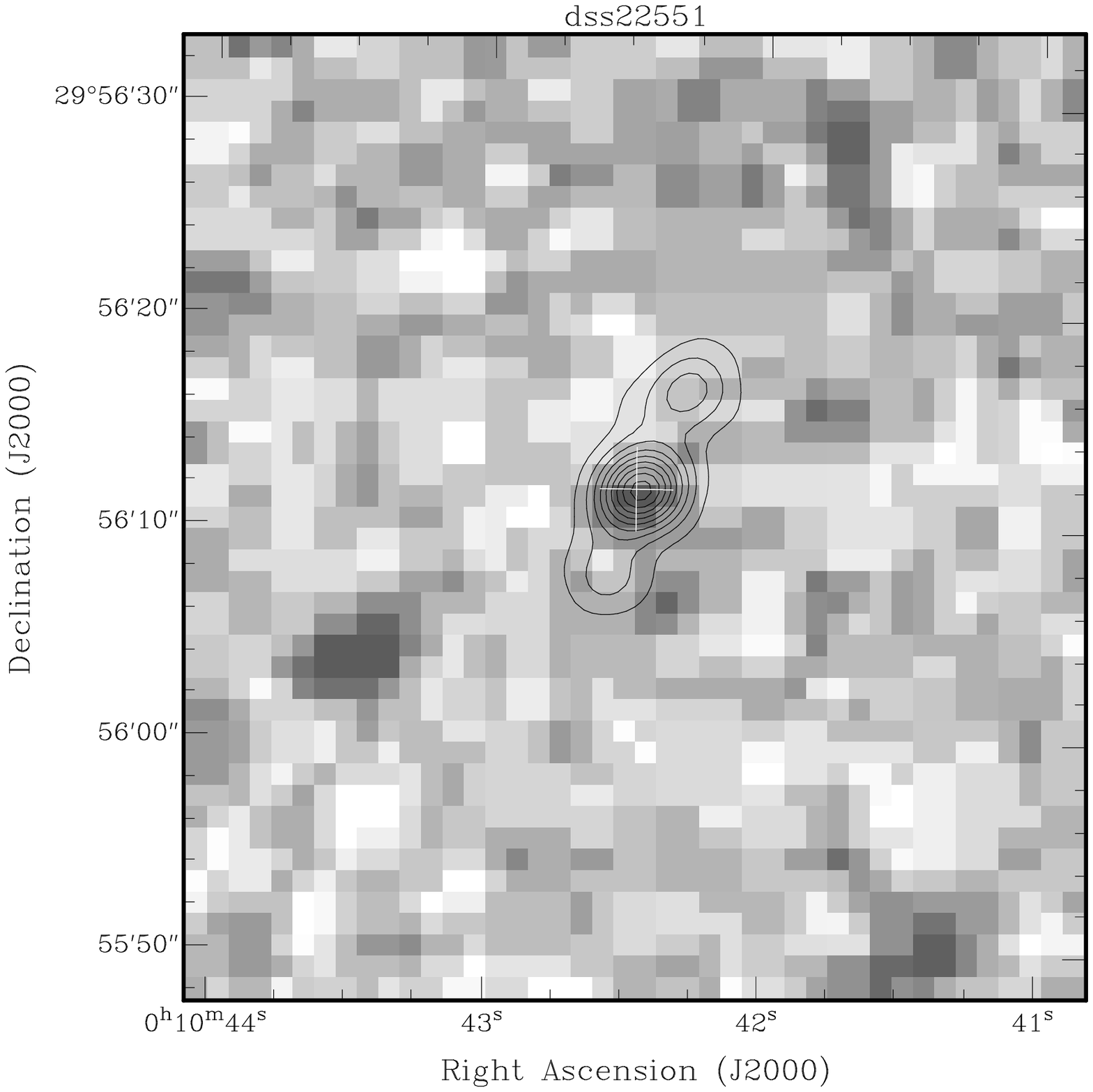,width=4.0cm,clip=}\label{b}}
}
\caption{\label{overlays}Optical counterparts for sources 9CJ0002+3456 to 9CJ0010+2956. Crosses mark maximum radio flux density and are 4\,arcsec top to bottom. Contours: \ref{a}, 43\,GHz contours at 40,60,80\,\% of peak (12.9\,mJy/beam); \ref{b}, 22\,GHz contours at 15,25,35,45,55,65,75,85\,\% of peak (29.9\,mJy/beam).} 
 \end{center}
\end{figure*}

\begin{figure*}
\mbox{ 
\subfigure[9CJ0010+2650 (DSS2 \it{R}\normalfont)]{\epsfig{figure=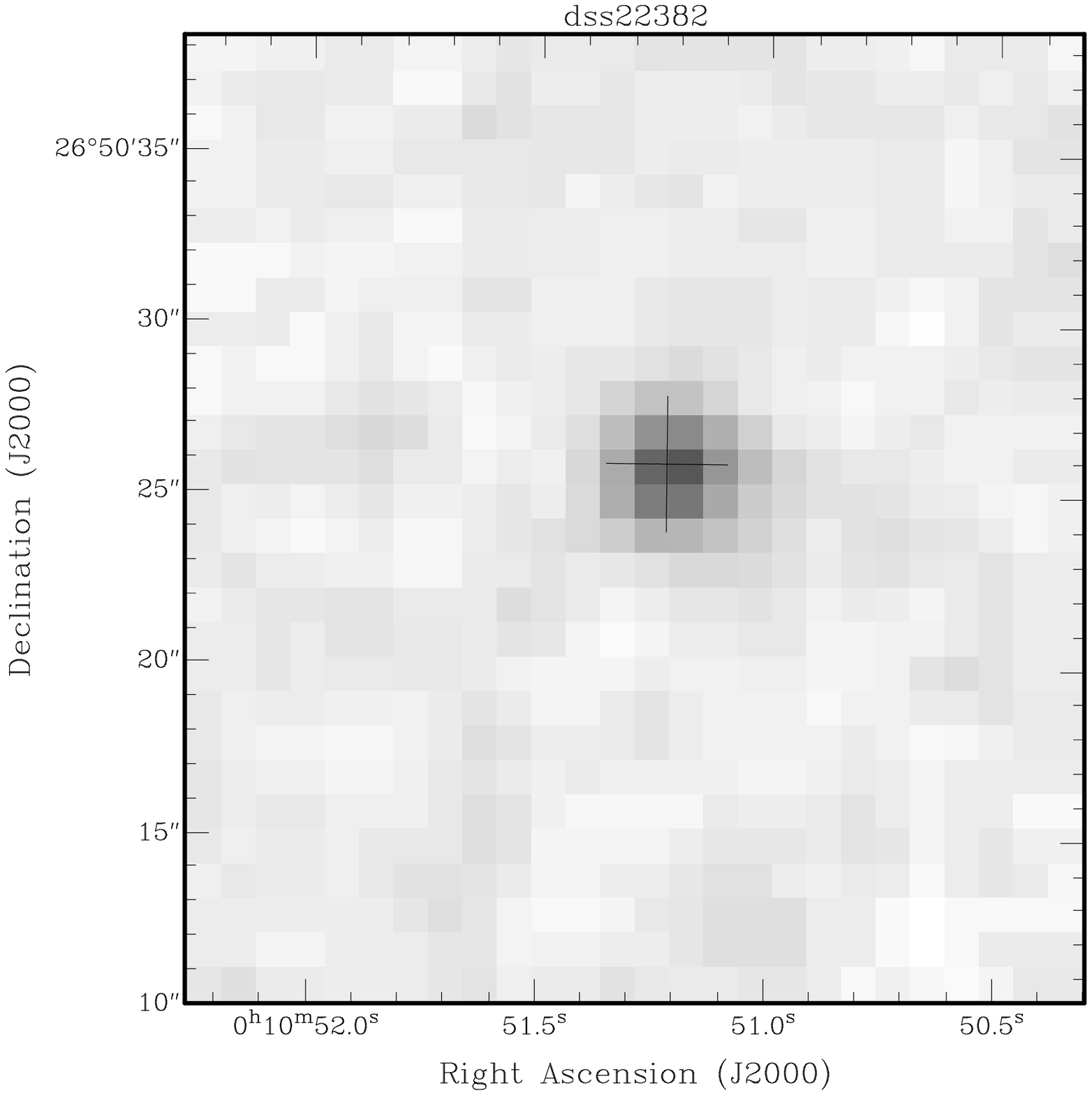 ,width=4.0cm,clip=}}\quad 
\subfigure[9CJ0011+3529 (P60 \it{R}\normalfont)]{\epsfig{figure=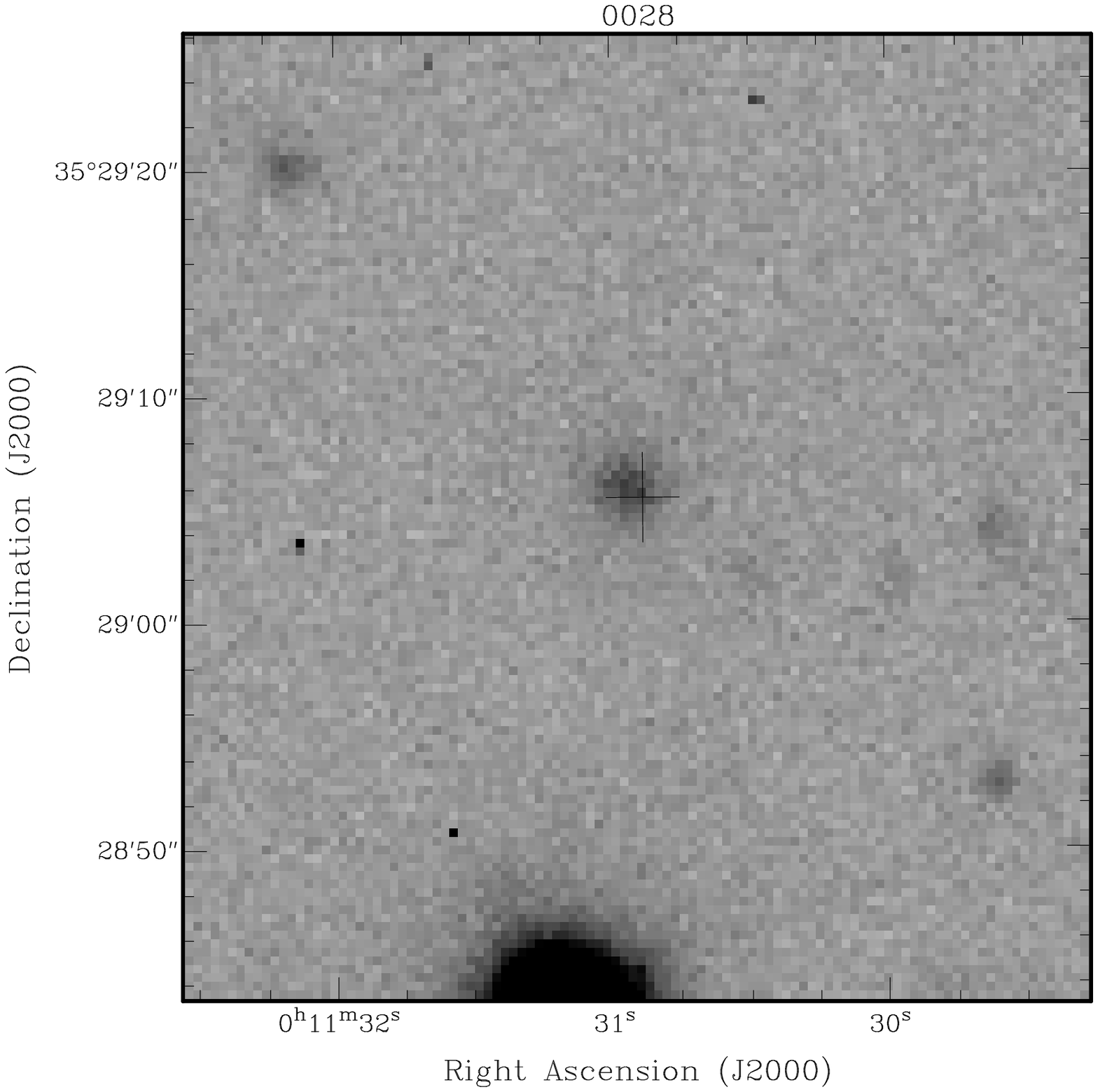 ,width=4.0cm,clip=}}\quad 
\subfigure[9CJ0011+2803 (DSS2 \it{R}\normalfont)]{\epsfig{figure=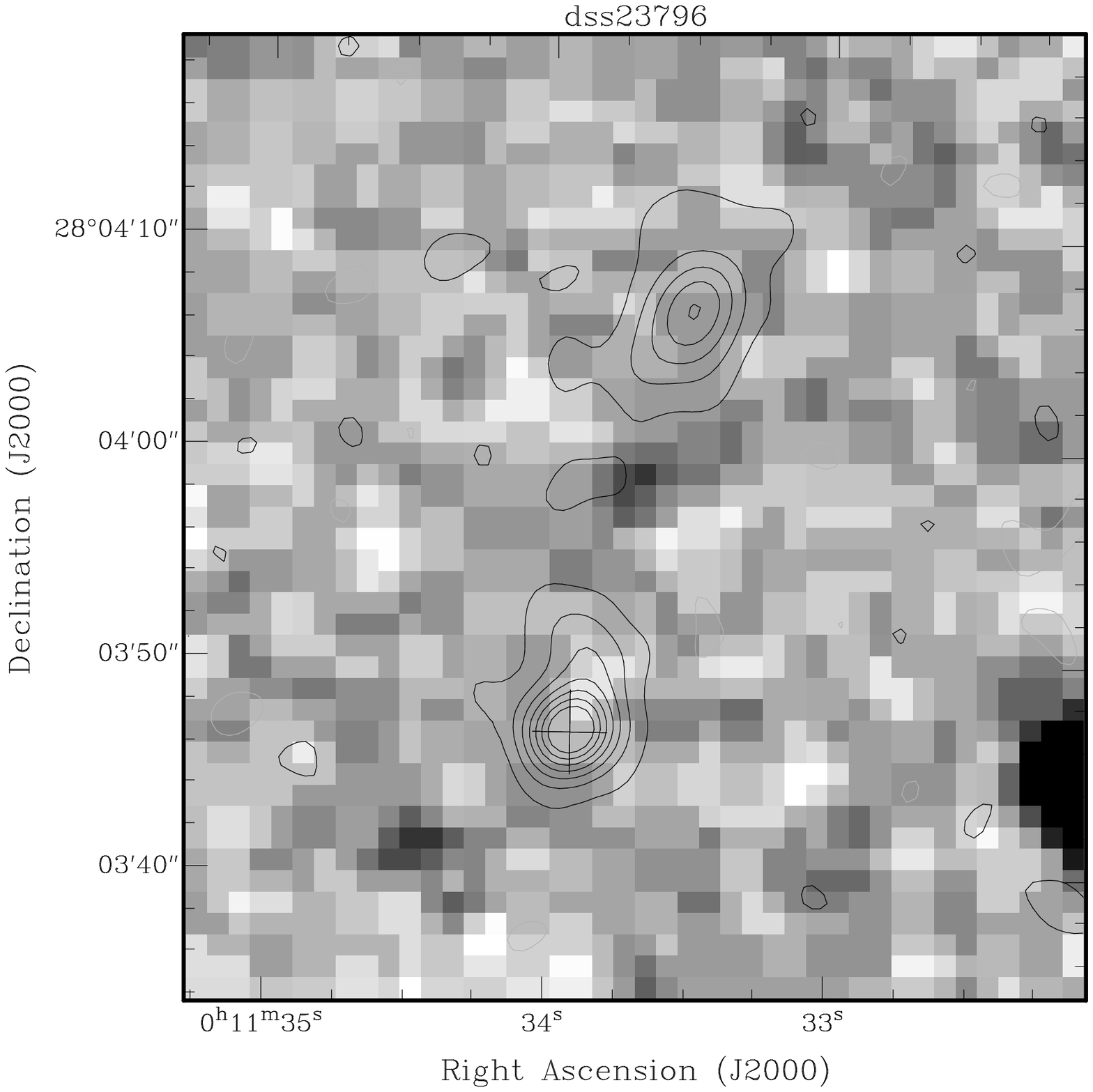 ,width=4.0cm,clip=}\label{c}}
}
\mbox{ 
\subfigure[9CJ0011+2928 (DSS2 \it{R}\normalfont)]{\epsfig{figure=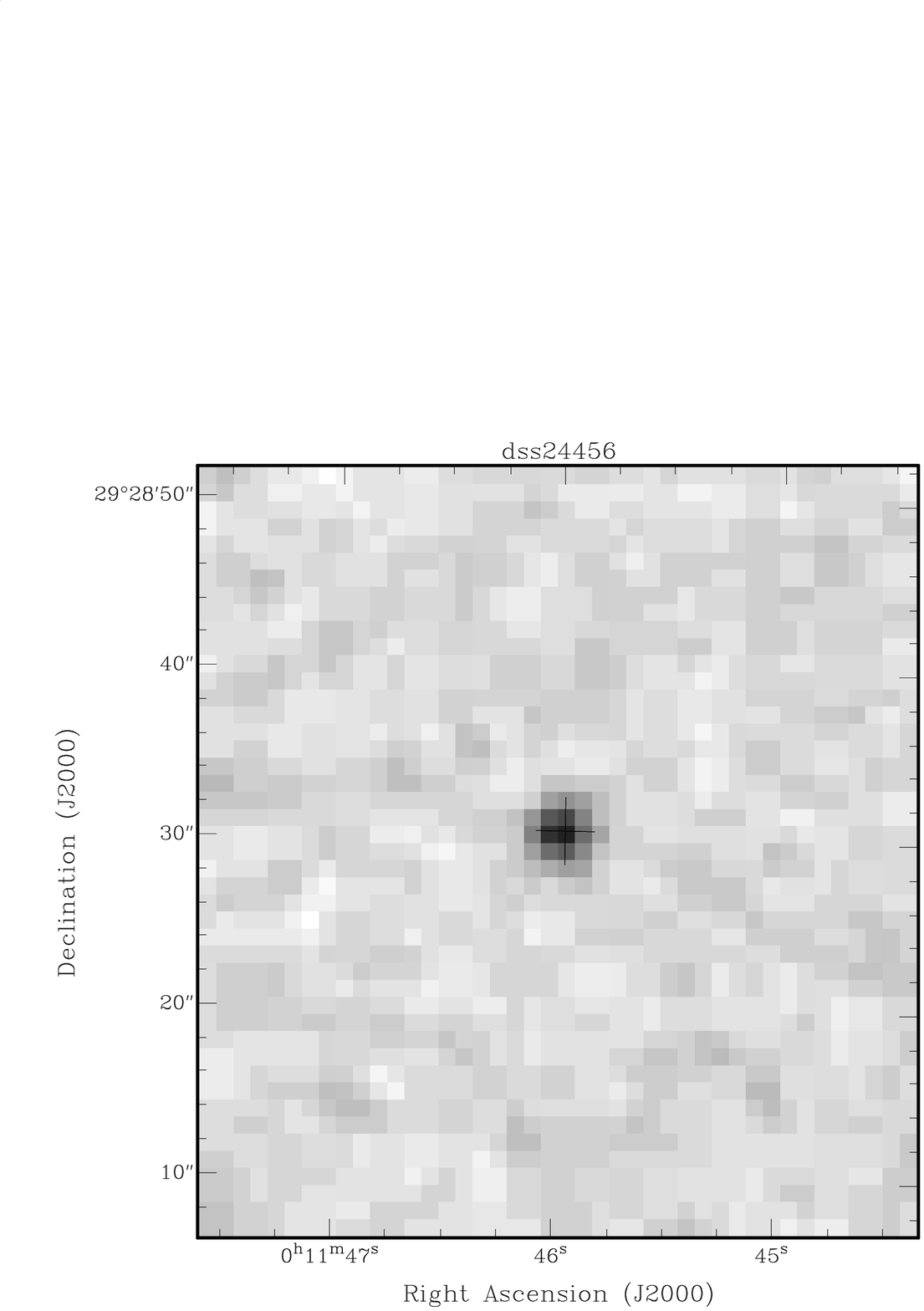 ,width=4.0cm,clip=}}\quad 
\subfigure[9CJ0012+2702 (DSS2 \it{R}\normalfont)]{\epsfig{figure=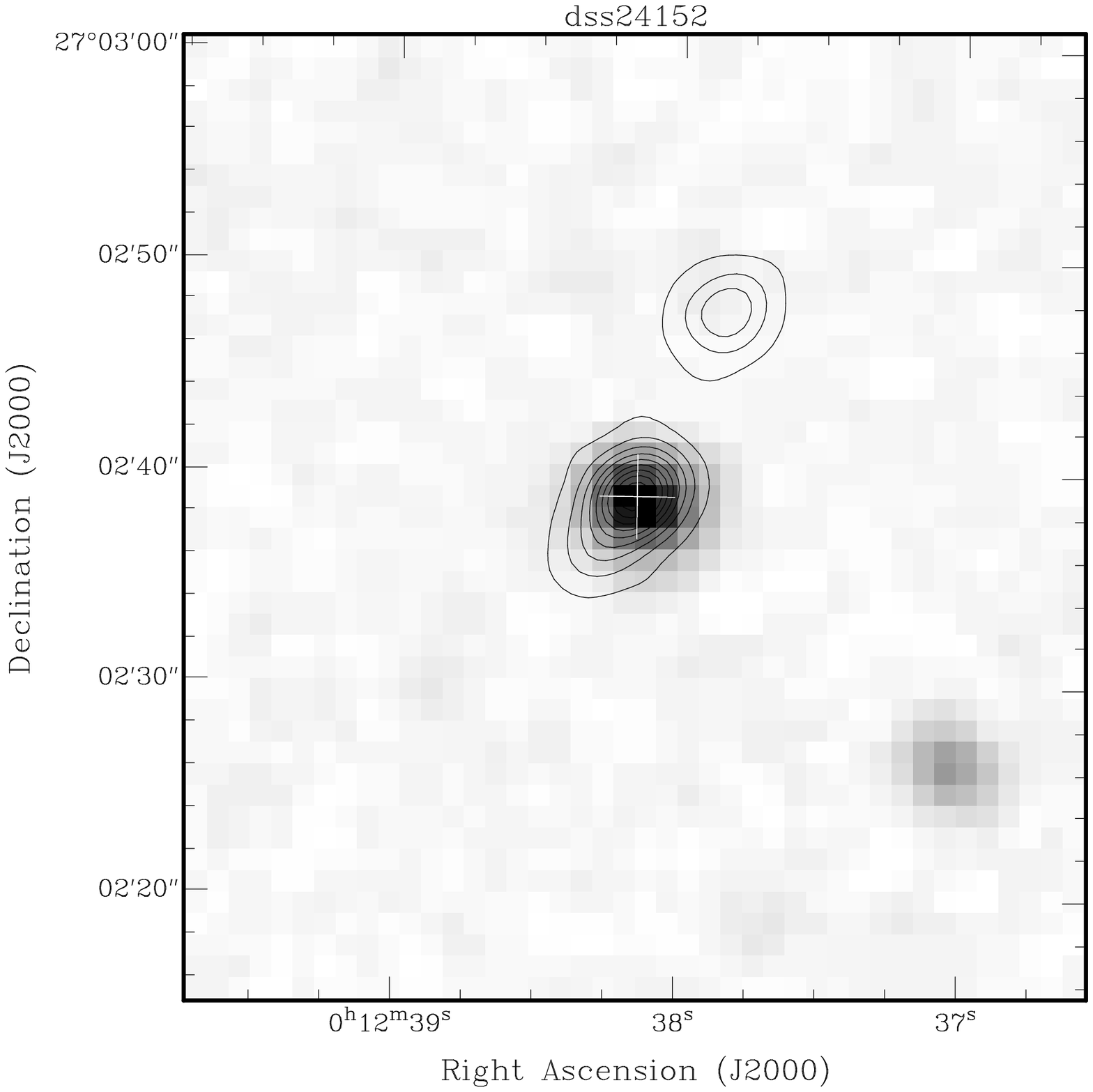 ,width=4.0cm,clip=}\label{d}}\quad 
\subfigure[9CJ0012+3353 (P60 \it{R}\normalfont)]{\epsfig{figure=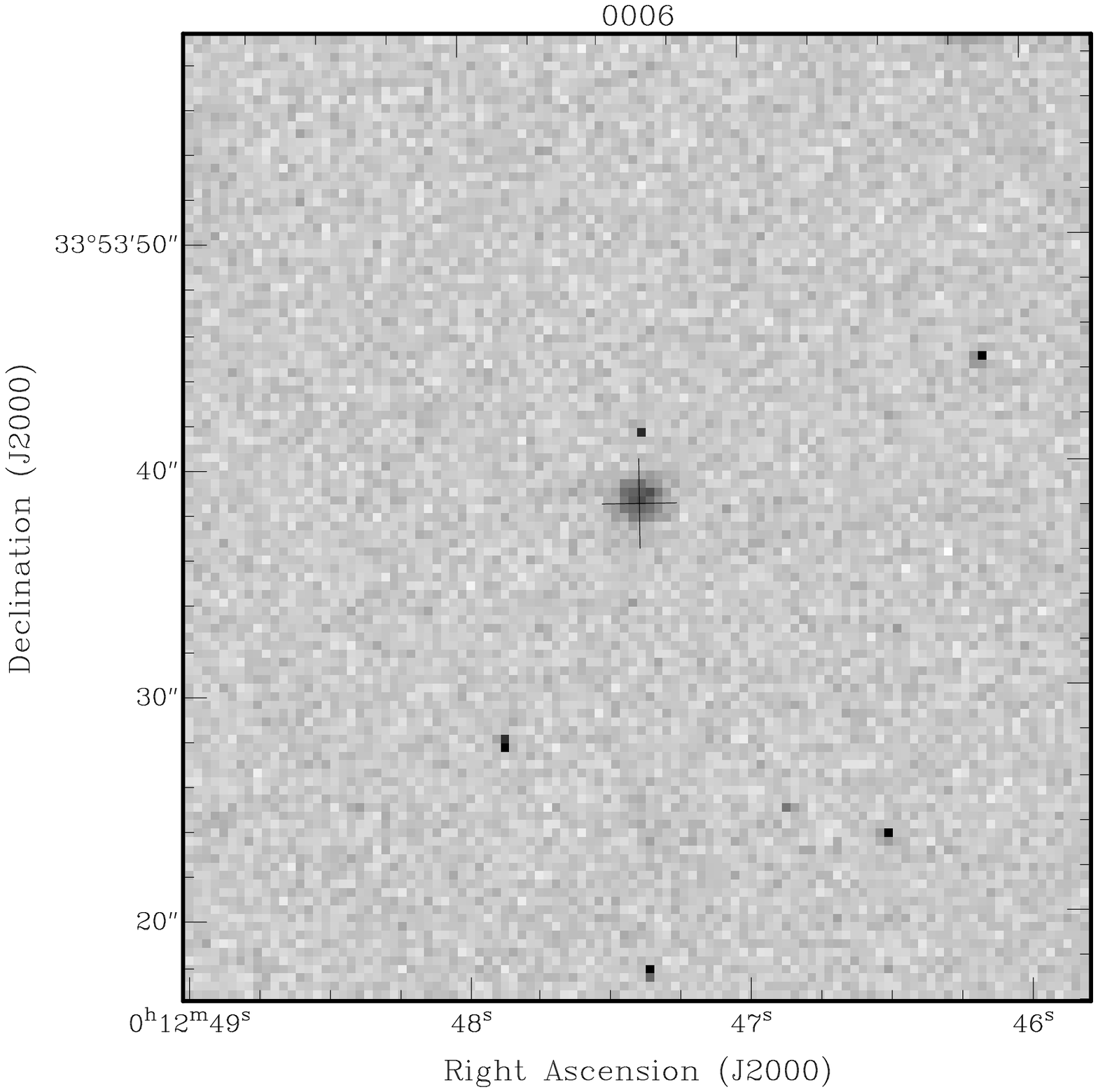 ,width=4.0cm,clip=}}
}
\mbox{ 
\subfigure[9CJ0012+3053 (P60 \it{R}\normalfont)]{\epsfig{figure=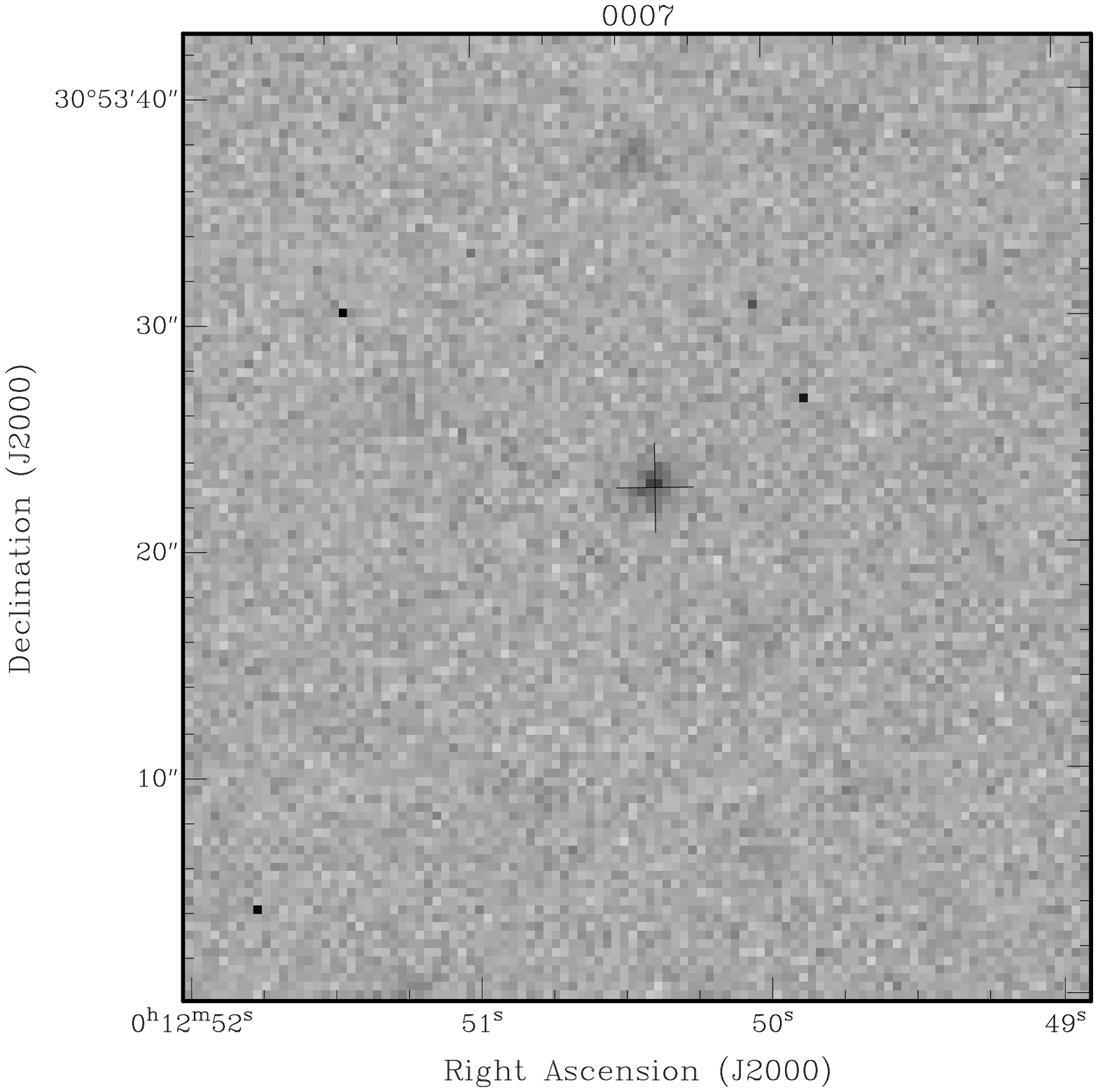 ,width=4.0cm,clip=}}\quad 
\subfigure[9CJ0013+2834 (P60 \it{R}\normalfont)]{\epsfig{figure=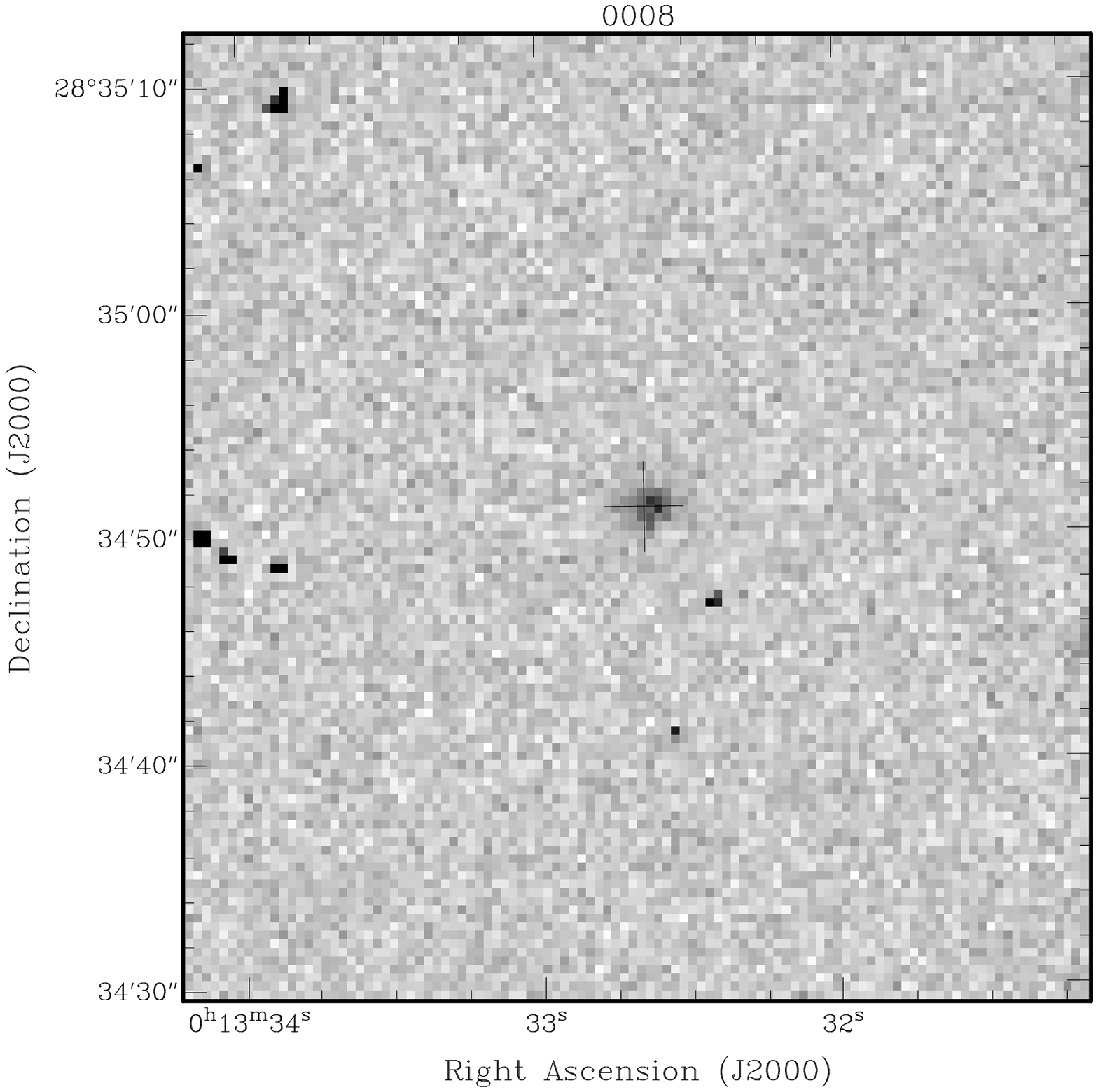 ,width=4.0cm,clip=}}\quad 
\subfigure[9CJ0013+2646 (P60 \it{R}\normalfont)]{\epsfig{figure=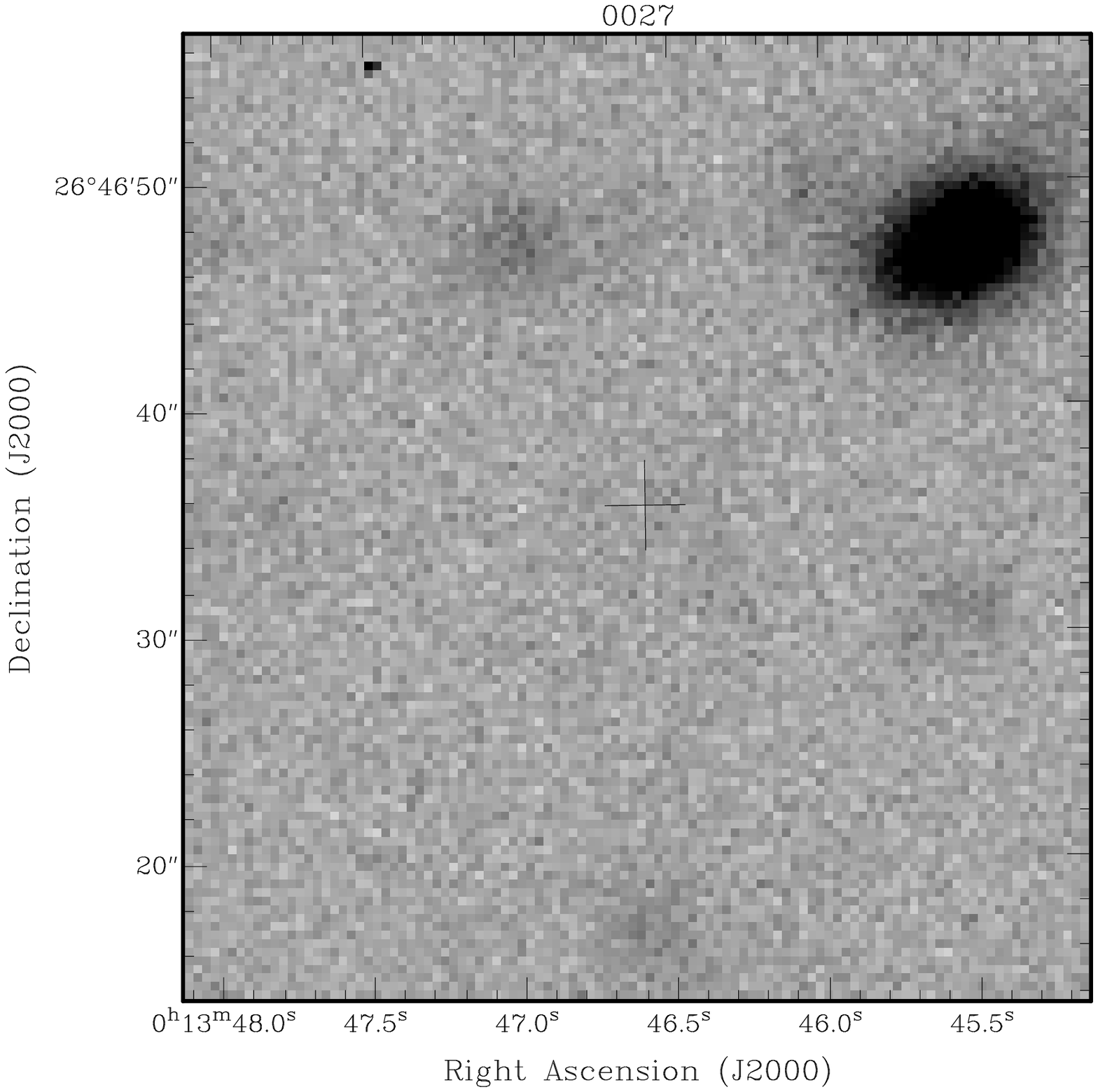 ,width=4.0cm,clip=}}
}
\mbox{ 
\subfigure[9CJ0014+2815 (P60 \it{R}\normalfont)]{\epsfig{figure=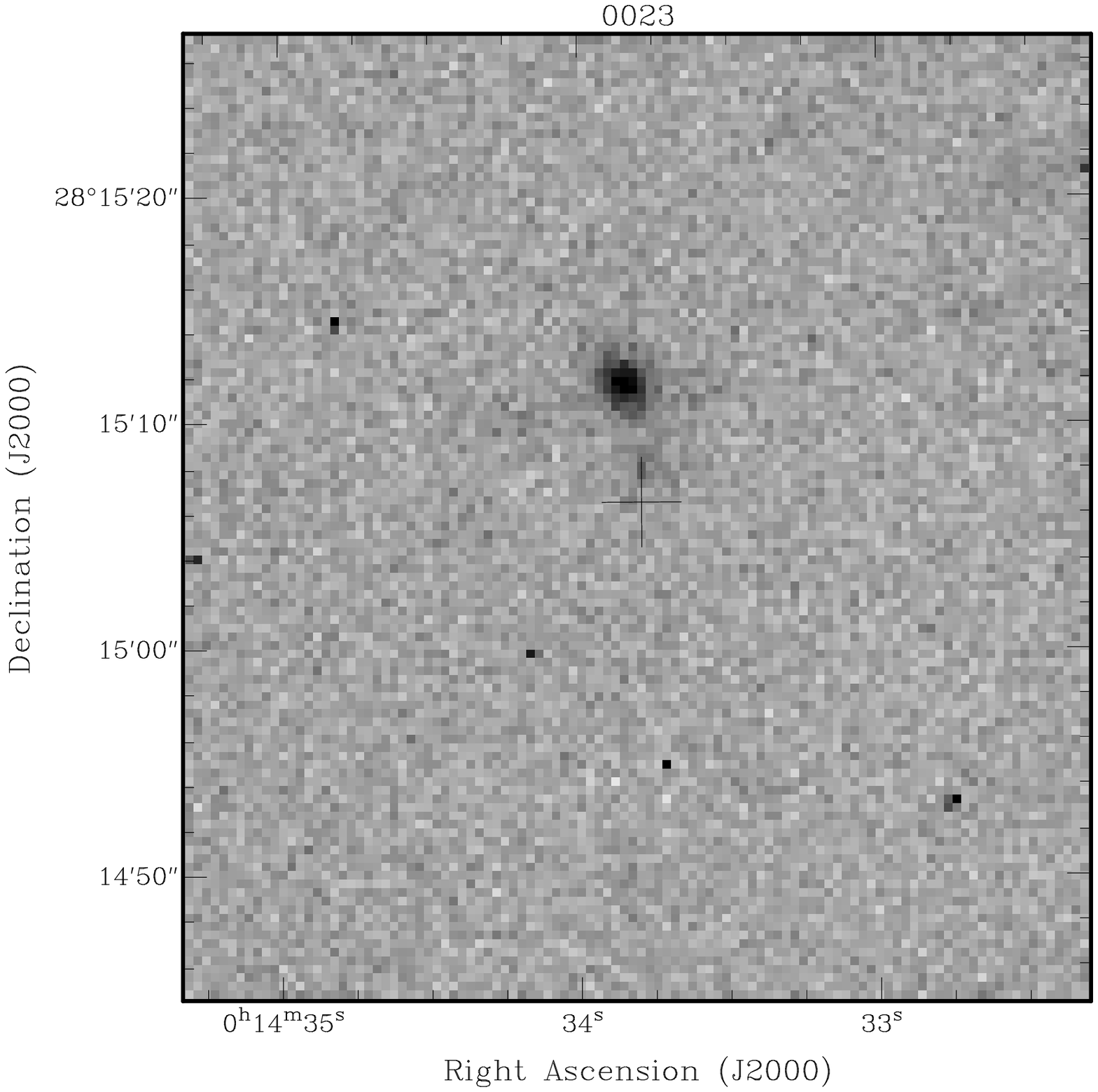 ,width=4.0cm,clip=}}\quad 
\subfigure[9CJ0014+2815 (P60 \it{R}\normalfont) Detail]{\epsfig{figure=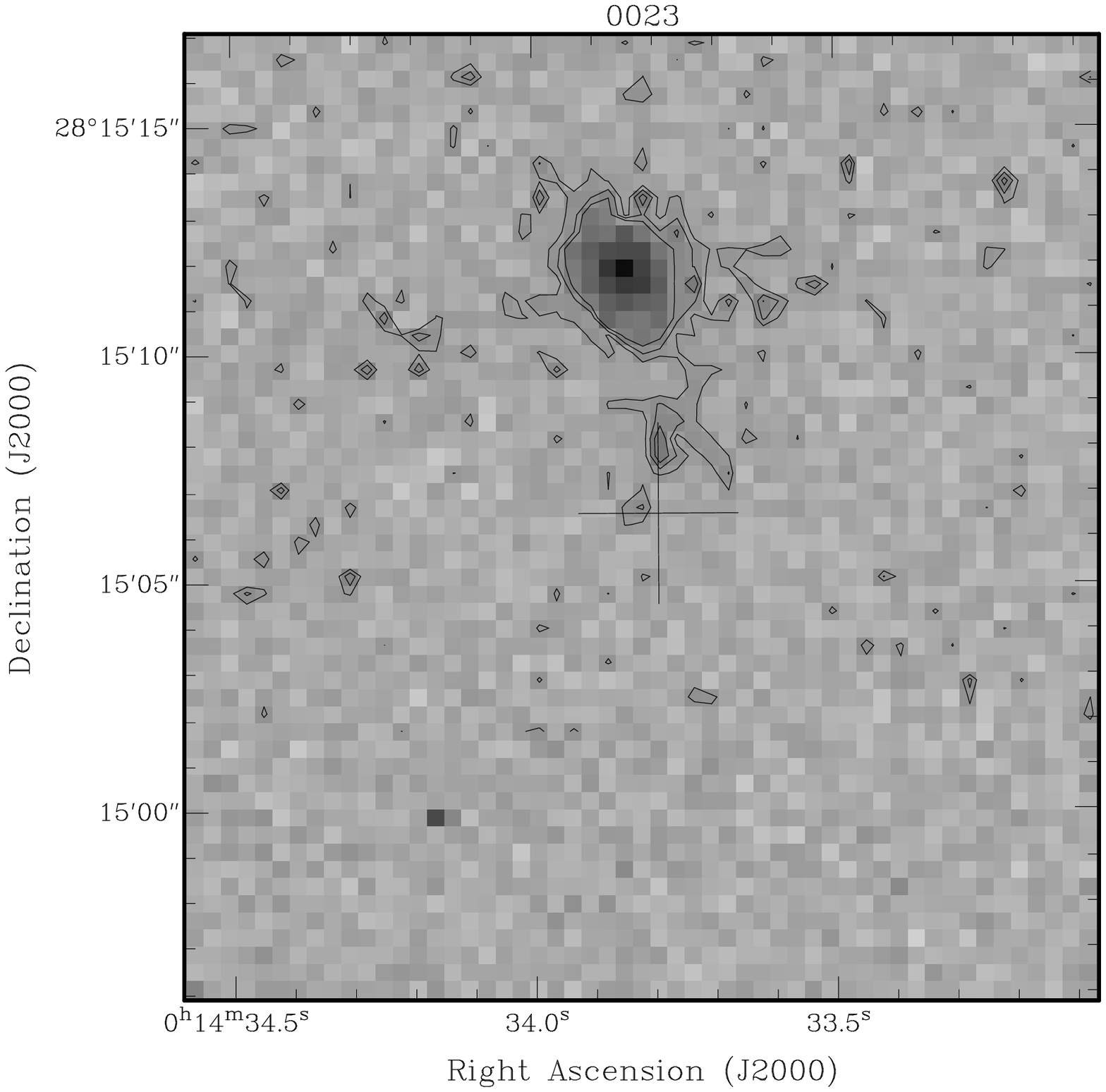 ,width=4.0cm,clip=}\label{e}}\quad 
\subfigure[9CJ0015+3216 (DSS2 \it{R}\normalfont) ]{\epsfig{figure=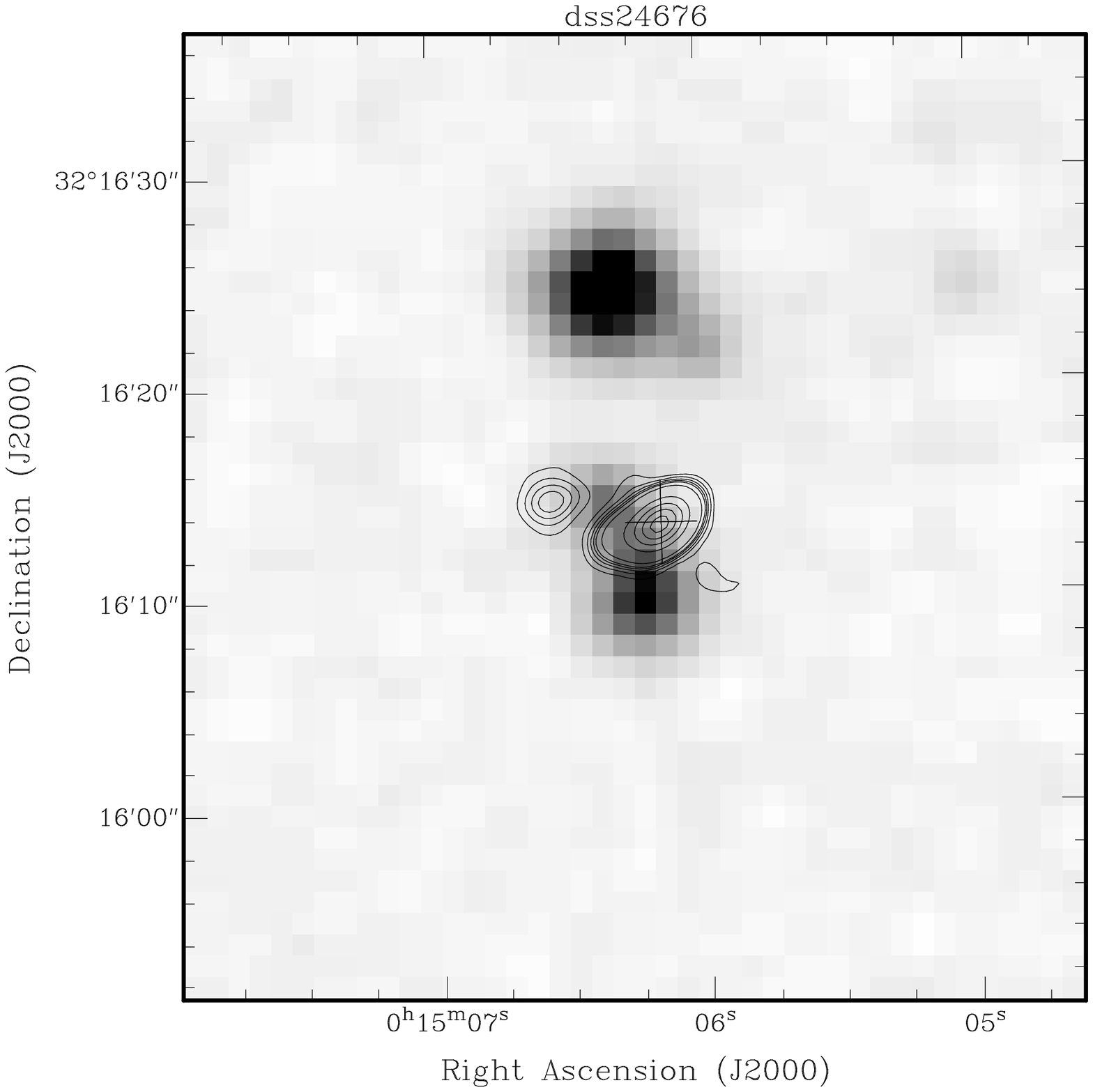 ,width=4.0cm,clip=}\label{f}}
}
\caption{Optical counterparts for sources 9CJ0010+2650 to 9CJ0015+3216. Crosses mark maximum radio flux density and are 4\,arcsec top to bottom. Contours: \ref{c}; 22\,GHz contours at 10-80 every 10\,\% of peak (15.2\,mJy/beam); \ref{d}, 22\,GHz contours at 10-90 every 10\,\% of peak (33.8\,mJy/beam); \ref{e}, P60 \it{R} \normalfont optical coutours; \ref{f}, 43\,GHz contours at 3,4,5,6,7,10,30,50,70,90\,\% of peak (220\,mJy/beam).}
\end{figure*}

\begin{figure*}
\mbox{ 
\subfigure[9CJ0015+3052 (DSS2 \it{R}\normalfont)]{\epsfig{figure=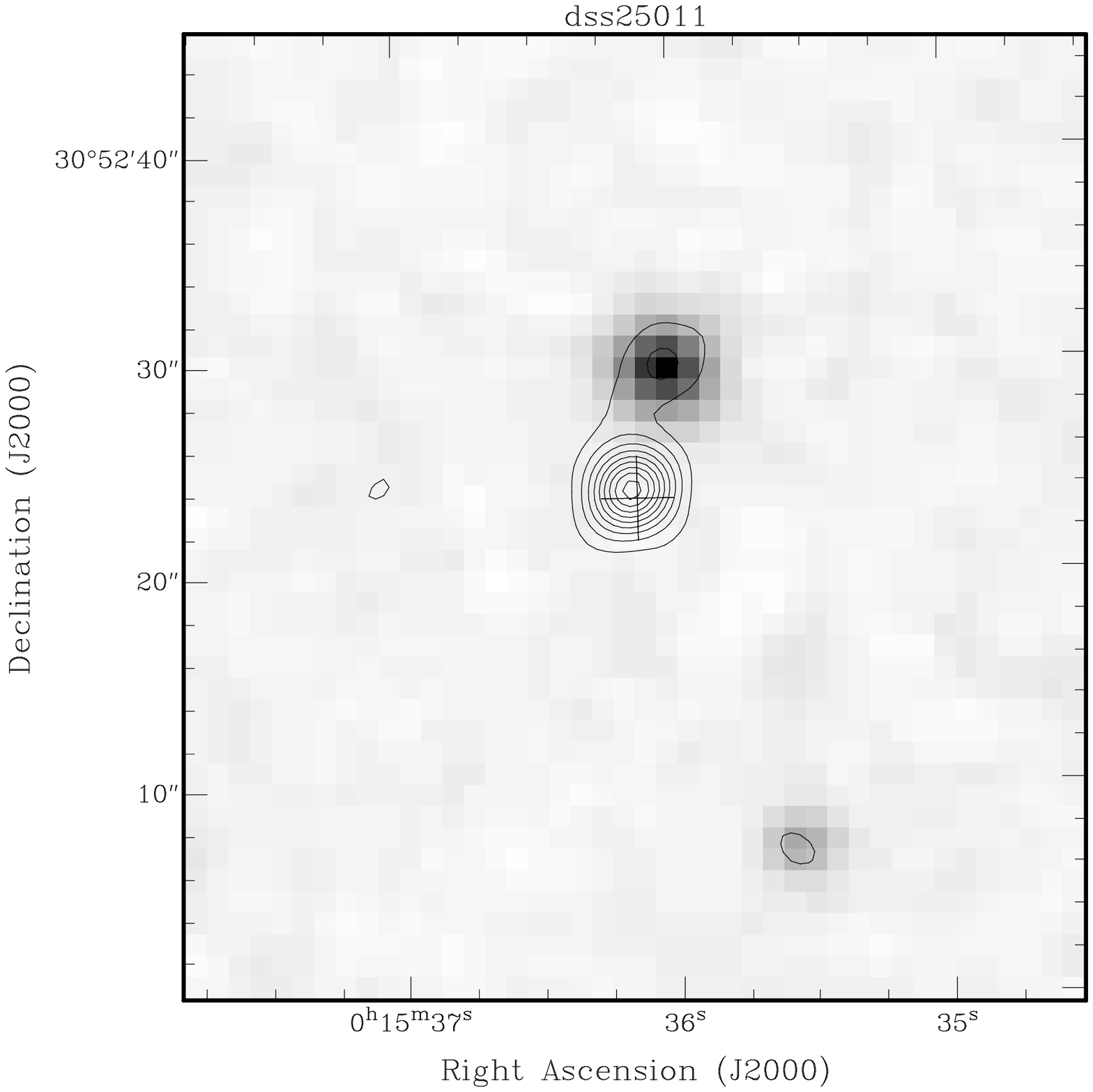 ,width=4.0cm,clip=}\label{g}}\quad 
\subfigure[9CJ0018+2921 (DSS2 \it{R}\normalfont)]{\epsfig{figure=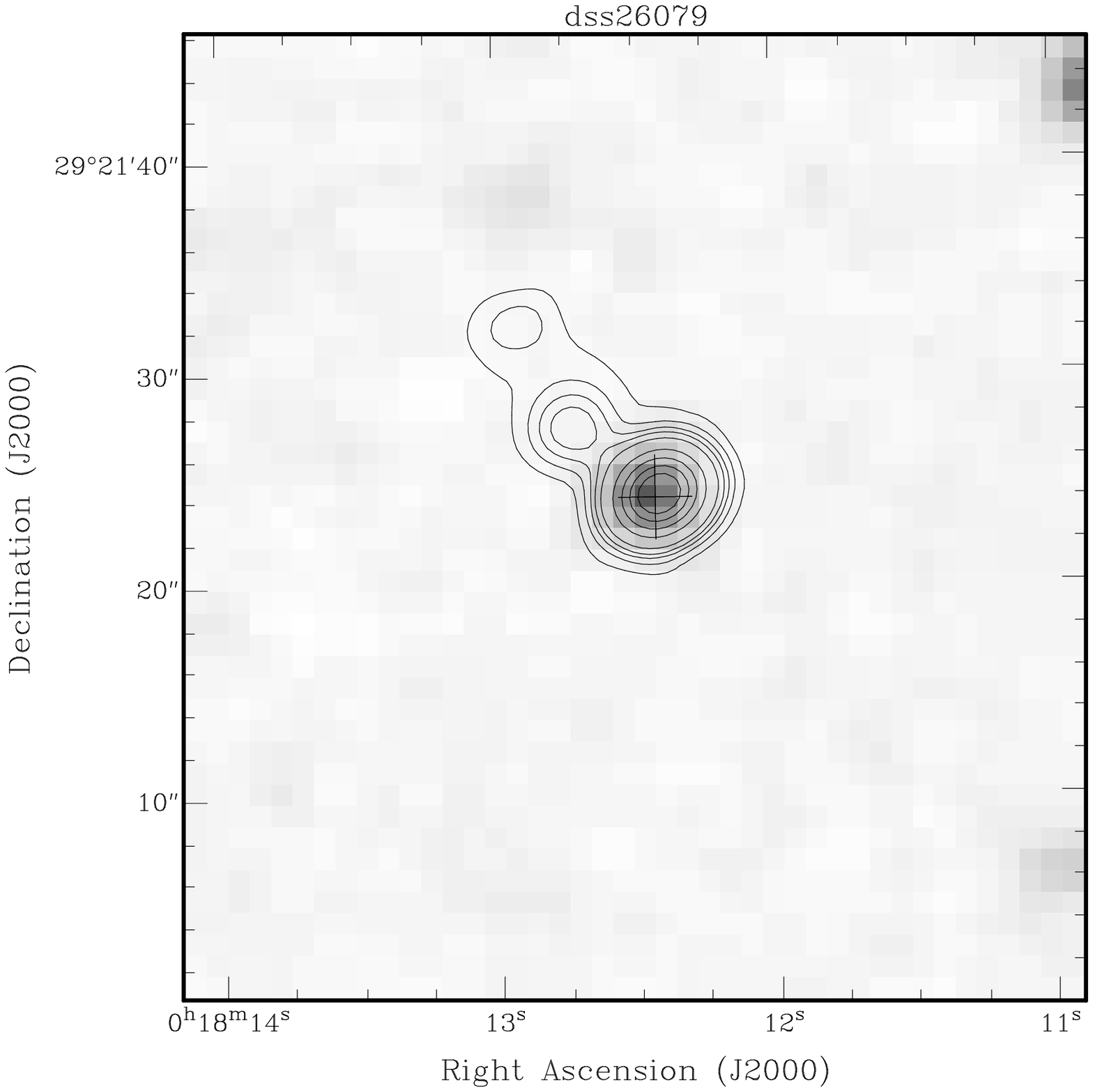 ,width=4.0cm,clip=}\label{h}}\quad 
\subfigure[9CJ0018+3105 (DSS2 \it{R}\normalfont)]{\epsfig{figure=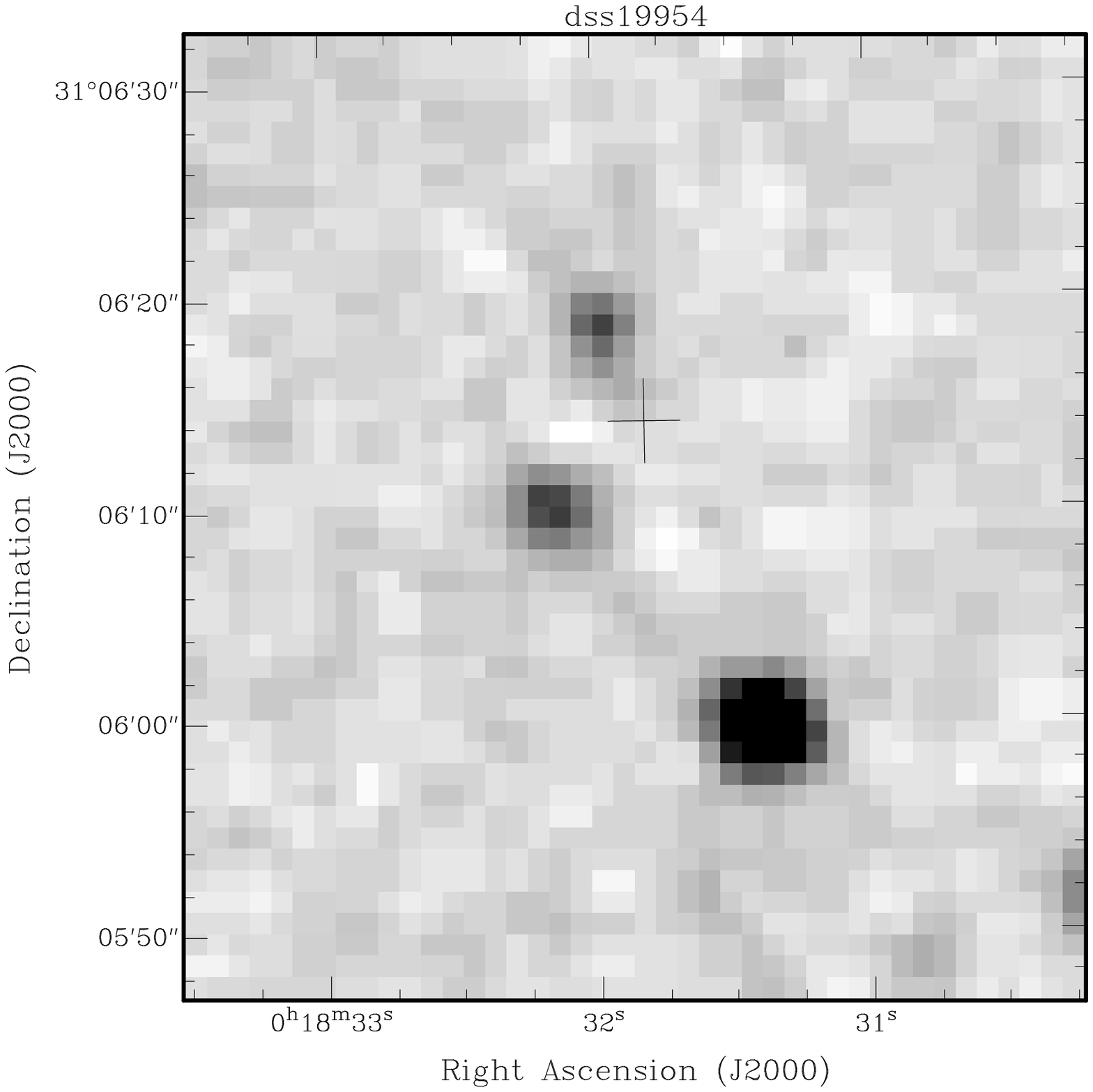 ,width=4.0cm,clip=}\label{i}}
} 
\mbox{ 
\subfigure[9CJ0018+3105, large (DSS2 \it{R}\normalfont)]{\epsfig{figure=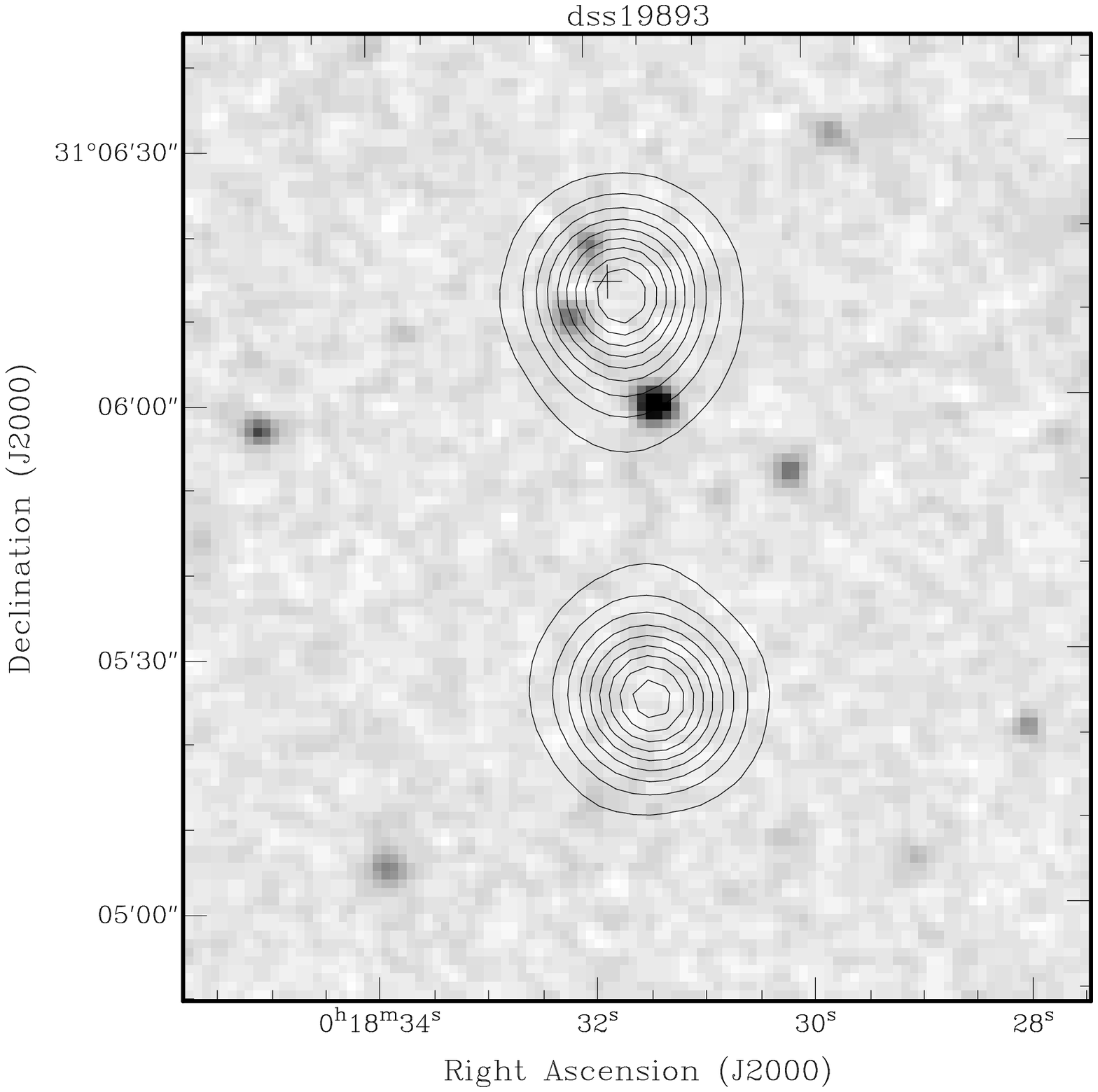 ,width=4.0cm,clip=}\label{j}}\quad 
\subfigure[9CJ0018+2907 (DSS2 \it{R}\normalfont)]{\epsfig{figure=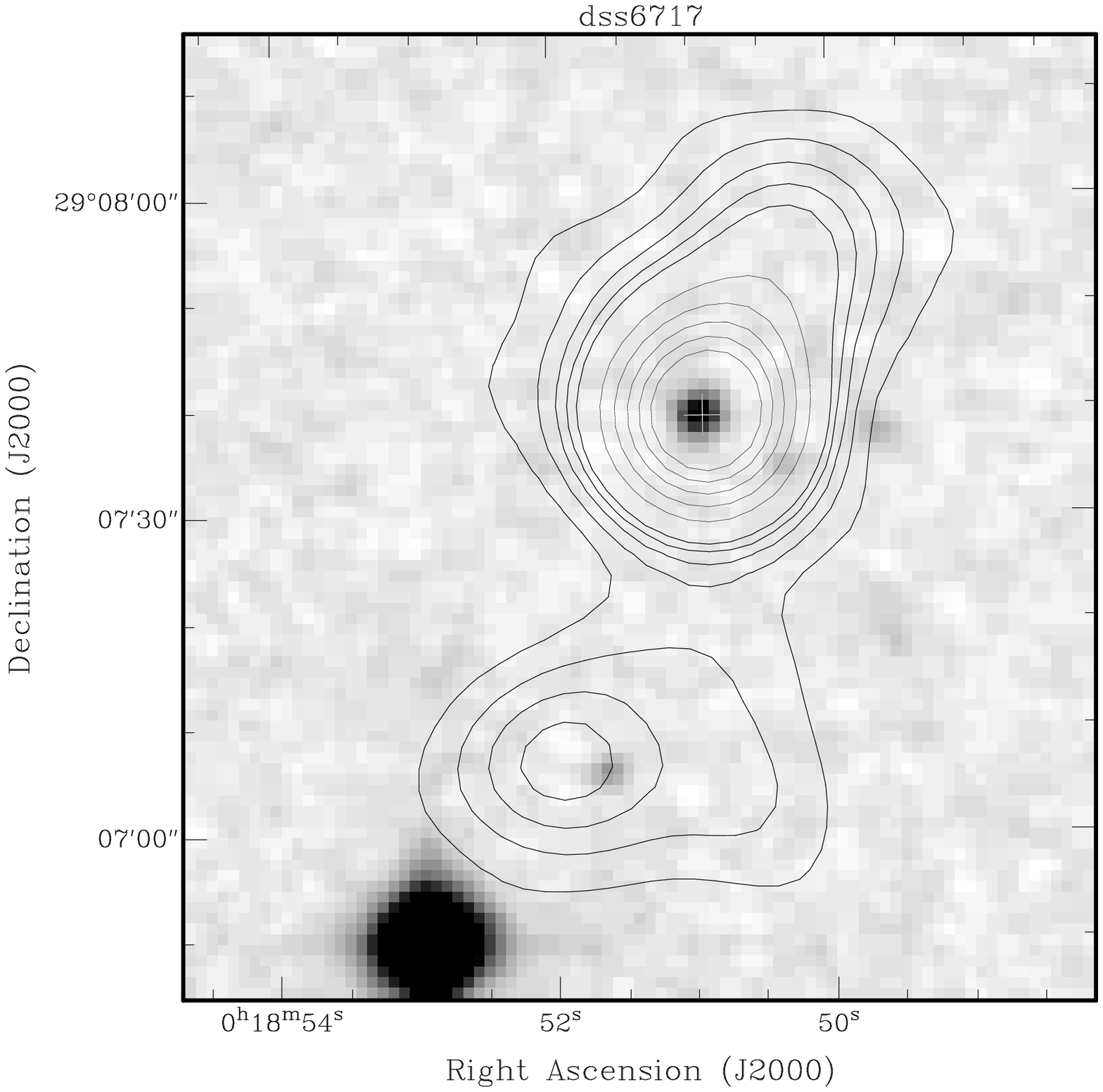 ,width=4.0cm,clip=}\label{k}}\quad 
\subfigure[9CJ0019+2817 (DSS2 \it{R}\normalfont)]{\epsfig{figure=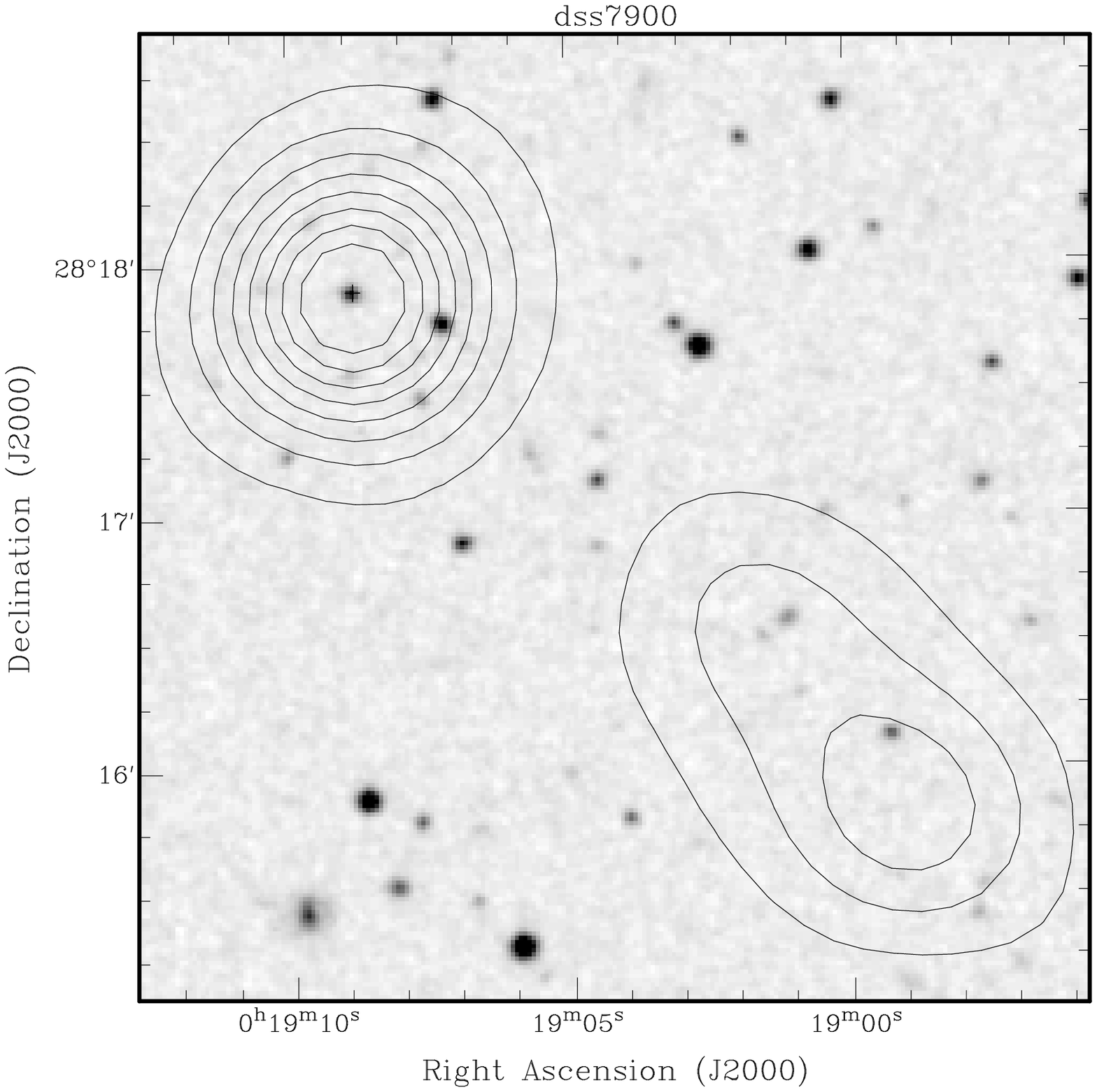 ,width=4.0cm,clip=}\label{l}}
} 
\mbox{ 
\subfigure[9CJ0019+2956 (P60 \it{R}\normalfont)]{\epsfig{figure=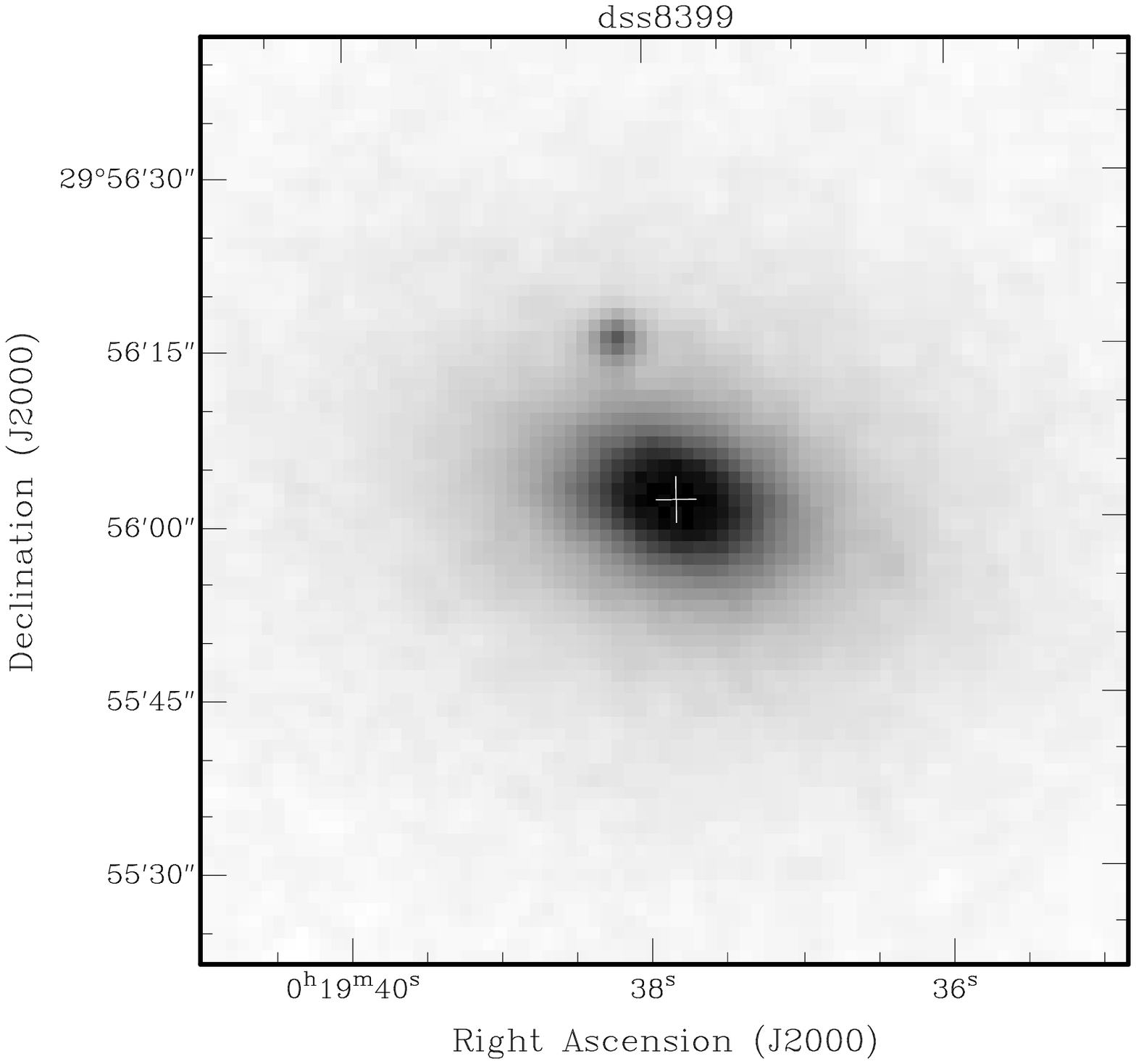 ,width=4.0cm,clip=}}\quad 
\subfigure[9CJ0019+2647 (DSS2 \it{R}\normalfont)]{\epsfig{figure=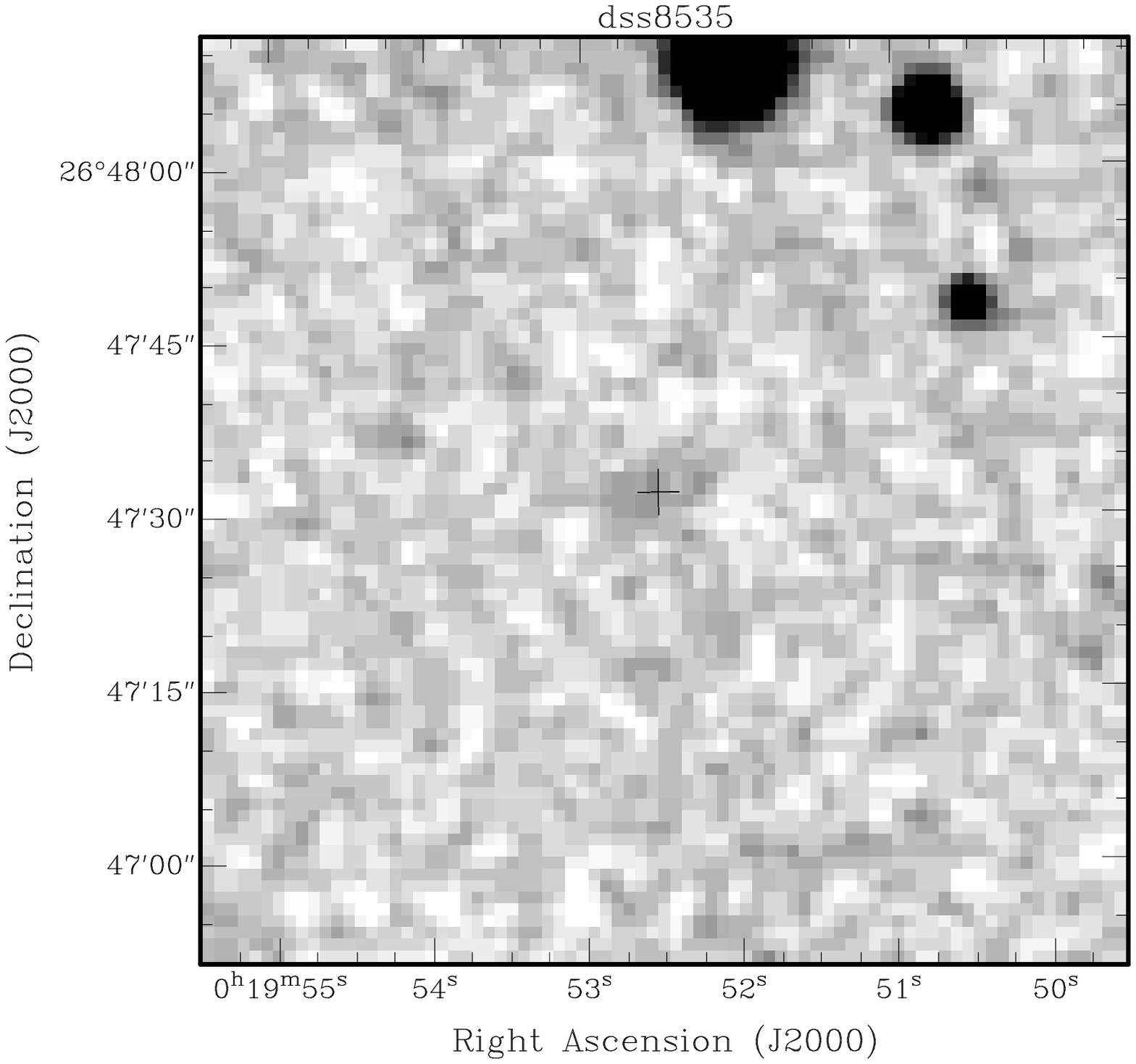 ,width=4.0cm,clip=}}\quad 
\subfigure[9CJ0019+3320 (DSS2 \it{R}\normalfont)]{\epsfig{figure=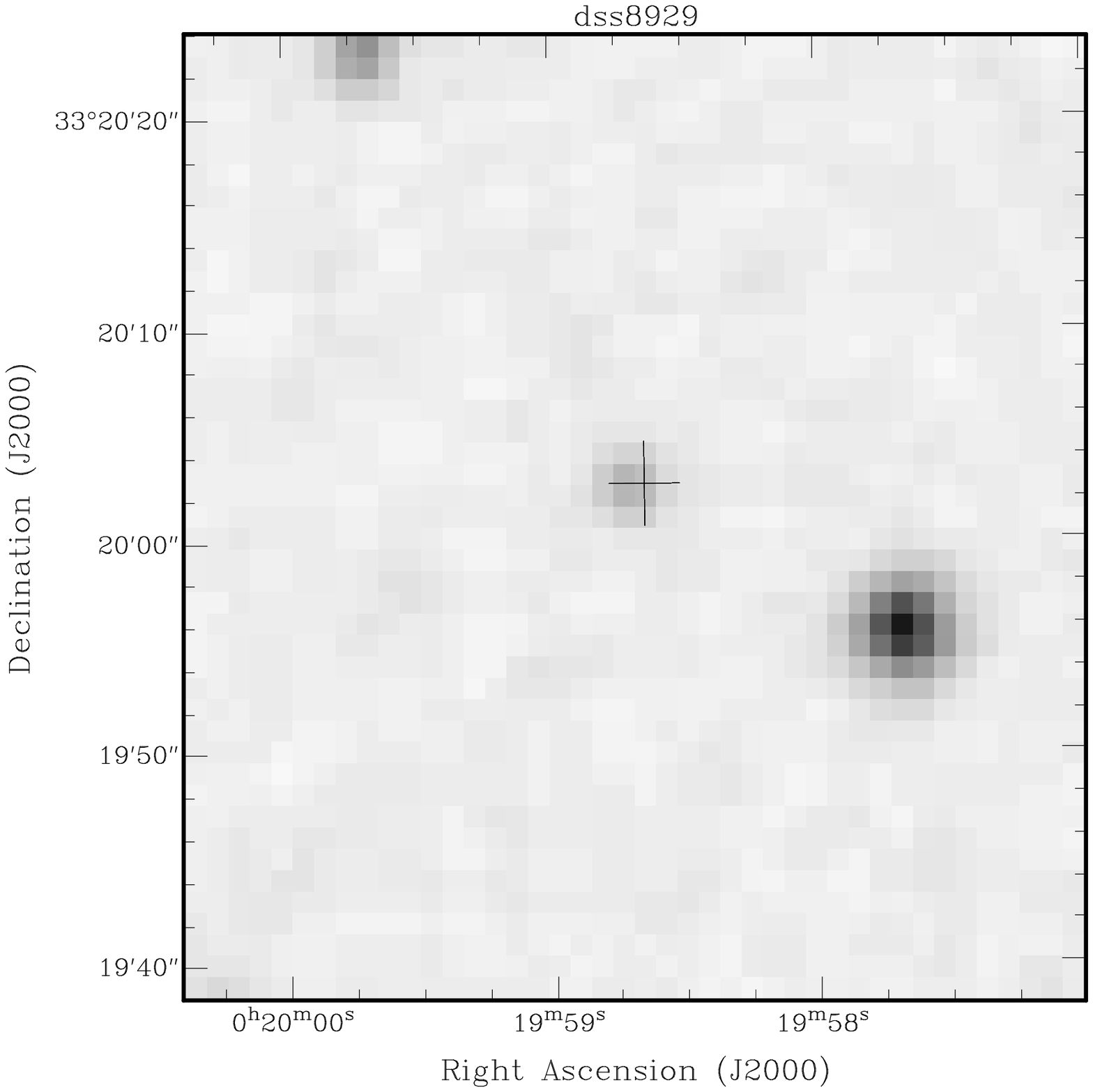 ,width=4.0cm,clip=}}
} 
\mbox{ 
\subfigure[9CJ0020+3152 (P60 \it{R}\normalfont)]{\epsfig{figure=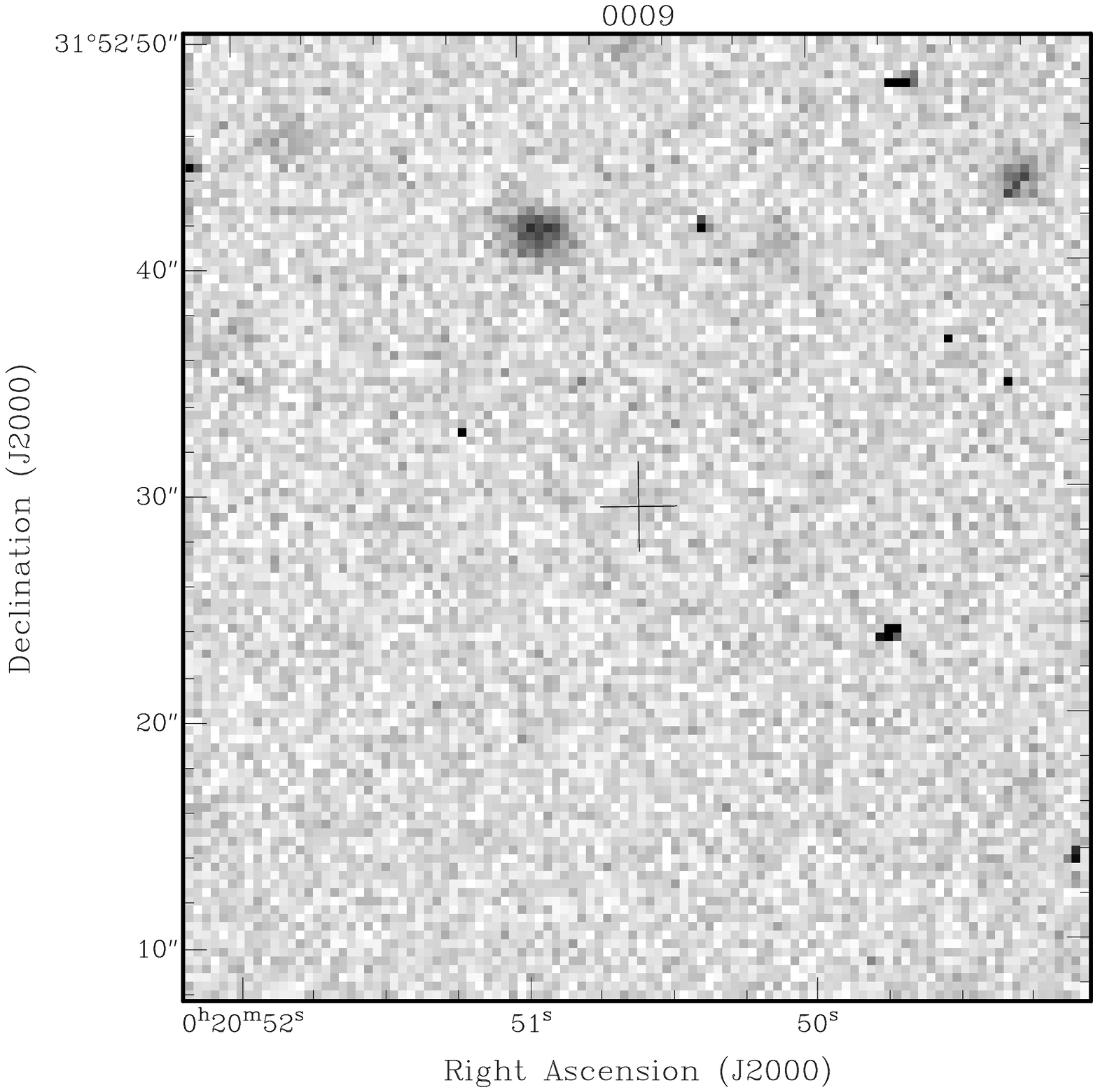 ,width=4.0cm,clip=}}\quad 
\subfigure[9CJ0021+2711 (DSS2 \it{R}\normalfont)]{\epsfig{figure=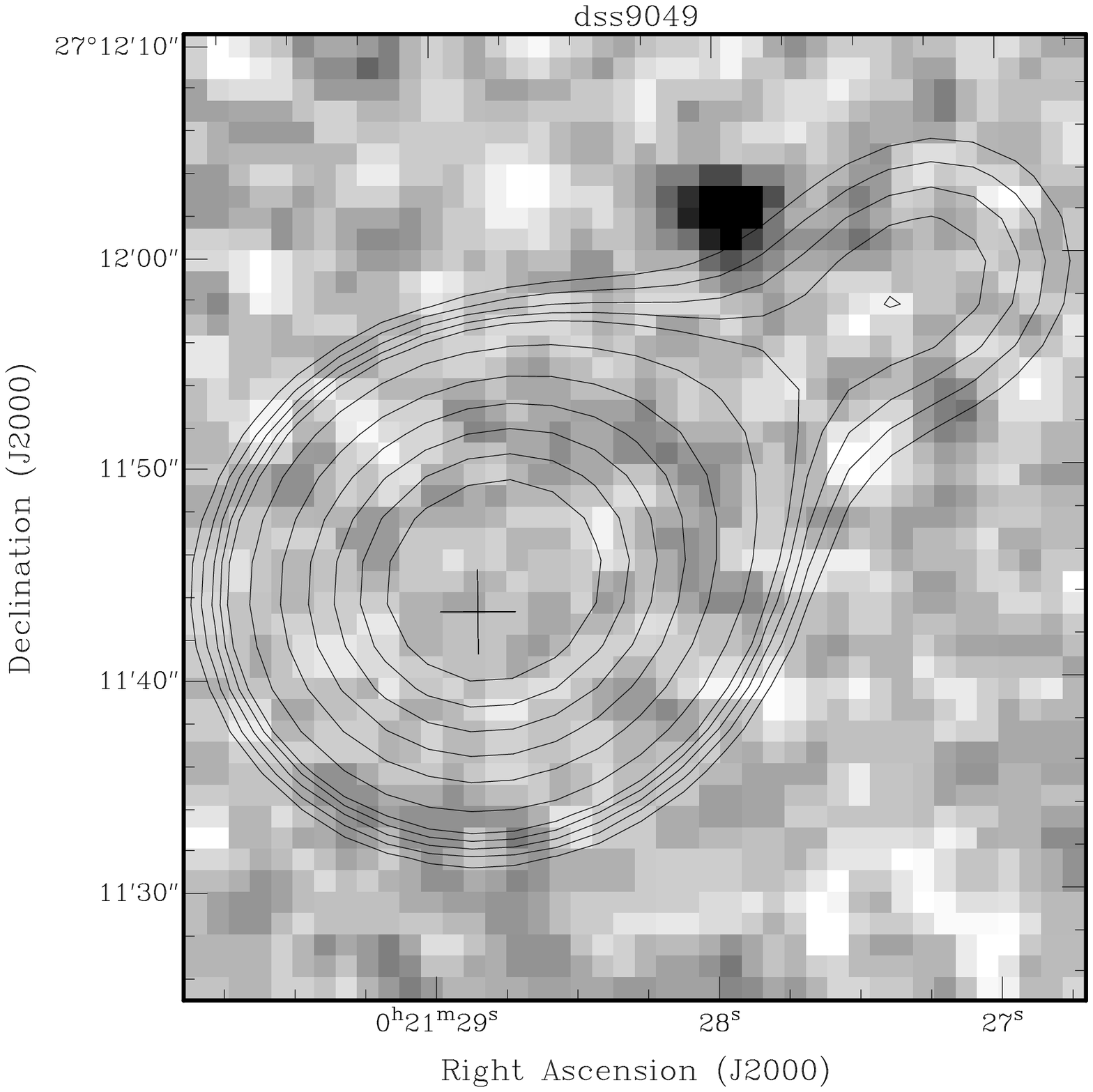 ,width=4.0cm,clip=}\label{m}}\quad 
\subfigure[9CJ0021+3226 (P60 \it{R}\normalfont)]{\epsfig{figure=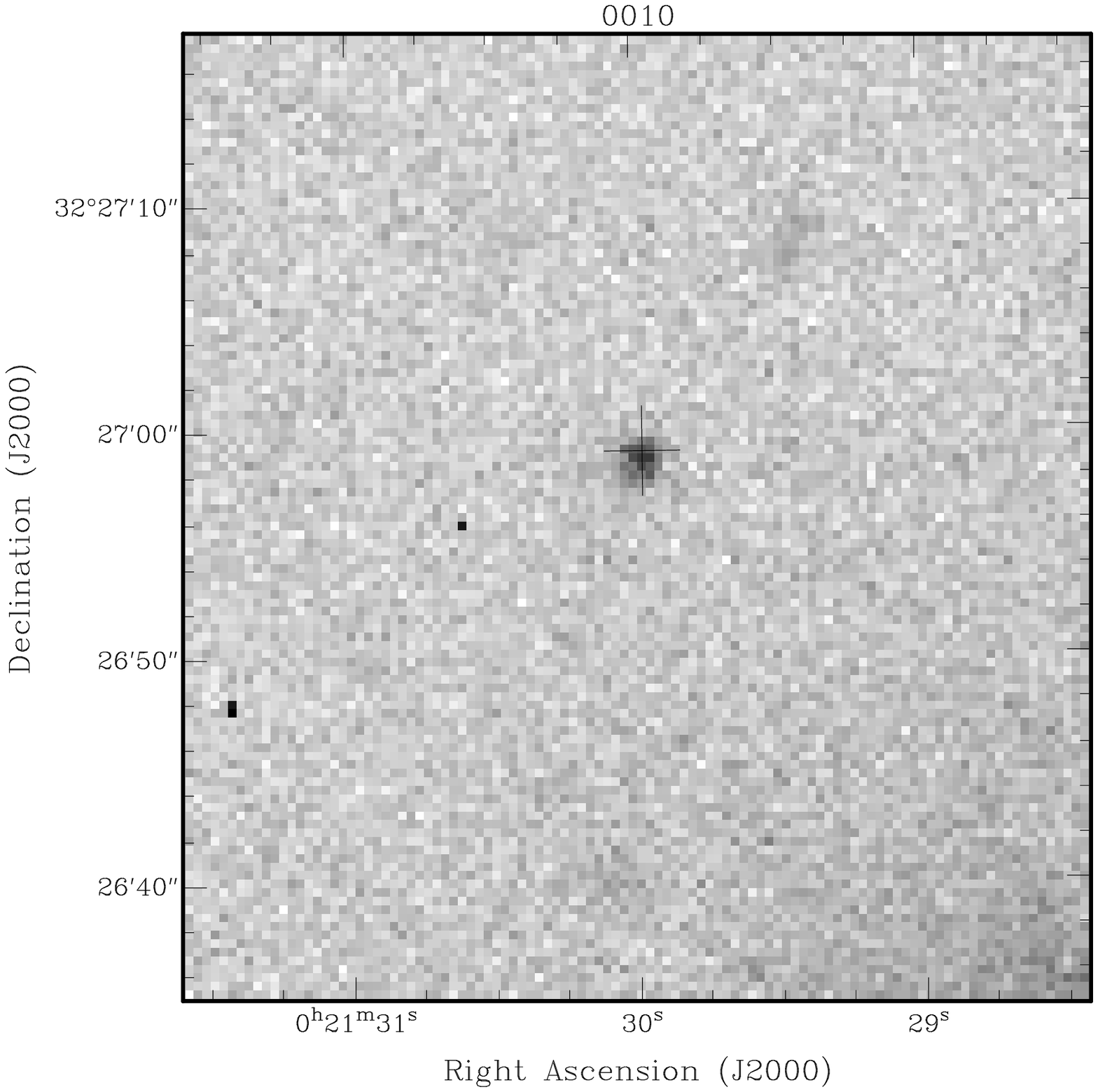 ,width=4.0cm,clip=}}
}\caption{ Optical counterparts for sources 9CJ0015+3052 to 9CJ0021+3226. Crosses mark maximum radio flux density and are 4\,arcsec top to bottom. Contours: \ref{g}, 22\,GHz contours at 15-95 every 10\,\% of peak (20.3\,mJy/beam); \ref{h}, 22\,GHz contours 6-15 every 3\,\% and 20-80 every 15\,\% of peak (63.9\,mJy/beam); \ref{j}, 4.8\,GHz contours 5-25 every 5\,\% and 35 to 95 every 10 \% of peak (50.0\,mJy/beam); \ref{k}, 4.8\,GHz contours at 7-21 every 3\,\% and 30-70 every 10\,\% of peak (27.1\,mJy/beam); \ref{l}, 1.4\,GHz contours 15-85 every 10\,\% of peak(25.8\,mJy/beam); \ref{m}, 4.8\,GHz contours 16-24 every 2\,\% and 30-80 every 10\,\% of peak (88.5\,mJy/beam).}
\end{figure*}

\begin{figure*}
\mbox{ 
\subfigure[9CJ0022+3250 (P60 \it{R}\normalfont)]{\epsfig{figure=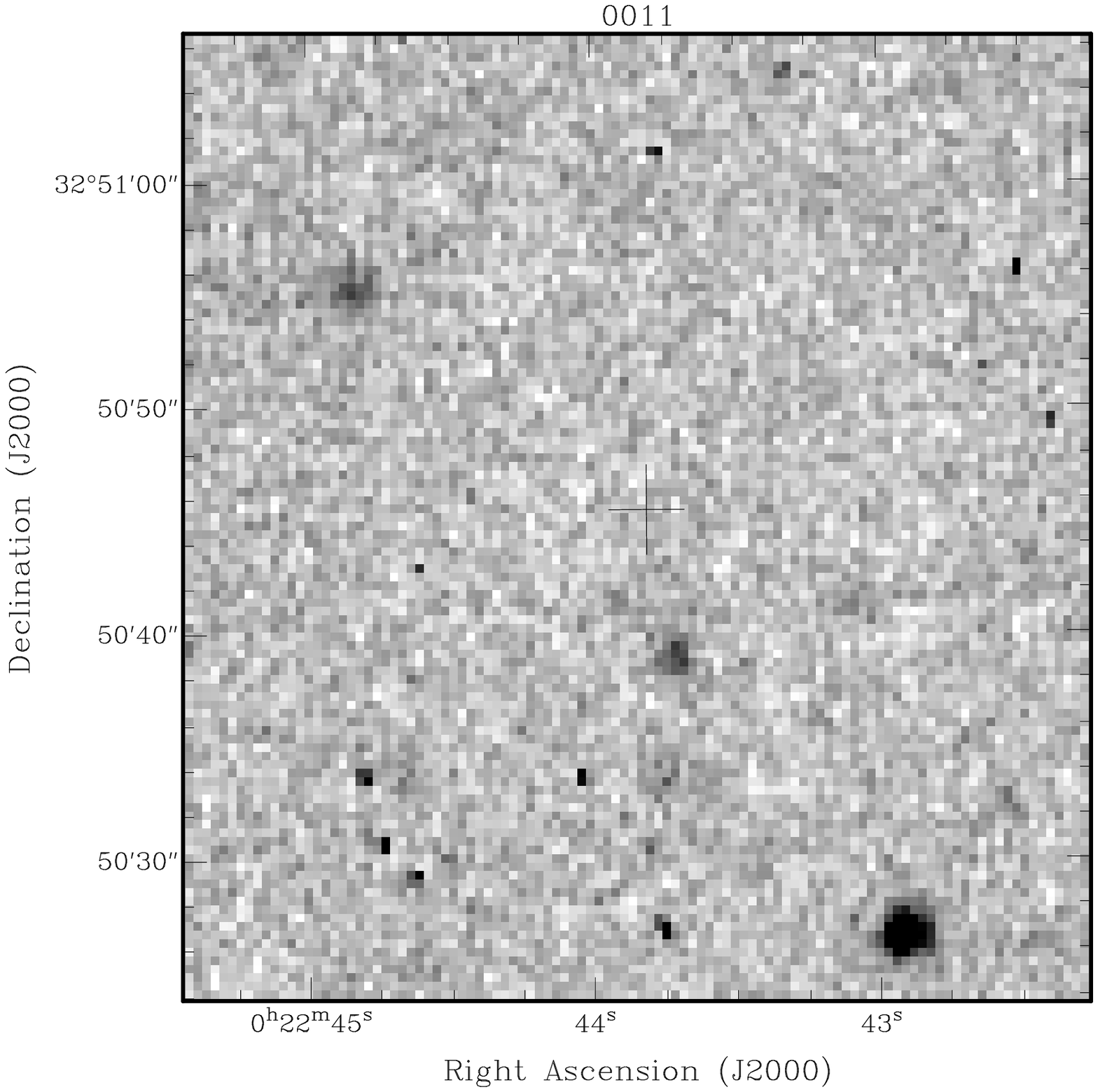 ,width=4.0cm,clip=}}\quad 
\subfigure[9CJ0023+3114 (P60 \it{R}\normalfont)]{\epsfig{figure=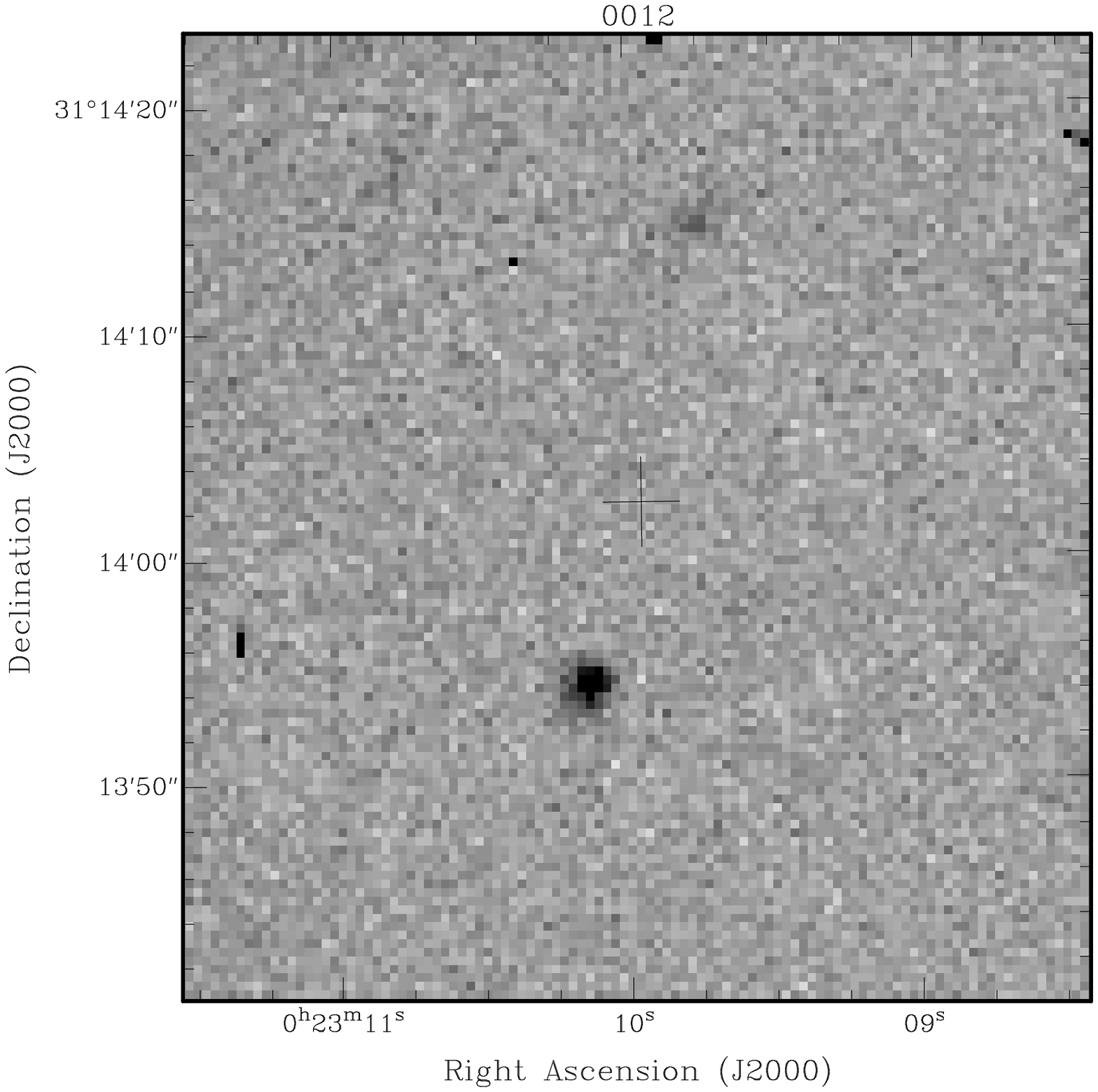 ,width=4.0cm,clip=}}\quad 
\subfigure[9CJ0023+2734 (P60 \it{R}\normalfont)]{\epsfig{figure=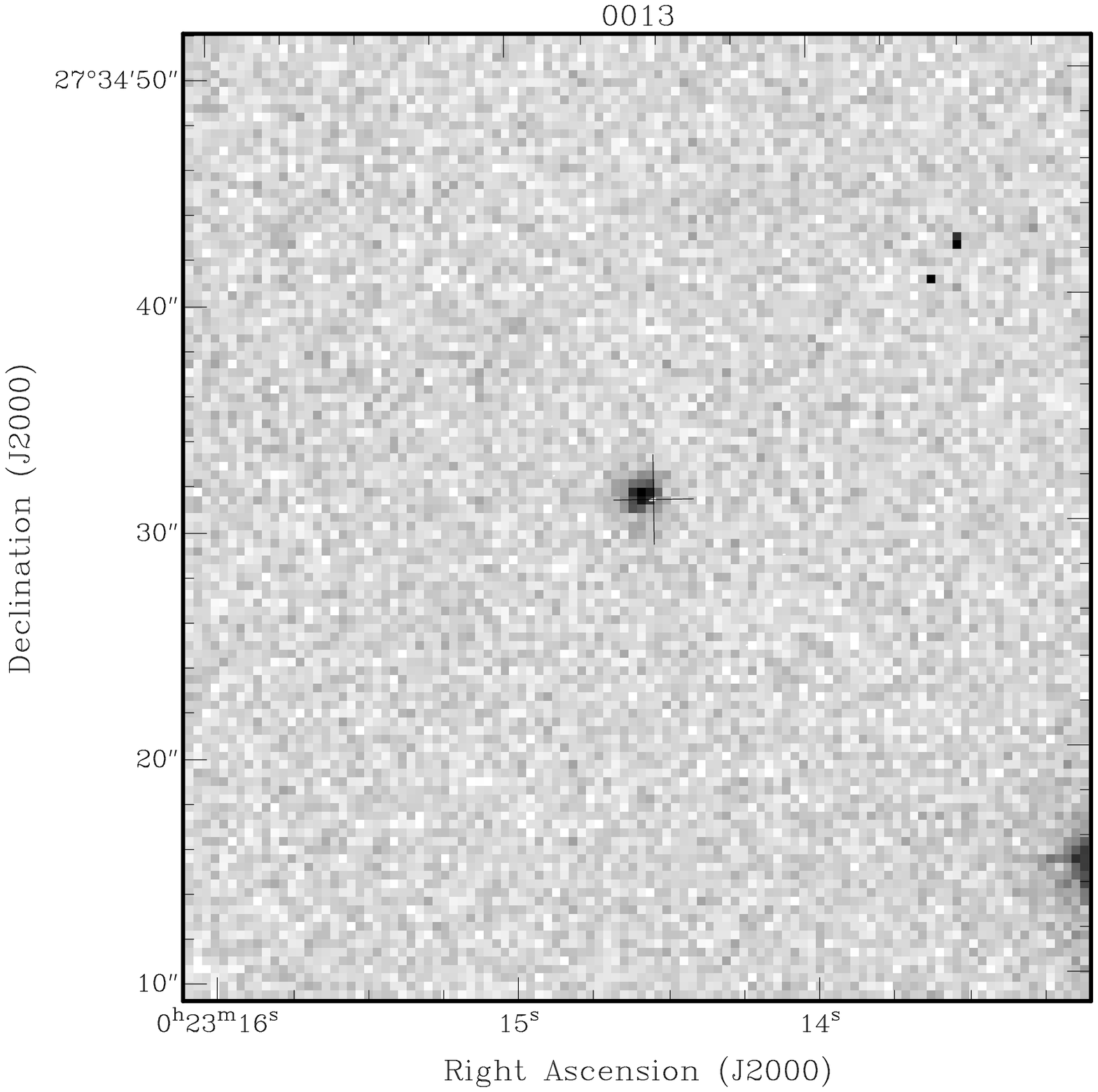 ,width=4.0cm,clip=}}
} 
\mbox{
\subfigure[9CJ0023+2734 (P60 \it{R}\normalfont) Detail]{\epsfig{figure=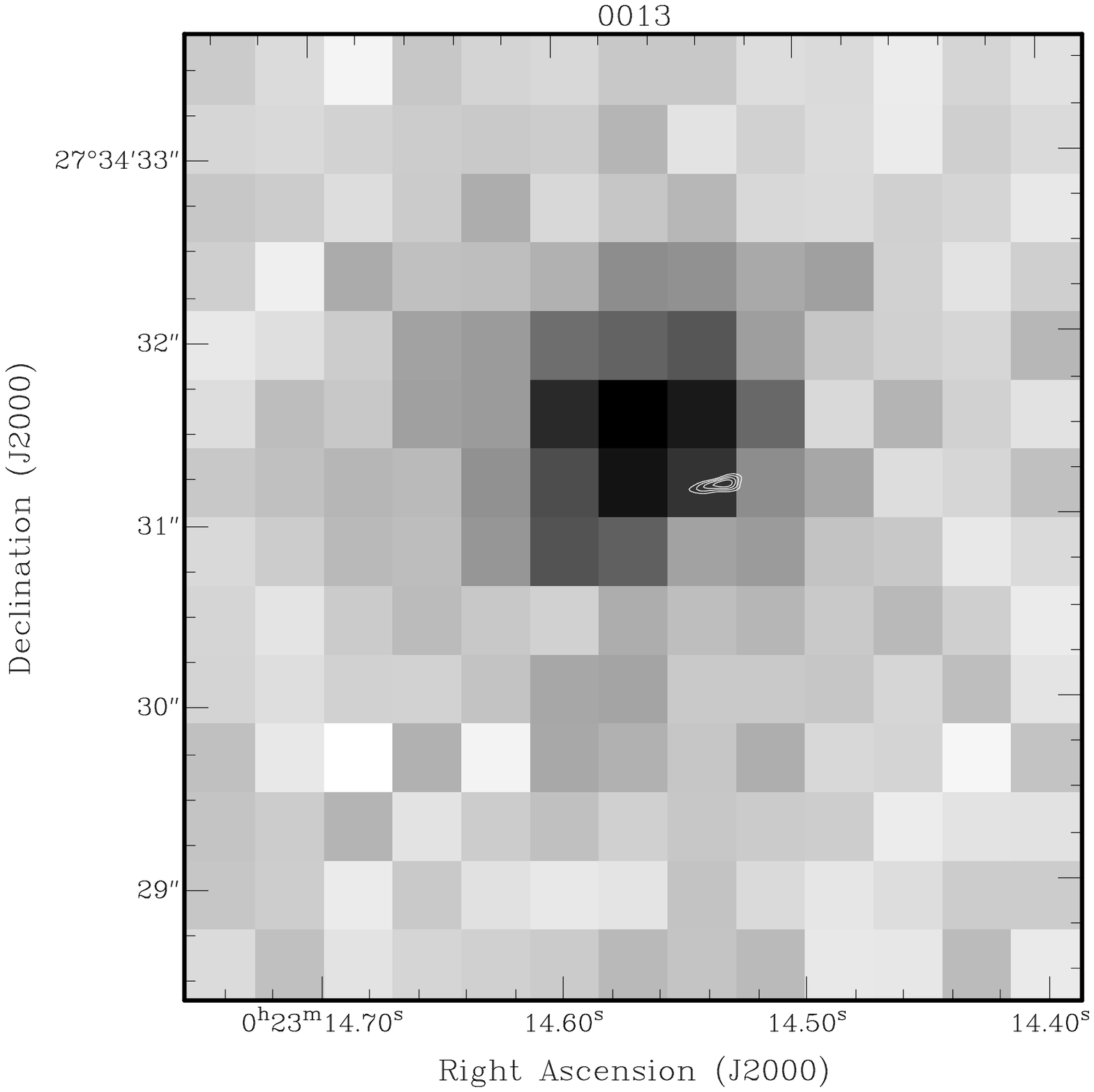 ,width=4.0cm,clip=}\label{o}}\quad 
\subfigure[9CJ0023+2539 (P60 \it{R}\normalfont)]{\epsfig{figure=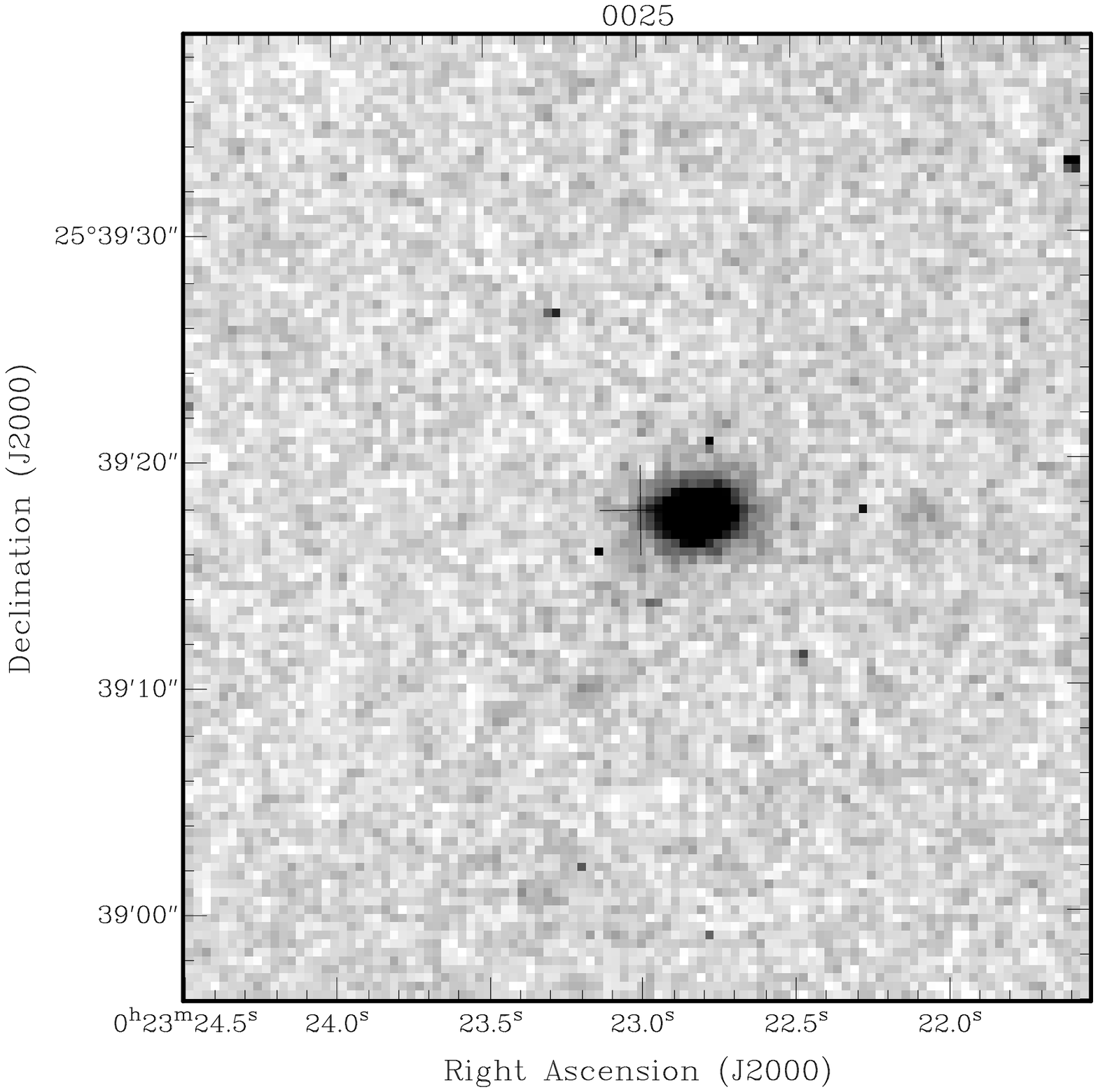 ,width=4.0cm,clip=}}\quad 
\subfigure[9CJ0024+2911 (P60 \it{R}\normalfont)]{\epsfig{figure=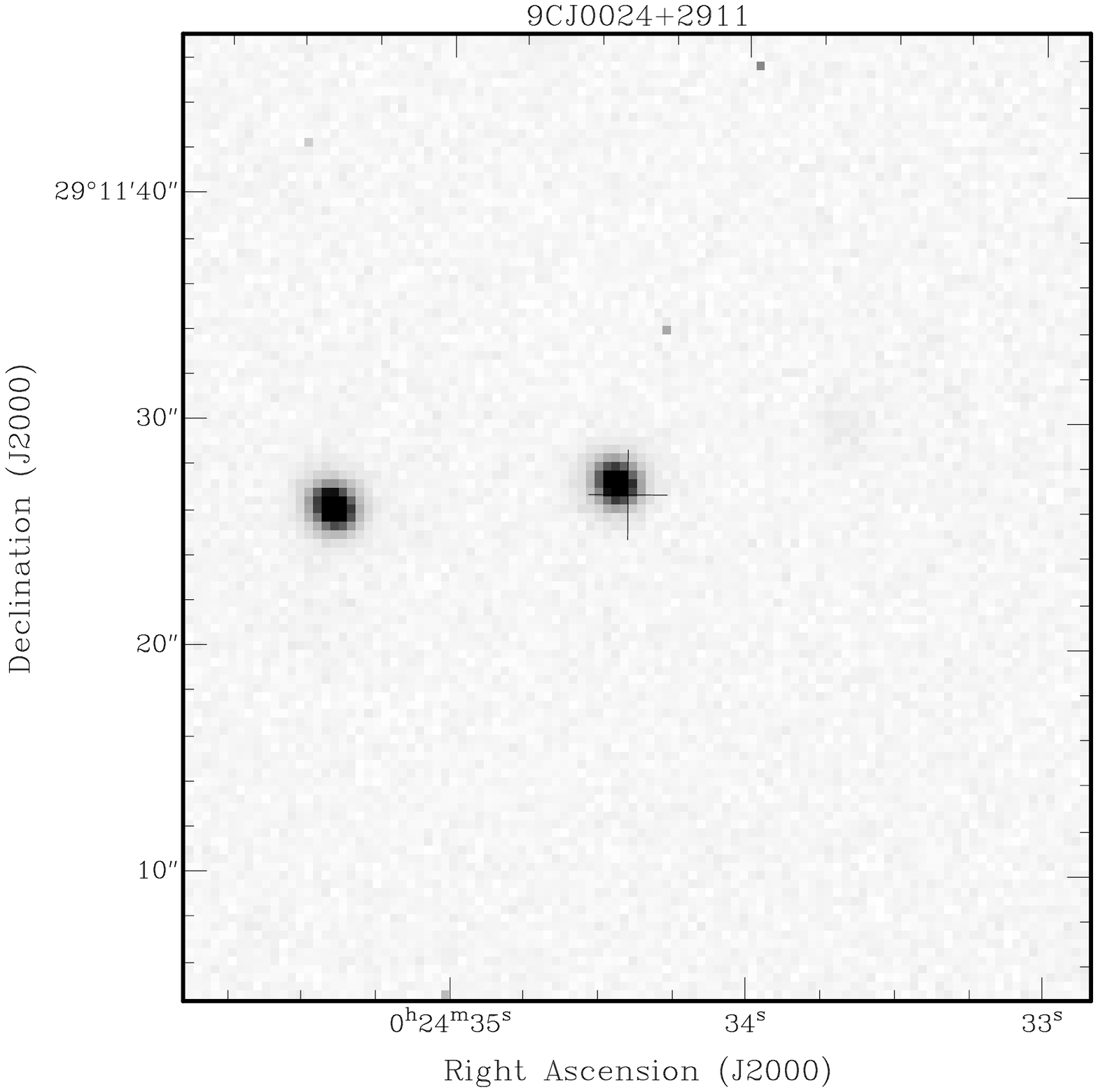 ,width=4.0cm,clip=}}
} 
\mbox{
\subfigure[9CJ0027+2830 (DSS2 \it{R}\normalfont)]{\epsfig{figure=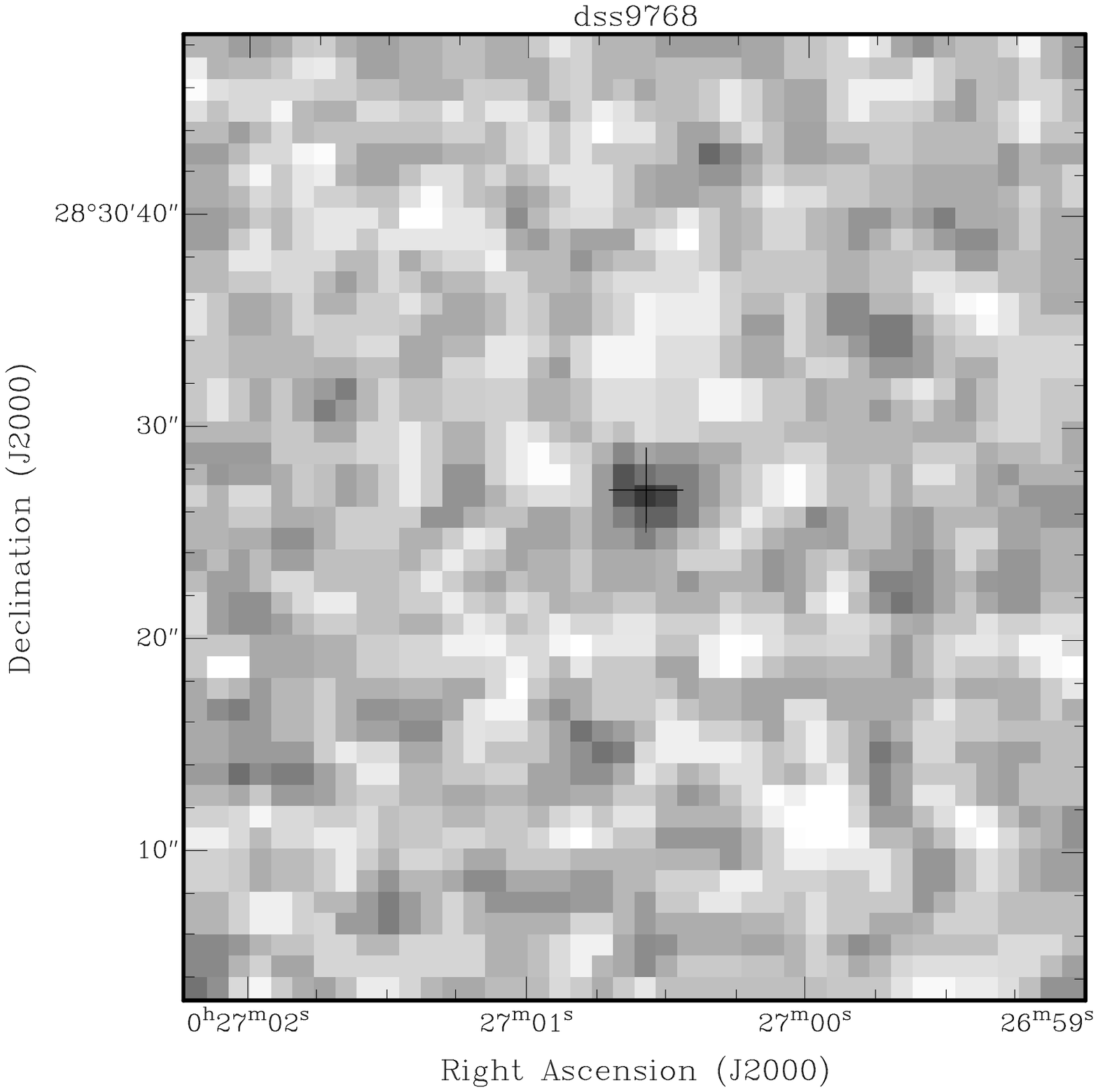 ,width=4.0cm,clip=}}\quad 
\subfigure[9CJ0028+3103 (DSS2 \it{R}\normalfont)]{\epsfig{figure=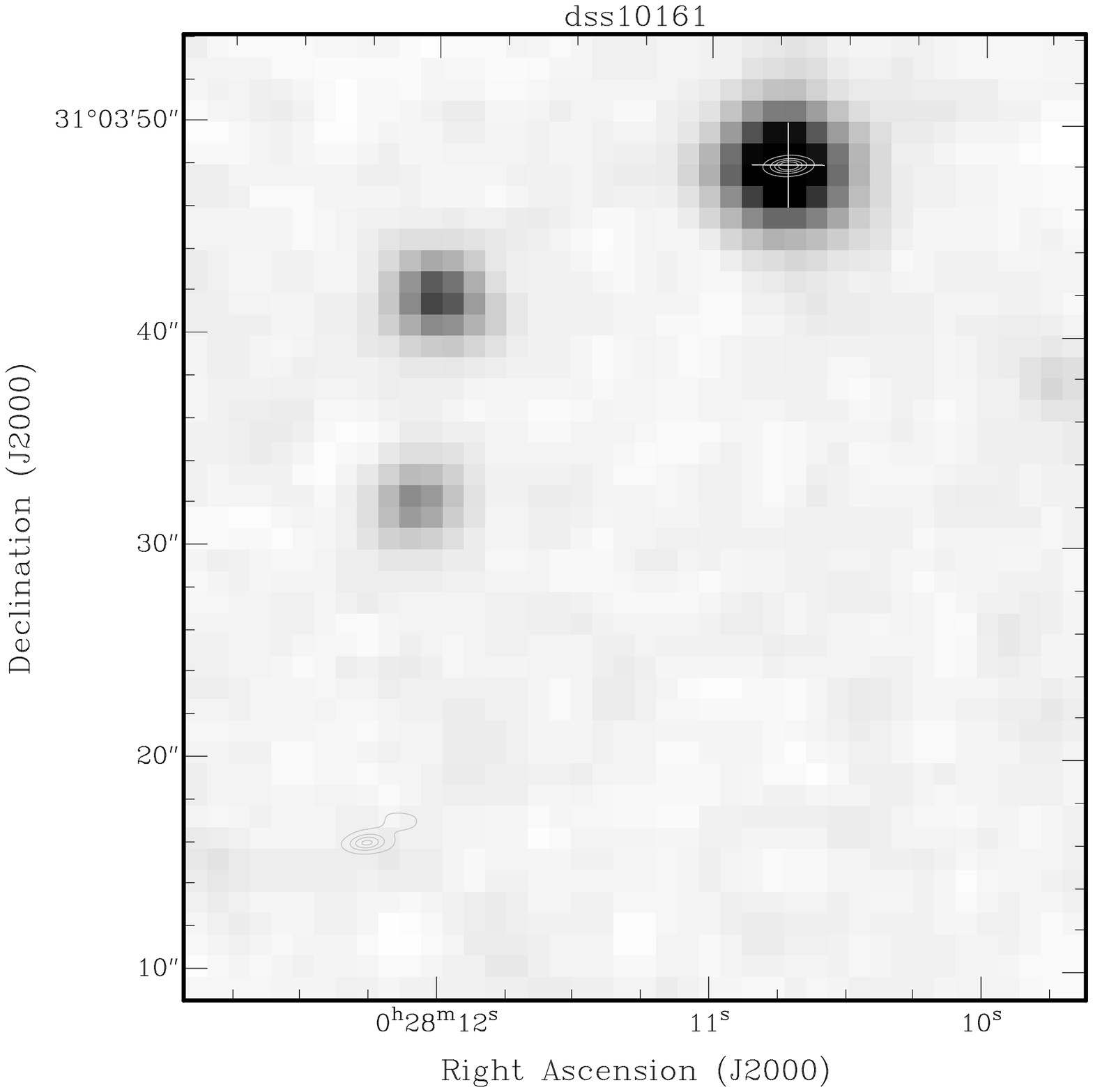 ,width=4.0cm,clip=}\label{p}}\quad 
\subfigure[9CJ0028+2914 (DSS2 \it{R}\normalfont)]{\epsfig{figure=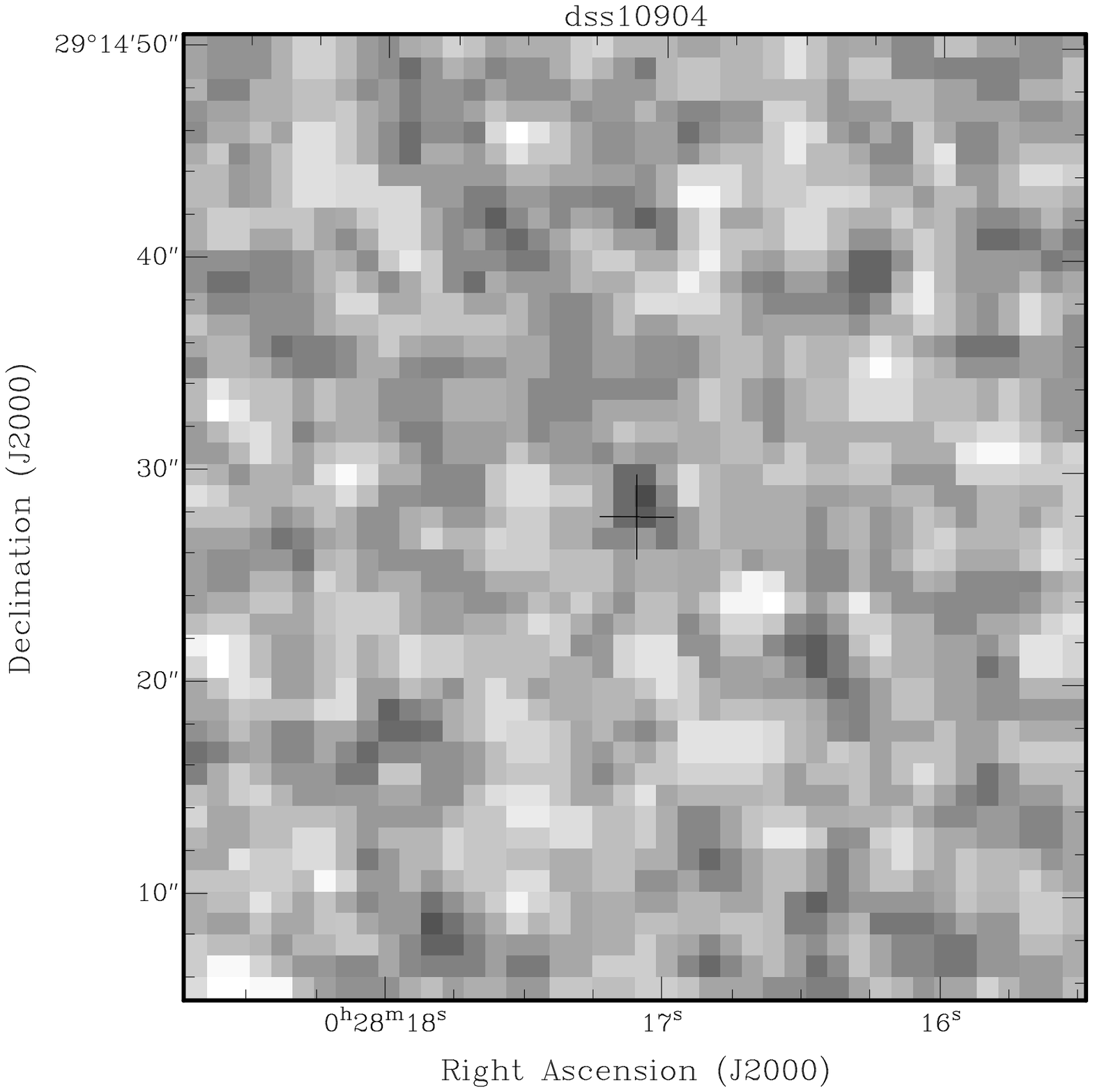 ,width=4.0cm,clip=}}
} 
\mbox{
\subfigure[9CJ0028+2954 (P60 \it{R}\normalfont)]{\epsfig{figure=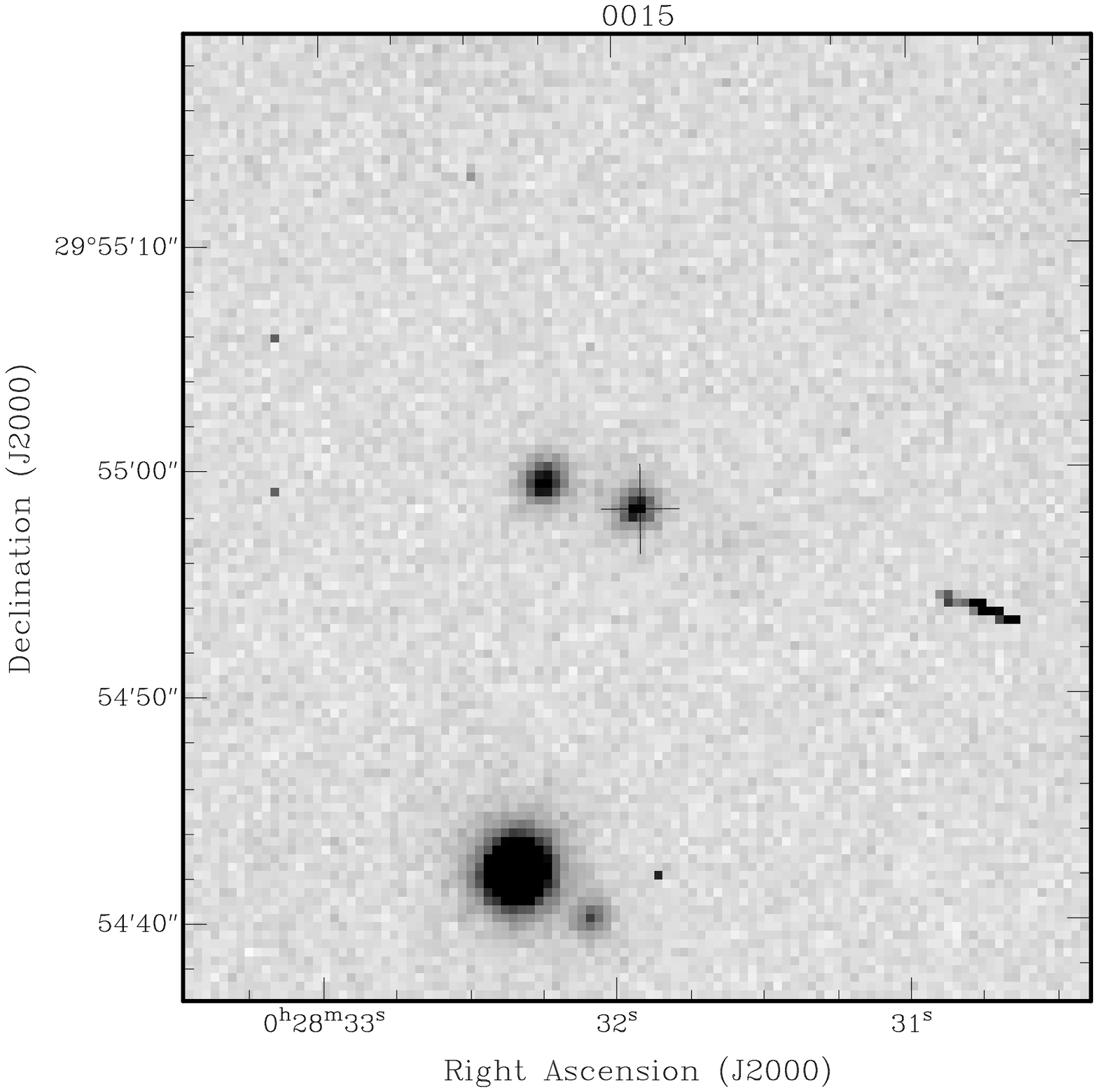 ,width=4.0cm,clip=}}\quad 
\subfigure[9CJ0028+2408 (DSS2 \it{R}\normalfont)]{\epsfig{figure=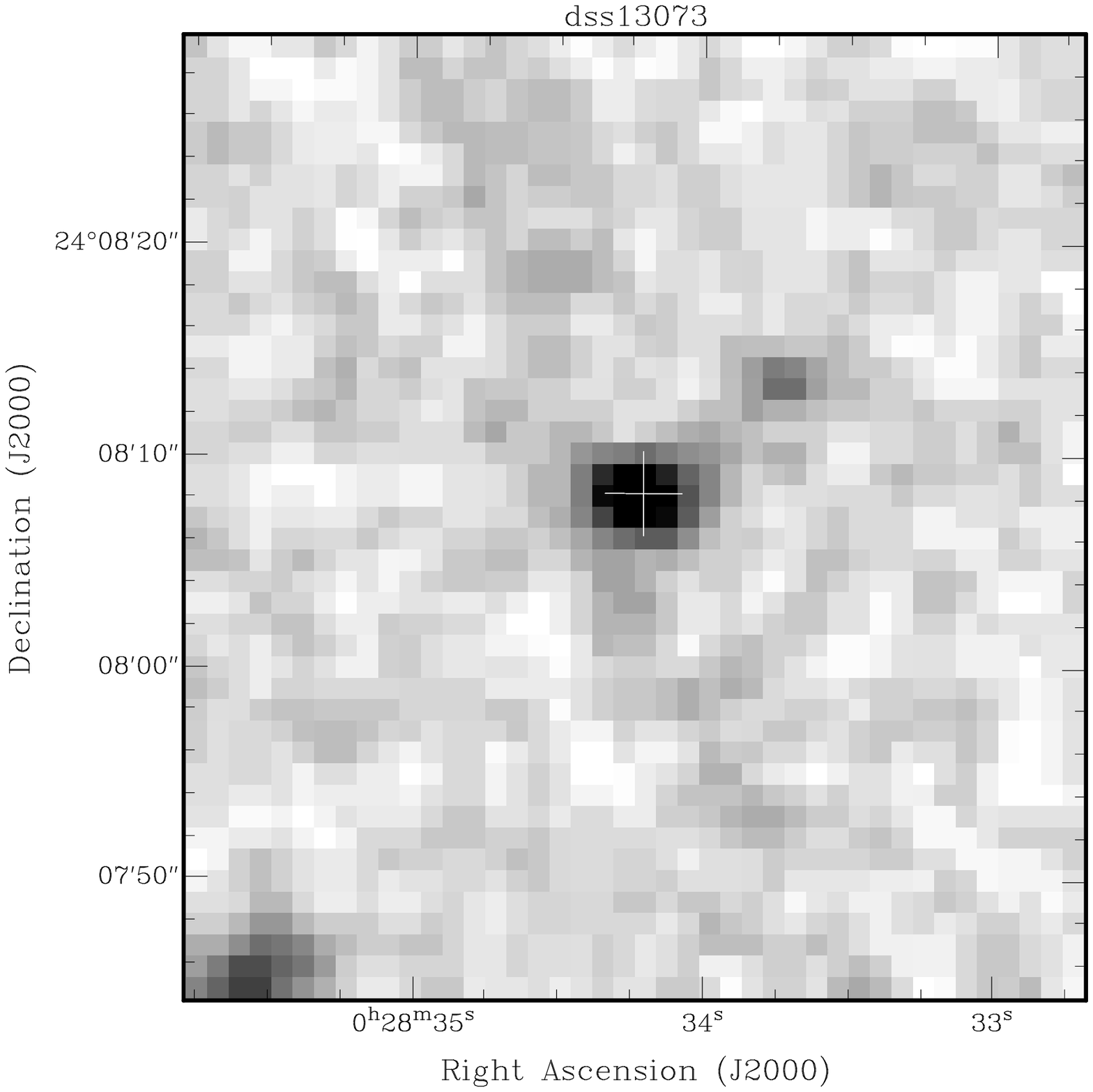 ,width=4.0cm,clip=}}\quad 
\subfigure[9CJ0029+3244 (P60 \it{R}\normalfont)]{\epsfig{figure=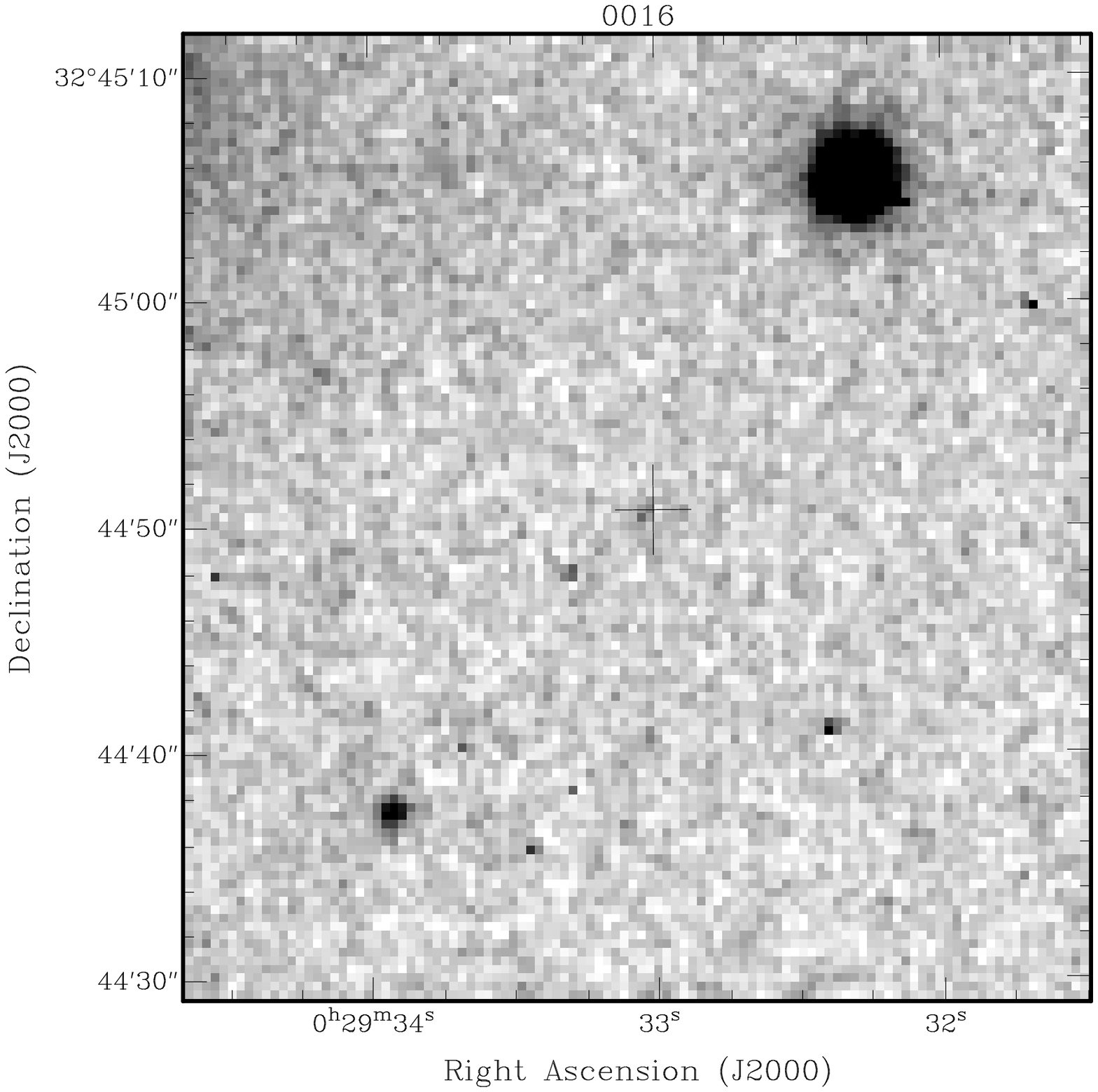 ,width=4.0cm,clip=}}
}\caption{ Optical counterparts for sources 9CJ0022+3250 to 9CJ0029+3244. Crosses mark maximum radio flux density and are 4\,arcsec top to bottom. Contours: \ref{o}, Detail showing 43\,GHz radio contours for this 0.25\,arcsec source: 60-90 every 10\,\% of peak (9.6\,mJy/beam); \ref{p}, 4.8\,GHz contours at 10-80 every 20\,\% of peak (26.2\,mJy/beam).}\end{figure*}

\begin{figure*}
\mbox{
\subfigure[9CJ0030+2957 (P60 \it{R}\normalfont)]{\epsfig{figure=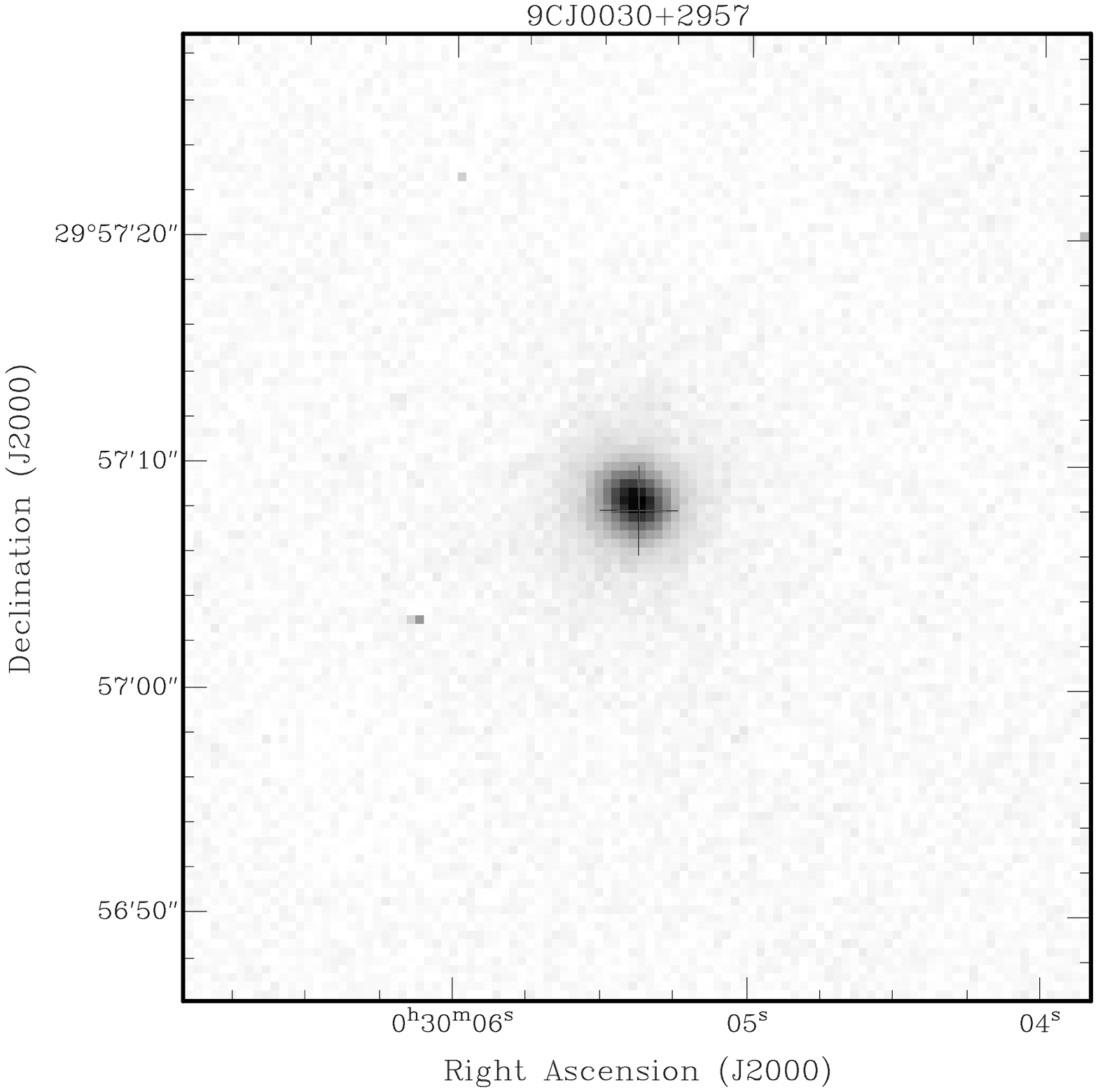 ,width=4.0cm,clip=}}\quad 
\subfigure[9CJ0030+3415 (P60 \it{R}\normalfont)]{\epsfig{figure=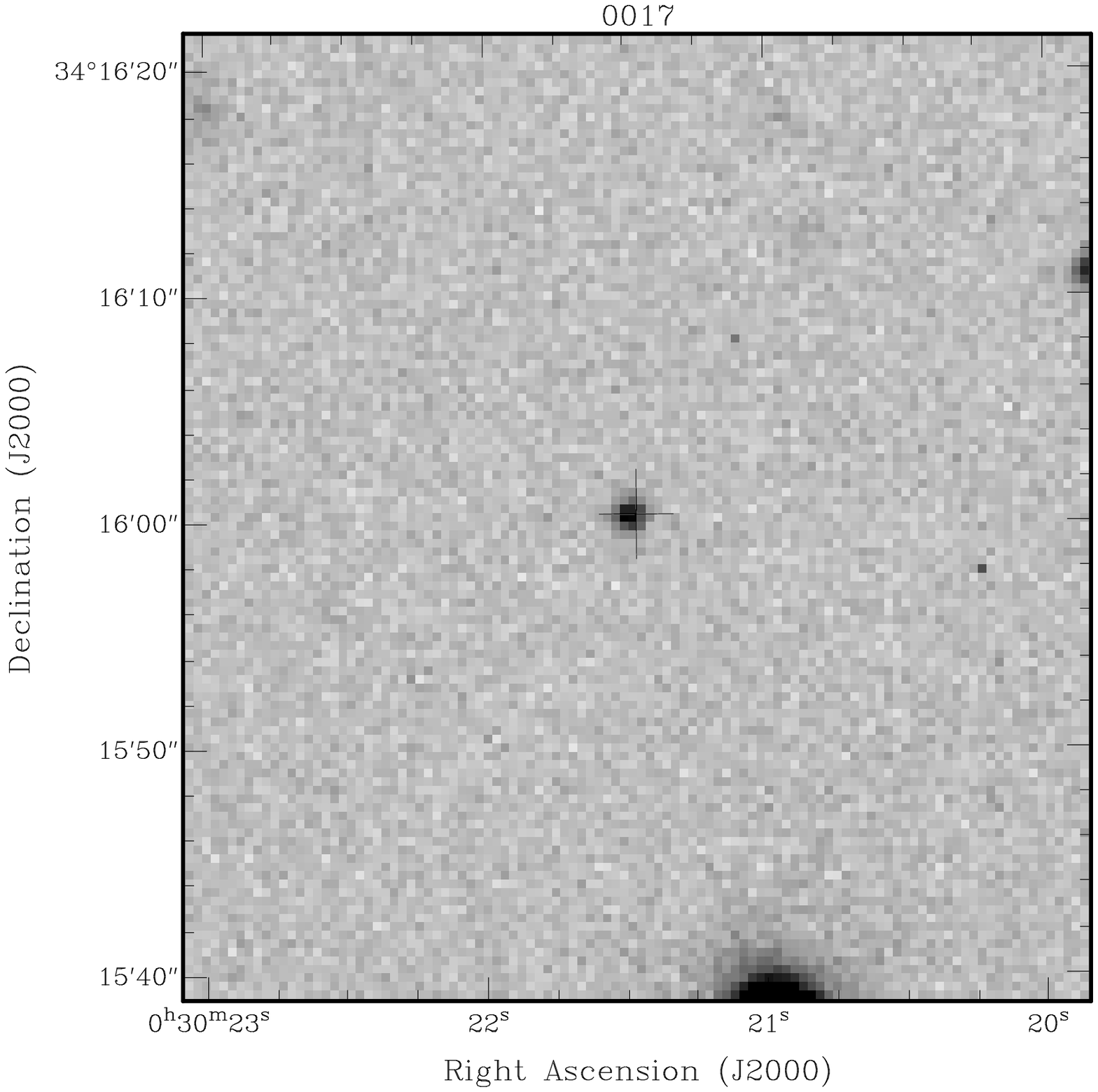 ,width=4.0cm,clip=}}\quad 
\subfigure[9CJ0030+2833 (P60 \it{R}\normalfont)]{\epsfig{figure=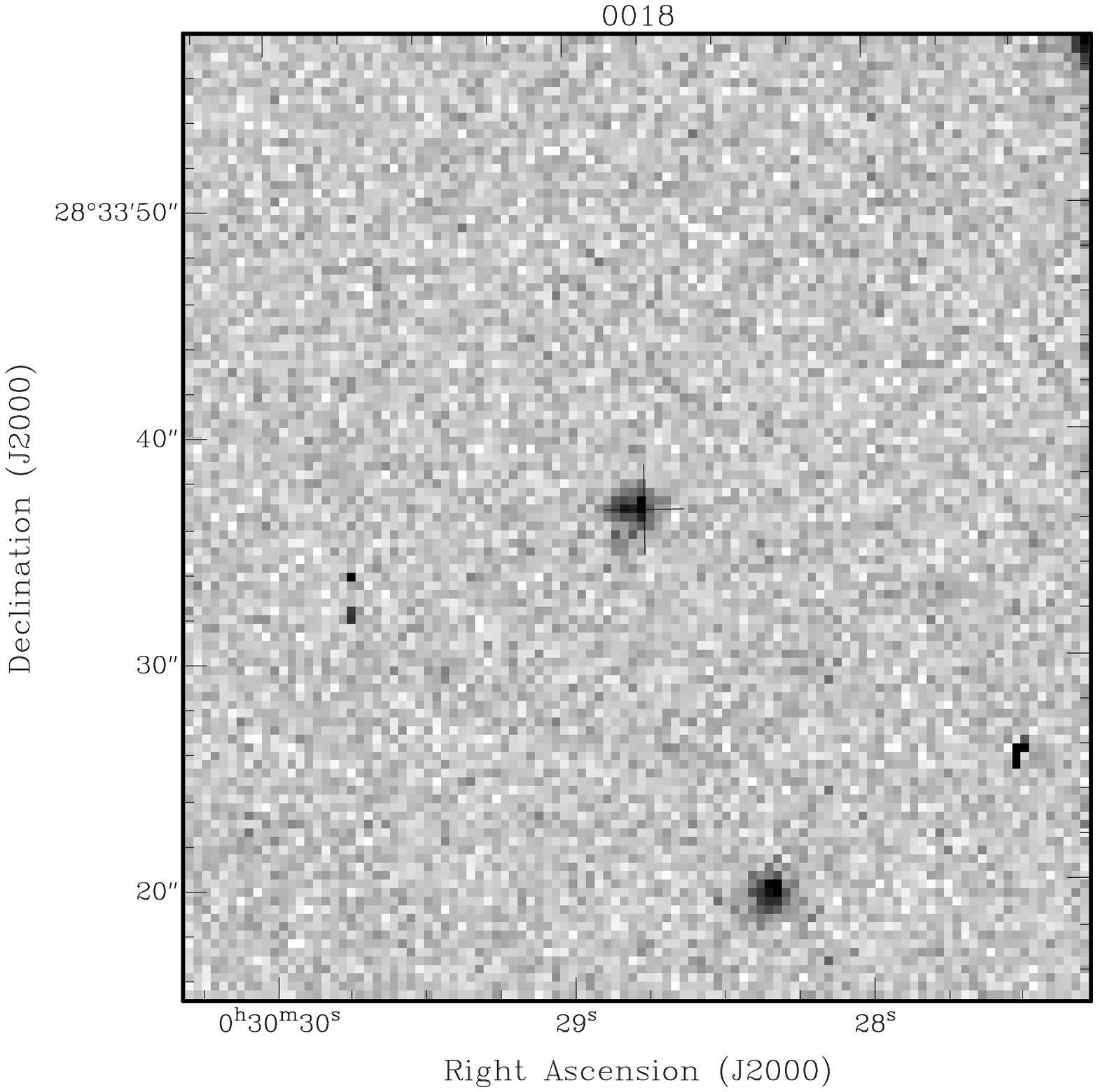 ,width=4.0cm,clip=}}
} 
\mbox{
\subfigure[9CJ0030+2809 (P60 \it{R}\normalfont)]{\epsfig{figure=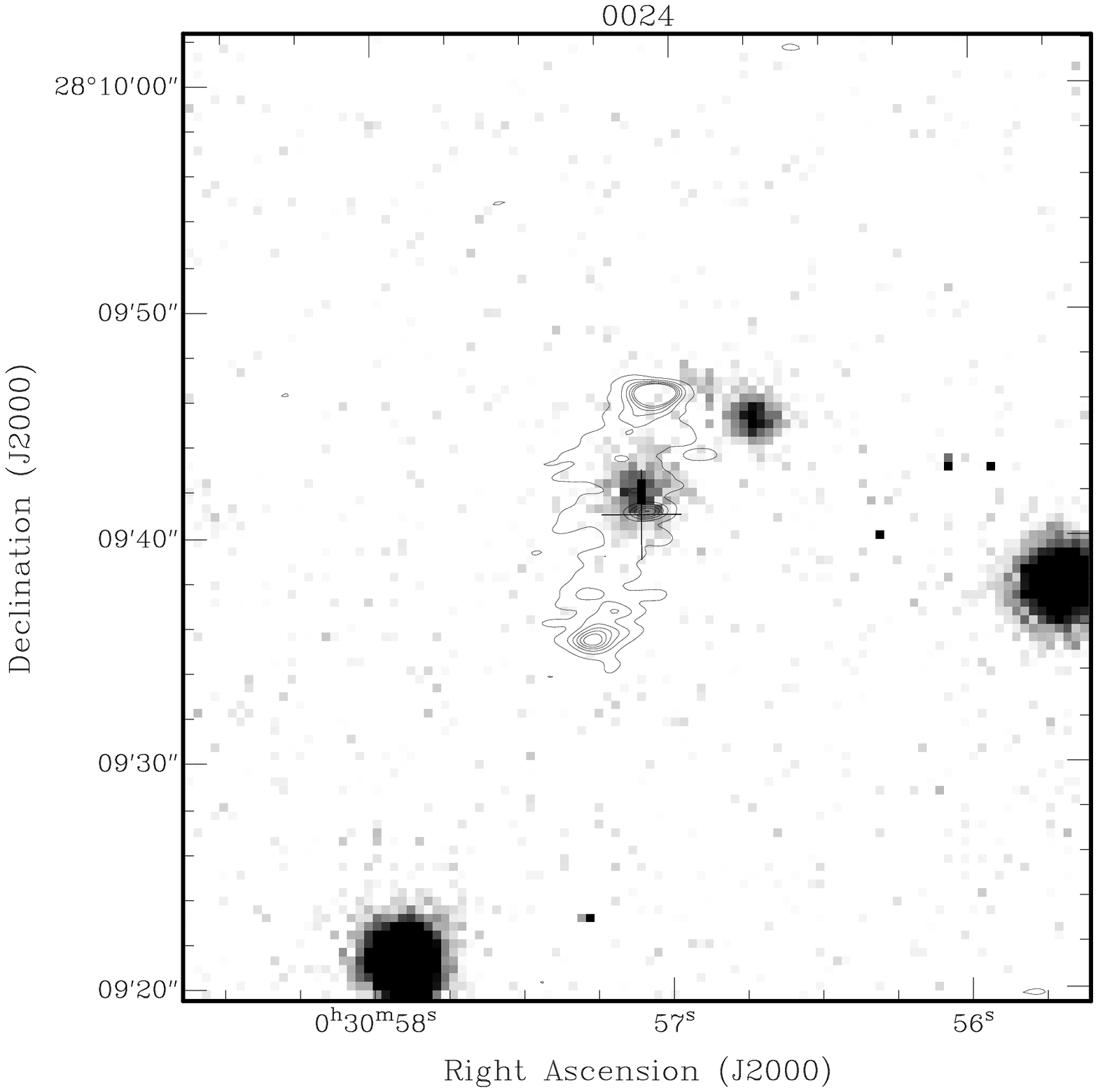 ,width=4.0cm,clip=}\label{q}}\quad 
\subfigure[9CJ0030+2809 (P60 \it{R}\normalfont) Detail]{\epsfig{figure=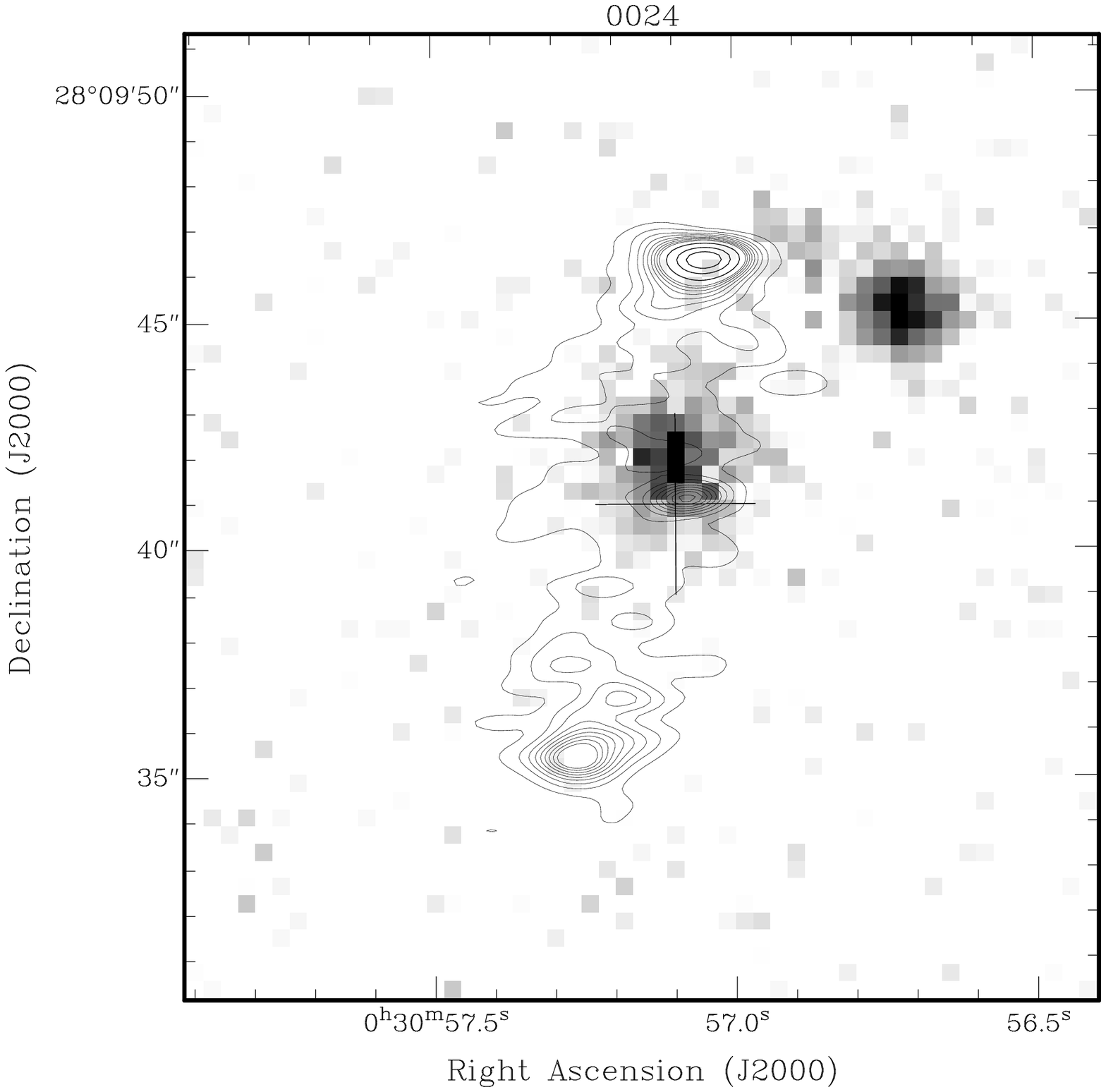 ,width=4.0cm,clip=}\label{r}}\quad 
\subfigure[9CJ0031+3016 (P60 \it{R}\normalfont)]{\epsfig{figure=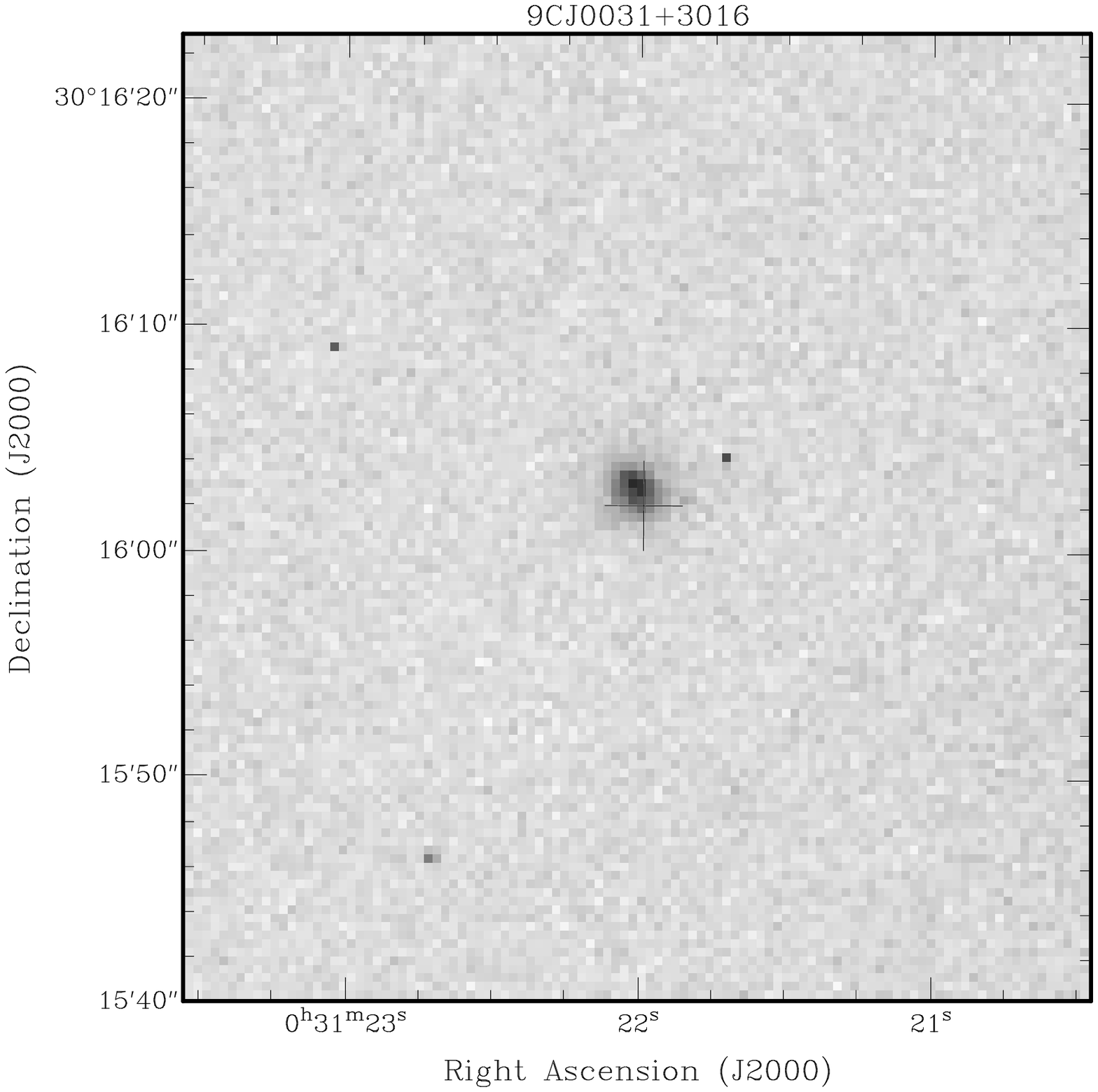 ,width=4.0cm,clip=}}
} 
\mbox{
\subfigure[9CJ0032+2758 (P60 \it{R}\normalfont)]{\epsfig{figure=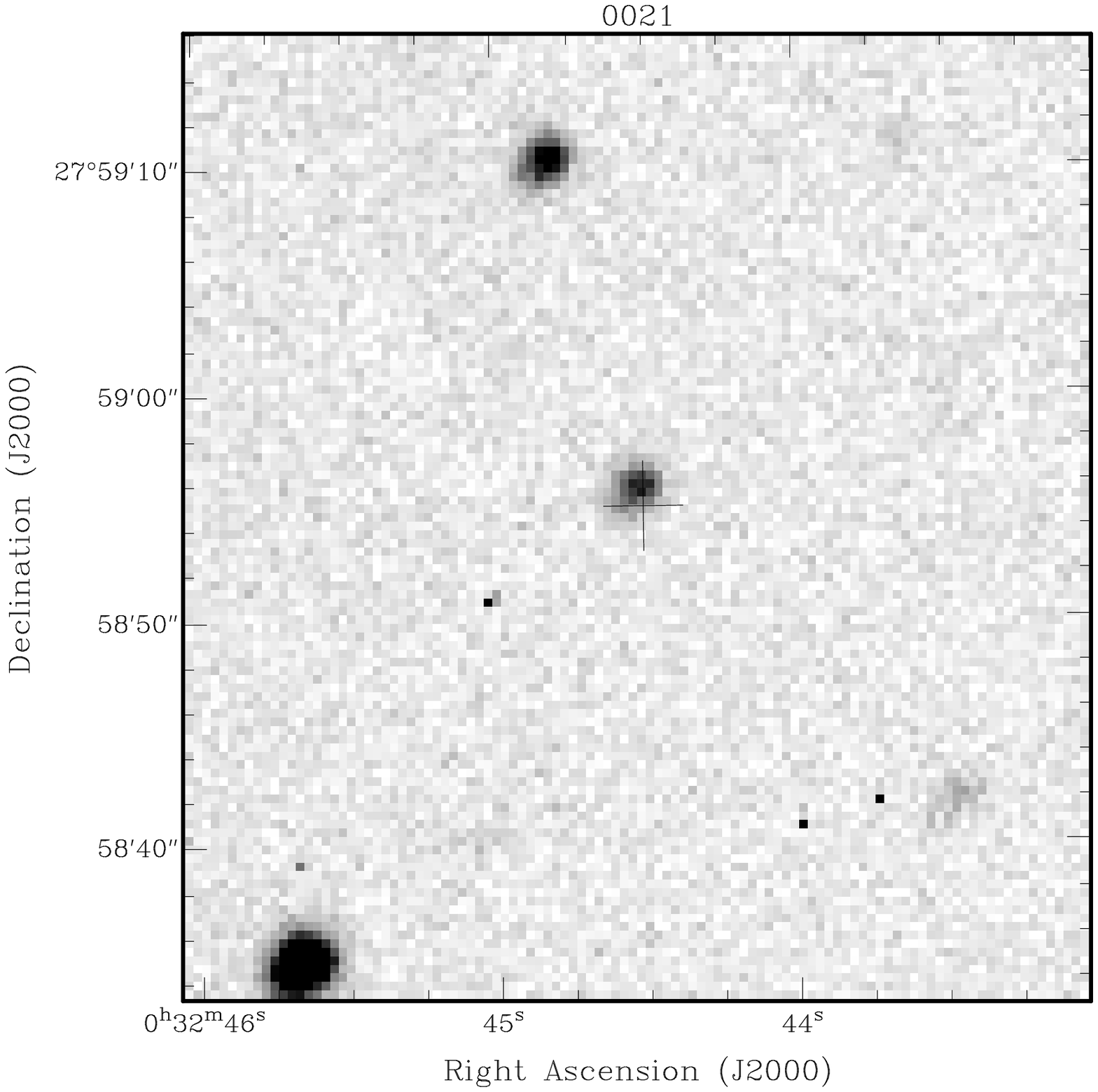 ,width=4.0cm,clip=}}\quad 
\subfigure[9CJ0033+2752 (DSS2 \it{R}\normalfont)]{\epsfig{figure=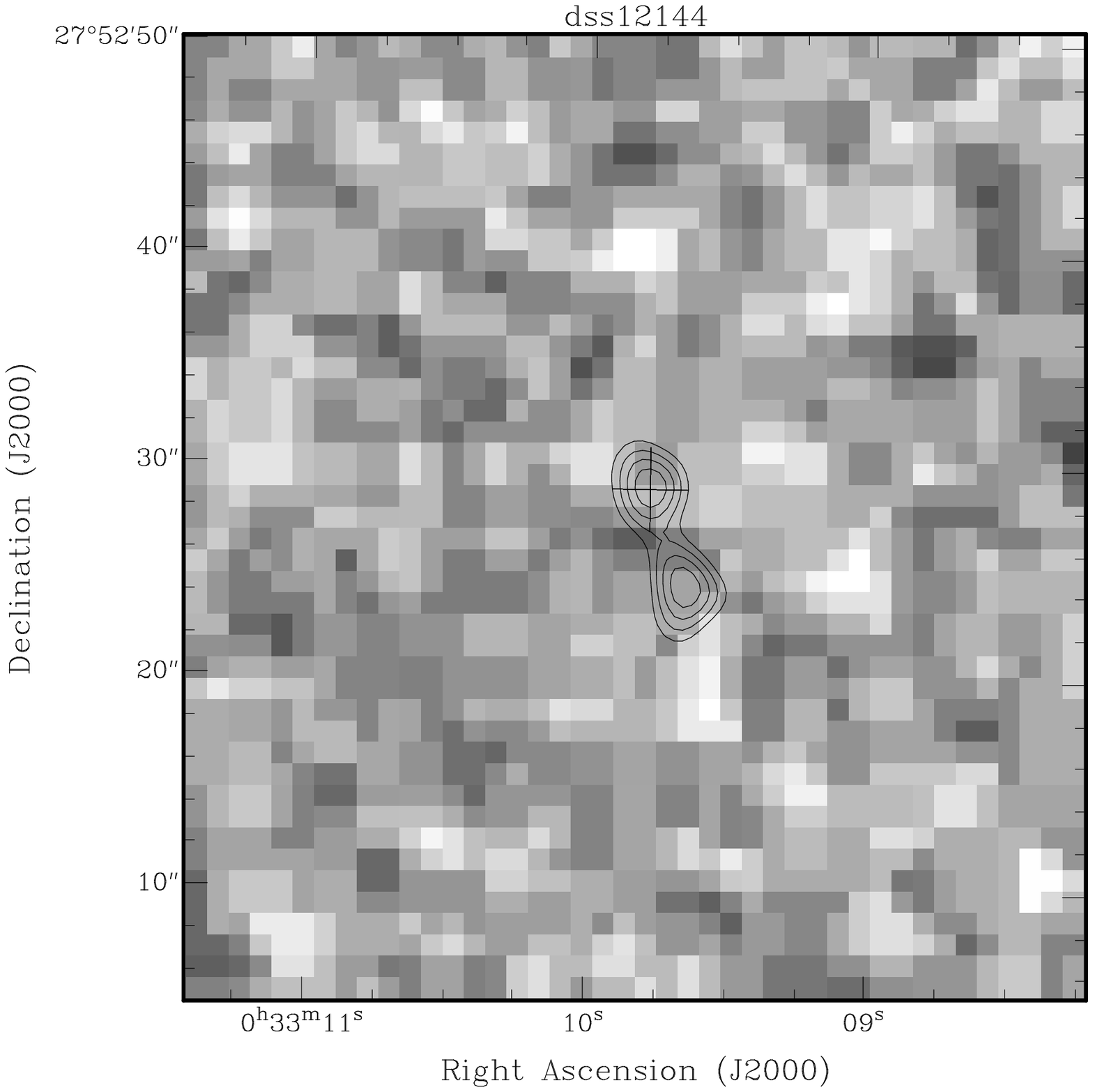 ,width=4.0cm,clip=}\label{s}}\quad 
\subfigure[ 9CJ0034+2754 (DSS2 \it{R}\normalfont)]{\epsfig{figure=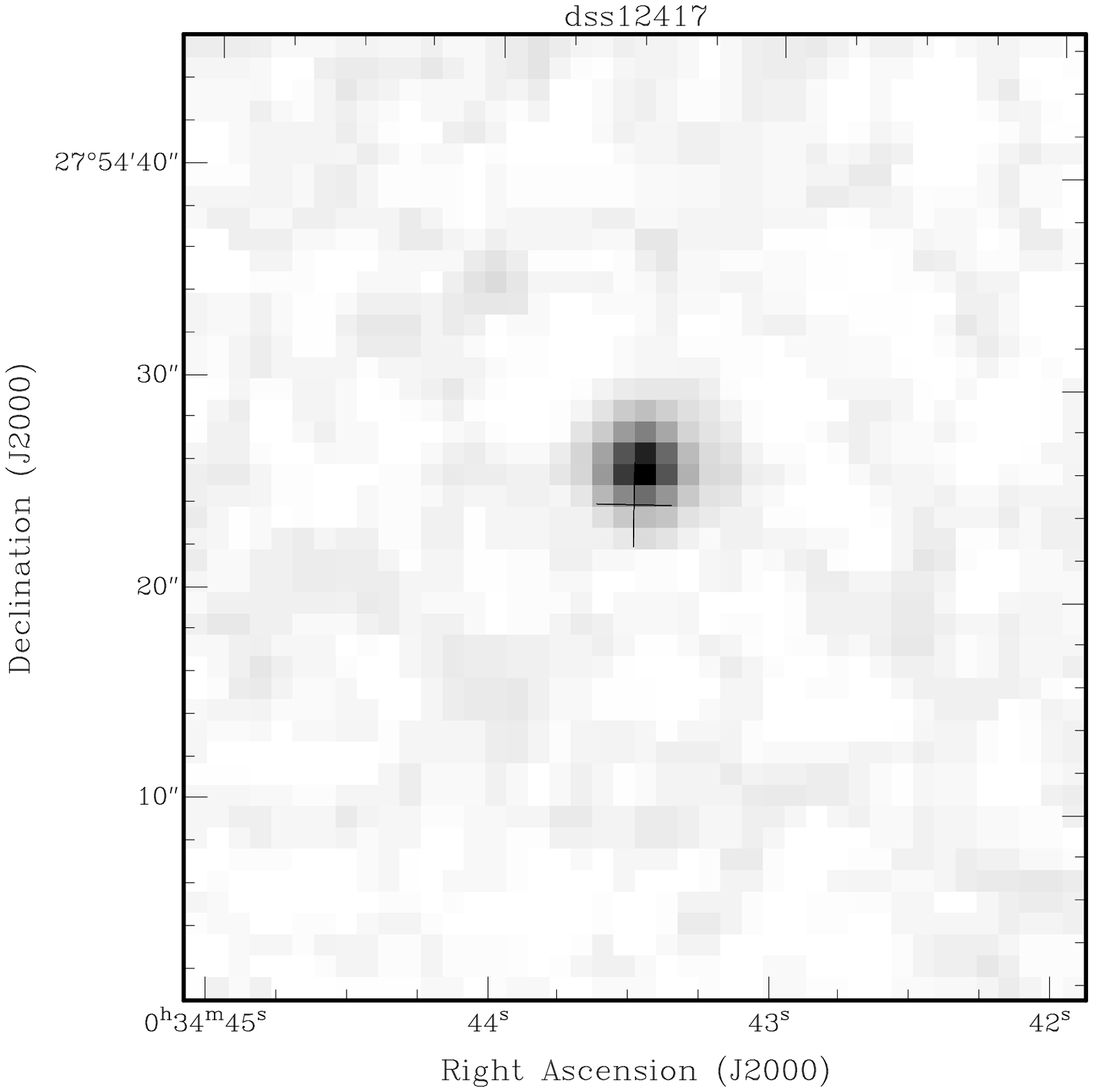 ,width=4.0cm,clip=}}
} 
\mbox{
\subfigure[9CJ0036+2620 (P60 \it{R}\normalfont)]{\epsfig{figure=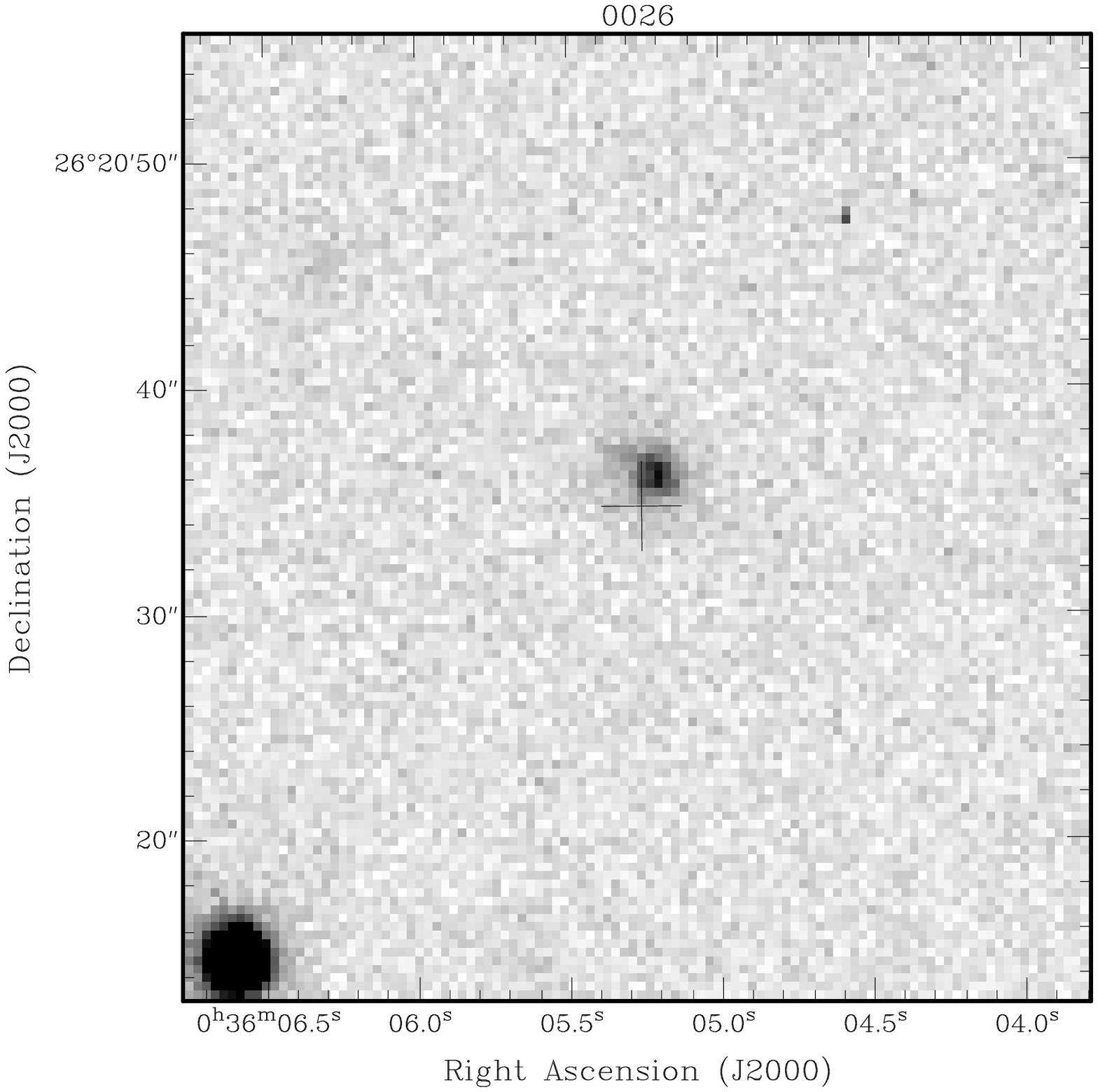 ,width=4.0cm,clip=}}\quad 
\subfigure[9CJ0917+3446 (P60 \it{R}\normalfont)]{\epsfig{figure=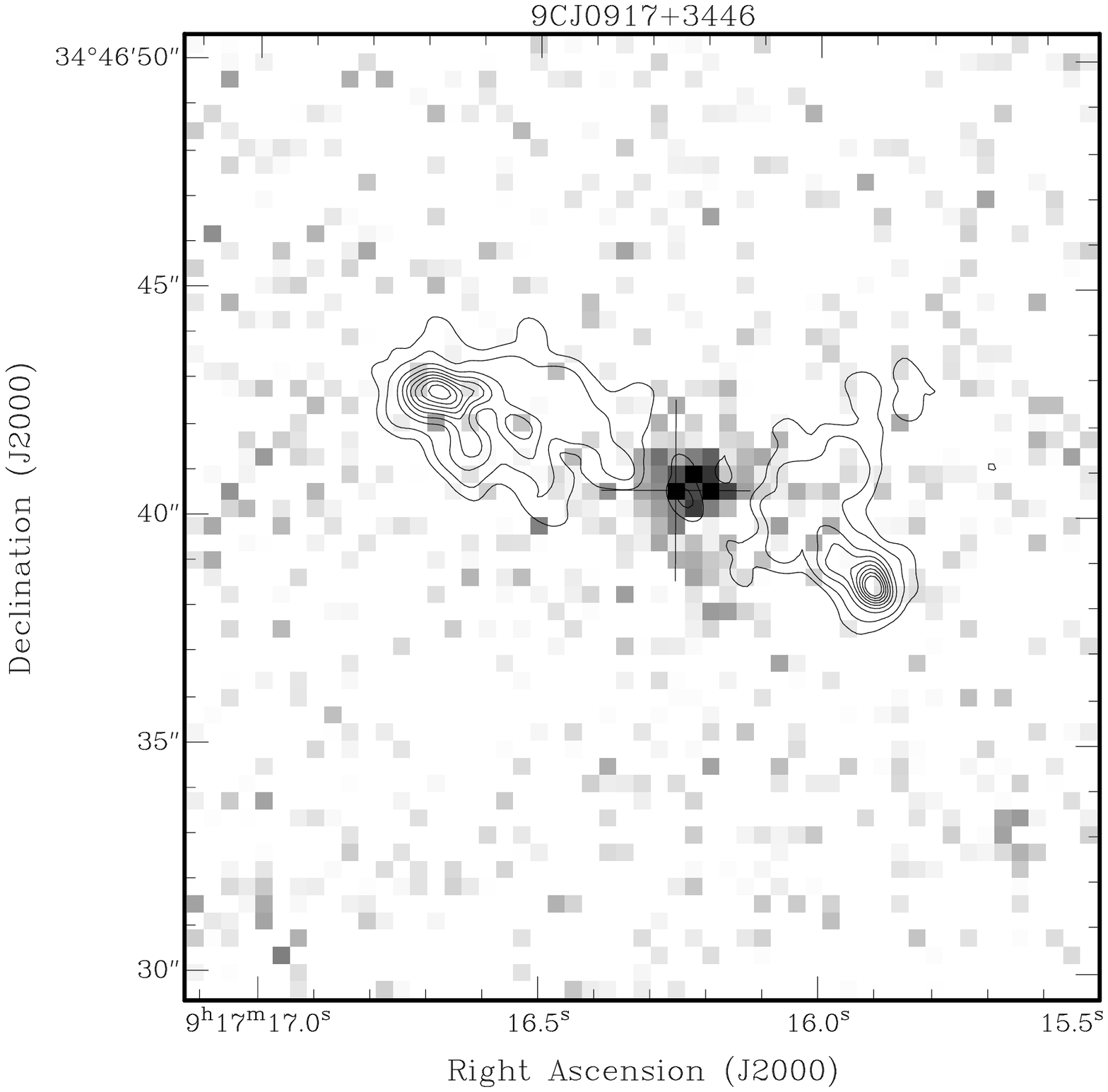 ,width=4.0cm,clip=}\label{t}}\quad 
\subfigure[9CJ0919+3324 (P60 \it{R}\normalfont)]{\epsfig{figure=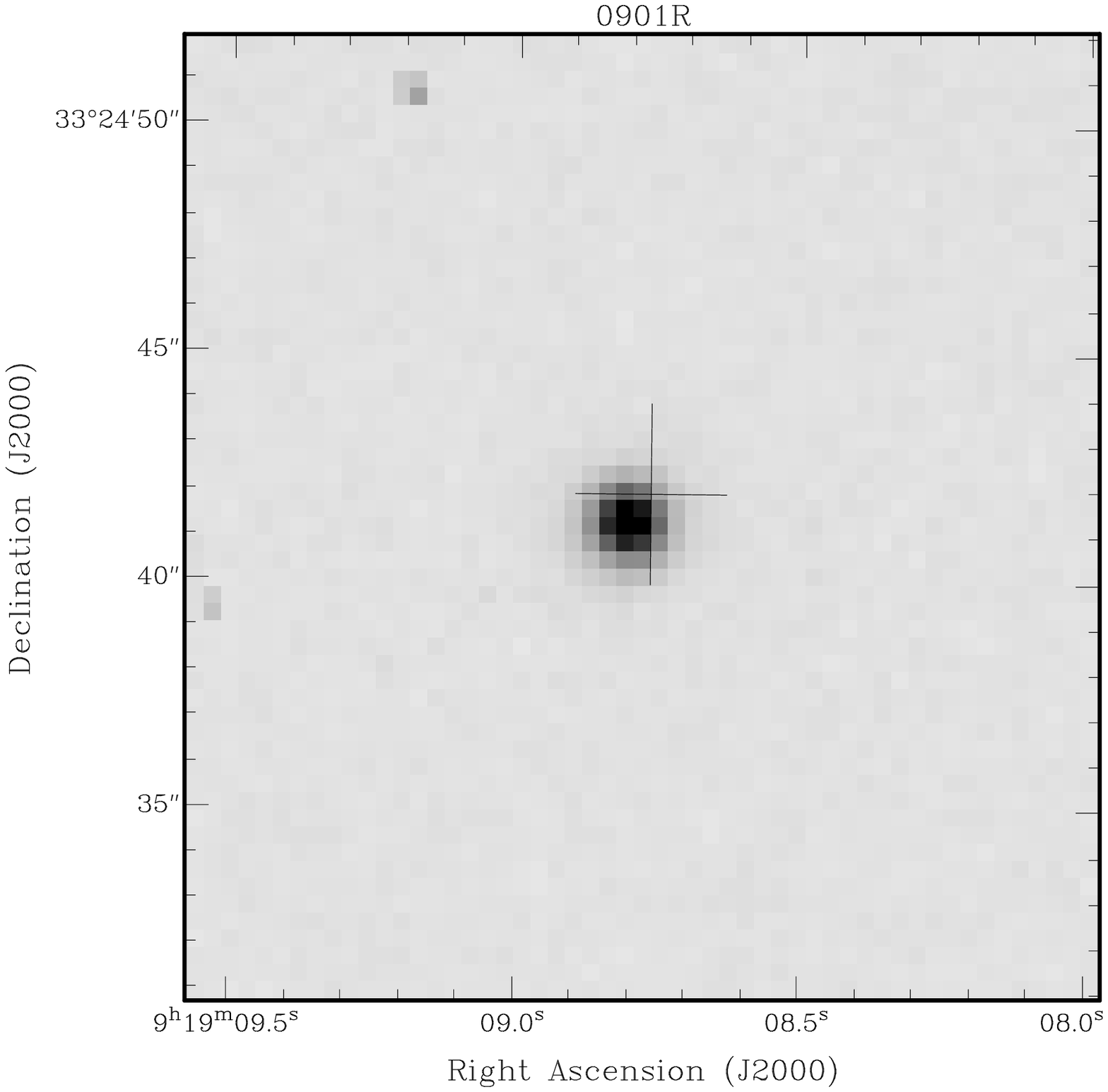 ,width=4.0cm,clip=}}
}\caption{Optical counterparts for sources 9CJ0030+2957 to 9CJ0919+3324. Crosses mark maximum radio flux density and are 4\,arcsec top to bottom.Contours: \ref{q}, 4.8\,GHz contours 5-30 every 5\,\% of peak (18.4\,mJy/beam); \ref{r}, 4.8\,GHz contours 5-30 every 3\,\% and 40-80 every 20\,\% of peak; \ref{s}, 22\,GHz contours 60-90 every 10\,\% of peak (8.1\,mJy/beam); \ref{t}, 4.8\,GHz contours at 6\,\% and 10-85 every 15\,\% of peak (23.1\,mJy/beam).}\end{figure*}

\begin{figure*}
\mbox{
\subfigure[9CJ0923+3107 (P60 \it{R}\normalfont)]{\epsfig{figure=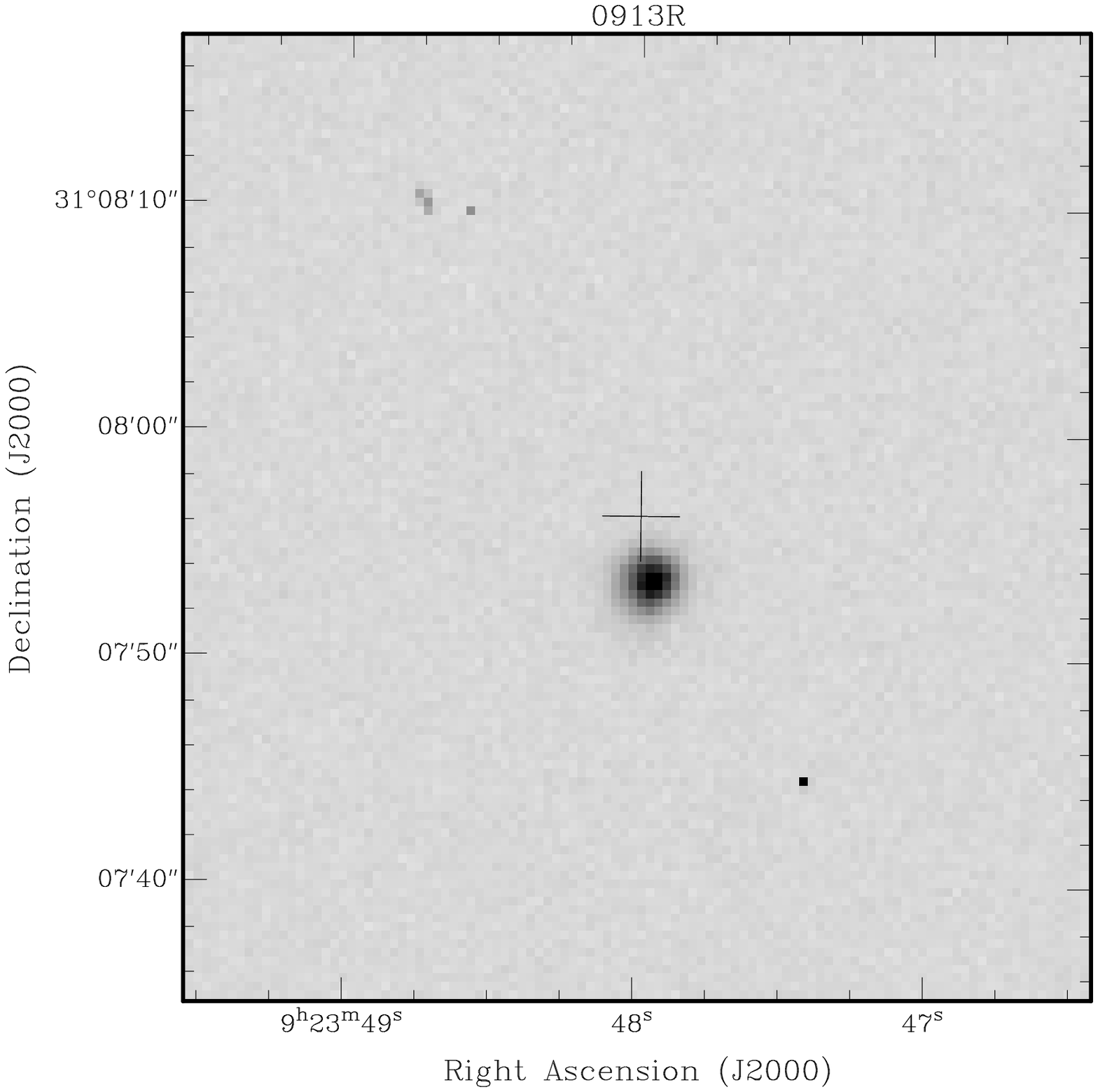 ,width=4.0cm,clip=}}\quad 
\subfigure[9CJ0923+2815 (P60 \it{R}\normalfont)]{\epsfig{figure=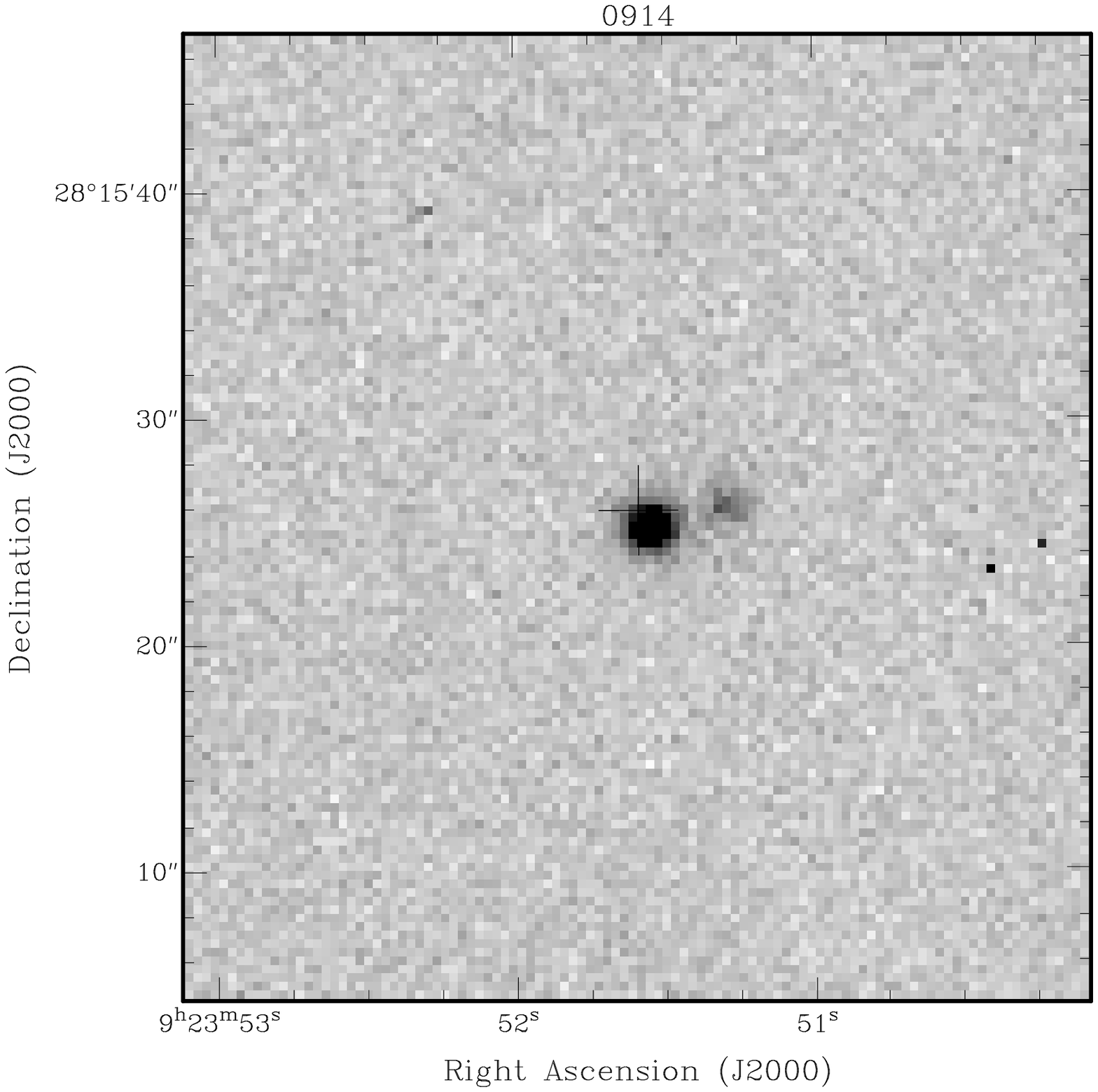 ,width=4.0cm,clip=}}\quad 
\subfigure[9CJ0925+3159 (P60 \it{R}\normalfont)]{\epsfig{figure=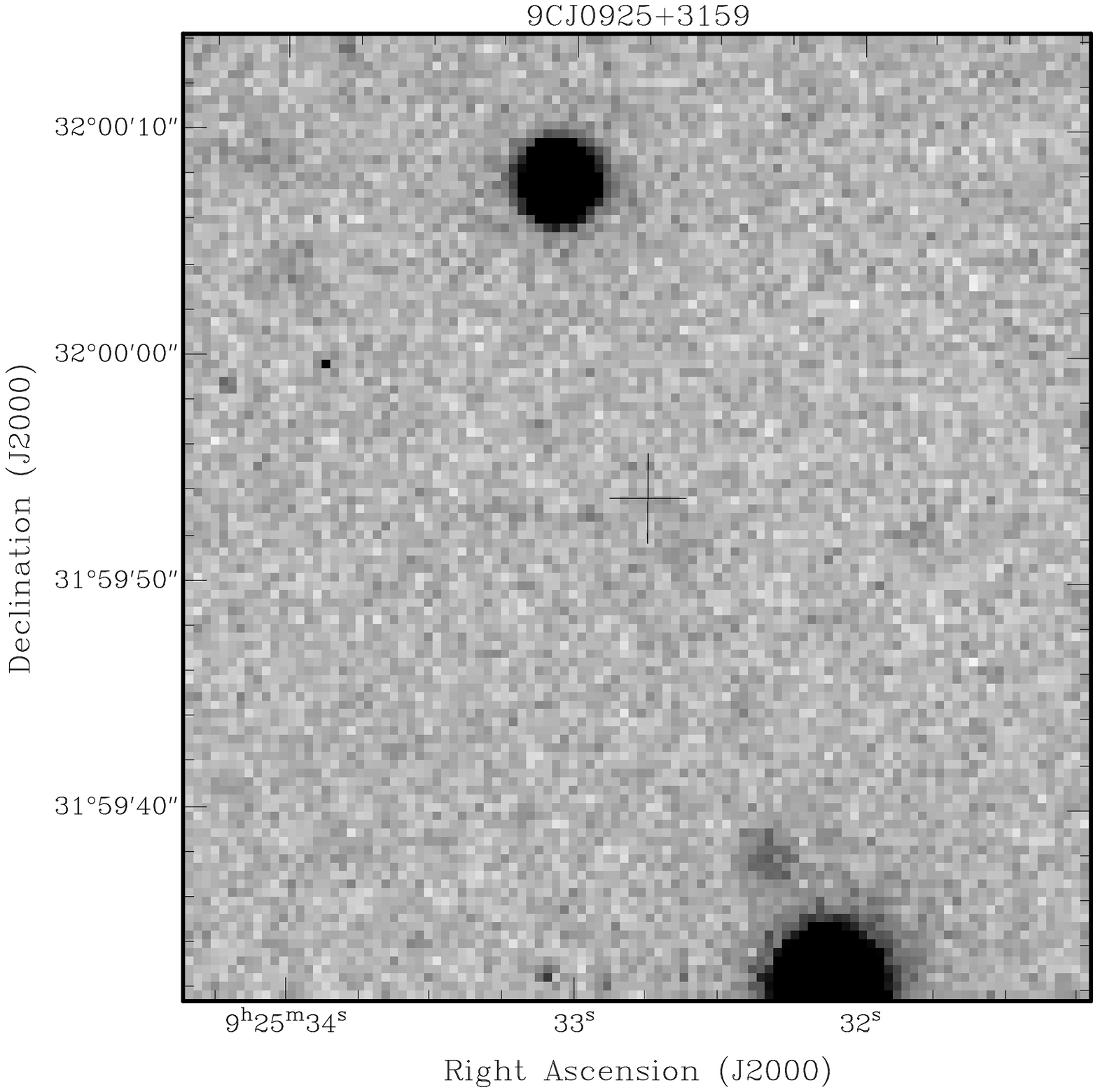 ,width=4.0cm,clip=}\label{u}}
}
\mbox{
\subfigure[9CJ0925+3159 (P60 \it{R}\normalfont) Detail]{\epsfig{figure=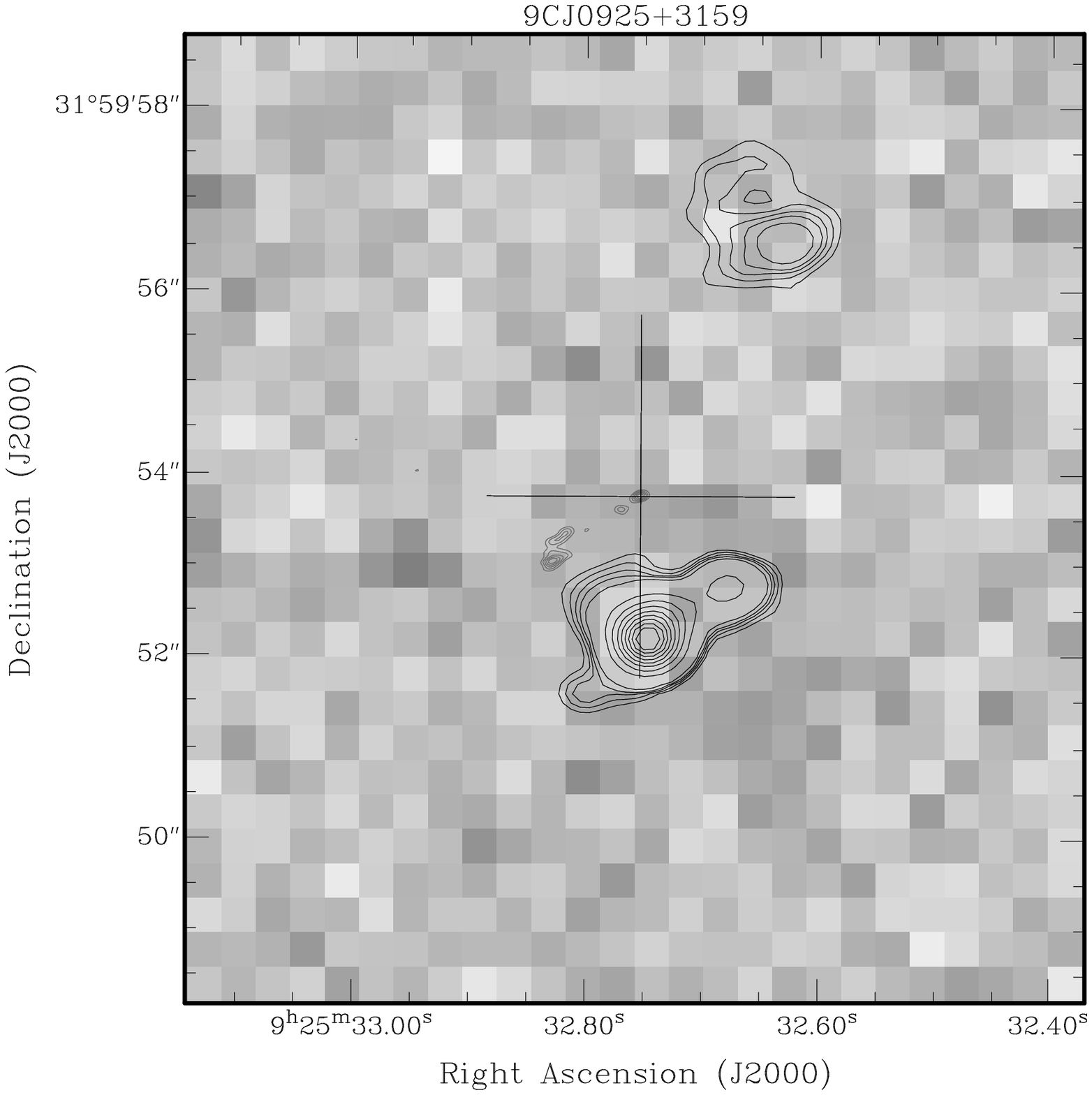 ,width=4.0cm,clip=}\label{v}}\quad 
\subfigure[9CJ0925+3127 (P60 \it{R}\normalfont)]{\epsfig{figure=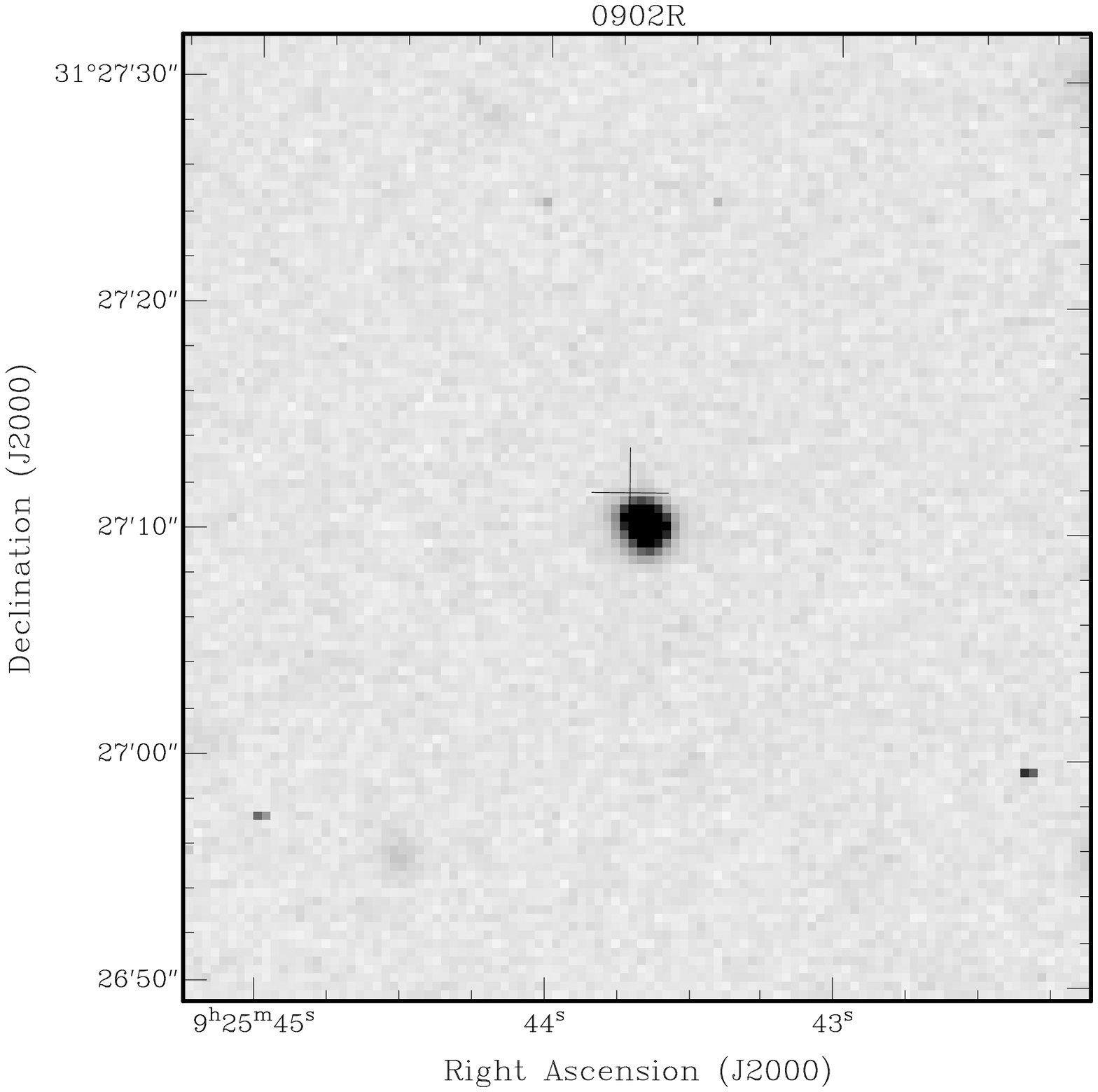 ,width=4.0cm,clip=}}\quad
\subfigure[9CJ0926+2758 (P60 \it{R}\normalfont)]{\epsfig{figure=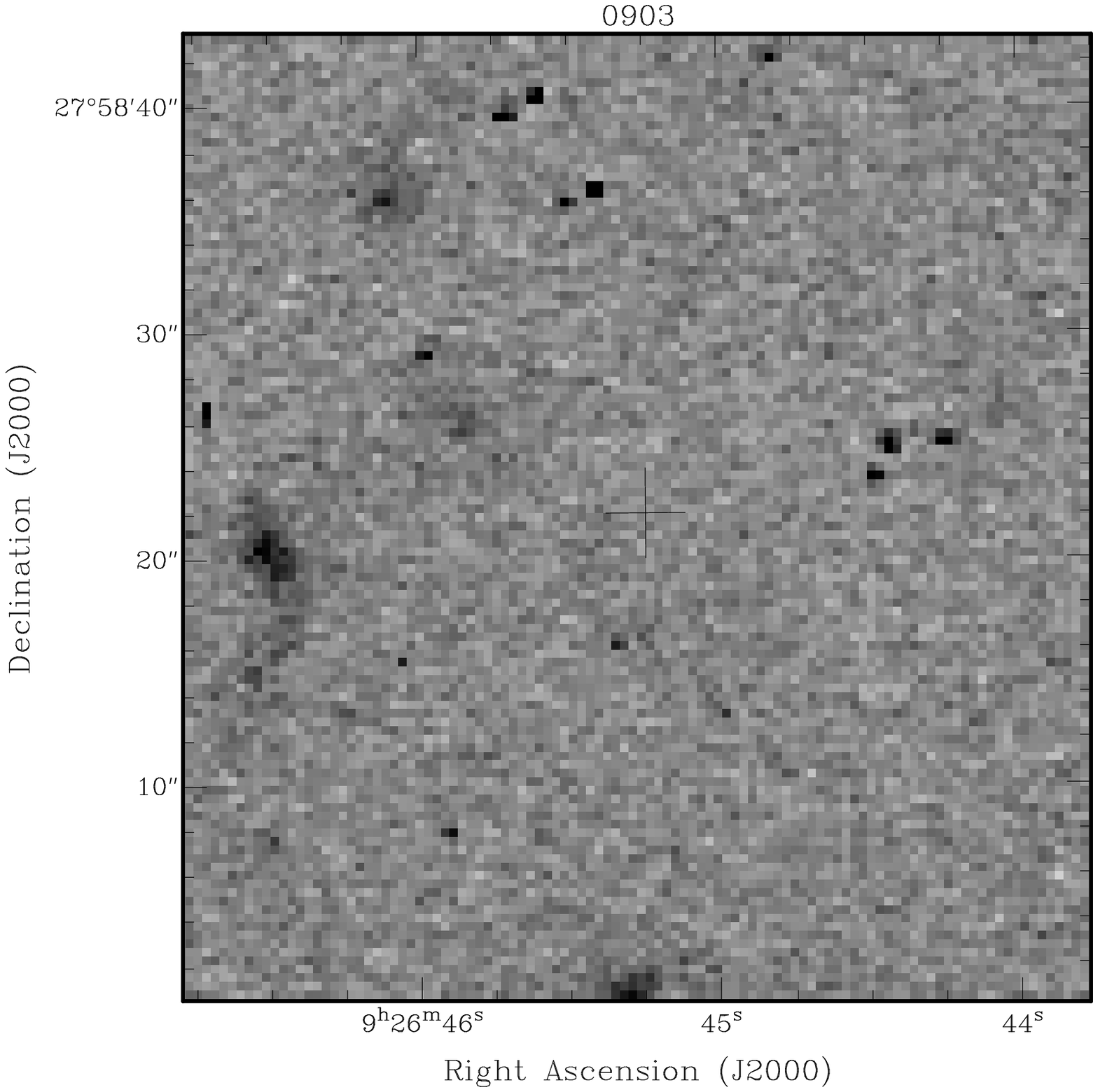 ,width=4.0cm,clip=}}
}
\mbox{
\subfigure[9CJ0927+2954 (P60 \it{R}\normalfont)]{\epsfig{figure=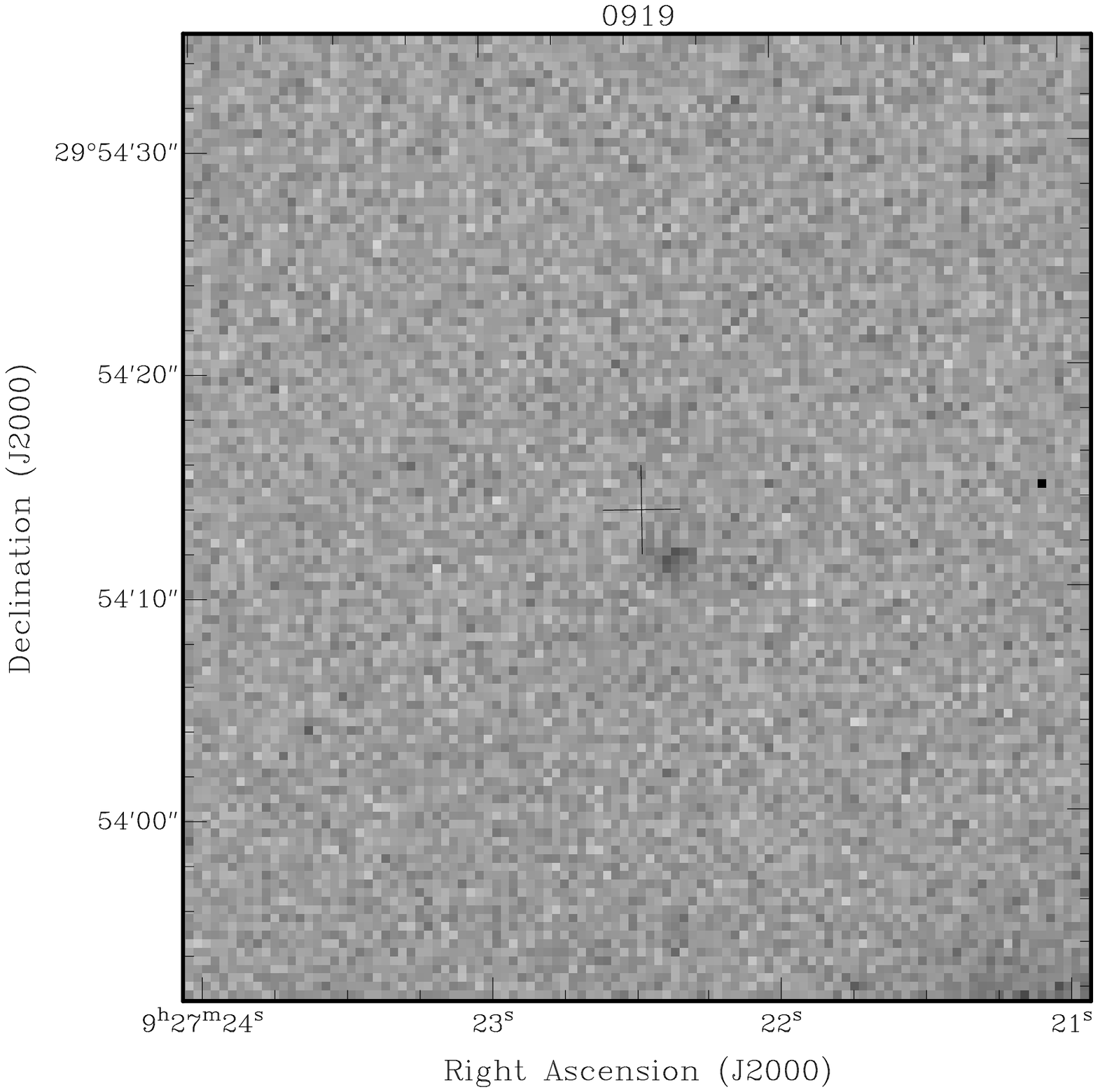 ,width=4.0cm,clip=}}\quad 
\subfigure[9CJ0927+3034 (DSS2 \it{R}\normalfont)]{\epsfig{figure=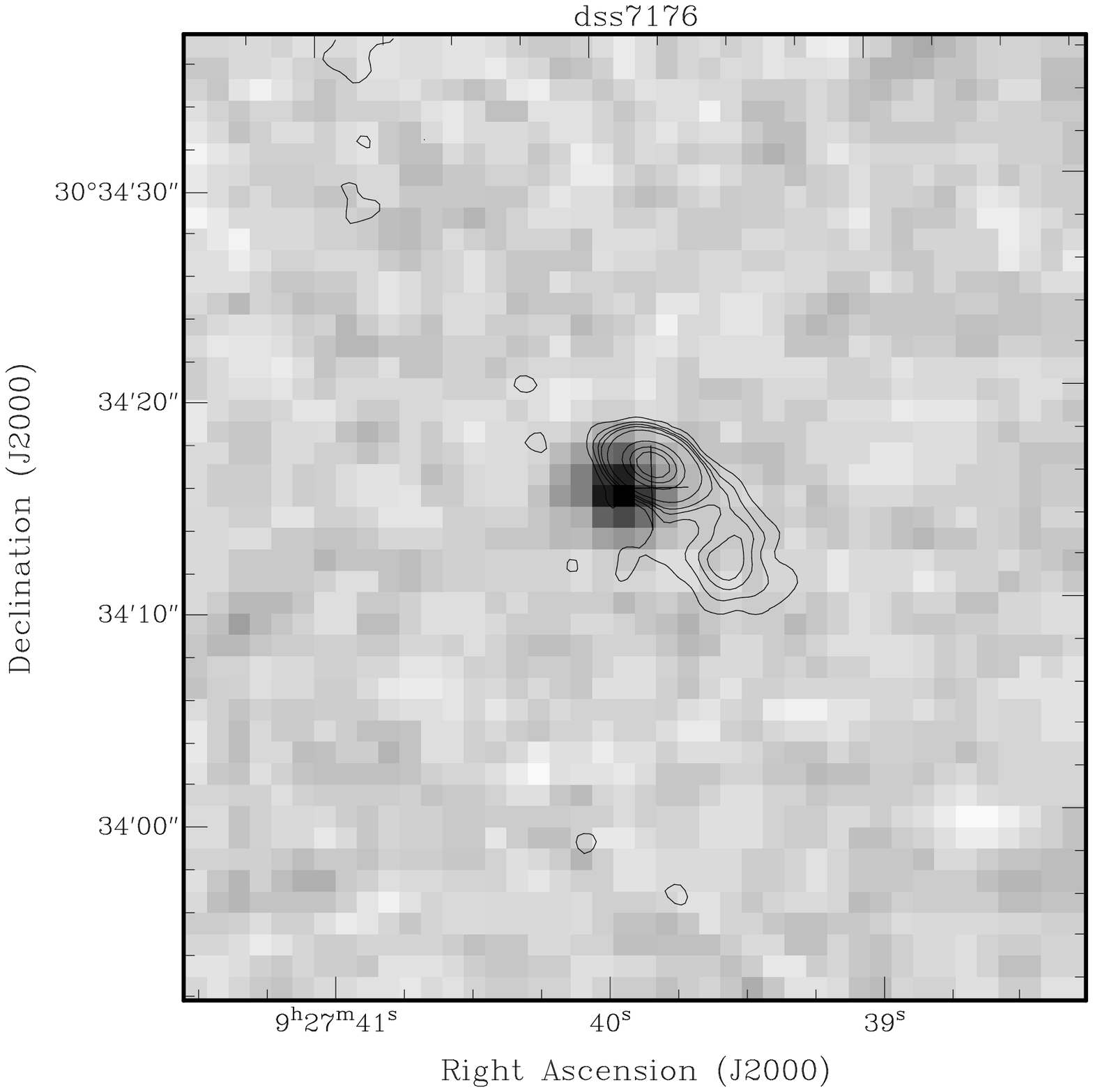 ,width=4.0cm,clip=}\label{w}}\quad 
\subfigure[9CJ0928+2904 (P60 \it{R}\normalfont)]{\epsfig{figure=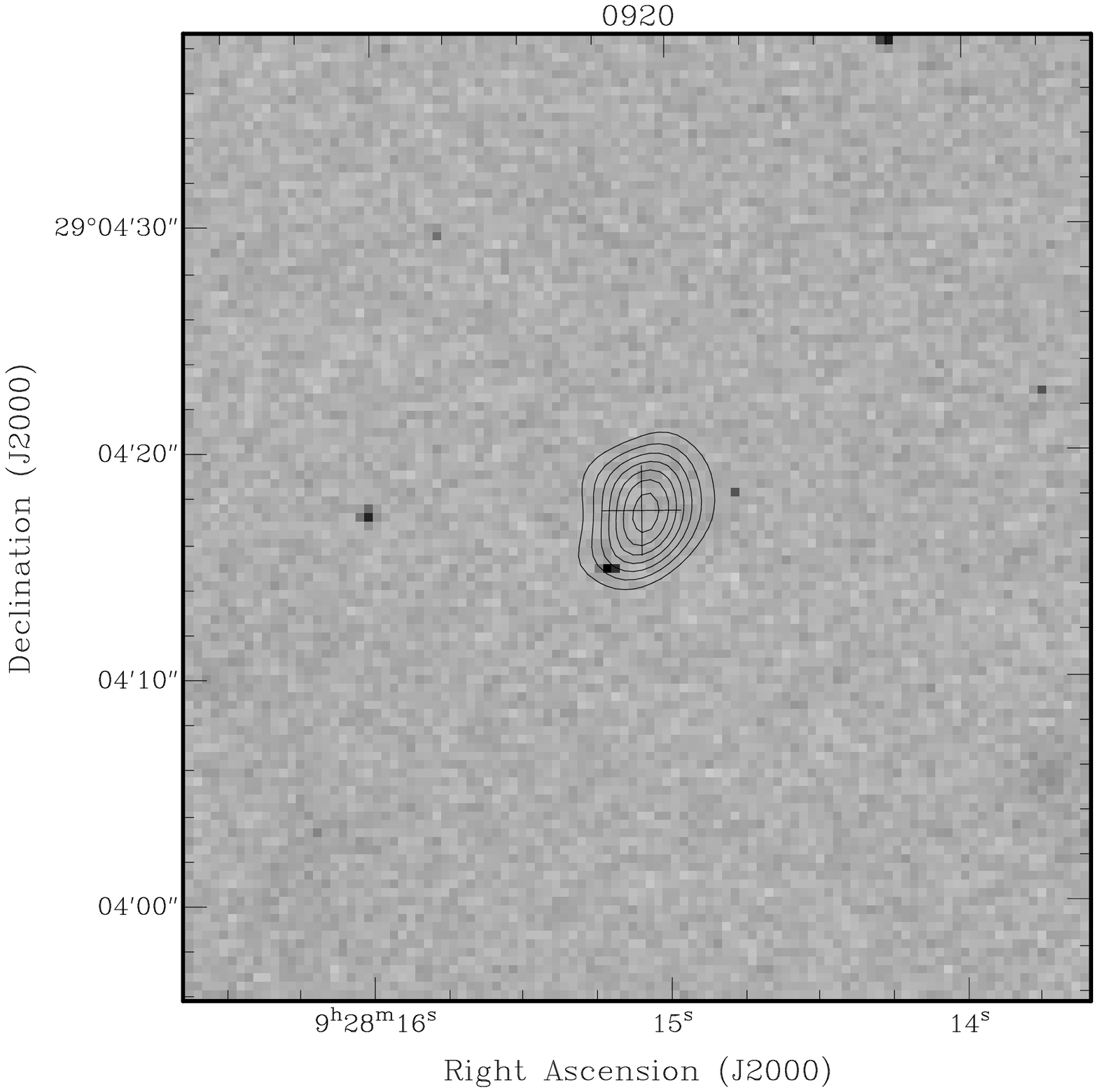 ,width=4.0cm,clip=}\label{x}}
}
\mbox{
\subfigure[9CJ0930+3503 (P60 \it{R}\normalfont)]{\epsfig{figure=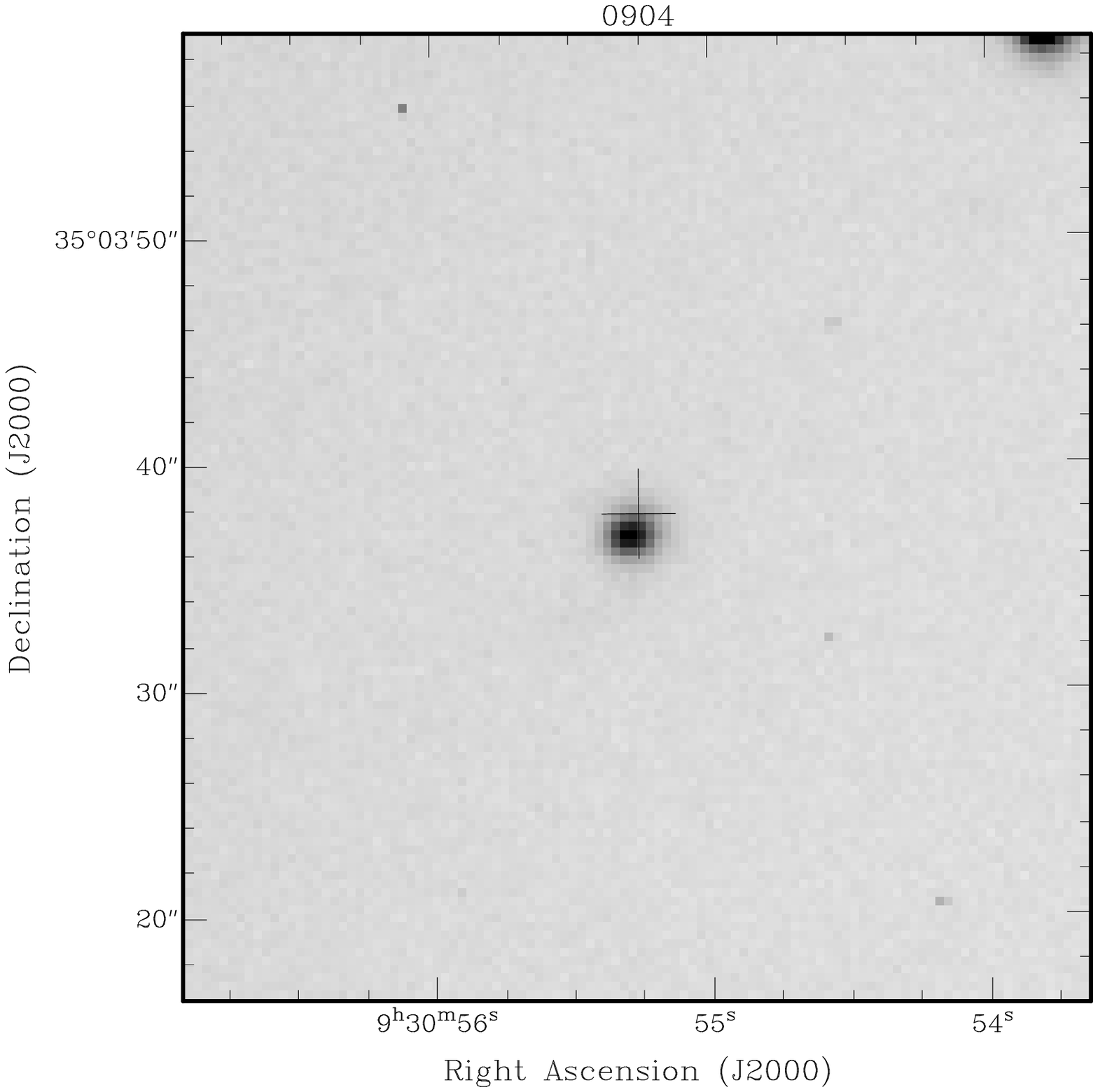 ,width=4.0cm,clip=}}\quad 
\subfigure[9CJ0931+2750 (P60 \it{R}\normalfont)]{\epsfig{figure=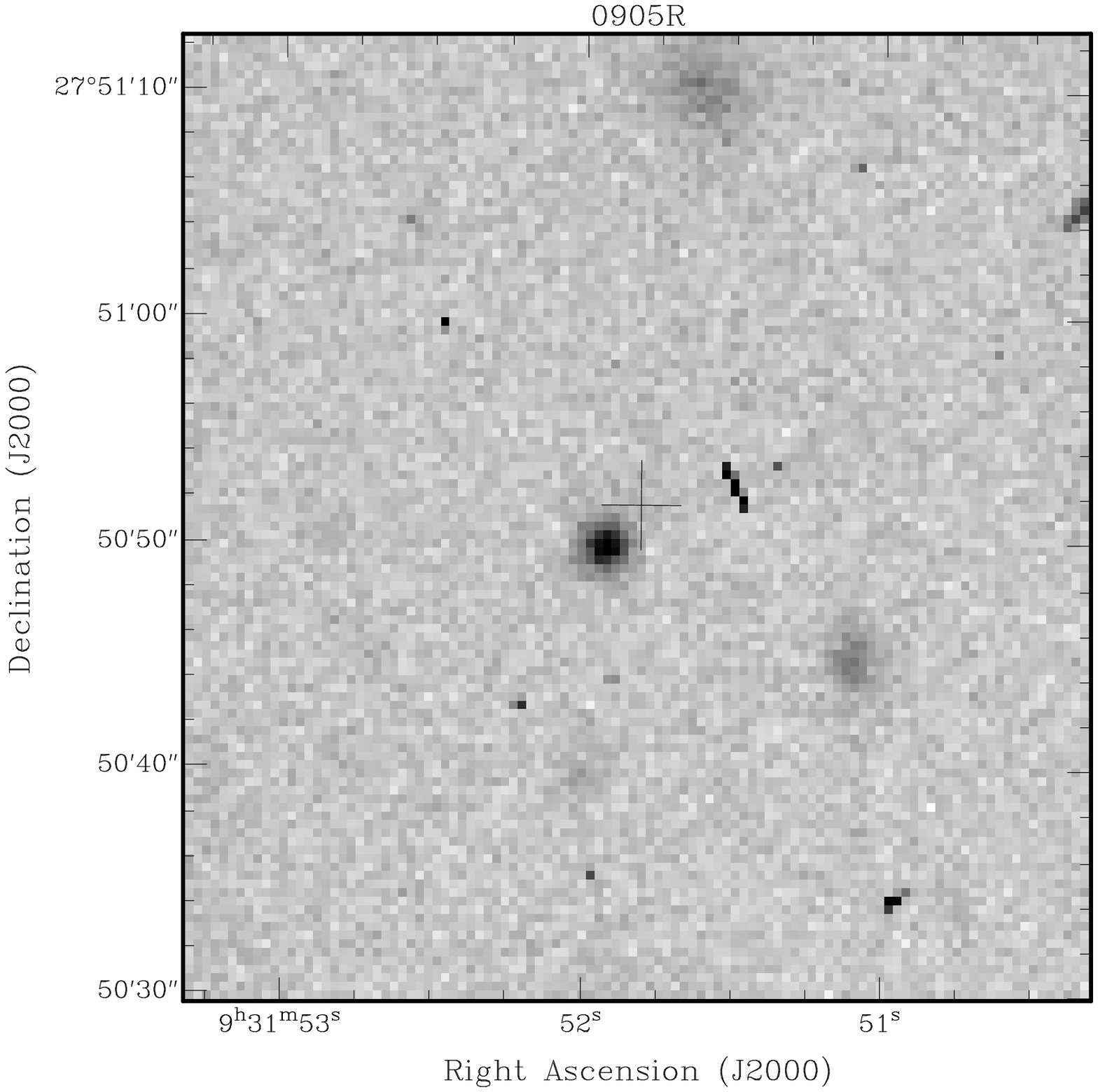 ,width=4.0cm,clip=}}\quad 
\subfigure[9CJ0932+2837 (P60 \it{R}\normalfont)]{\epsfig{figure=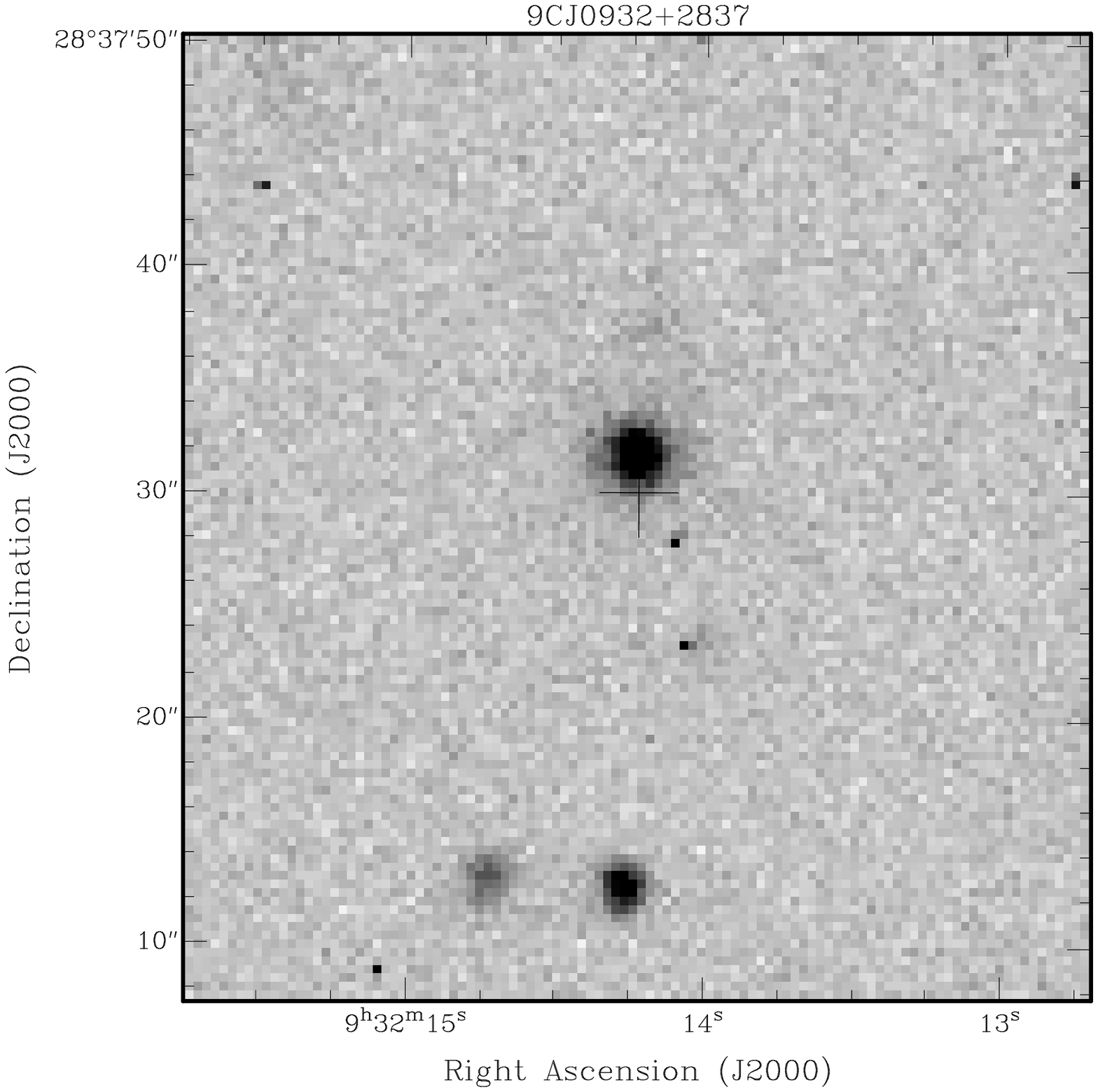 ,width=4.0cm,clip=}}}
\caption{ Optical counterparts for sources 9CJ0923+3107 to 9CJ0932+2837. Crosses mark maximum radio flux density and are 4\,arcsec top to bottom. Contours: \ref{v}, close-up showing radio contours at 4.8\,GHz (black, larger lobes -- 2.5-5 every 0.5\,\% and 10-90 every 10\,\% of peak (84.9\,mJy/beam)) and 22\,GHz (grey, above and to the left of the lower lobe -- 60-90 every 10\,\% of peak (8.1\,mJy/beam)); \ref{w}, 1.4\,GHz contours 4,5,6,7 and 10-70 every 20\,\% of peak (41.1\,mJy/beam); \ref{x}, 22\,GHz contours 35-95 every 10\,\% of peak (11.8\,mJy/beam).}\end{figure*}

\begin{figure*}
\mbox{
\subfigure[9CJ0932+3339 (DSS2 \it{R}\normalfont)]{\epsfig{figure=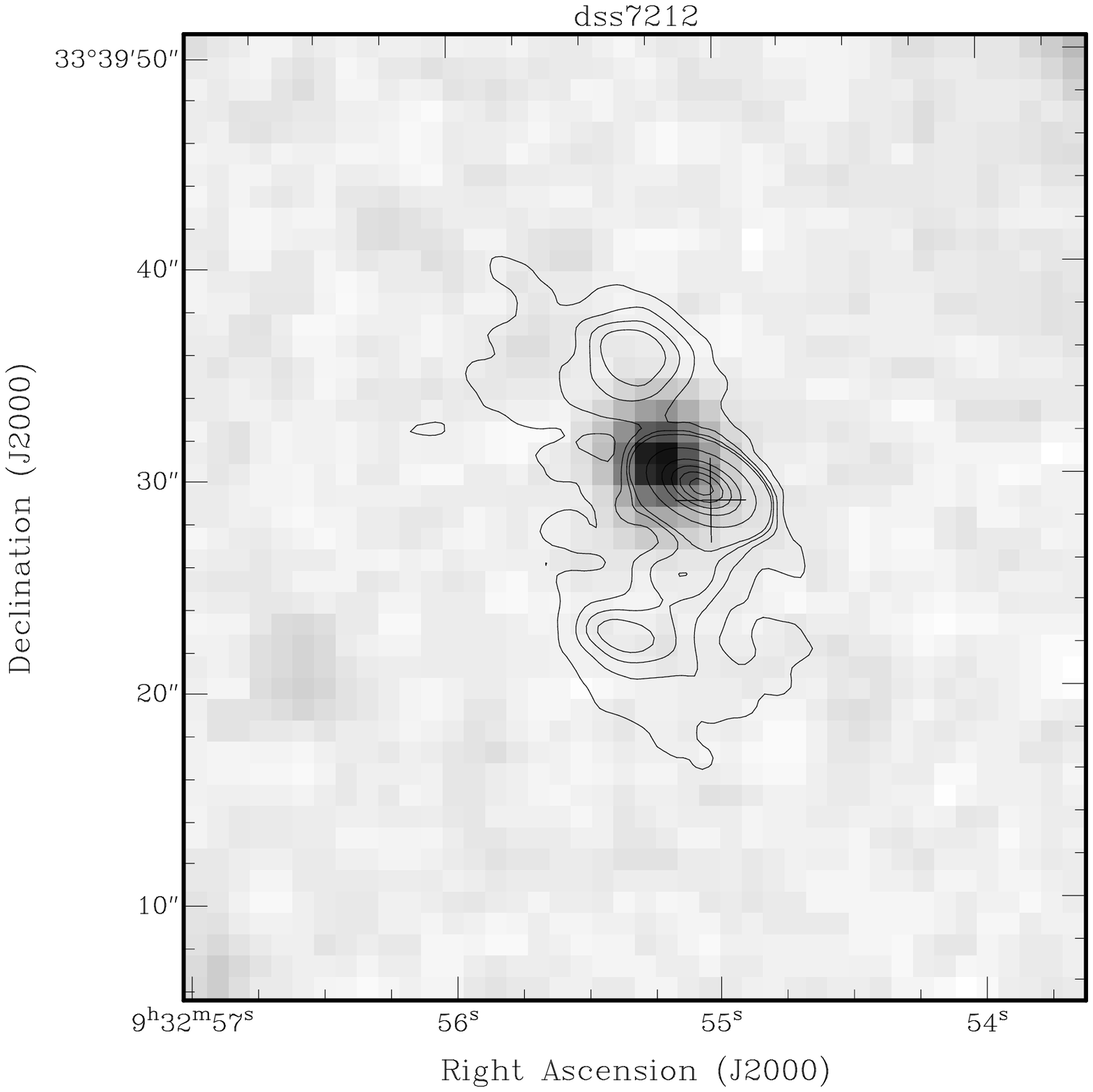 ,width=4.0cm,clip=}\label{aa}}\quad 
\subfigure[9CJ0933+2845 (P60 \it{R}\normalfont)]{\epsfig{figure=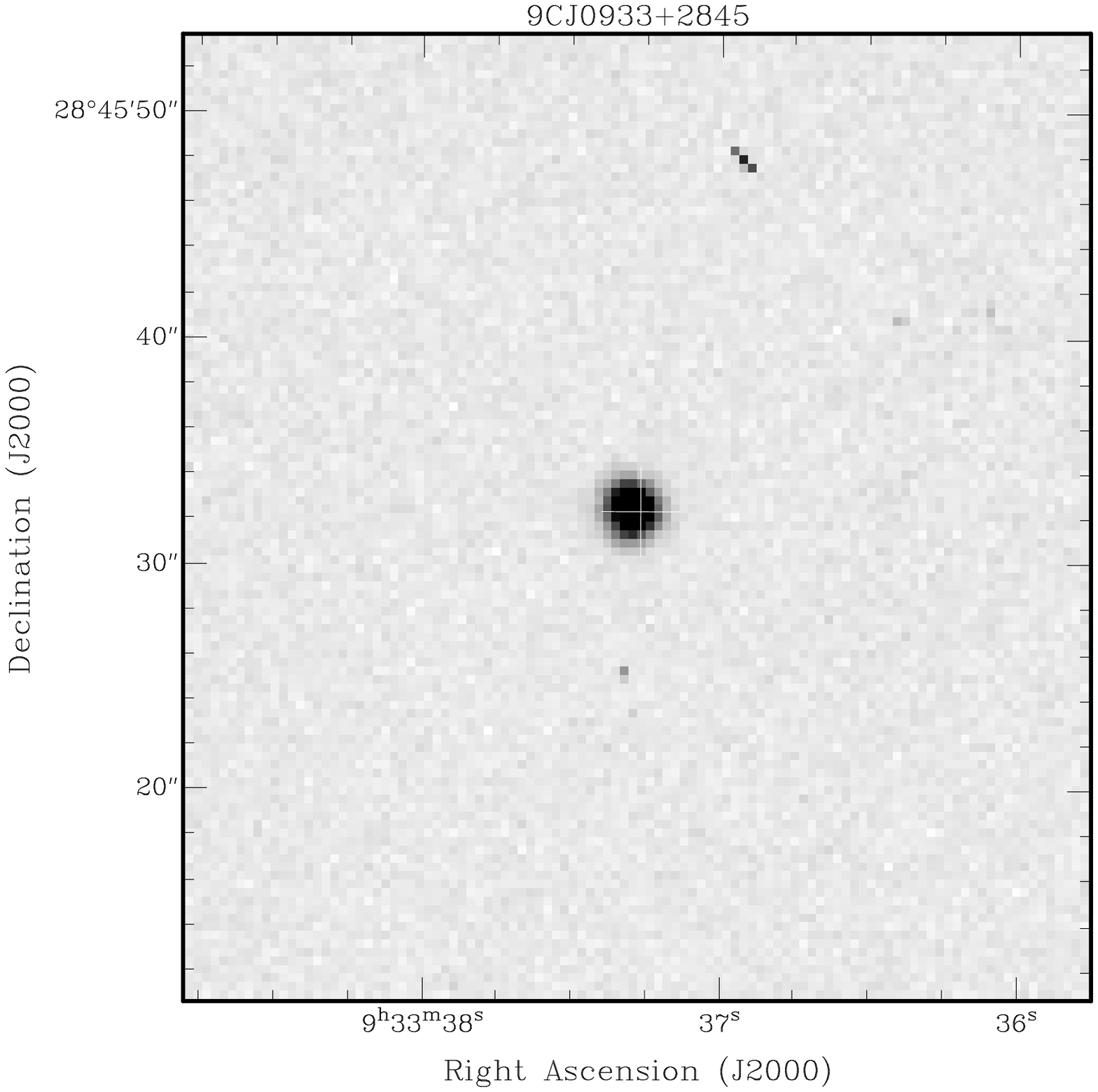 ,width=4.0cm,clip=}}\quad 
\subfigure[9CJ0933+3254 (P60 \it{R}\normalfont)]{\epsfig{figure=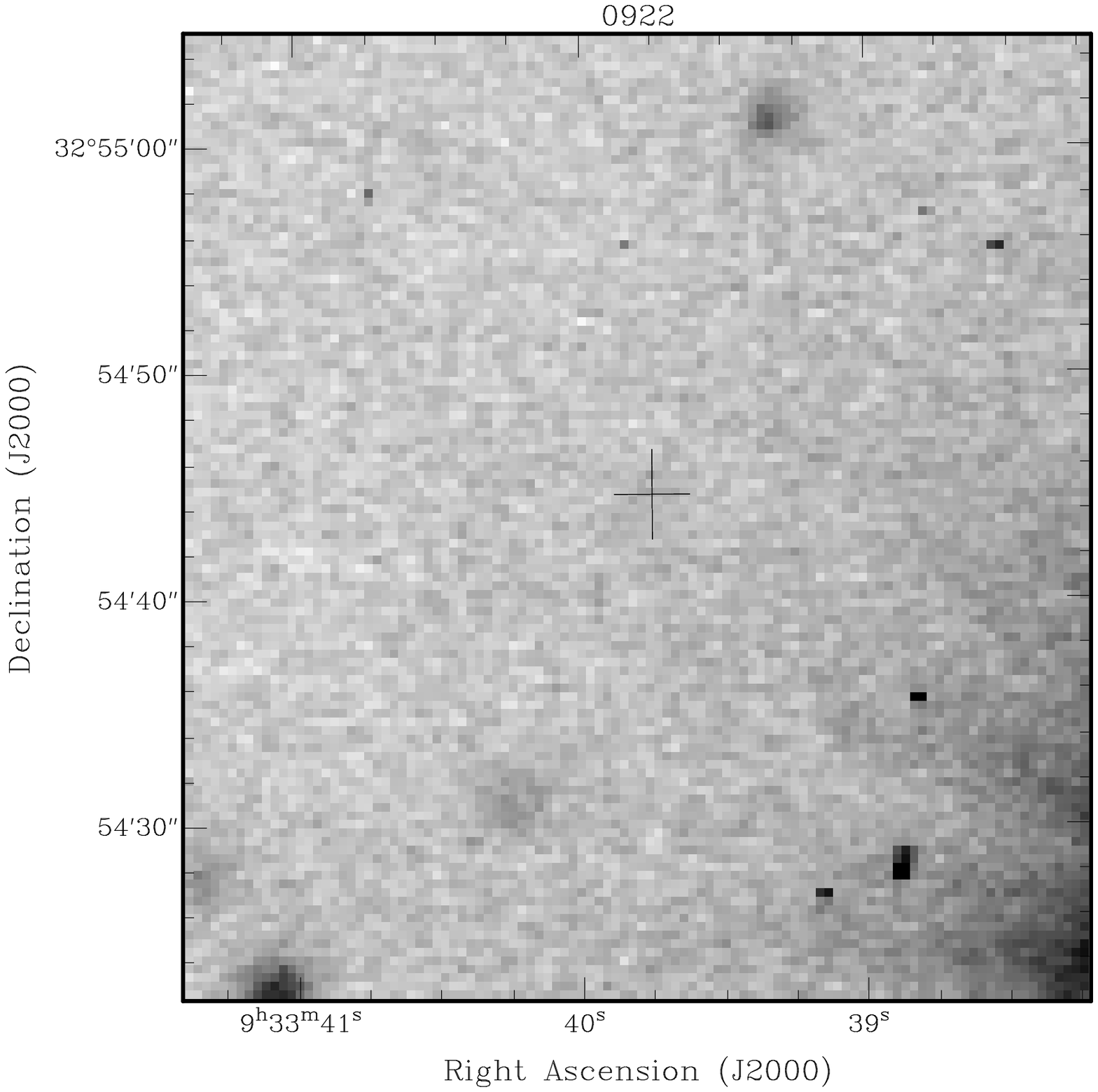 ,width=4.0cm,clip=}}}
 \mbox{
\subfigure[9CJ0934+2756 (P60 \it{R}\normalfont)]{\epsfig{figure=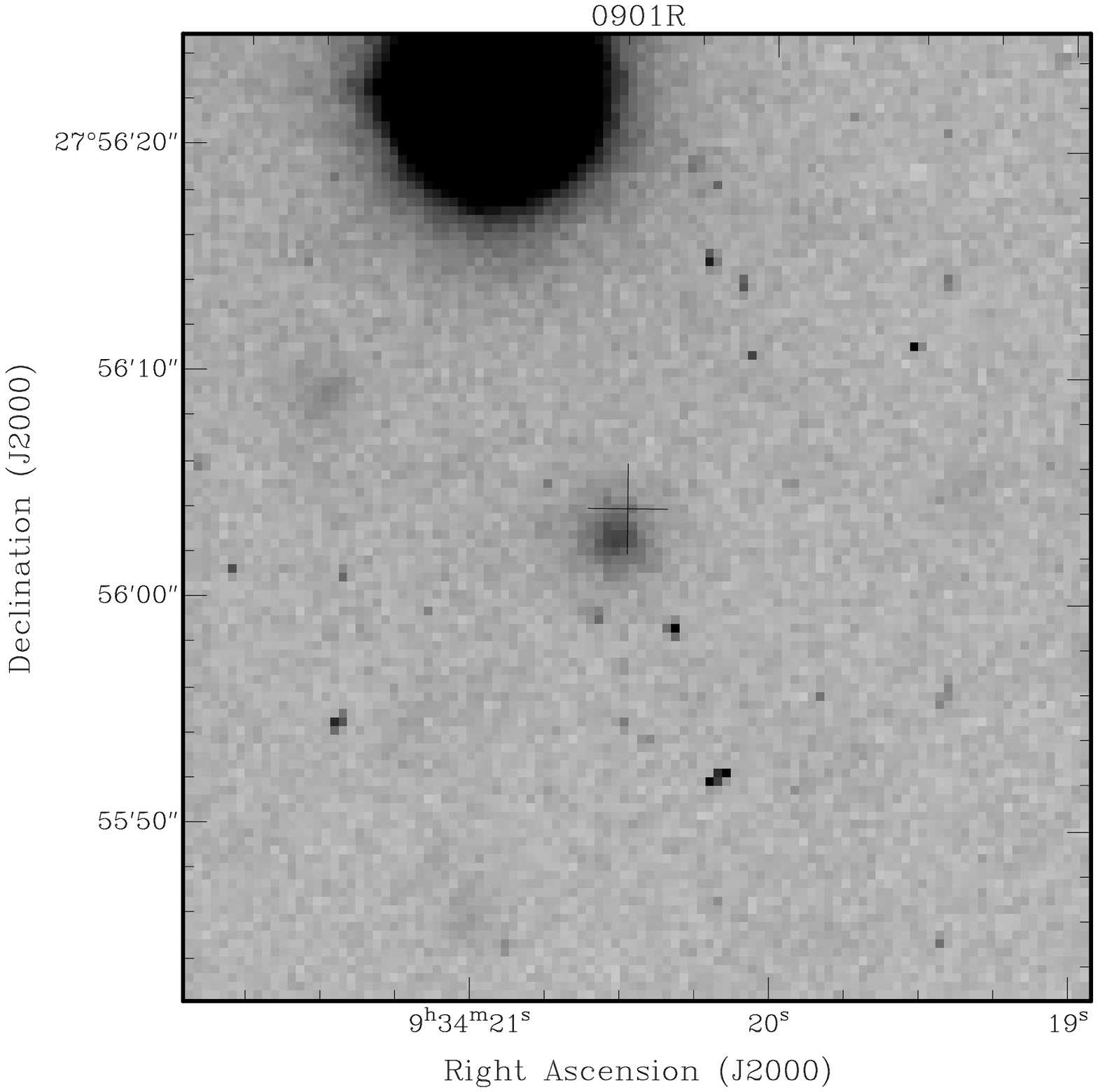 ,width=4.0cm,clip=}}\quad 
\subfigure[9CJ0934+3050 (P60 \it{R}\normalfont)]{\epsfig{figure=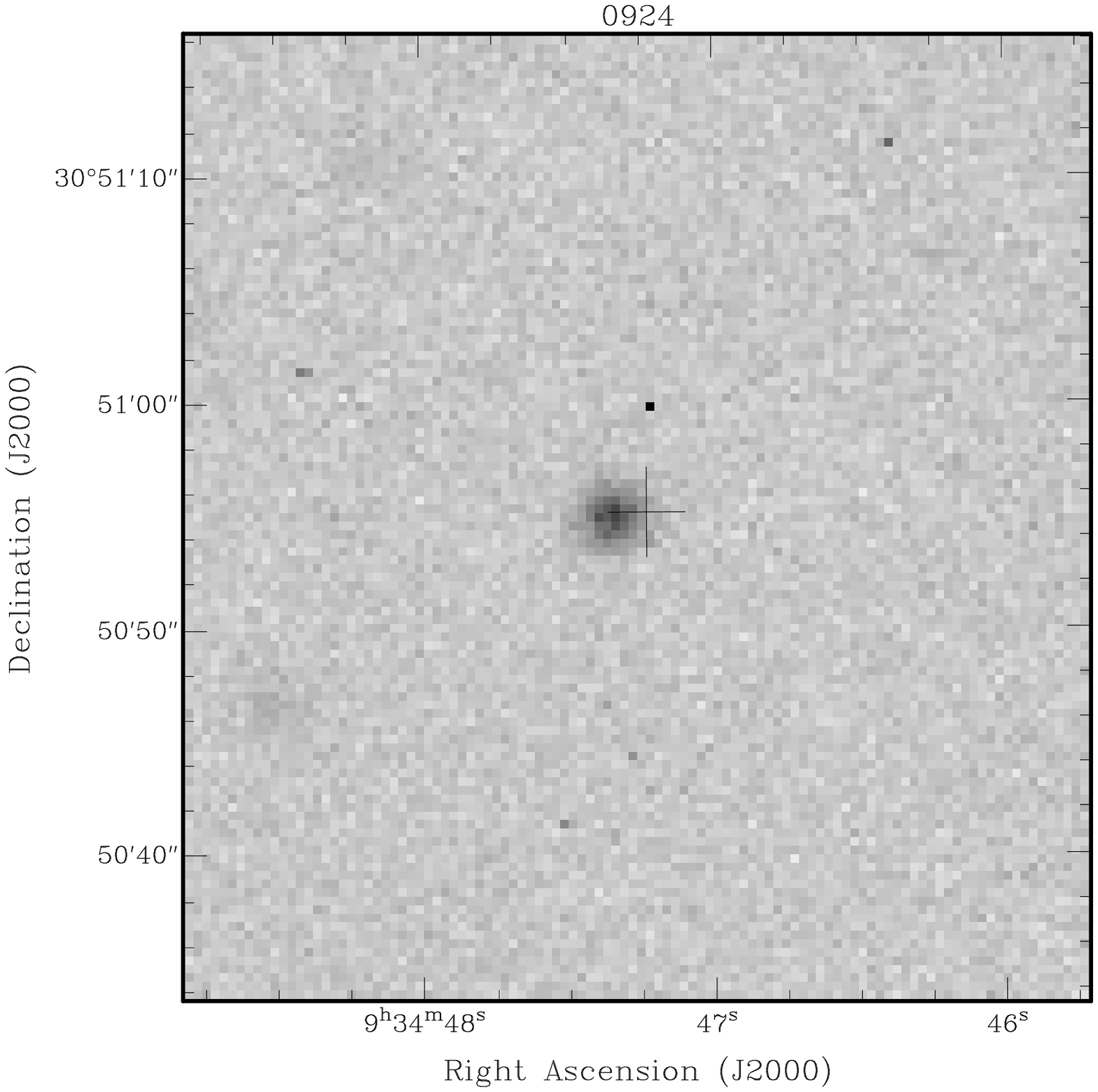 ,width=4.0cm,clip=}}\quad 
\subfigure[9CJ0935+2917 (P60 \it{R}\normalfont)]{\epsfig{figure=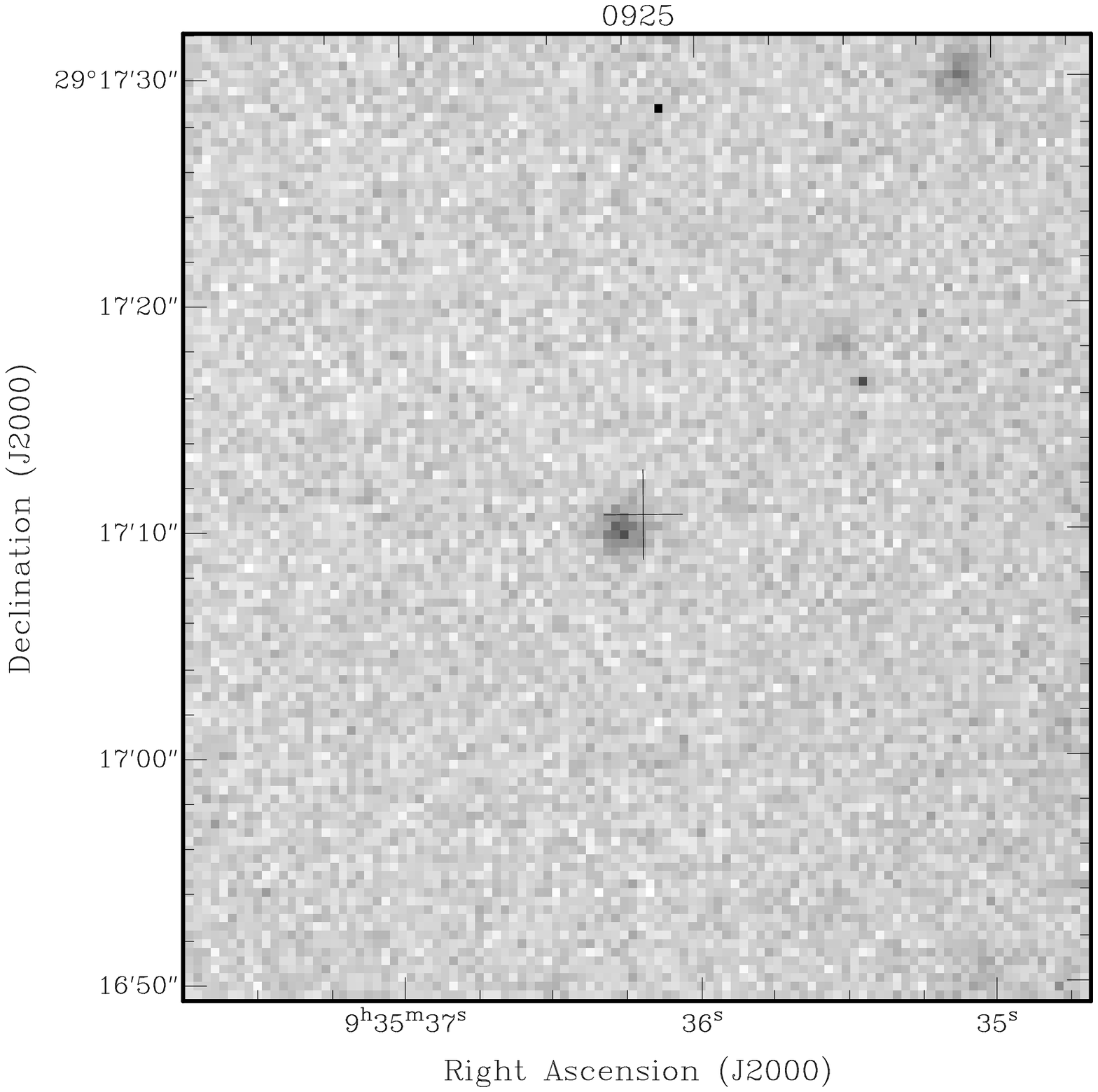 ,width=4.0cm,clip=}}} 
\mbox{
\subfigure[9CJ0936+3207 (P60 \it{R}\normalfont)]{\epsfig{figure=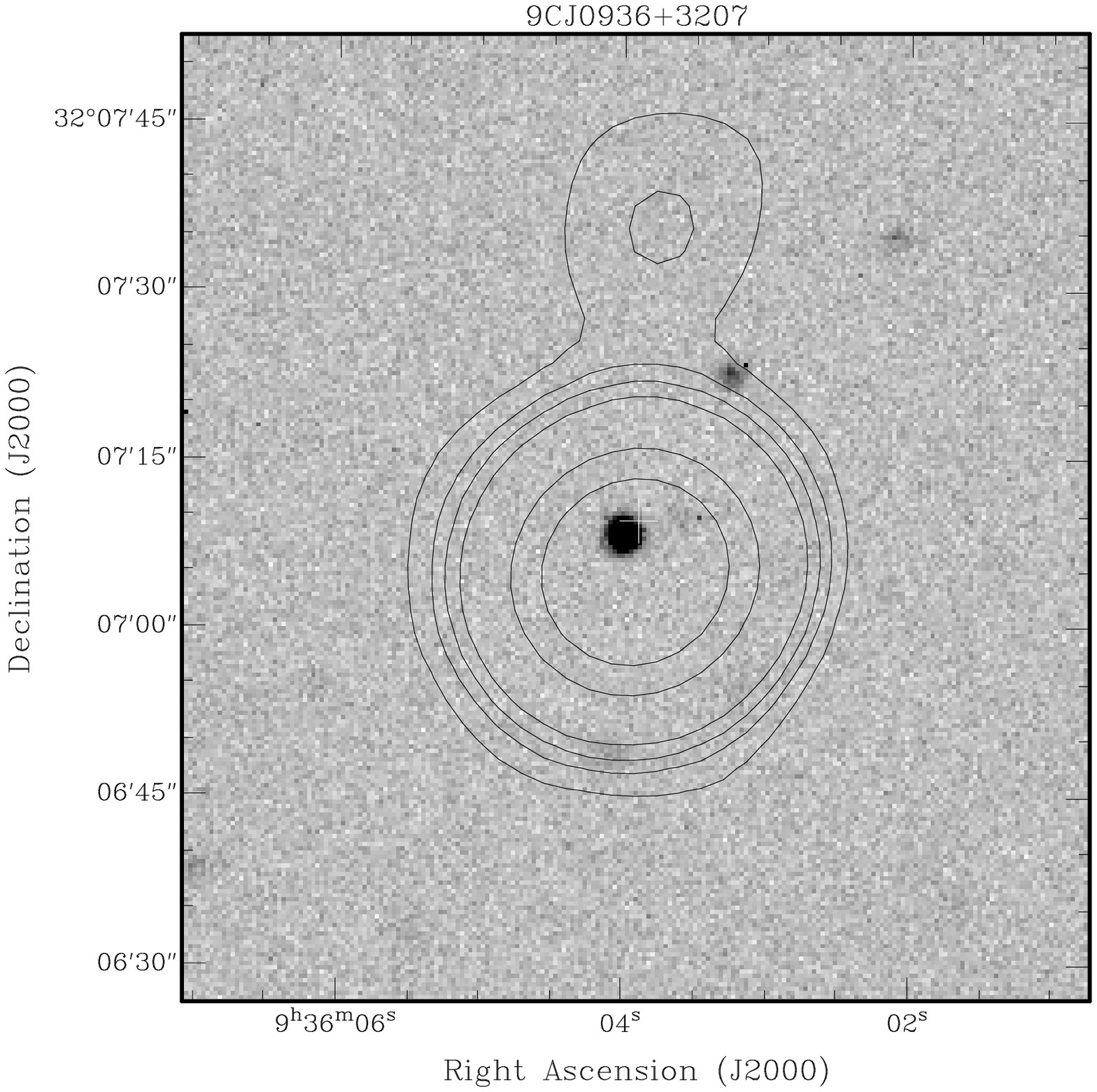 ,width=4.0cm,clip=}\label{ab}}\quad 
\subfigure[9CJ0936+3313 (P60 \it{R}\normalfont)]{\epsfig{figure=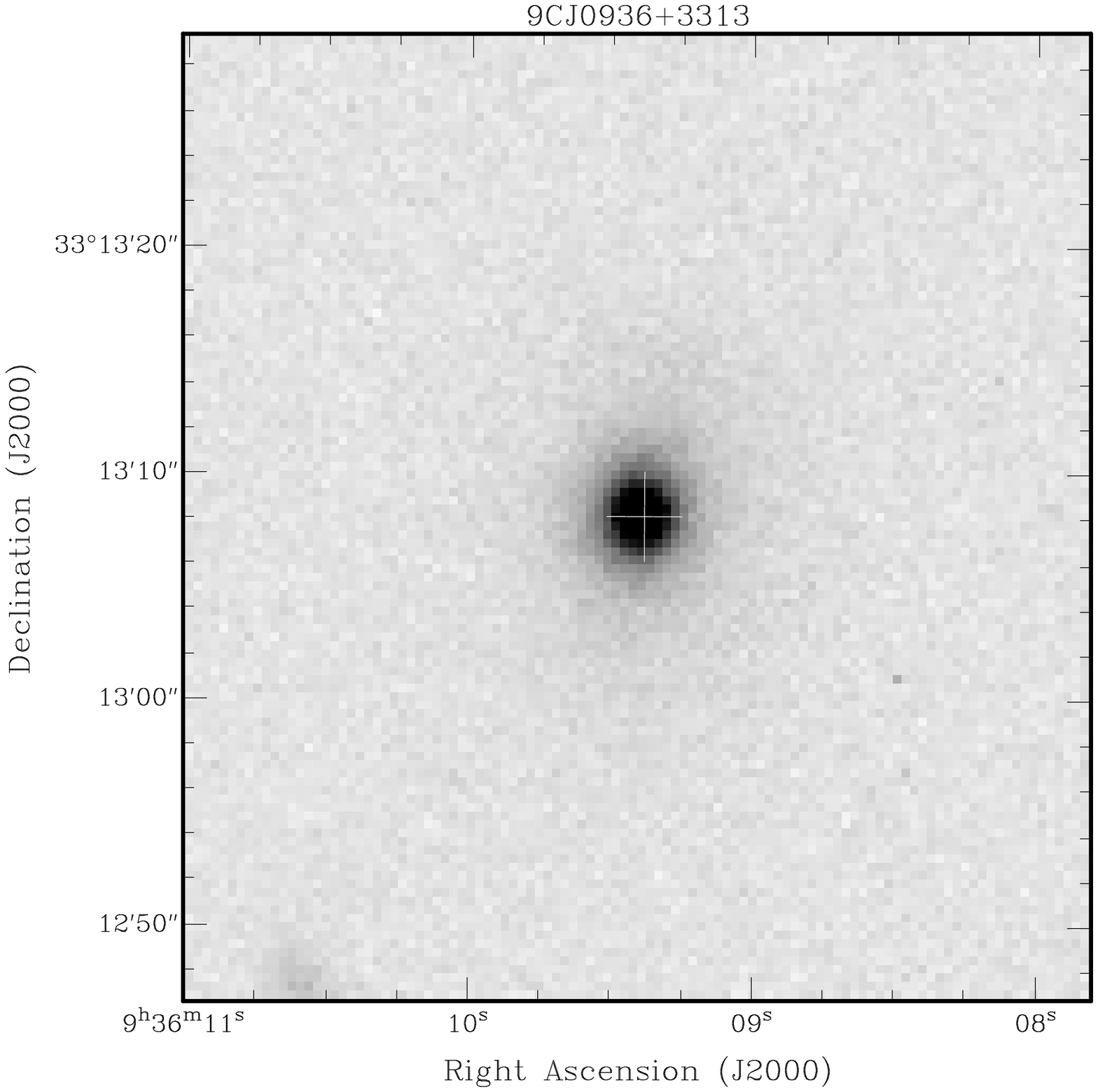 ,width=4.0cm,clip=}}\quad 
\subfigure[9CJ0936+2624 (P60 \it{R}\normalfont)]{\epsfig{figure=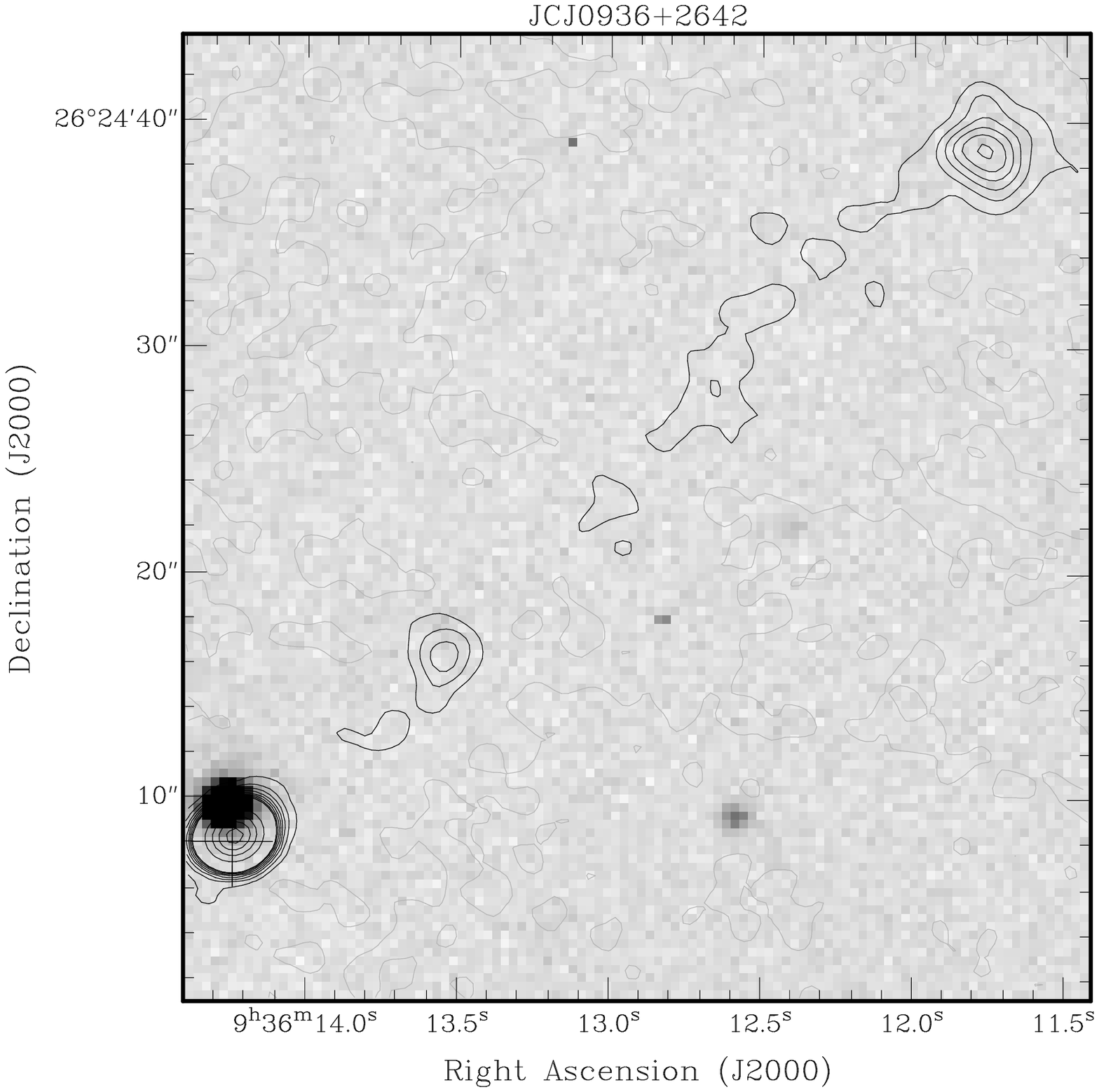 ,width=4.0cm,clip=}\label{ac}}
} 
\mbox{
\subfigure[9CJ0936+2624 (P60 \it{R}\normalfont)]{\epsfig{figure=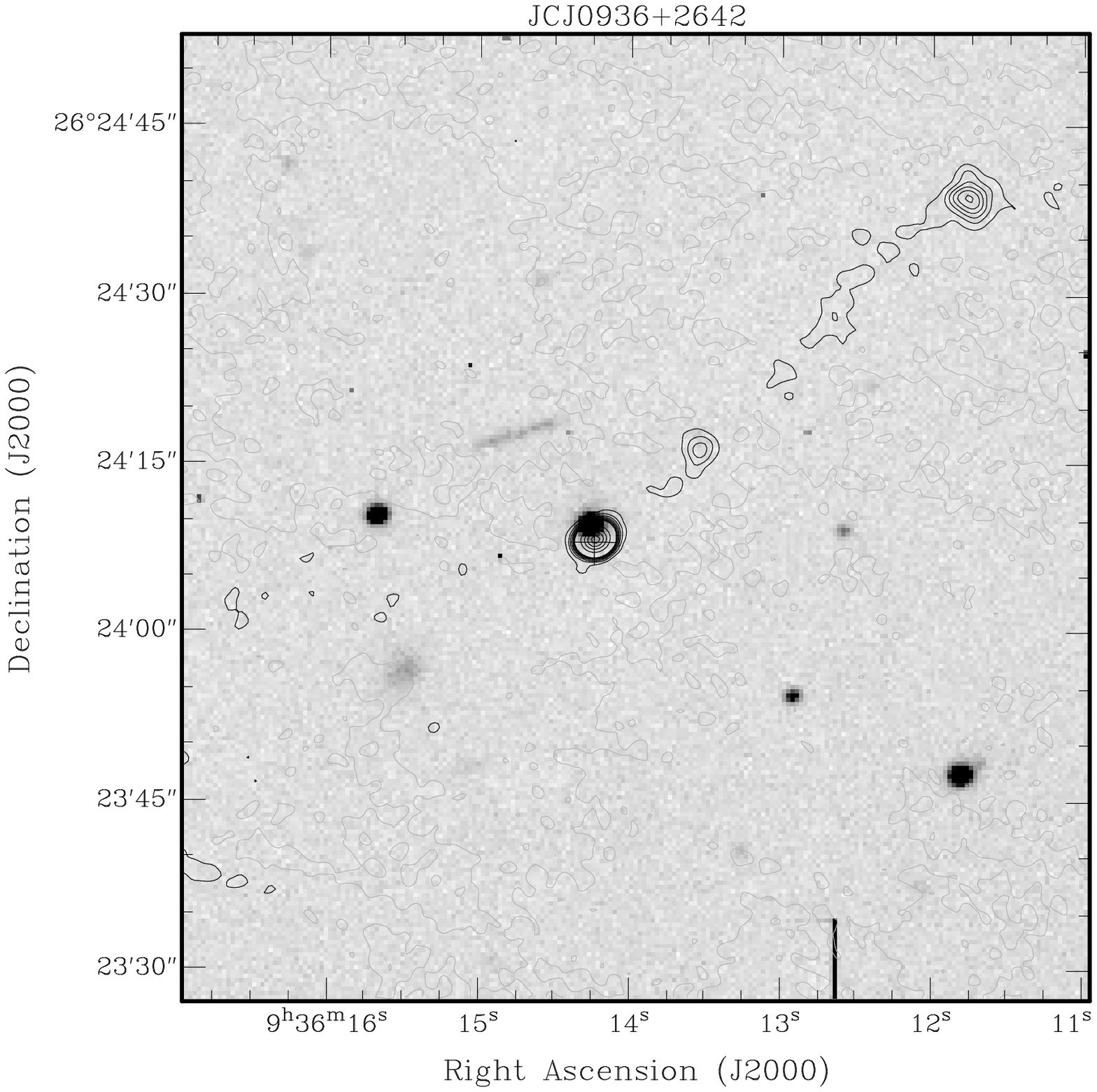 ,width=4.0cm,clip=}\label{ad}}\quad 
\subfigure[9CJ0937+3206 (P60 \it{R}\normalfont)]{\epsfig{figure=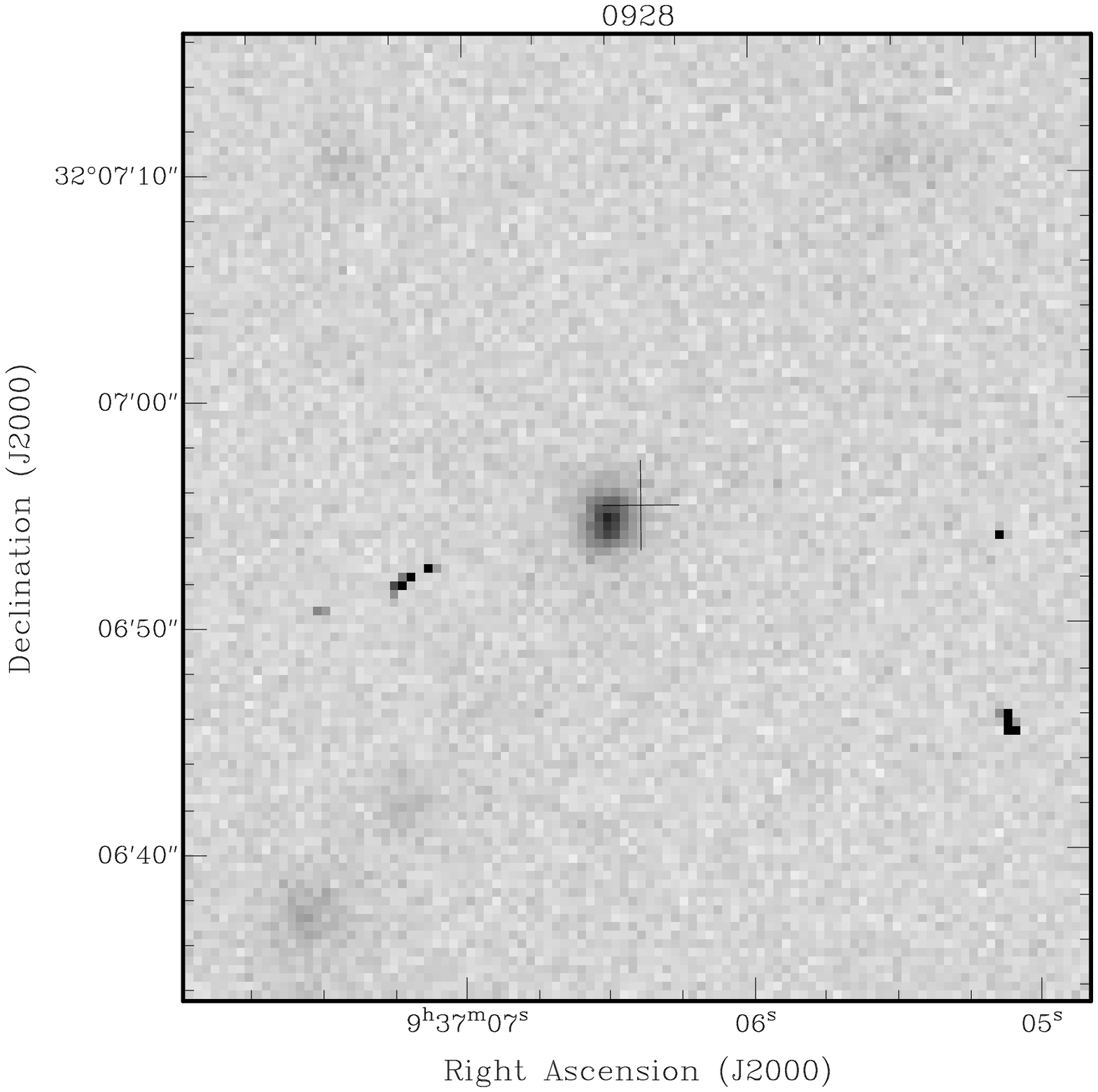 ,width=4.0cm,clip=}}\quad 
\subfigure[9CJ0939+2908 (P60 \it{R}\normalfont)]{\epsfig{figure=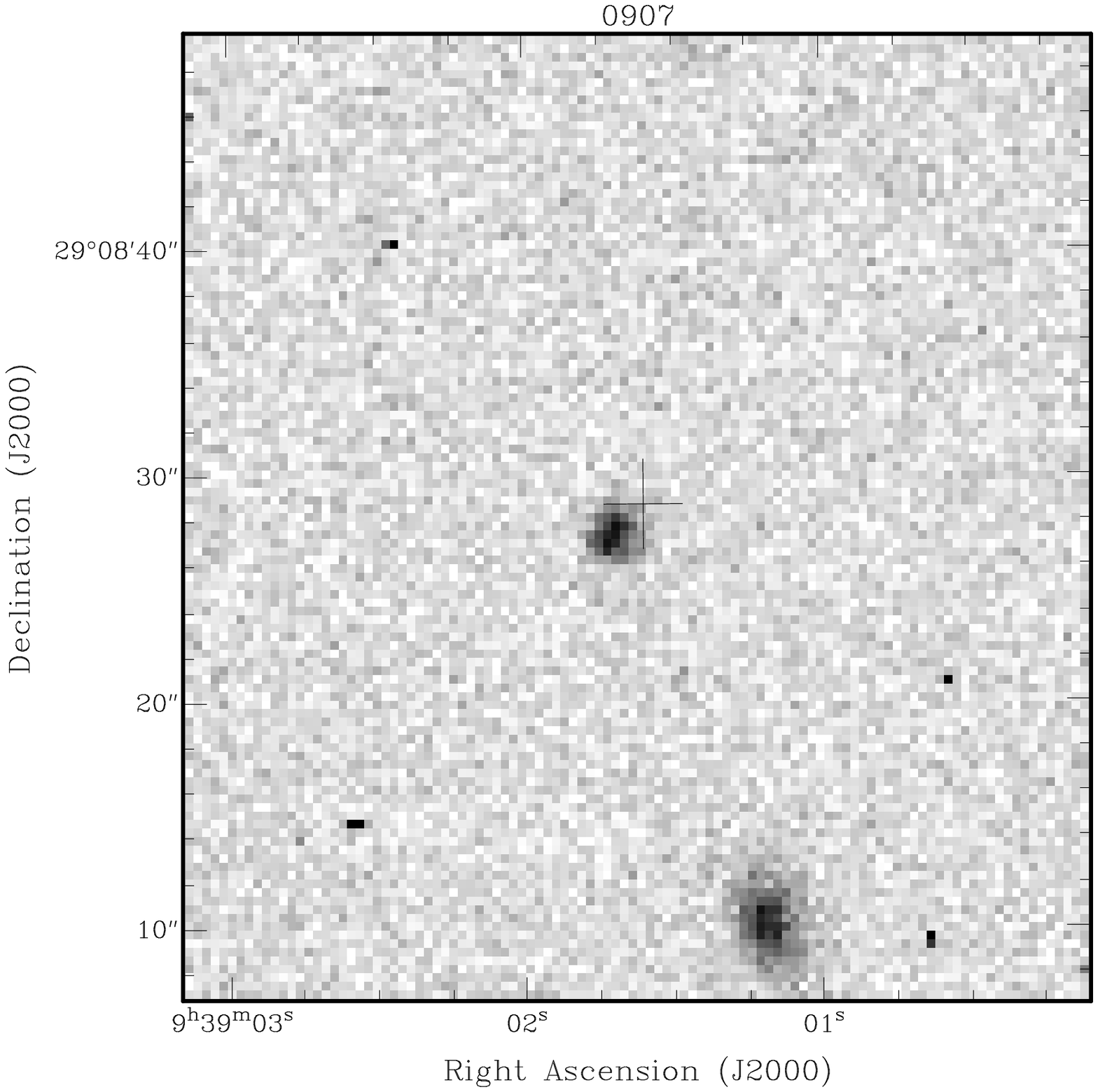 ,width=4.0cm,clip=}}}\caption{Optical counterparts for sources 9CJ0932+3339 to 9CJ0939+2908. Crosses mark maximum radio flux density and are 4\,arcsec top to bottom. Contours: \ref{aa}, 1.4\,GHz contours at 1-2.5 every 0.5\,\% and 10-90 every 10\,\% of peak (182\,mJy/beam); \ref{ab}, 4.8\,GHz contours 3,5,7,10,30 and 50\,\% of peak (39.6\,mJy/beam); \ref{ac} and \ref{ad}, in both 1.4\,GHz contours 2-9 every 1\,\% and 10-90 every 20\,\% of peak (53.1\,mJy/beam).}\end{figure*}

\begin{figure*}
\mbox{
\subfigure[9CJ0937+3411 (P60 \it{R}\normalfont)]{\epsfig{figure=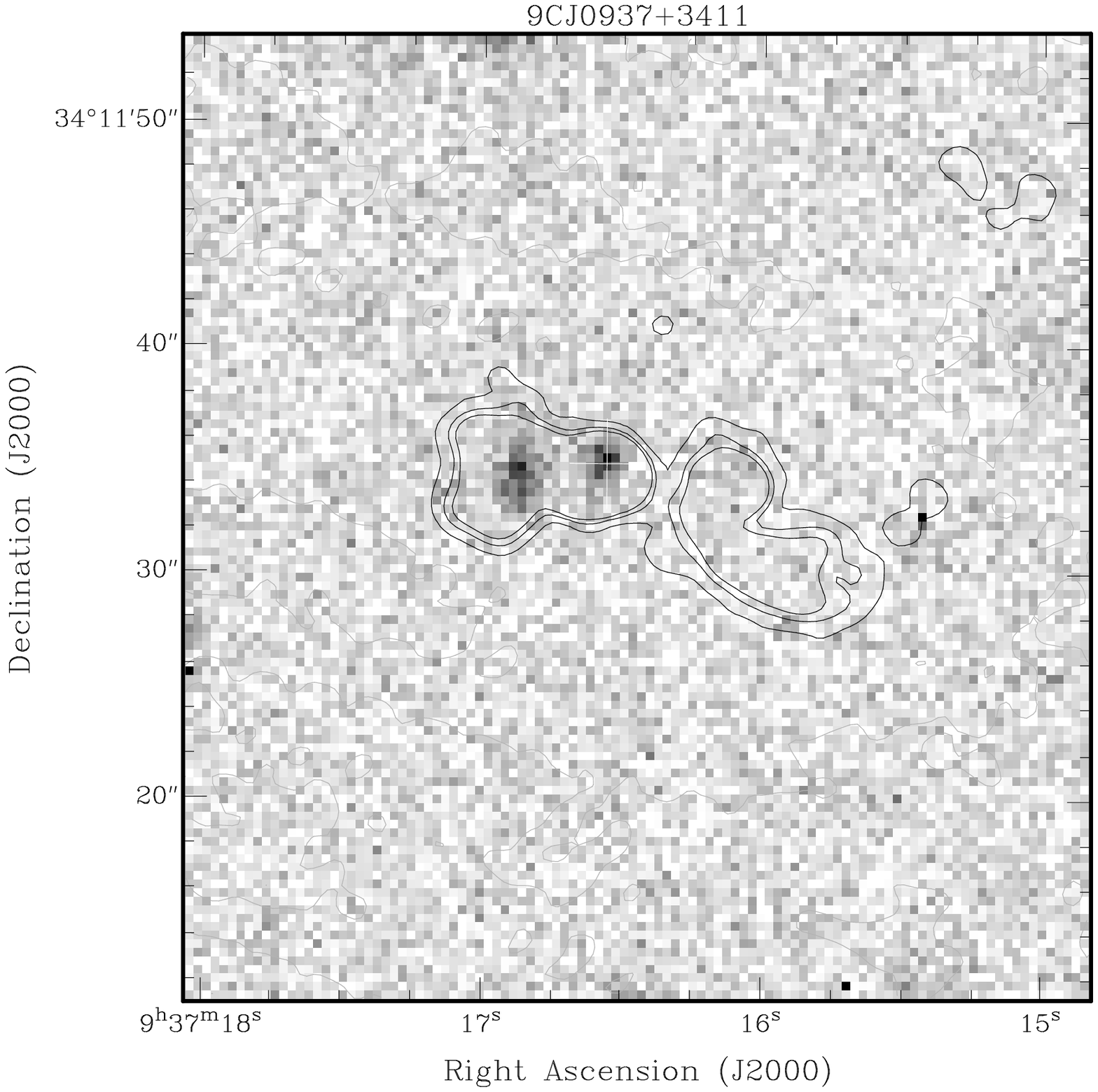 ,width=4.0cm,clip=}\label{ae}}\quad 
\subfigure[9CJ0937+3411 (P60 \it{R}\normalfont)]{\epsfig{figure=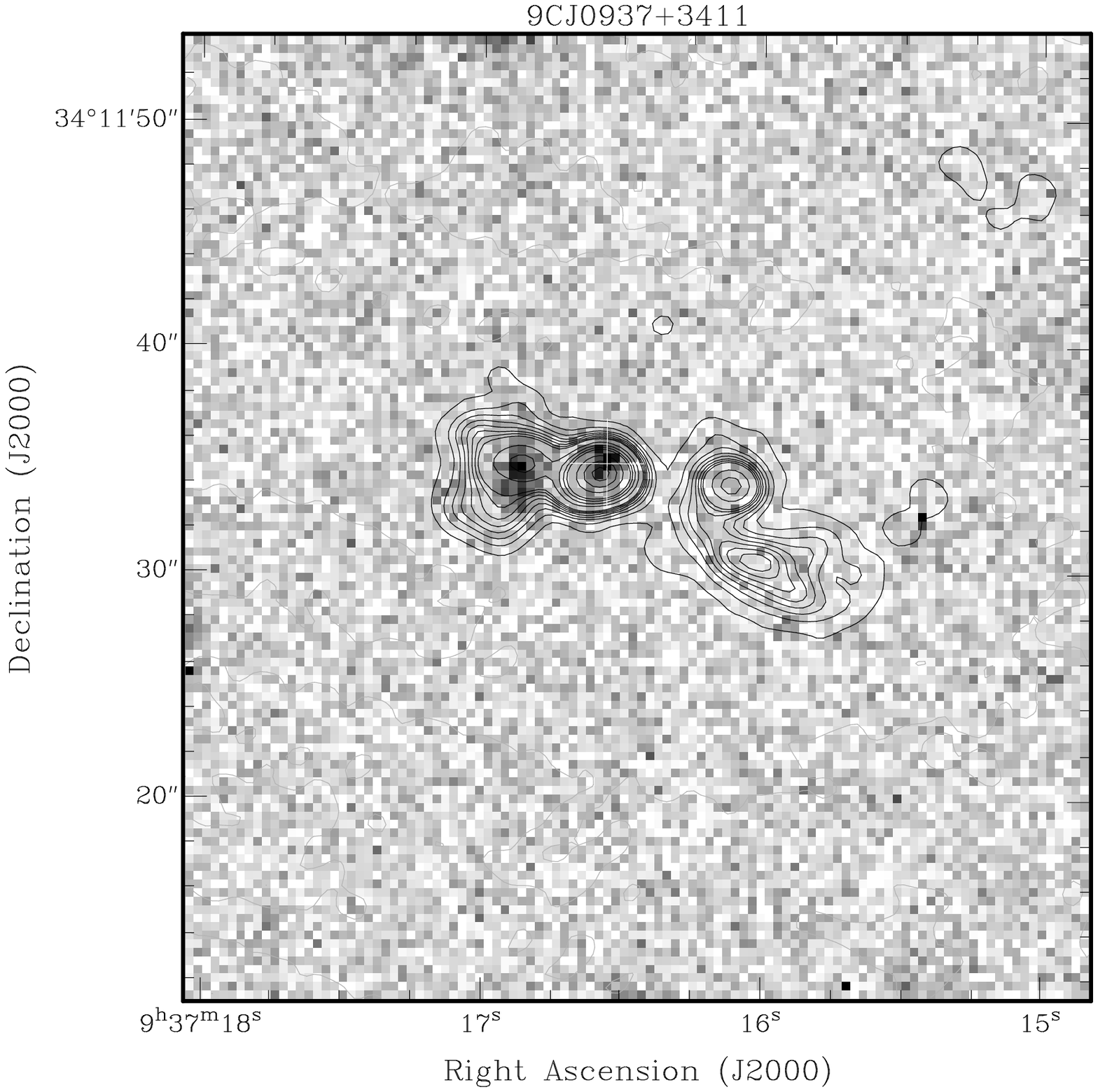 ,width=4.0cm,clip=}\label{af}}\quad 
\subfigure[9CJ0940+2626 (P60 \it{R}\normalfont)]{\epsfig{figure=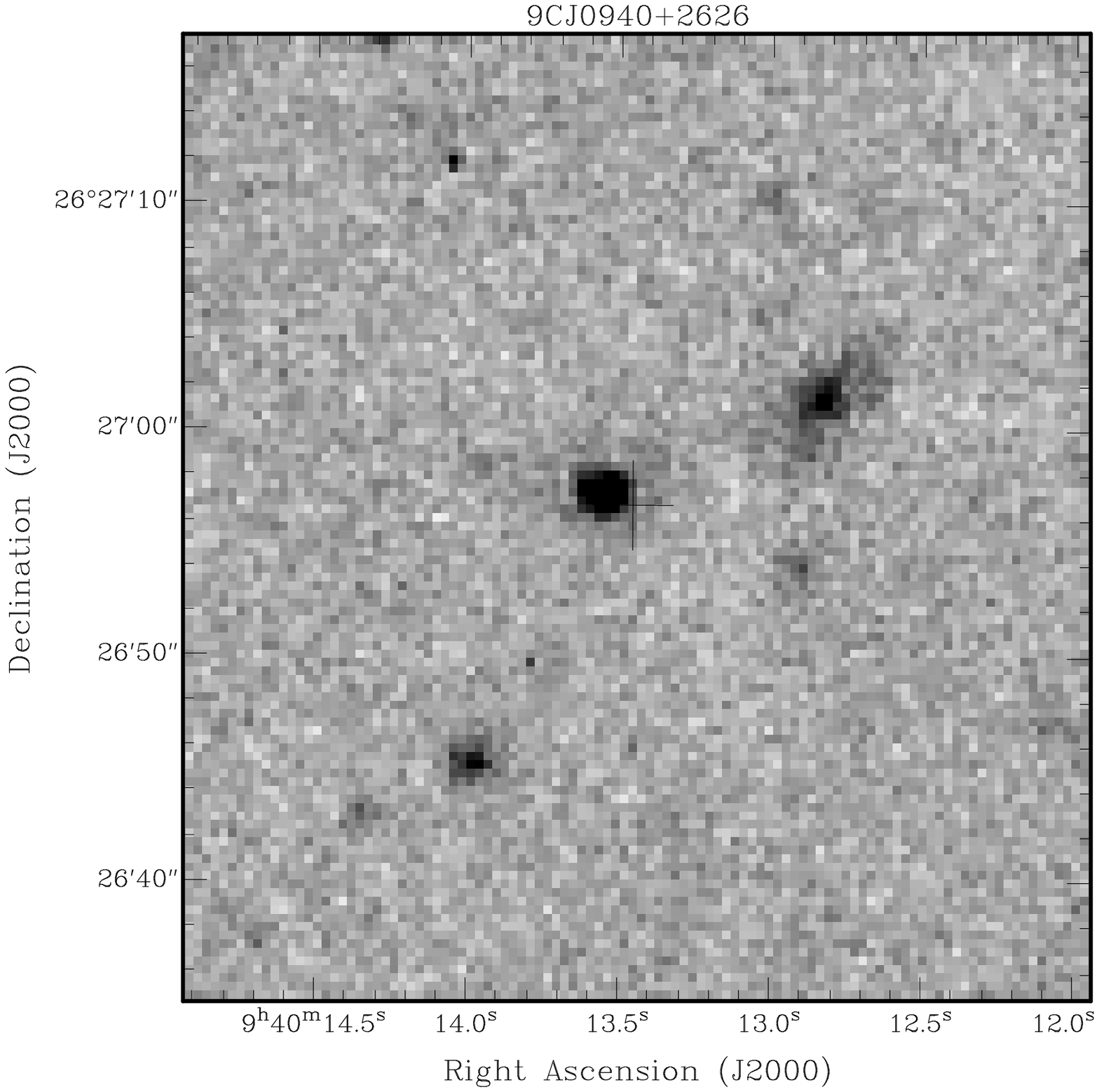 ,width=4.0cm,clip=}}
}
\mbox{
\subfigure[9CJ0940+2603 (DSS2 \it{R}\normalfont)]{\epsfig{figure=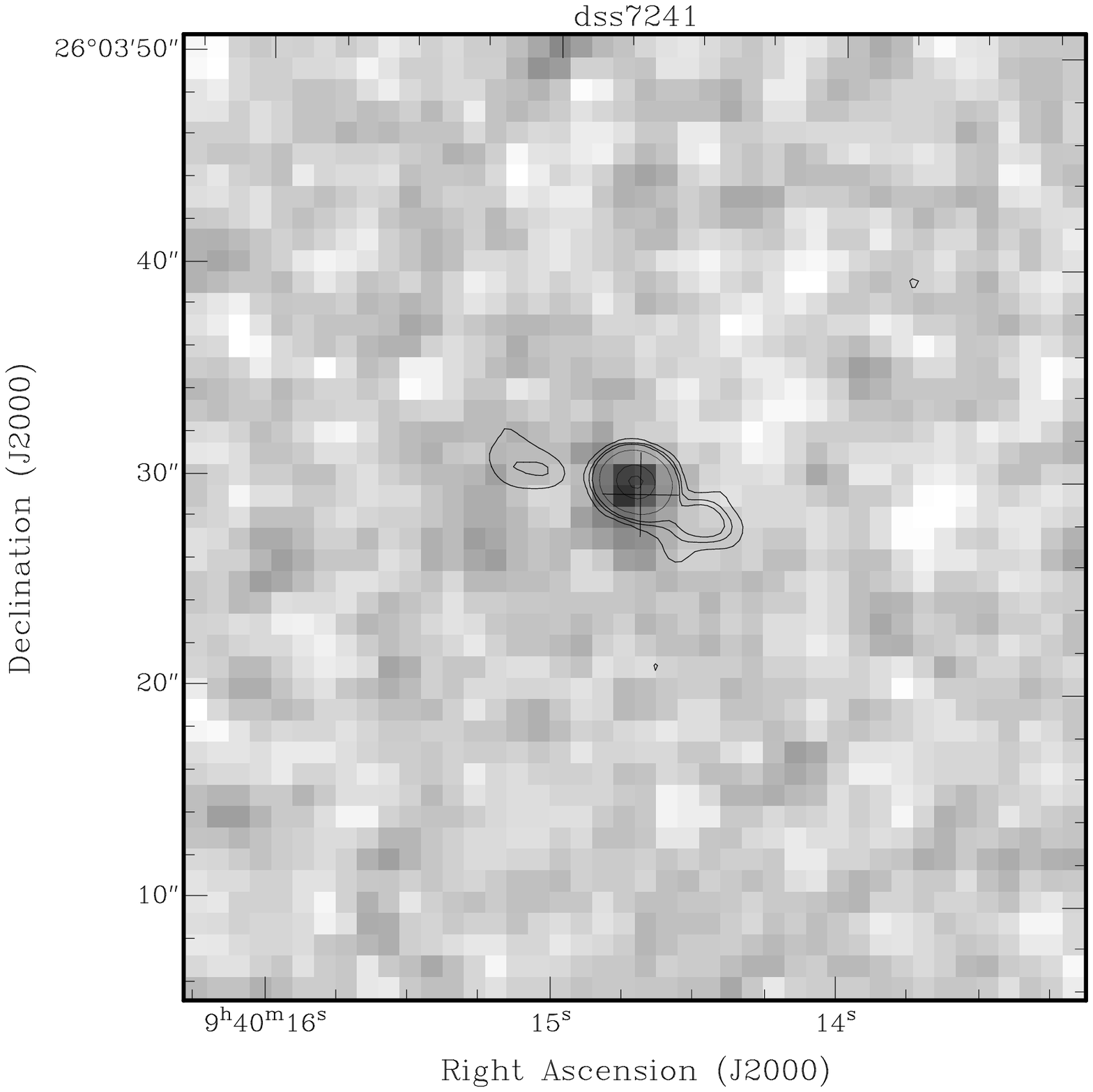 ,width=4.0cm,clip=}\label{ag}}\quad 
\subfigure[9CJ0940+3015 (DSS2 \it{R}\normalfont)]{\epsfig{figure=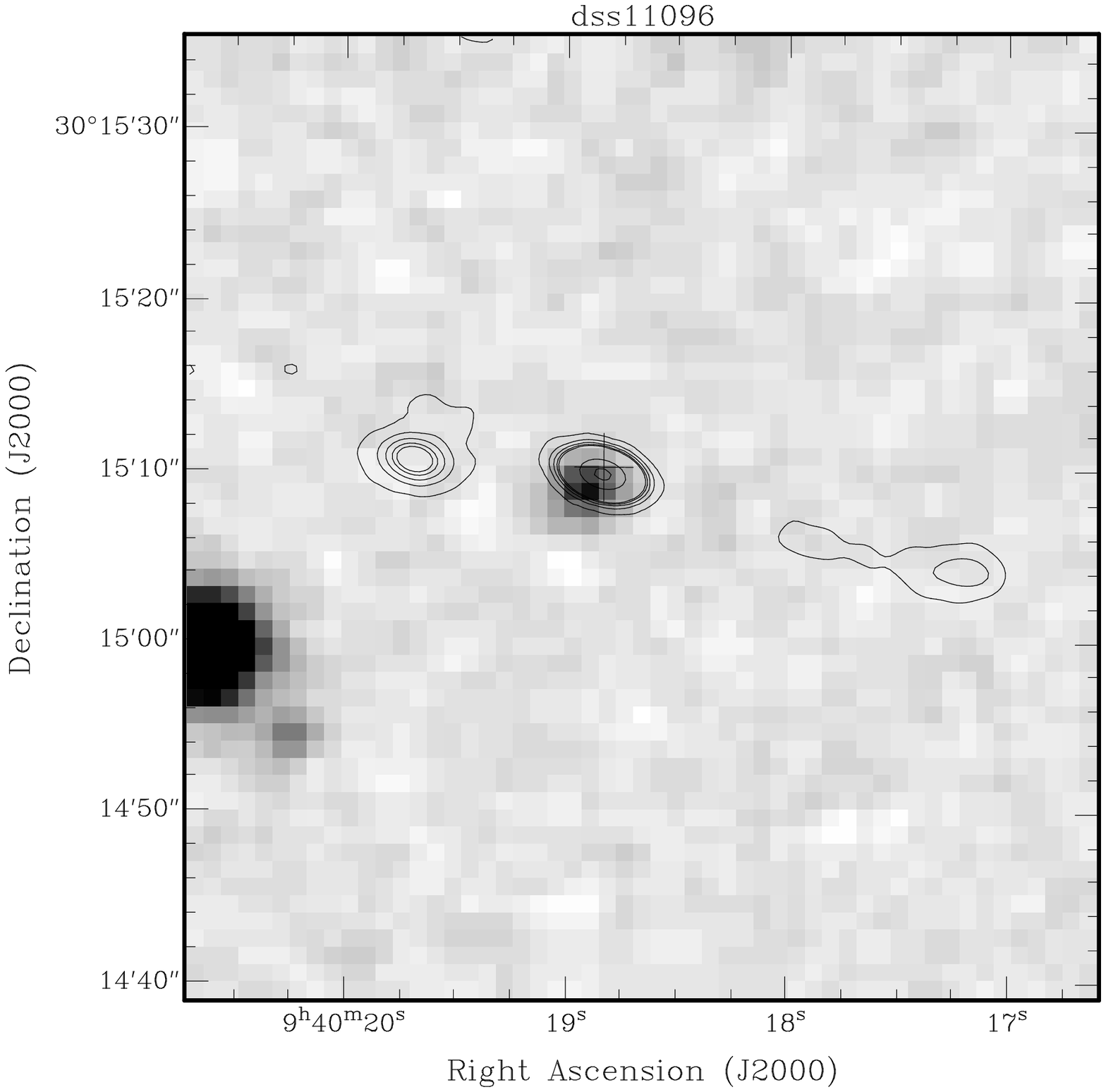 ,width=4.0cm,clip=}\label{ah}}\quad 
\subfigure[9CJ0941+2728 (P60 \it{R}\normalfont)]{\epsfig{figure=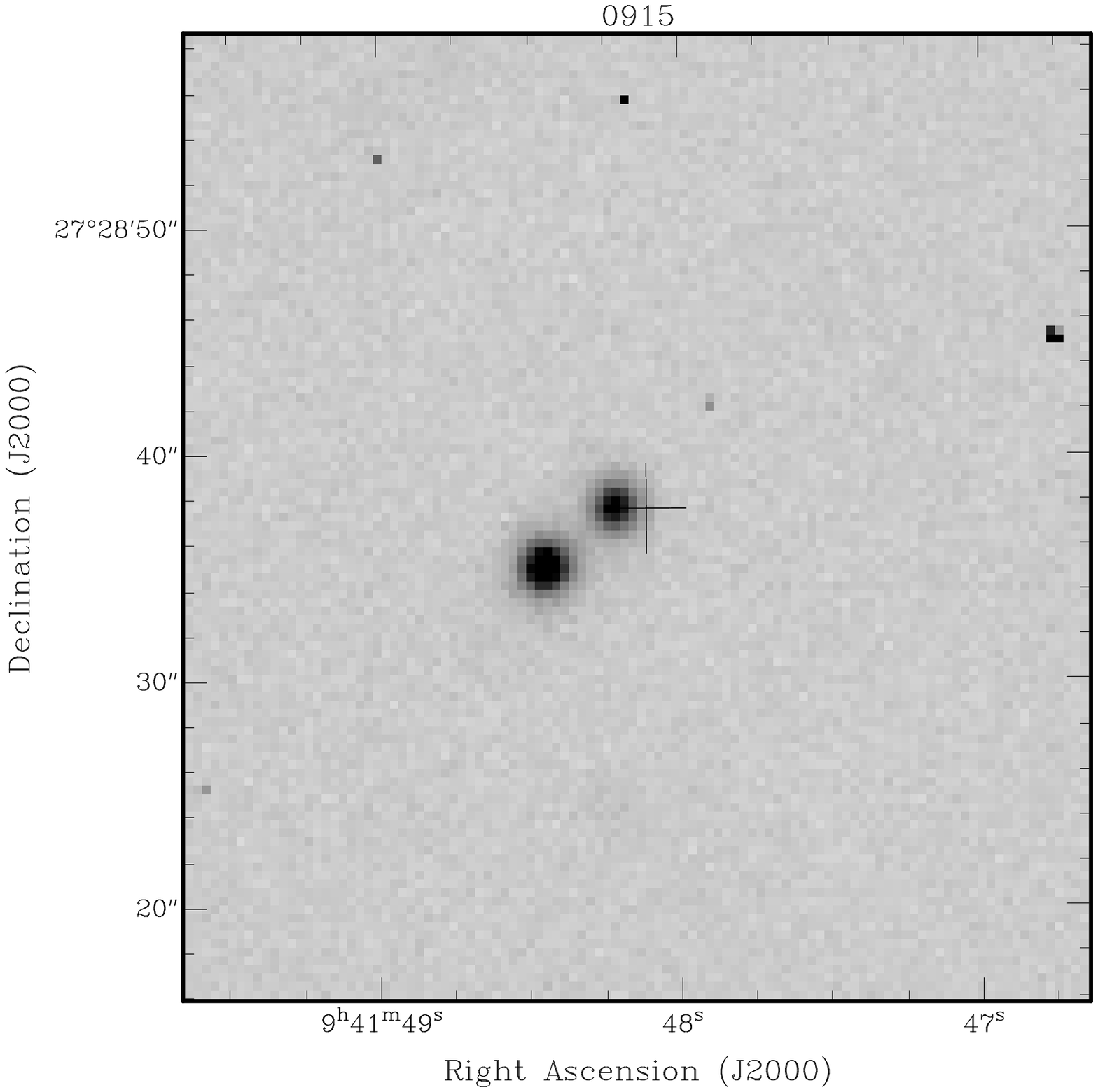 ,width=4.0cm,clip=}}
} 
\mbox{
\subfigure[9CJ0942+3309 (P60 \it{R}\normalfont)]{\epsfig{figure=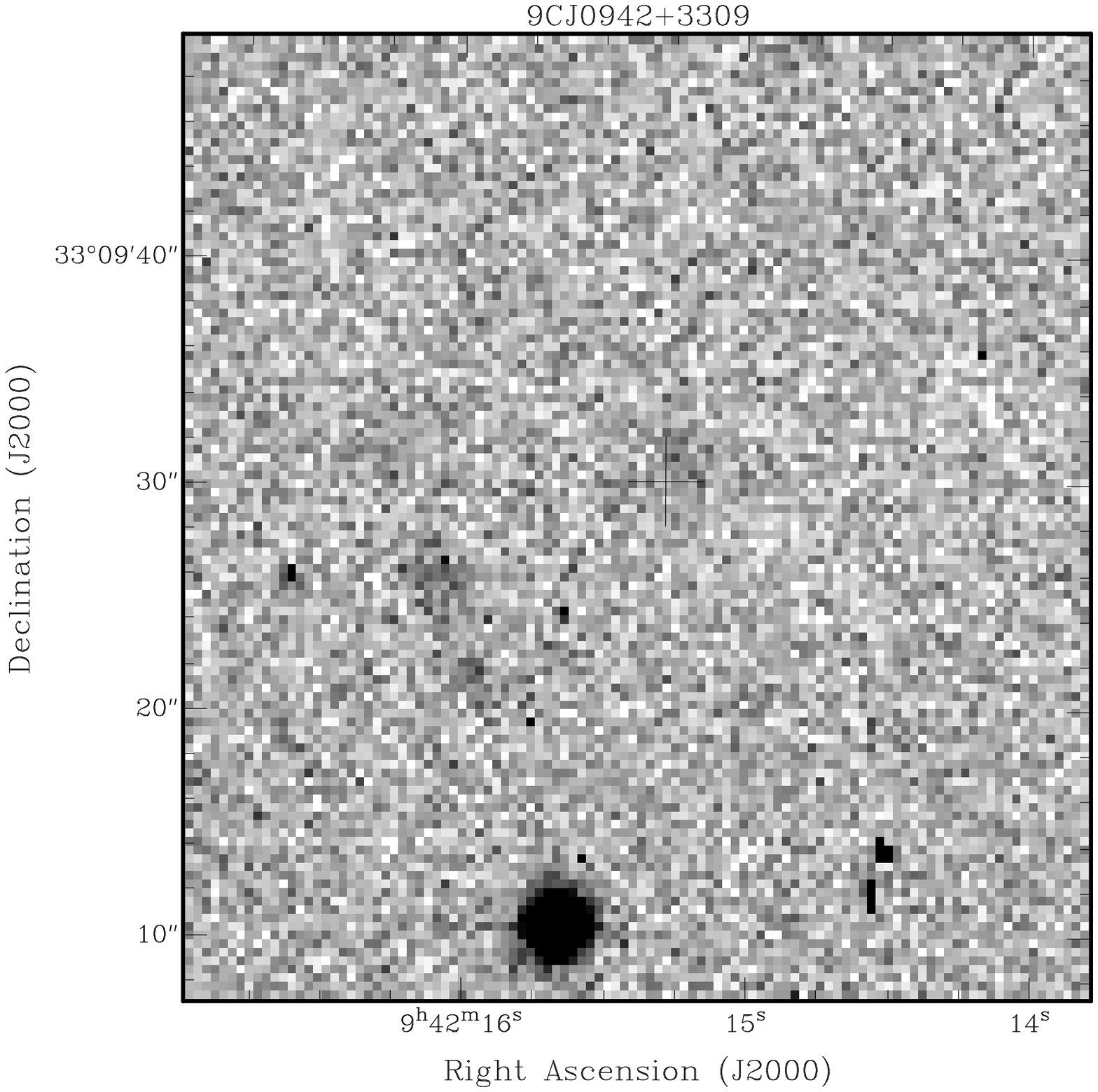 ,width=4.0cm,clip=}}\quad 
\subfigure[9CJ0943+3614 (P60 \it{R}\normalfont)]{\epsfig{figure=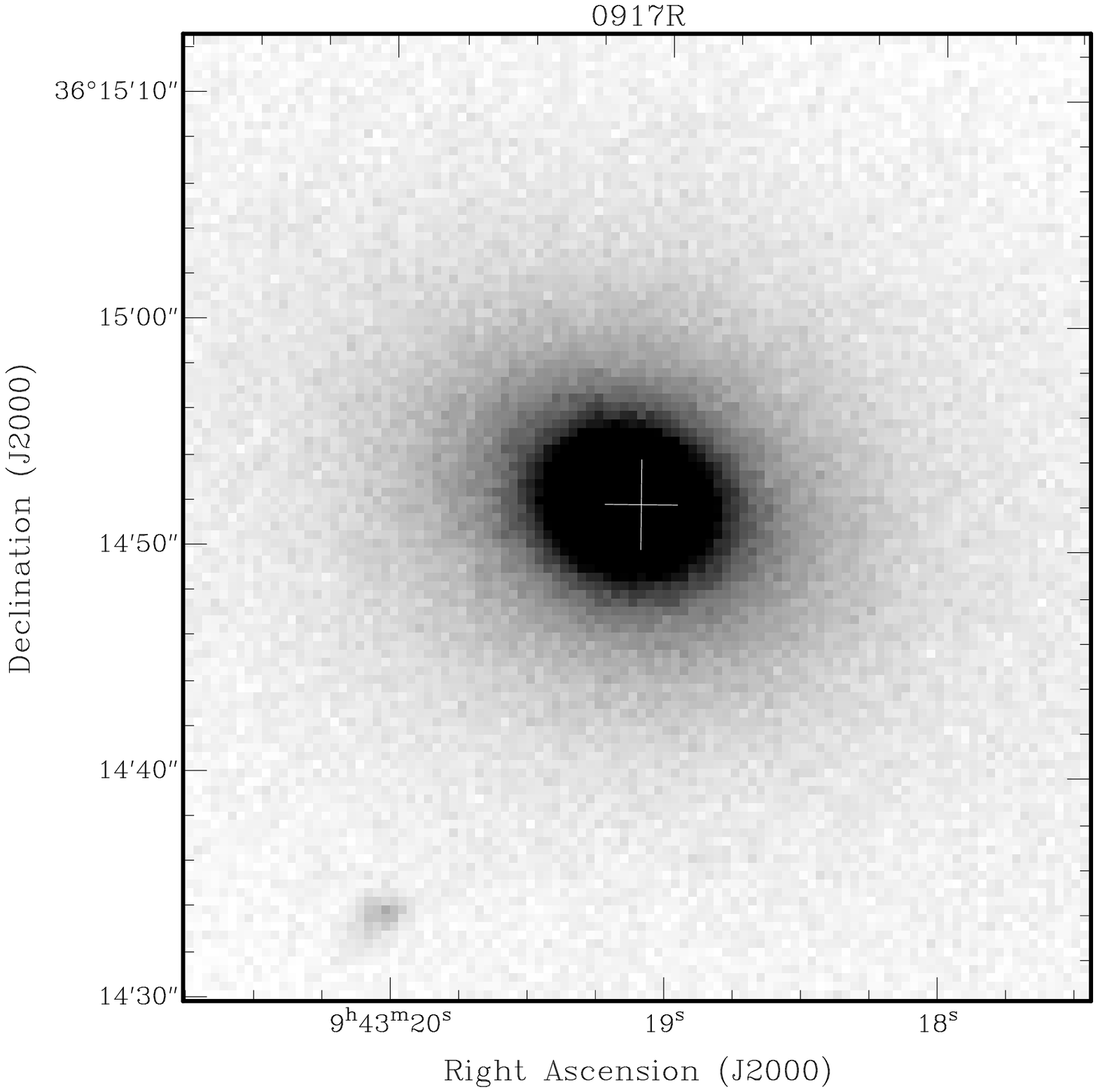 ,width=4.0cm,clip=}}\quad 
\subfigure[9CJ0944+2554 (P60 \it{R}\normalfont)]{\epsfig{figure=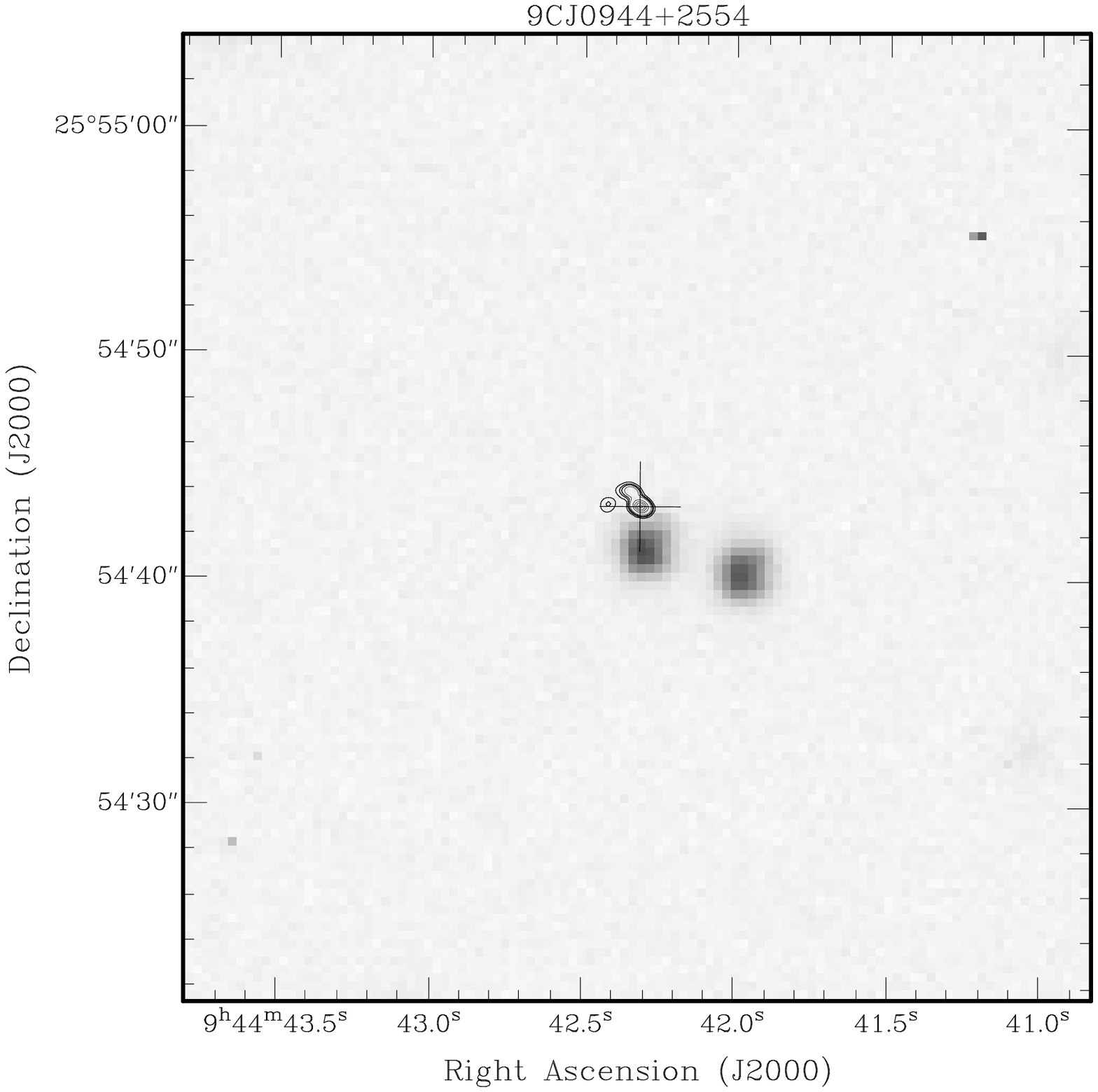 ,width=4.0cm,clip=}\label{ai}}
}
\mbox{
\subfigure[9CJ0944+2554 (P60 \it{R}\normalfont) Detail]{\epsfig{figure=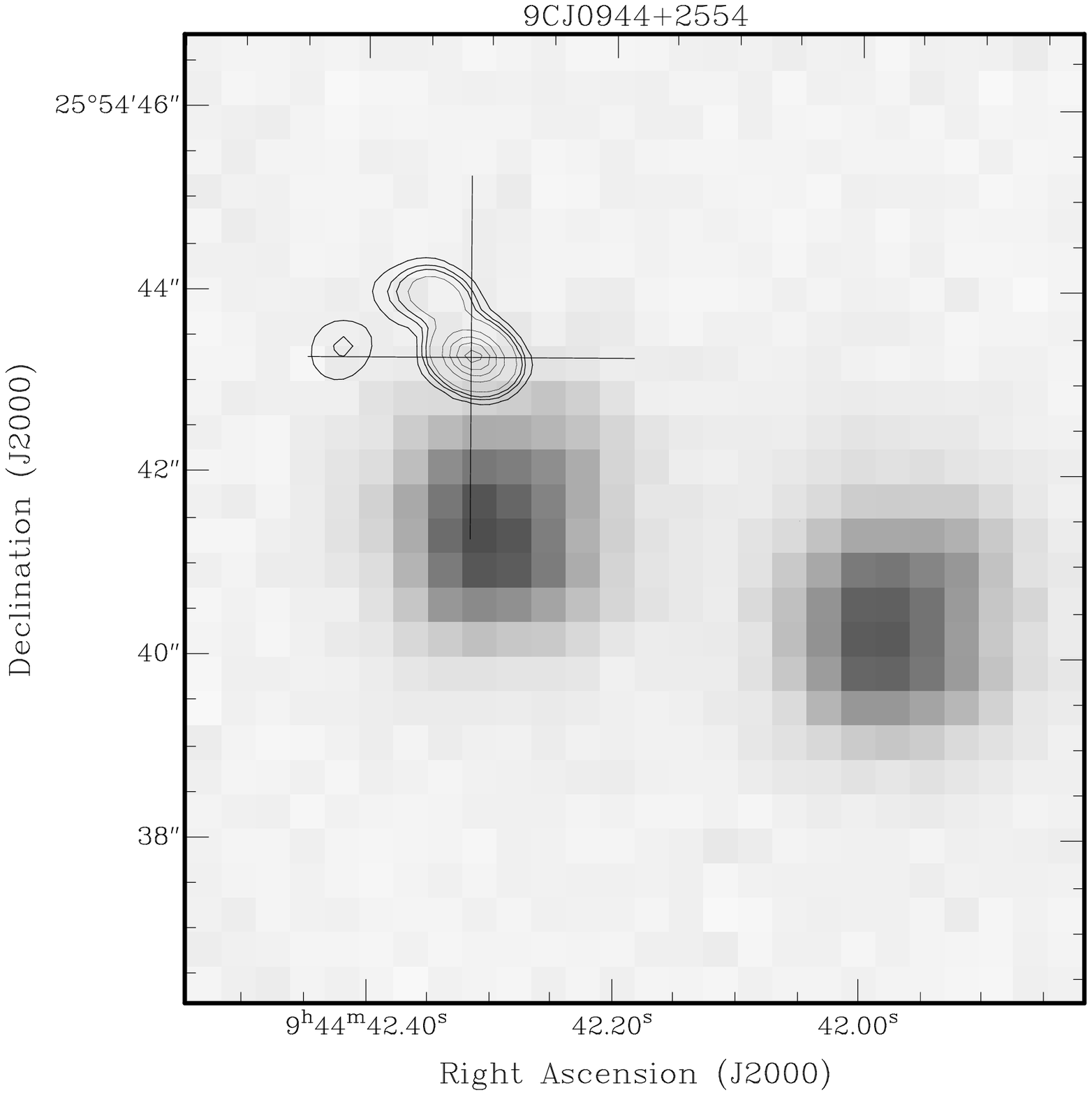 ,width=4.0cm,clip=}\label{aj}}\quad 
\subfigure[9CJ0945+2640 (P60 \it{R}\normalfont)]{\epsfig{figure=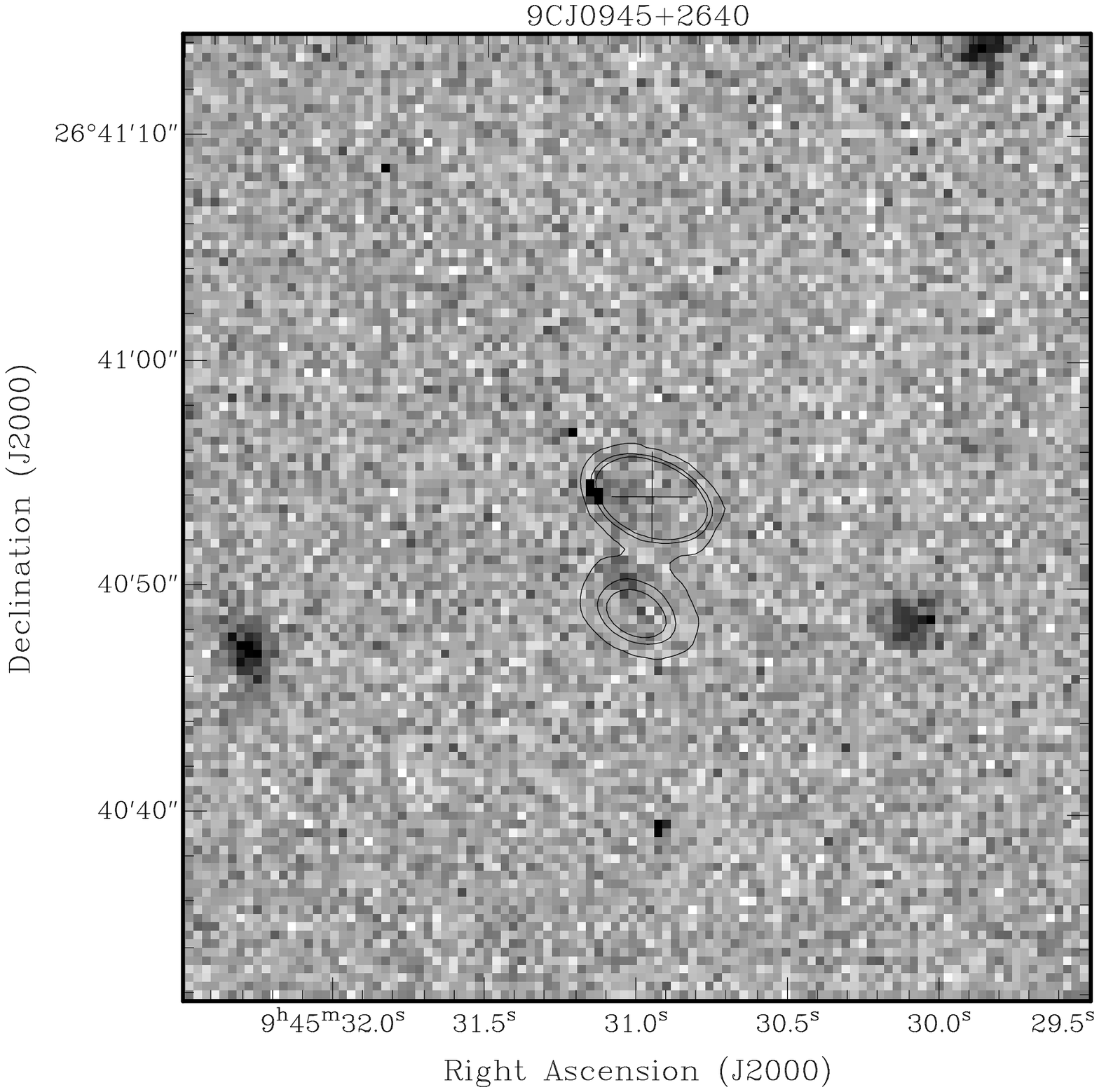 ,width=4.0cm,clip=}\label{ak}}\quad 
\subfigure[9CJ0945+2640 (P60 \it{R}\normalfont)]{\epsfig{figure=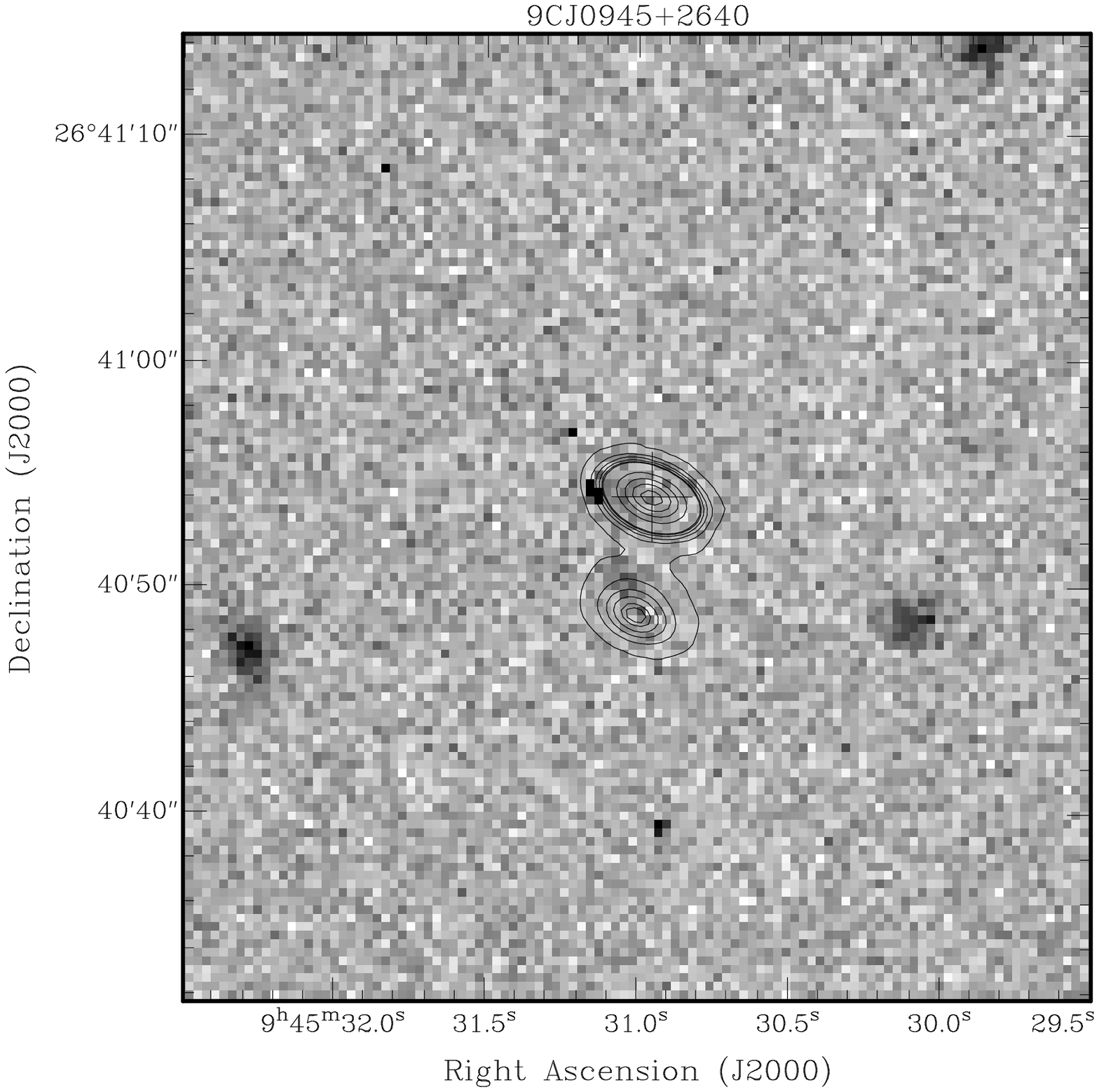 ,width=4.0cm,clip=}\label{al}}}\caption{Optical counterparts for sources 9CJ0937+3411 to 9CJ0945+2640. Crosses mark maximum radio flux density and are 4\,arcsec top to bottom. Contours: \ref{ae}, 1.4\,GHz contours at 1,2,3,4,5\,\% of peak (52.4\,mJy/beam); \ref{af}, 1.4\,GHz radio contours: 1-9 every 1\,\% and 10,15,25,25 and 30-90 every 20\,\% of peak (52.4\,mJy/beam); \ref{ag}, 1.4\,GHz contours 3,4,5,10,50,90\,\% of peak (292\,mJy/beam));\ref{ah}, 1.4\,GHz contours 2.5-9.5 every 1\,\% and 10,50,90\,\% of peak (102\,mJy/beam); \ref{ai} and \ref{aj}, 4.8\,GHz contours at 4,5,6\,\% and 10-90 every 20\,\% of peak (243\,mJy/beam); \ref{ak} 1.4\,GHz at 1-9 every 2\,\% and 10-90 every 20\,\% of peak (485\,mJy/beam) -- an optical counterpart is just seen; \ref{al}, 1.4\,GHz at 1-9 every 2\,\% and 10-90 every 20\,\% of peak (485\,mJy/beam).}\end{figure*}

\begin{figure*}
\mbox{
\subfigure[9CJ0945+2729 (P60 \it{R}\normalfont)]{\epsfig{figure=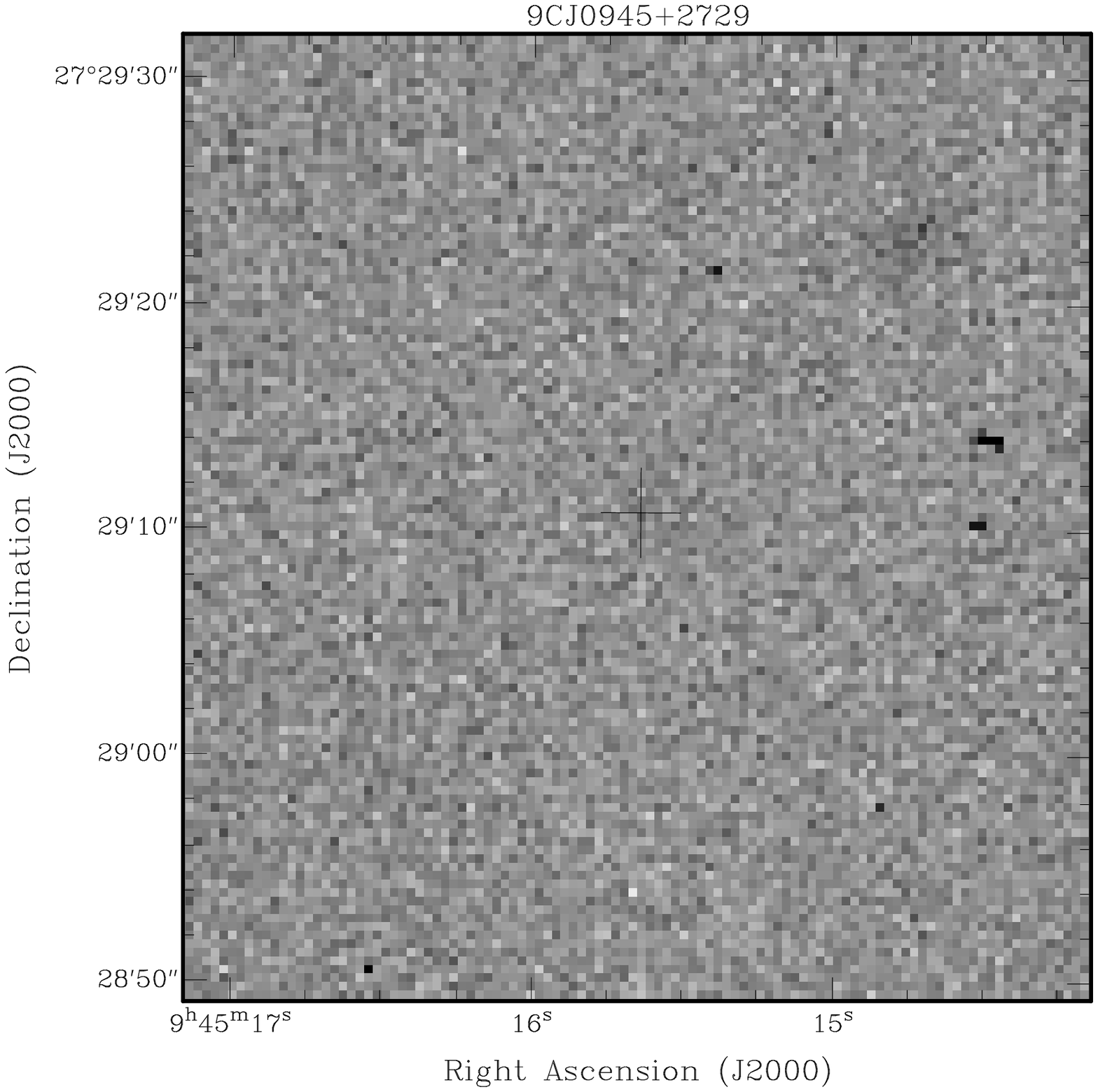 ,width=4.0cm,clip=}}\quad 
\subfigure[9CJ0945+3534 (DSS2 \it{R}\normalfont)]{\epsfig{figure=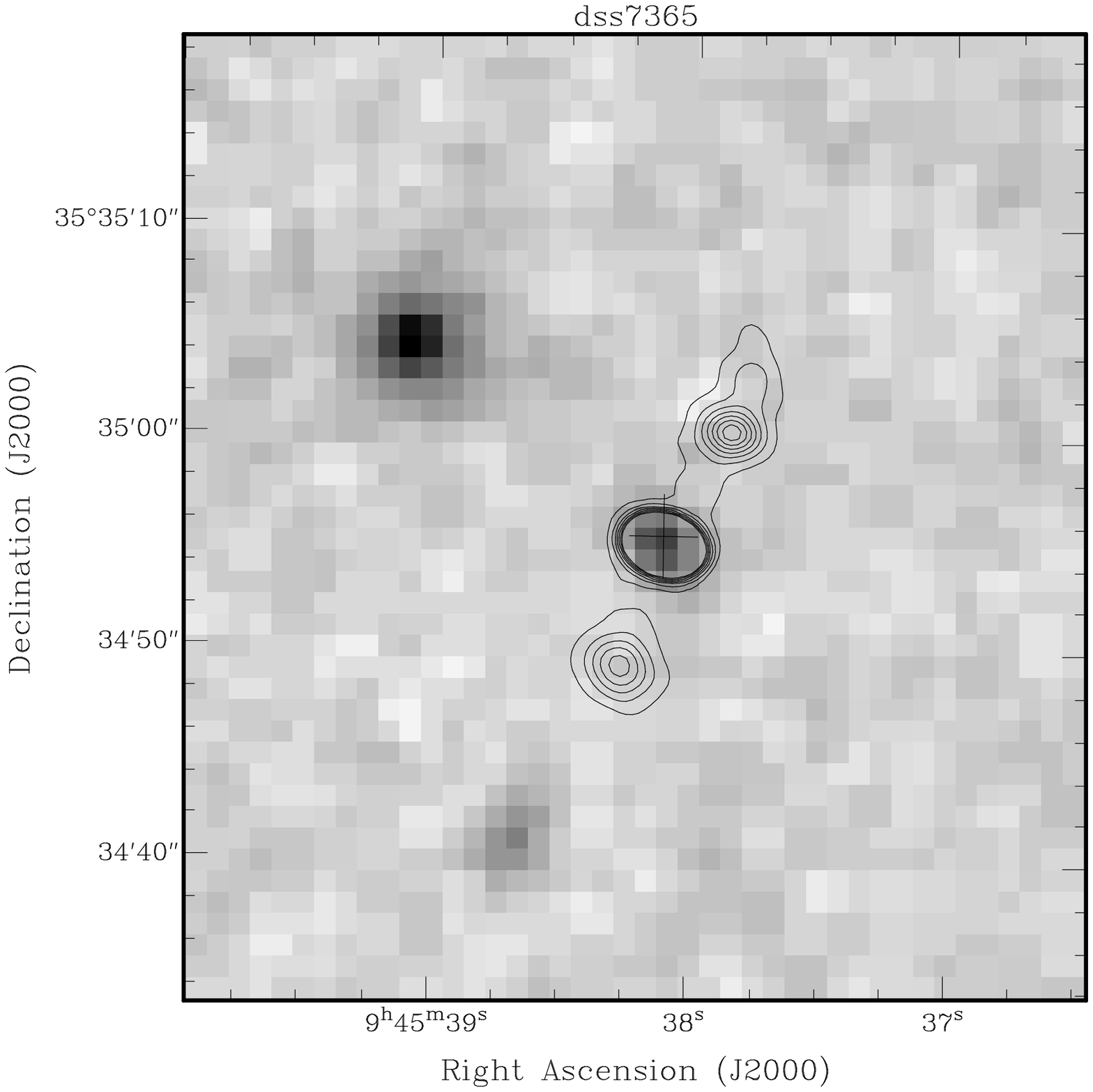 ,width=4.0cm,clip=}\label{am}}\quad 
\subfigure[9CJ0945+3534 (DSS2 \it{R}\normalfont) Detail]{\epsfig{figure=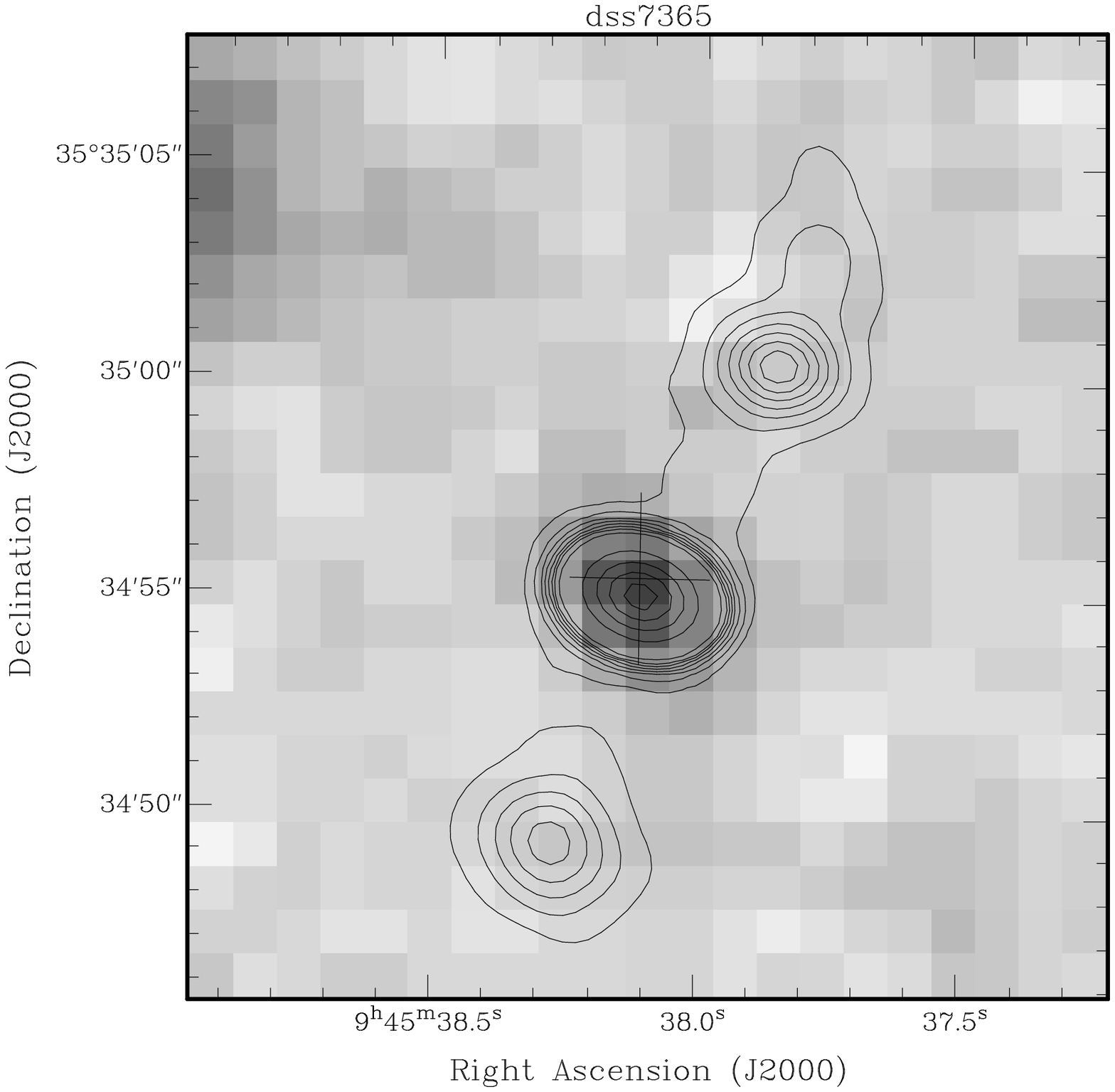 ,width=4.0cm,clip=}\label{an}}
} 
\mbox{
\subfigure[9CJ0945+3003 (P60 \it{R}\normalfont)]{\epsfig{figure=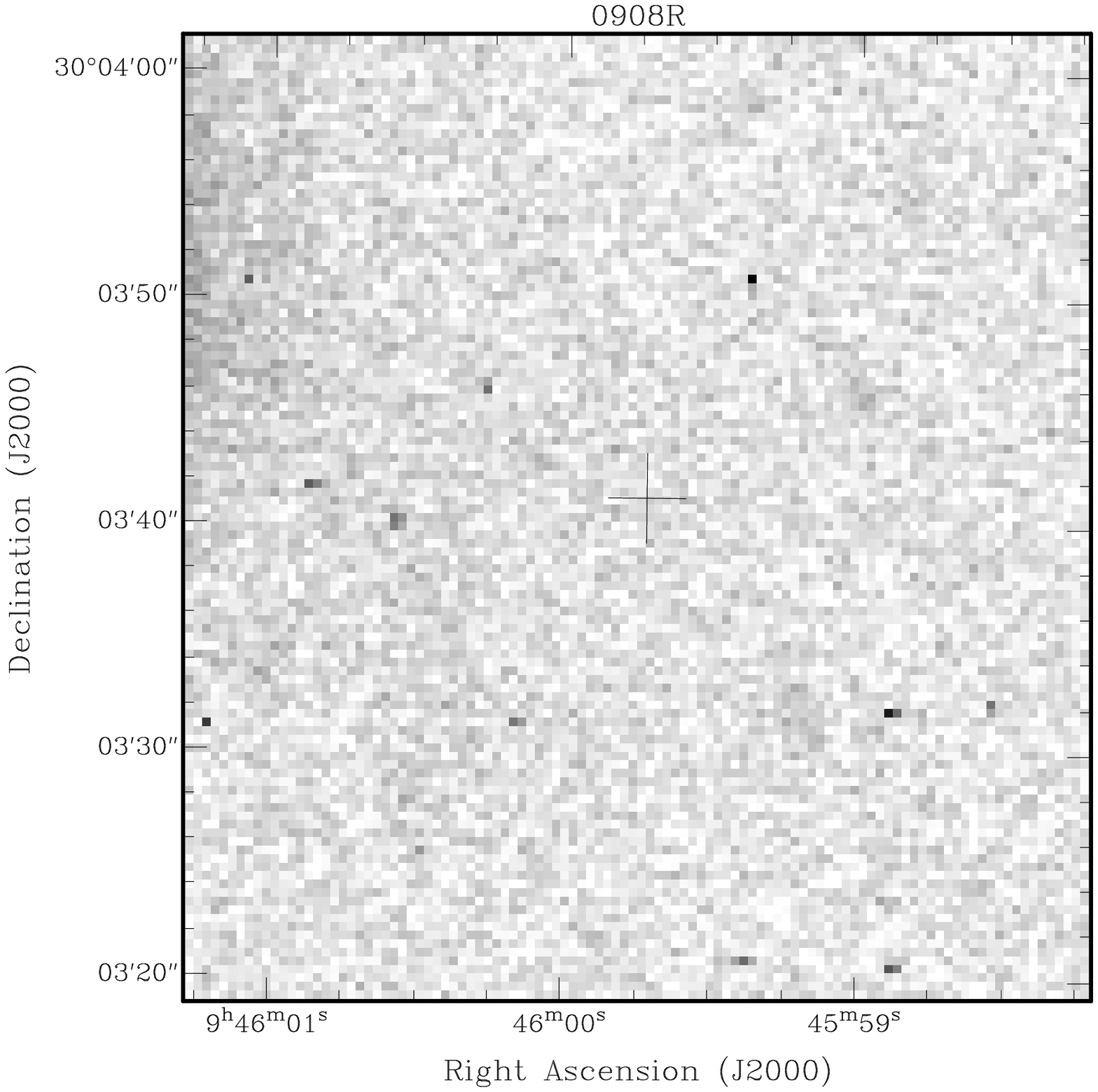 ,width=4.0cm,clip=}}\quad 
\subfigure[9CJ0949+2920 (P60 \it{R}\normalfont)]{\epsfig{figure=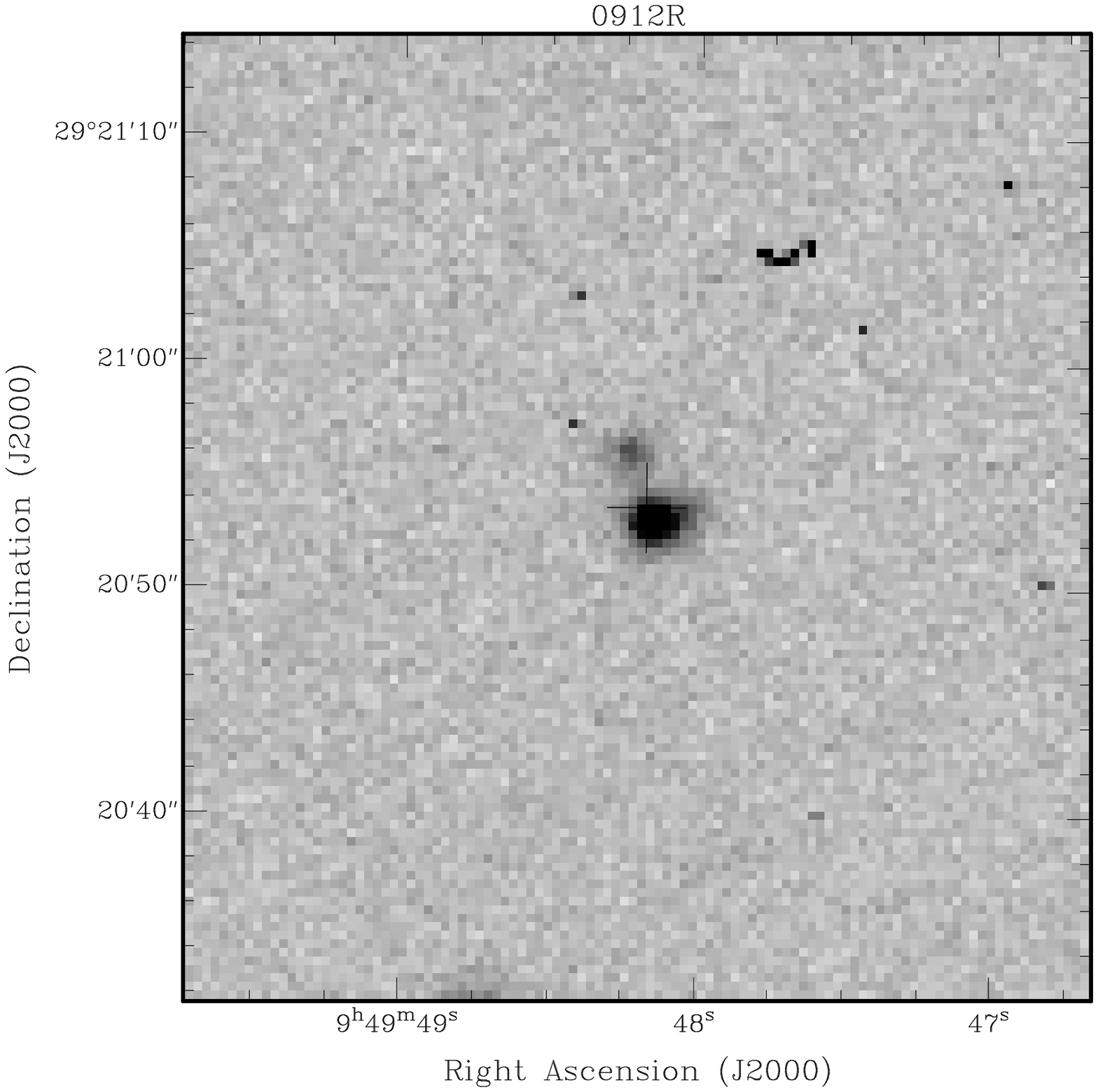 ,width=4.0cm,clip=}}\quad 
\subfigure[9CJ0952+2828 (22\,GHz radio map)]{\epsfig{figure=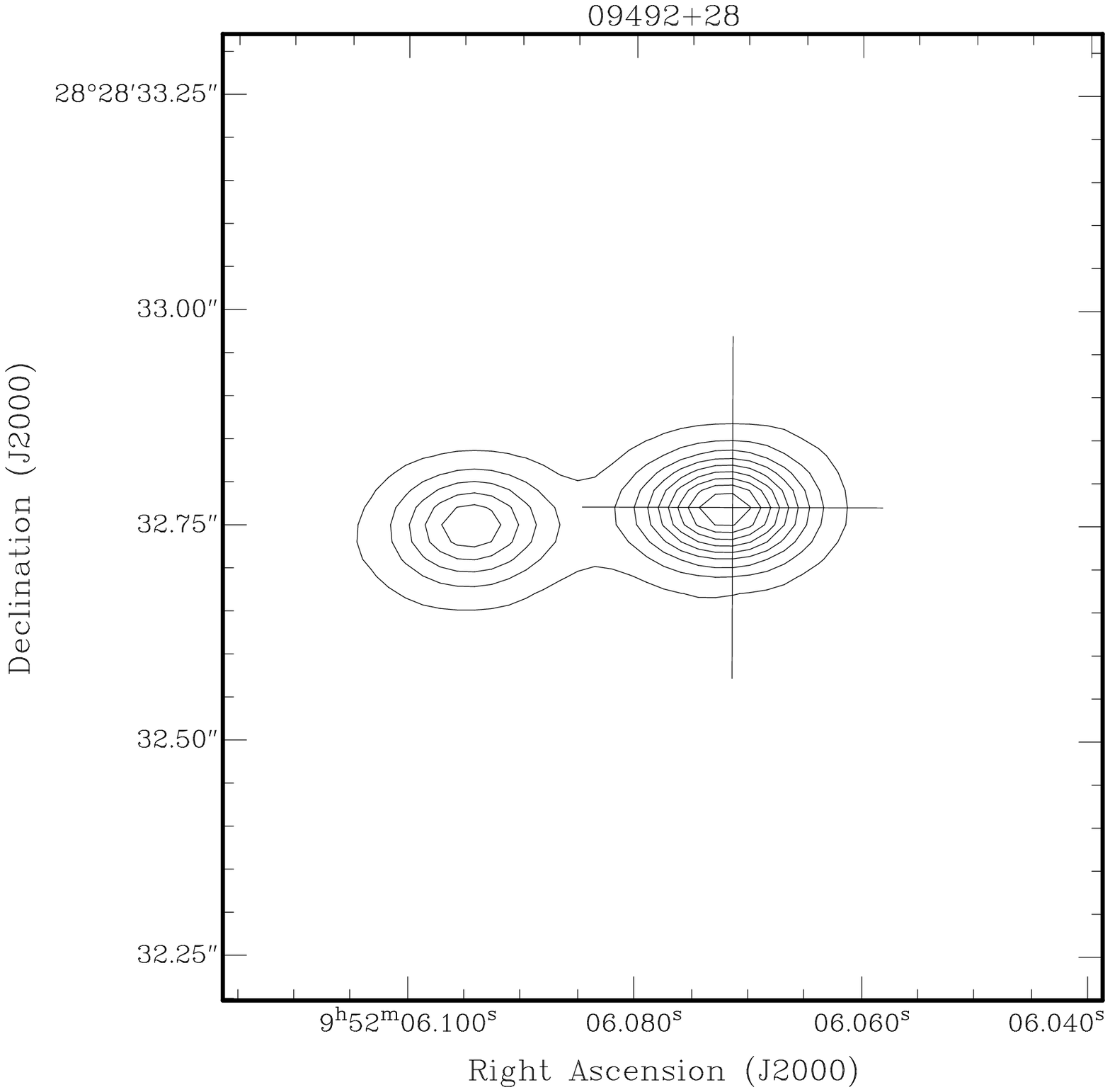 ,width=4.0cm,clip=}\label{ao}}
} 
\mbox{
\subfigure[9CJ0952+2828 (DSS2 \it{R}\normalfont)]{\epsfig{figure=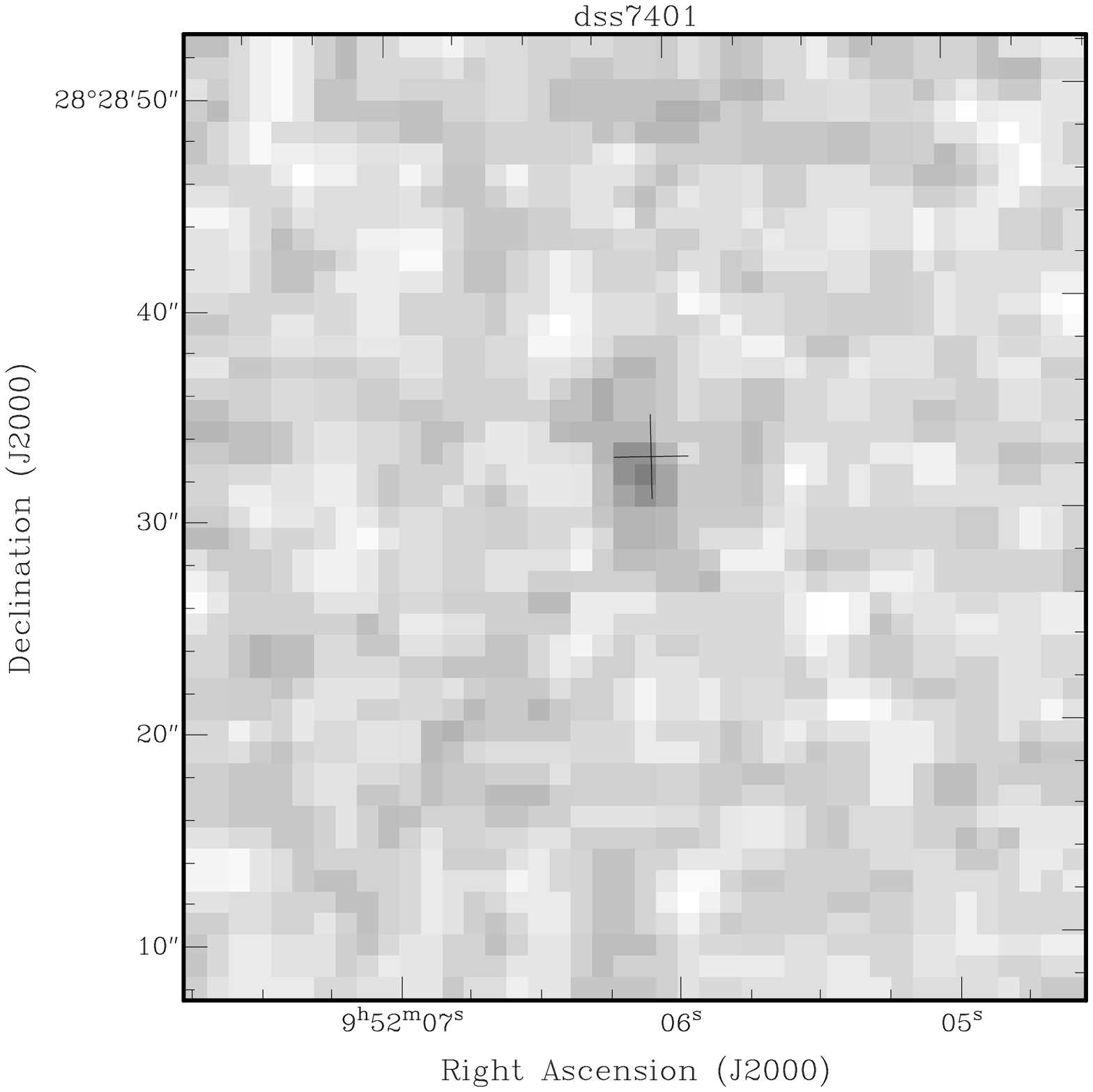 ,width=4.0cm,clip=}\label{ap}}\quad 
\subfigure[9CJ0952+3512 (DSS2 \it{R}\normalfont)]{\epsfig{figure=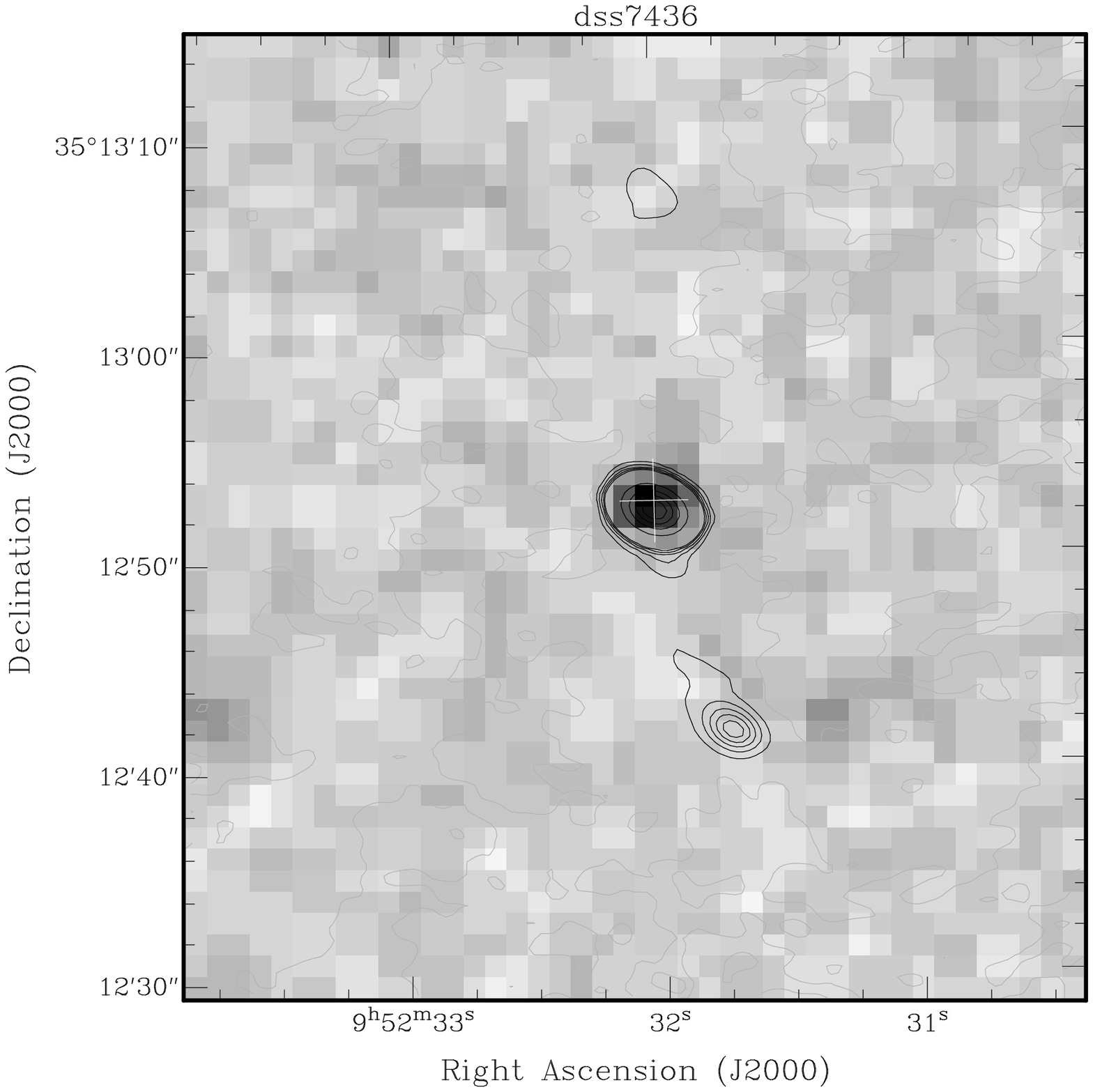 ,width=4.0cm,clip=}\label{aq}}\quad 
\subfigure[9CJ0953+3225 (P60 \it{R}\normalfont)]{\epsfig{figure=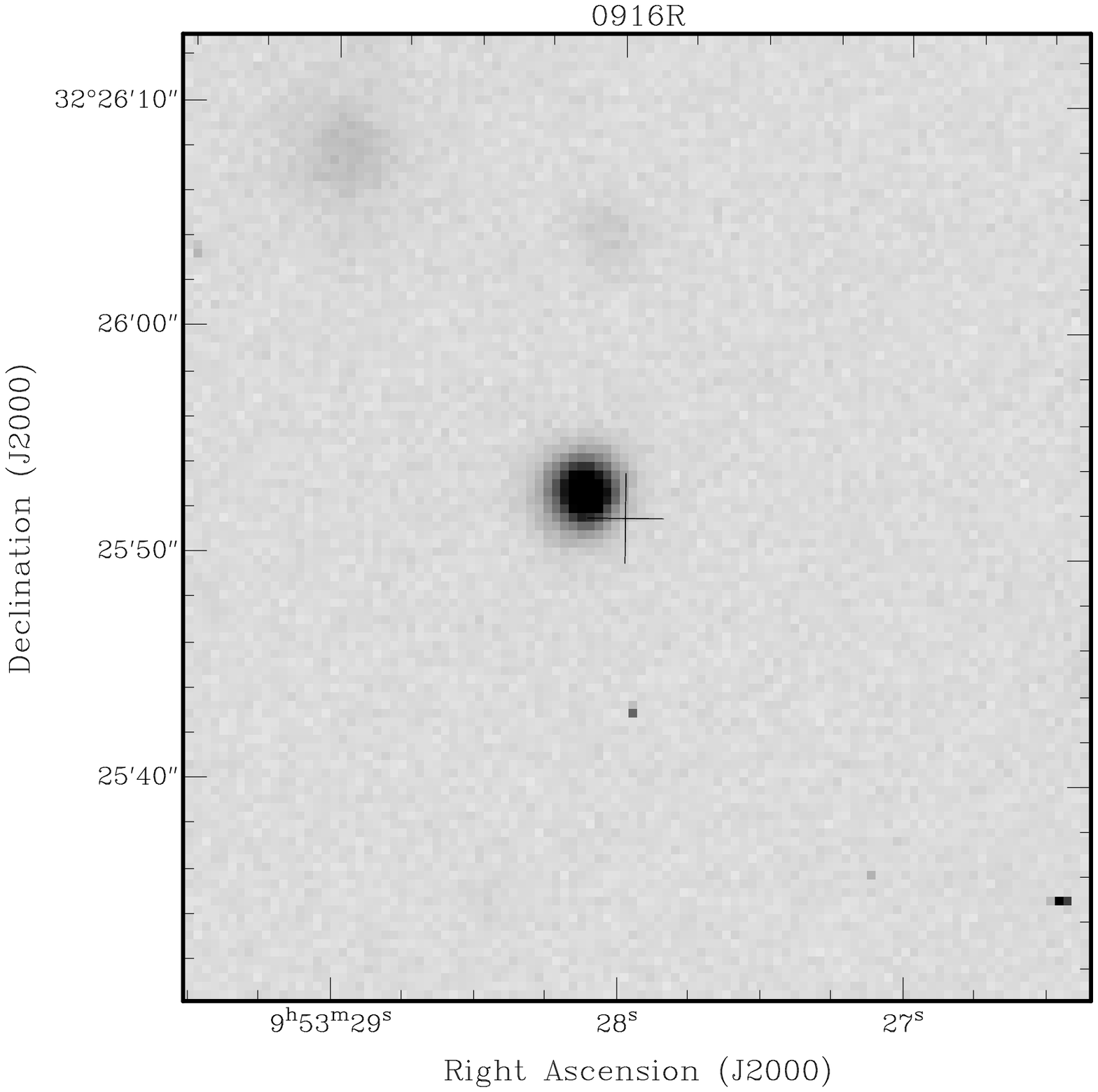 ,width=4.0cm,clip=}}
} 
\mbox{
\subfigure[9CJ0954+2639 (DSS2 \it{R}\normalfont)]{\epsfig{figure=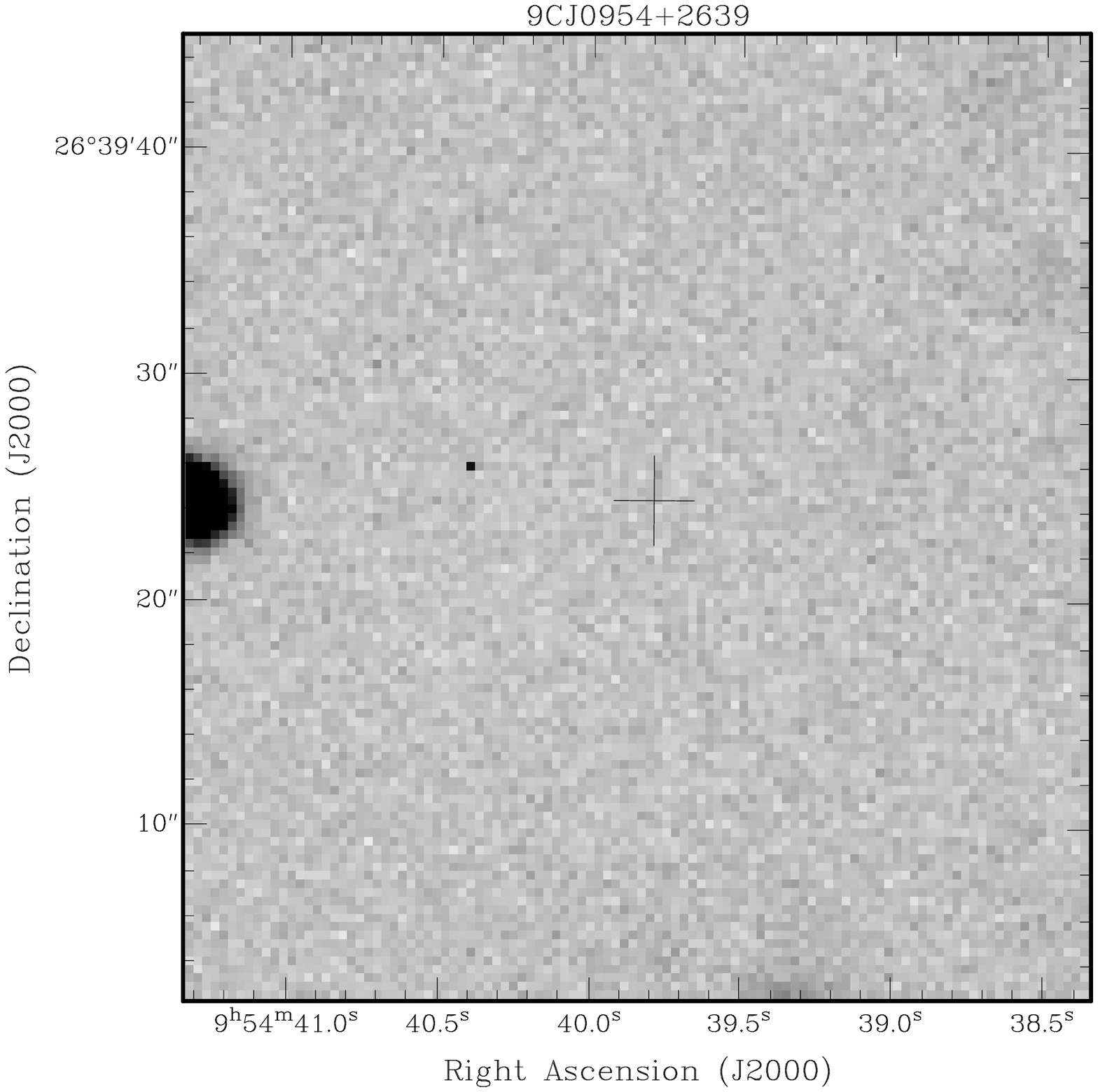 ,width=4.0cm,clip=}}\quad 
\subfigure[9CJ0954+2639 (DSS2 \it{O}\normalfont)]{\epsfig{figure=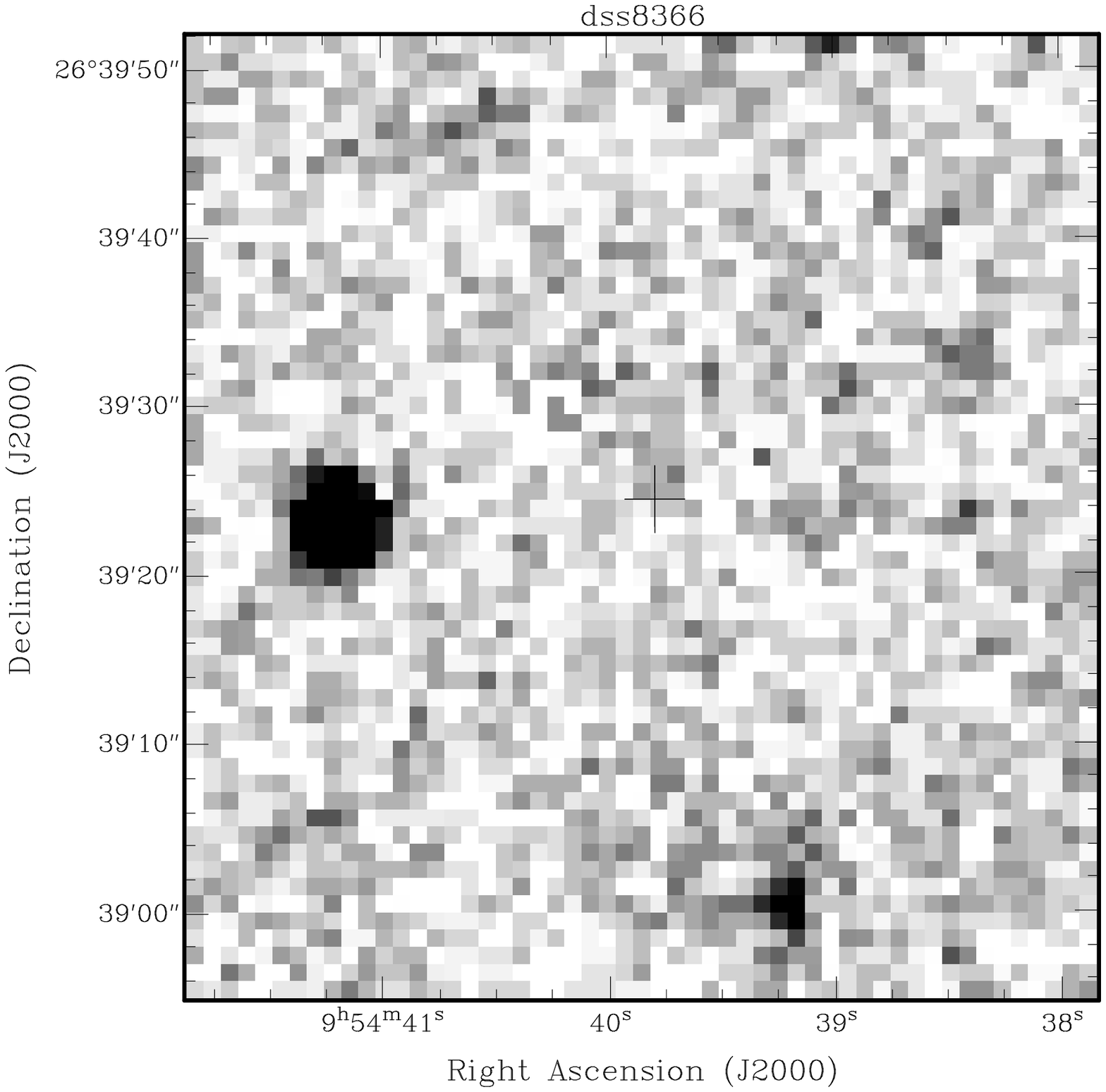 ,width=4.0cm,clip=}}\quad 
\subfigure[9CJ0955+3335 (P60 \it{R}\normalfont)]{\epsfig{figure=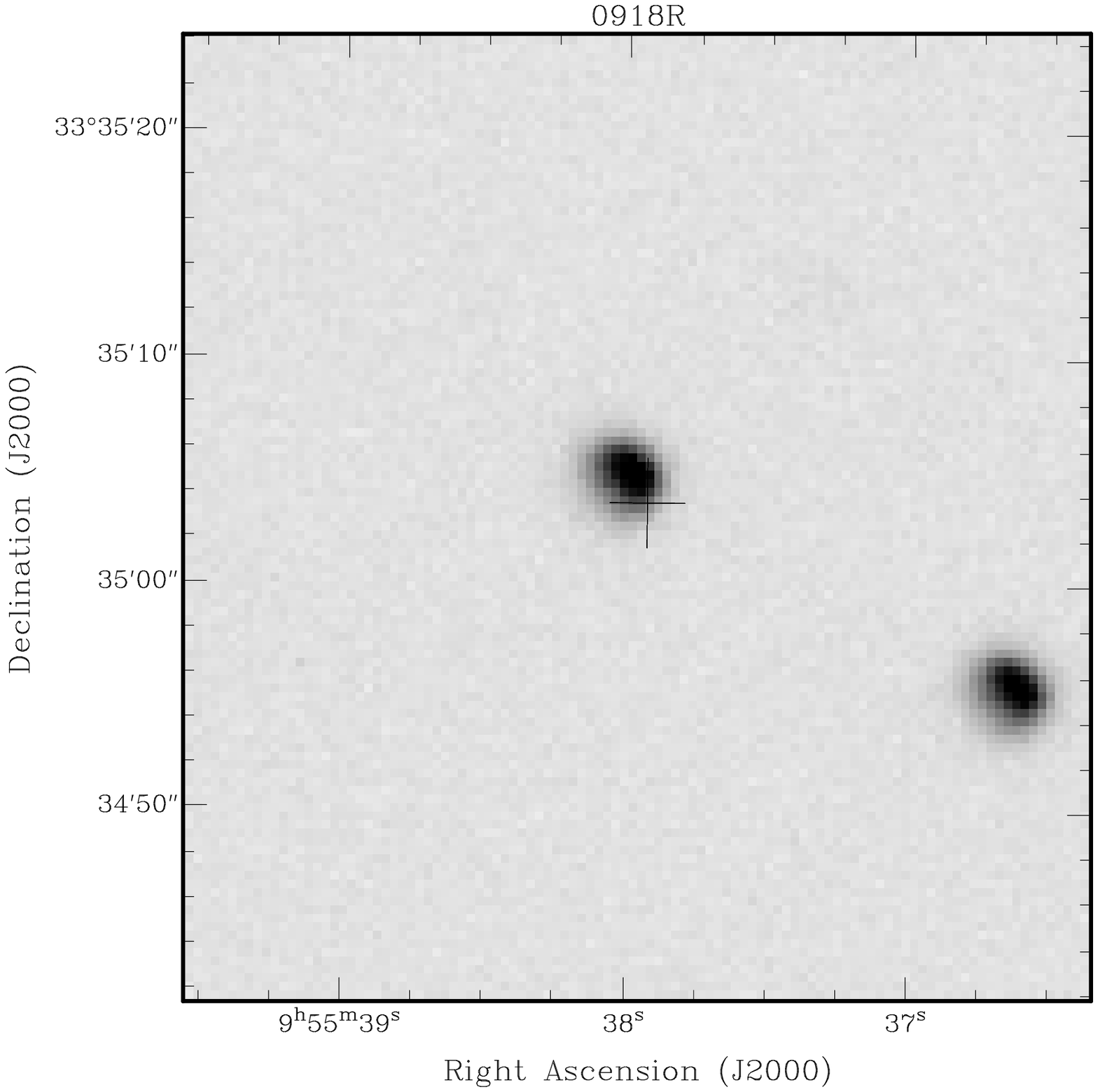 ,width=4.0cm,clip=}}}\caption{Optical counterparts for sources 9CJ0945+2729 to 9CJ0955+3335. Crosses mark maximum radio flux density and are 4\,arcsec top to bottom. Contours: \ref{am} and \ref{an} 1.4\,GHz contours 1.5-7.5 every 1\,\% and 10-90 every 10\,\% of peak (252\,mJy/beam). Fig. \ref{an} zoomed in to show radio map better; \ref{ao}, 22\,GHz contours 10-90 every 10\,\% of peak (67\,mJy/beam) of this 0.03\,arcsec source; \ref{ap} optical image without radio contours; \ref{aq}, 1.4\,GHz contours at 0.5-3 every 0.5\,\% and 20-90 every 20\,\% of peak (345\,mJy/beam).}\end{figure*}

\begin{figure*}
\mbox{
\subfigure[9CJ0957+3422 (DSS2 \it{R}\normalfont)]{\epsfig{figure=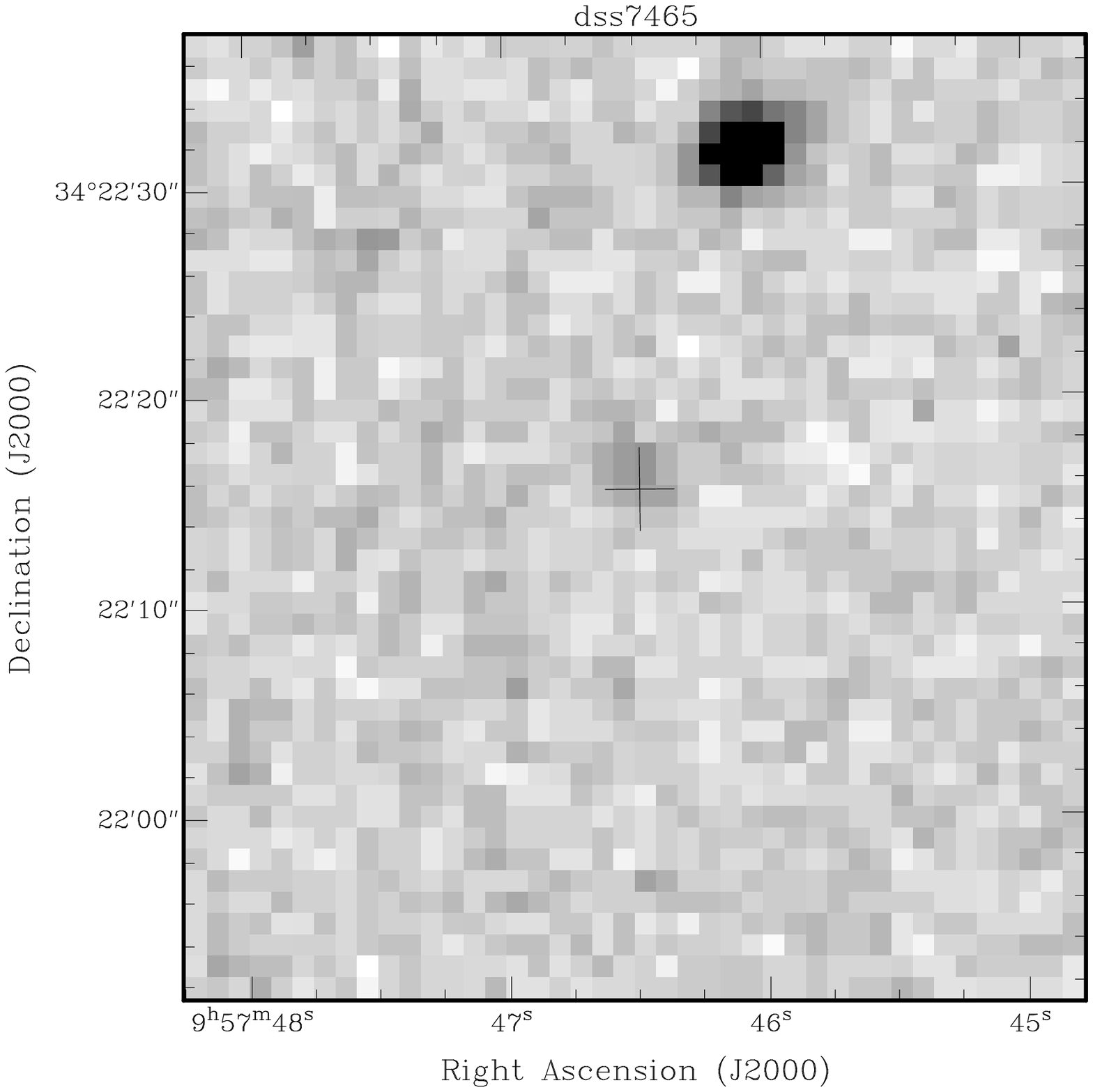 ,width=4.0cm,clip=}}\quad 
\subfigure[9CJ0958+2640 (DSS2 \it{R}\normalfont)]{\epsfig{figure=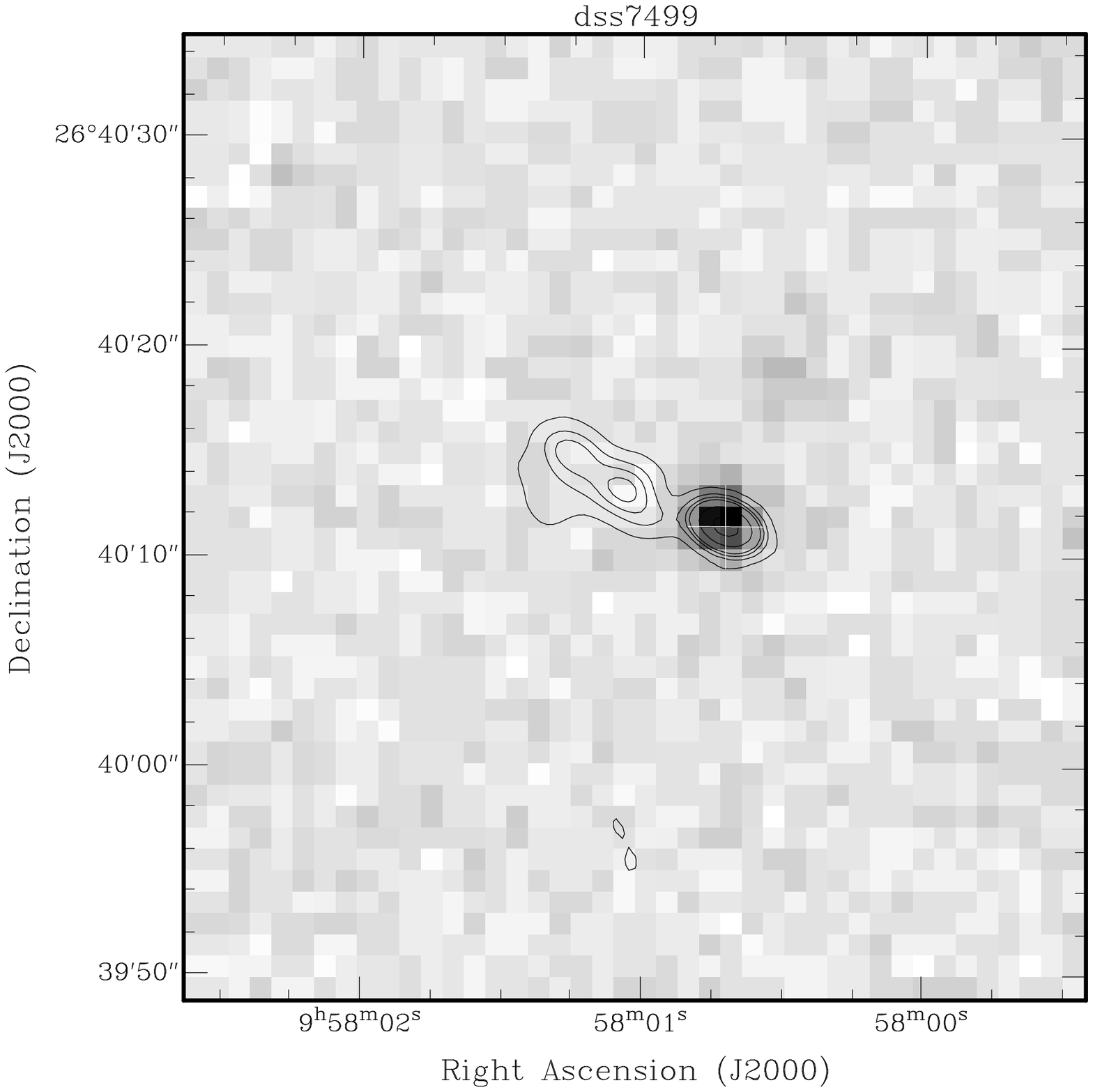 ,width=4.0cm,clip=}\label{ar}}\quad 
\subfigure[9CJ0958+3224 (DSS2 \it{R}\normalfont)]{\epsfig{figure=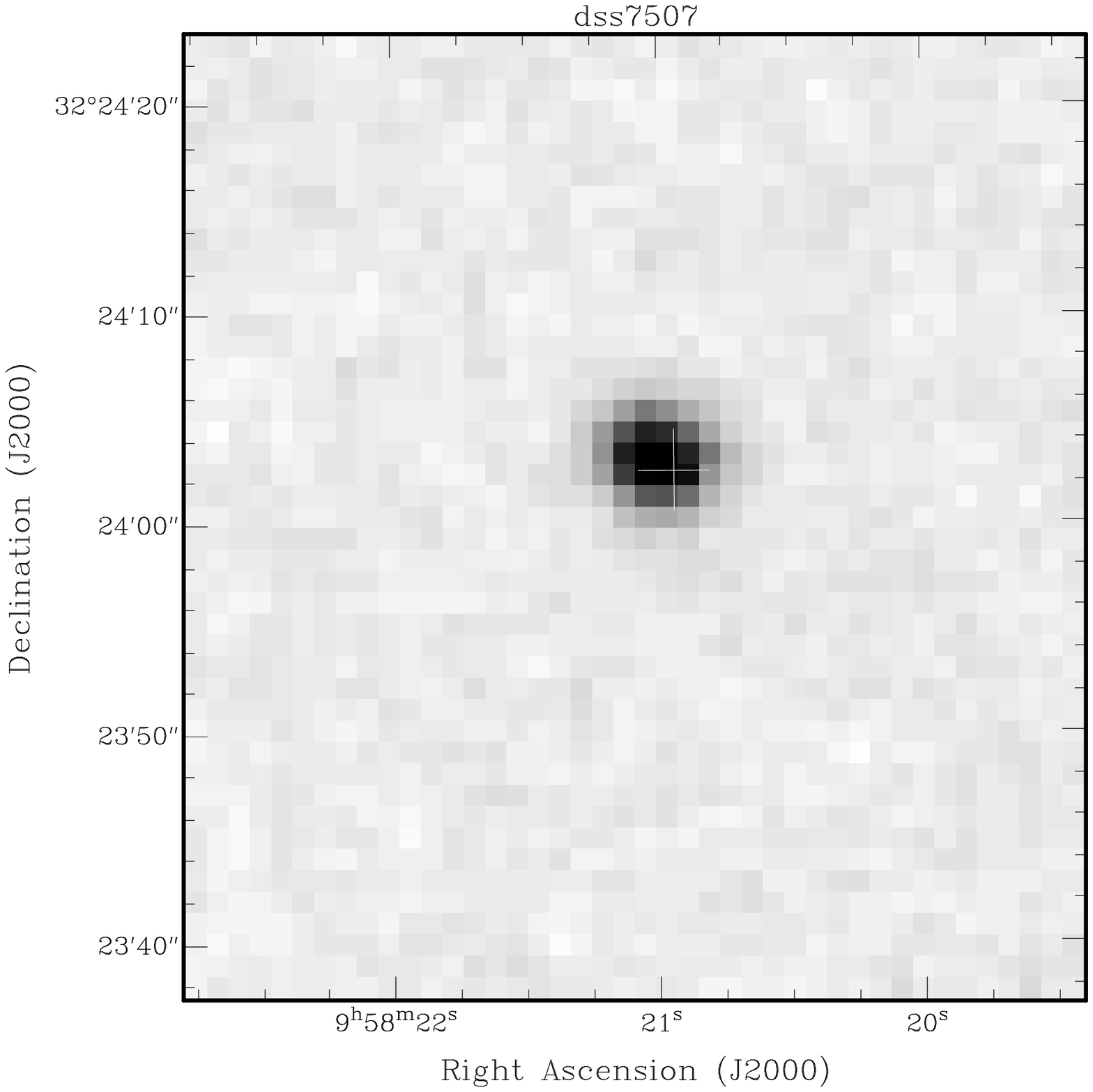 ,width=4.0cm,clip=}}
} 
\mbox{
\subfigure[9CJ0958+2948 (DSS2 \it{R}\normalfont)]{\epsfig{figure=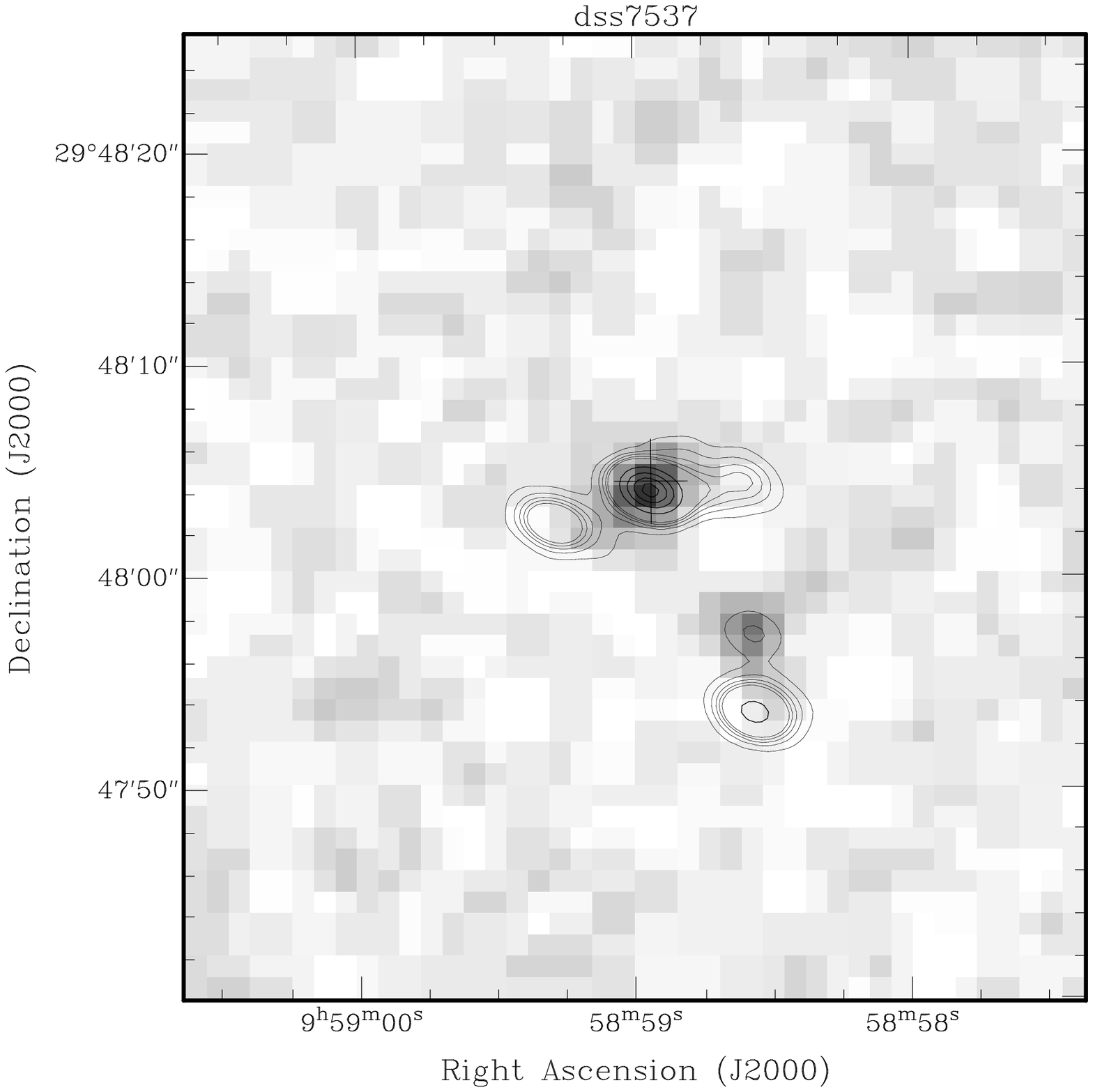 ,width=4.0cm,clip=}\label{9CJ0958+2948}}\quad 
\subfigure[9CJ0959+2512 (P60 \it{R}\normalfont)]{\epsfig{figure=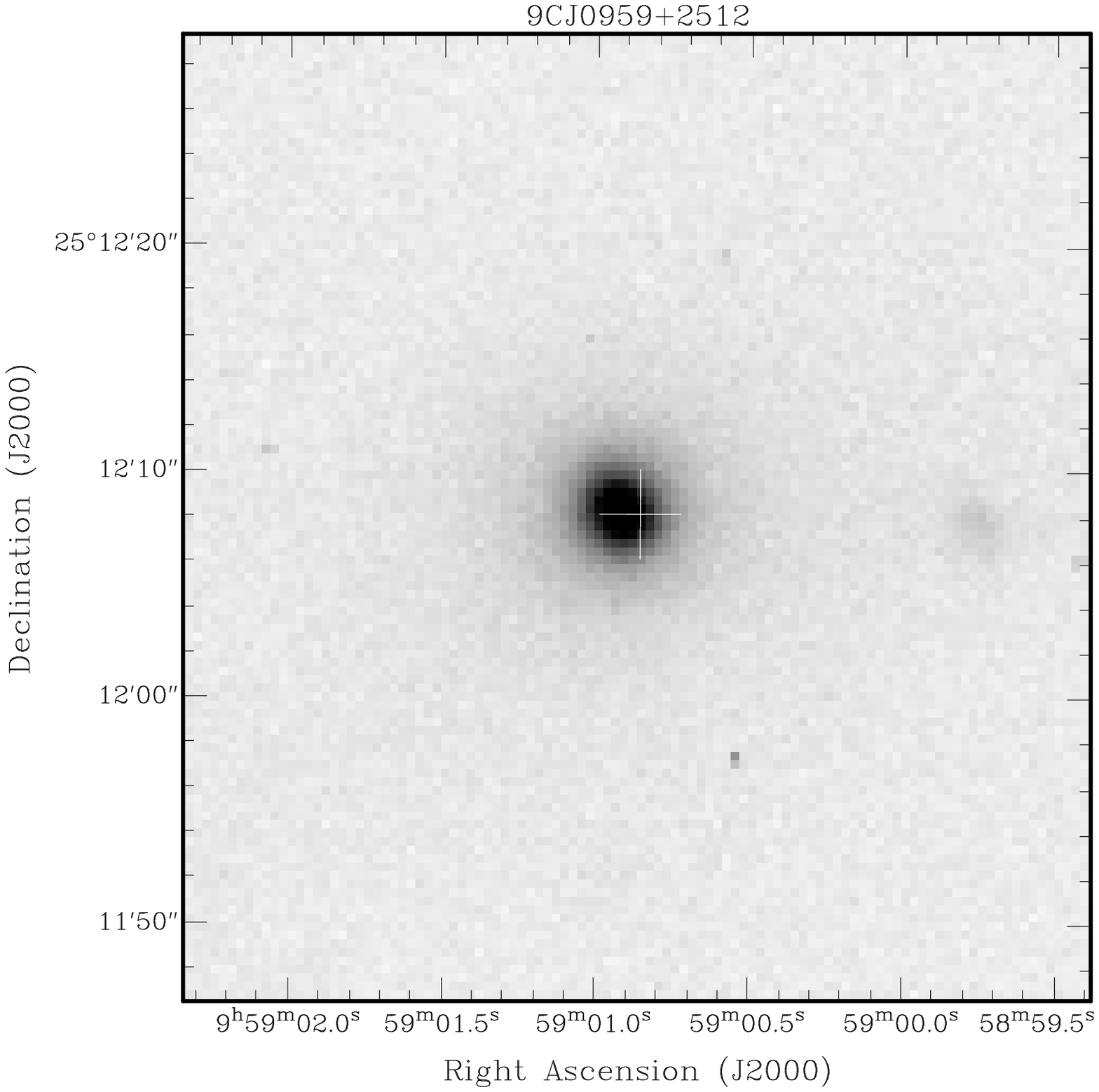 ,width=4.0cm,clip=}}\quad 
\subfigure[9CJ0959+2645 (P60 \it{R}\normalfont)]{\epsfig{figure=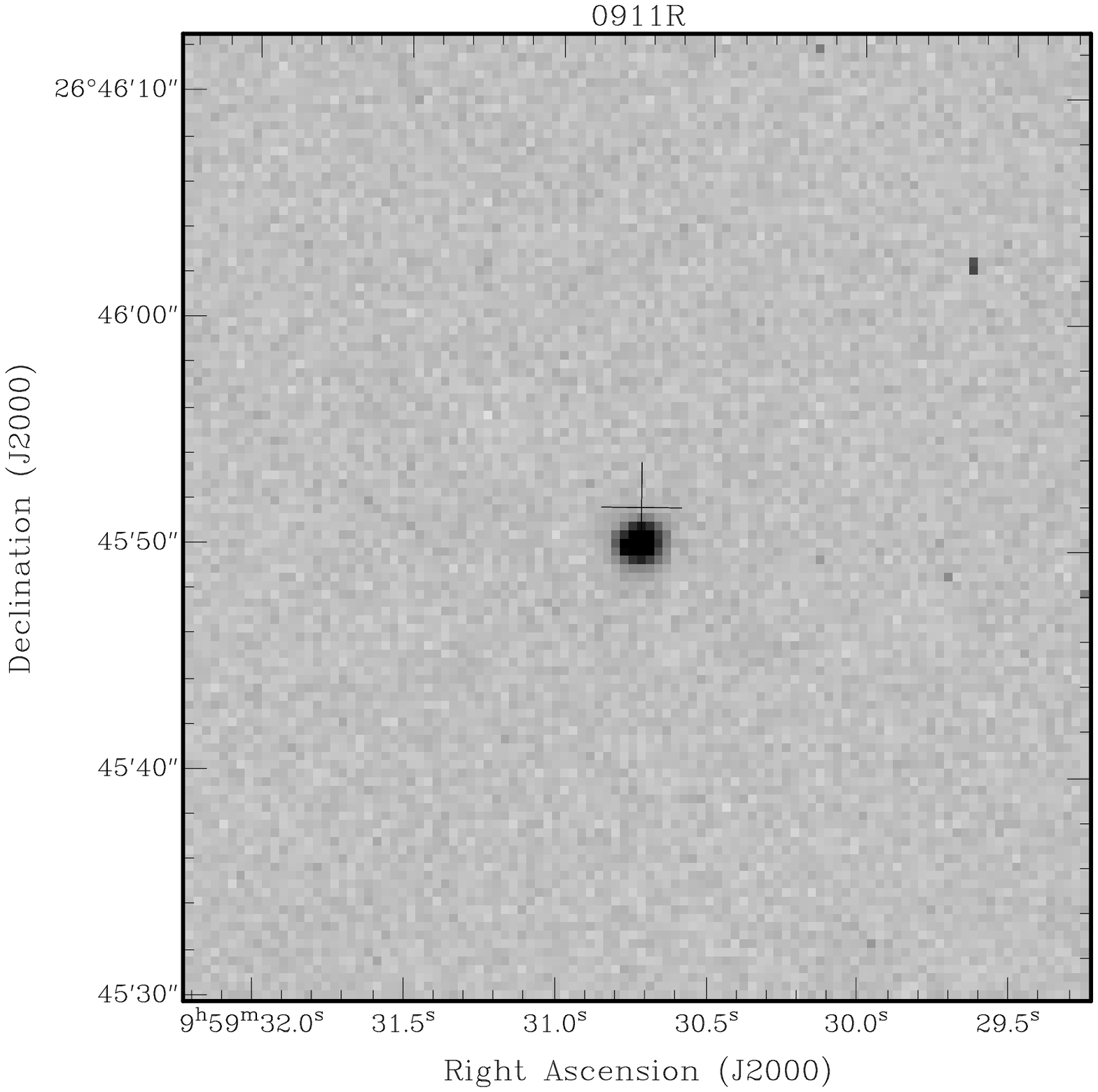 ,width=4.0cm,clip=}}
} 
\mbox{
\subfigure[9CJ1000+2752 (P60 \it{R}\normalfont)]{\epsfig{figure=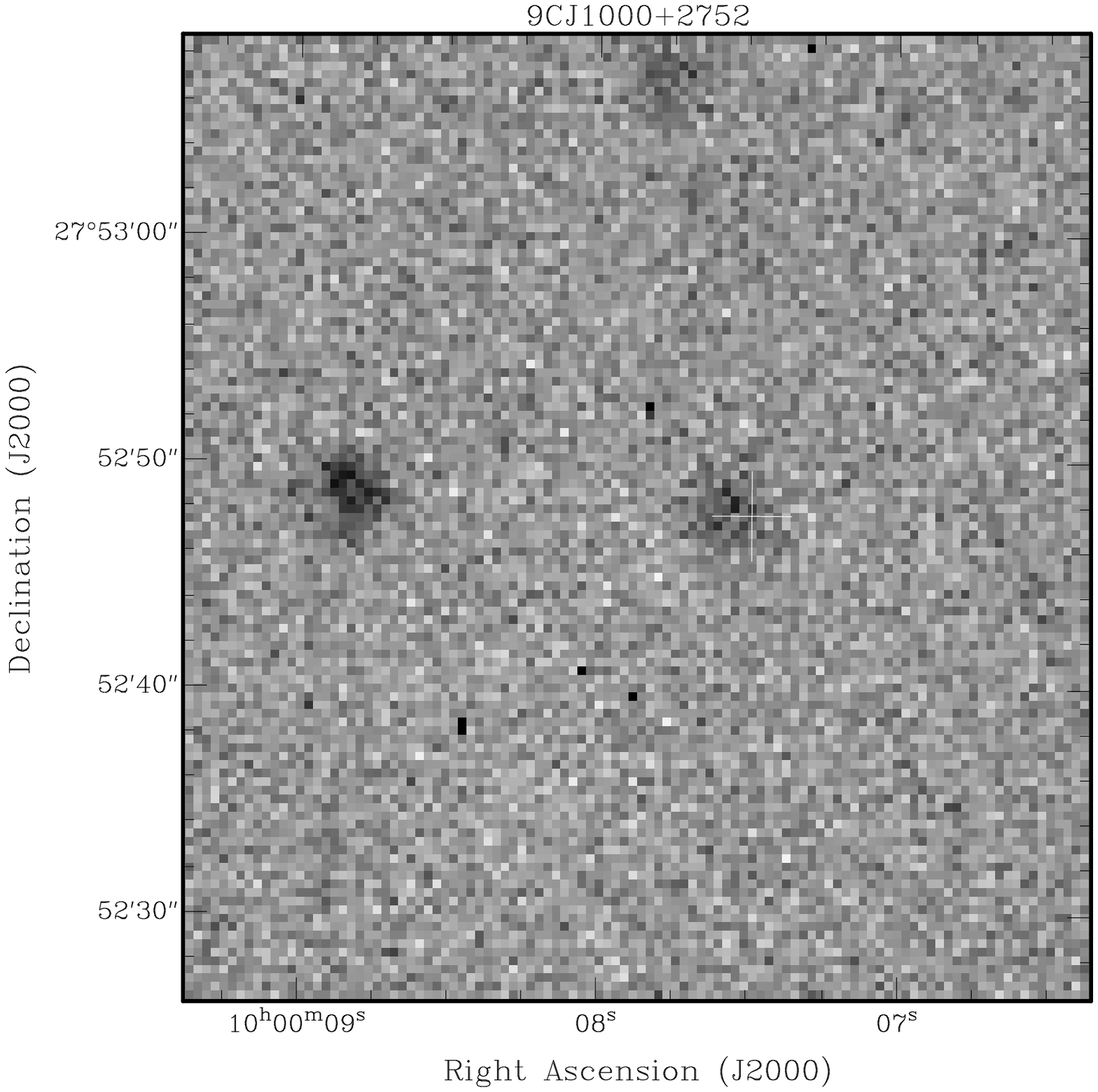 ,width=4.0cm,clip=}}\quad 
\subfigure[9CJ1002+3409 (P60 \it{R}\normalfont)]{\epsfig{figure=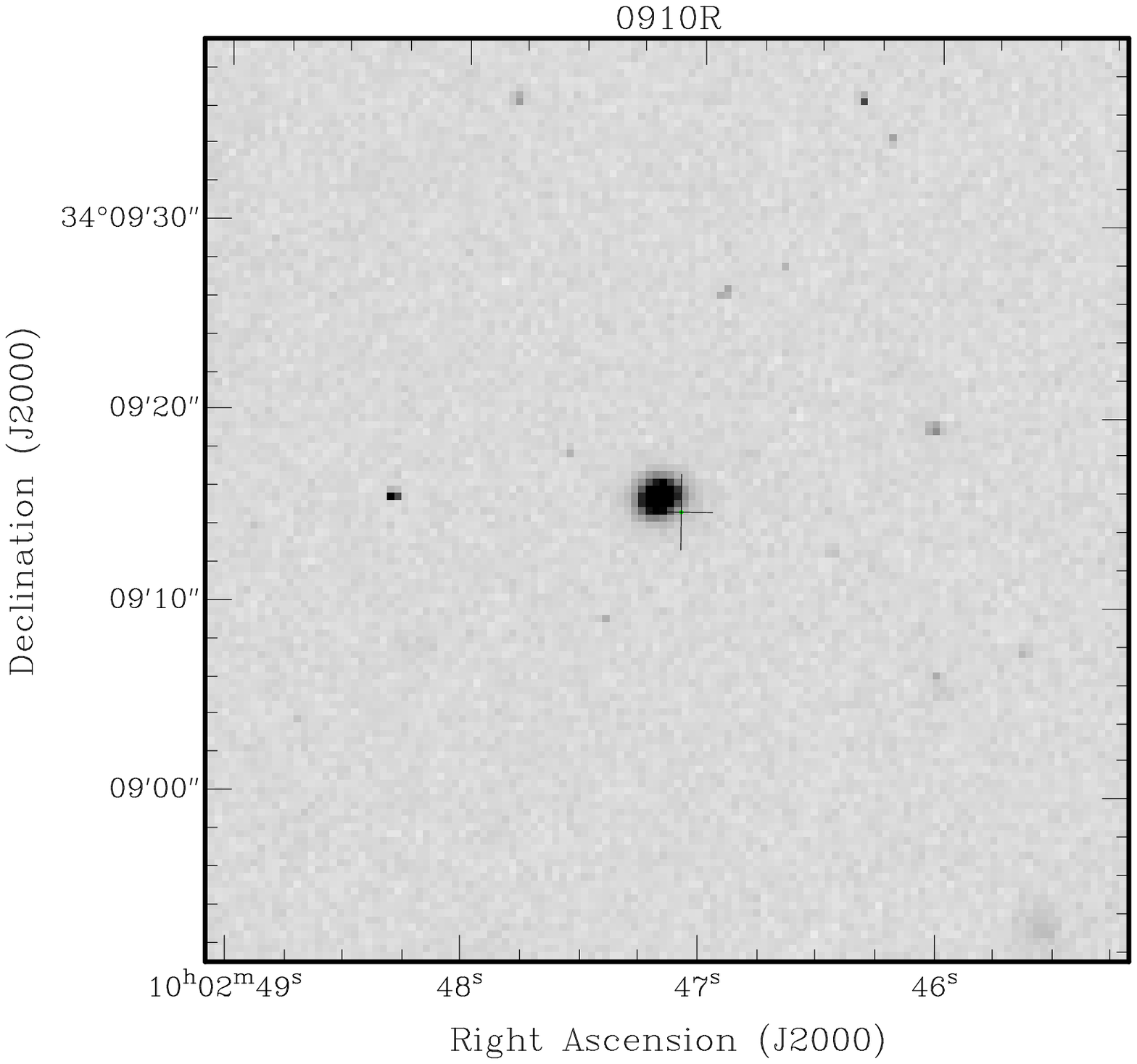 ,width=4.0cm,clip=}}\quad 
\subfigure[9CJ1003+3347 (DSS2 \it{R}\normalfont)]{\epsfig{figure=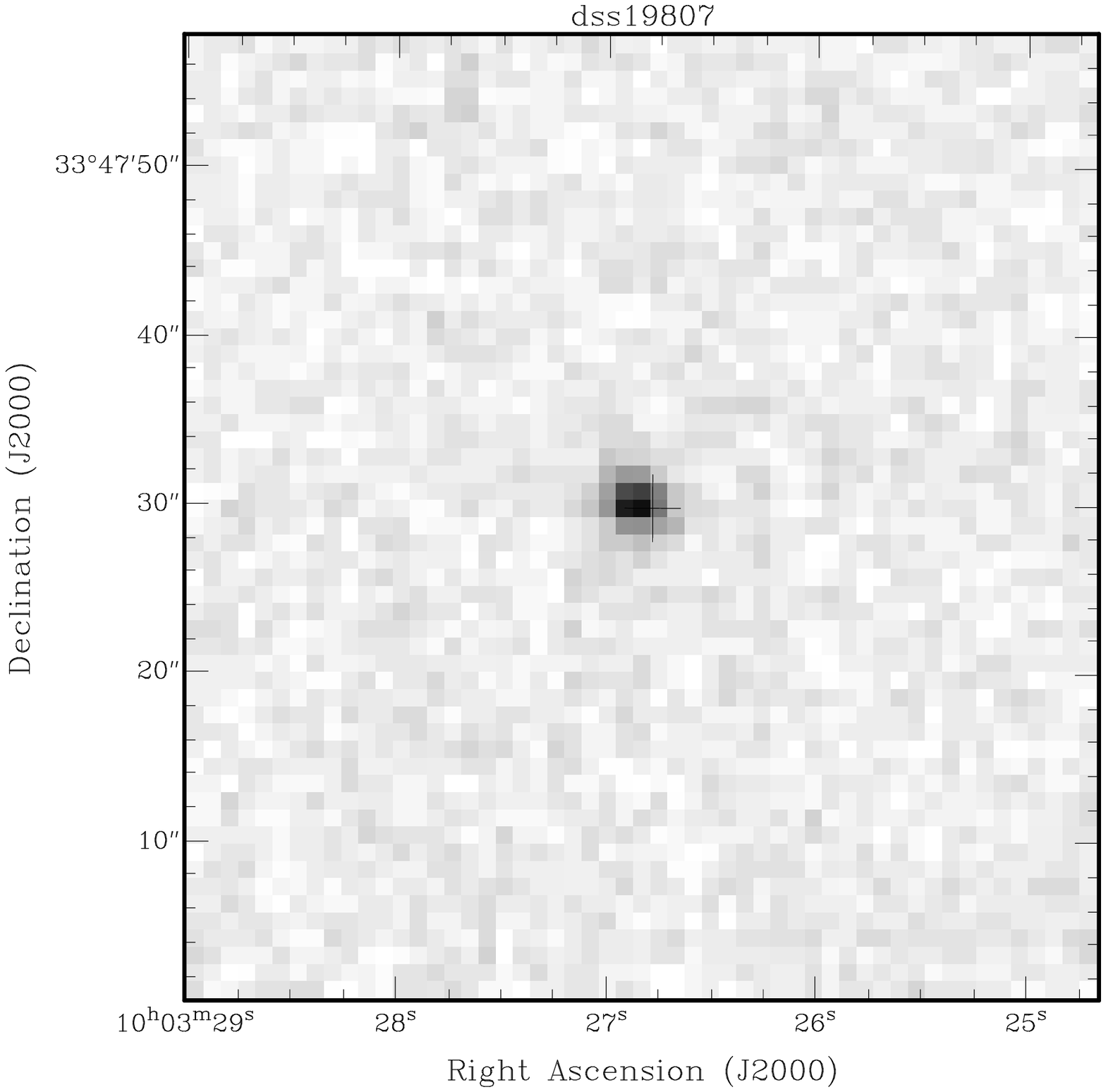 ,width=4.0cm,clip=}}
} 
\mbox{
\subfigure[9CJ1004+3010 (DSS2 \it{R}\normalfont)]{\epsfig{figure=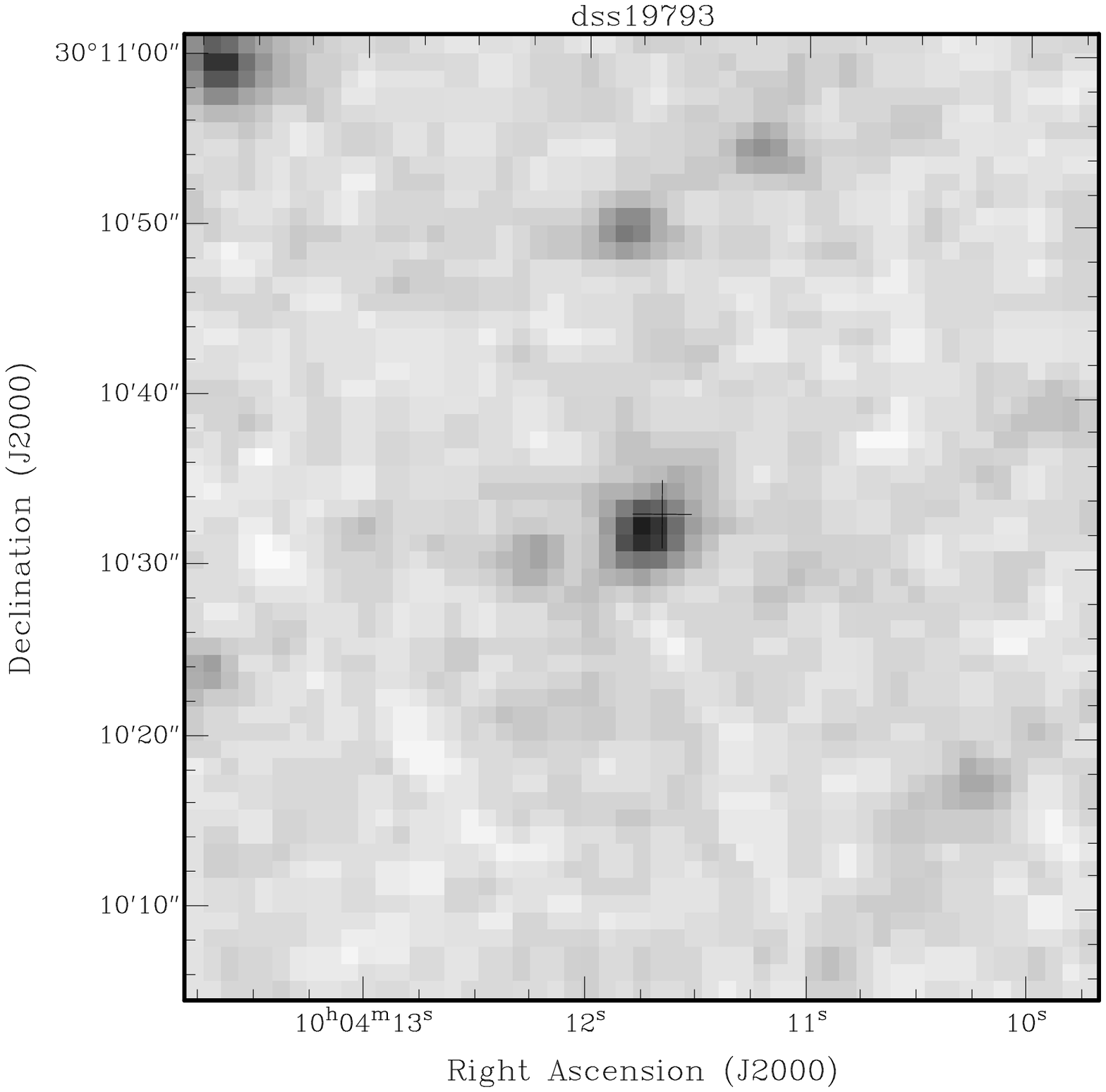 ,width=4.0cm,clip=}}\quad 
\subfigure[9CJ1501+4211 (DSS2 \it{R}\normalfont)]{\epsfig{figure=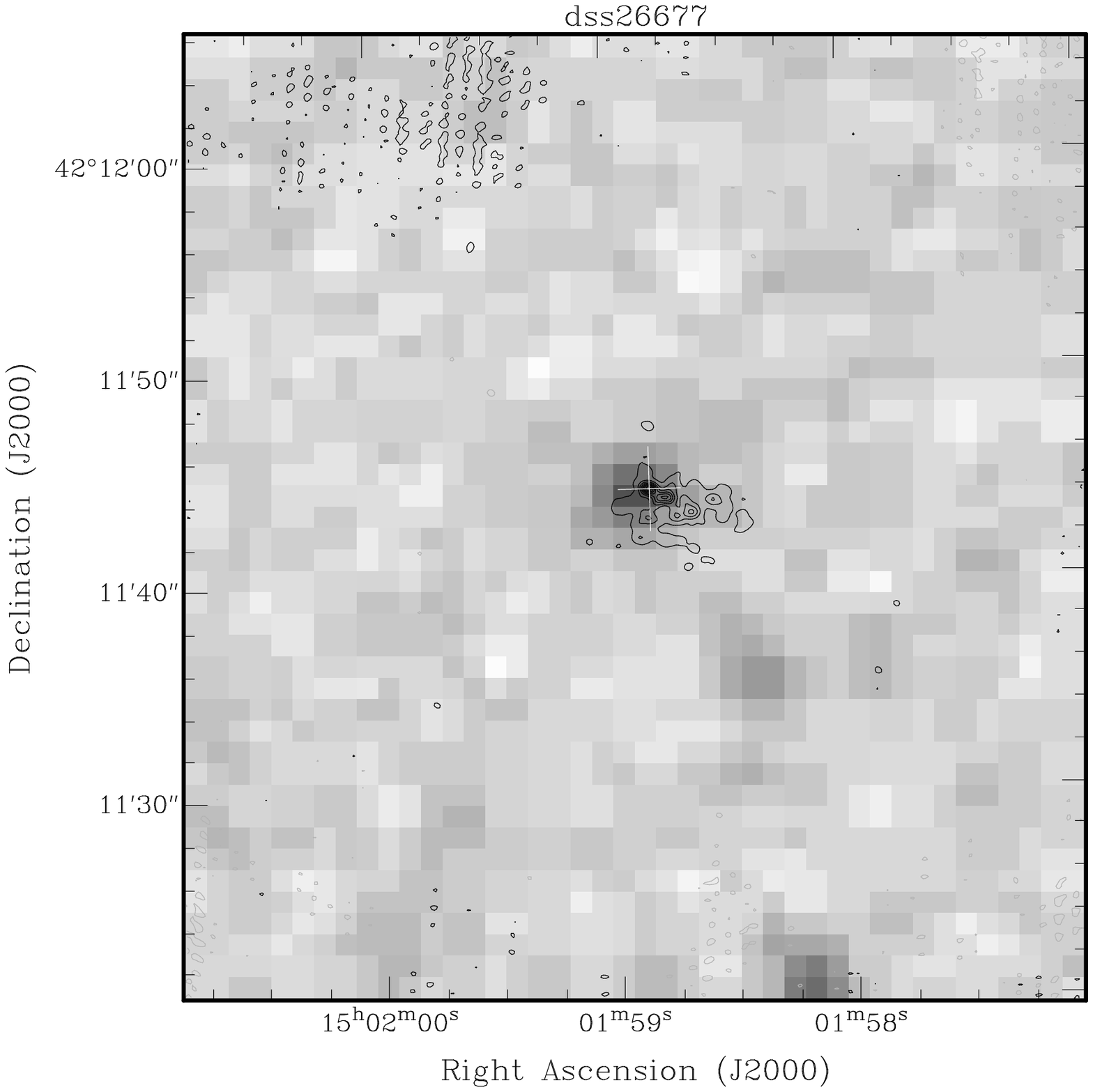 ,width=4.0cm,clip=}\label{as}}\quad 
\subfigure[9CJ1501+4211 (DSS2 \it{R}\normalfont) Detail]{\epsfig{figure=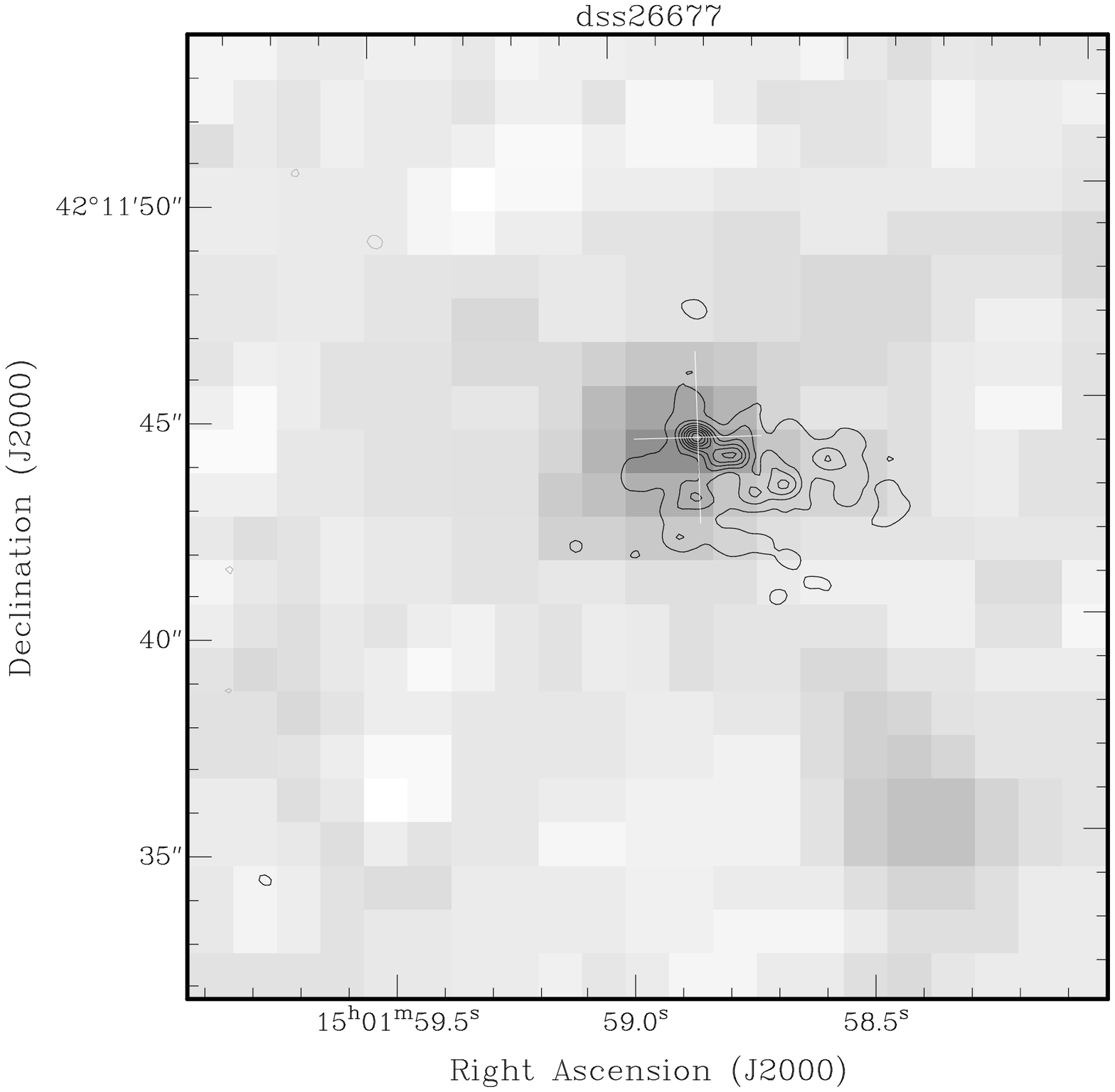 ,width=4.0cm,clip=}\label{at}}}\caption{Optical counterparts for sources 9CJ0957+3422 to 9CJ1501+4211. Crosses mark maximum radio flux density and are 4\,arcsec top to bottom. Contours: \ref{ar}, 1.4\,GHz contours 5,10,15,20,40,80\,\% of peak (91\,mJy/beam); \ref{9CJ0958+2948}, 1.4\,GHz contours at 3-11 every 2\,\% and 30-90 every 20\,\% of peak (89.9\,mJy/beam); \ref{as} and \ref{at}, 4.8\,GHz contours 10-90 every 10\,\% of peak (7.4\,mJy/beam).}\end{figure*}

\begin{figure*}
\mbox{
\subfigure[9CJ1459+4442 (P60 \it{R}\normalfont)]{\epsfig{figure=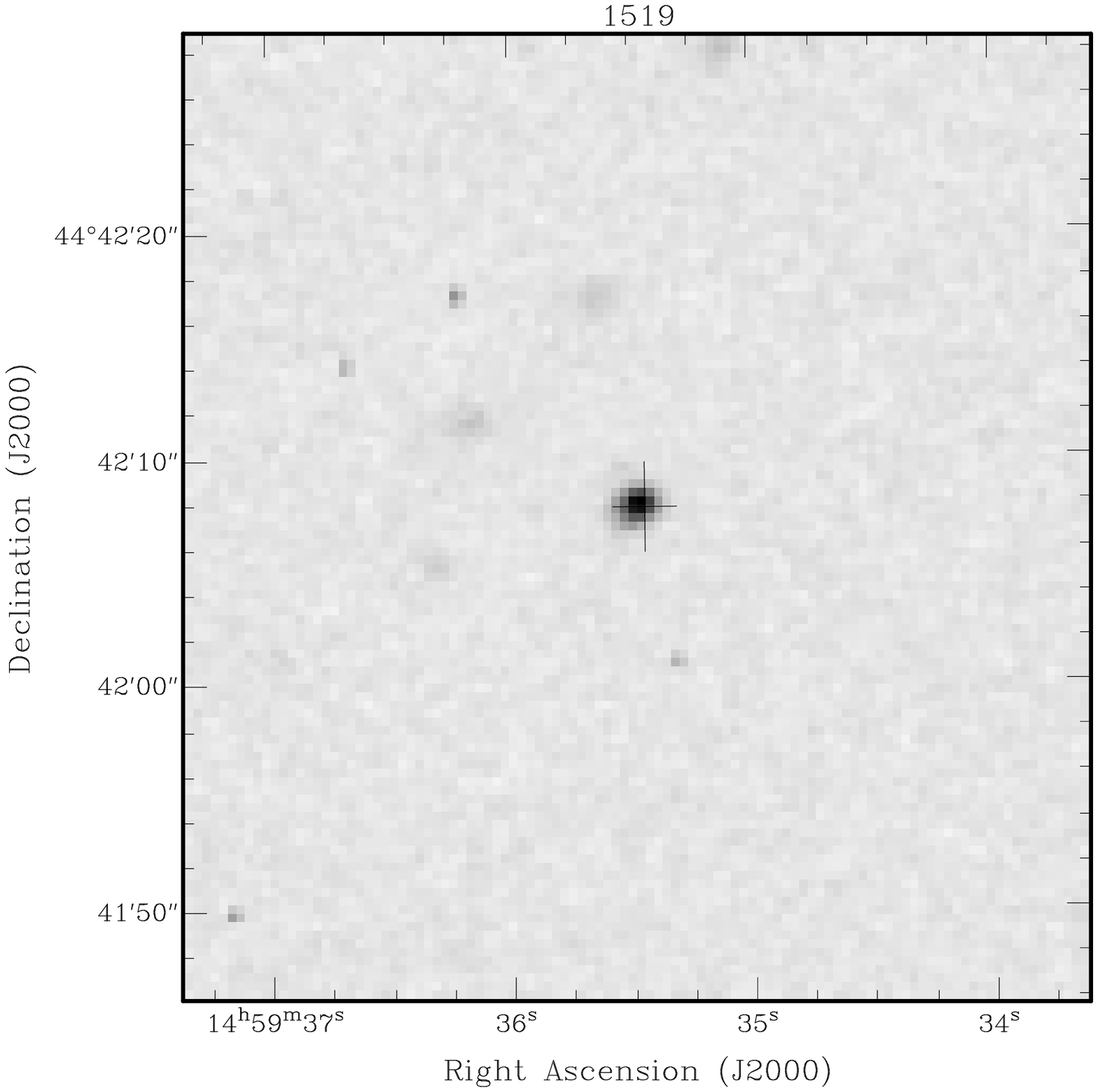 ,width=4.0cm,clip=}}\quad 
\subfigure[9CJ1501+4537 (P60 \it{R}\normalfont)]{\epsfig{figure=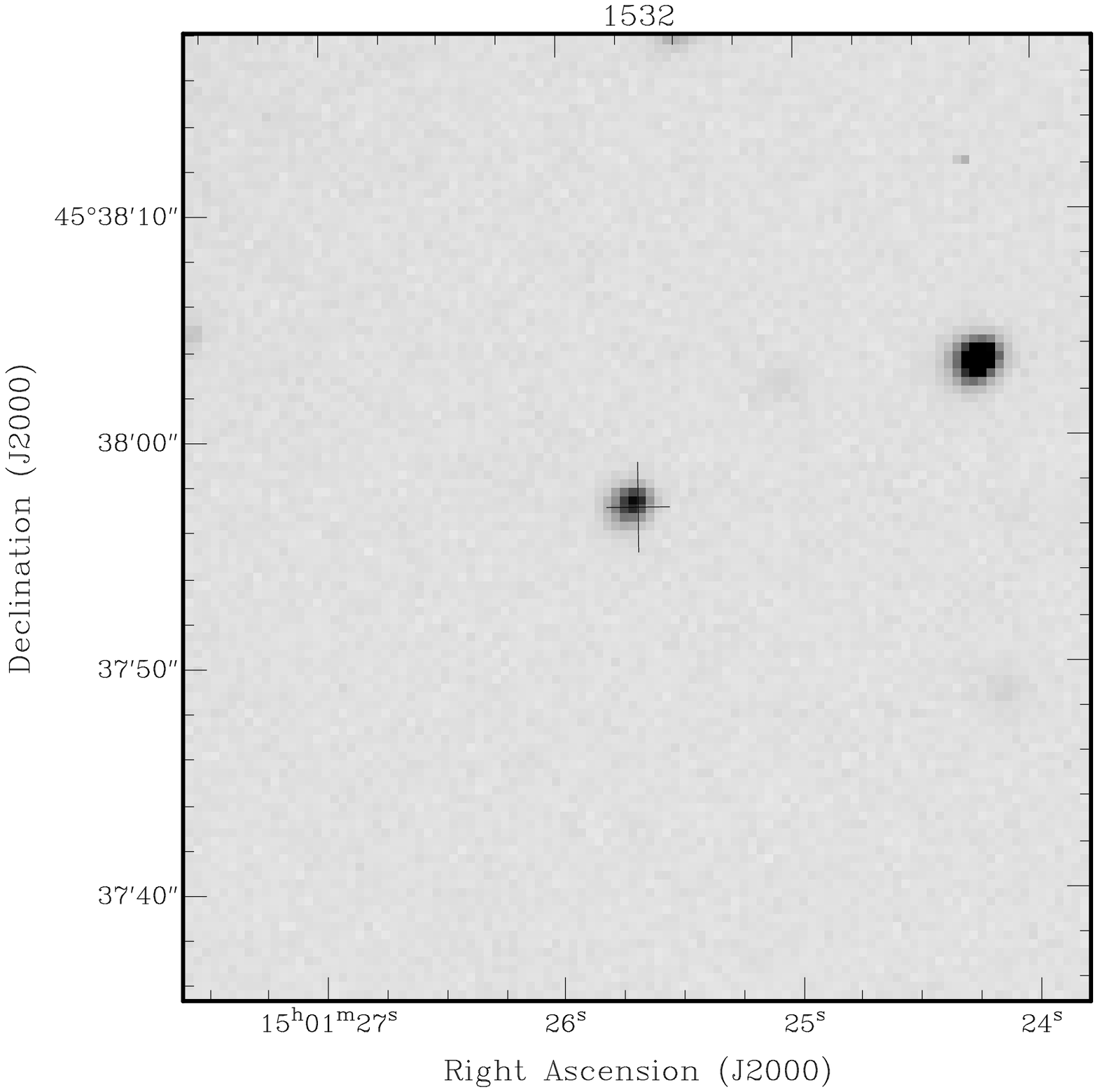 ,width=4.0cm,clip=}}\quad 
\subfigure[9CJ1502+3956 (DSS2 \it{R}\normalfont)]{\epsfig{figure=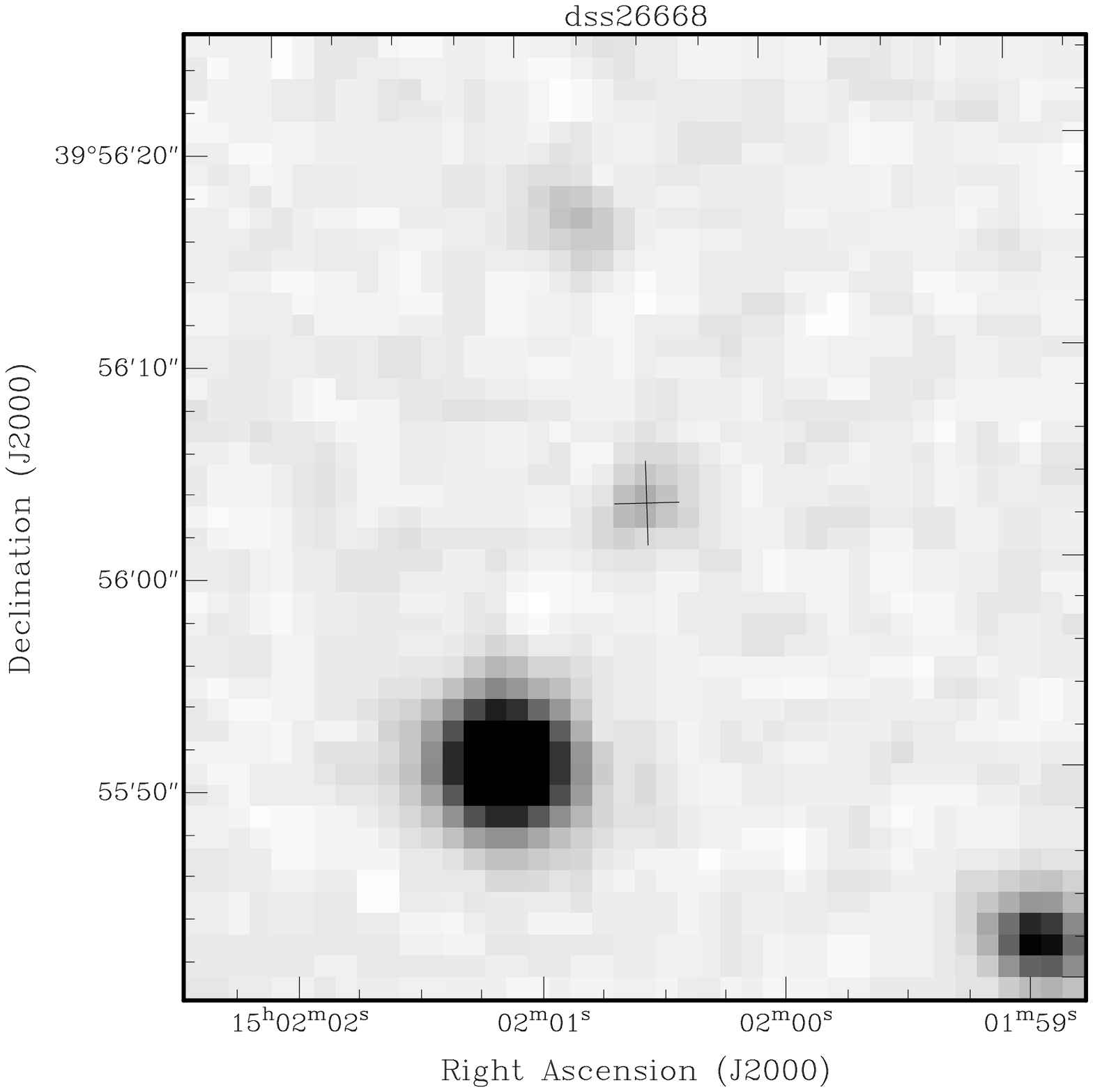 ,width=4.0cm,clip=}}
} 
\mbox{
\subfigure[9CJ1502+3947 (DSS2 \it{R}\normalfont)]{\epsfig{figure=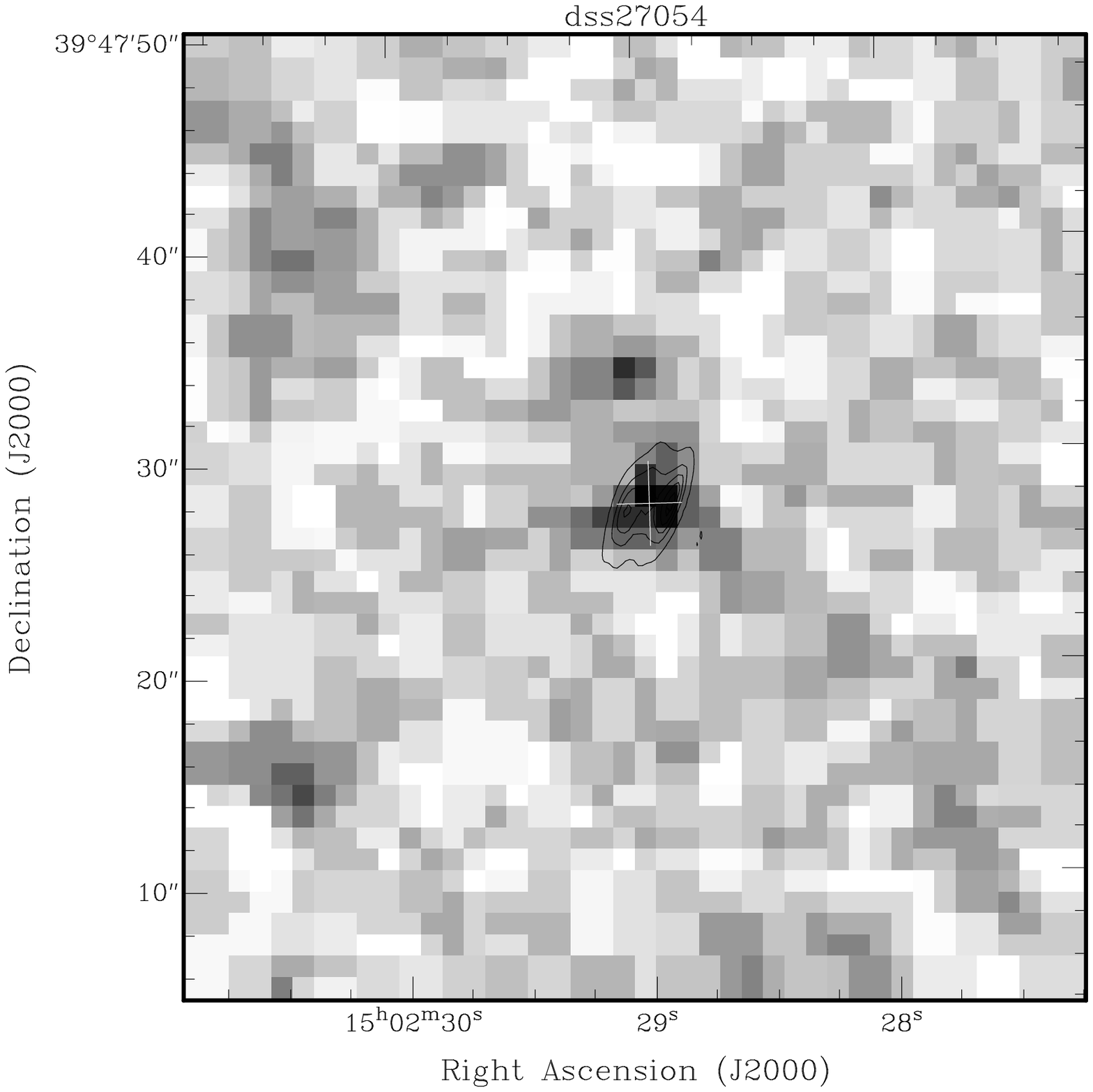 ,width=4.0cm,clip=}\label{au}}\quad 
\subfigure[9CJ1502+3947 (DSS2 \it{R}\normalfont) Detail]{\epsfig{figure=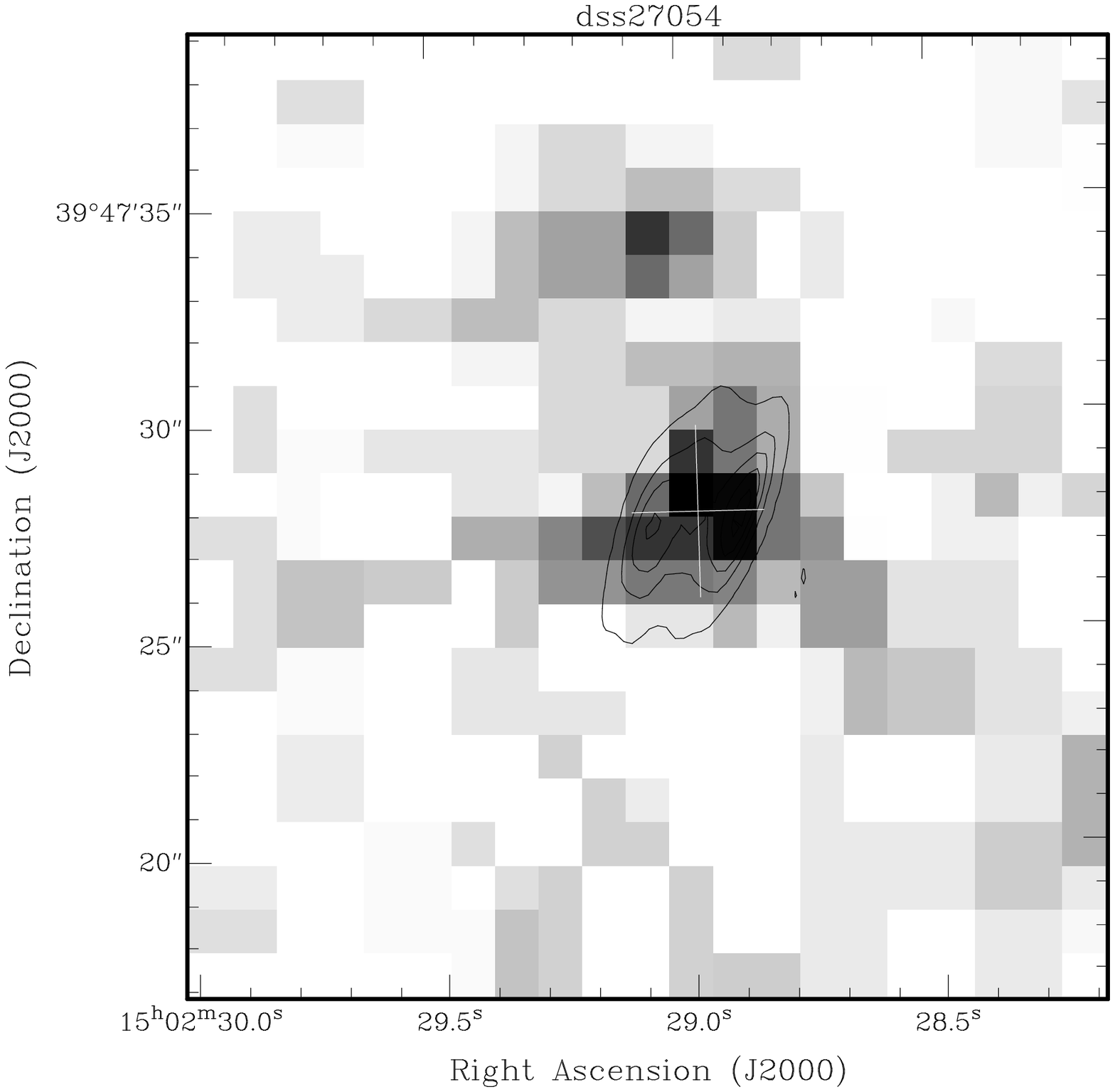 ,width=4.0cm,clip=}\label{av}}\quad 
\subfigure[9CJ1502+3753 (P60 \it{R}\normalfont)]{\epsfig{figure=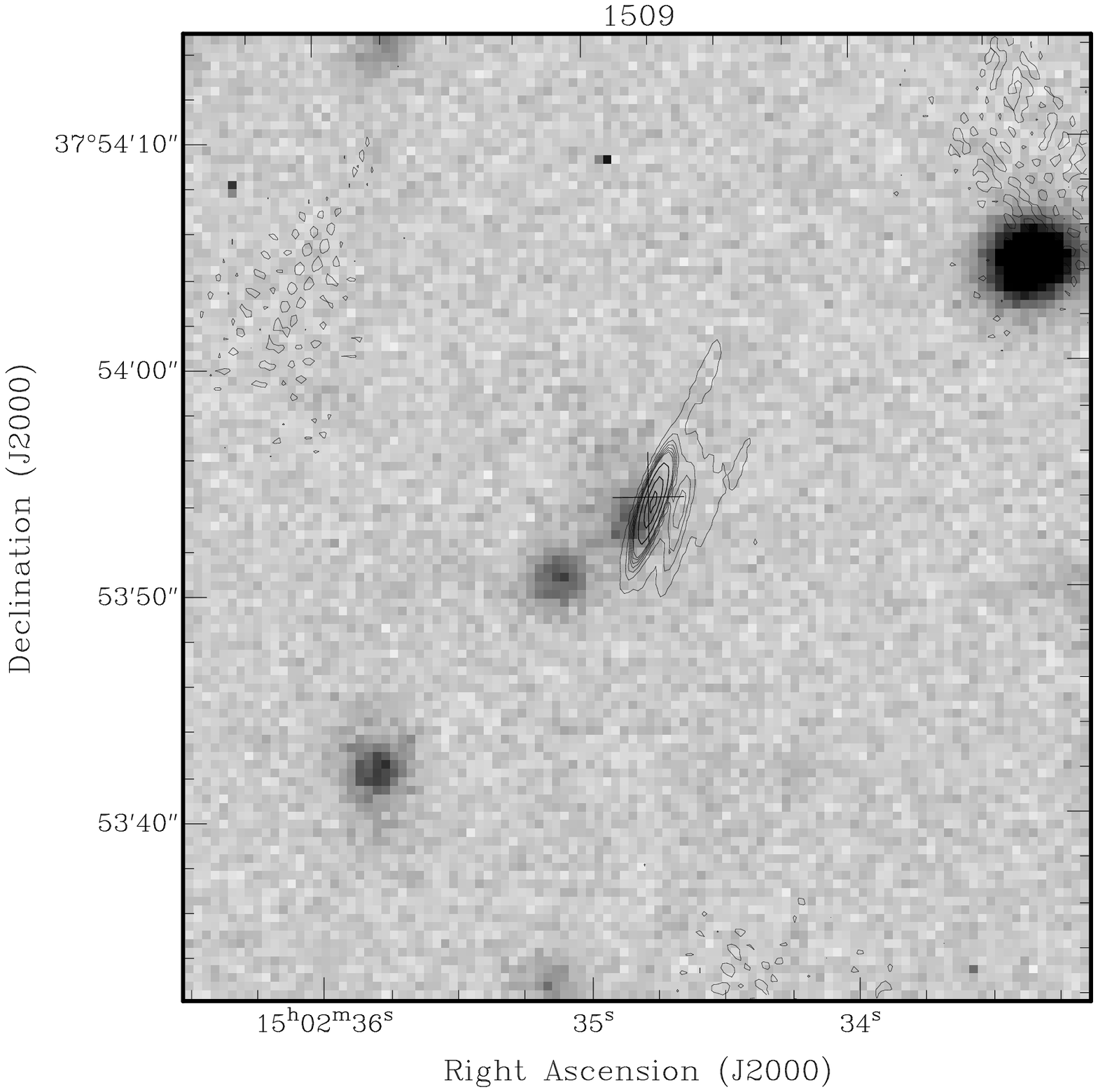 ,width=4.0cm,clip=}\label{aw}}
} 
\mbox{
\subfigure[9CJ1502+3753 (P60 \it{R}\normalfont) Detail]{\epsfig{figure=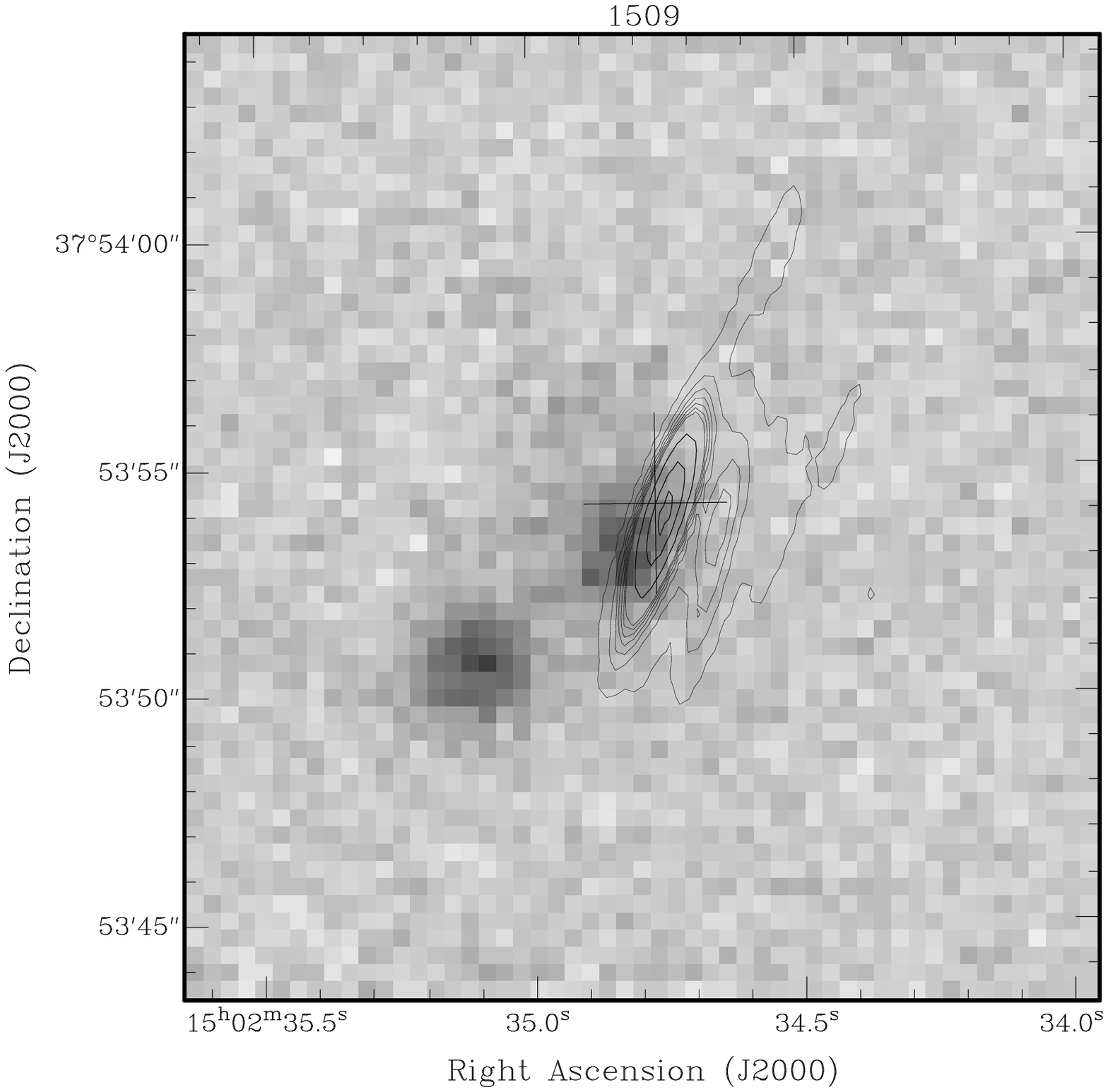 ,width=4.0cm,clip=}\label{ax}}\quad 
\subfigure[9CJ1503+4528 (DSS2 \it{R}\normalfont)]{\epsfig{figure=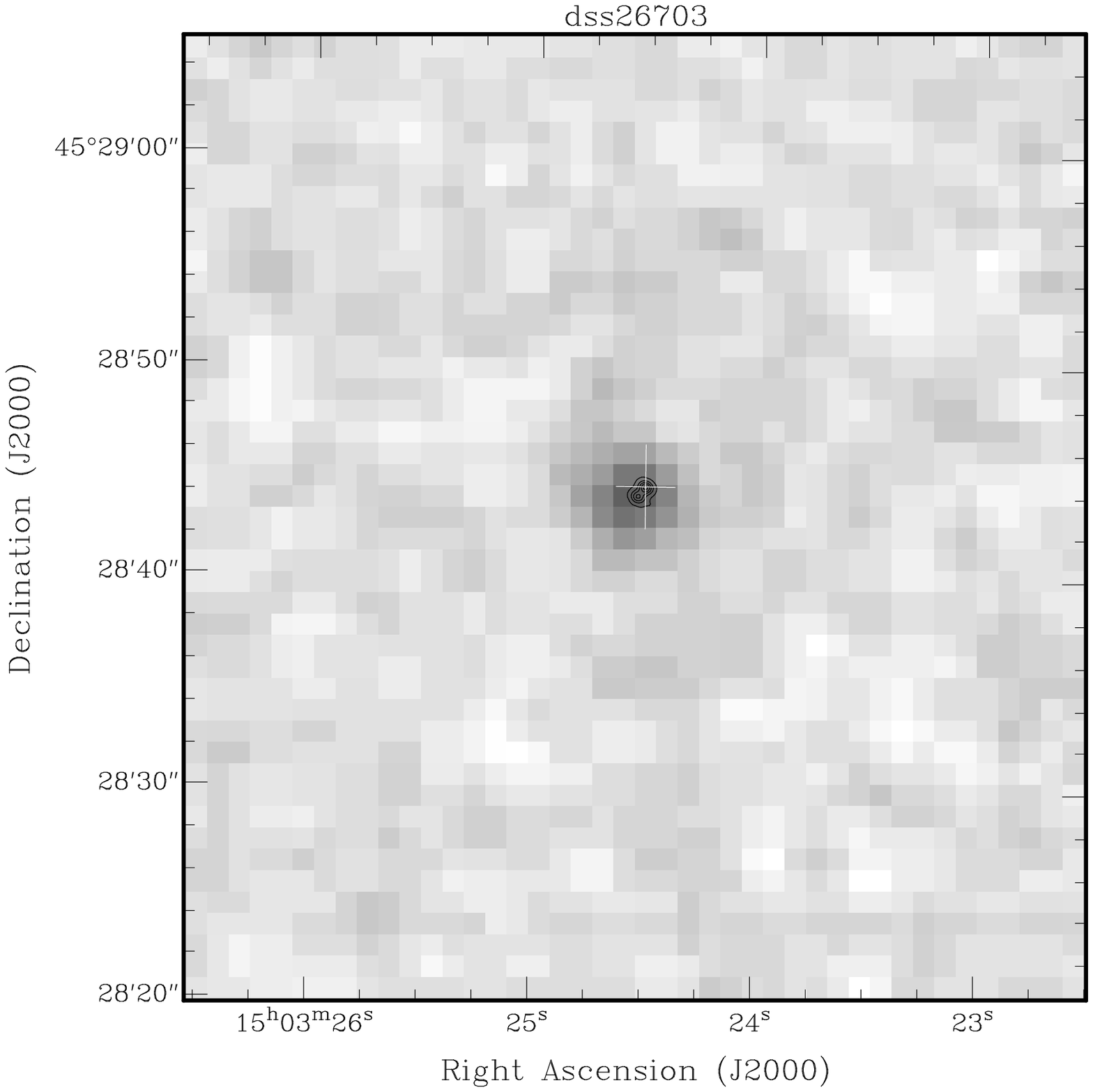 ,width=4.0cm,clip=}\label{ay}}\quad 
\subfigure[9CJ1503+4528 (DSS2 \it{R}\normalfont) Detail]{\epsfig{figure=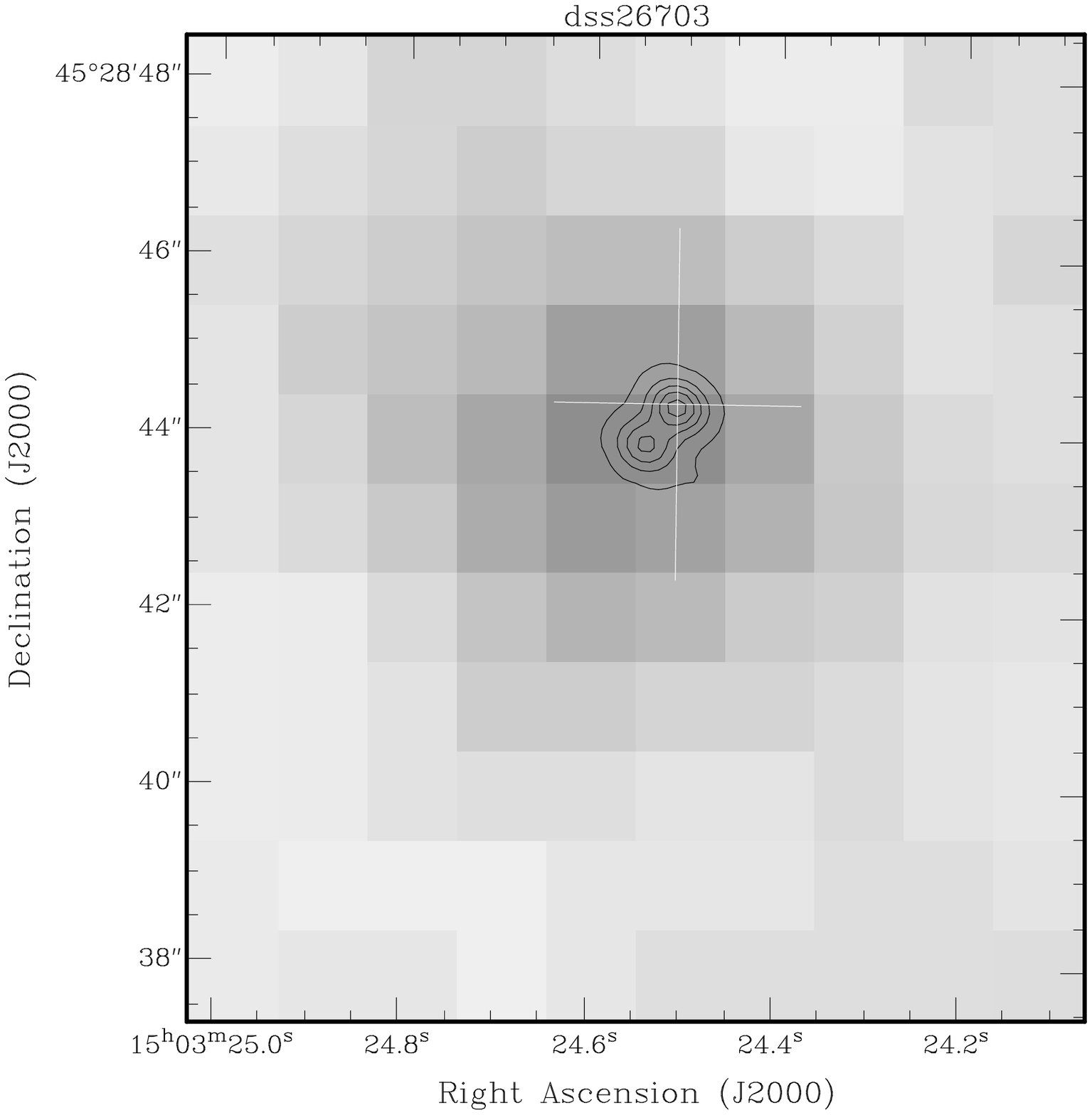 ,width=4.0cm,clip=}\label{az}}
}
\mbox{
\subfigure[9CJ1505+3702 (DSS2 \it{R}\normalfont)]{\epsfig{figure=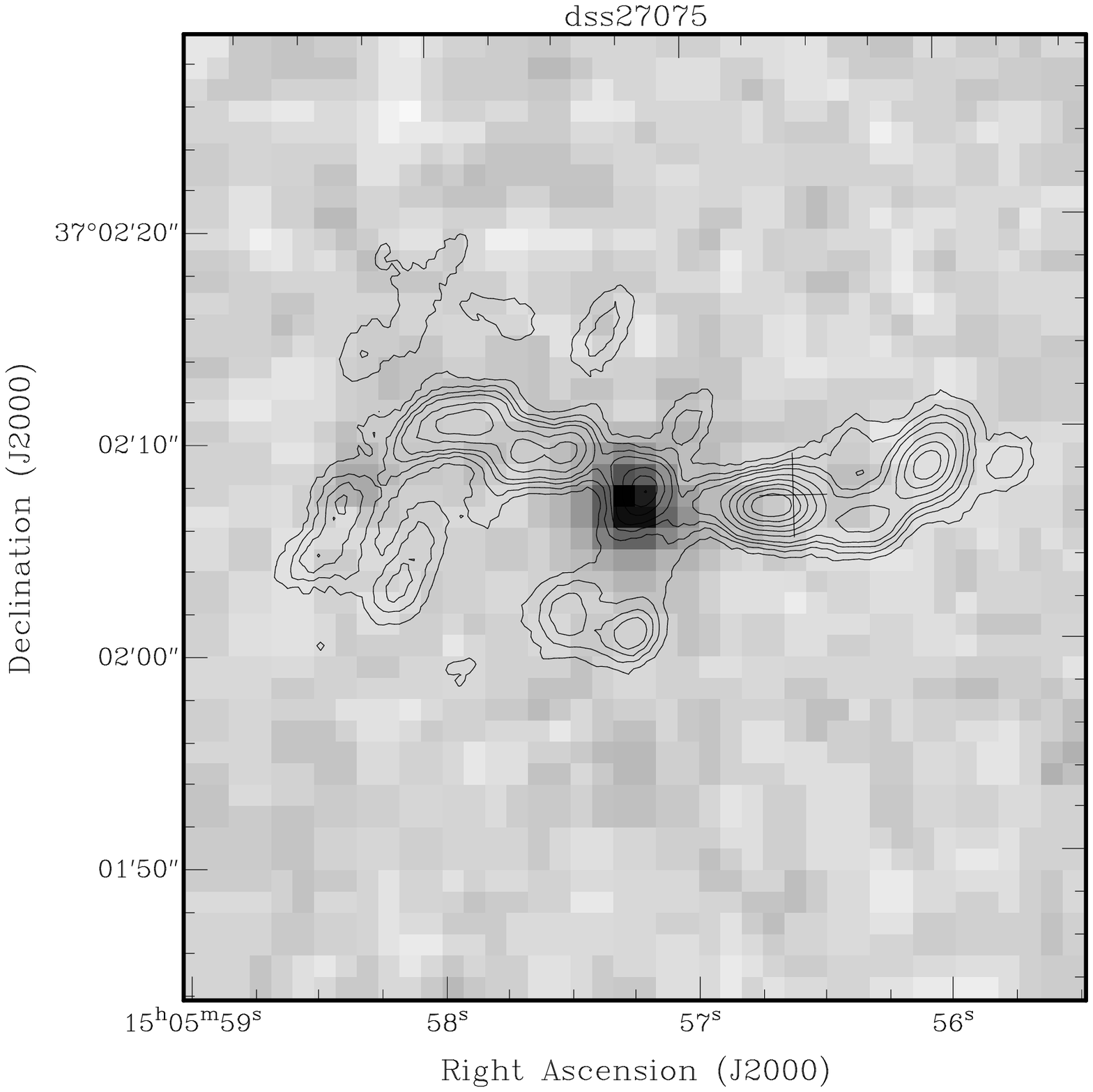 ,width=4.0cm,clip=}\label{ba}}\quad 
\subfigure[9CJ1506+4359 (DSS2 \it{R}\normalfont)]{\epsfig{figure=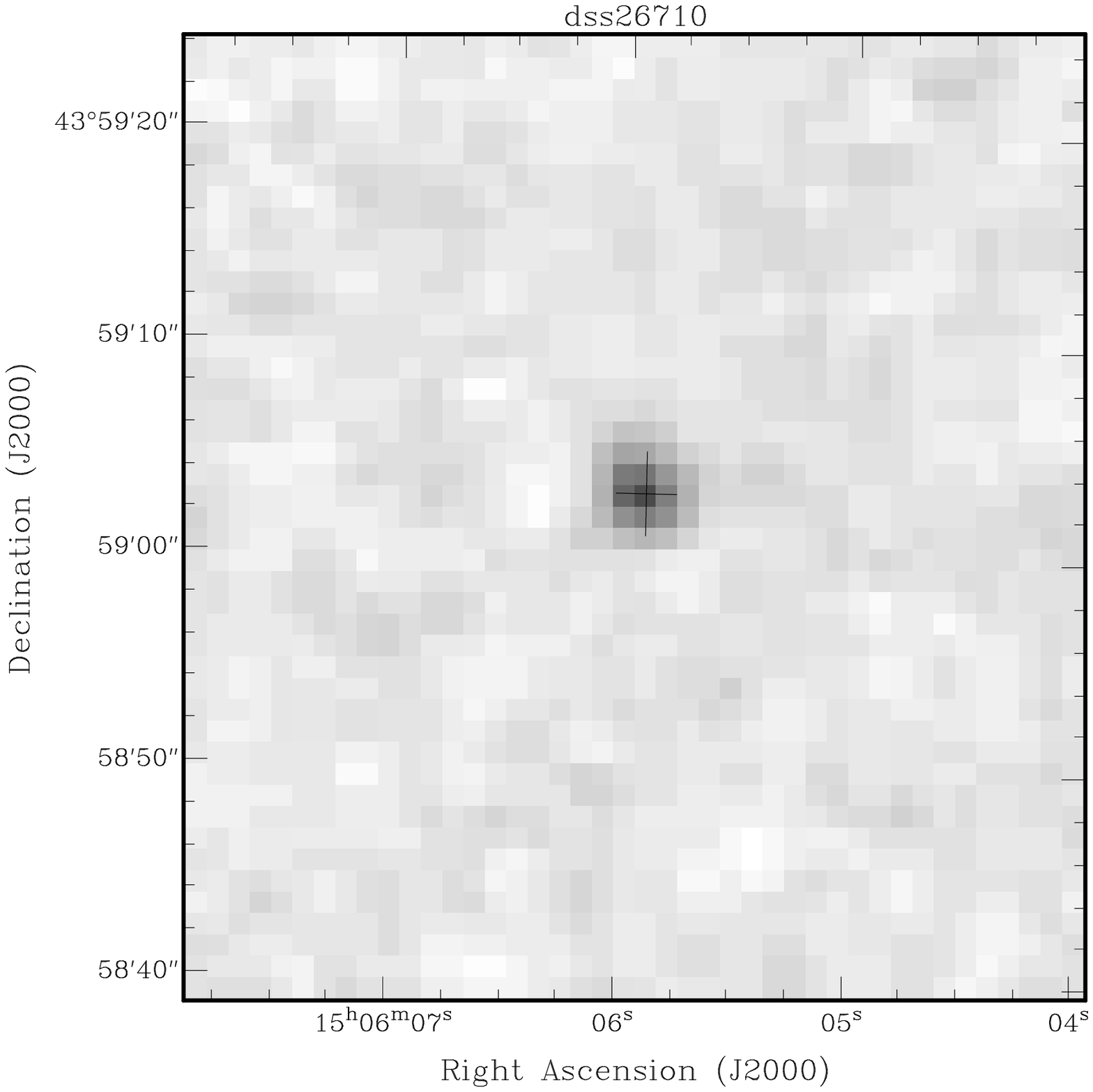 ,width=4.0cm,clip=}}\quad 
\subfigure[9CJ1506+3730 (P60 \it{R}\normalfont)]{\epsfig{figure=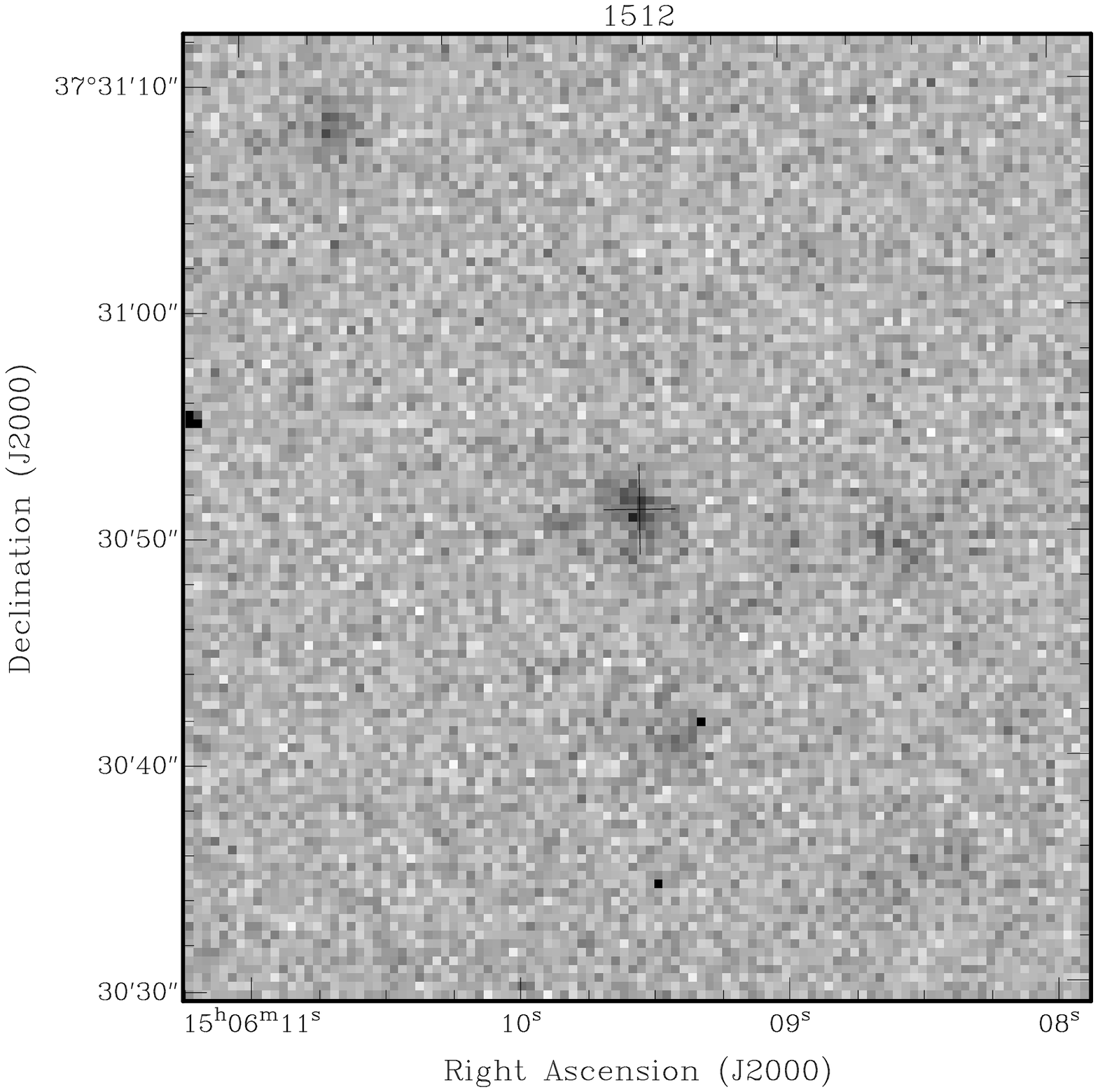 ,width=4.0cm,clip=}}}\caption{ Optical counterparts for sources 9CJ1459+4442 to 9CJ1506+3730. Crosses mark maximum radio flux density and are 4\,arcsec top to bottom. Contours: \ref{au} and \ref{av}, 4.8\,GHz contours 10-90 every 20\,\% of peak (27.4\,mJy/beam); \ref{aw} and \ref{ax}, 4.8\,GHz contours 3-15 every 2\,\% and 30,60,90\,\% of peak (61.2\,mJy/beam); \ref{ay} and \ref{az}, 4.8\,GHz contours 10-90 every 20\,\% of peak (99.6\,mJy/beam); \ref{ba}, 4.8\,GHz contours 25,30,35\,\% and 40-90 every 10\,\% of peak (6.8\,mJy/beam).}\end{figure*}

\begin{figure*}
\mbox{
\subfigure[9CJ1506+4239 (DSS2 \it{R}\normalfont)]{\epsfig{figure=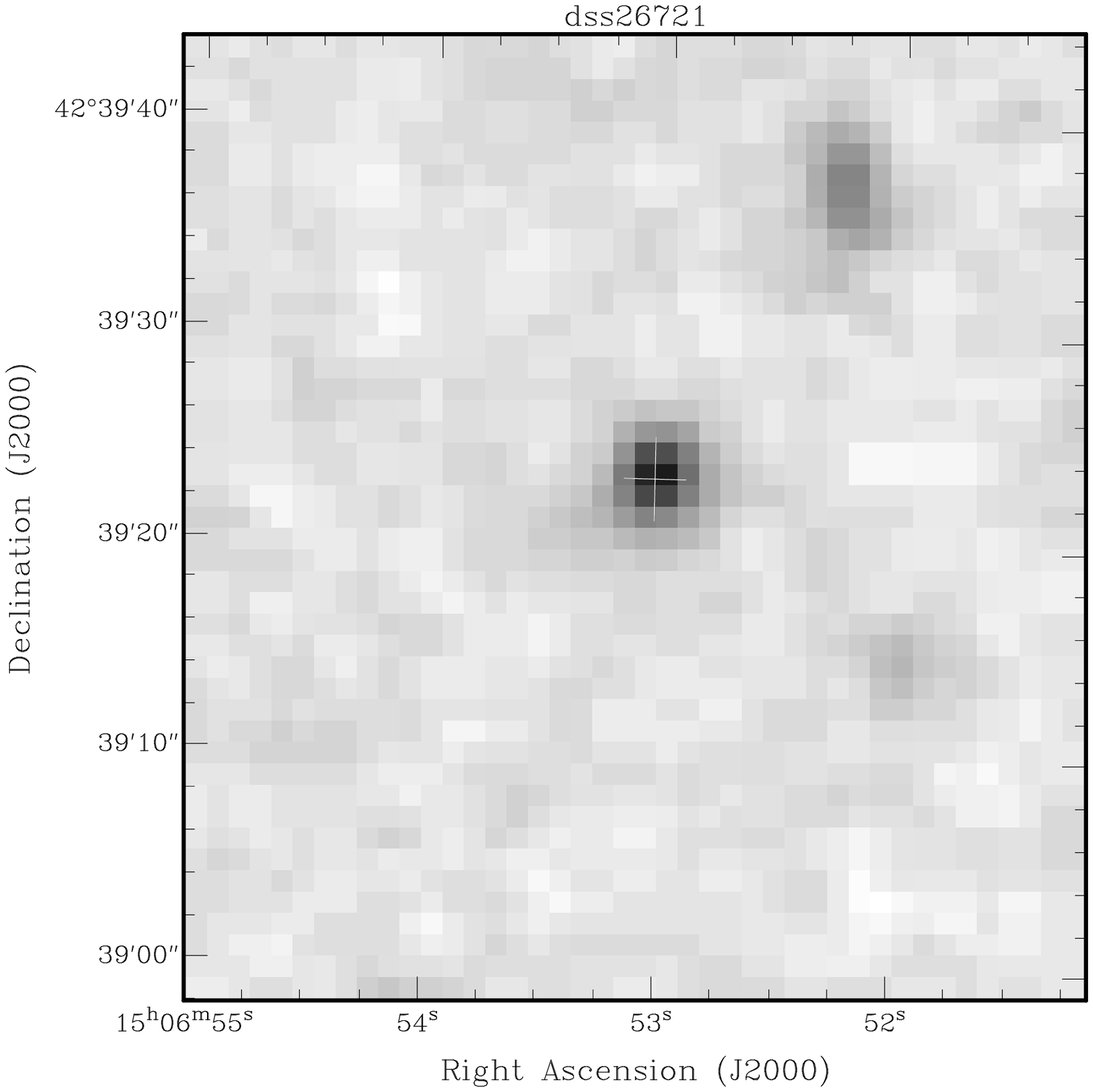 ,width=4.0cm,clip=}}\quad 
\subfigure[9CJ1508+4127 (DSS2 \it{R}\normalfont)]{\epsfig{figure=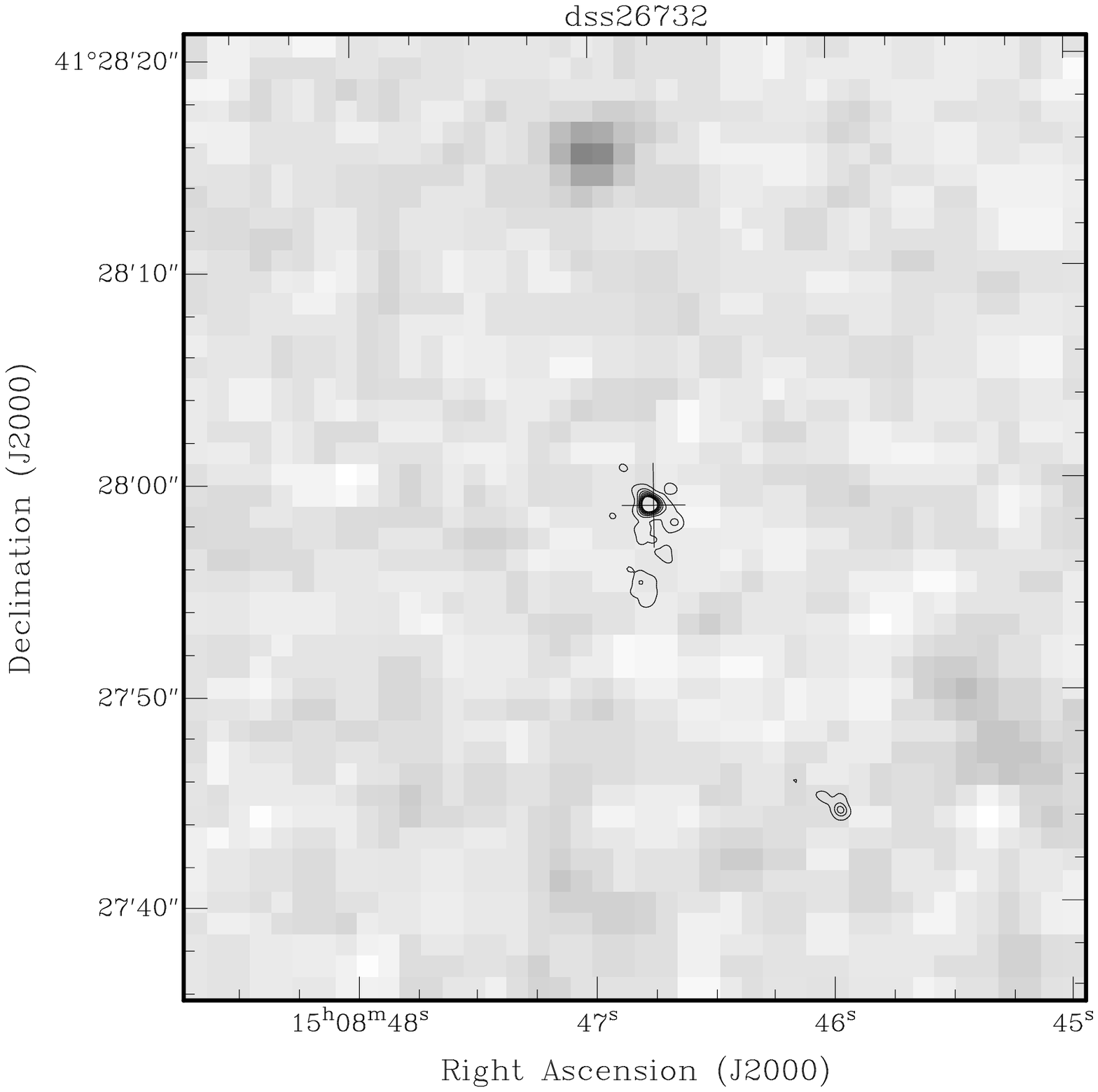 ,width=4.0cm,clip=}\label{bb}}\quad 
\subfigure[9CJ1508+4127 (DSS2 \it{R}\normalfont) Detail]{\epsfig{figure=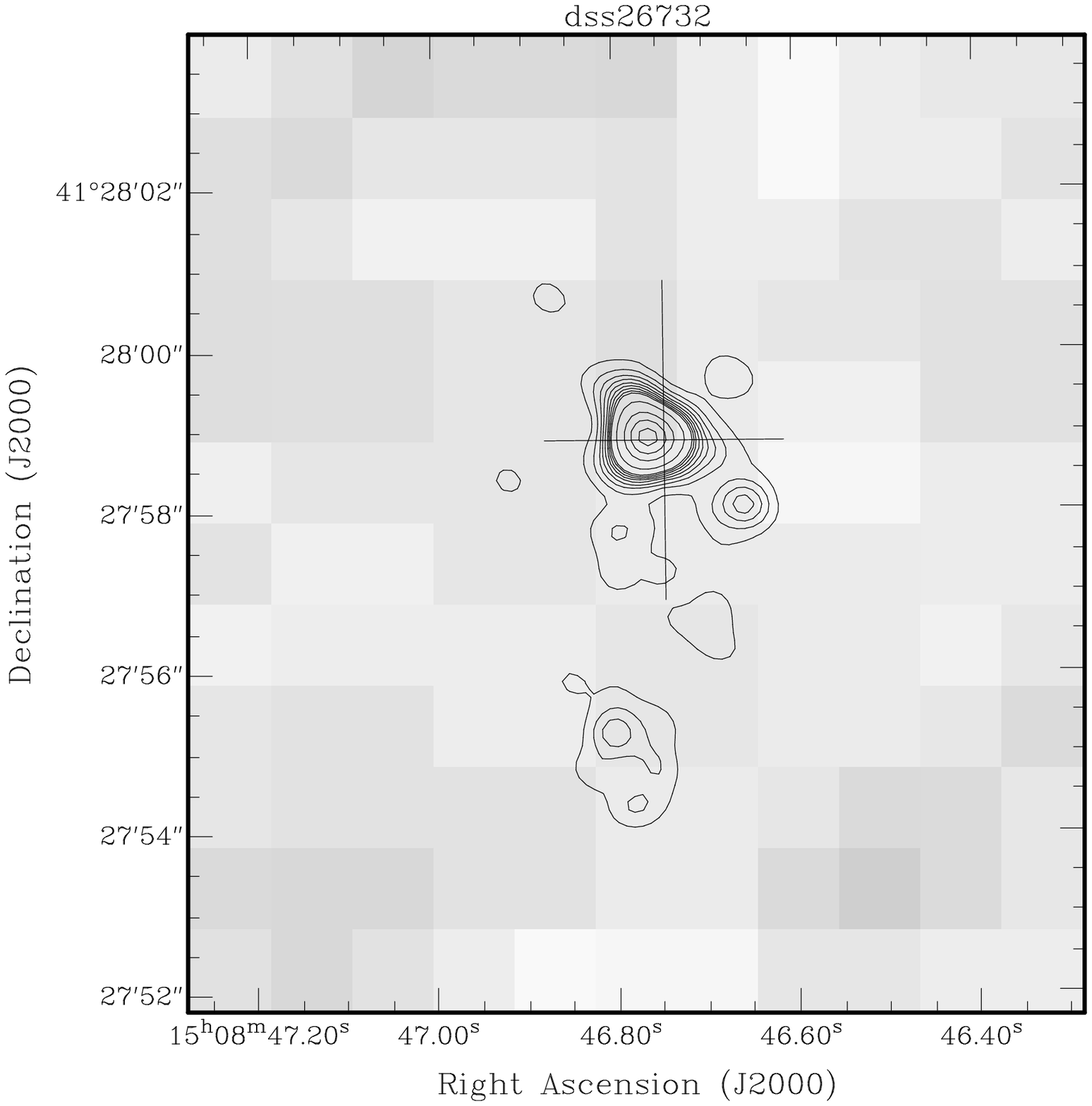 ,width=4.0cm,clip=}\label{bc}}
} 
\mbox{
\subfigure[9CJ1510+4138 (P60 \it{R}\normalfont)]{\epsfig{figure=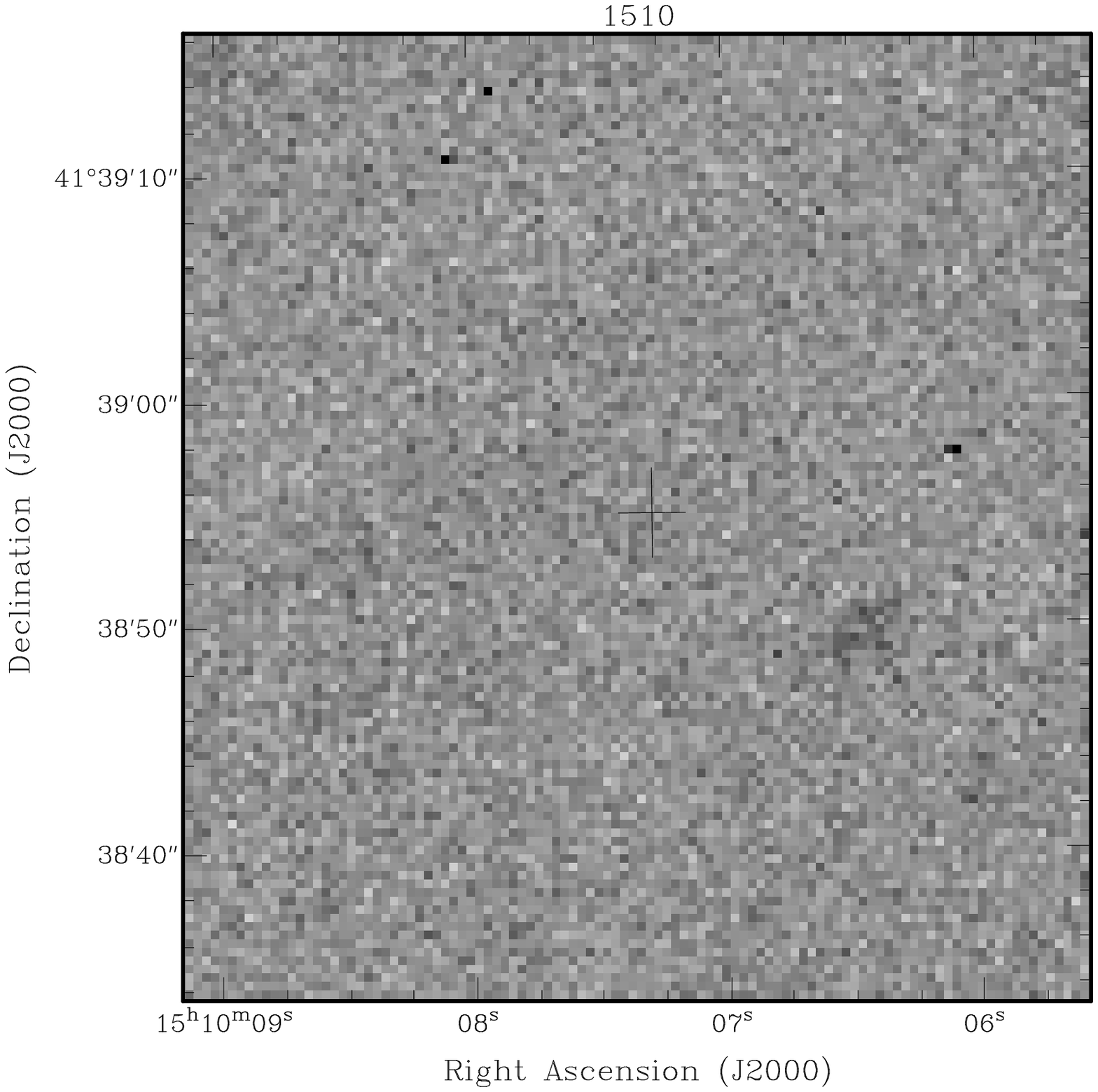 ,width=4.0cm,clip=}}\quad 
\subfigure[9CJ1510+3750 (DSS2 \it{R}\normalfont)]{\epsfig{figure=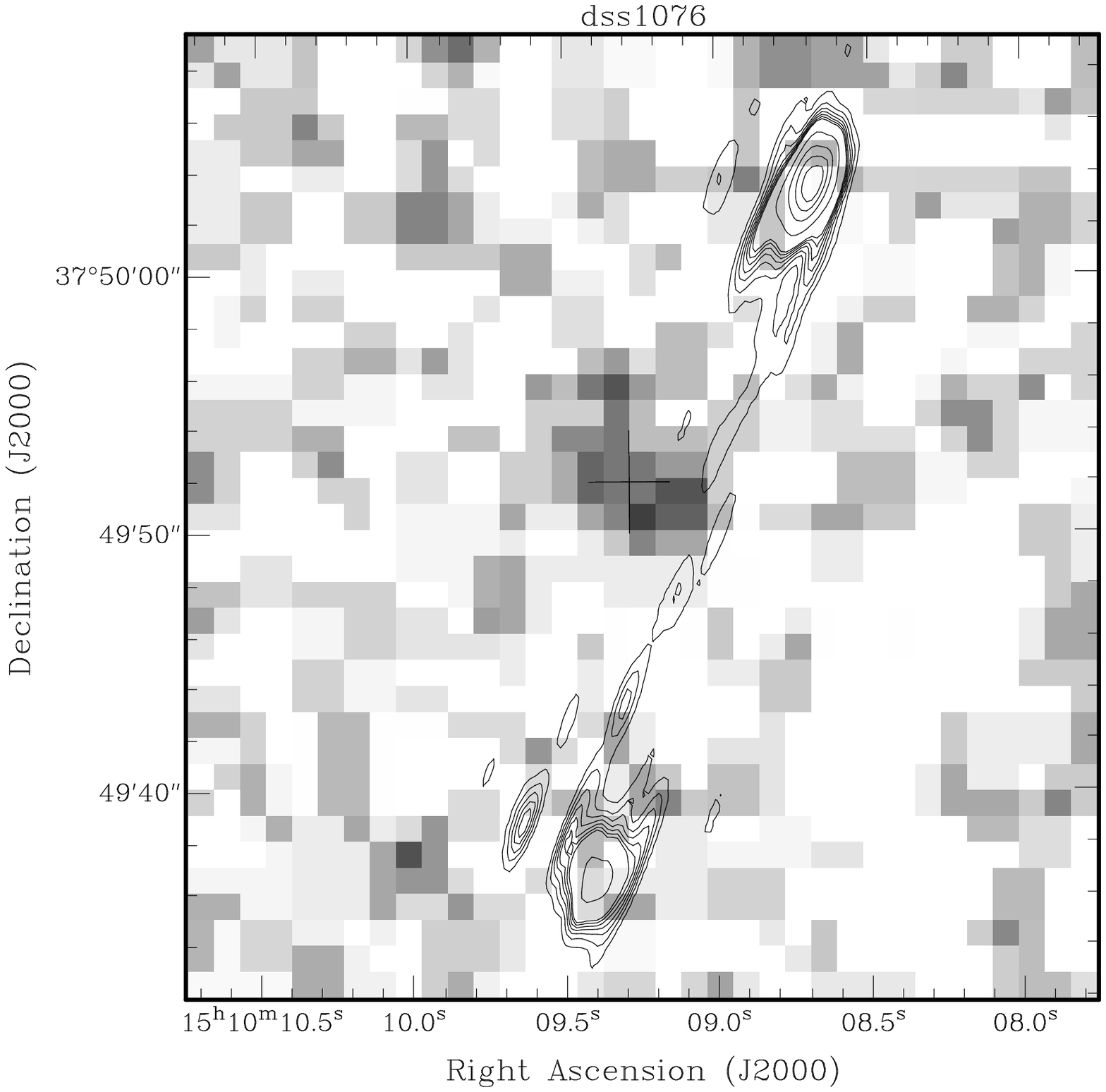 ,width=4.0cm,clip=}\label{bd}}\quad 
\subfigure[9CJ1510+4221 (P60 \it{R}\normalfont)]{\epsfig{figure=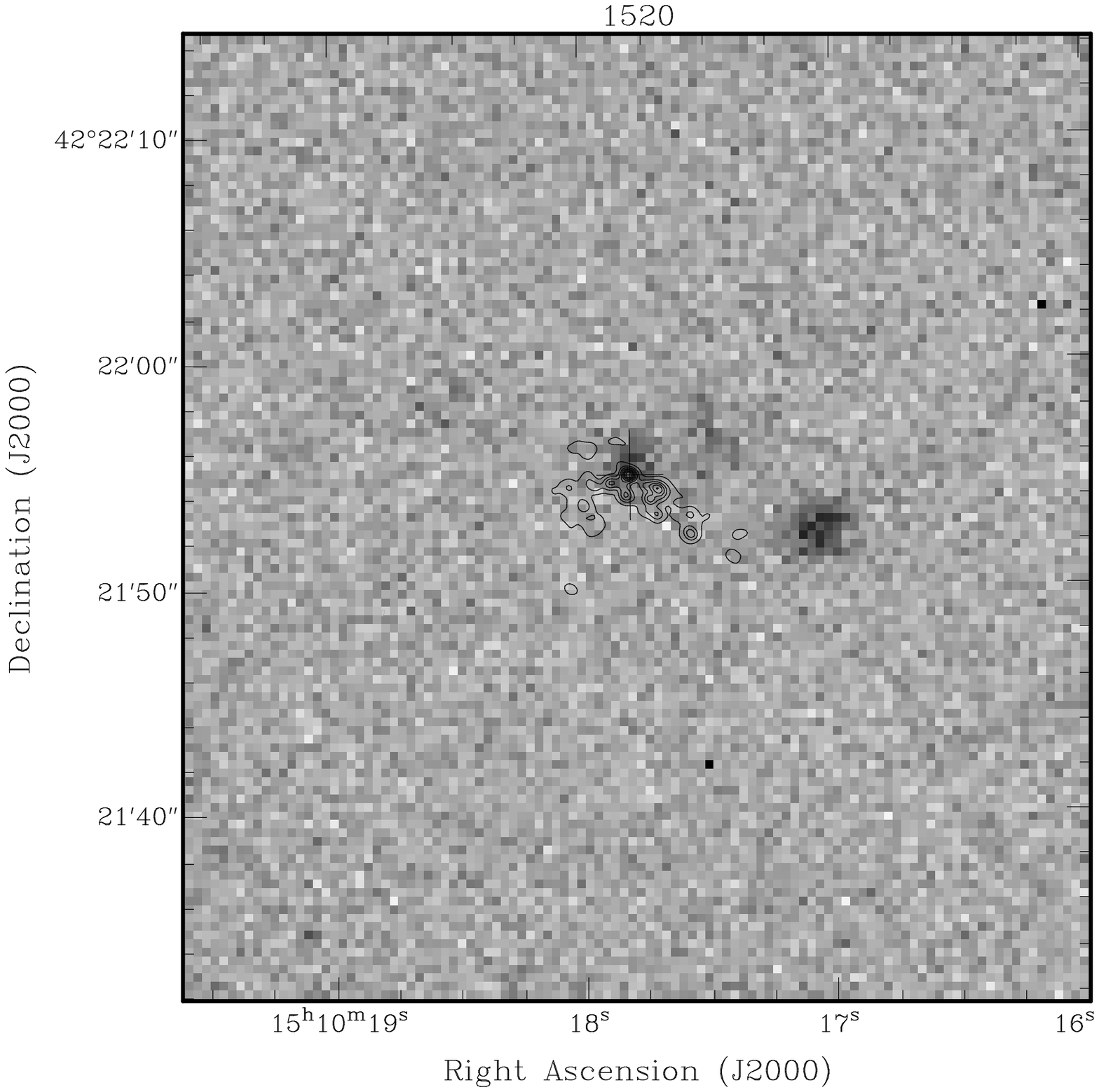 ,width=4.0cm,clip=}\label{be}}
} 
\mbox{
\subfigure[9CJ1510+4221 (P60 \it{R}\normalfont) Detail]{\epsfig{figure=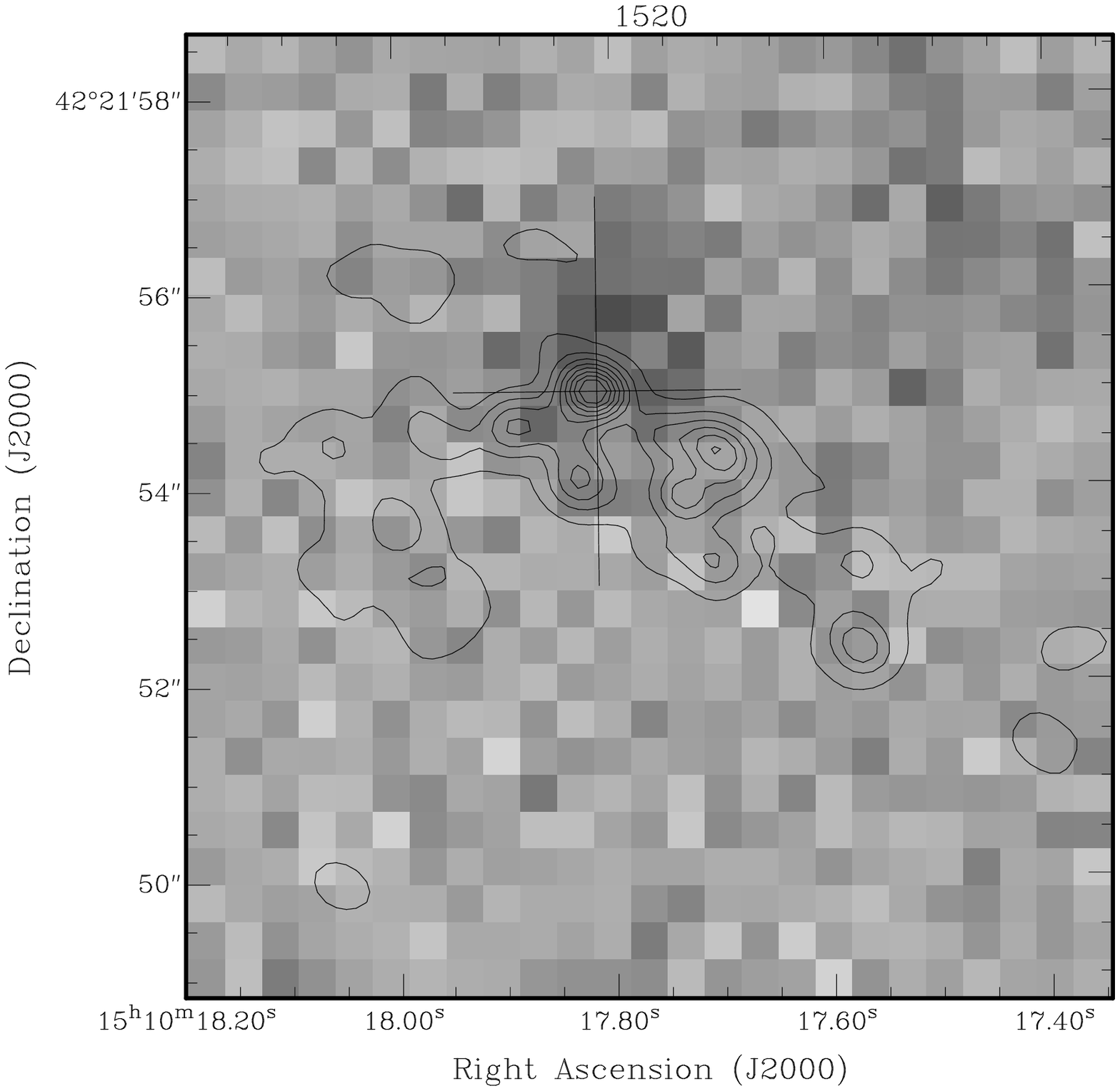 ,width=4.0cm,clip=}\label{bf}}\quad 
\subfigure[9CJ1511+4430 (DSS2 \it{R}\normalfont)]{\epsfig{figure=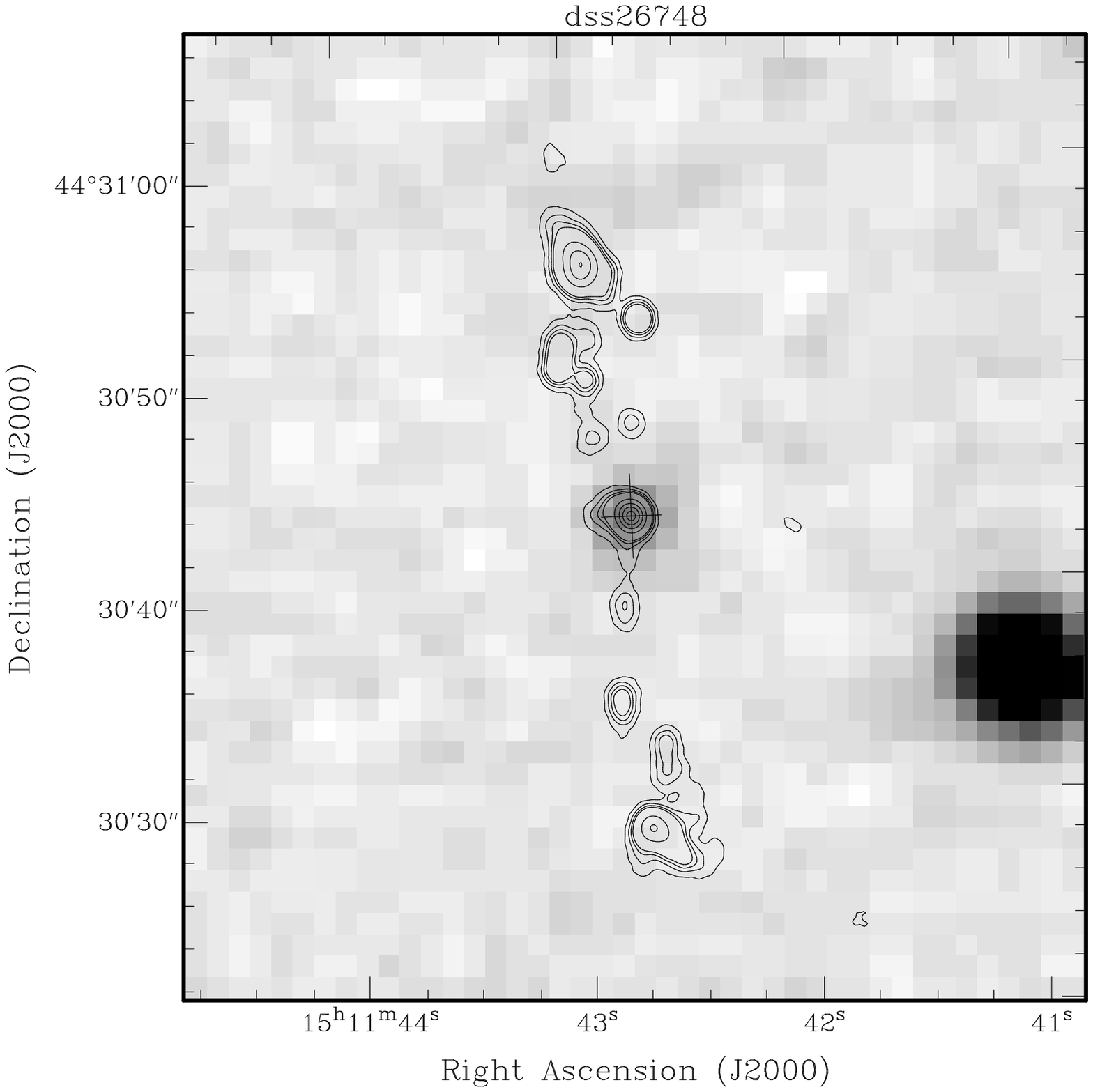,width=4.0cm,clip=}\label{9cJ1511+4430}}\quad 
\subfigure[9CJ1512+4540 (DSS2 \it{R}\normalfont)]{\epsfig{figure=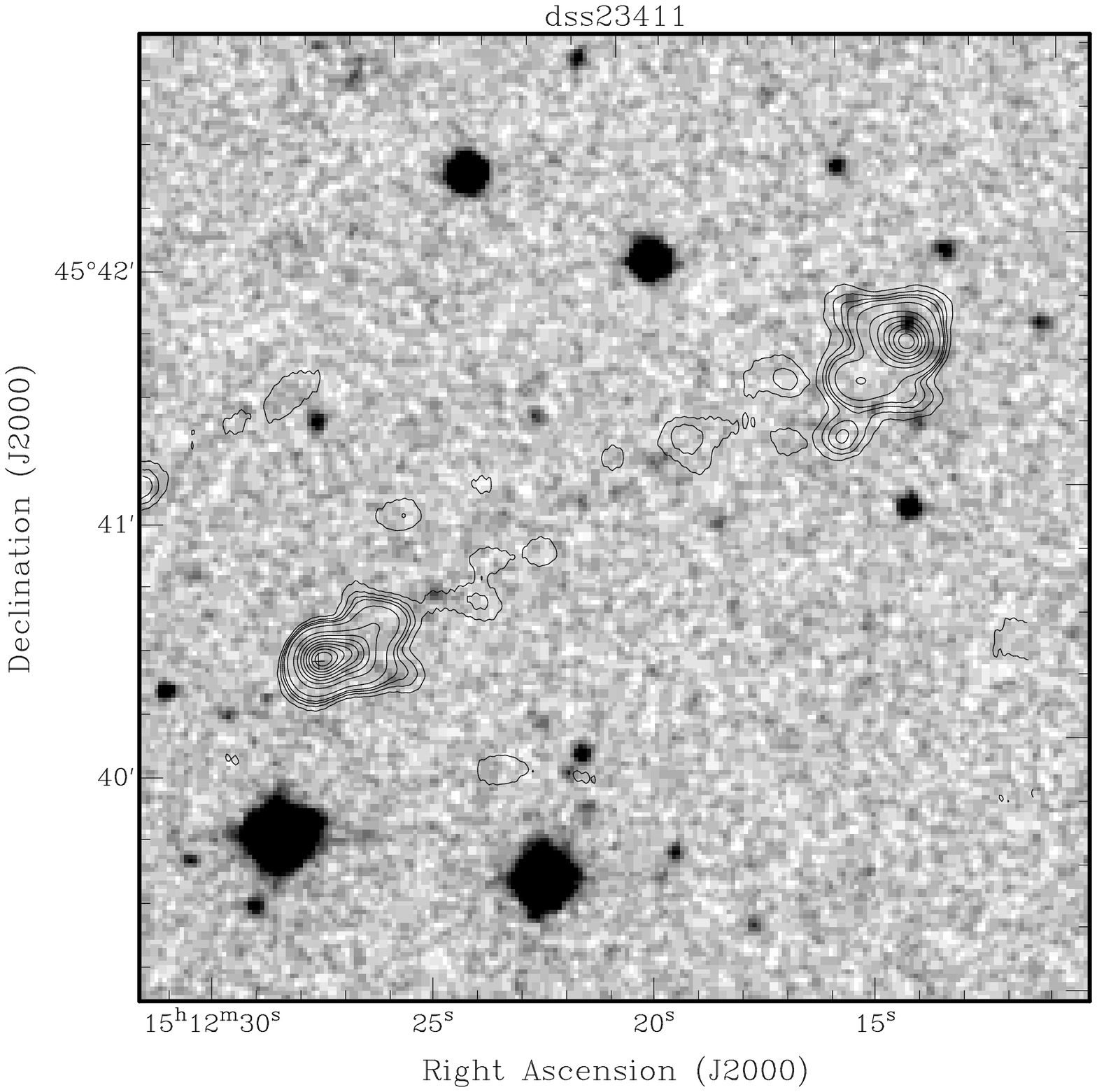,width=4.0cm,clip=}\label{bg}}
} 
\mbox{
\subfigure[9CJ1514+3650 (DSS2 \it{R}\normalfont)]{\epsfig{figure=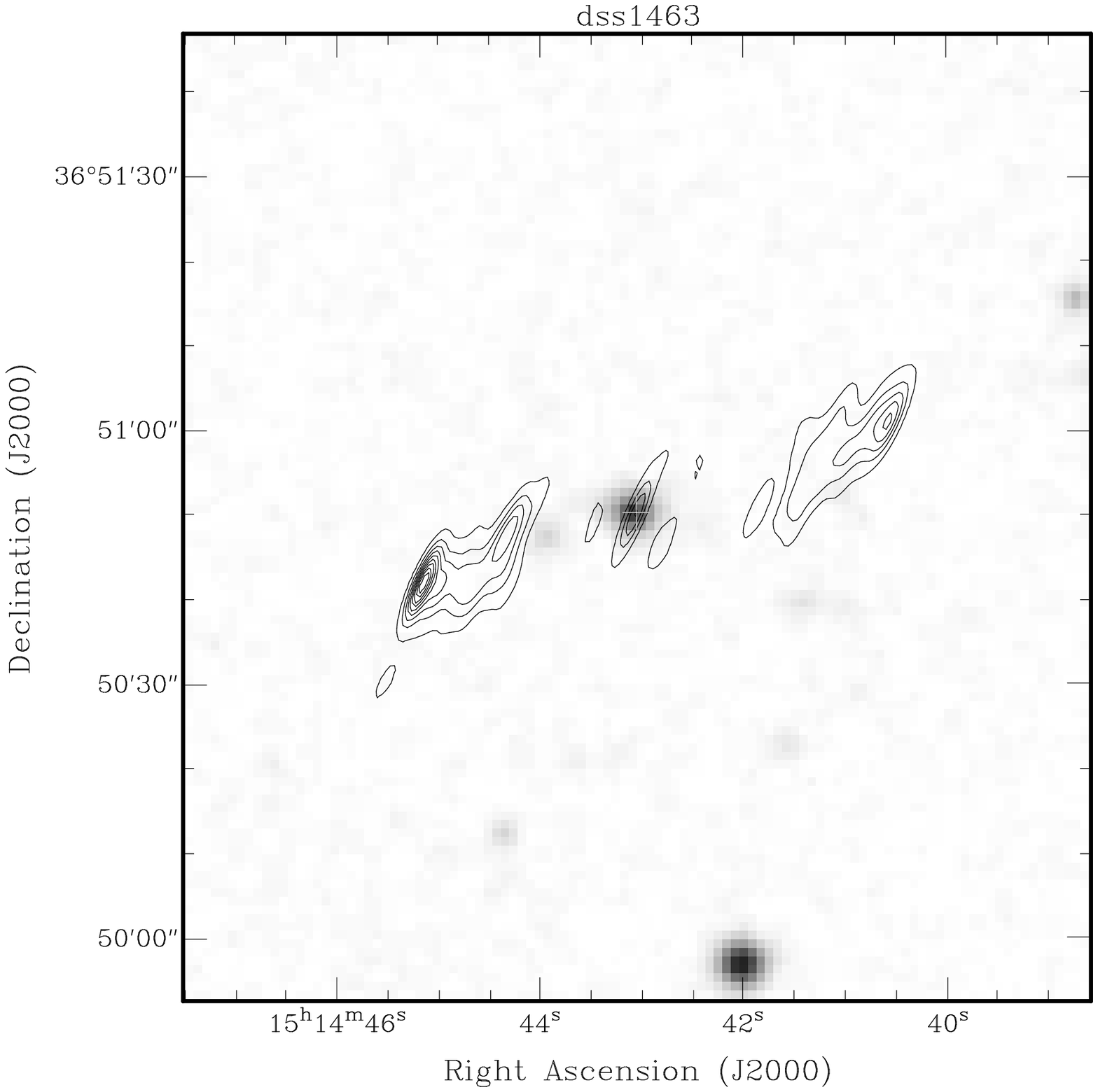,width=4.0cm,clip=}\label{bh}}\quad 
\subfigure[9CJ1516+4349 (P60 \it{R}\normalfont)]{\epsfig{figure=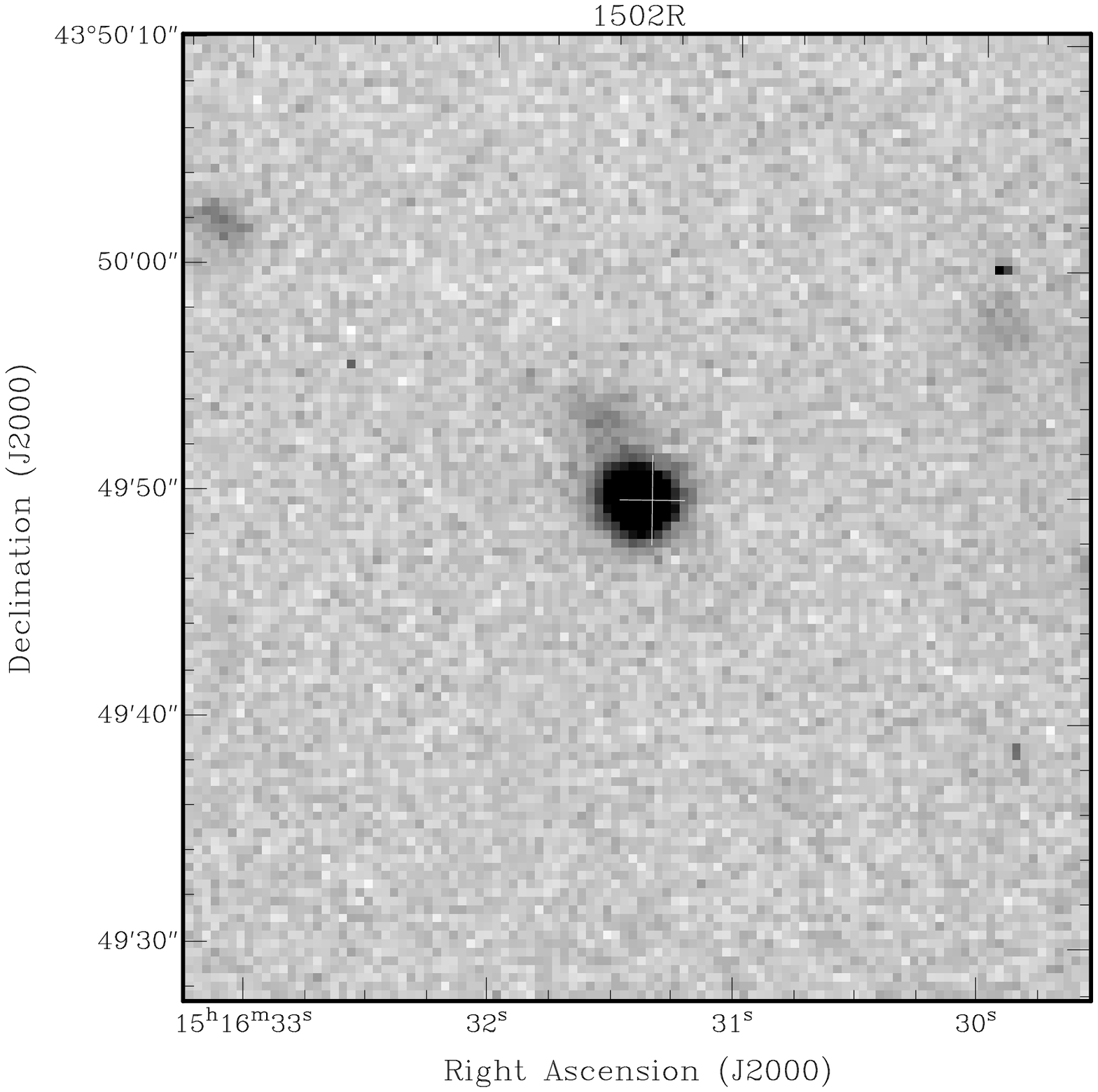,width=4.0cm,clip=}}\quad 
\subfigure[9CJ1516+3650 (DSS2 \it{R}\normalfont)]{\epsfig{figure=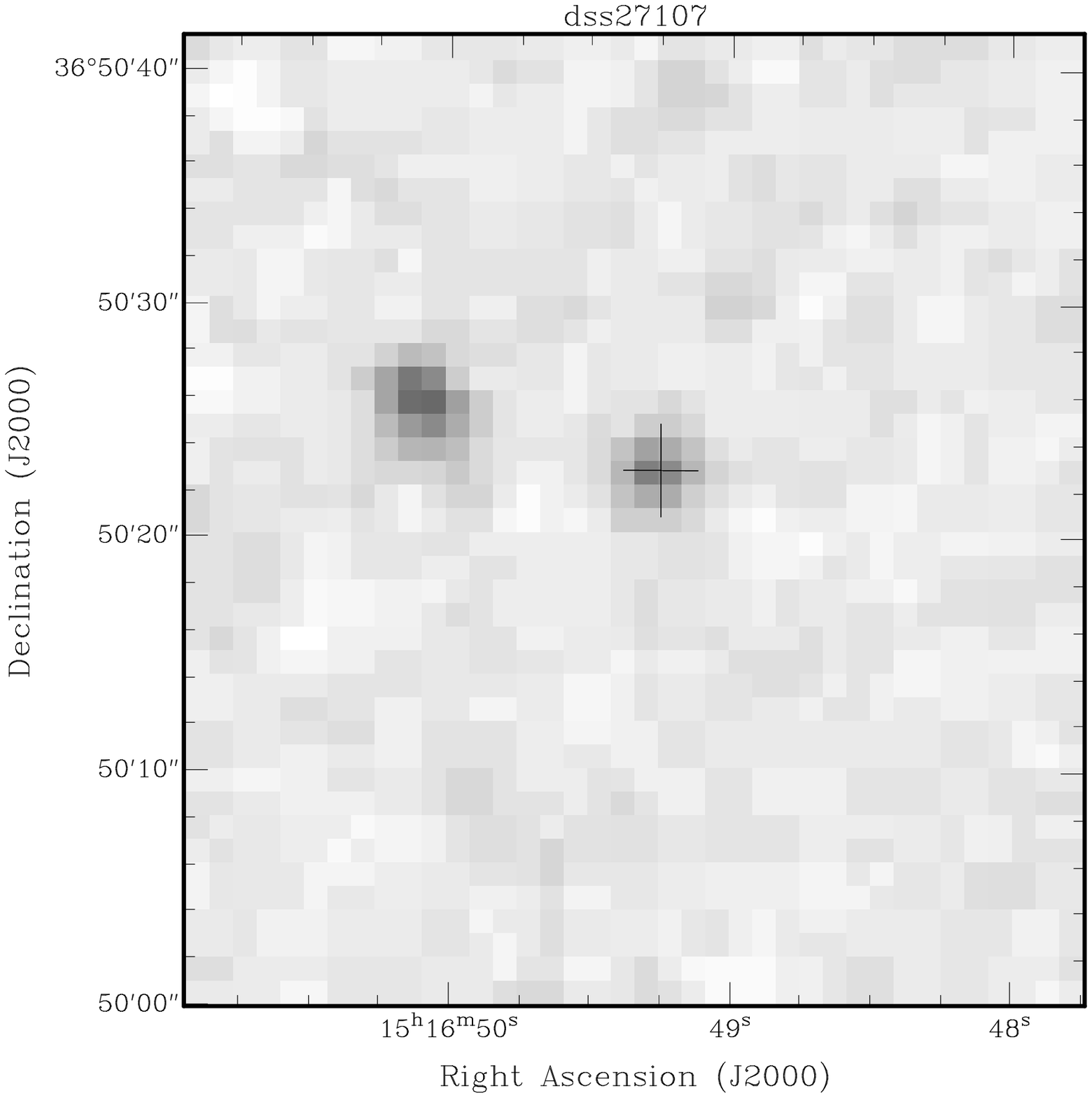,width=4.0cm,clip=}}}\caption{Optical counterparts for sources 9CJ1506+4239 to 9CJ1516+3650. Crosses mark maximum radio flux density and are 4\,arcsec top to bottom. Contours: \ref{bb}, 4.8\,GHz contours 4-40 every 4\,\%; \ref{bc}, 4.8\,GHz contours 4-20 every 2\,\% and 30-90 every 10\,\% of peak (46.8\,mJy/beam); \ref{bd}, 4.8\,GHz contours 3.5-9.5 every 1\,\% and 20-80 every 20\,\% of peak (61.5\,mJy); \ref{be} and \ref{bf}, 4.8\,GHz contours 5-85 every 10\,\% of peak (13.4\,mJy/beam); \ref{9cJ1511+4430}, 4.8\,GHz contours at 6-12 every 2\,\% and 30-90 every 20\,\% of peak (27.8\,mJy/beam); \ref{bg}, 4.8\,GHz contours 4-7 every 1\,\% and 10-90 every 10\,\% of peak (53.5\,mJy/beam); \ref{bh}, 1.4\,GHz contours 10-90 every 10\,\% of peak (97.0\,mJy/beam).}\end{figure*}

\begin{figure*}
\mbox{
\subfigure[9CJ1516+4159 (DSS2 \it{R}\normalfont)]{\epsfig{figure=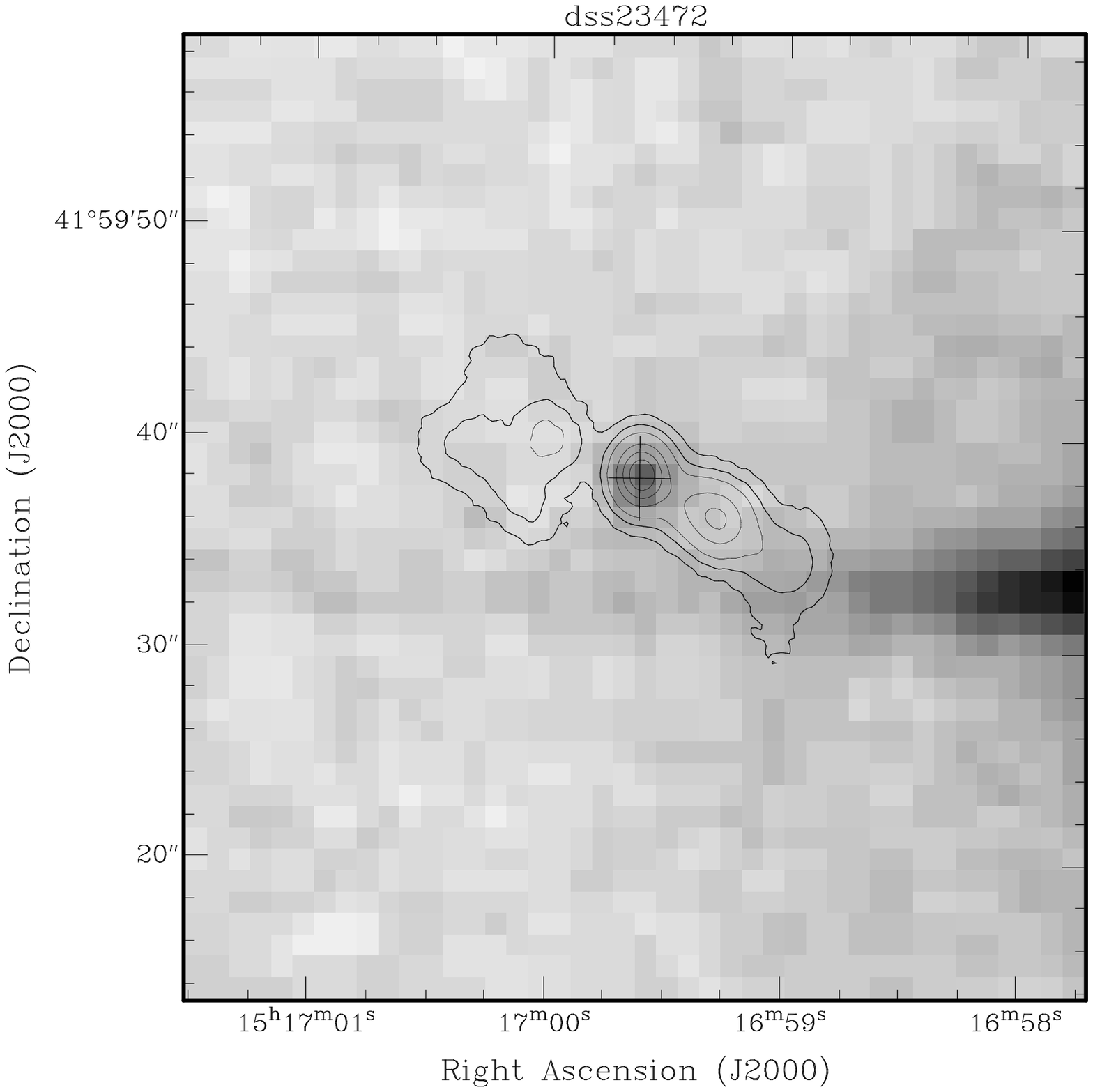,width=4.0cm,clip=}\label{bi}}\quad 
\subfigure[9CJ1517+3936 (P60 \it{R}\normalfont)]{\epsfig{figure=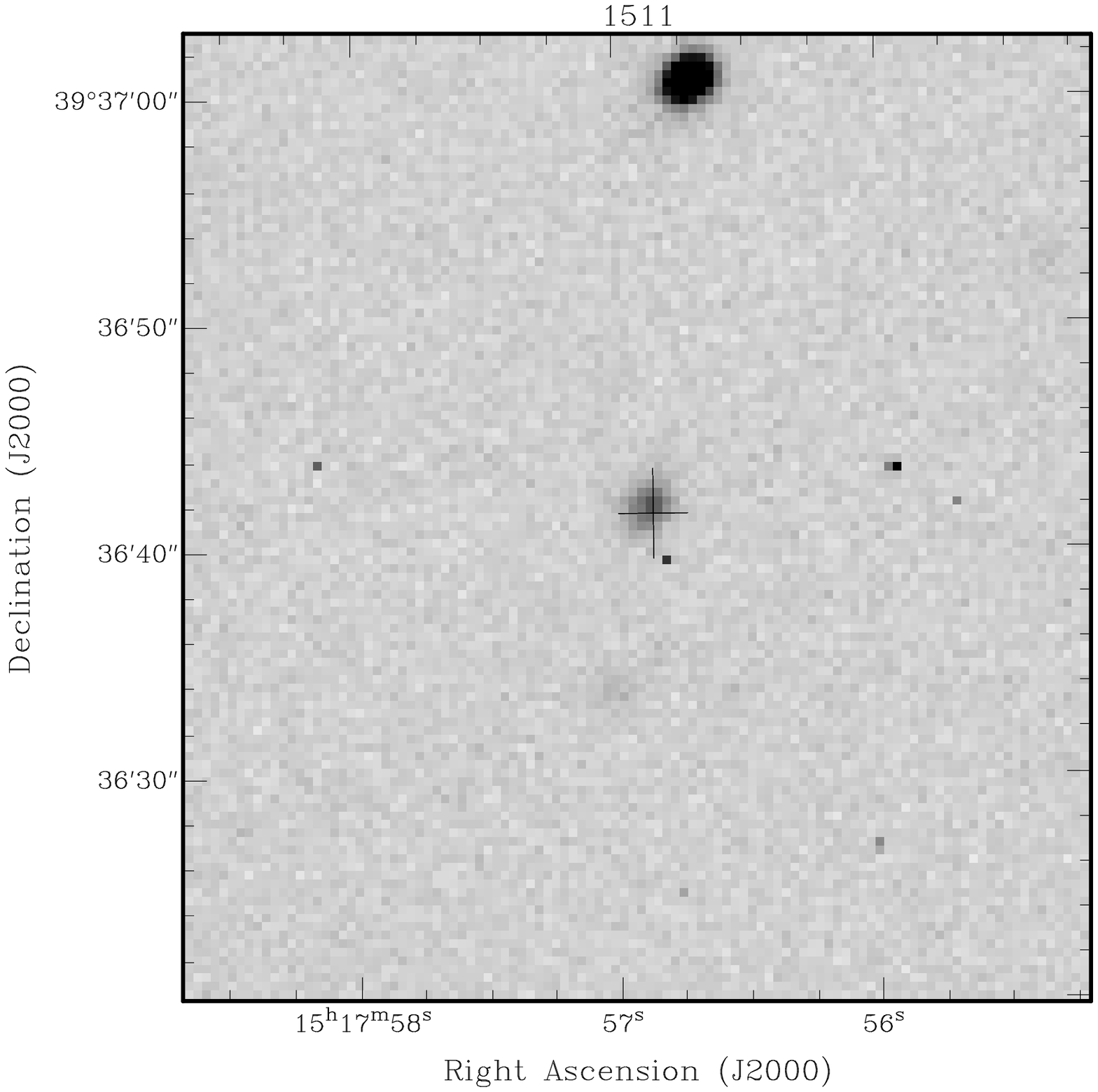,width=4.0cm,clip=}}\quad 
\subfigure[9CJ1518+4131 (DSS2 \it{R}\normalfont)]{\epsfig{figure=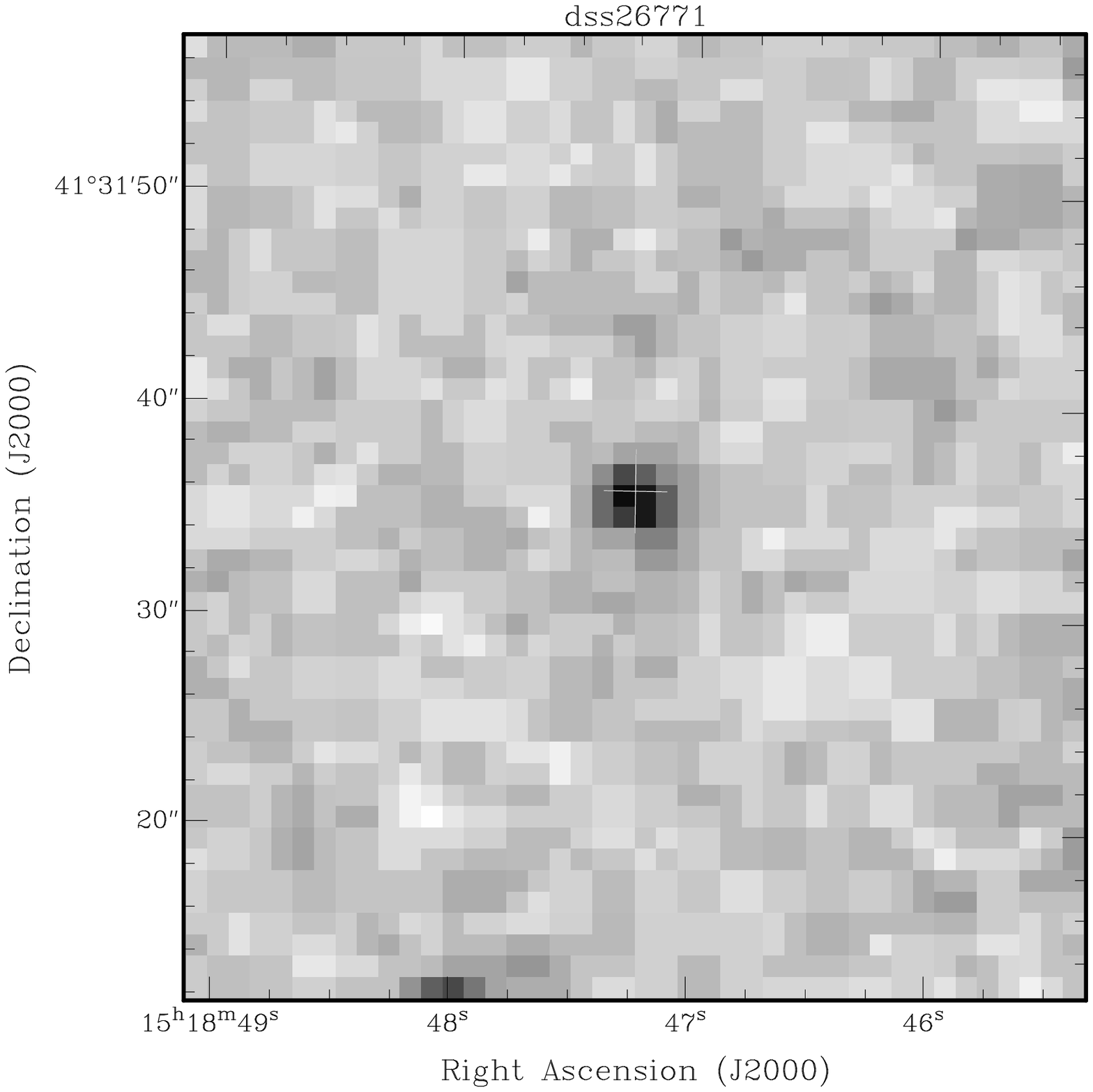,width=4.0cm,clip=}}
} 
\mbox{
\subfigure[9CJ1519+4254 (P60 \it{R}\normalfont)]{\epsfig{figure=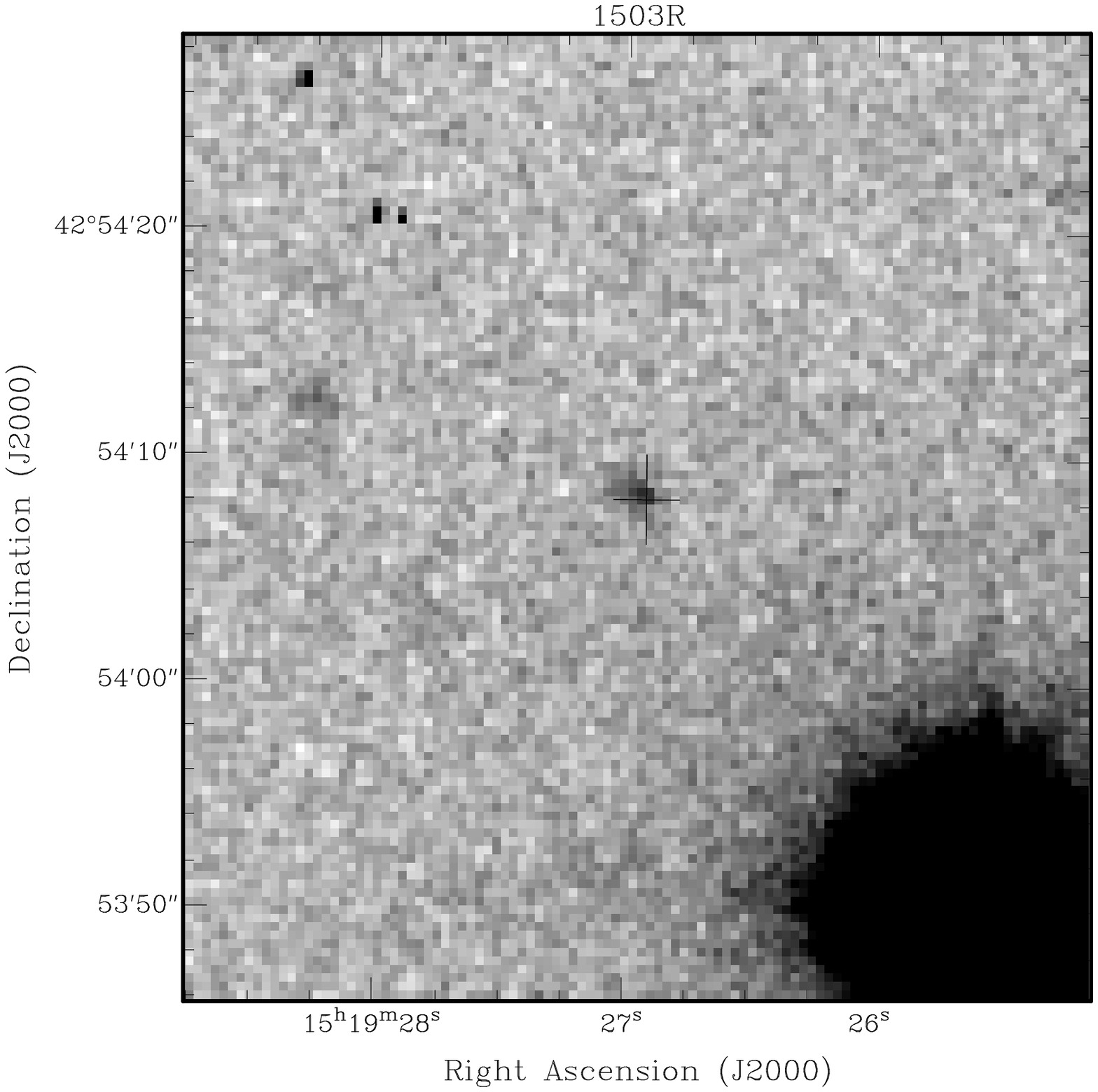,width=4.0cm,clip=}}\quad 
\subfigure[9CJ1519+3844 (DSS2 \it{R}\normalfont)]{\epsfig{figure=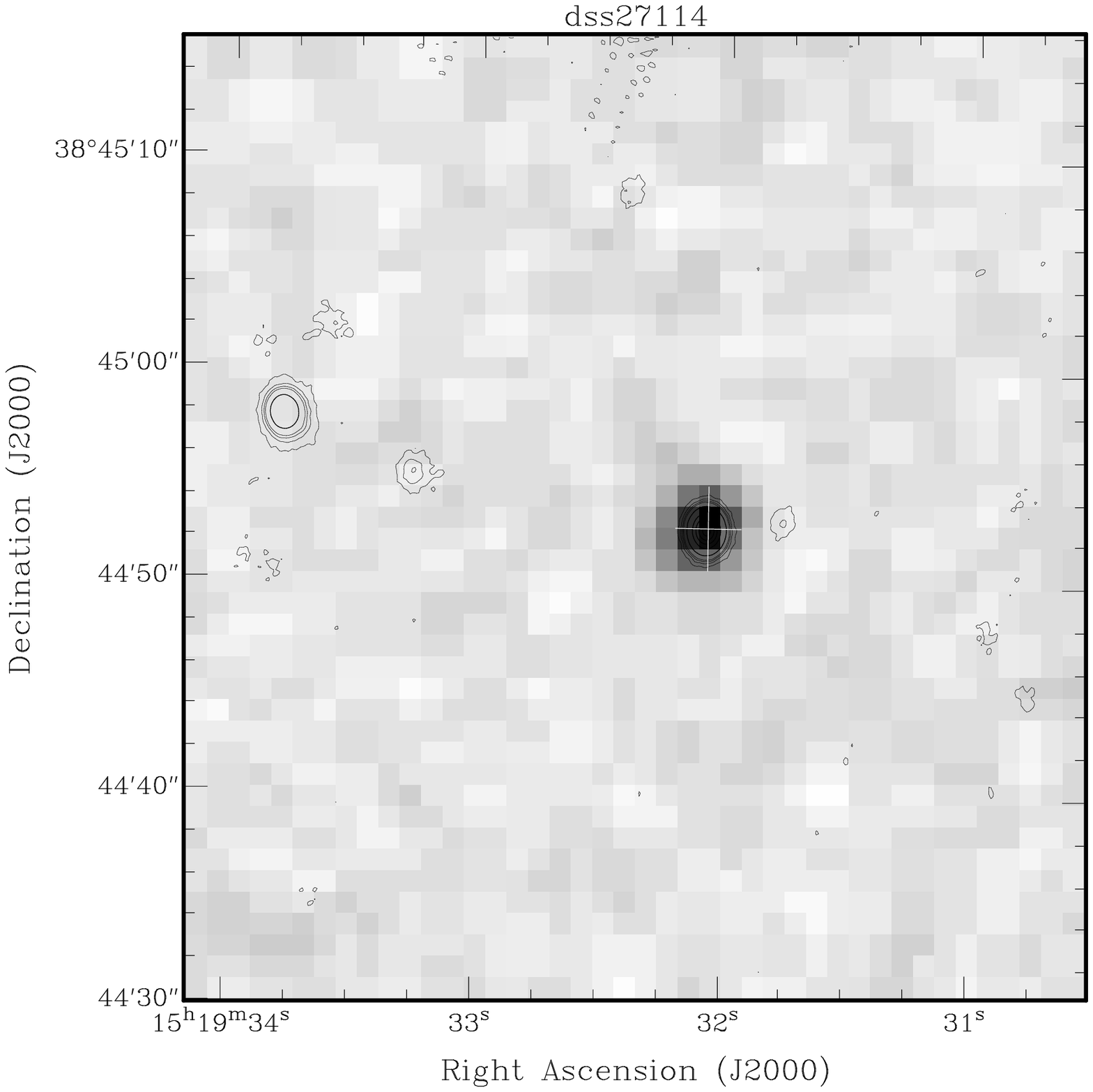,width=4.0cm,clip=}\label{bj}}\quad 
\subfigure[9CJ1519+3913 (DSS2 \it{R}\normalfont)]{\epsfig{figure=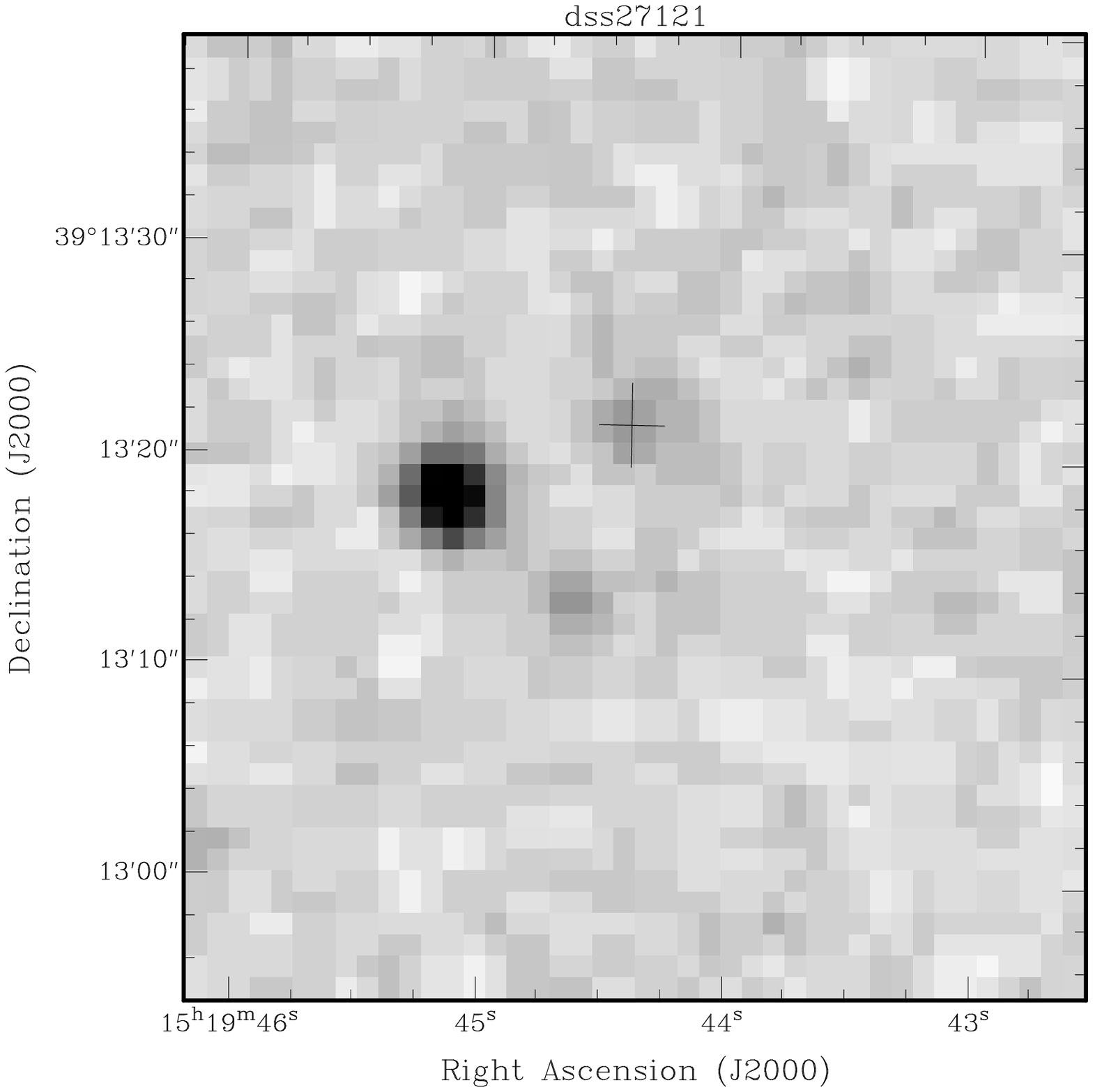,width=4.0cm,clip=}}
} 
\mbox{
\subfigure[9CJ1520+3843 (DSS2 \it{R}\normalfont)]{\epsfig{figure=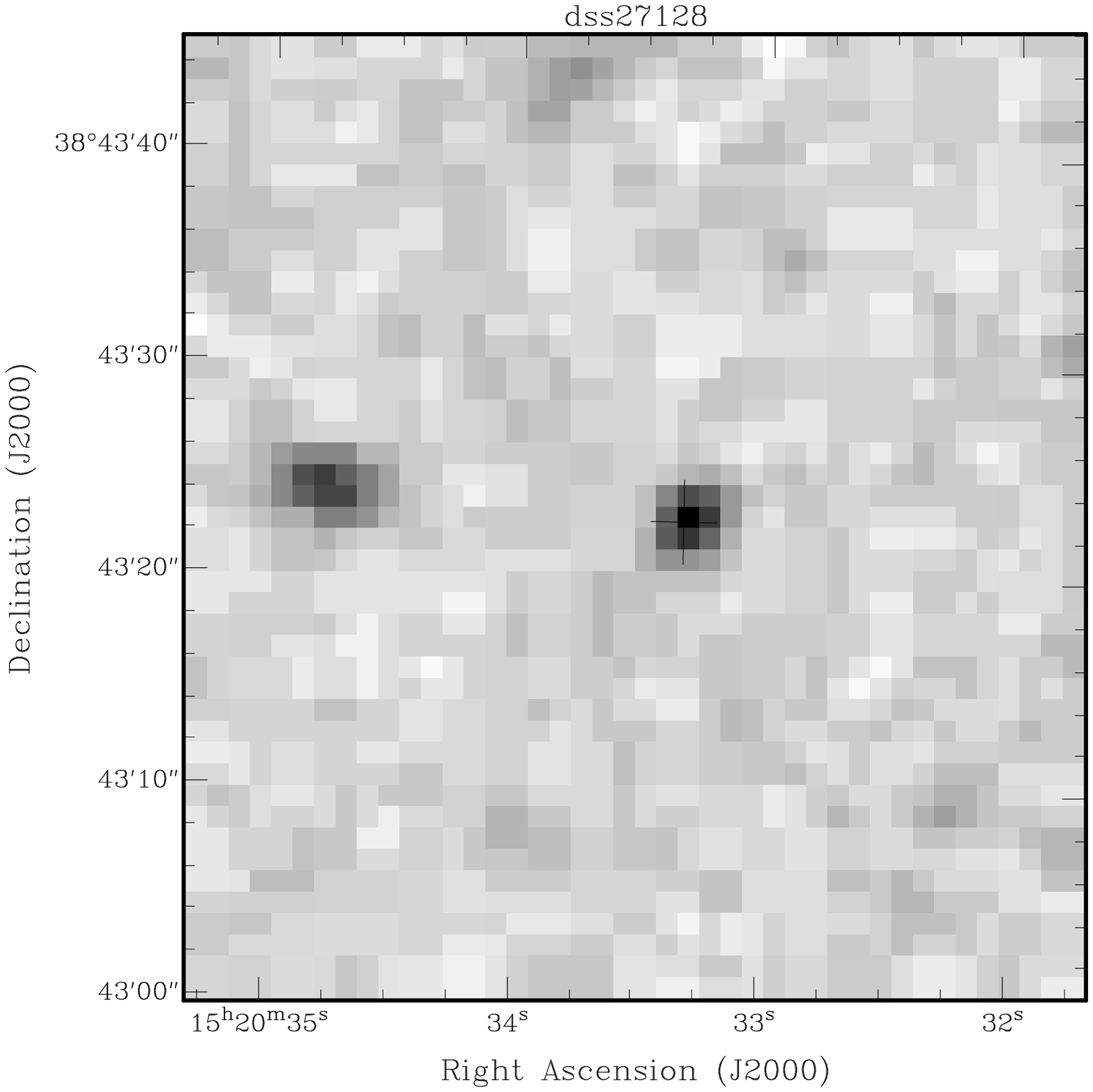,width=4.0cm,clip=}}\quad 
\subfigure[9CJ1520+4211 (DSS2 \it{R}\normalfont)]{\epsfig{figure=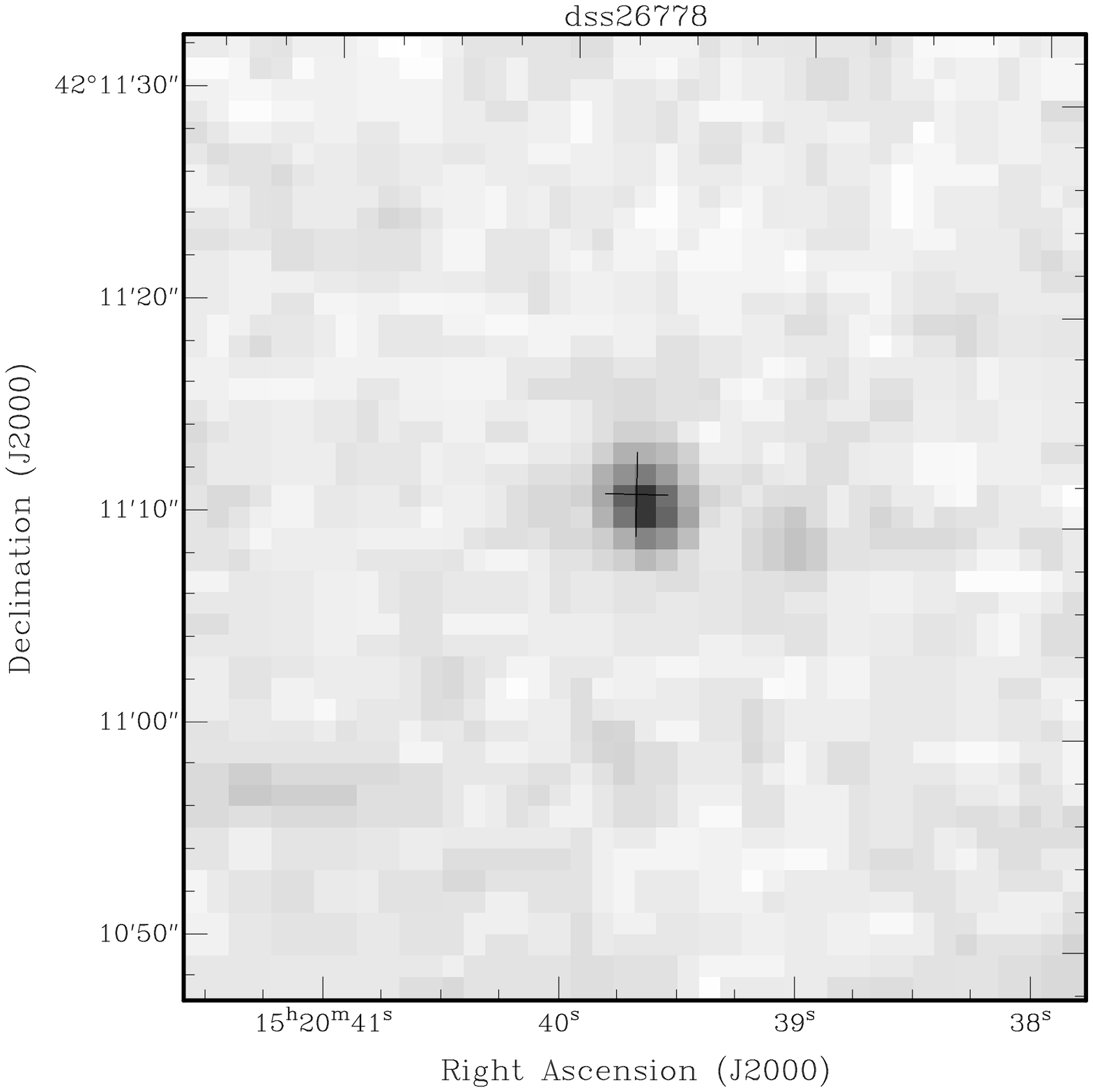,width=4.0cm,clip=}}\quad 
\subfigure[9CJ1521+4336 (DSS2 \it{R}\normalfont)]{\epsfig{figure=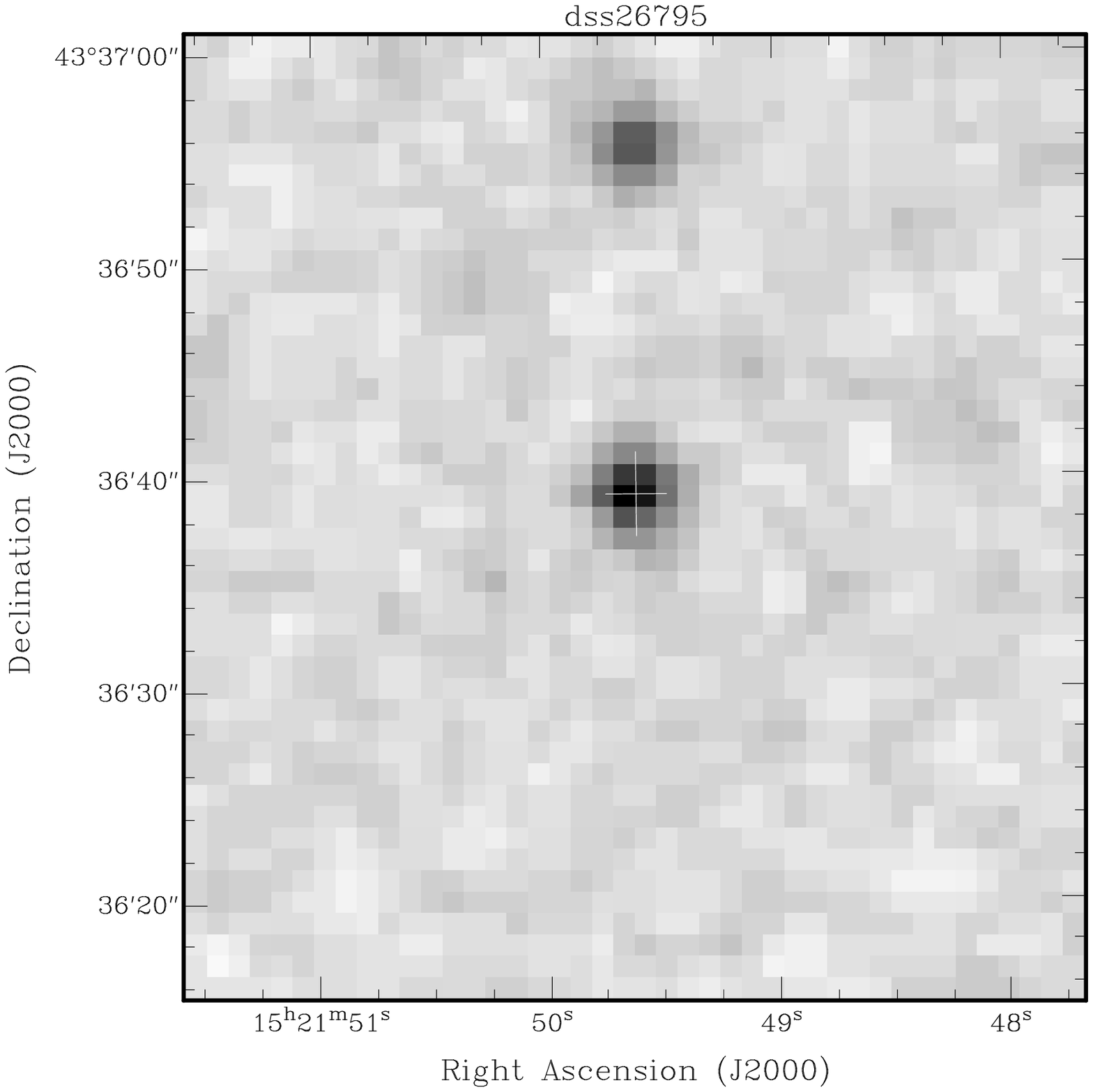,width=4.0cm,clip=}}
} 
\mbox{
\subfigure[9CJ1523+4156 (DSS2 \it{R}\normalfont)]{\epsfig{figure=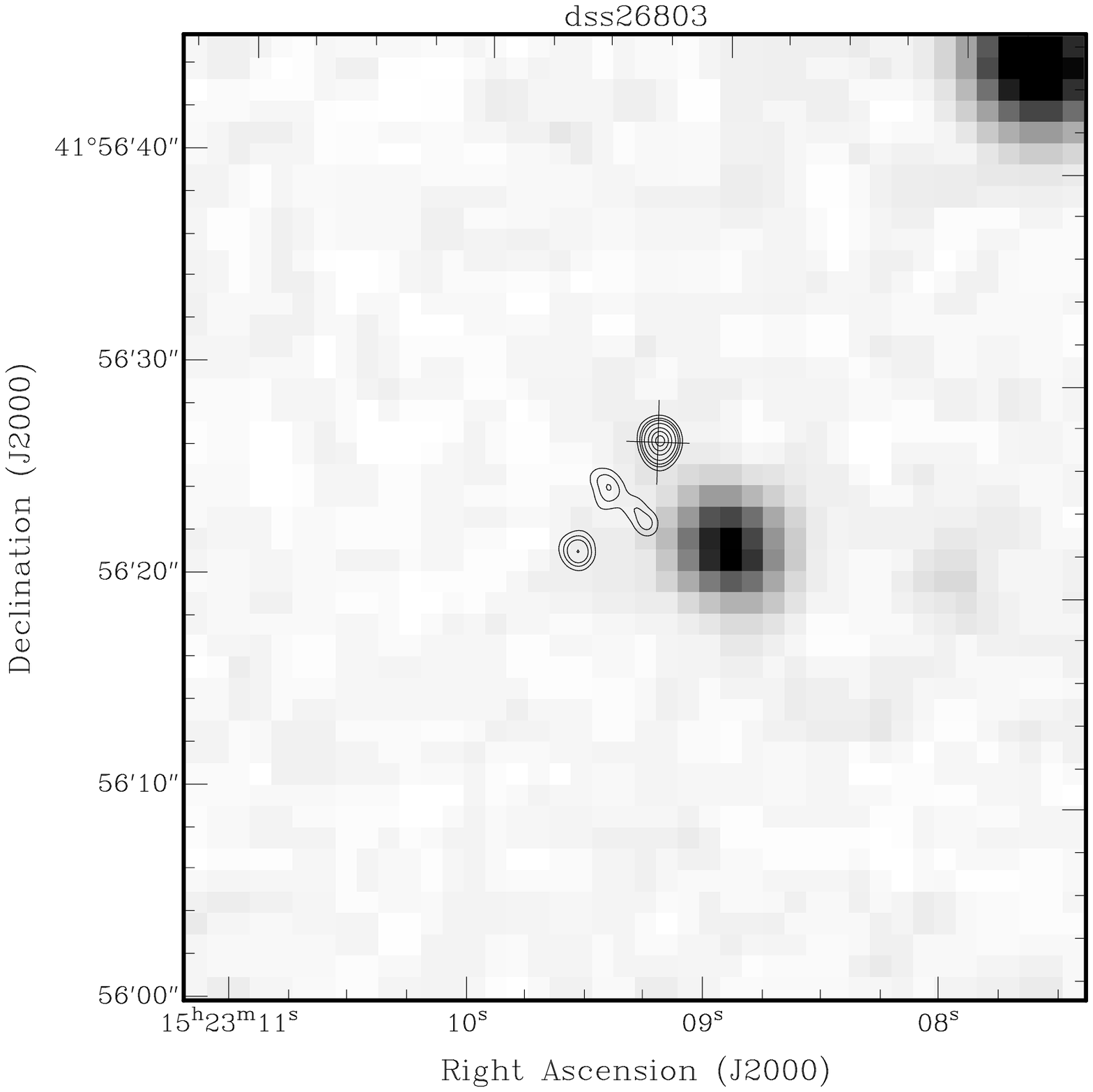,width=4.0cm,clip=}\label{bk}}\quad 
\subfigure[9CJ1525+4201 (DSS2 \it{R}\normalfont)]{\epsfig{figure=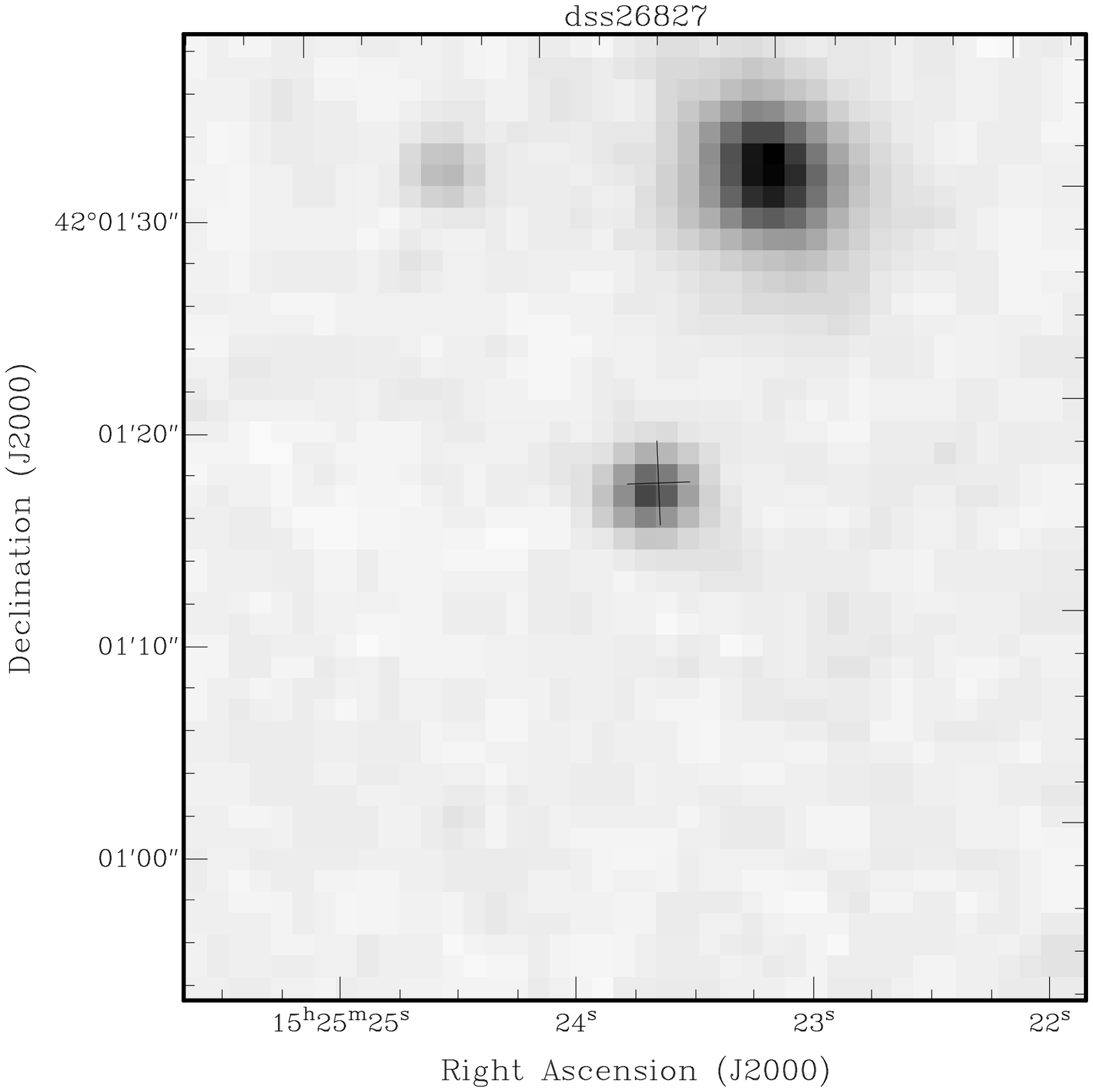,width=4.0cm,clip=}}\quad 
\subfigure[9CJ1526+3712 (P60 \it{R}\normalfont)]{\epsfig{figure=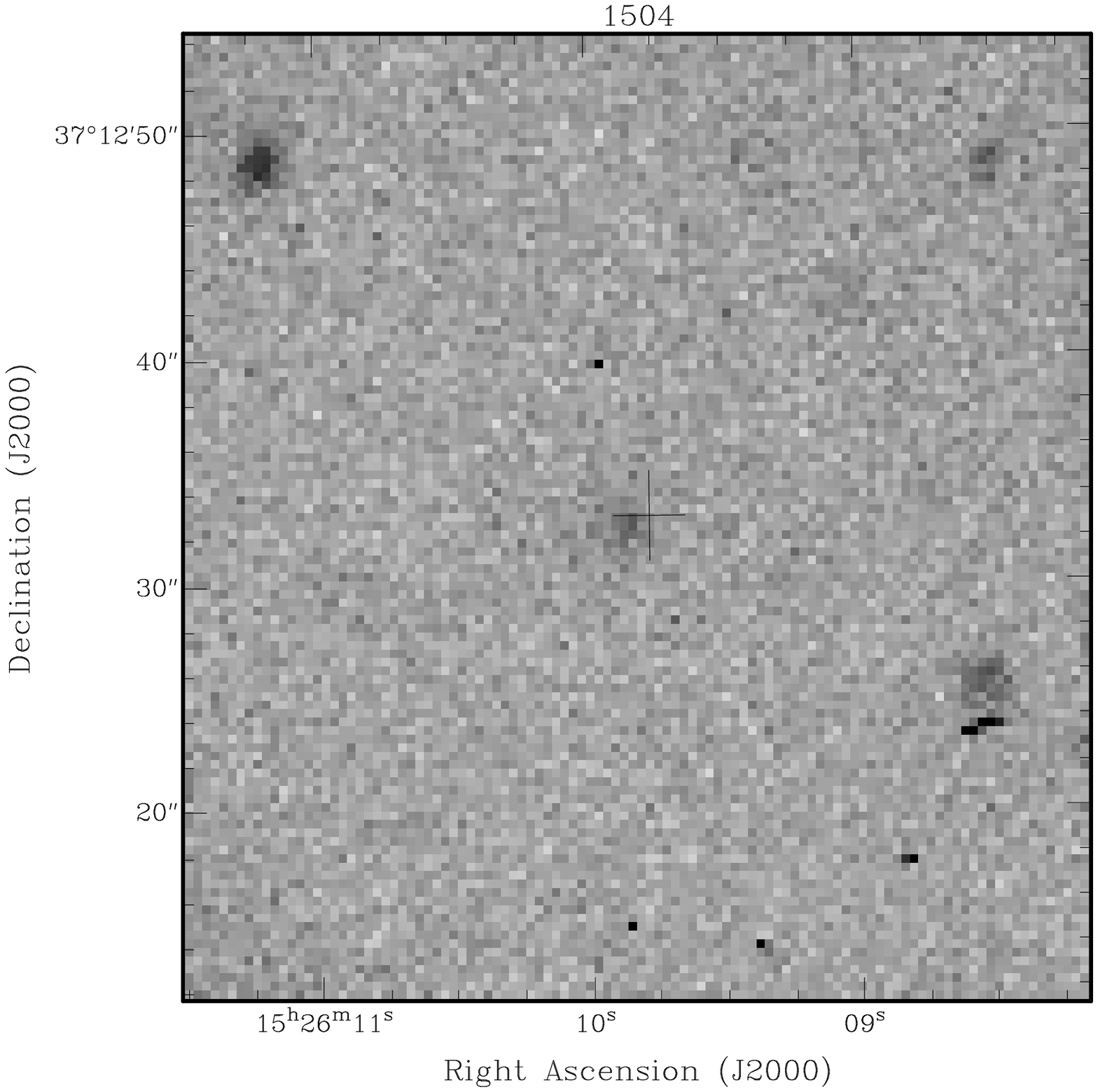,width=4.0cm,clip=}}}\caption{ Optical counterparts for sources 9CJ1516+4159 to 9CJ1526+3712. Crosses mark maximum radio flux density and are 4\,arcsec top to bottom. Contours: \ref{bi}, 4.8\,GHz contours at 7,12 and 20-90 every 10\,\% of peak (31.0\,mJy/beam);\ref{bj}, 4.8\,GHz contours at 2-5 every 1\,\% and 10-90 every 20\,\% of peak (45.6\,mJy/beam); \ref{bk}, 4.8\,GHz contours at 10,15,20 and 30-90 every 20\,\% of peak (86.1\,mJy/beam).}\end{figure*}

\begin{figure*}
\mbox{
\subfigure[9CJ1526+4201 (P60 \it{R}\normalfont)]{\epsfig{figure=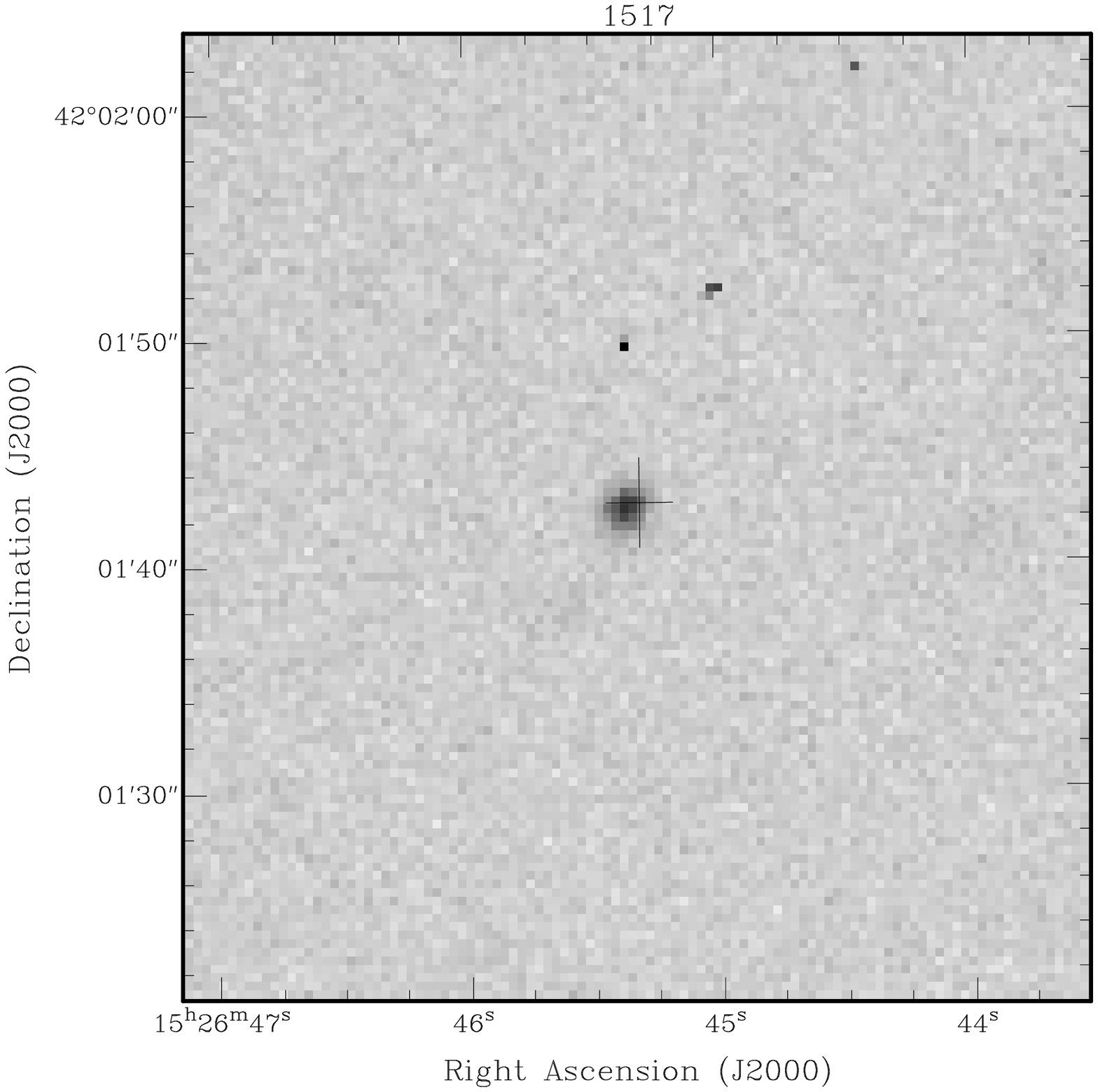 ,width=4.0cm,clip=}}\quad 
\subfigure[9CJ1528+4219 (DSS2 \it{R}\normalfont)]{\epsfig{figure=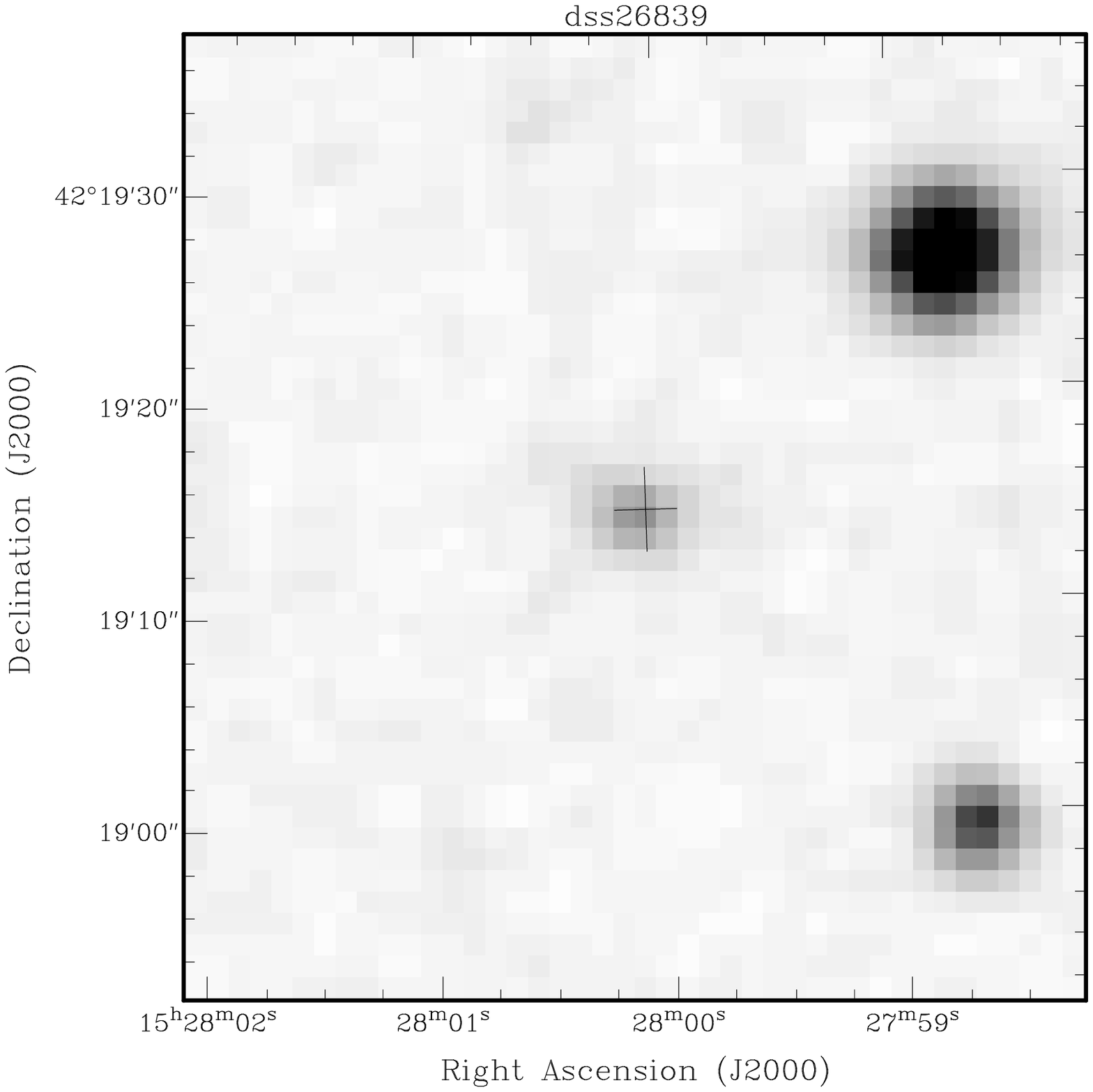 ,width=4.0cm,clip=}}\quad 
\subfigure[9CJ1528+4233 (P60 \it{R}\normalfont)]{\epsfig{figure=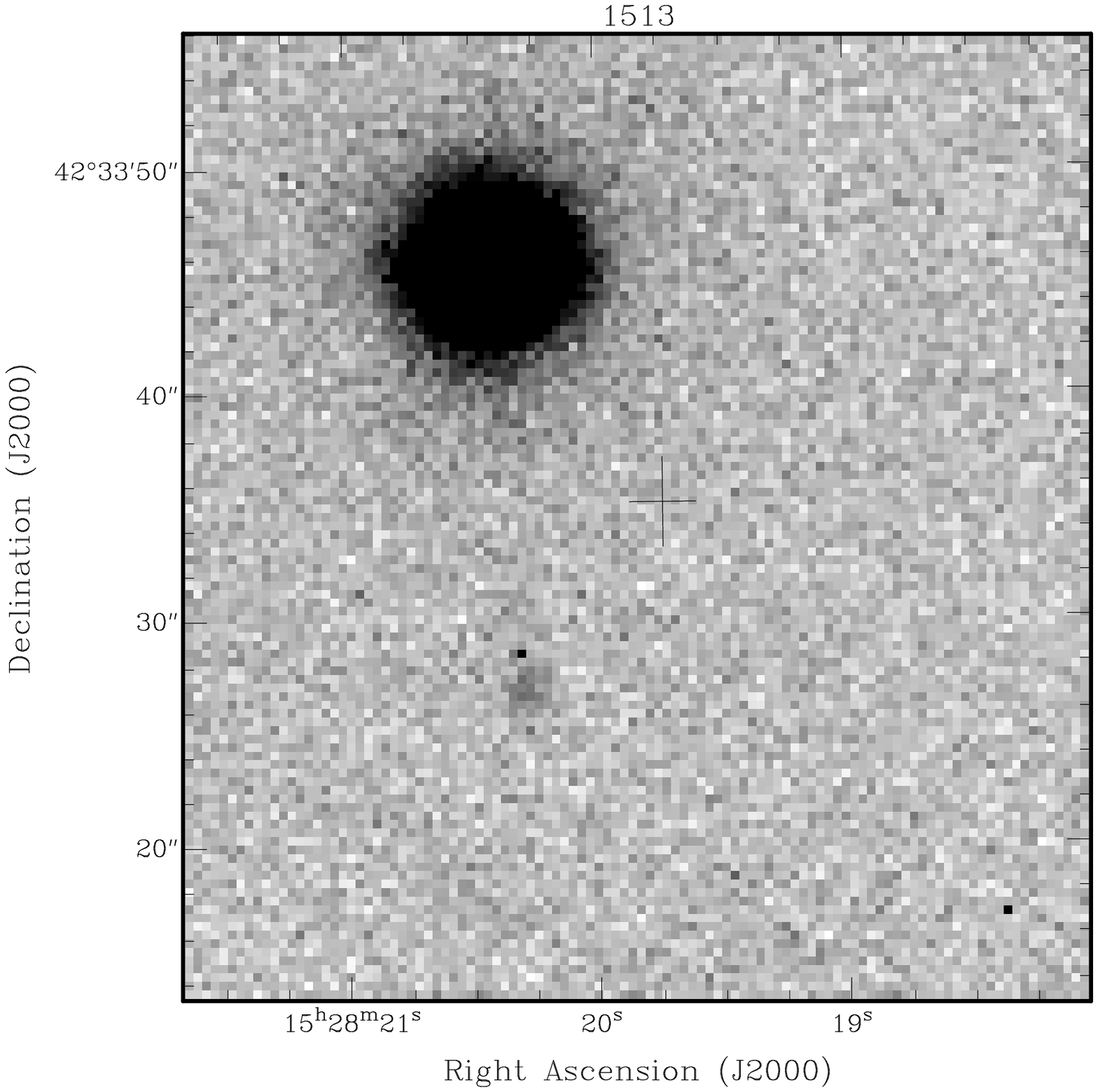 ,width=4.0cm,clip=}}
} 
\mbox{
\subfigure[9CJ1528+4233 (DSS2 \it{O}\normalfont]{\epsfig{figure=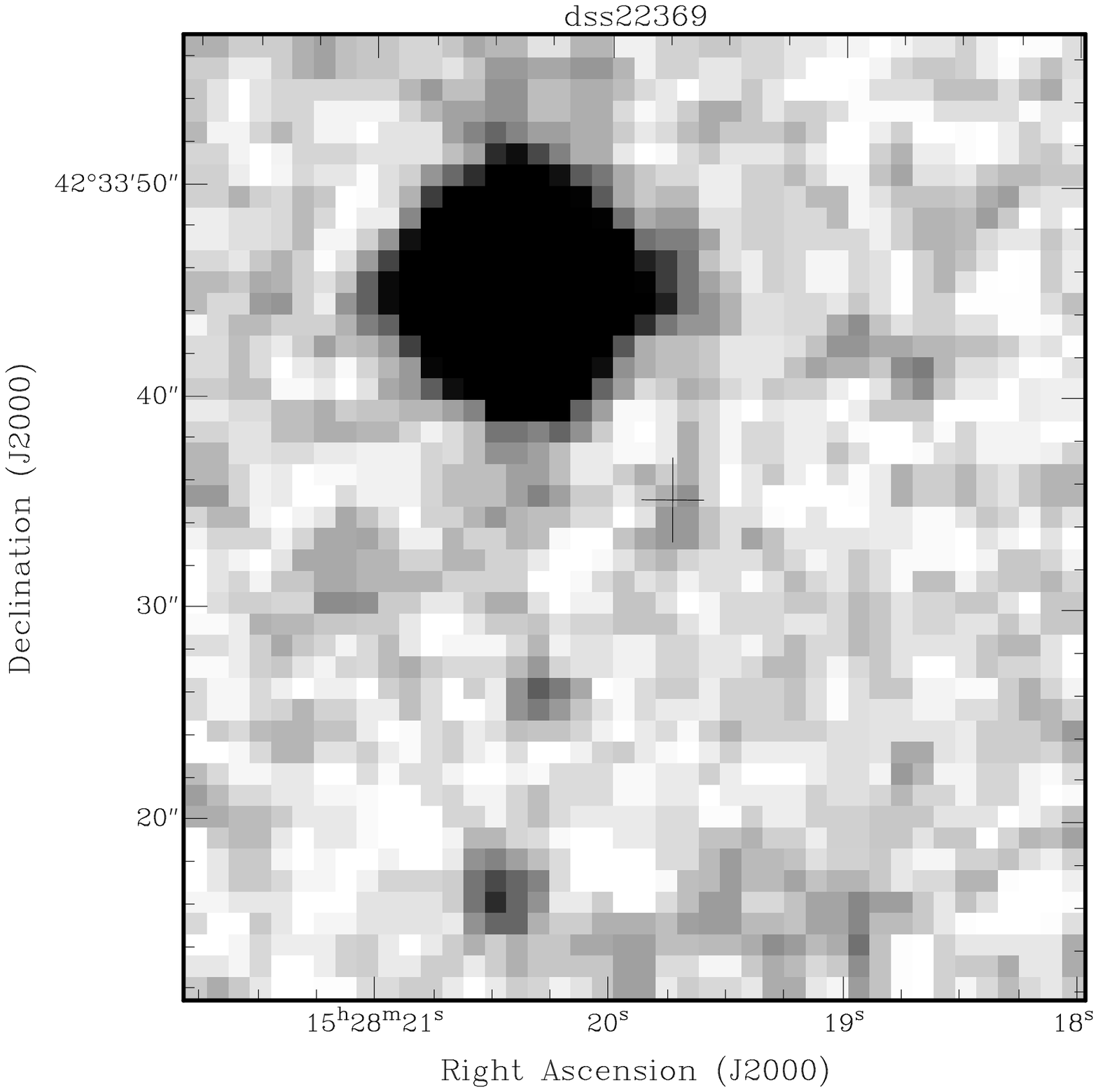 ,width=4.0cm,clip=}}\quad 
\subfigure[9CJ1528+3738 (P60 \it{R}\normalfont)]{\epsfig{figure=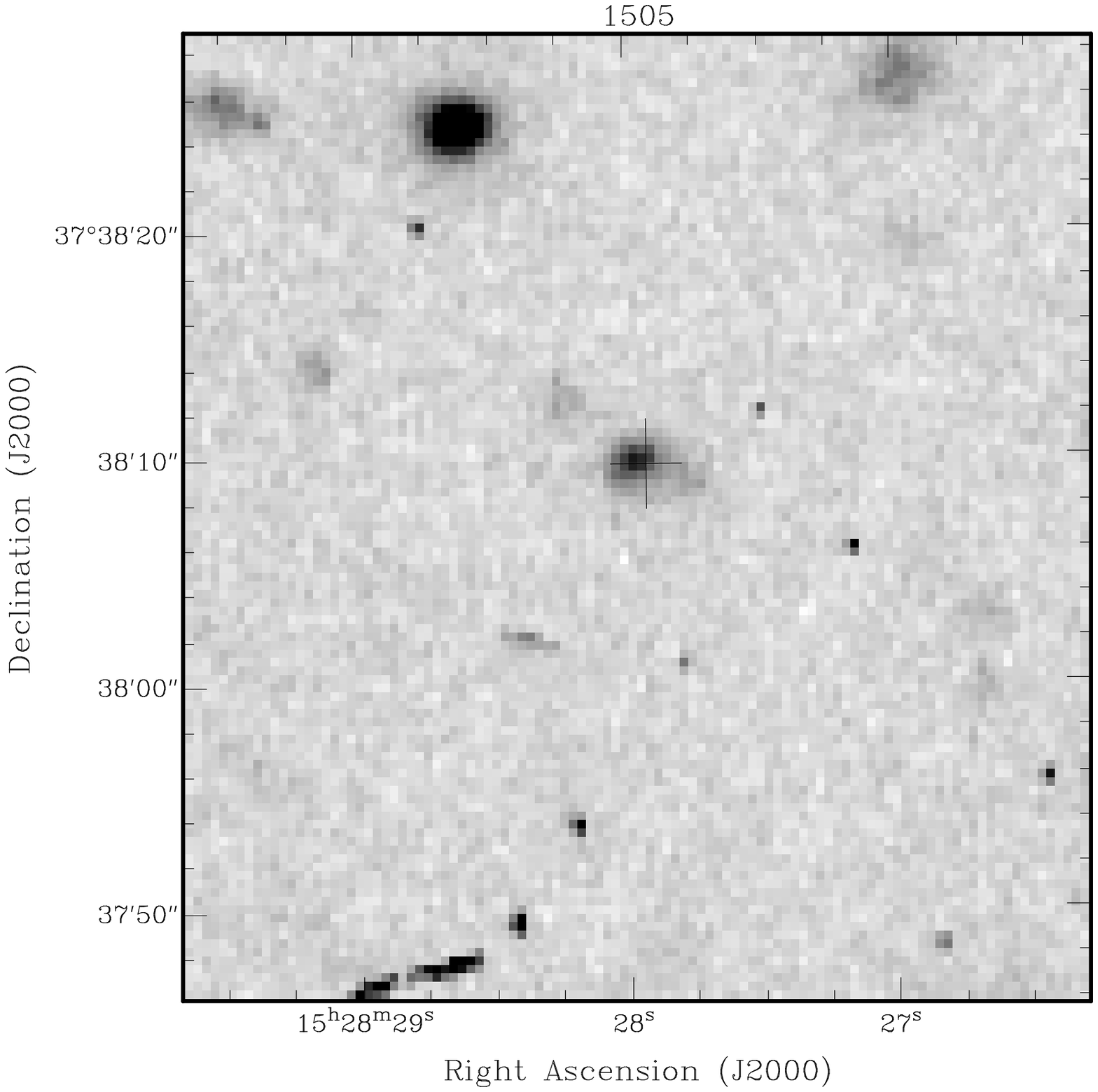 ,width=4.0cm,clip=}\label{bl}}\quad 
\subfigure[9CJ1528+3738 (P60 \it{R}\normalfont) Detail]{\epsfig{figure=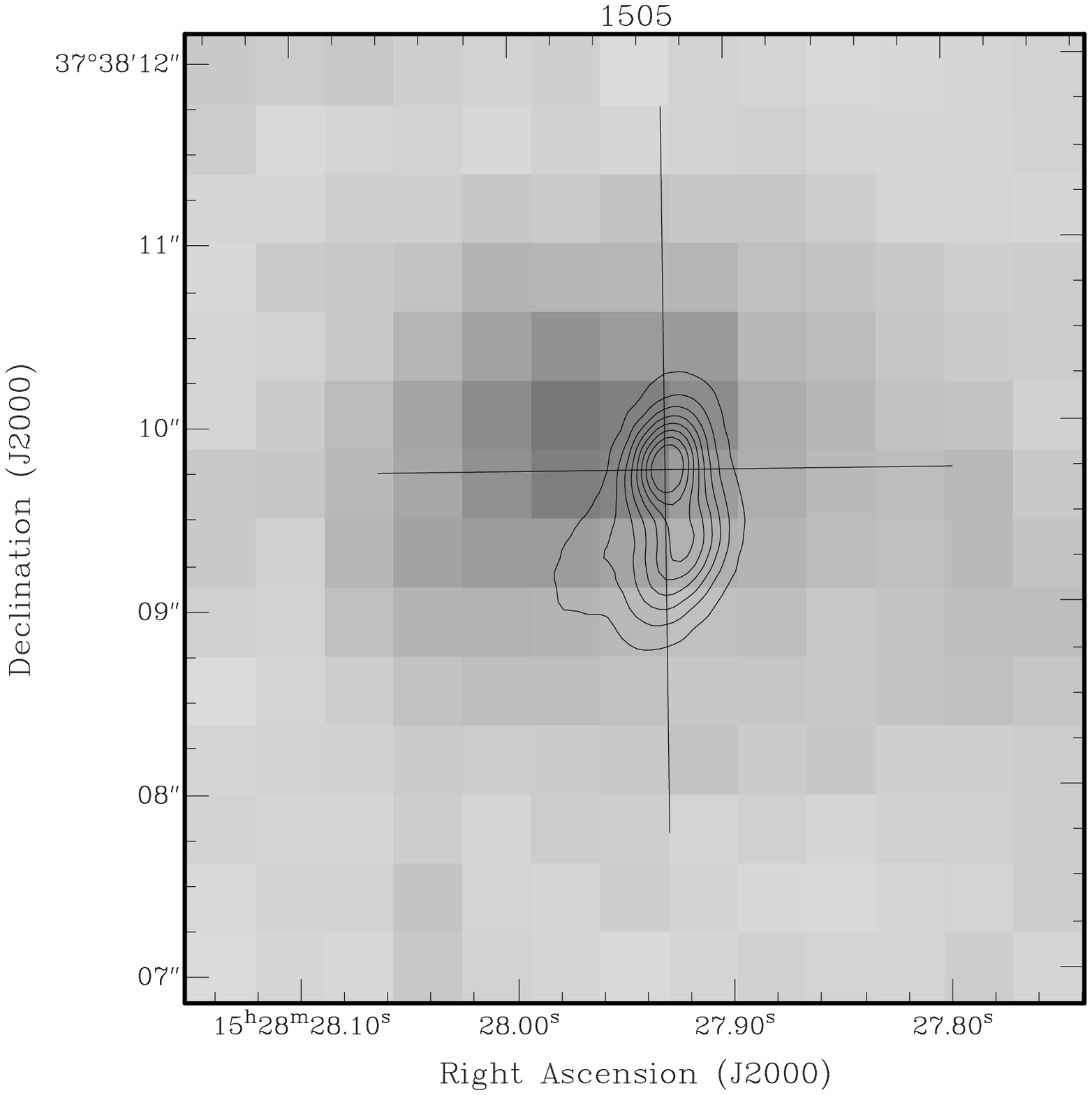 ,width=4.0cm,clip=}\label{bm}}
} 
\mbox{
\subfigure[9CJ1528+3816 (P60 \it{R}\normalfont)]{\epsfig{figure=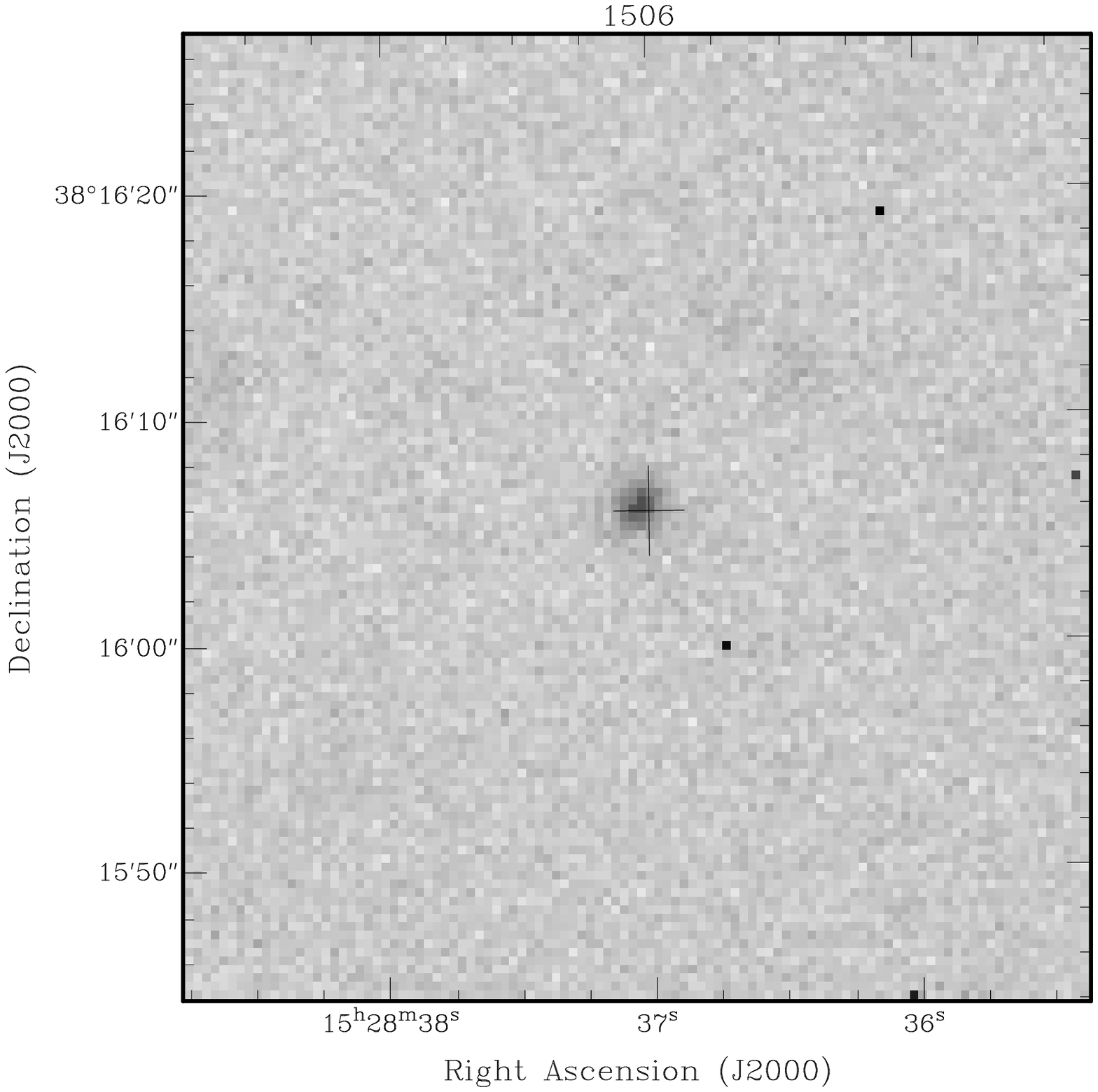 ,width=4.0cm,clip=}}\quad 
\subfigure[9CJ1528+4522 (P60 \it{R}\normalfont)]{\epsfig{figure=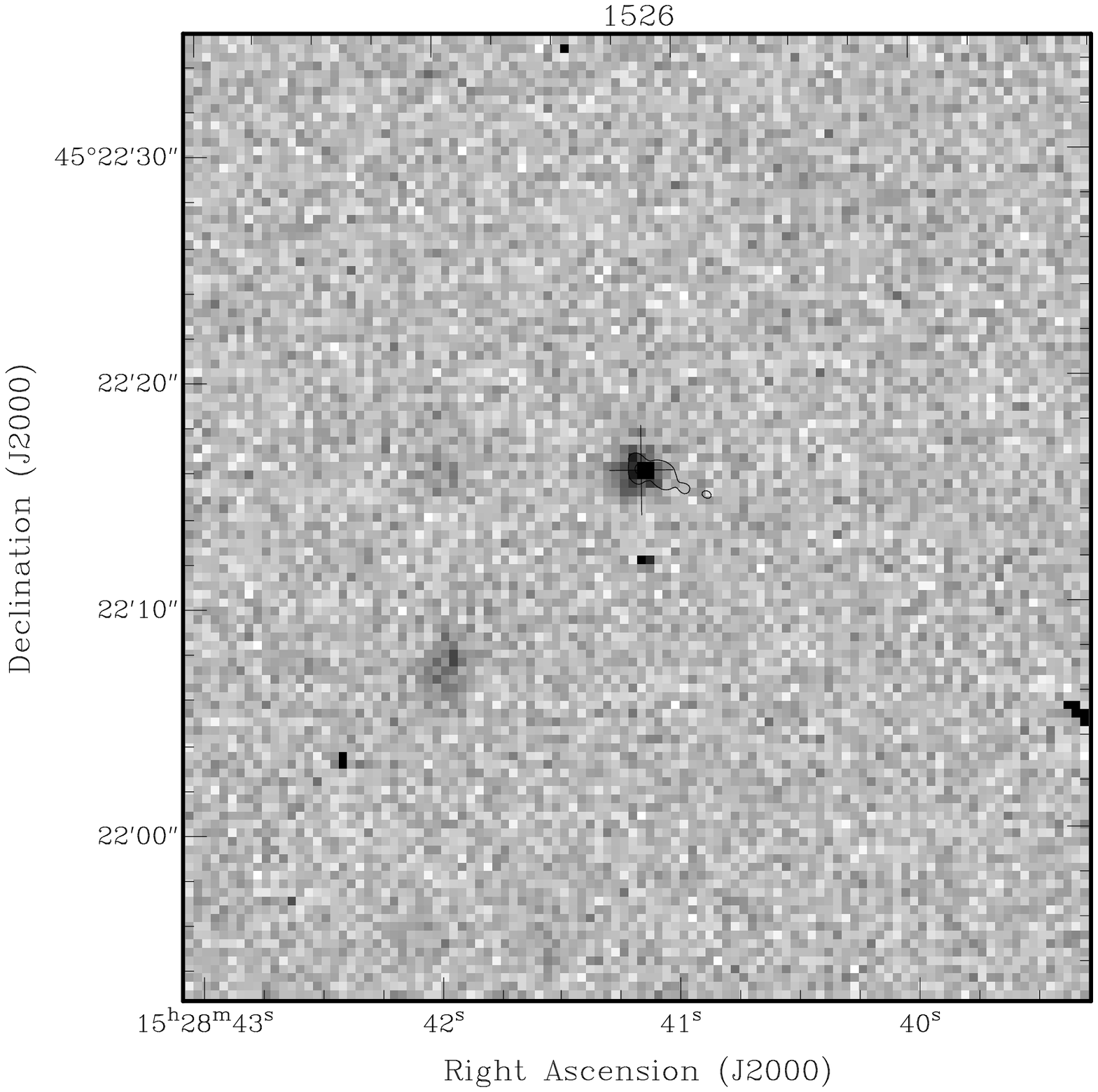 ,width=4.0cm,clip=}\label{bn}}\quad 
\subfigure[9CJ1528+4522 (P60 \it{R}\normalfont) Detail]{\epsfig{figure=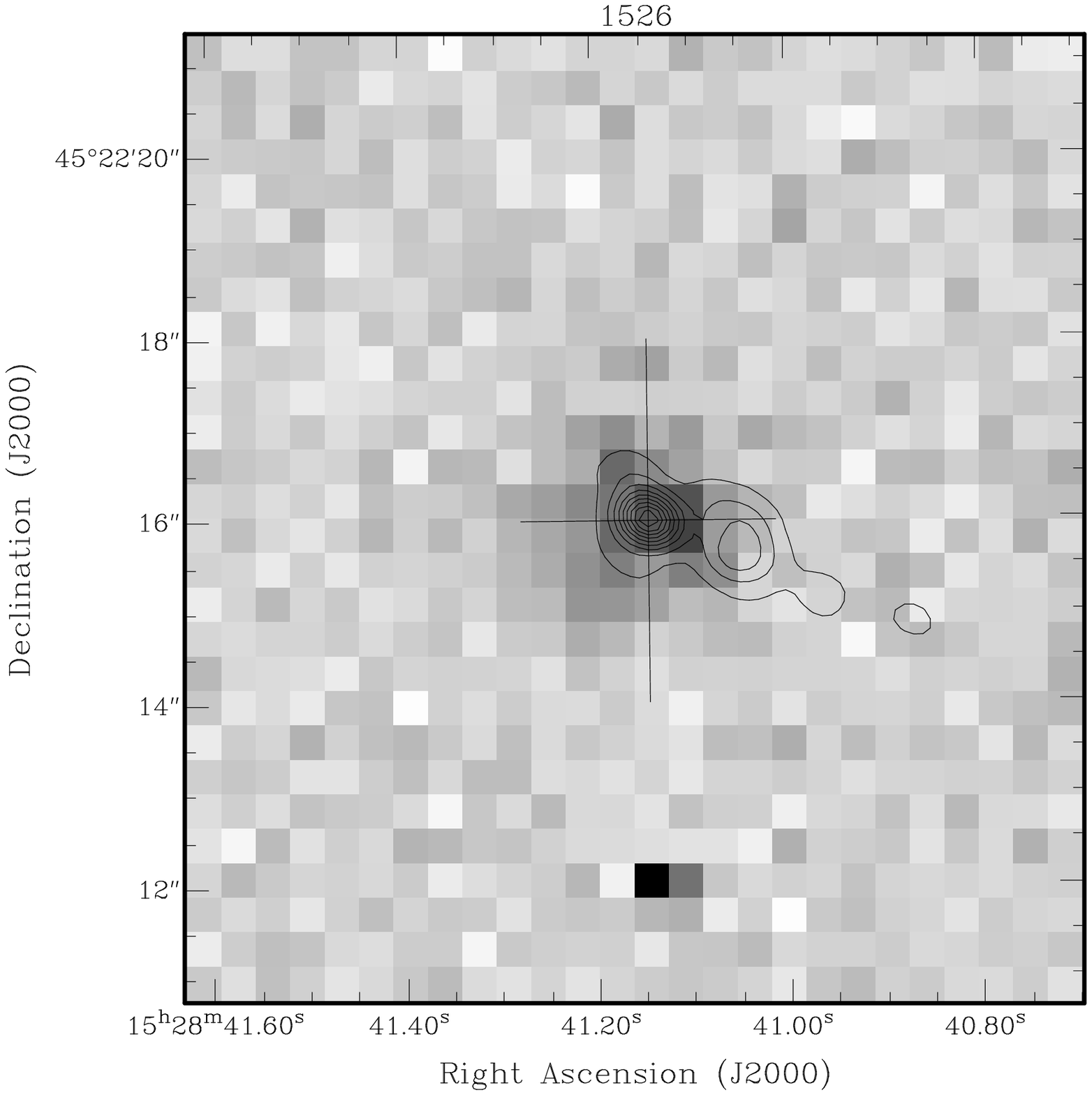 ,width=4.0cm,clip=}\label{bo}}
} 
\mbox{
\subfigure[9CJ1529+4538 (DSS2 \it{R}\normalfont)]{\epsfig{figure=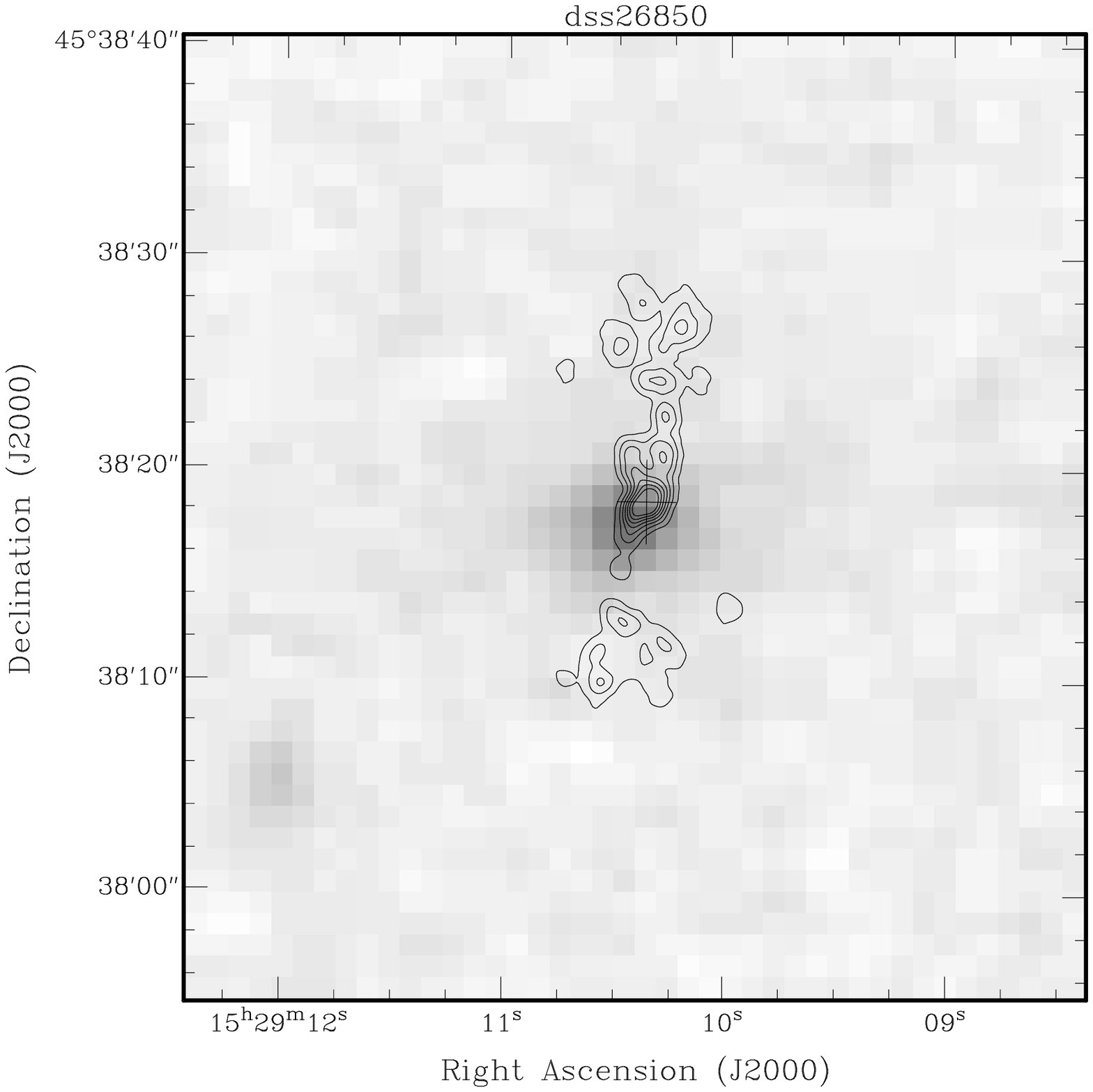 ,width=4.0cm,clip=}\label{bp}}\quad 
\subfigure[9CJ1529+3945 (DSS2 \it{R}\normalfont)]{\epsfig{figure=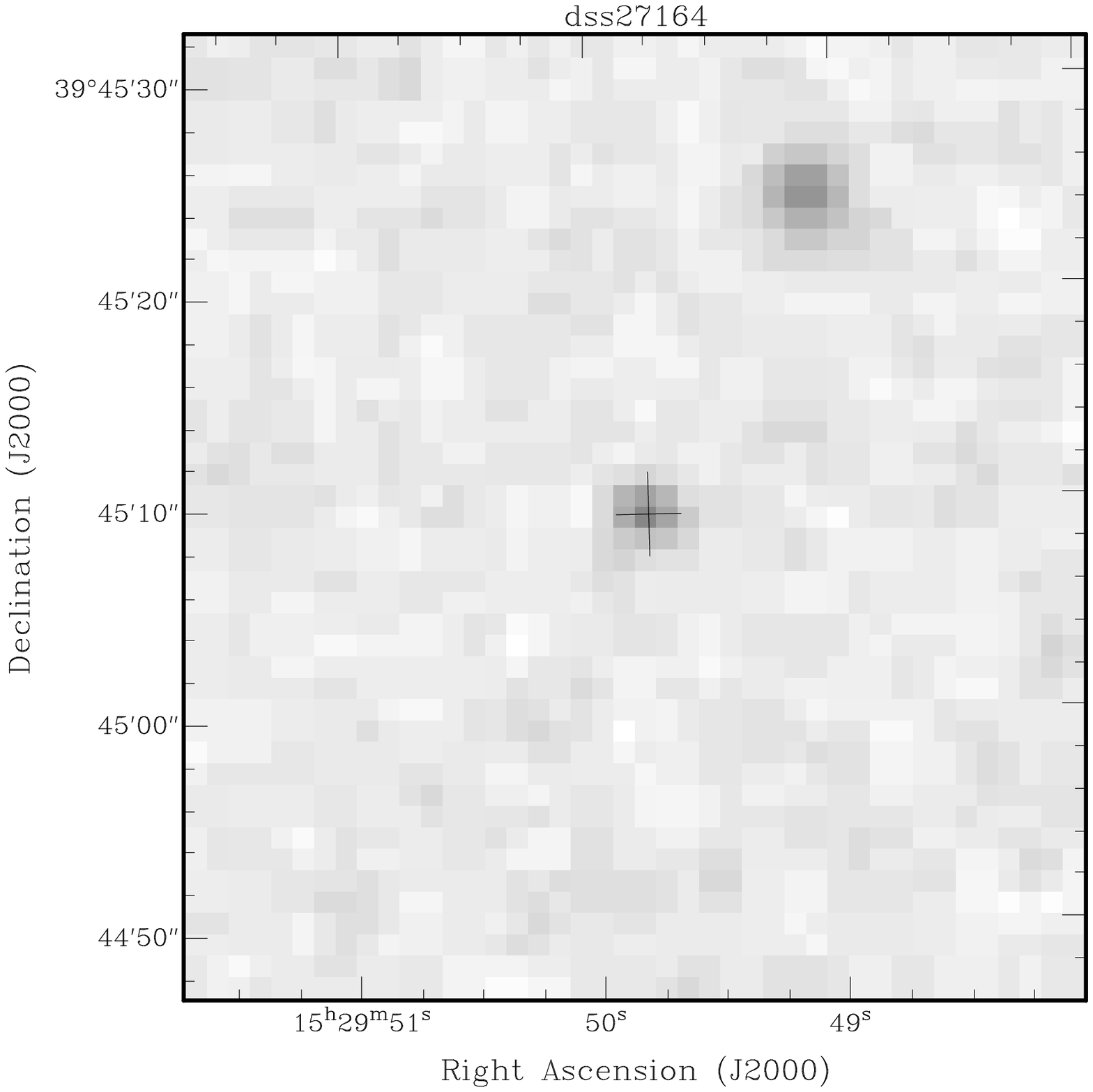 ,width=4.0cm,clip=}}\quad 
\subfigure[9CJ1530+3758 (P60 \it{R}\normalfont)]{\epsfig{figure=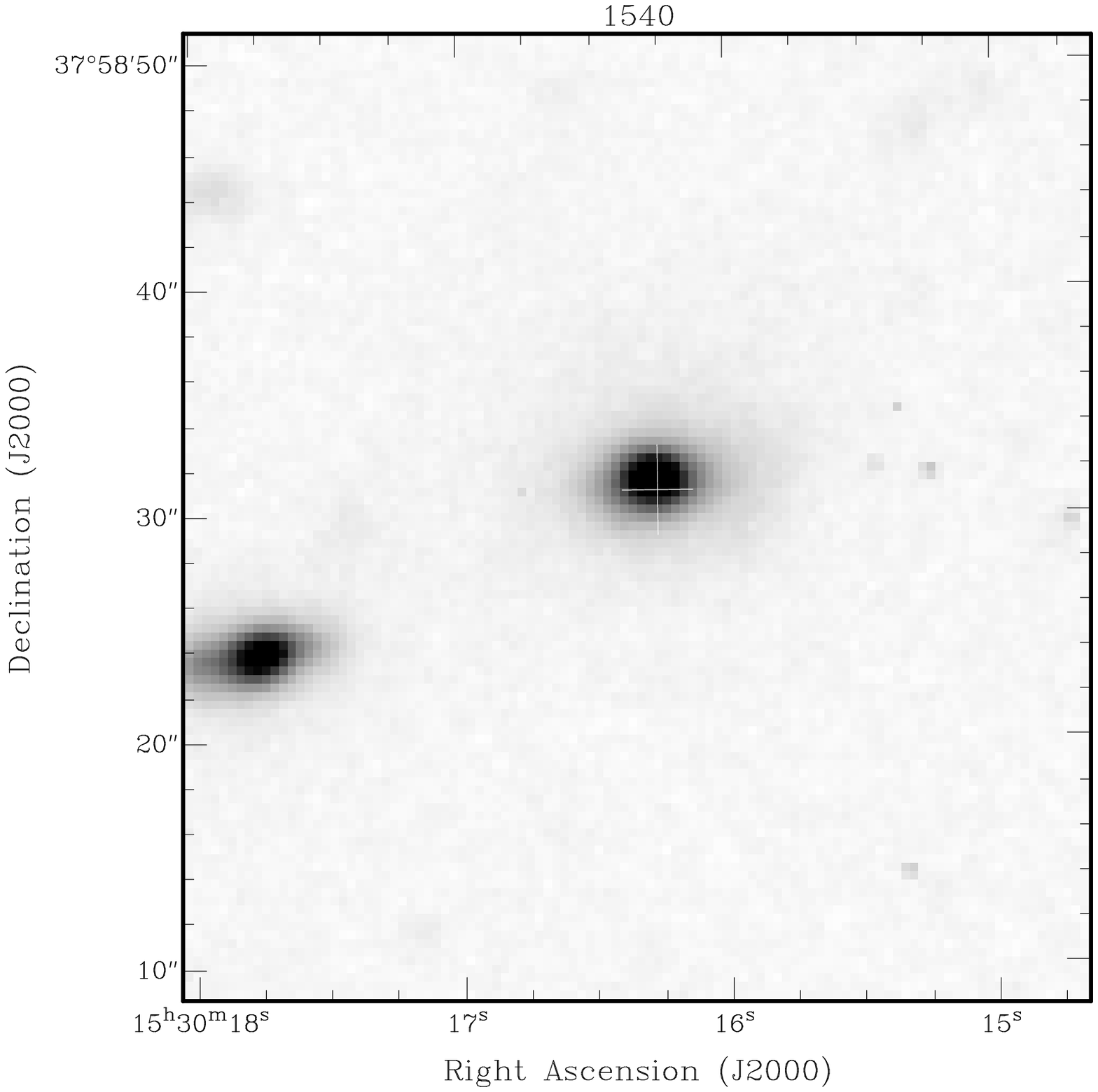 ,width=4.0cm,clip=}}}\caption{ Optical counterparts for sources 9CJ1526+4201 to 9CJ1530+3758. Crosses mark maximum radio flux density and are 4\,arcsec top to bottom. Contours: \ref{bm} 22\,GHz contours at 15-90 every 15\,\% of peak (48.6\,mJy/beam); \ref{bn}, 4.8\,GHz contours at 10,50 and 90\,\% of peak (29.3\,mJy/beam); \ref{bo}, 4.8\,GHz contours at 10-90 every 10\,\% of peak (29.3\,mJy/beam); \ref{bp}, 4.8\,GHz contours at 20-80 every 20\,\% of peak (9.9\,mJy/beam).}\end{figure*}

\begin{figure*}
\mbox{
\subfigure[9CJ1531+4356 (P60 \it{R}\normalfont)]{\epsfig{figure=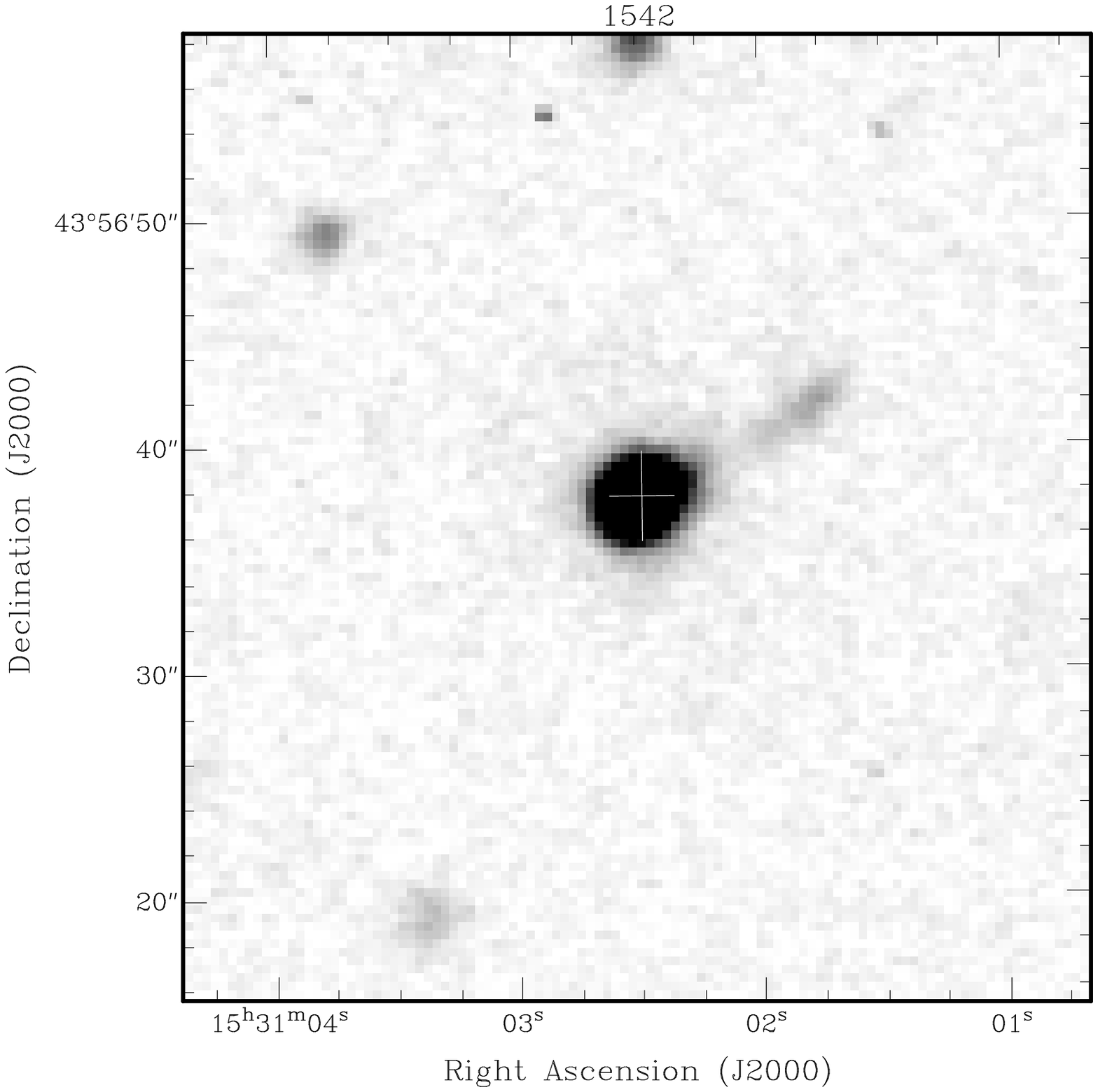  ,width=4.0cm,clip=}}\quad 
\subfigure[9CJ1531+4048 (DSS2 \it{R}\normalfont)]{\epsfig{figure=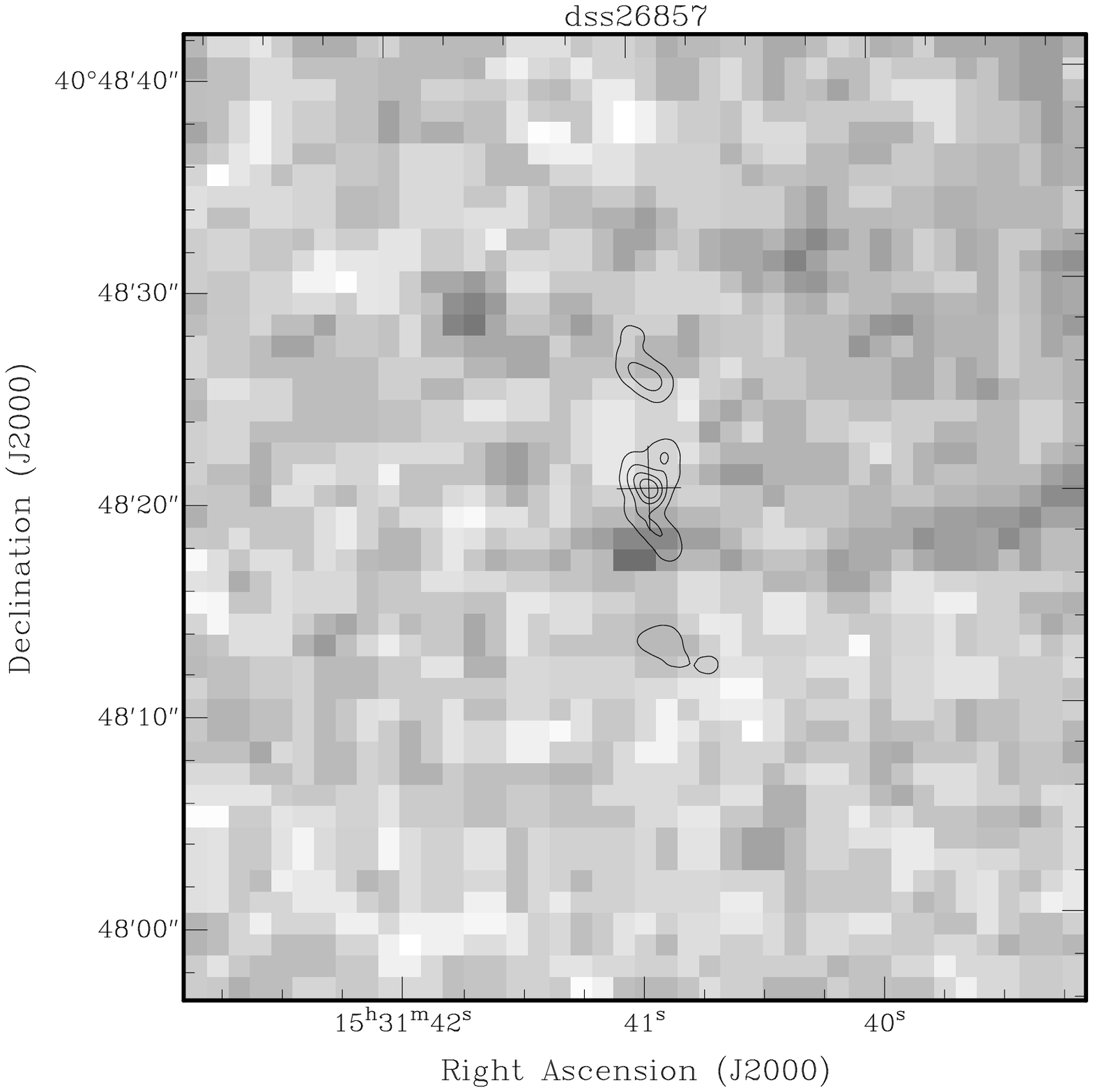 ,width=4.0cm,clip=}\label{bq}}\quad 
\subfigure[9CJ1531+4048 (DSS2 \it{O}\normalfont)]{\epsfig{figure=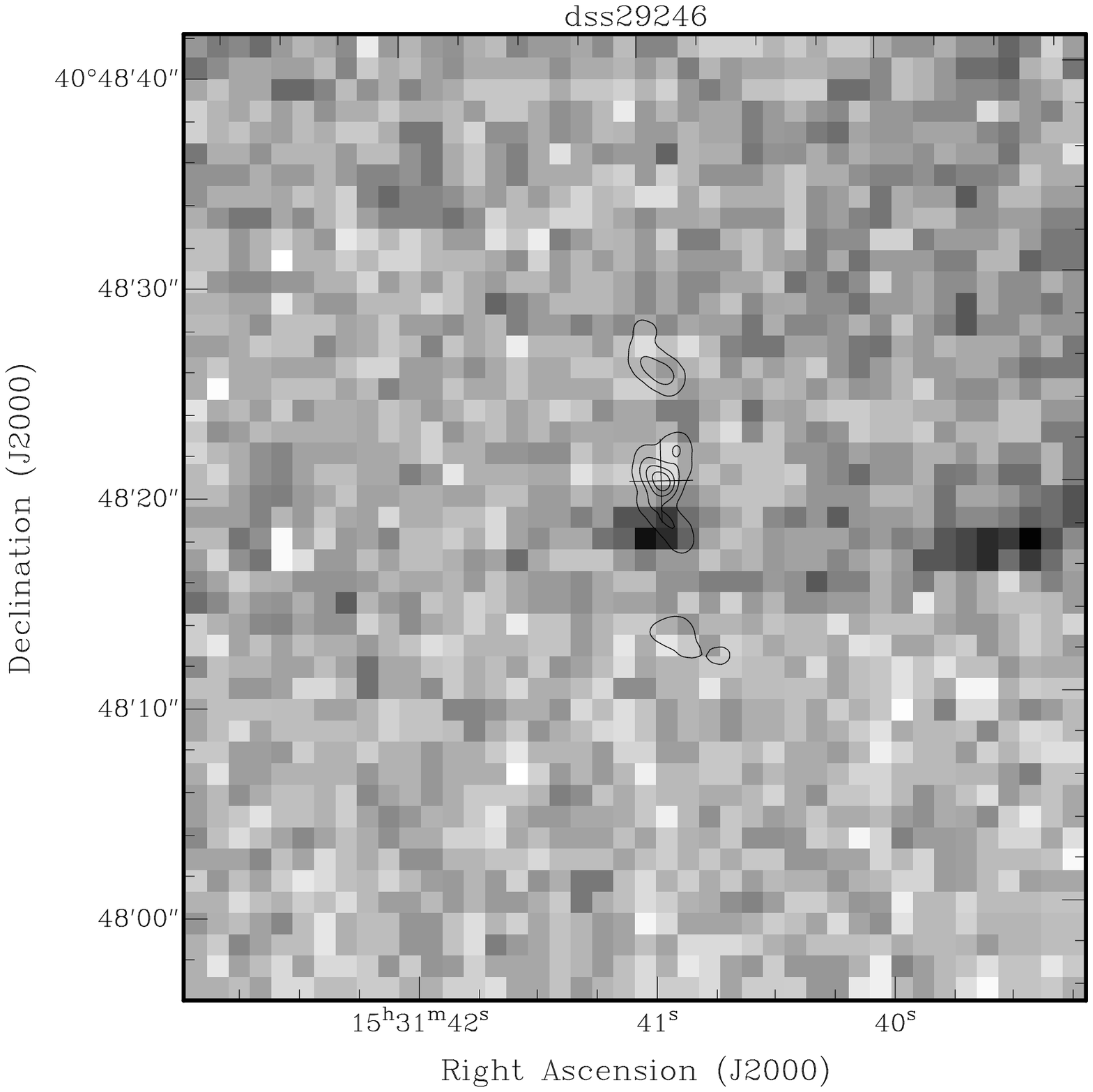 ,width=4.0cm,clip=}\label{br}}
} 
\mbox{
\subfigure[9CJ1533+4107 (DSS2 \it{R}\normalfont)]{\epsfig{figure=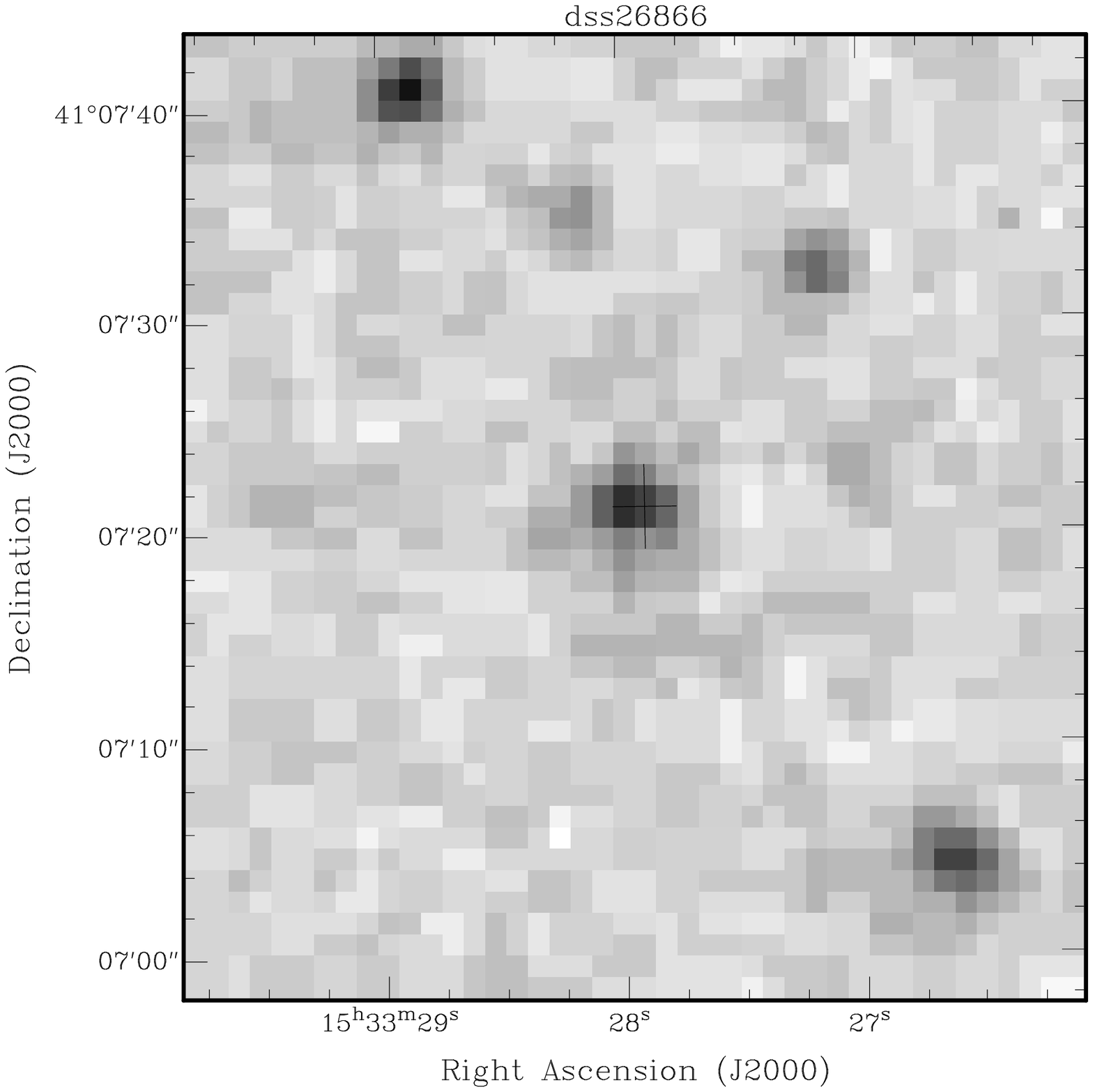 ,width=4.0cm,clip=}}\quad 
\subfigure[9CJ1538+4225 (P60 \it{R}\normalfont)]{\epsfig{figure=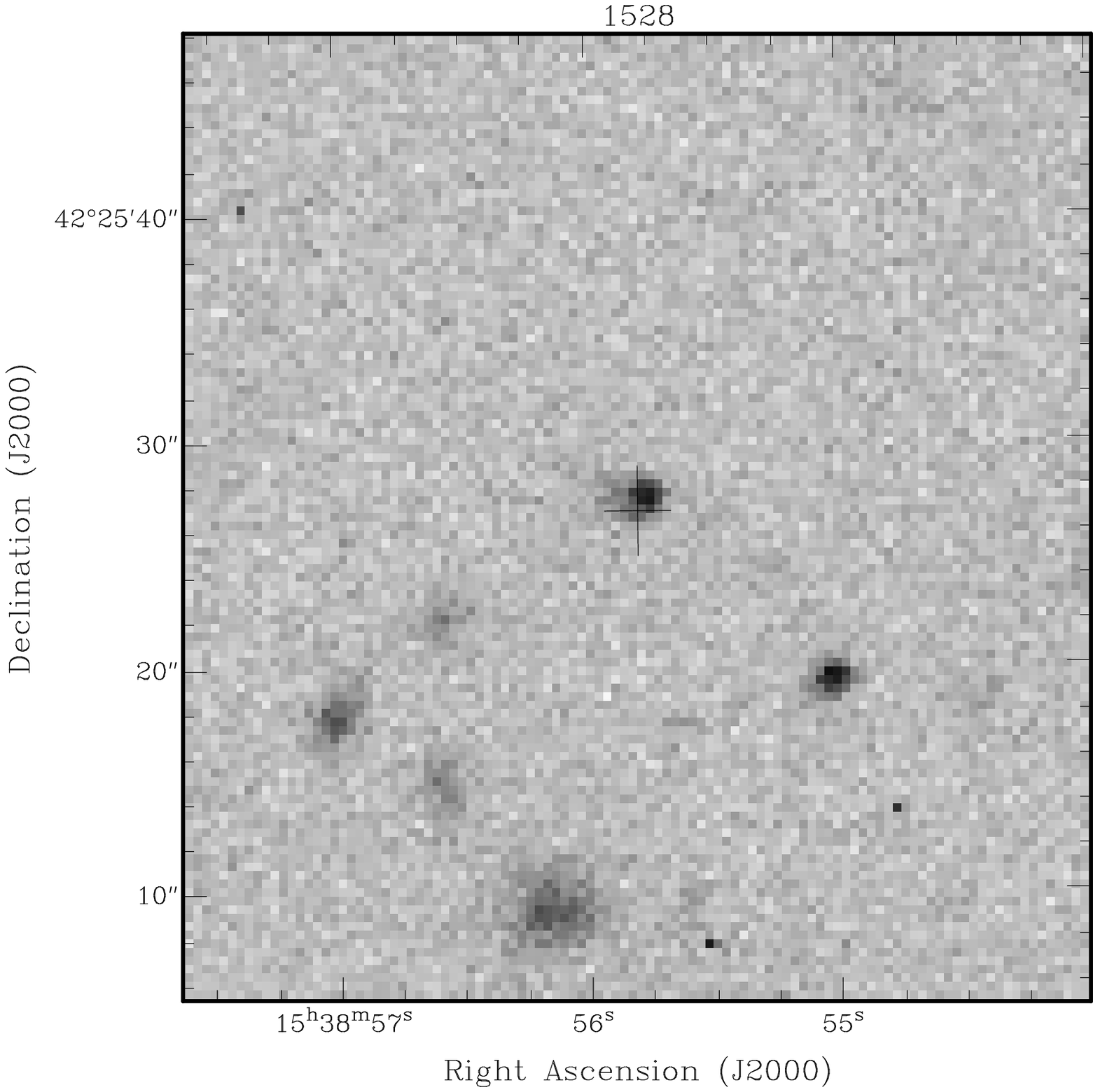 ,width=4.0cm,clip=}}\quad 
\subfigure[9CJ1539+4217 (DSS2 \it{R}\normalfont)]{\epsfig{figure=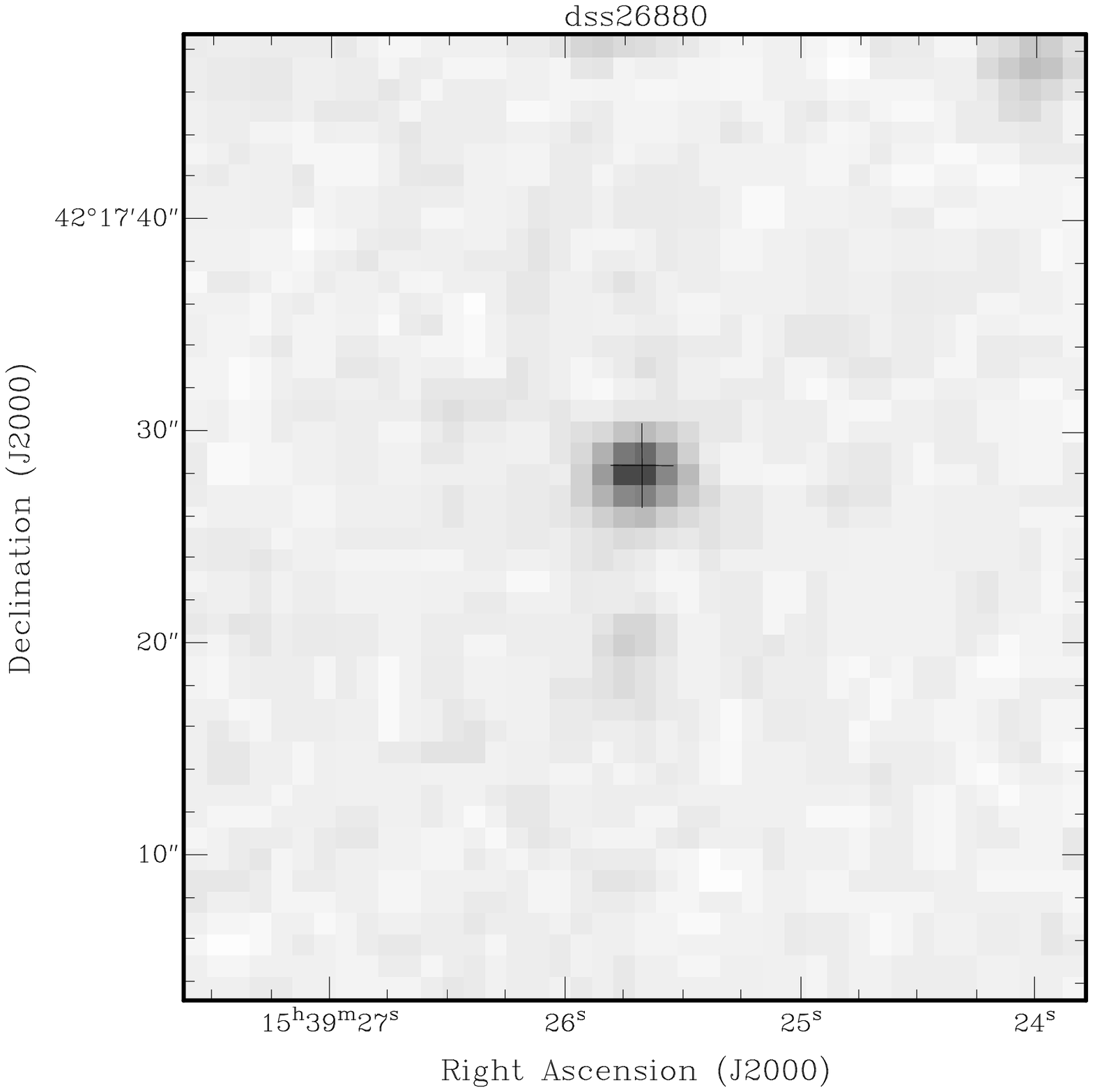 ,width=4.0cm,clip=}}
} 
\mbox{
\subfigure[9CJ1540+4138 (DSS2 \it{R}\normalfont)]{\epsfig{figure=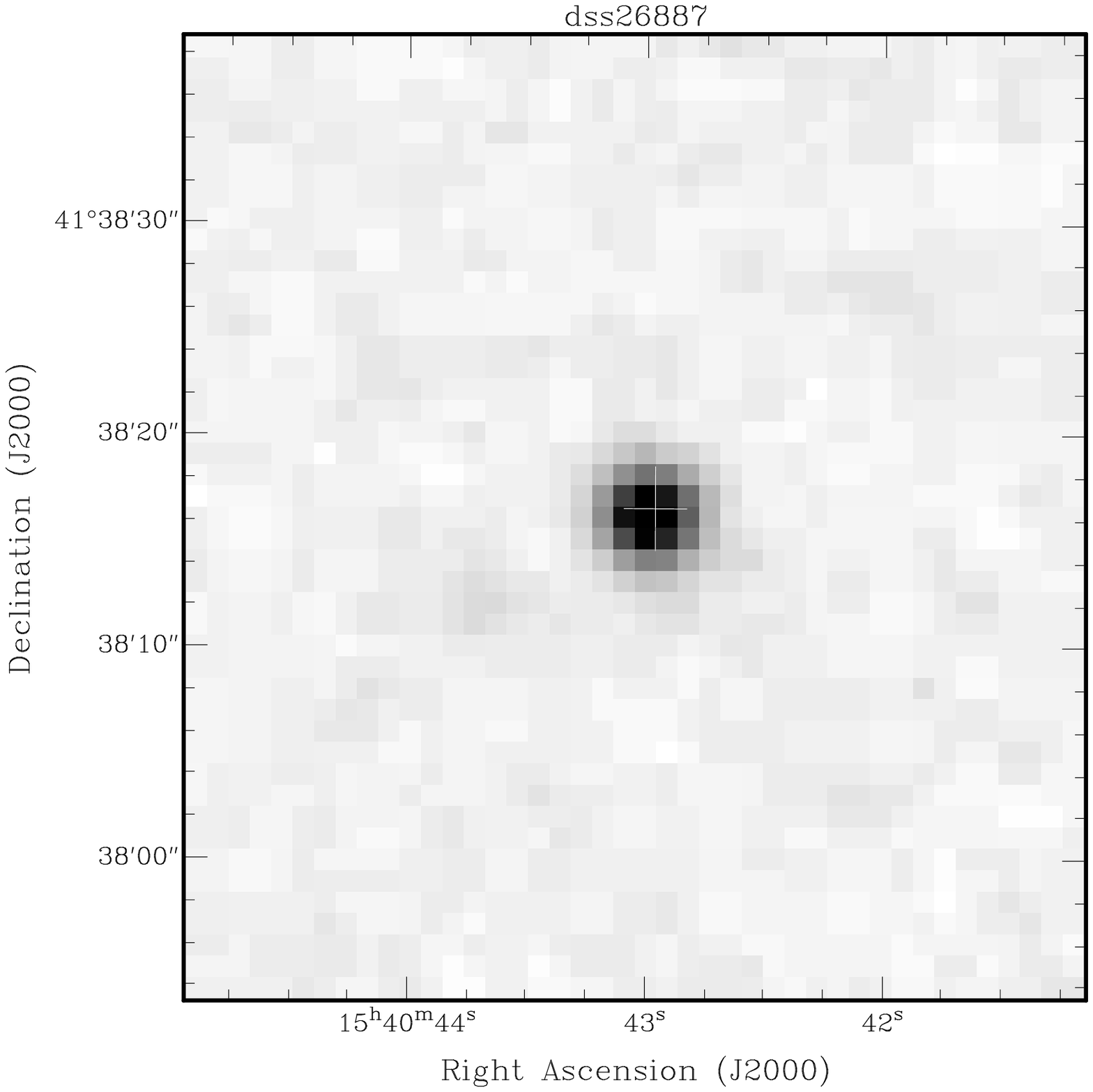 ,width=4.0cm,clip=}}\quad 
\subfigure[9CJ1541+4114 (DSS2 \it{R}\normalfont)]{\epsfig{figure=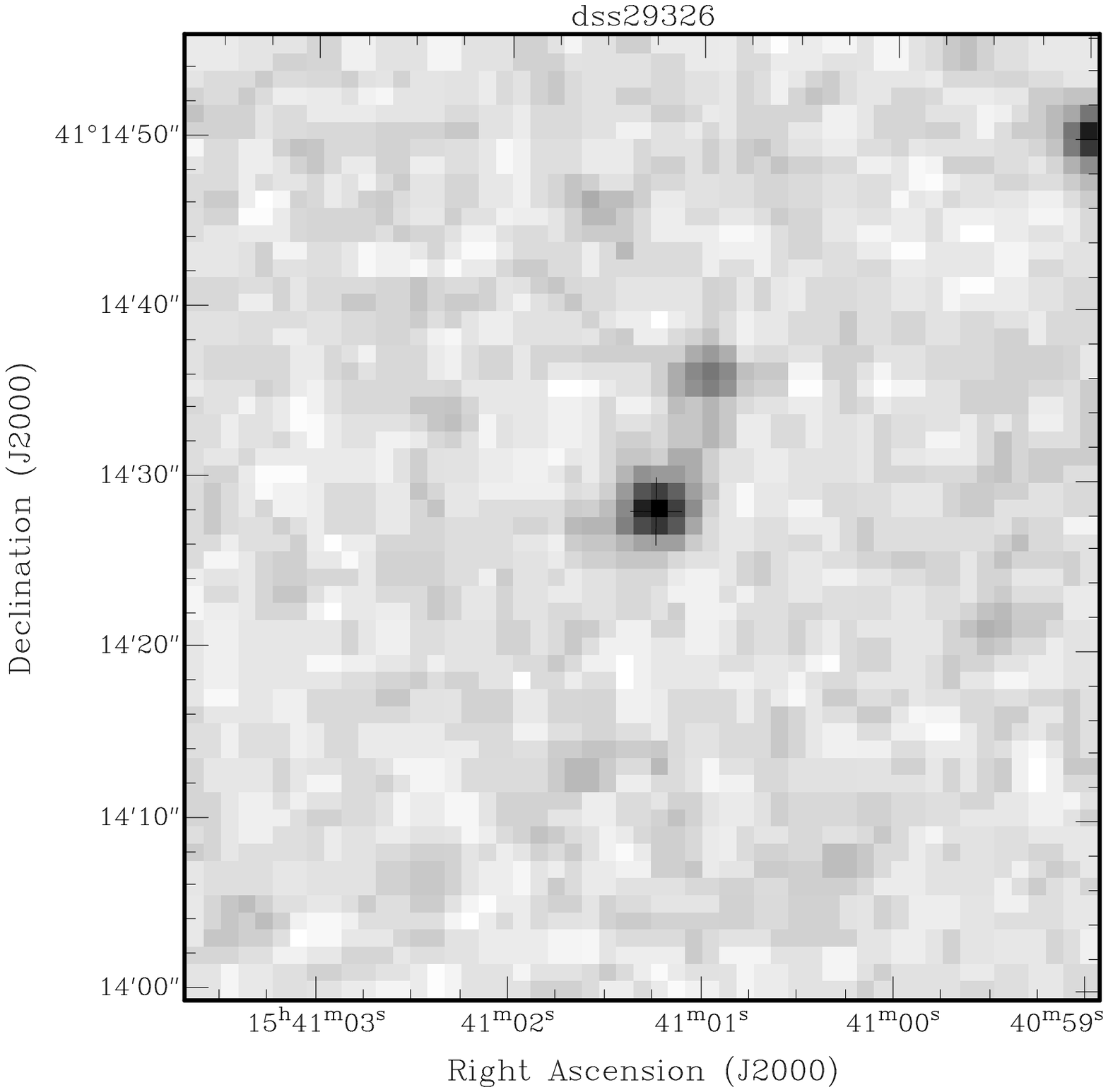 ,width=4.0cm,clip=}}\quad 
\subfigure[9CJ1541+4114 (DSS2 \it{R}\normalfont)]{\epsfig{figure=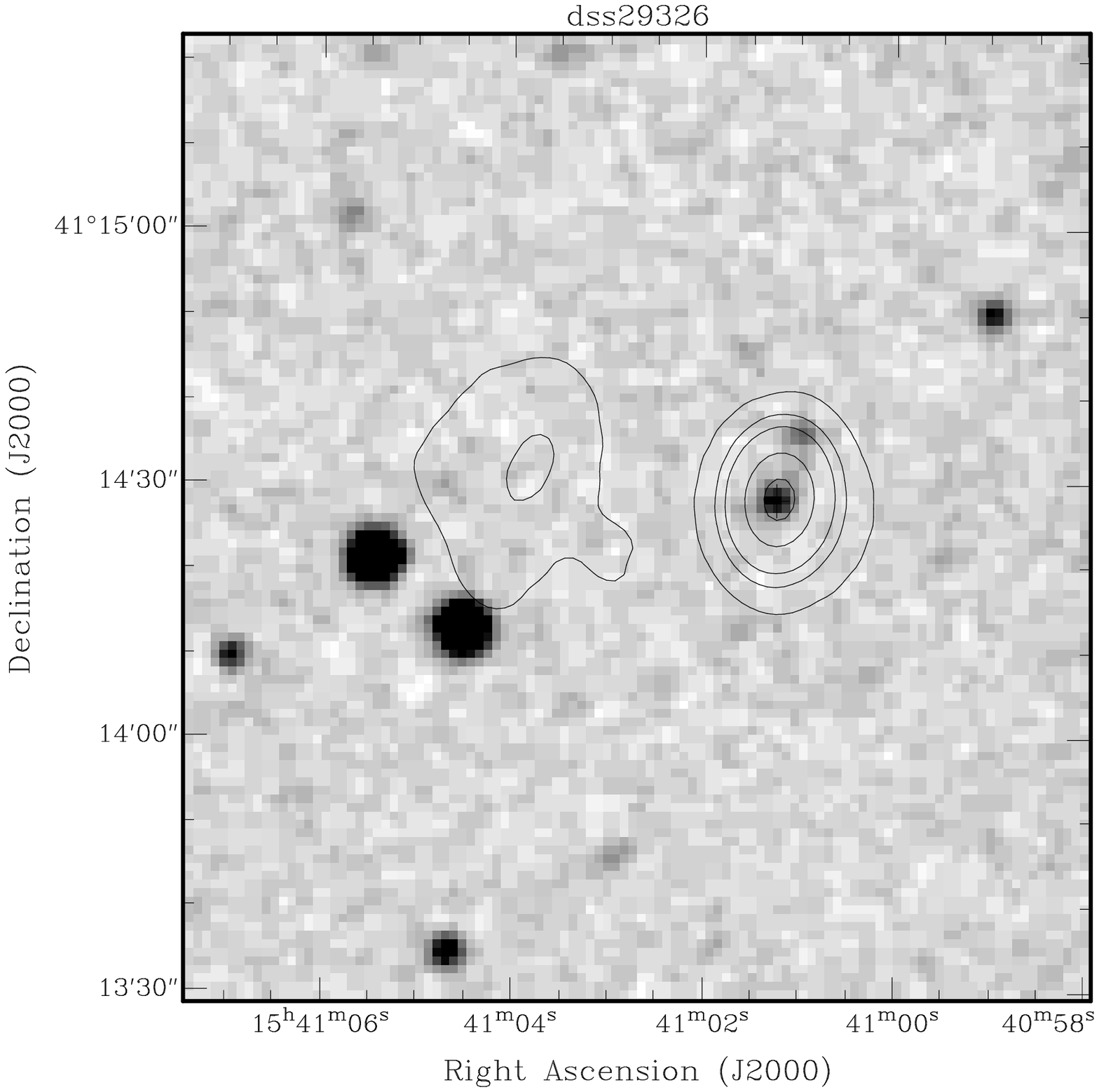 ,width=4.0cm,clip=}\label{bs}}
} 
\mbox{
\subfigure[9CJ1541+4456 (DSS2 \it{R}\normalfont)]{\epsfig{figure=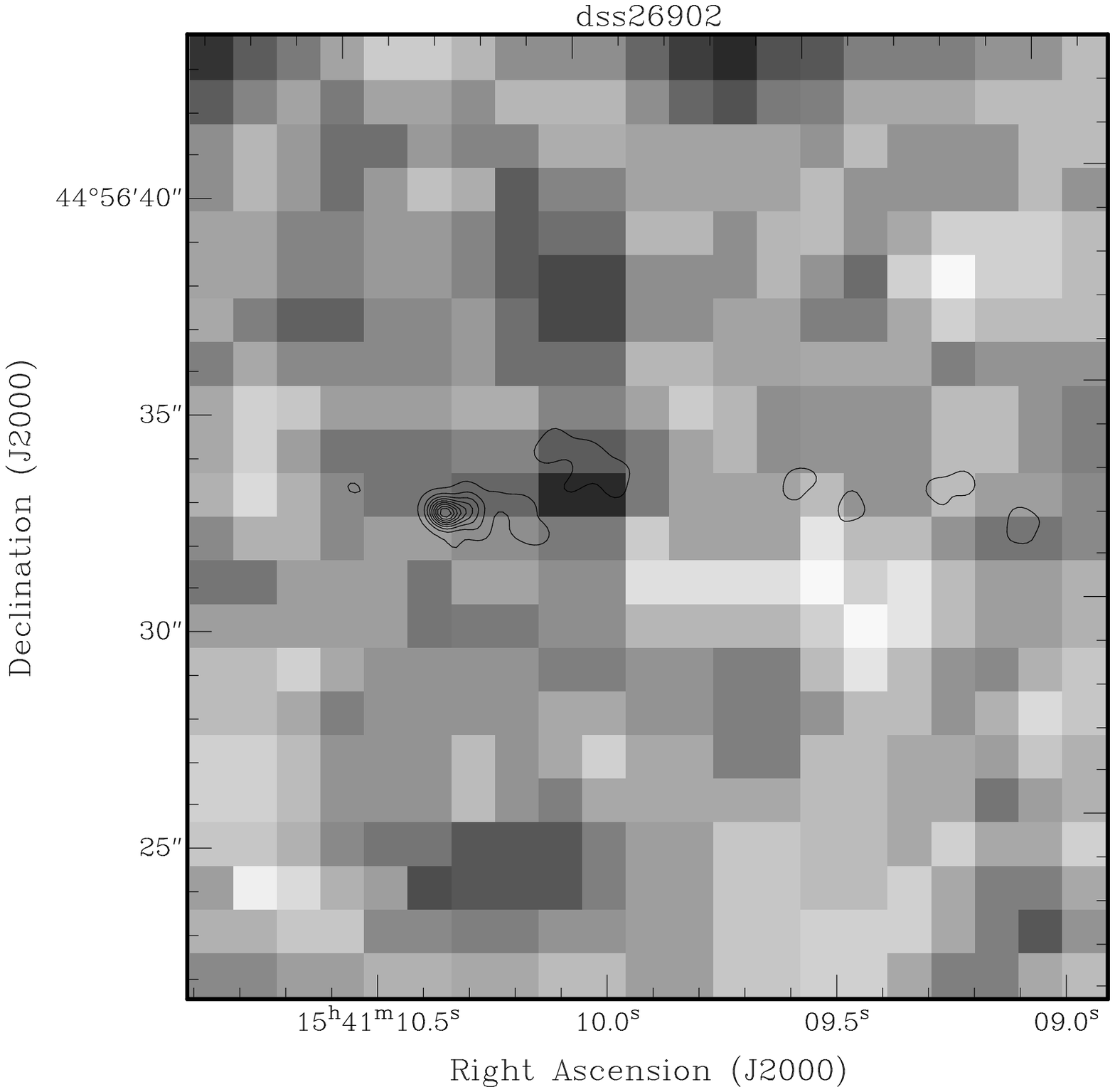 ,width=4.0cm,clip=}\label{bt}}\quad 
\subfigure[9CJ1545+4130 (P60 \it{R}\normalfont)]{\epsfig{figure=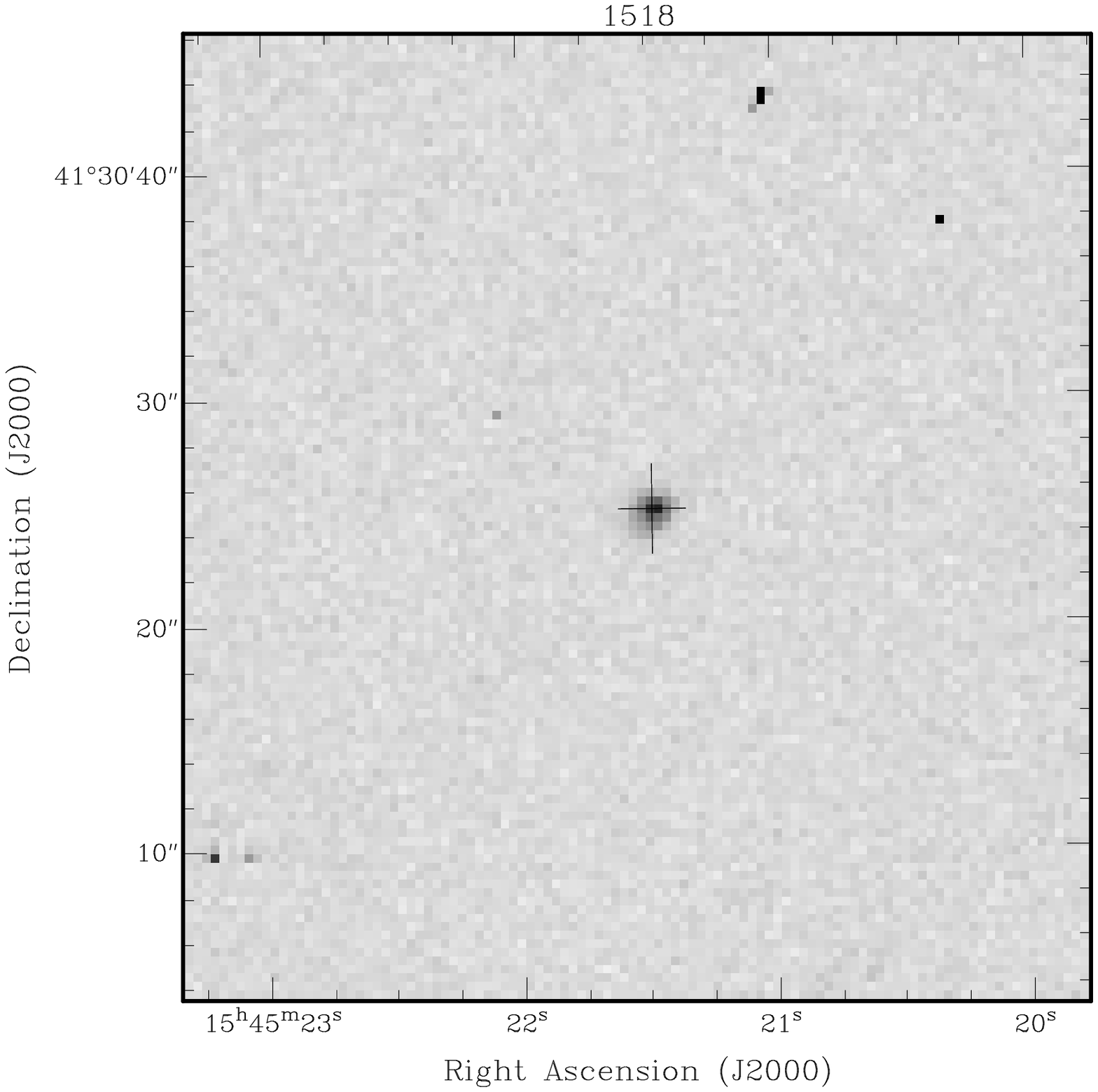 ,width=4.0cm,clip=}}\quad 
\subfigure[9CJ1546+4257 (DSS2 \it{R}\normalfont)]{\epsfig{figure=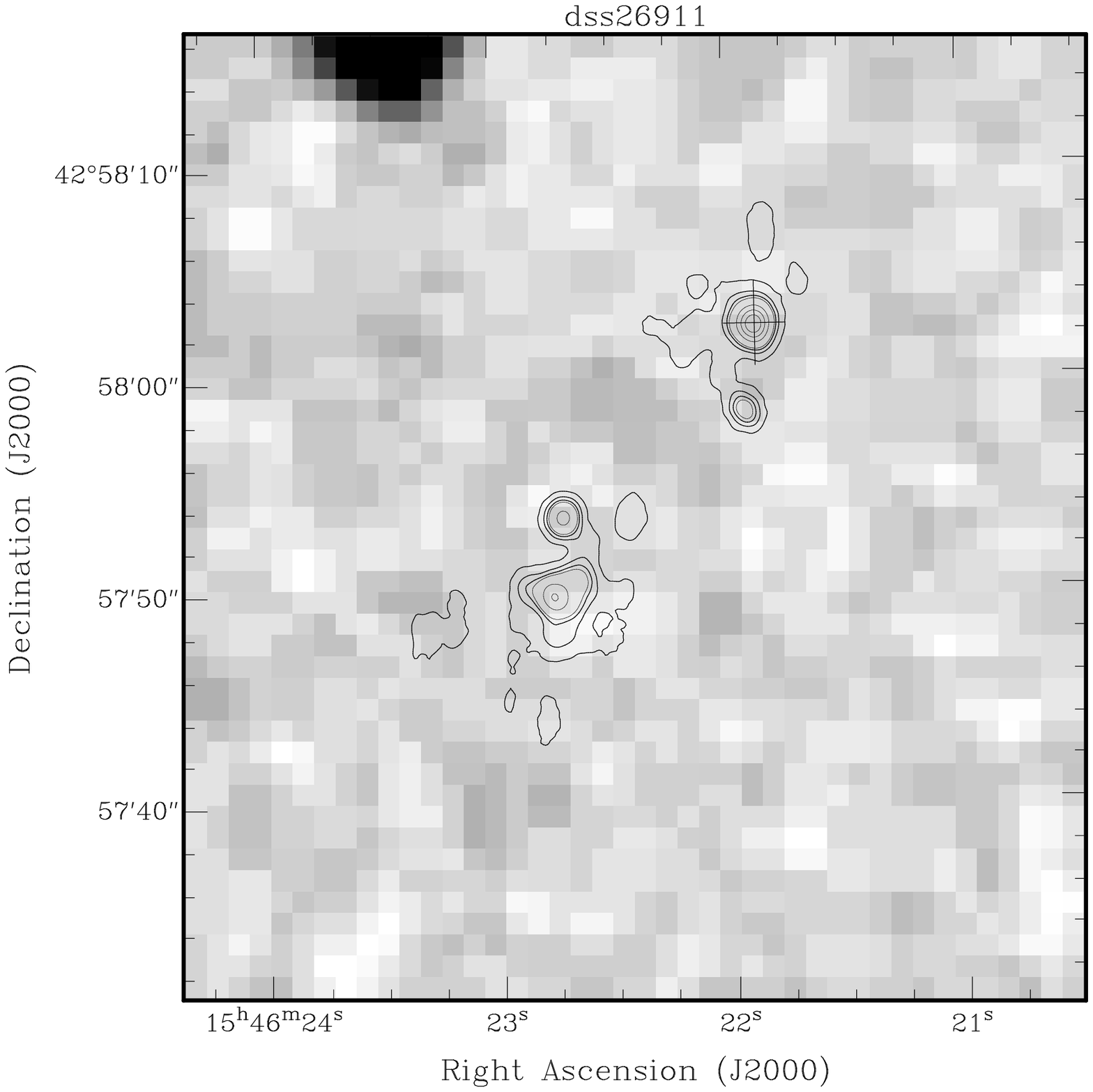 ,width=4.0cm,clip=}\label{bu}}
}\caption{ Optical counterparts for sources 9CJ1531+4356 to 9CJ1546+4257. Crosses mark maximum radio flux density and are 4\,arcsec top to bottom. Contours: \ref{bq} and \ref{br}, 4.8\,GHz contours at 20-80 every 20\,\% of peak (17.5\,mJy/beam); \ref{bs}, 4.8\,GHz, 10,20,30,60,90\,\% of peak (24.7\,mJy/beam); \ref{bt}, 4.8\,GHz 10-90 every 10\,\% of peak (29.6\,mJy/beam); \ref{bu}, 4.8\,GHz contours at 6,11,16 and 20-80 every 20\,\% of peak (24.5\,mJy/beam).}\end{figure*}

\begin{figure*}
\mbox{
\subfigure[9CJ1547+4208 (P60 \it{R}\normalfont)]{\epsfig{figure=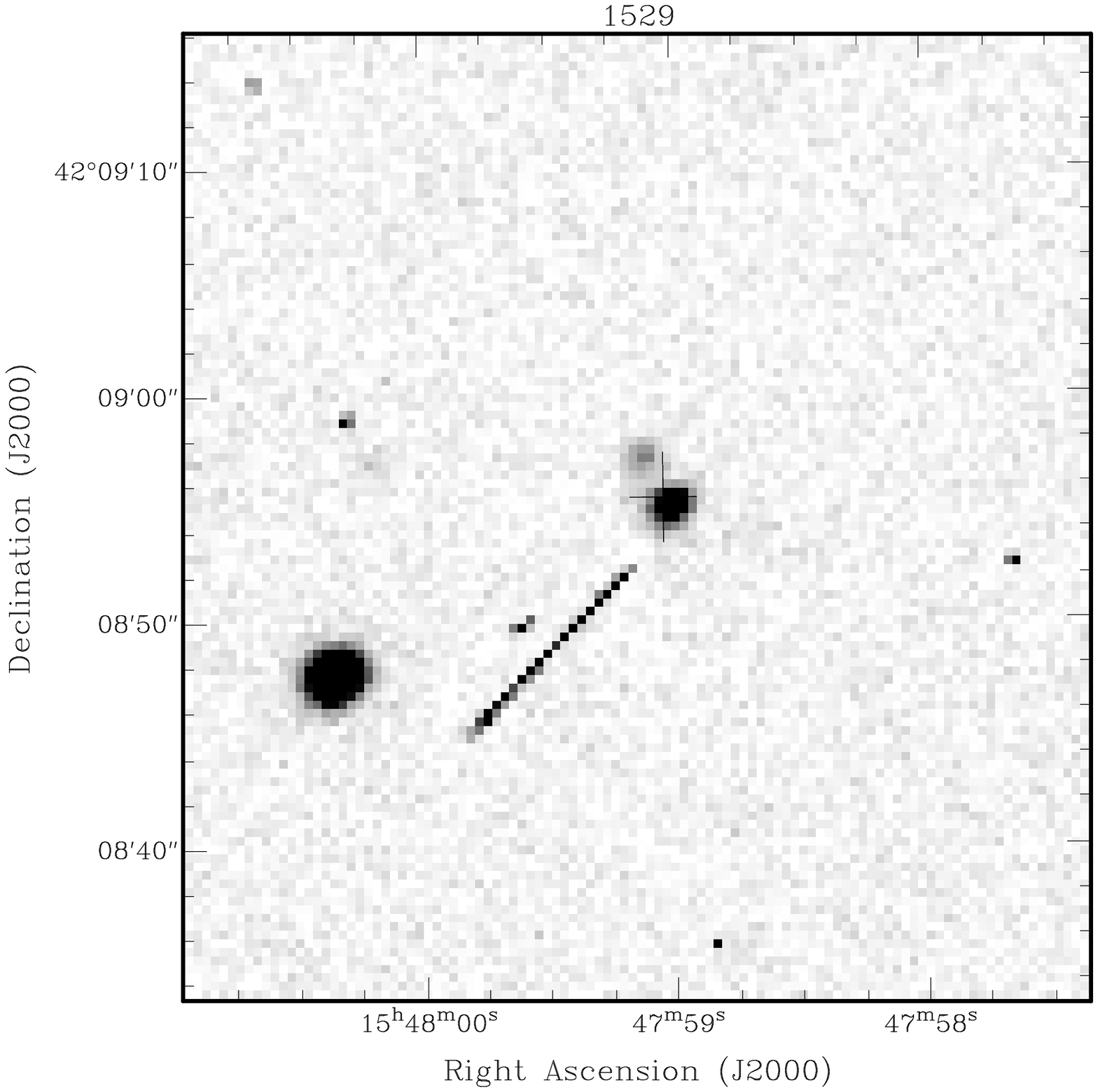 ,width=4.0cm,clip=}}\quad 
\subfigure[9CJ1548+4031 (DSS2 \it{R}\normalfont)]{\epsfig{figure=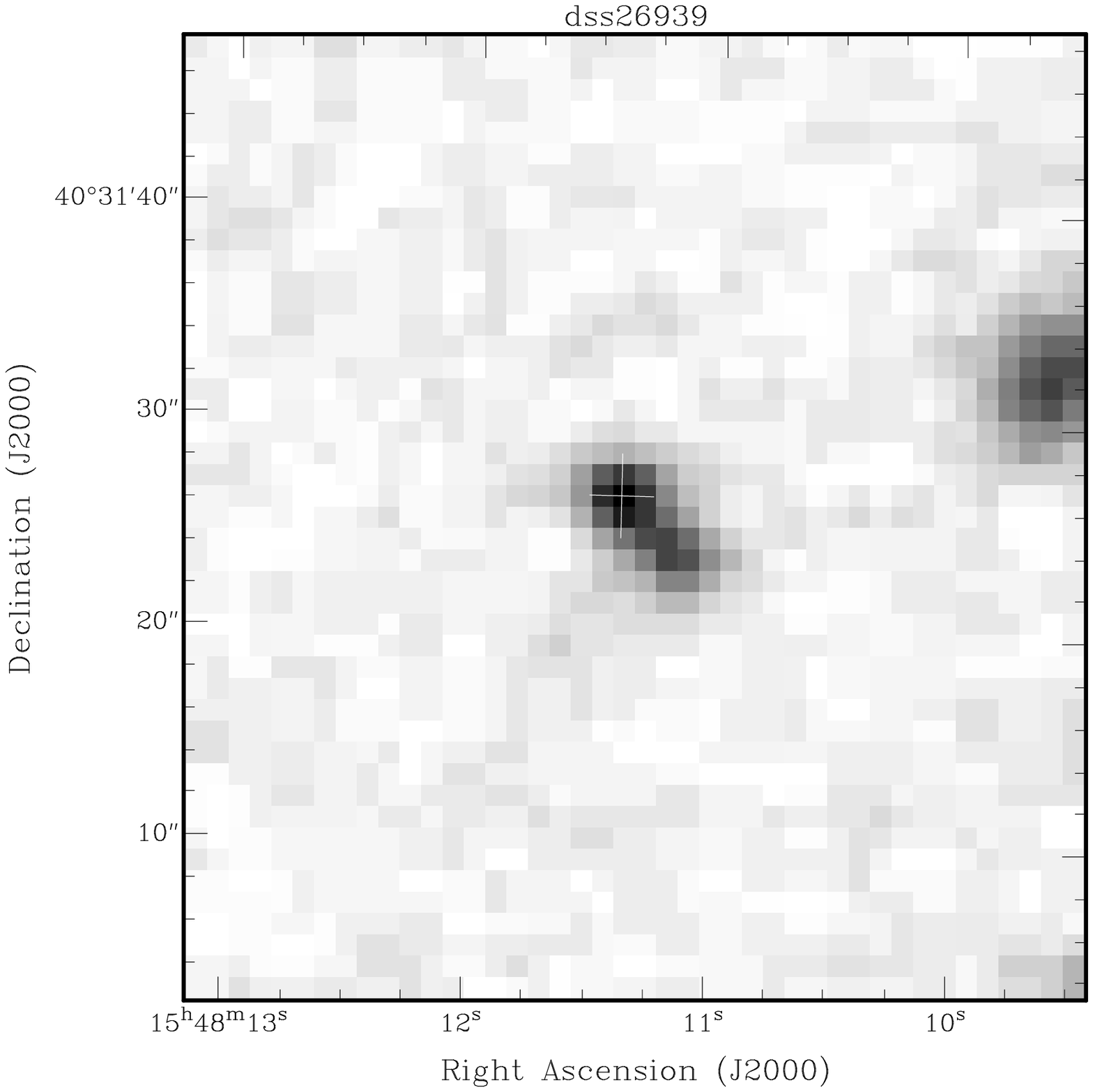 ,width=4.0cm,clip=}}\quad 
\subfigure[9CJ1550+4536 (DSS2 \it{R}\normalfont)]{\epsfig{figure=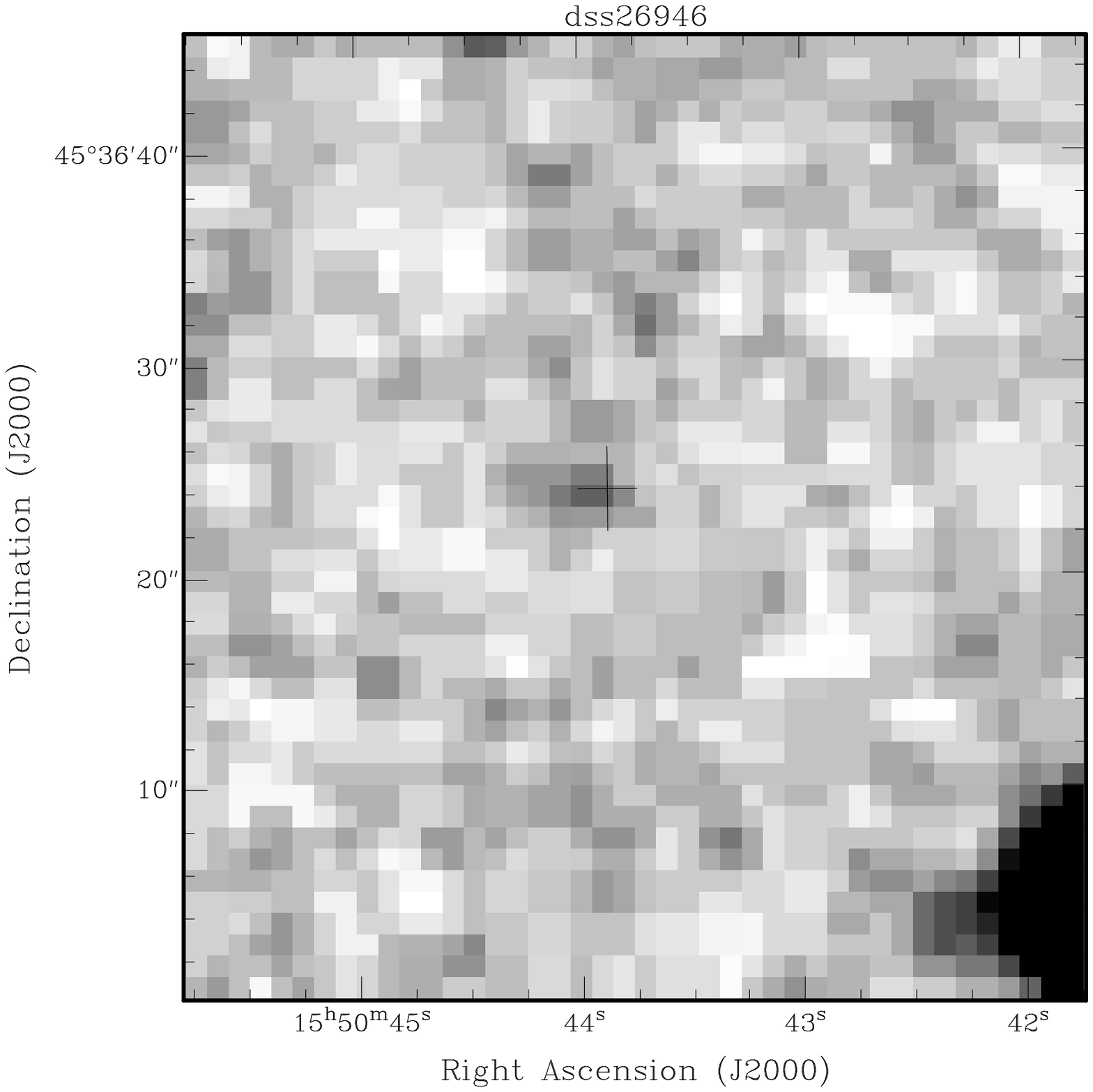 ,width=4.0cm,clip=}}
}
\mbox{
\subfigure[9CJ1550+4536 (DSS2 \it{O}\normalfont)]{\epsfig{figure=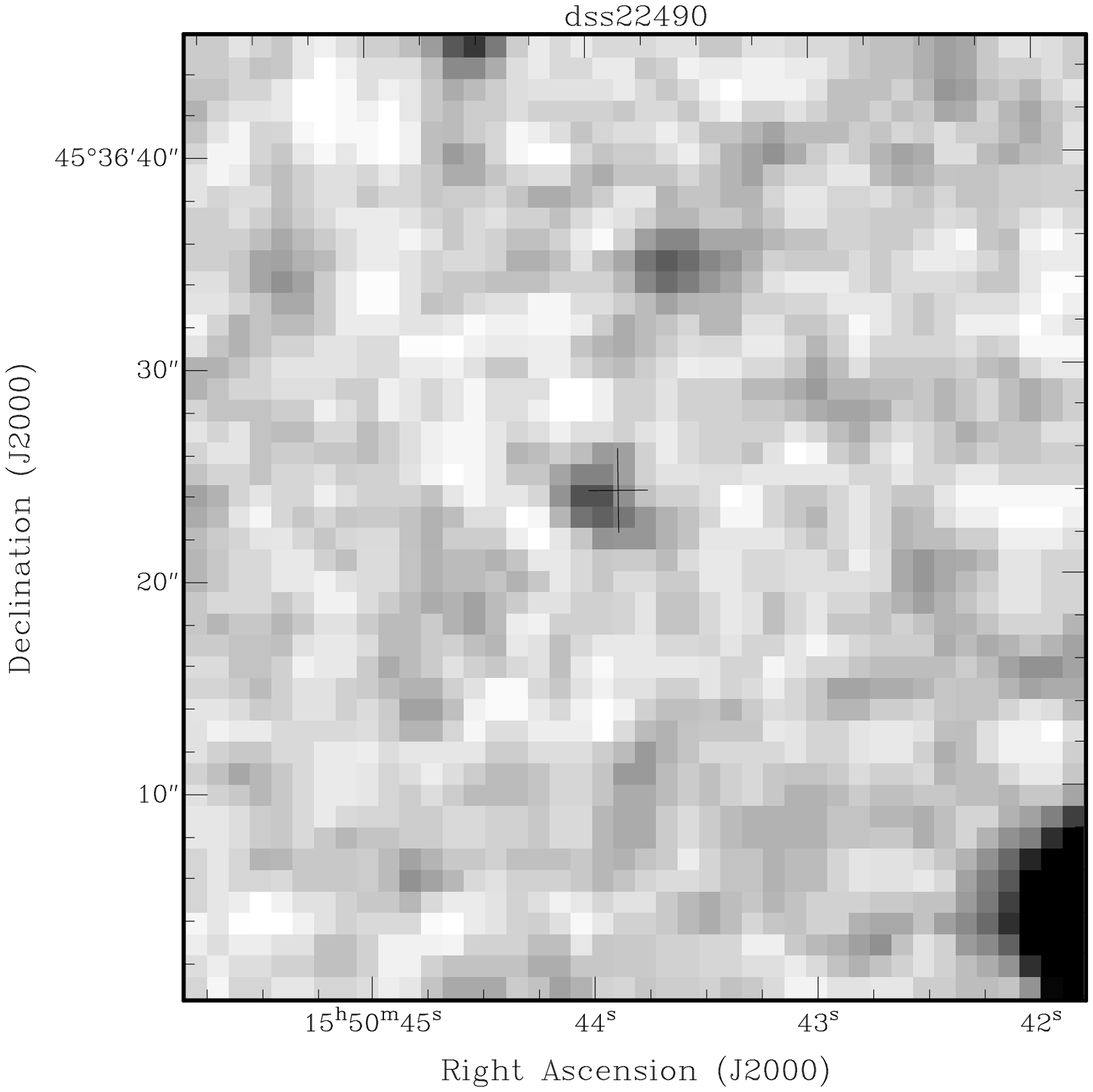 ,width=4.0cm,clip=}}\quad 
\subfigure[9CJ1550+4545 (DSS2 \it{R}\normalfont)]{\epsfig{figure=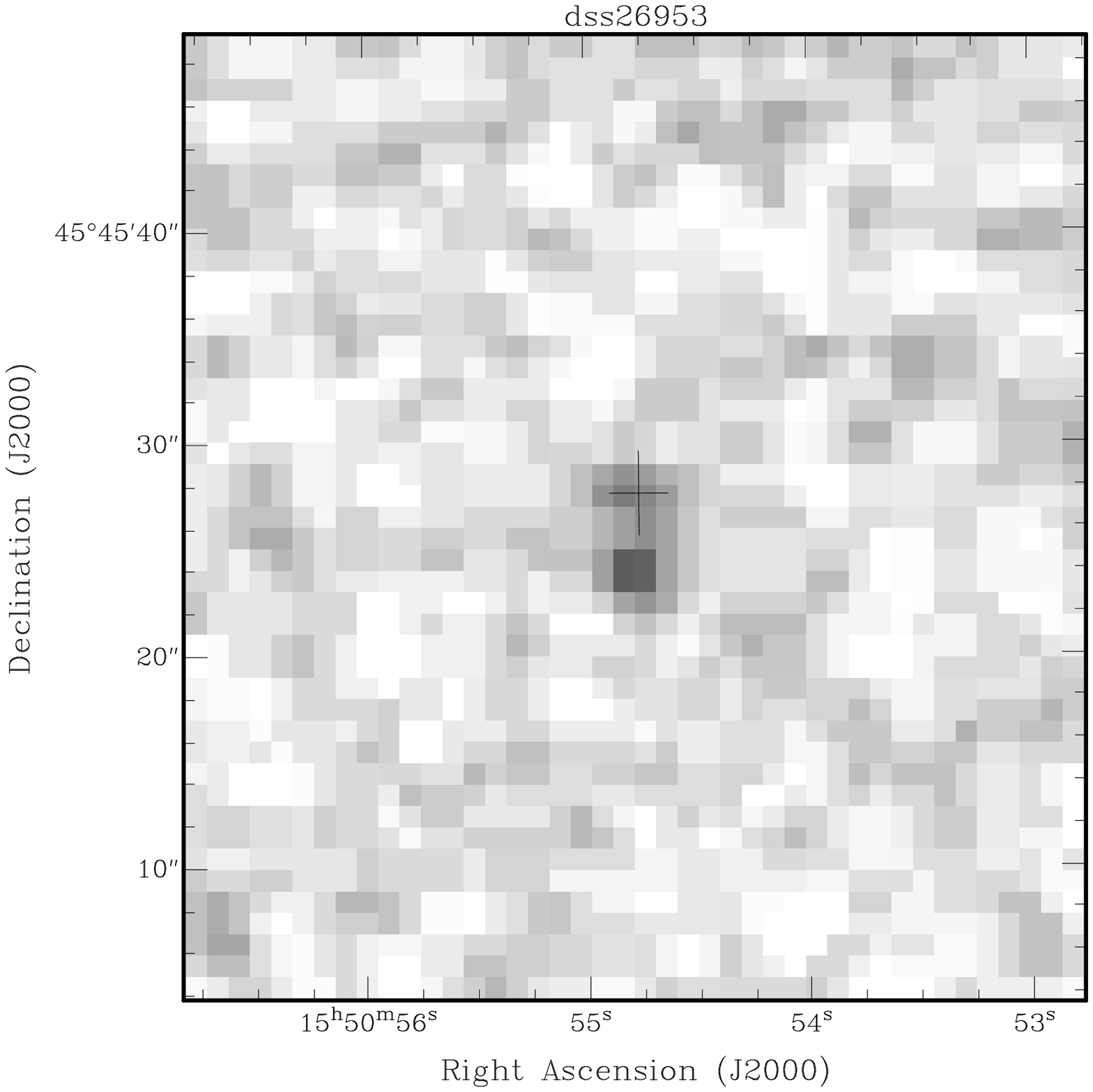 ,width=4.0cm,clip=}}\quad 
\subfigure[9CJ1553+4039 (DSS2 \it{R}\normalfont)]{\epsfig{figure=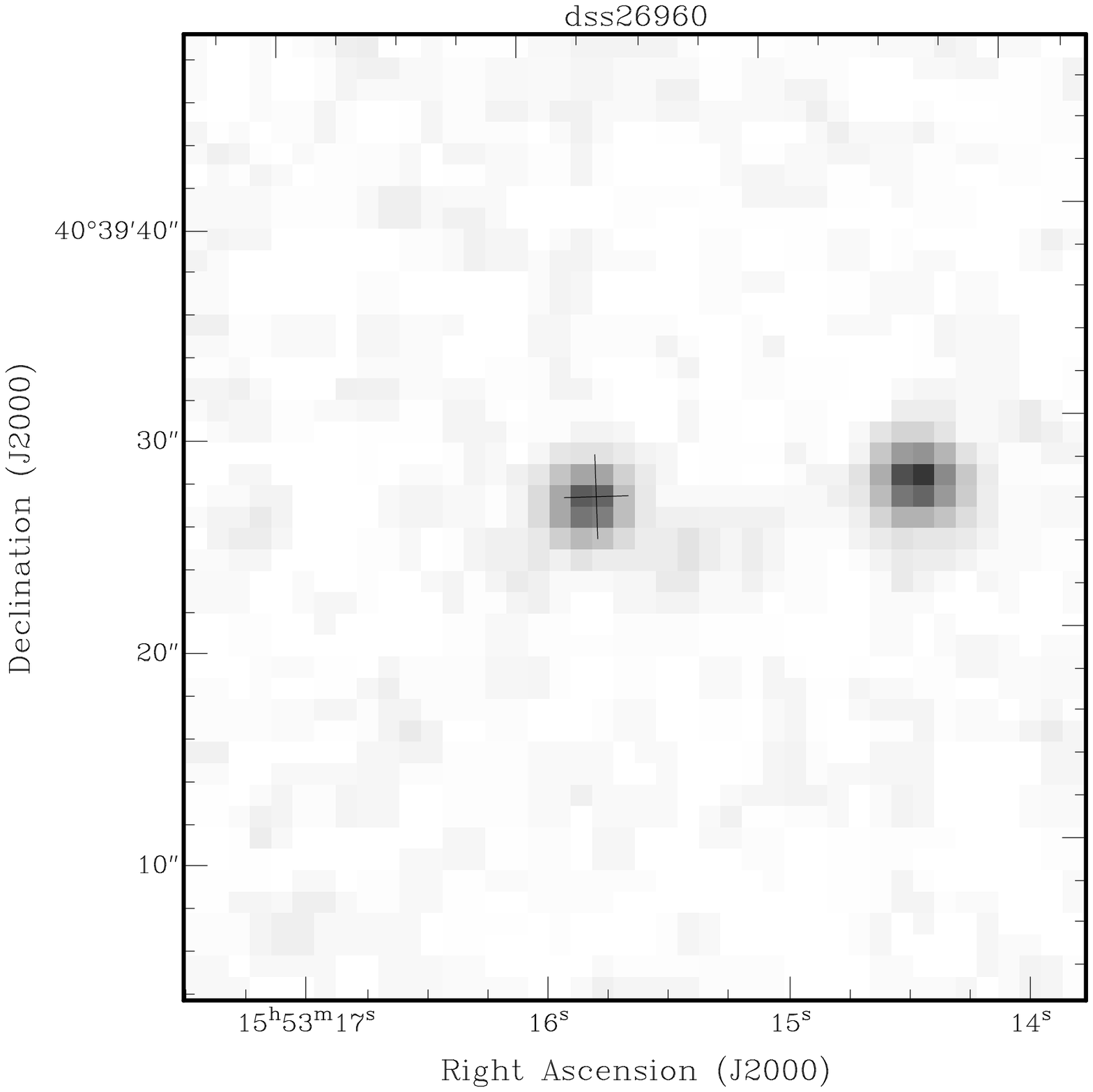 ,width=4.0cm,clip=}}
} 
\mbox{
\subfigure[9CJ1554+4350 (P60 \it{R}\normalfont)]{\epsfig{figure=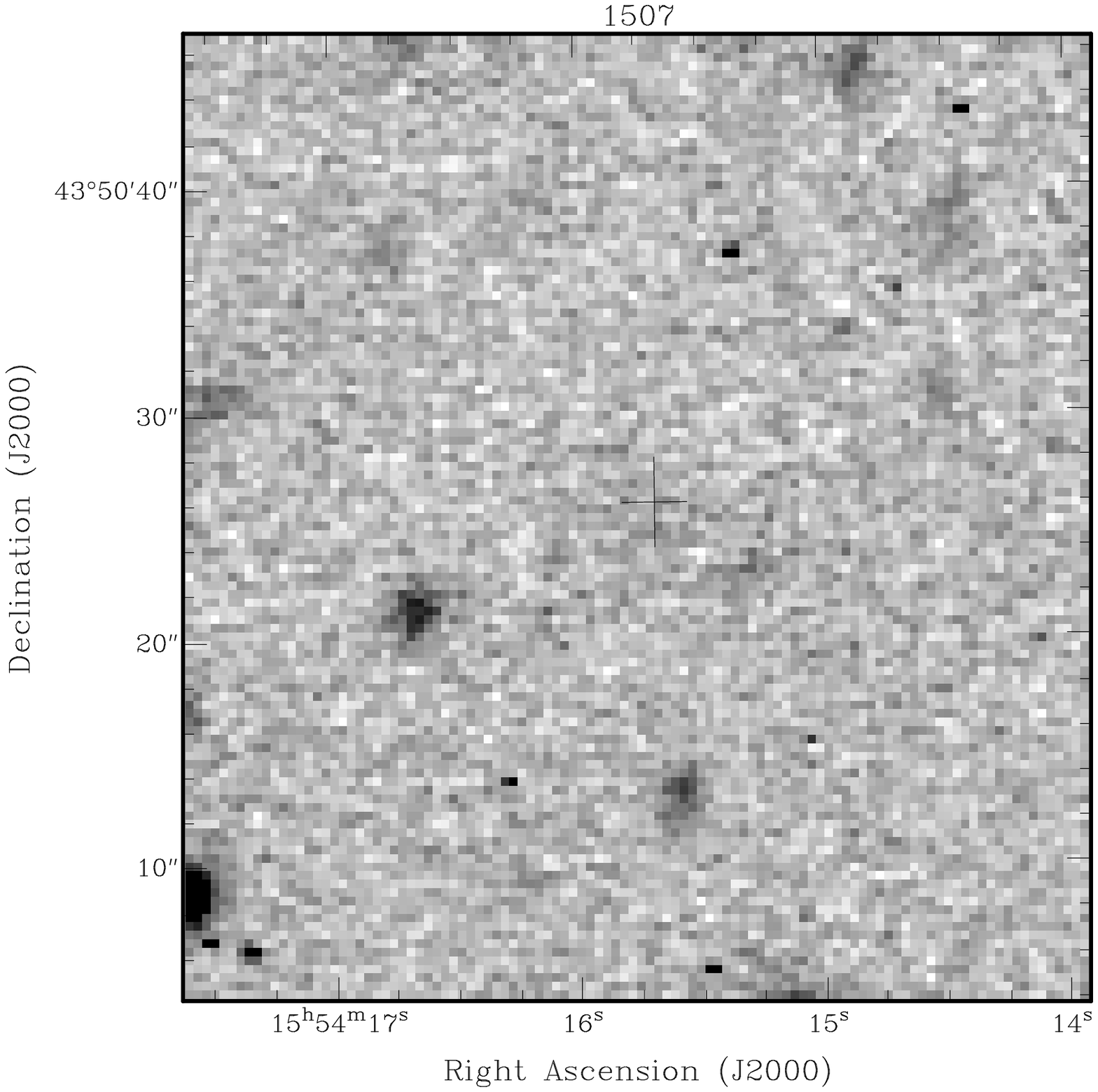 ,width=4.0cm,clip=}}\quad 
\subfigure[9CJ1554+4348 (P60 \it{R}\normalfont)]{\epsfig{figure=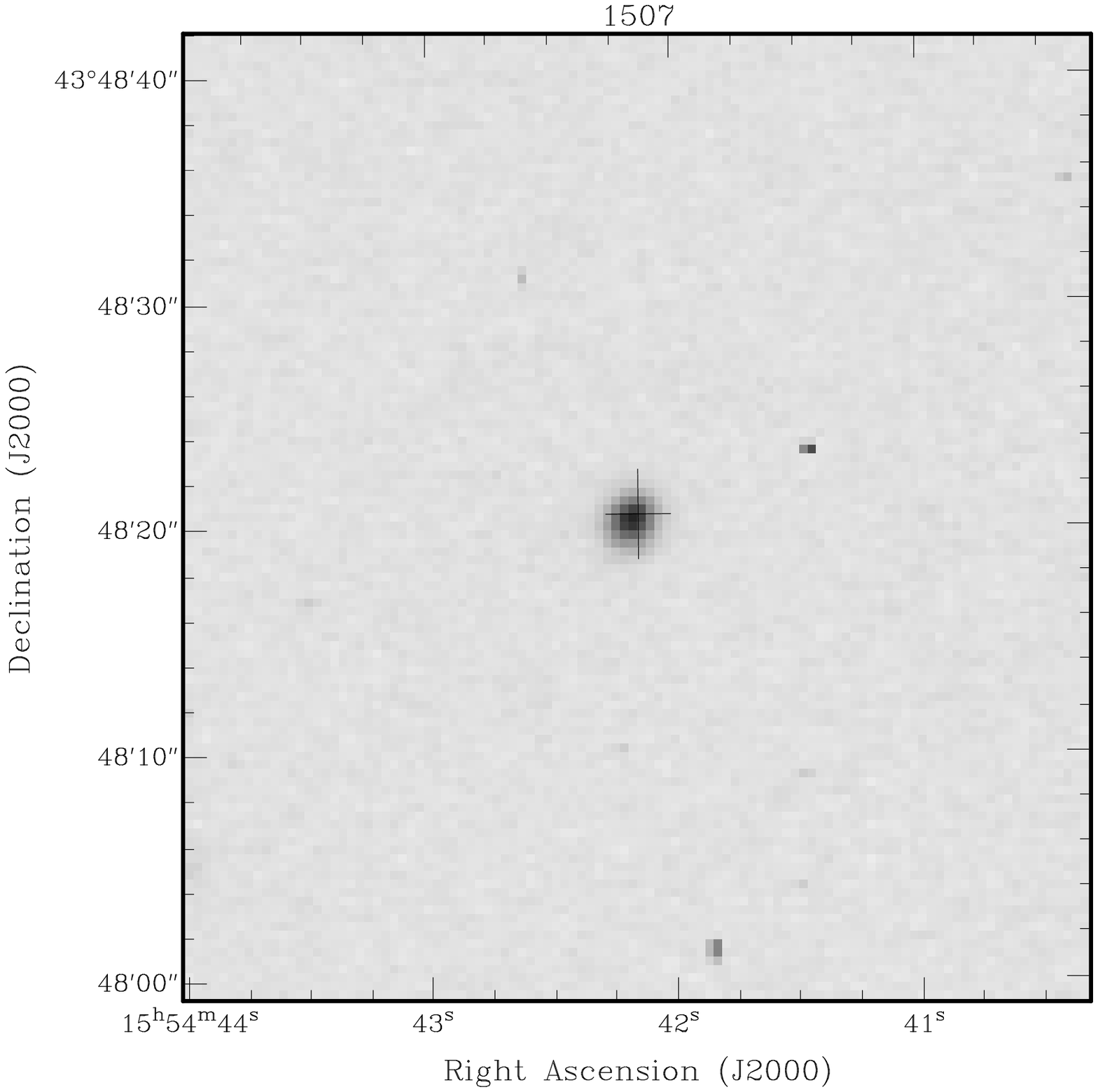 ,width=4.0cm,clip=}}\quad 
\subfigure[9CJ1556+4257 (DSS2 \it{R}\normalfont)]{\epsfig{figure=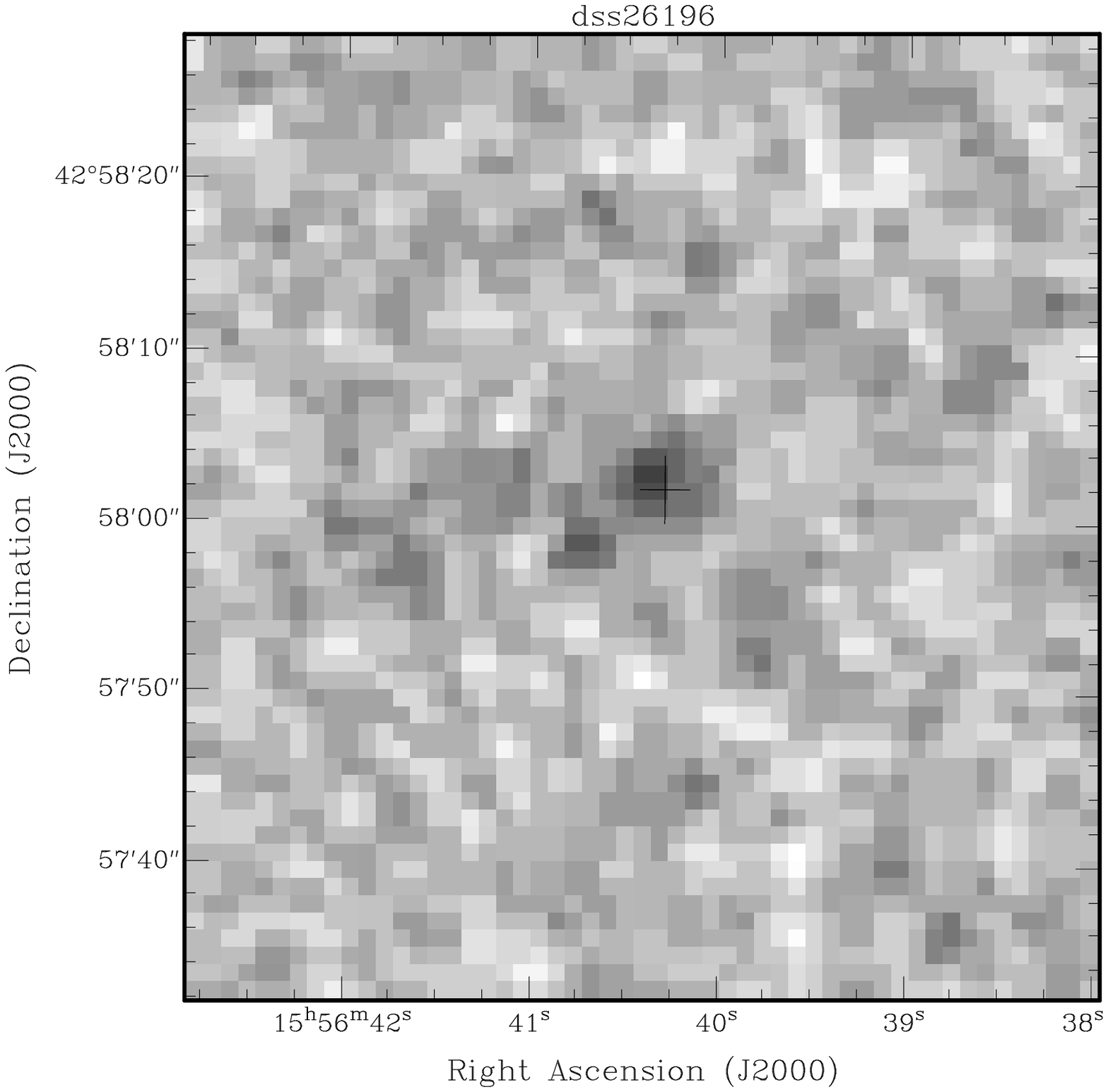 ,width=4.0cm,clip=}}
} 
\mbox{
\subfigure[9CJ1556+4259 (DSS2 \it{R}\normalfont)]{\epsfig{figure=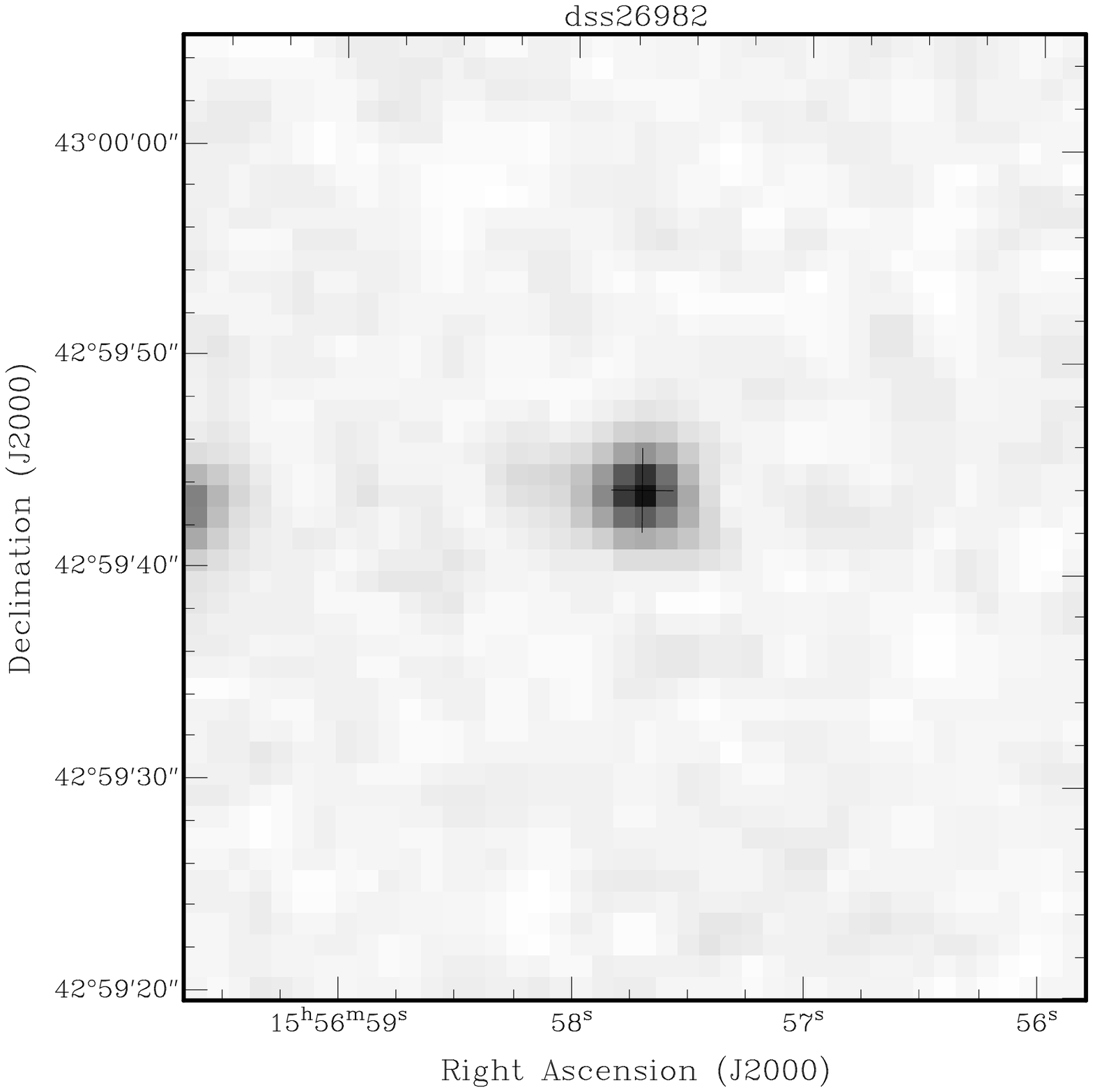 ,width=4.0cm,clip=}\label{id's}}\quad 
\subfigure[9CJ1557+4522 (P60 \it{R}\normalfont)]{\epsfig{figure=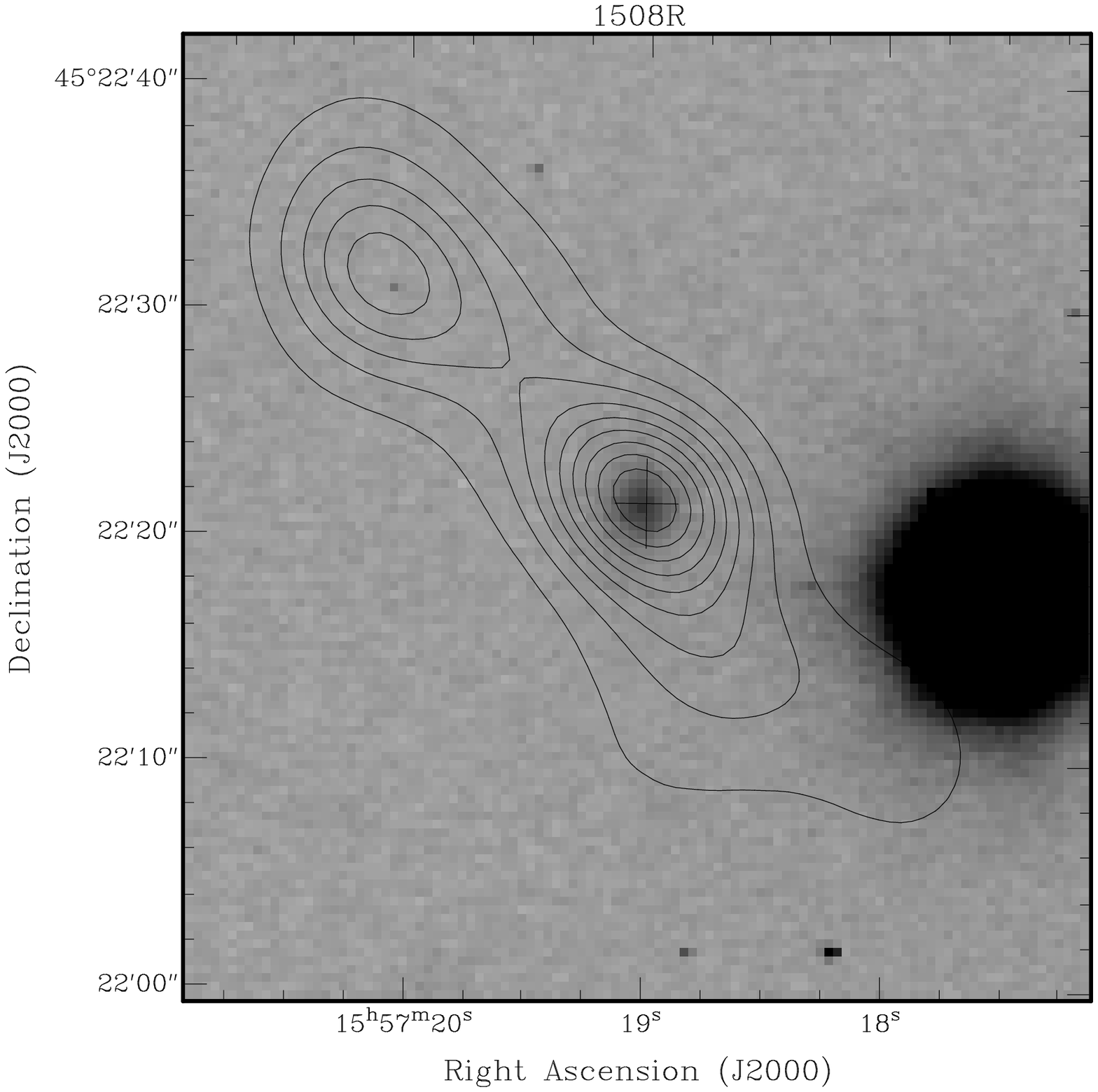 ,width=4.0cm,clip=}\label{bv}}\quad 
\subfigure[9CJ1557+4007 (DSS2 \it{R}\normalfont)]{\epsfig{figure=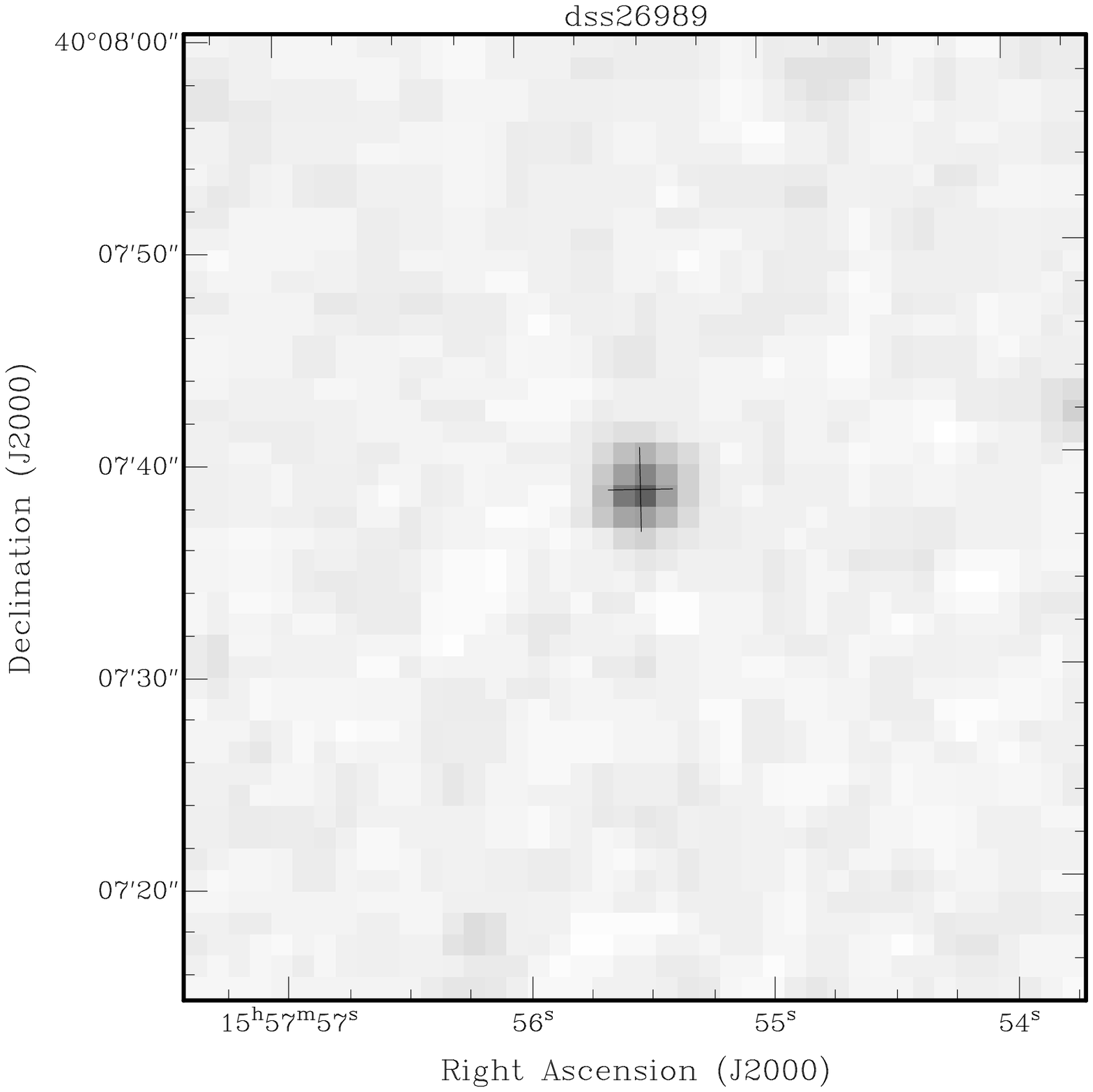 ,width=4.0cm,clip=}}
}\caption{ Optical counterparts for sources 9CJ1547+4208 to 9CJ1557+4007. Crosses mark maximum radio flux density and are 4\,arcsec top to bottom. Contours: \ref{bv}, 4.8\,GHz contours 10-90 every 10\,\% of peak (87.6\,mJy/beam).}\end{figure*}

\begin{figure*}
\mbox{
\subfigure[9CJ1558+4146 (DSS2 \it{R}\normalfont)]{\epsfig{figure=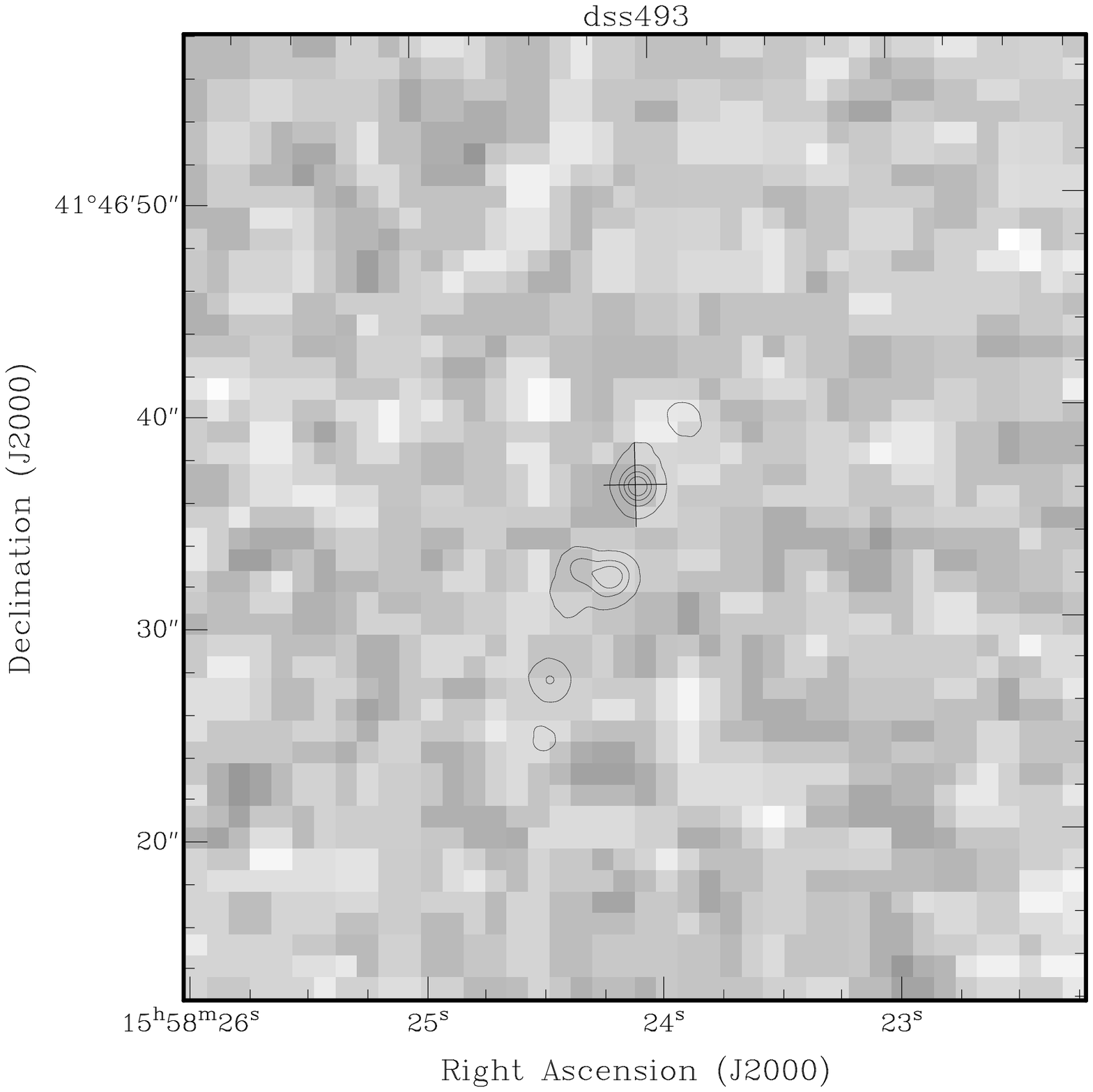 ,width=4.0cm,clip=}}\quad 
\subfigure[9CJ2351+3018 (DSS2 \it{R}\normalfont)]{\epsfig{figure=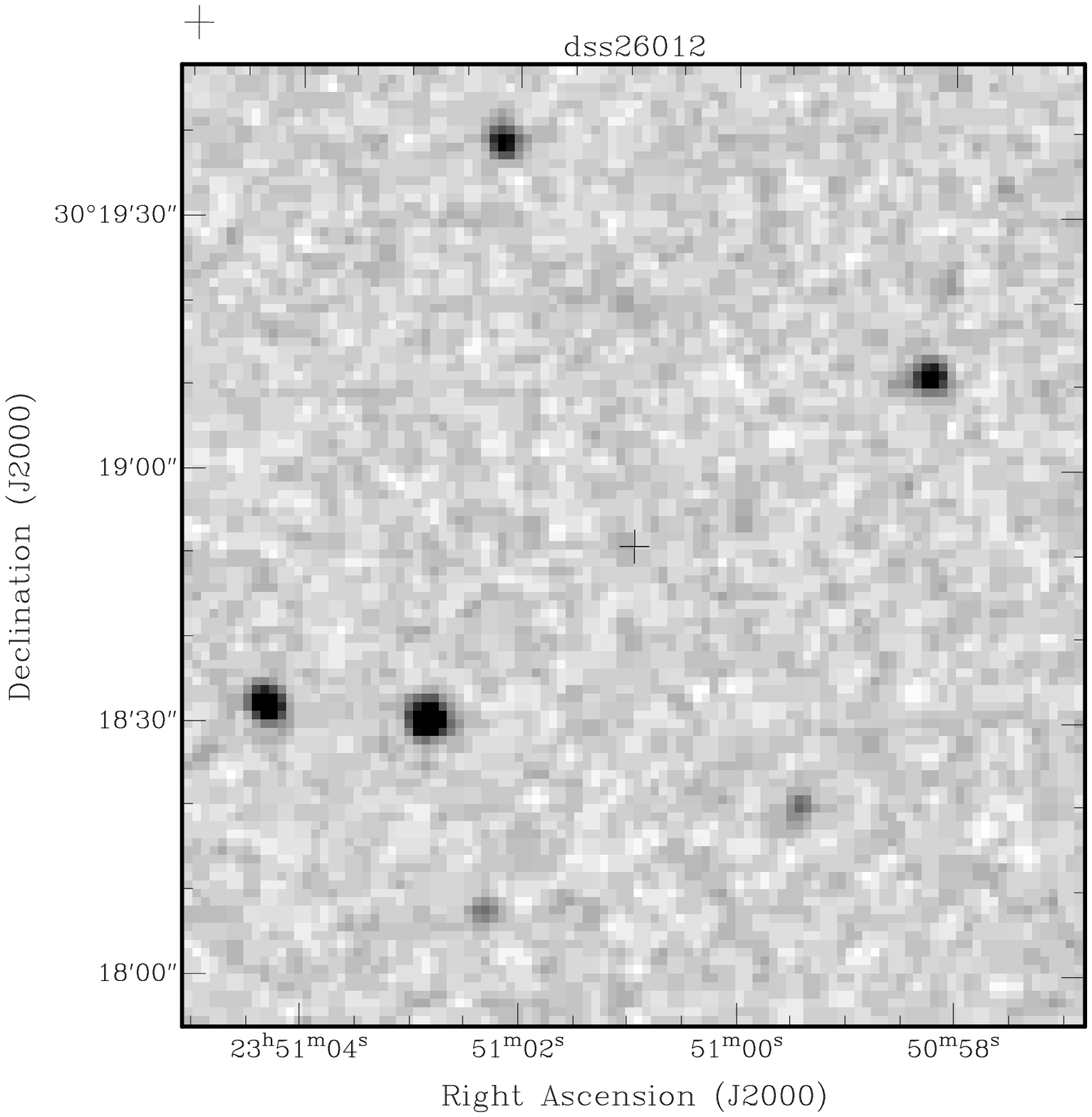 ,width=4.0cm,clip=}}\quad 
\subfigure[9CJ2351+3019 (DSS2 \it{R}\normalfont)]{\epsfig{figure=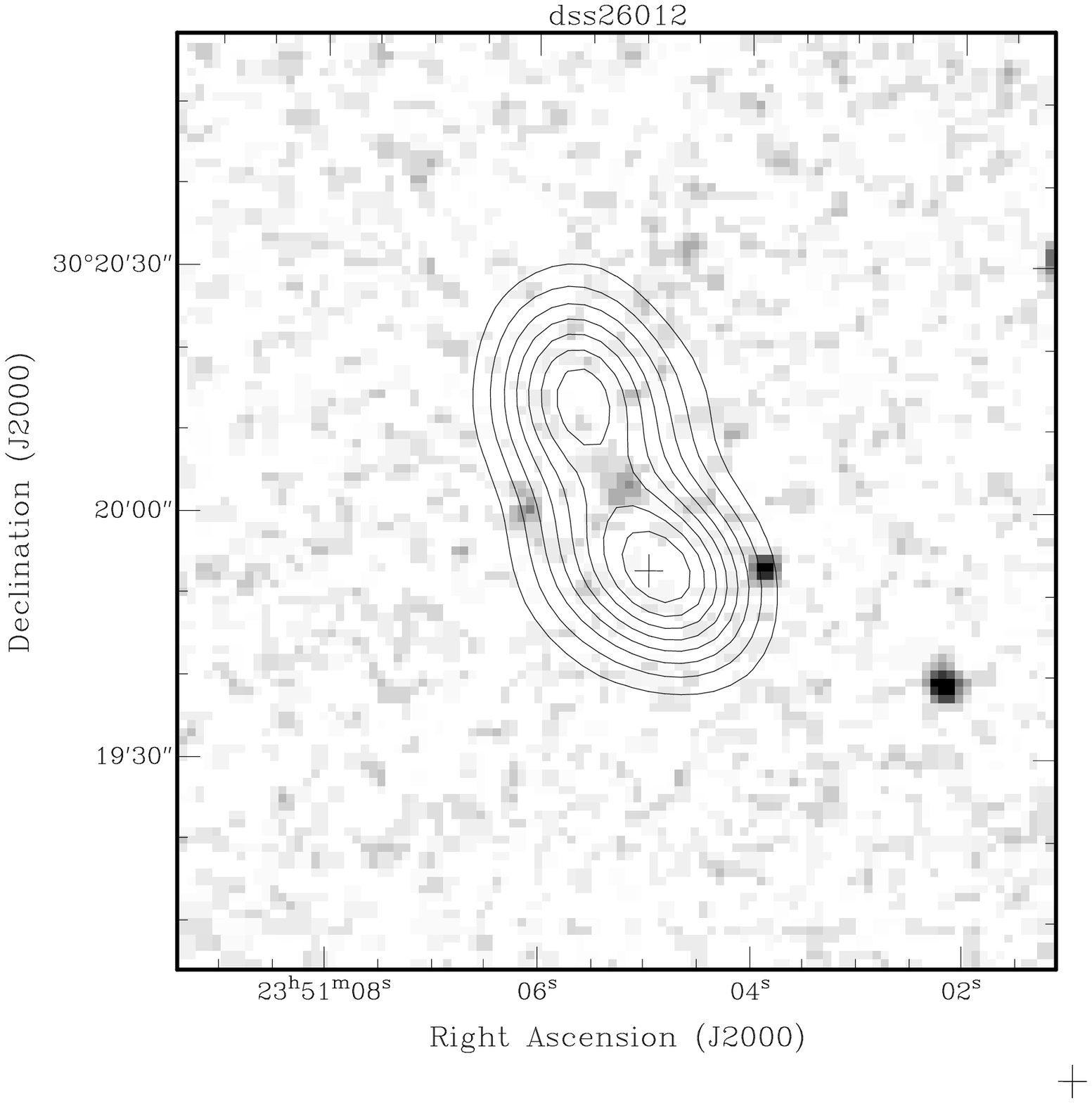 ,width=4.0cm,clip=}\label{bw}}
} 
\mbox{
\subfigure[9CJ2352+3035 (DSS2 \it{R}\normalfont)]{\epsfig{figure=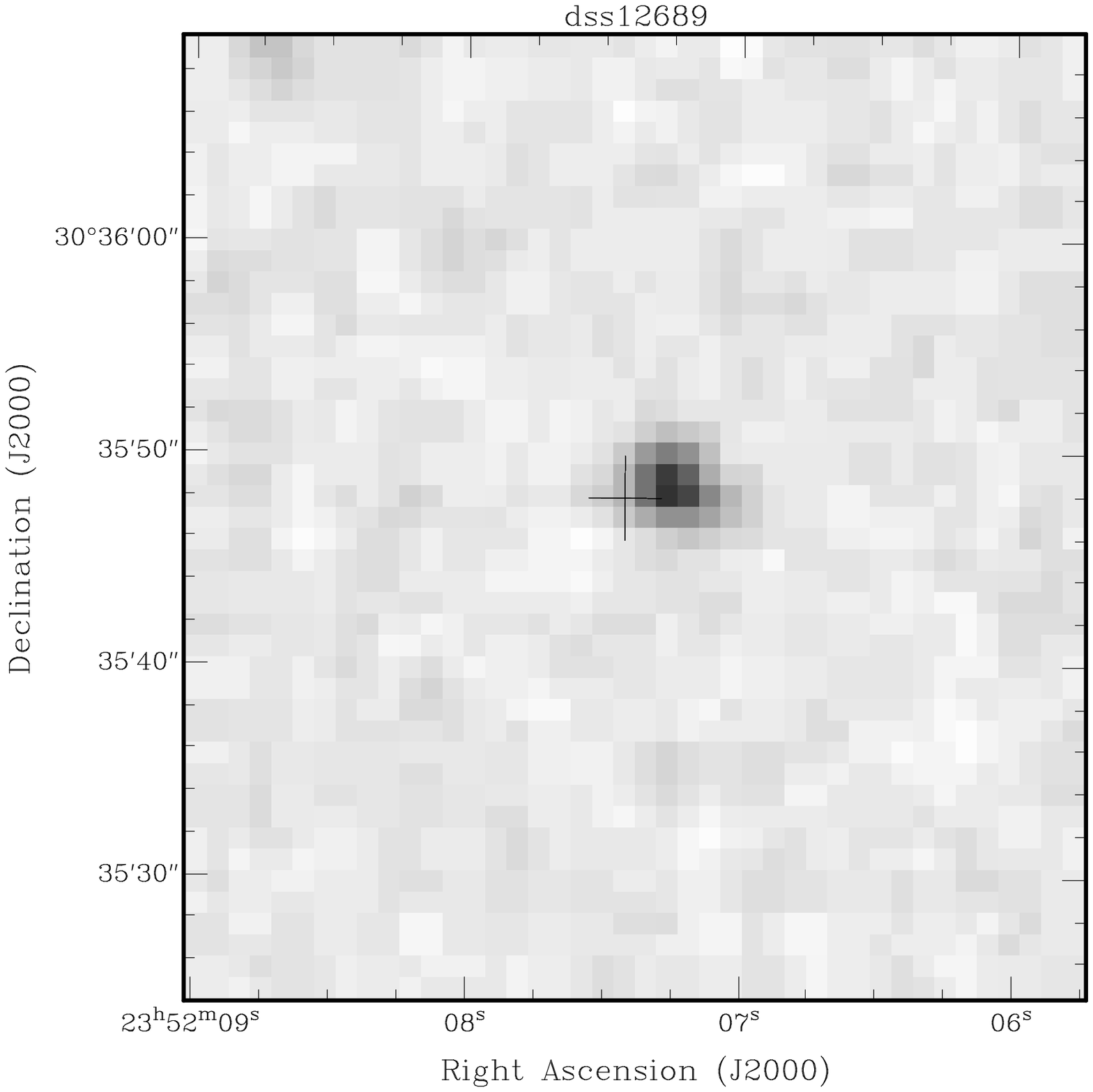 ,width=4.0cm,clip=}}\quad 
\subfigure[9CJ2353+3136 (P60 \it{R}\normalfont)]{\epsfig{figure=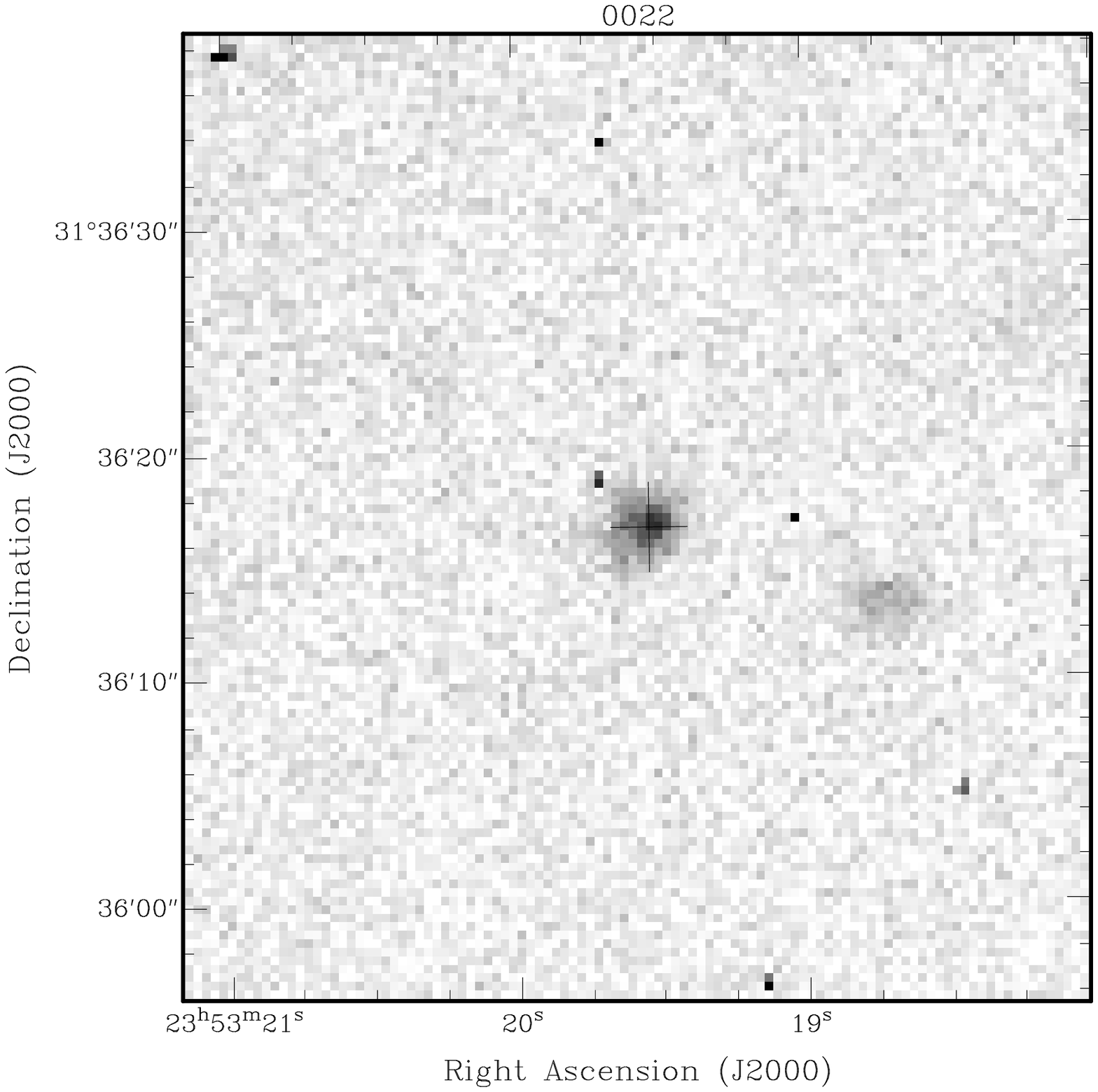 ,width=4.0cm,clip=}}\quad
\subfigure[9CJ2358+2402 (P60 \it{R}\normalfont)]{\epsfig{figure=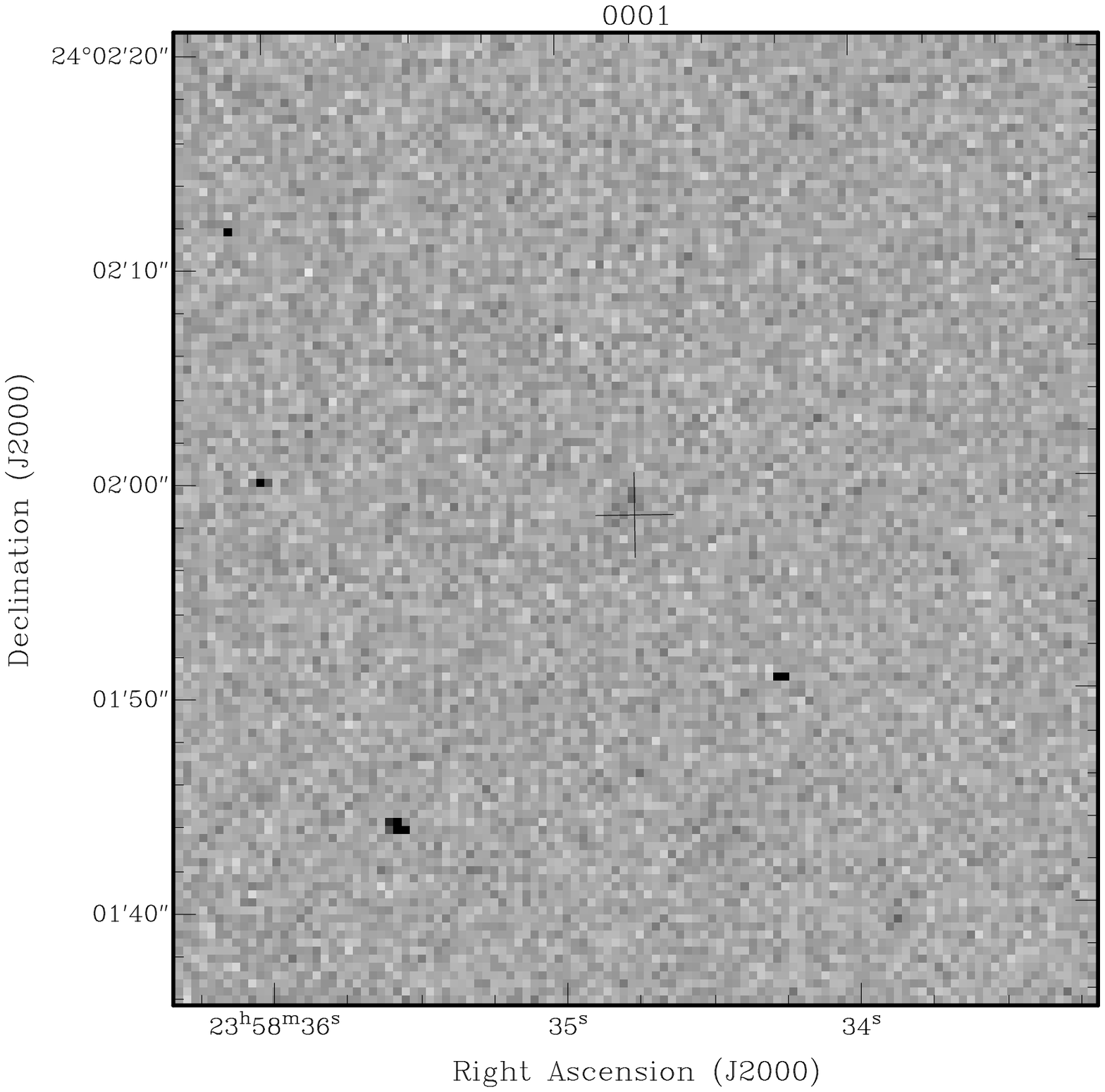 ,width=4.0cm,clip=}}
}
\mbox{
\subfigure[9CJ2359+3543 (DSS2 \it{R}\normalfont)]{\epsfig{figure=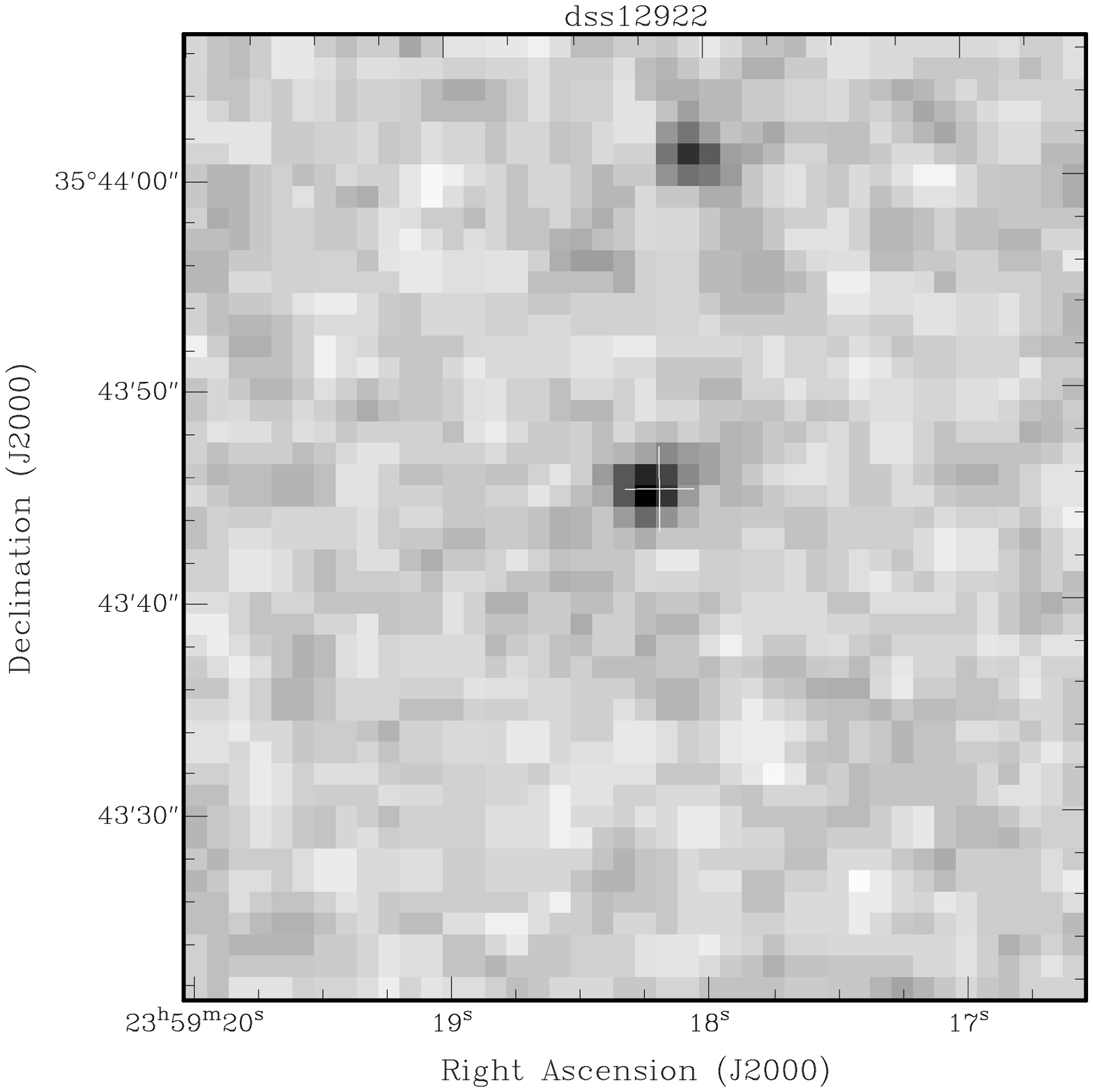 ,width=4.0cm,clip=}}
\subfigure[9CJ2359+2352 (P60 \it{R}\normalfont)]{\epsfig{figure=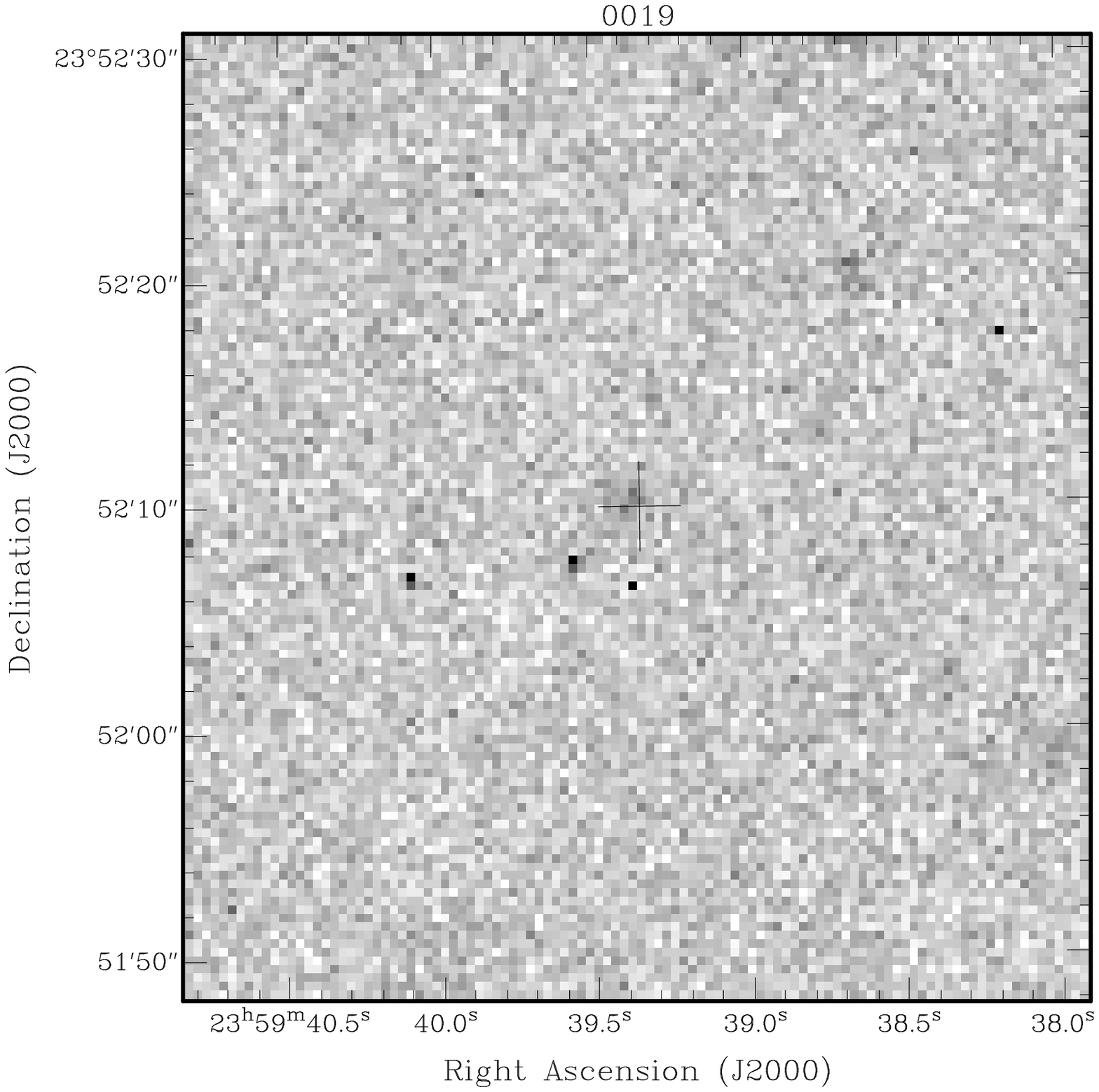 ,width=4.0cm,clip=}}
}
\caption{ \label{overlays_end}Optical counterparts for sources 9CJ1558+4146 to 9CJ2359+2352. Crosses mark maximum radio flux density and are 4\,arcsec top to bottom. Contours: \ref{bw}, 4.8\,GHz contours 10-90 every 10\,\% of peak (48.1\,mJy/beam).}\end{figure*}

The radio and optical results are presented in table 2 and figures \ref{spectra} to \ref{overlays_end}. The positions given in table 2 are taken from the VLA maps; for each source the most accurate position was obtained using the highest frequency map with a good signal to noise ratio. For compact or slightly extended sources the position given is that of the peak (as found by the Gaussian-fitting \texttt{AIPS} task \texttt{jmfit}). For extended sources with an obvious core, the position is that of the core. For extended sources with no core the position, in general, is that of the brightest peak on the map. However for three classical double sources with no cores (9CJ1510+3750 - 29\,arcsec in angular extent, 9CJ1512+4540 - 165\,arcsec in angular extent and 9CJ1556+4257 - 160\,arcsec in angular extent) the position quoted is for the optical counterpart, situated midway between the two hotspots in each case. 

The radio spectra from the simultaneous measurements are shown in figures \ref{spectra} to \ref{spectra_end}. The OVRO 31-GHz points are plotted on the same graphs as circles rather than crosses. The OVRO 40-m telescope has a beam size of $\sim 1\,$acrmin, slightly larger than the  1.4\,GHz synthesised beam of the VLA in its D configuration. Power on angular scales between the resolution of the VLA at high frequency (3\,arcsec or so in D array) and 1\,arcmin will be missed from the VLA measurements but present in the OVRO measurements. The OVRO points thus appear relatively high for several extended sources: notably 9CJ0011+2803, this 25\,arcsec double radio source was observed with the VLA in its D configuration, so it is unresolved at 1.4\,GHz but is resolved at 4.8\,GHz and at 15\,GHz with the RT (which has resolution of 20\,arcsec). The 22 and 43\,GHz flux densities appear to be significantly lower than the unresolved flux densities predict. There are several sources with OVRO flux densities that are lower than the VLA flux densities predict; none of which are steep spectrum sources, and all of which are dominated by an unresolved component. There are several sources for which the OVRO measurement is high but the source is unresolved: this suggests that the discrepancy may by due, at least in part, to variability over the six months or so during which observations were made.

The optical images with the radio contours overlaid are shown in figures \ref{overlays} to \ref{overlays_end}. For clarity, contours are not shown for the unresolved radio sources. In all cases a cross, 4\,arcsec in size, shows the radio position given in table 2. Where contours are shown, they have been chosen to best describe the radio-source morphology, emphasising the extended structures. In a few cases the DSS2 \it{O}\normalfont-band image provides a higher signal to noise detection of a counterpart and we present it alongside the \it{R}\normalfont-band image. Where descriptive, smaller or larger scale images are also shown. Images are in RA order, except where a pair of images of the same source would otherwise occur across a page break.

The astrometry of the P60 images was carried out by comparison with DSS2 images and is accurate to better than 0.5\,arcsec. The error on the position of the peak in the radio source, given by \texttt{jmfit}, is generally better than about 0.1\,arcsec. However, many compact radio sources were phase calibrated on themselves, so the astrometry could be affected by mean phase errors persistent over the course of the snap-shot observation. Additionally, the 2\,arcsec beam of the VLA in D-config and at 43\,GHz will not resolve sources with extended structures as large as $\sim 1$\,arcsec which could pull the position of the peak (as fit by \texttt{jmfit}) away from the core position. For these reasons optical identifications have been made where the mis-match between the radio and optical positions is as large as 6\,arcsec, although for all but three sources it is better than 2\,arcsec. The probability of a random alignment within 2\,arcsec is $\sim3$\,percent on the deepest P60 images (and, of course, is even smaller on less deep images), it is therefore expected that a handful of the IDs will prove to be chance coincidences. The three sources with optical IDs more distant than 2\,arcsec from the radio position are 9CJ0005+3139 (this has a 6\,arcsec discrepancy - intriguingly the optical source in extended in the direction of elongation of the radio source); 9CJ0923+3107 (3\,arcsec discrepancy) and 9CJ1523+4156 (5\,arcsec difference). Including these three, optical identifications have been made for a total of 155 (88\,percent) of the 176 radio sources.

\newpage\clearpage
\addtocounter{page}{5}
\addtocounter{table}{1}
\begin{figure*}
\includegraphics*[bb=50 60 510 760, width=15cm]{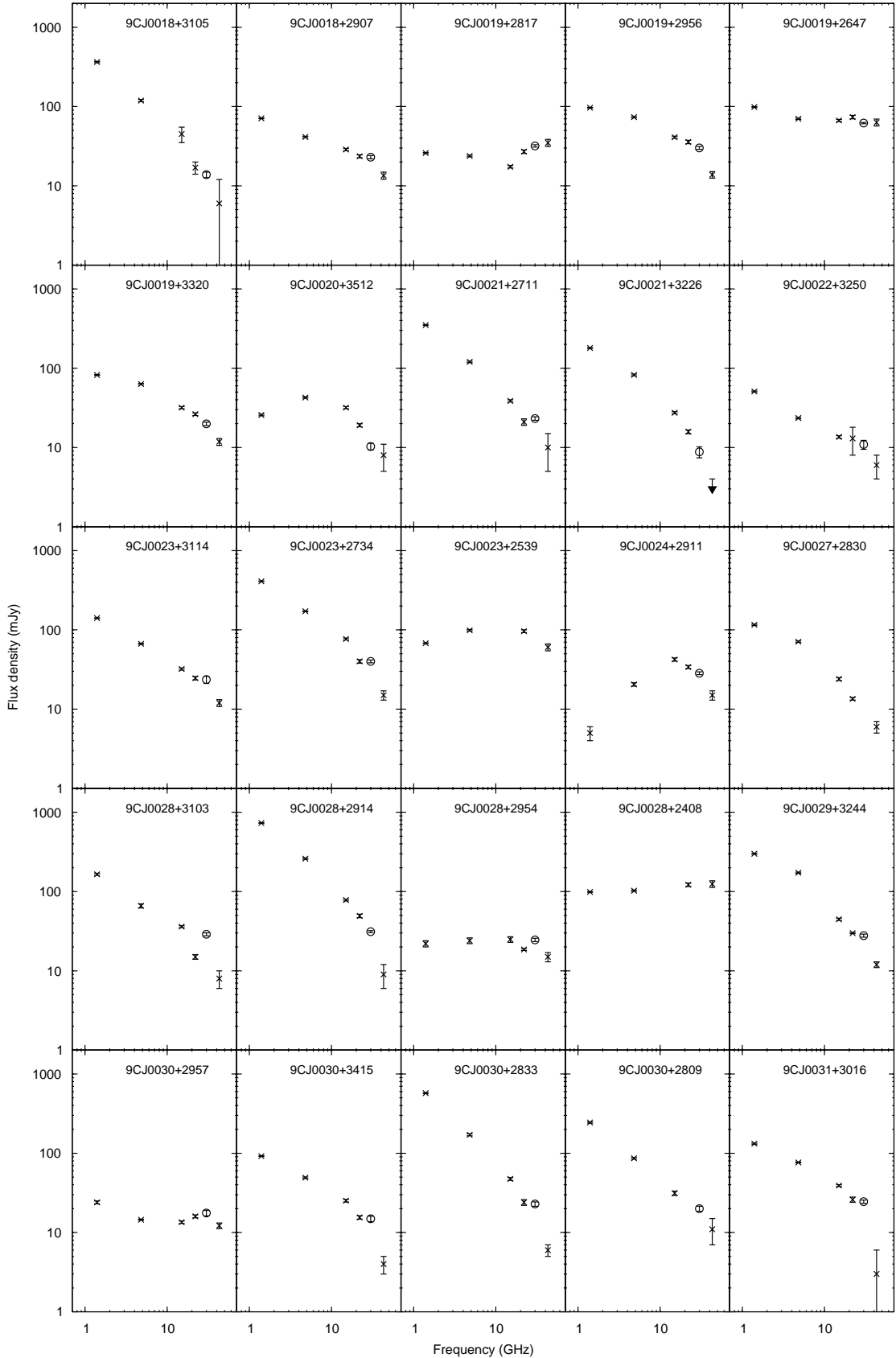}\caption{Radio spectra for sources 9CJ0018+3105 to 9CJ0031+3016.}
\end{figure*}
\begin{figure*}
\includegraphics*[bb=50 60 510 760, width=15cm]{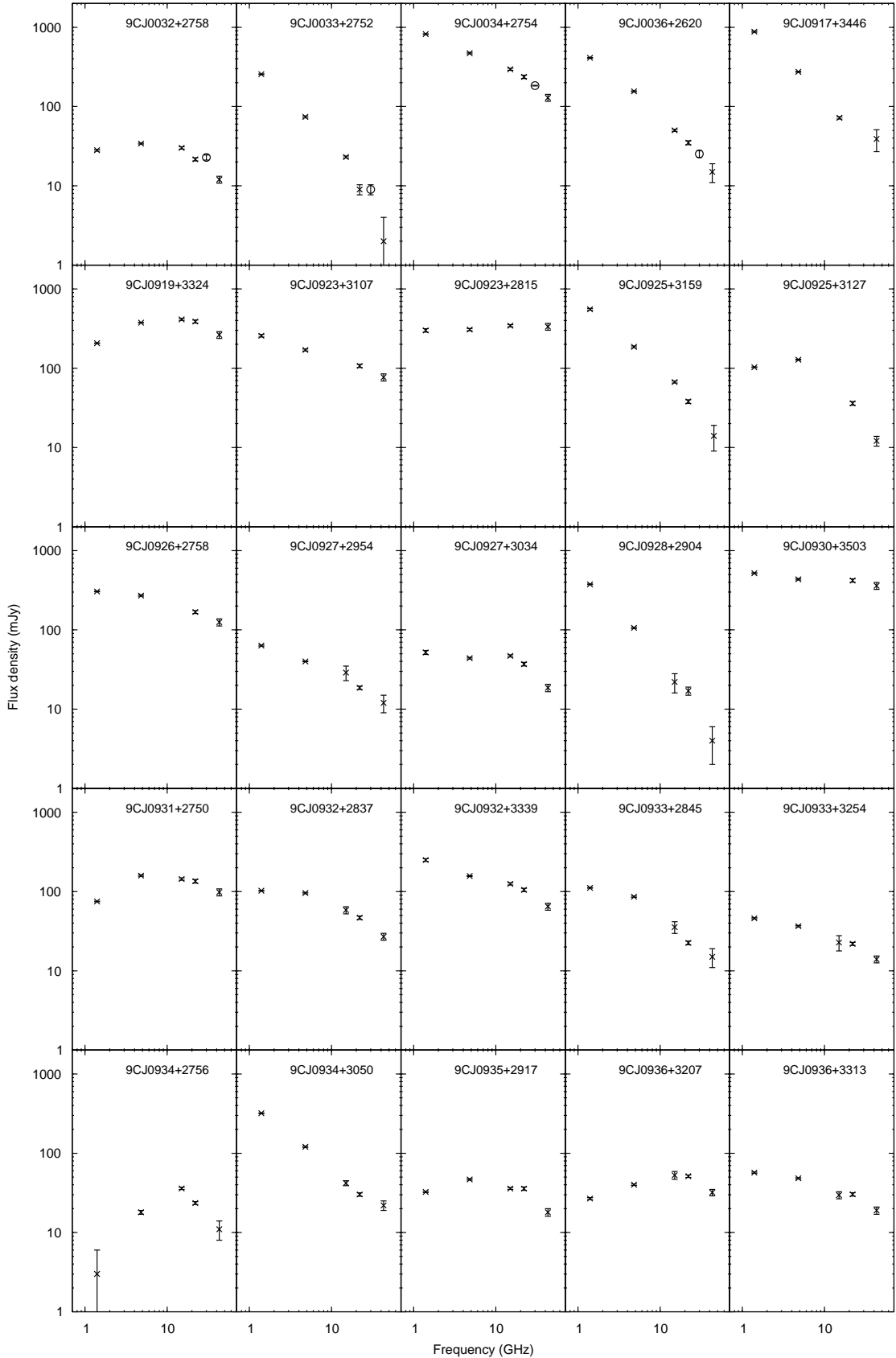}\caption{Radio spectra for sources 9CJ0032+2758 to 9CJ0936+3313.}
\end{figure*}
\begin{figure*}
\includegraphics*[bb=50 60 510 760, width=15cm]{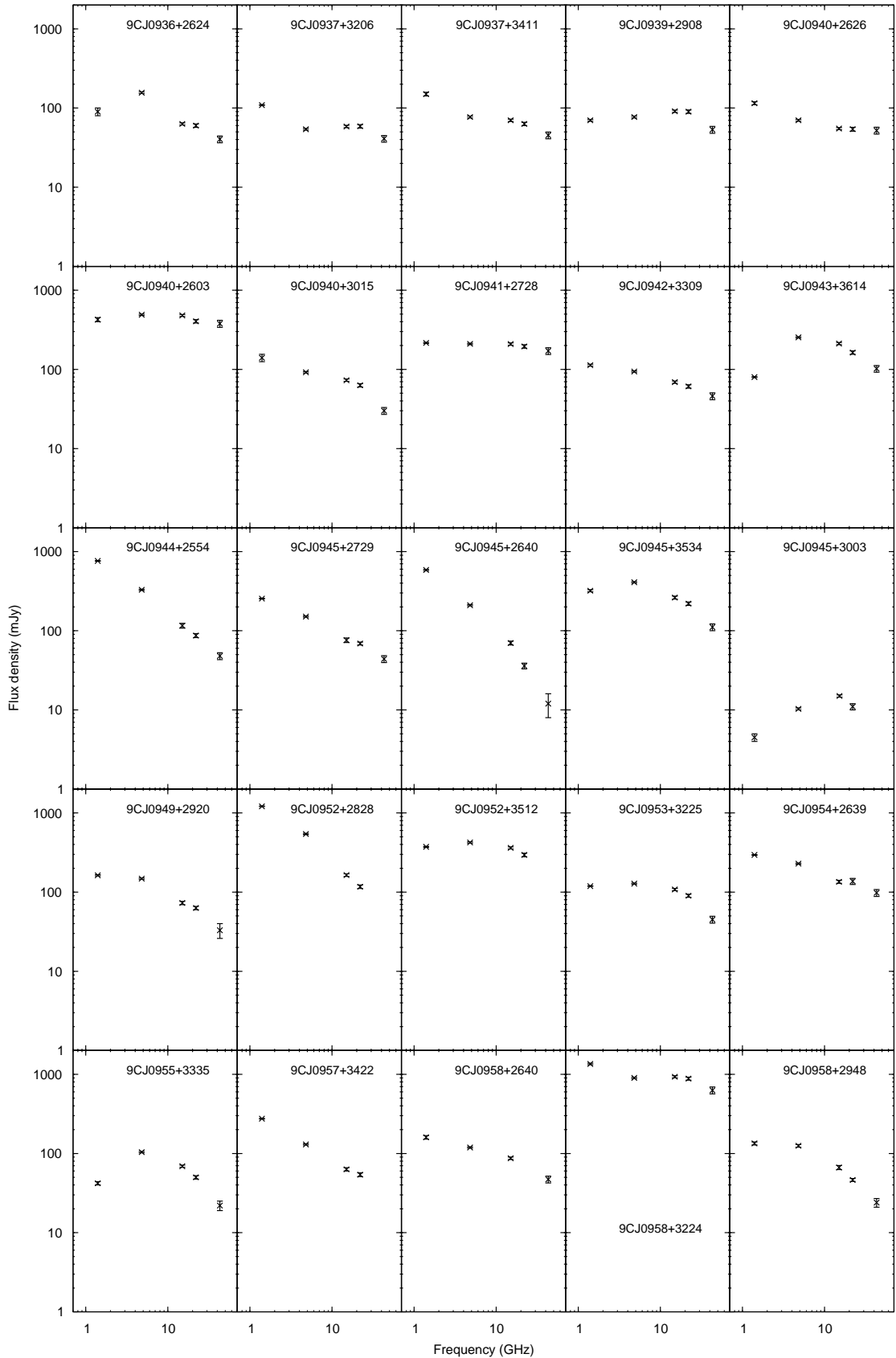}\caption{Radio spectra for sources 9CJ0936+2624 to 9CJ0958+2948.}
\end{figure*}
\begin{figure*}
\includegraphics*[bb=50 60 510 760, width=15cm]{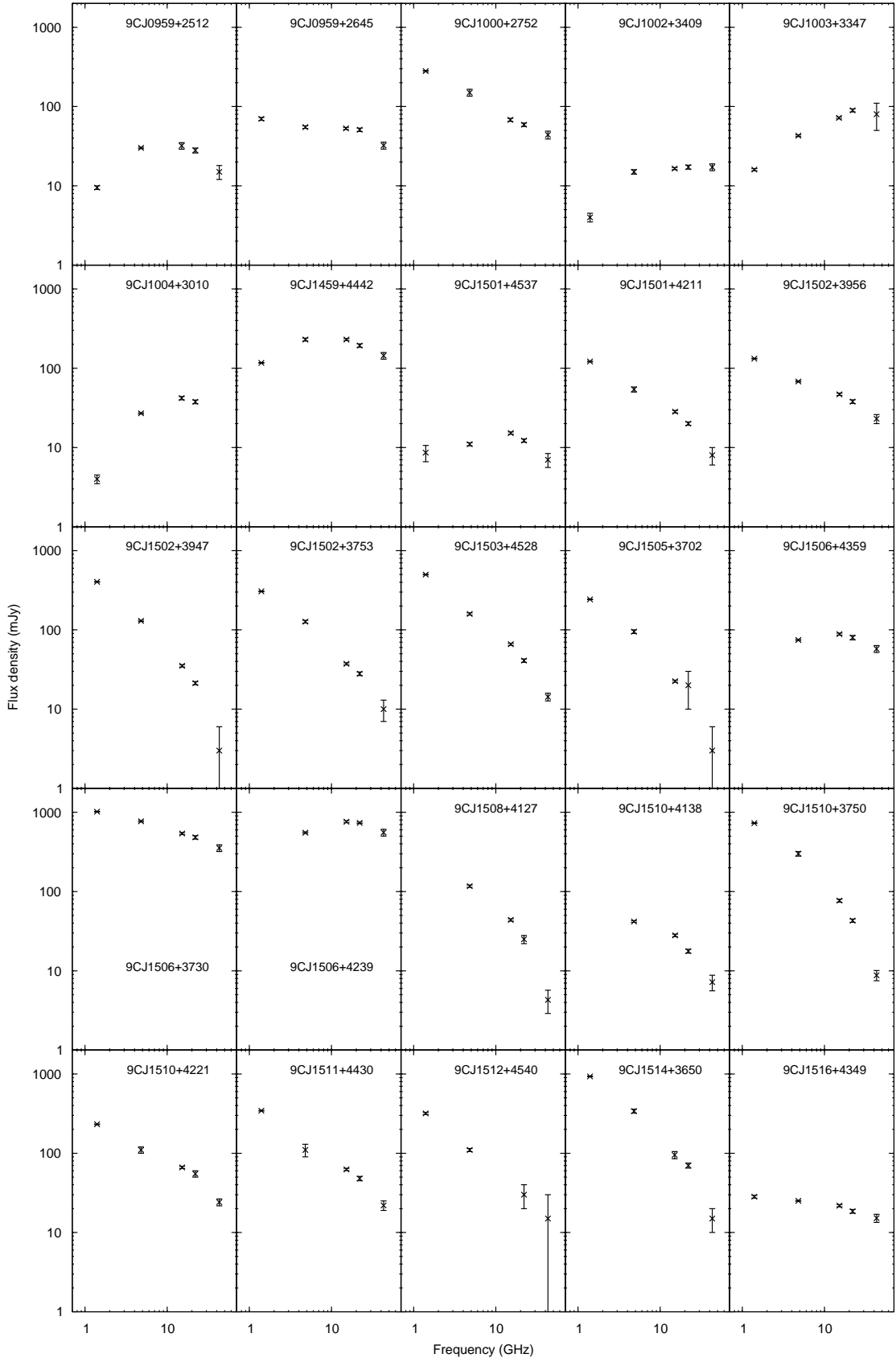}\caption{Radio spectra for sources 9CJ0959+2512 to 9CJ1516+4349.}
\end{figure*}
\begin{figure*}
\includegraphics*[bb=50 60 510 760, width=15cm]{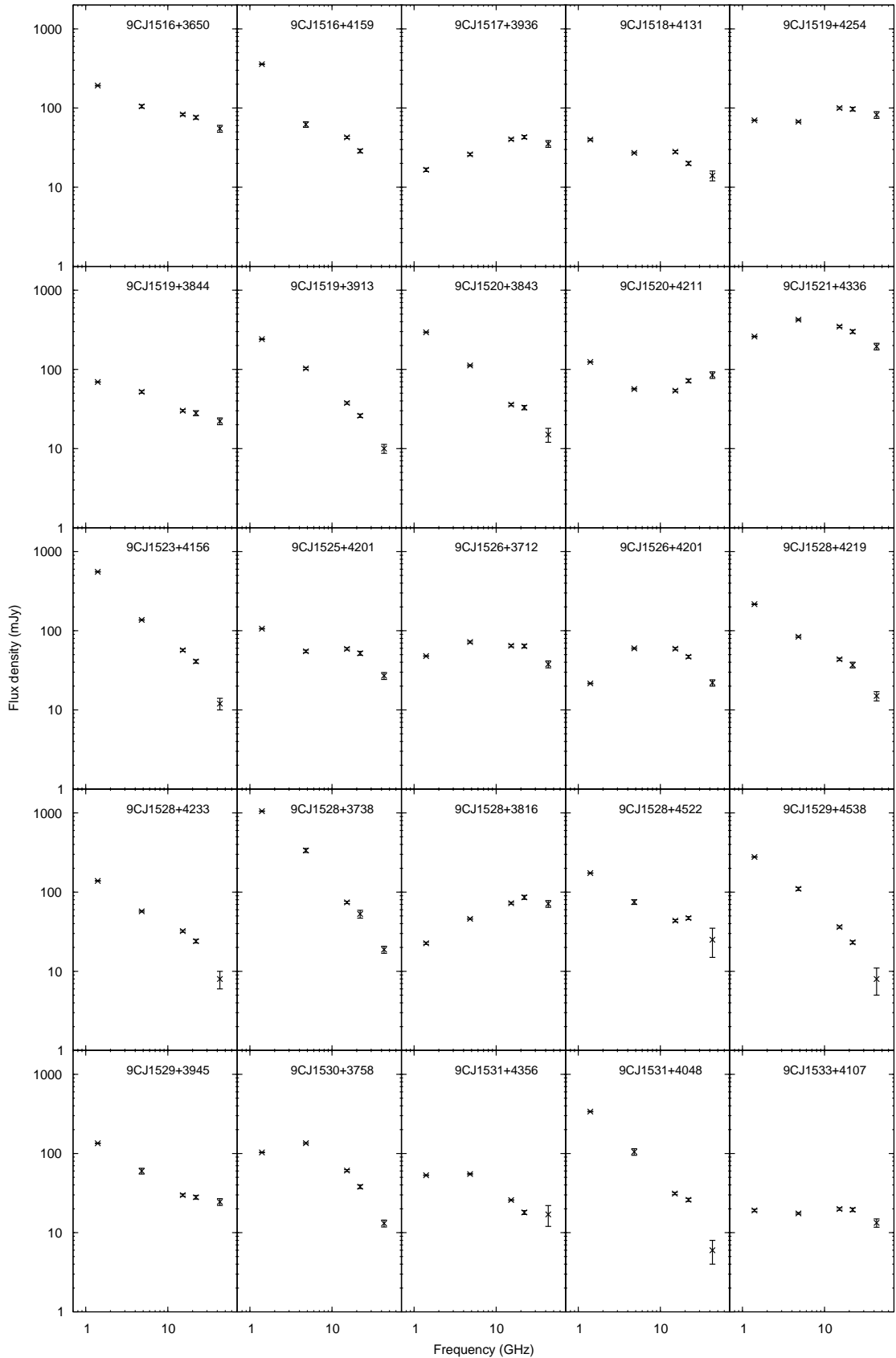}\caption{Radio spectra for sources 9CJ1516+3650 to 9CJ1533+4107.}
\end{figure*}
\begin{figure*}
\includegraphics*[bb=50 60 510 760, width=15cm]{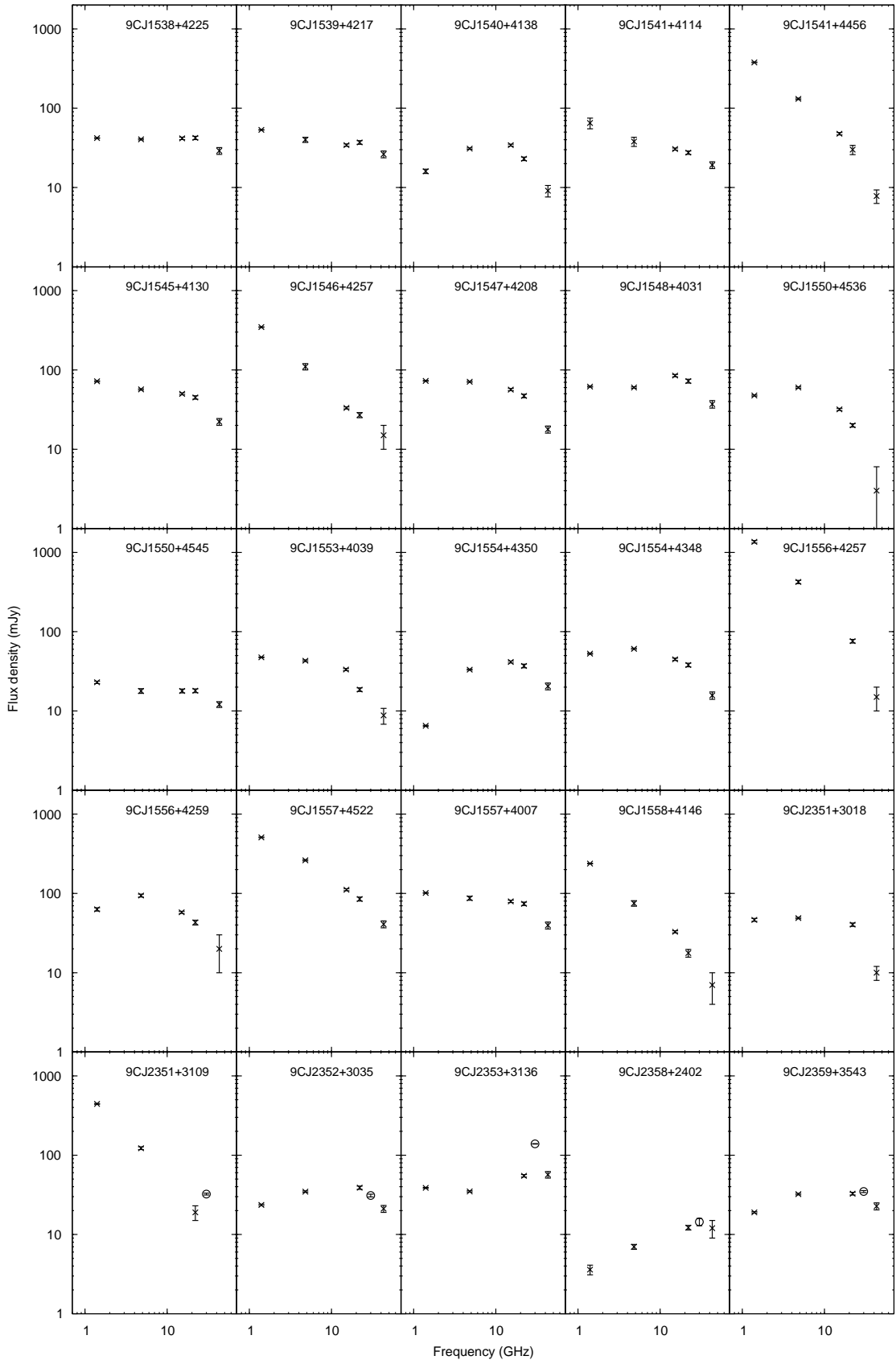}\caption{Radio spectra for sources 9CJ1538+4225 to 9CJ2359+3543.}
\end{figure*}

\clearpage 

\renewcommand{\thesubfigure}{(\roman{subfigure})}

\newpage\clearpage
\newpage\clearpage
\newpage\clearpage
\newpage\clearpage
\newpage\clearpage
 \newpage\clearpage

\subsection {Notes on individual sources}  
   
The map of 9CJ0958+2948 (see Fig. \ref{9CJ0958+2948}) shows four radio components, the brightest two of which form a core/jet source centred on the 9C position and with an optical counterpart. The remaining two components are offset about 7\,arcsec to the south-west and not aligned with the other components. They are very close together and we assume that they are associated. The total flux density of this second radio source is only 40\,mJy at 1.4\,GHz and 25\,mJy at 4.8\,GHz and we leave it out of the sample since its expected 15\,-GHz flux would be too low to allow its inclusion.

\begin{figure}
\centerline{\epsfig{figure=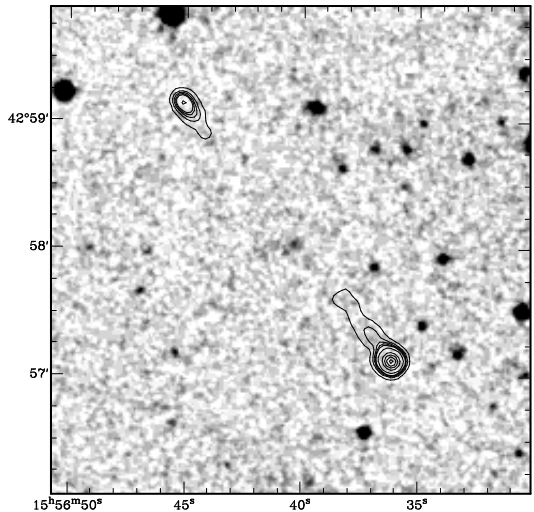 ,width=7cm,clip=}}
\caption{\label{uni} POSS-II image of the field of 9CJ1556+4257 with contours from the FIRST survey map overlaid. Radio contours are at 1,2,3,4,5,10,30,50,70 and 90\,percent of the peak flux density of 0.88 Jy/beam.}
\end{figure}

The two sources 9CJ1556+4257 and 9CJ1556+4259 appear in the same radio map; inspection of the maps in the Faint Images of the Radio Sky at Twenty-cm \citep[FIRST,][]{becker} survey reveals that they are physically distinct objects. 9CJ1556+4257 is a 160\,-arcsec classical double radio source (see figure \ref{uni}), with an optical counterpart mid-way between the two lobes. 9CJ1556+4259 is a point-like radio source, with a peaked spectrum and an optical counterpart (see figure \ref{id's}).

\begin{figure}
\centerline{\epsfig{figure=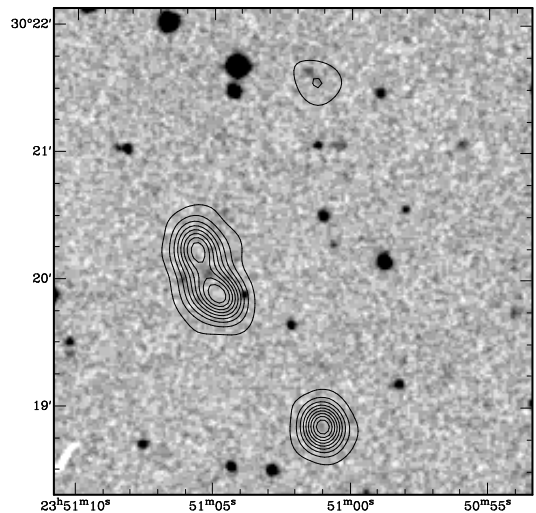,bb=50 50 541 525,width=7cm,clip=}}
\caption{\label{multi}9CJ2351+3018 and 9CJ2351+3019. 4.8\,GHz radio contours at 10,20,30,40,50,60,70,80,and 90\,percent of the peak flux density of 48.1 mJy/beam.}
\end{figure}

The two sources 9CJ2351+3018 and 9CJ2351+3019 are only 1\,arcmin apart on the sky and appear as one source in the 9C catalogue. Figure \ref{multi} shows the 4.8\,GHz radio contours of 9CJ2351+3018, the point source to the bottom of the image, and 9CJ2351+3019, the double source in the centre left. Both sources have optical counterparts and we therefore assume that they are separate objects. There is a third radio source which is faint (10\,mJy) and also has a possible optical ID; this source is not seen at all at higher frequencies and is therefore not considered any further. The OVRO 40m dish flux for 9CJ2351+3019 appears to be slightly high, but this may be accounted for by the inclusion of some flux from 9CJ2351+3018 in the $\sim 1$ arcmin beam.

\section{Sample statistics}
\label{sec:stats}
Here we consider the population of our flux-limited samples, A and B.

\subsection{Radio properties}

\label{sec:radio_stats}We classify our radio sources as compact or extended on the basis of their radio size, compact sources being those with angular extent $< 2$\,arcsec. 68\,percent of the sources in sample A are compact and 67 percent of those in sample B are compact.

As the radio flux densities have been measured simultaneously we can reliably classify objects using their radio spectral index, $\alpha$ (we take $ S \propto \nu^{-\alpha}$). We define three different spectral classes using the spectral index between 1.4 and 4.8\,GHz. The classes are: steep spectrum sources with $\alpha^{4.8}_{1.4} \ge 0.5$, flat spectrum sources with $-0.1 \le \alpha^{4.8}_{1.4} < 0.5 $, and rising spectrum sources with $\alpha^{4.8}_{1.4} < -0.1 $. We use -0.1, rather than zero, as the flat/rising cut-off to reduce contamination of the rising spectrum class by objects with poorly defined spectral peaks. These criteria mean that we classify sources that actually peak at about 1\,GHz  (truly ``gigahertz'' peaked spectrum objects) as steep or (more probably) flat spectrum objects. The sources with spectra that rise between 1.4 and 4.8\,GHz must have a peak at some frequency: in fact all but two (which have spectra still rising at 43\,GHz) of the ``rising'' spectrum sources show a peak in the range 4.8 to 22\,GHz, and are presumably just more extreme versions of gigahertz peaked spectrum (GPS) sources; we refer to them as GPS sources in this work, taking the definition of a GPS source to be a source with a radio spectrum peaking at $\sim5$\,GHz \it{or above}\normalfont.

Four sources have no 1.4\,GHz measurements so have been classified using the spectral index between 4.8 and 15\,GHz and by looking in the FIRST catalogues. They are: 9CJ1506+4359, 9CJ1506+4239, 9CJ1508+4127 and  9CJ1510+4138. 9CJ1506+4359 has a FIRST flux density of 65\,mJy at 1.4\,GHz, when measurement errors are considered the FIRST/4.8\,GHz spectral index straddles the value of $-0.1$ so we conservatively classify it as flat spectrum. 9CJ1506+4239, has a FIRST flux density of 308\,mJy, so both between FIRST and 4.8\,GHz and simultaneous 4.8 and 15\,GHz it has a rising spectrum. 9CJ1508+4127 is a steep spectrum source (with a FIRST flux density of 350\,mJy) and 9CJ1510+4138, does not appear in FIRST but we classify it as flat spectrum. 
\suppressfloats

\begin{table}
 \centering
 \caption{\label{radio_props}Numbers of radio sources of different radio spectral and size types in samples A and B. Percentages are given as the fractions within each radio  size class that fall into each spectral class.}
 \begin{tabular}{@{}crrrrrrrr@{}}
 \hline
Sample A & \multicolumn{2}c{Steep} & \multicolumn{2}c{Flat} & \multicolumn{2}c{Rising} & Total\\ 
\hline
Compact & 23  & (27\,\%) & 40 & (48\,\%) & 21 & (25\,\%) & 84 \\
Extended & 33 & (83\,\%) & 6 & (15\,\%) & 1 & (3\,\%) & 40 \\
Total & 56 & (45\,\%) & 46 & (37\,\%) & 22 & (18\,\%) & 124 \\
\hline 
Sample B & \multicolumn{2}c{Steep} & \multicolumn{2}c{Flat} & \multicolumn{2}c{Rising} & Total\\ 
 \hline
Compact &  10  & (21\,\%) & 22 & (47\,\%) & 15 & (32\,\%) & 47\\
Extended & 13 & (57\,\%) & 6 & (26\,\%) & 4 & (17\,\%) & 23 \\
Total & 23 & (33\,\%) & 28 & (40\,\%) & 19 & (27\,\%) & 70 \\
\hline
\end{tabular}
\end{table}

In table \ref{radio_props} we compare the proportions of sources in each spectral and radio size class in samples A and B. From this it is clear that there is a greater proportion of rising spectrum sources in the sample with the higher flux limit. This trend is confirmed by taking the subsample of B with flux densities $\ge 150$\,mJy at 15\,GHz; 7 (39 percent) of the 18 sources have rising spectra. This behaviour is expected since we are measuring $\alpha$ below the selection frequency. There is also a strong tendency for the flat and rising spectrum sources to be compact in the radio; conversely 83\,percent of the extended radio sources in sample A are steep spectrum. This trend is already well established in previous work at lower frequencies: see e.g \citet[][PW hereafter]{P1}.

\begin{figure*}
\centerline{ \epsfig{figure=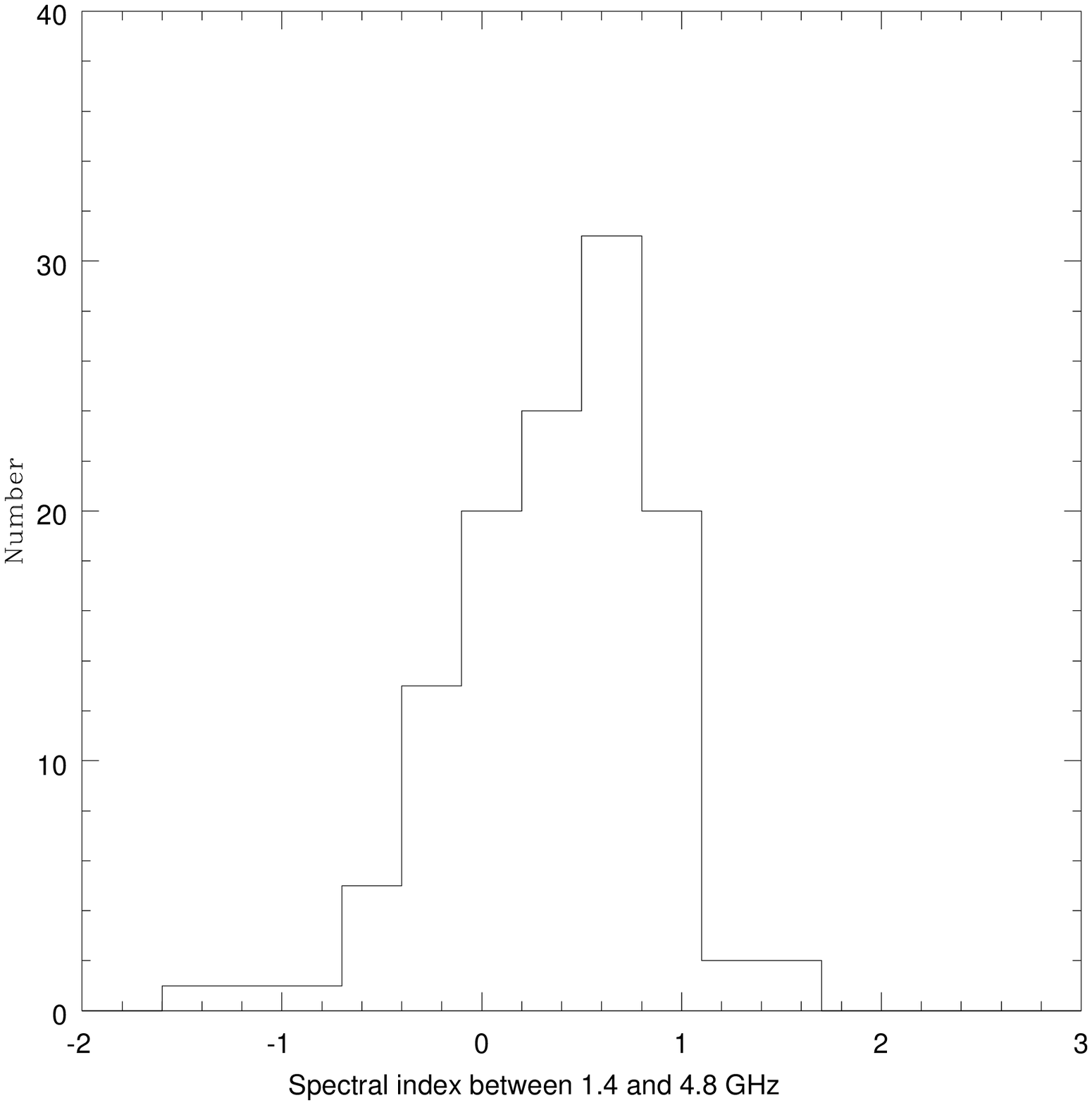,width=7cm,clip=}
\qquad \epsfig{figure=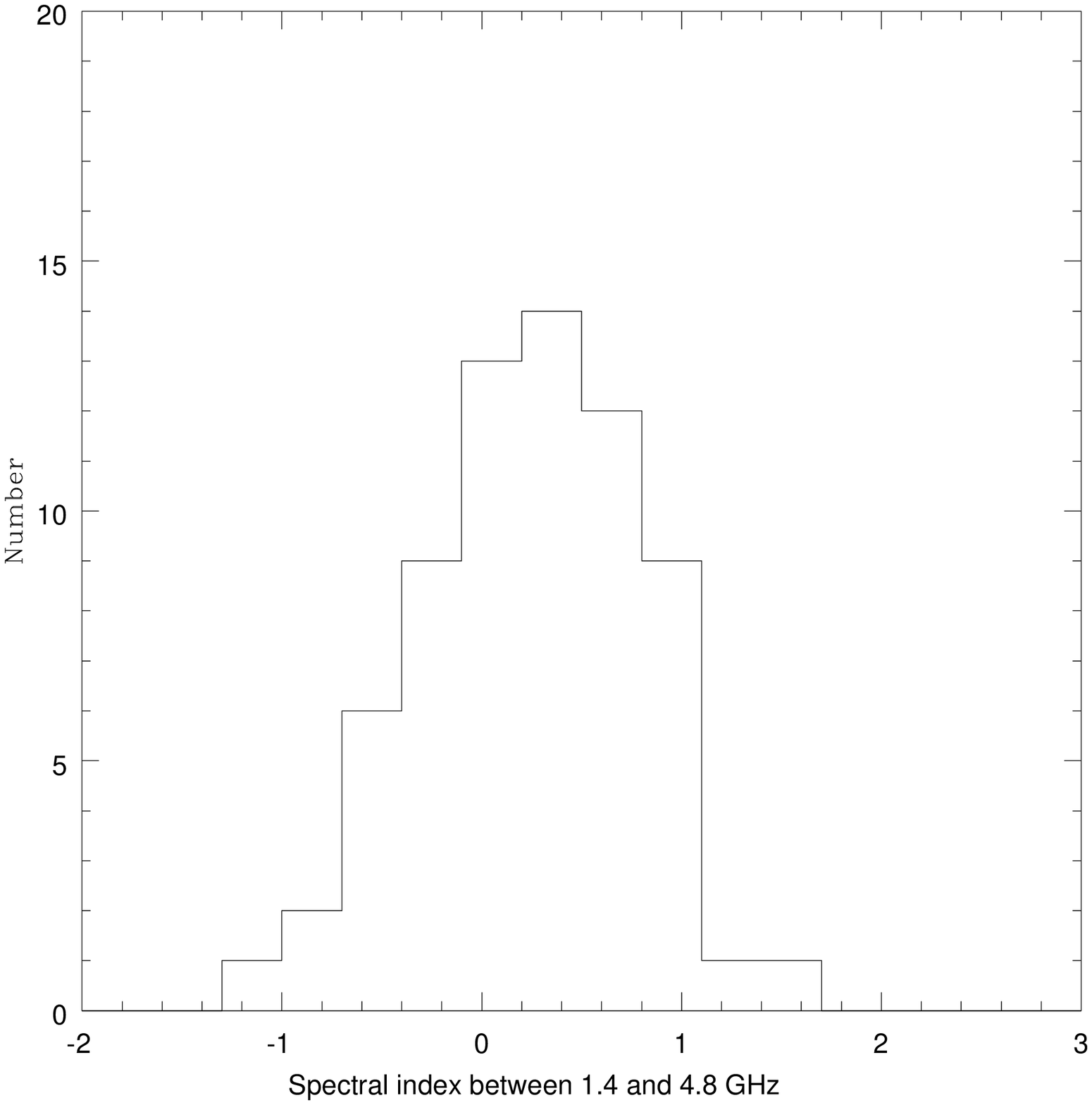,width=7cm,clip=}}
\vspace{0.5cm}
\centerline{ \epsfig{figure=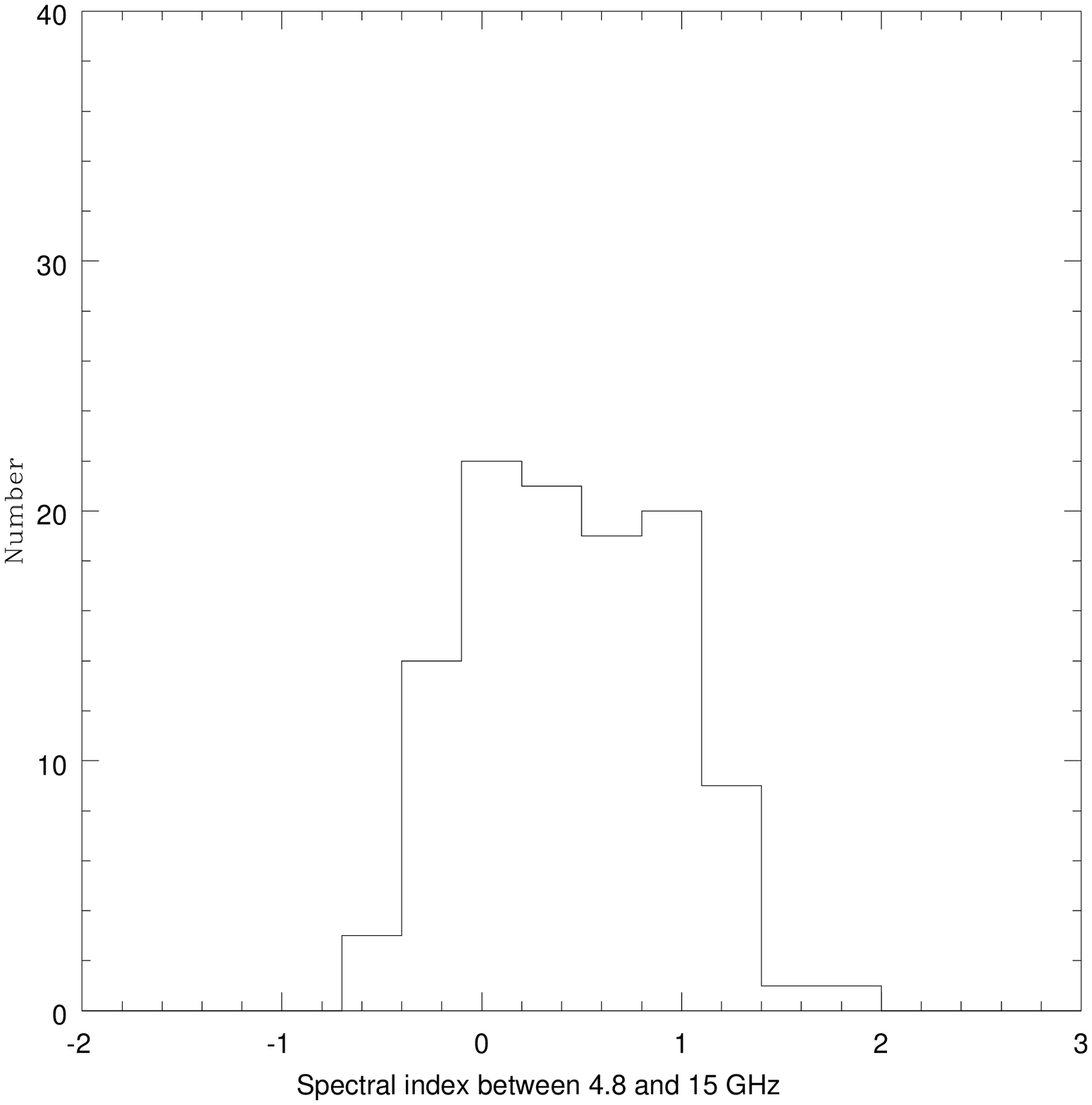,width=7cm,clip=}
\qquad \epsfig{figure=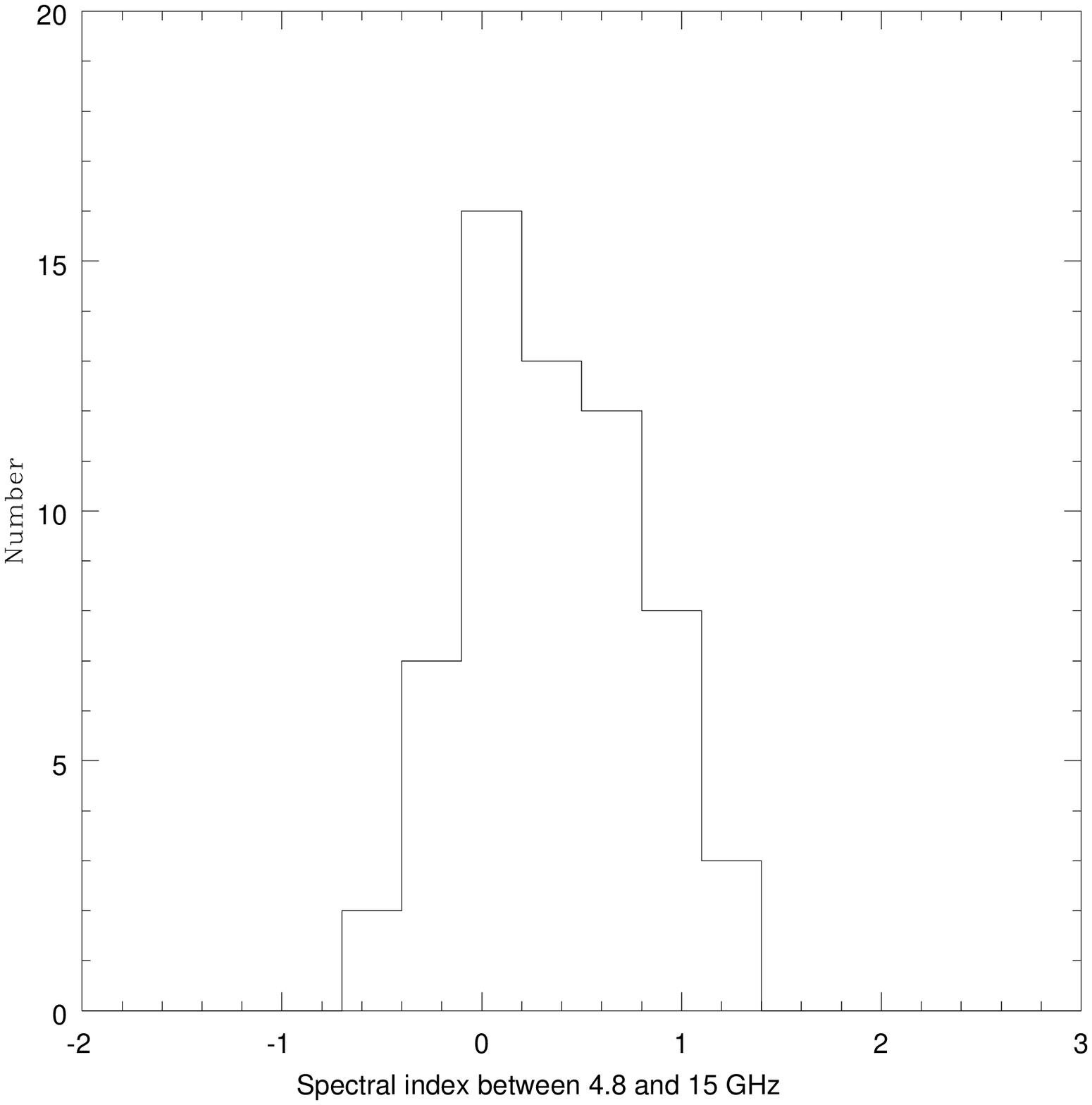,width=7cm,clip=}}
\vspace{0.5cm}
\centerline{ \epsfig{figure=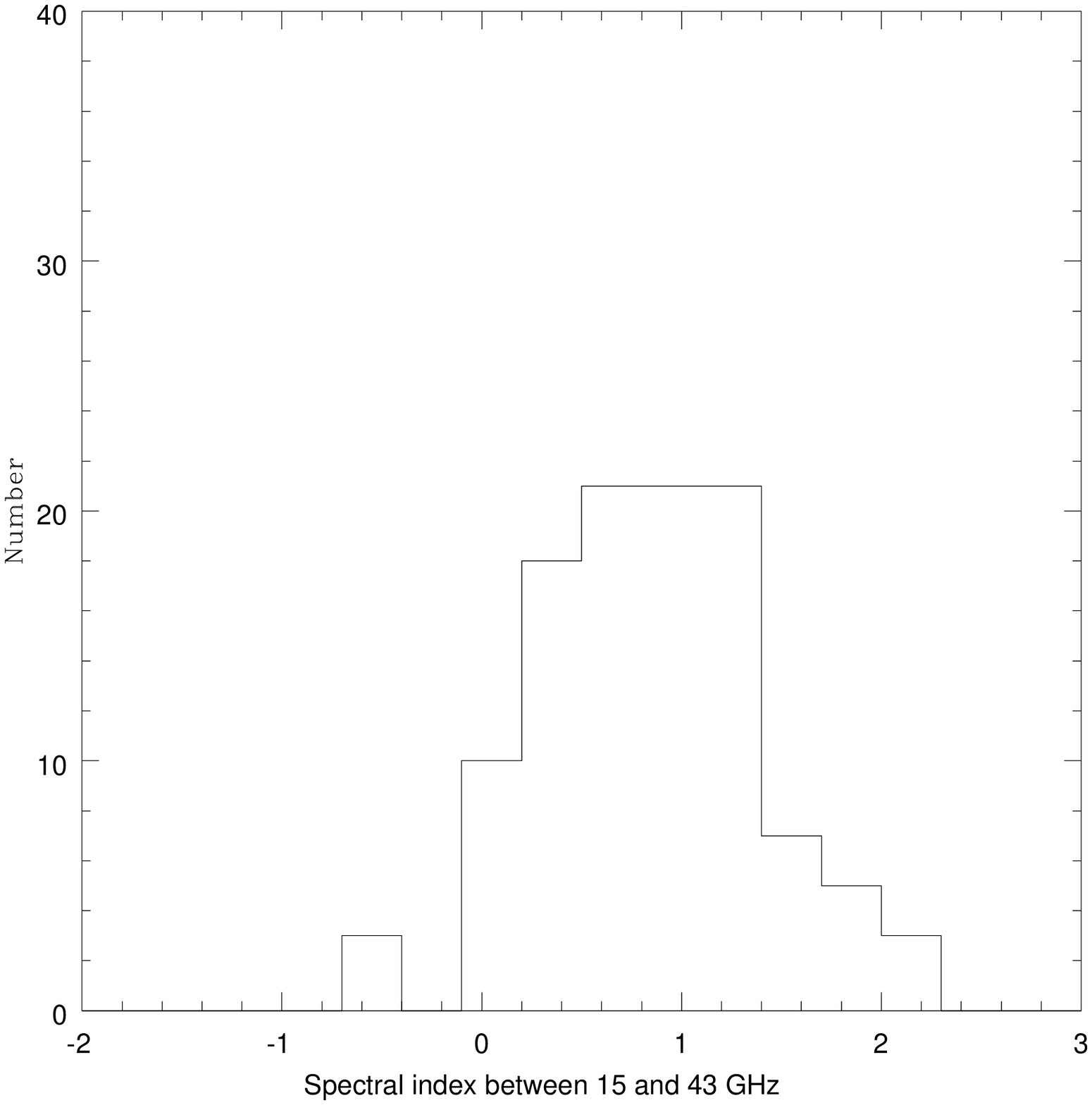,width=7cm,clip=}
\qquad \epsfig{figure=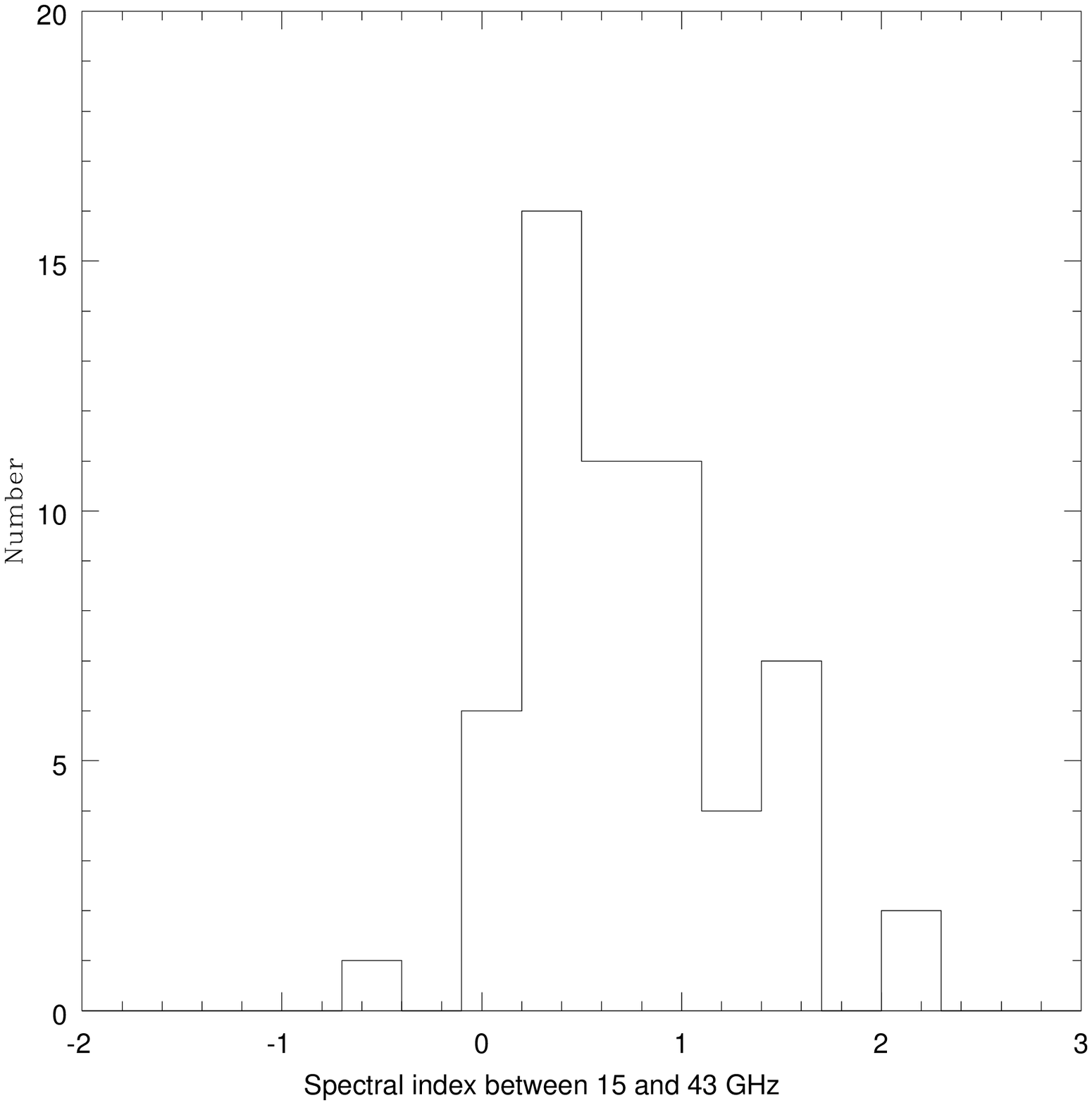,width=7cm,clip=}}\caption{\label{spectral_trend}Distribution of spectral index for Sample A, left and B, right. Top to bottom: $\alpha^{4.8}_{1.4}$; $\alpha^{15}_{4.8}$; $\alpha^{43}_{15}$ }
\end{figure*}

Figure \ref{spectral_trend} shows the distribution of spectral index measured between different frequencies for samples A and B. There is a clear trend for the spectral index taken between 15 and 43\,GHz to be steeper than that calculated at lower frequencies. The median values and 25th and 75th percentile values are given in table \ref{alphas}.

\begin{table}
 \centering
 \caption{\label{alphas}The 25th percentile, median and 75th percentile values of spectral index for samples A and B at different frequencies.}
 \begin{tabular}{@{}rrcrrcr@{}}
 \hline
&  \multicolumn{3}c{Sample A} & \multicolumn{3}c{Sample B} \\
& 25\,\% & Median & 75\,\% & 25\,\% & Median & 75\,\% \\
 \hline
\vspace{0.2cm}
$\alpha^{4.8}_{1.4}$ & 0.05 & 0.44 & 0.76 & -0.12 & 0.24 & 0.64\\
\vspace{0.2cm}
$\alpha^{15}_{4.8}$ & 0.06 & 0.39 & 0.95 & 0.02 & 0.27 & 0.70\\
$\alpha^{43}_{15}$ & 0.42 & 0.87 & 1.20 & 0.38 & 0.67 & 1.03 \\
\hline
\end{tabular}
\end{table}

For each sample the median values of $\alpha^{4.8}_{1.4}$ and $\alpha^{15}_{4.8}$ are very similar. This is perhaps surprising since 12 of the 22 peaked spectrum sources in sample A peak at 5\,GHz, but is explained by the fact that although their spectra are starting to fall above 5\,GHz, so that the left-hand, low-$\alpha$ tail of the distribution shrinks, they have not turned over sufficiently between 4.8 and 15\,GHz to be ``steep'' spectrum sources in this frequency range. On the other hand between 15 and 43\,GHz the spectra of these 5-GHz peakers have become steep, hence the large change in median spectral index when it is calculated between 15 and 43\,GHz. Sources in sample B tend to have lower (less steeply falling) values of $\alpha$, due to the increased fraction of peaked sources in this sample, as discussed earlier.

\subsection{Optical properties}

\begin{figure}
\centerline{\epsfig{figure=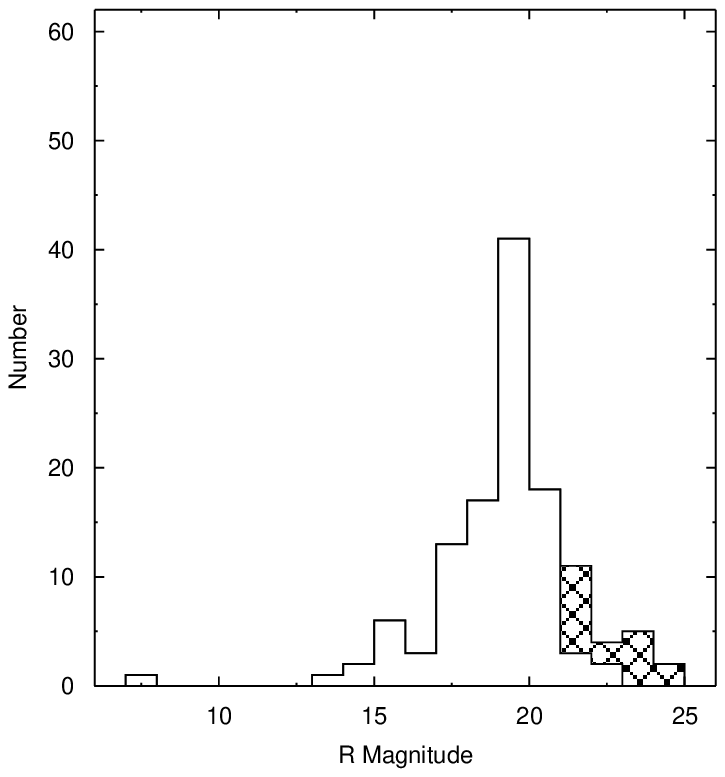, width=7cm,clip=}}\vspace{0.5cm}
\centerline{\epsfig{figure=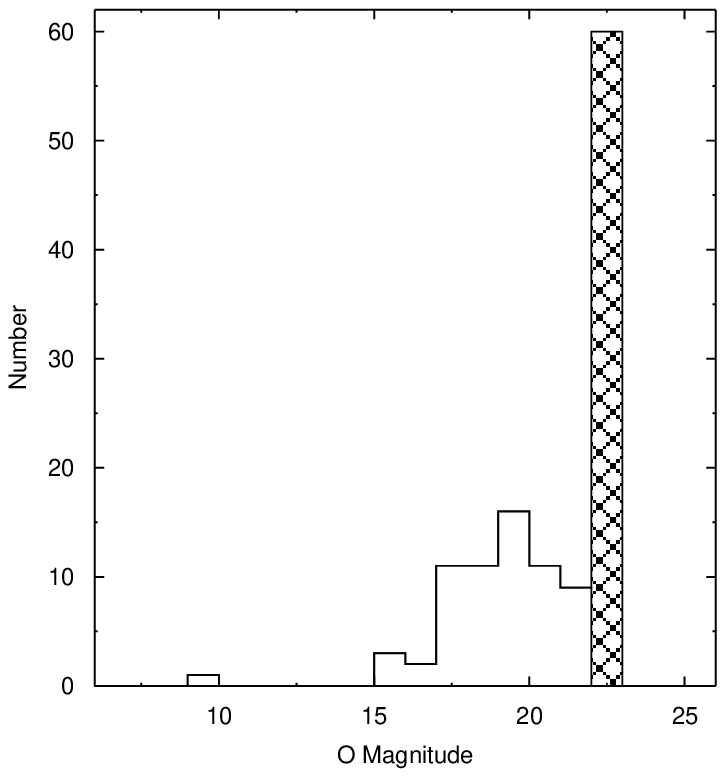, width=7cm,clip=}}\vspace{0.5cm}
\caption{\label{optical_magsA} Distribution of optical magnitudes for Sample A in \it{R}\normalfont-band (top) and \it{O}\normalfont-band (bottom). Unseen objects are assigned an optical flux density one magnitude fainter that the detection limit (cross-hatched area).}
\end{figure}

\begin{figure}
\centerline{\epsfig{figure=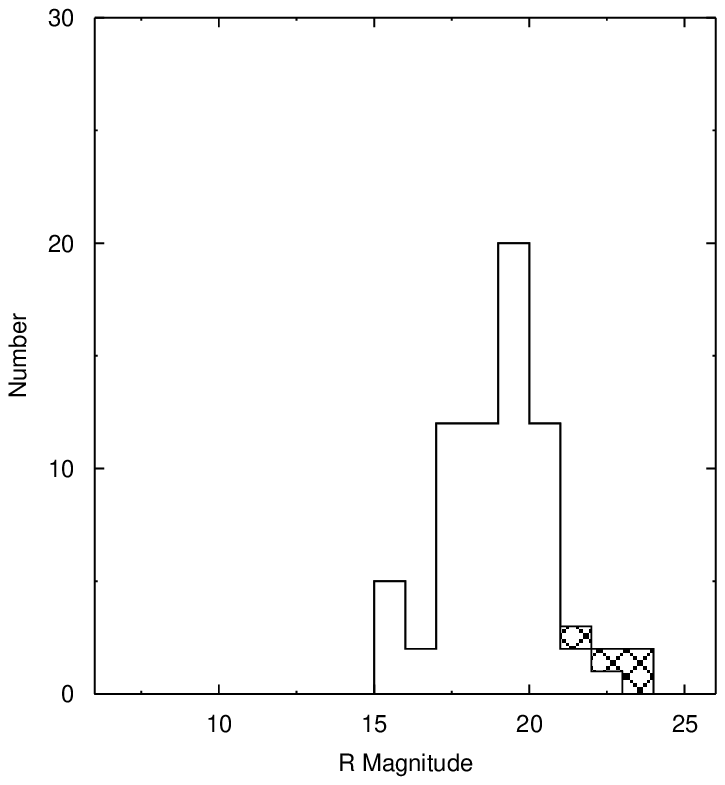, width=7cm,clip=}}\vspace{0.5cm}
\centerline{\epsfig{figure=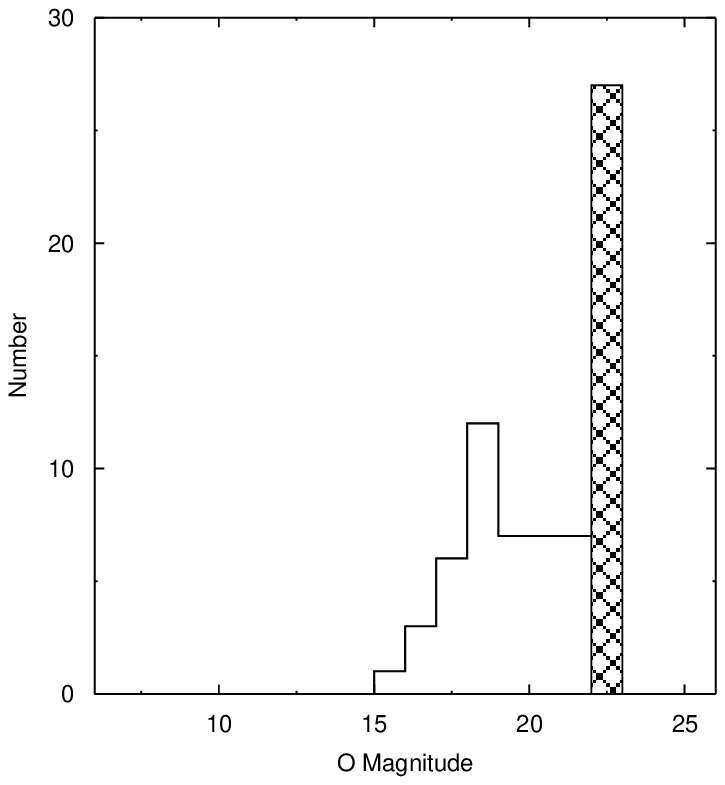, width=7cm,clip=}}\vspace{0.5cm}
\caption{\label{optical_magsB} Distribution of optical magnitudes for Sample B in \it{R}\normalfont-band (top) and \it{O}\normalfont-band (bottom). Unseen objects are assigned an optical flux density one magnitude fainter that the detection limit (cross-hatched area).} 

\end{figure}

Figures \ref{optical_magsA} and \ref{optical_magsB} show the distribution of optical magnitudes for samples A and B, where the \it{O}\normalfont-band magnitudes have been taken from the APM catalogue. Sources with no optical ID are assigned a magnitude which is one magnitude fainter than the detection limit and are shown by the cross-hatched bins; we take the detection limit for the blue plates to be the completeness limit: $O=21.5$. In sample A the median \it{R}\normalfont-band magnitude is 19.4, and the median\it{O}\normalfont-band magnitude is  21.6. In sample B the median \it{R}\normalfont-band magnitude is 19.3, and the median \it{O}\normalfont-band magnitude is 20.9; these are slightly lower values than for sample A, as expected since this is the sample with the higher flux density limit so the objects are expected to be at lower redshift. 

Using the \it{R}\normalfont-band optical images (P60 or DSS2 as appropriate), P60 or APM \it{R}\normalfont-band magnitudes and APM \it{O}\normalfont-band magnitudes we put the optical counterparts into four classes in table 2: those unseen in the images, those that appear extended (``G'': galaxies), those that are point-like and blue, with $O-R < 1.6$ \citep[``Q?'': potential quasars, this is a slightly more strict classification of objects as ``blue'' than][who take $O-R < 1.8$ in their selection of quasar candidates]{riley} and those that are point-like but do not have $O-R$ definitely less than 1.6 (``G?'', this will contain a mixture of unresolved galaxies and faint quasar candidates). Whilst this classification scheme is useful on a source by source basis, the differing quality of the optical images (in particular the relatively poor seeing of the DSS2 images compared to the P60 images) makes it difficult to draw comparisons between optical counterpart types for different radio classes since few steep spectrum sources were imaged with the P60.

To avoid these imaging biases we define optical classes on the basis of colour alone: counterparts with $O-R < 1.6$ are ``blue'' and we define objects with $O-R \geq 1.6$ to be ``red''. We calculate the median spectral indices for the ``red'' and ``blue'' counterparts, leaving out objects with inconclusive limits on the colour; e.g. all those unseen in \it{O}\normalfont-band but not bright enough in \it{R}\normalfont-band to have $O-R$ definitely greater than 1.6, and vice versa for those seen only in the blue filter. Objects not detected in either filter are also excluded. The median values of $\alpha$ are: 0.63 for the ``red'' objects in sample A; 0.30 for the ``blue'' objects in sample A; 0.49 for the ``red'' objects in sample B and 0.17 for the ``blue'' objects in sample B. Again, sample B has lower values of spectral index than sample A, and also, the ``blue'' counterparts in each sample have lower spectral indices than the ``red'' counterparts, which supports the standard interpretation that flat spectrum radio sources are quasars and are blue in the optical \citep[see, e.g.][]{riley}.

The full sample is classified into optical colour classes by combining the ``red'' objects and all those with inconclusive optical colours into a single class (``not blue''). Table \ref{radio_opticalA_bluered} shows the relative numbers of unseen, ``blue'' and ``not blue''  objects in the different radio classes of samples A and B. The expected trend is that a lower fraction of steep spectrum sources will be blue than the flat and rising sources. This appears to be the case for sample B: at least 74\,percent of the steep spectrum sources are ``not blue'' and at most 49\,percent of the flat or rising spectrum sources are ``not blue''. However, for sample A there are so many steep spectrum objects without optical IDs that there is no significant difference between the different radio classes. Again, given the uncertainties introduced by the unseen objects, Samples A and B are not statistically different in terms of the optical properties of each radio class.

\begin{table} 
\centering
 \caption{\label{radio_opticalA_bluered}Optical counterparts classified only by optical colour for samples A and B. Objects definitely bluer that $O-R = 1.6$ are classified as ``blue'', all others with IDs are ``not blue''. Percentages are given as the fractions within each radio class that fall into each optical colour class.} 
 \begin{tabular}{@{}l@{\hspace{0.4cm}}r@{\hspace{0.2cm}}r@{\hspace{0.4cm}}r@{\hspace{0.2cm}}r@{\hspace{0.4cm}}r@{\hspace{0.2cm}}r@{\hspace{0.4cm}}r@{}}
 \hline
SAMPLE A &  \multicolumn{2}l{Unseen} & \multicolumn{2}l{Blue} & \multicolumn{2}l{Not blue} & Total \\
 \hline
Steep spectrum & 13 & 23\,\% & 17 & 30\,\% & 26 & 46\,\% & 56 \\
Flat spectrum & 2 & 4\,\% & 22 & 48\,\% & 22 & 48\,\% & 46 \\
Rising spectrum & 2 & 9\,\% & 10 & 45\,\% & 10 & 45\,\% & 22\\
\hline
\hline
SAMPLE B &  \multicolumn{2}l{Unseen} & \multicolumn{2}l{Blue} & \multicolumn{2}l{Not blue} & Total\\ 
 \hline
Steep spectrum & 1 & 4\,\% & 5 & 22\,\%  & 17  & 74\,\% & 23\\
Flat spectrum & 3 & 11\,\% & 16 & 57\,\% & 9 & 32\,\% & 28\\
Rising spectrum & 0 & 0\,\% & 8 & 42\,\% & 11 & 58\,\% & 19\\
\hline
\end{tabular}

\end{table}

\section{Comparison with previous work}
\label{sec:compare}

The most extensively studied survey close in frequency to 9C is that of PW who looked at a sample of 168 radio sources defined at 2.7\,GHz and complete to 1.5 Jy. PW measured the (non-simultaneous) spectral index of each source between 1.4 and 2.7\,GHz and classified sources as extended steep spectrum (ESS: resolved and with $\alpha > 0.5$), compact steep spectrum (CSS: unresolved and with $\alpha > 0.5$) and flat spectrum ($\alpha < 0.5$). PW used data with a resolution of, at best, 2\,arcsec. We have reclassified their objects as flat or rising spectrum using the cut-off at $\alpha = -0.1$ as for the 9C sources.  In table \ref{PW_compare} the spectral and structural properties of the PW sample and samples A and B are compared. Clearly, the samples defined at 15\,GHz are strongly biased toward the flat and rising spectrum objects as compared with the PW sample, containing two to three times the fraction of rising spectrum sources. \citet{odea} shows that the fraction of GPS sources is still only $\sim10$\,percent in samples selected at frequencies as high as 5\,GHz, compared with the 20-30\,percent in 9C. Comparisons between the PW and 9C samples should be made with caution because of the very much higher flux density cut-ff in the PW sample. However, the trend for the fraction of GPS sources to increase as the flux limit is increased in the 9C samples suggests that a sample selected at 15\,GHz with a flux limit comparable to 1.5\,Jy would be even more rich in GPS sources than the samples we have studied here.

\begin{table}
\caption{\label{PW_compare} Percentages of different radio-source classes from the PW sample and the 15\,GHz samples.}
\begin{center}
\begin{tabular}{@{}lrrrr@{}} 
\hline
& \multicolumn{3}c{Selection frequency \& flux limit}\\
 & 2.7\,GHz 1.5\,Jy & 15\,GHz 60\,mJy  & 15\,GHz 25\,mJy \\ 
\hline  
 ESS & 48\,\% & 19\,\% & 27\,\% \\ 
\hline  
CSS & 22\,\% & 14\,\% & 19\,\% \\ 
\hline  
Flat & 21\,\% & 40\,\% & 37\,\% \\
\hline  
Rising & 9\,\% & 27\,\% & 18\,\% \\
\hline 
\end{tabular}
\end{center}
\end{table}

\section{Creating samples rich GPS sources}
\label{GPS}

\begin{figure}
\centerline{ \epsfig{figure=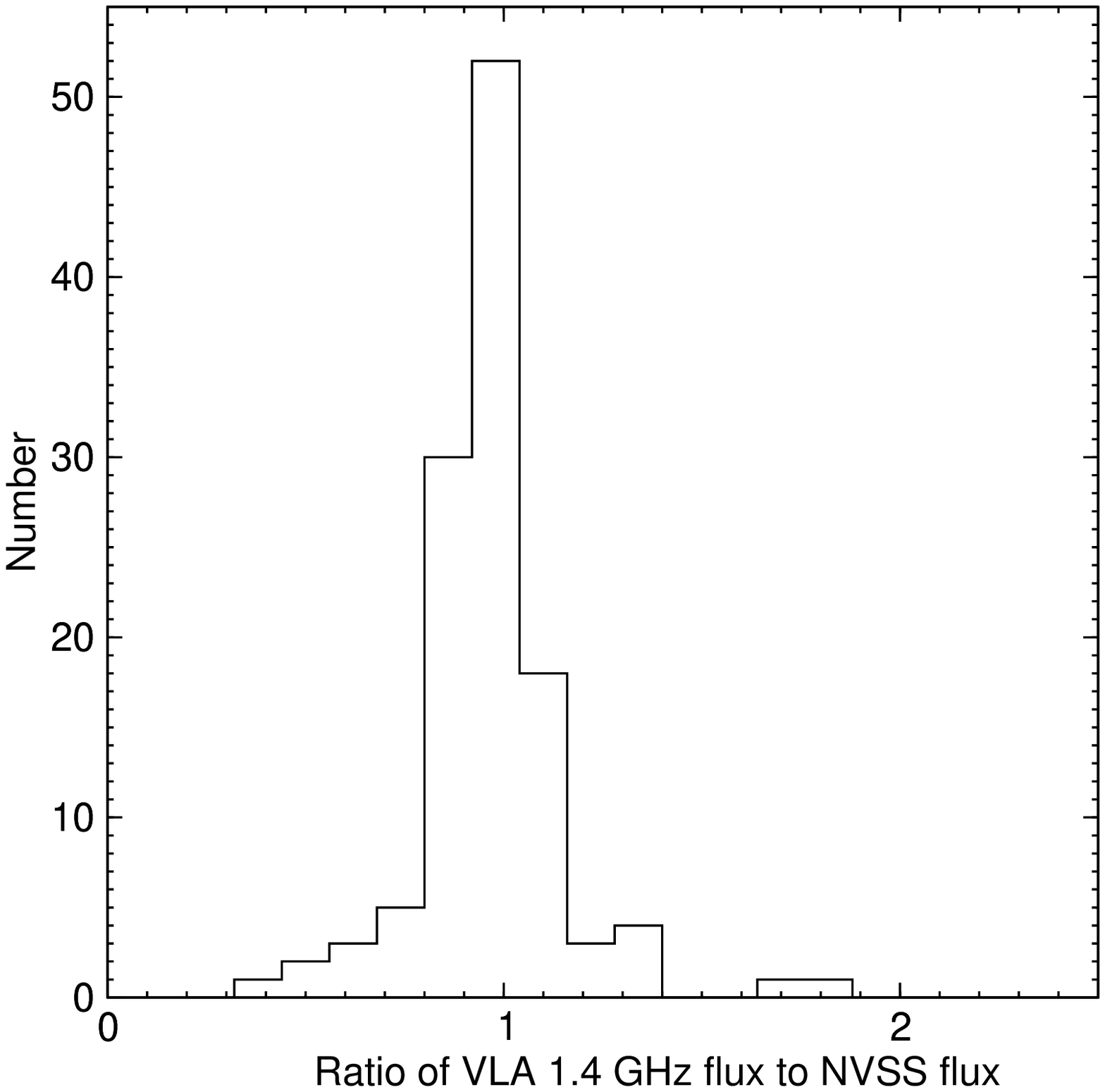,angle=0,width=7cm,clip=}}
\centerline{ \epsfig{figure=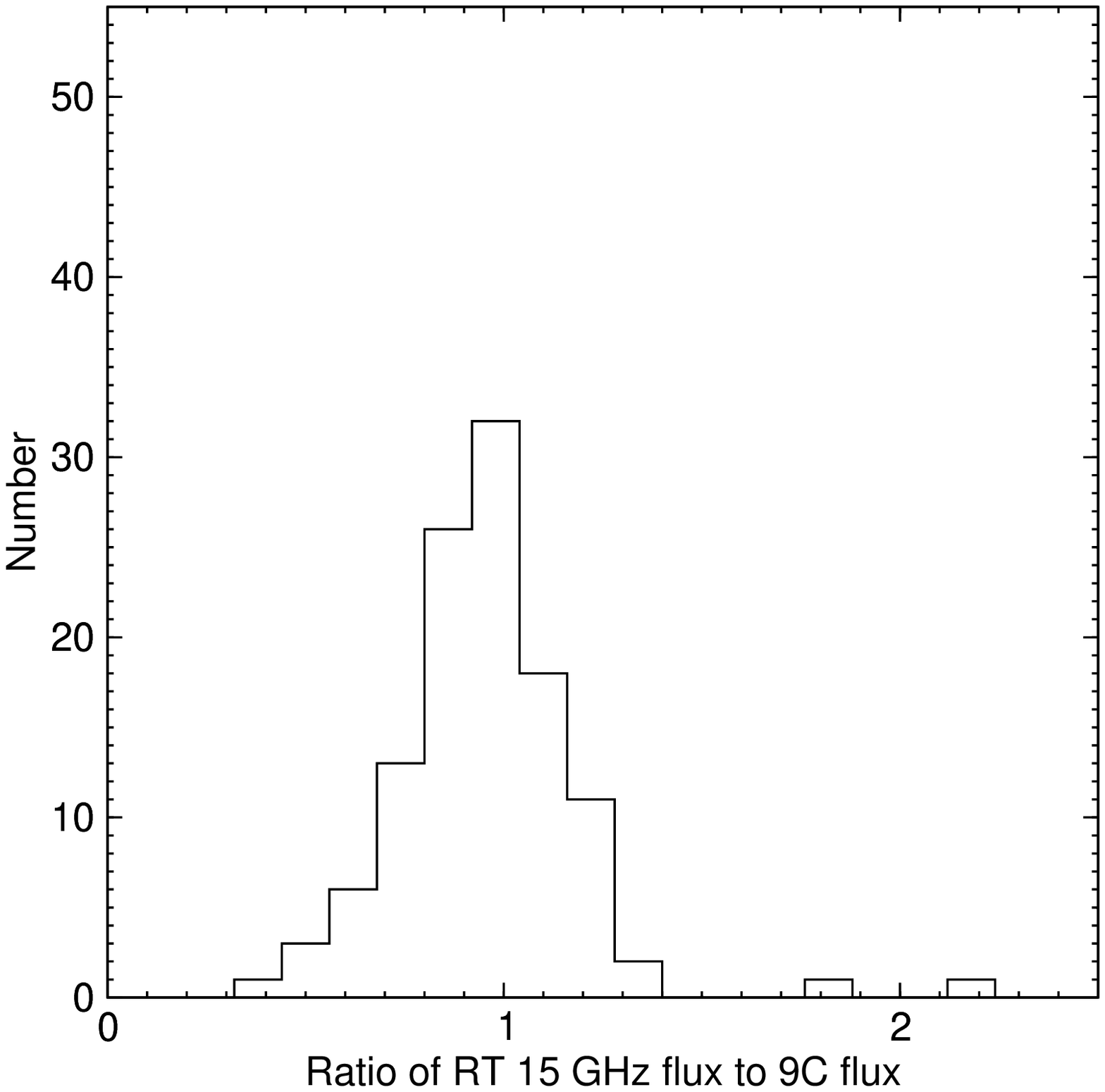,angle=0,width=7cm,clip=}}
\caption{\label{C} Histograms showing the ratios of flux densities from our follow-up work to the flux densities given in the surveys at  1.4\,GHz (NVSS; top) and 15\,GHz (9C; bottom).} 
\end{figure}

We have defined complete, flux-limited samples from a survey at 15\,GHz that contain higher fractions of GPS sources than samples selected at lower radio frequency. Future work on GPS sources will benefit if complete samples of such objects can be selected without the need for time-consuming multi-frequency measurements of all objects in the flux-limited samples. 

The efficiency of selecting GPS sources can be increased by removing sources with steeply falling spectra as indicated by existing survey measurements: in particular comparison of the old, non-simultaneous NRAO VLA Sky Survey \citep[NVSS, see][]{C1} flux density measurements at 1.4\,GHz and the 9C catalogue flux densities for each object gives an indication of the spectral type. Figure \ref{C} shows histograms of the ratio between the simultaneous flux density measurements to those from surveys -- the top figure shows this histogram for the 1.4-GHz data (follow-up VLA: NVSS) and the lower is for the 15-GHz data (follow-up RT: 9C). The 15\,GHz histogram is wider, indicating, as expected, that source variability is more prevalent at 15\,GHz than at lower radio frequency. NVSS was carried out between 1993 and 1996 and the 9C data points were taken between 1999 November and 2001 June, so the time offset between the survey data points and our simultaneous data points is around 6 years for the 1.4\,GHz data and  between about 1 and 2 years at 15\,GHz. To quantify the variability of the 9C samples on timescales from a few days to a year or so we are carrying out systematic RT observations at 15\,GHz of one third of the sources from the flux-limited samples (chosen randomly), to be presented elsewhere.

\begin{figure}
\centerline{ \epsfig{figure=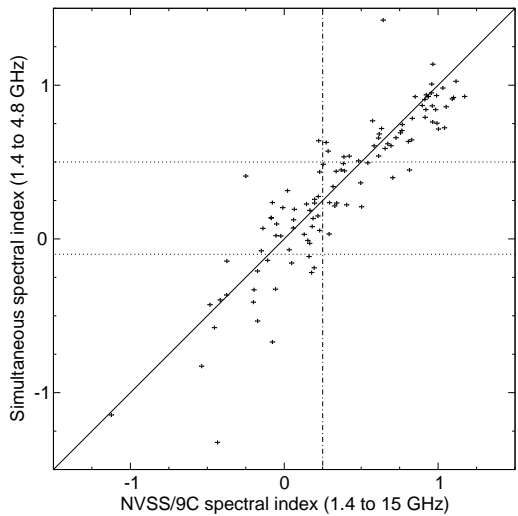,width=7cm,clip=}}
\caption{\label{oldLC} Simultaneous spectral index taken between 1.4 and 4.8\,GHz plotted against the spectral index calculated using the 1.4-GHz NVSS flux density and the 15-GHz 9C flux density. The horizontal dashed lines are at the cut-offs for the radio spectral classes ($+0.5$ and $-0.1$). The vertical dot-dashed line is at an NVSS/9C spectral index value of 0.25.}
\end{figure}

Figure \ref{oldLC} shows the correlation between spectral index calculated between 1.4 and 15\,GHz from NVSS and 9C and the spectral index calculated from the simultaneous measurements at 1.4 and 4.8\,GHz (i.e. the spectral index we use to classify the radio sources). The graph shows that the two measurements are similar, but there is significant scatter. The GPS sources with $\alpha^{4.8}_{1.4} < -0.1$ have a tendency for their NVSS/9C spectral indices to be higher than their simultaneous 1.4 to 4.8\,GHz spectral indices. This is due to the turnover or flattening of the spectra of these sources between 4.8 and 15\,GHz. Note however that, as suggested in section \ref{sec:radio_stats}, the sources have not turned over sufficiently to have steeply falling spectra when averaged between 1.4 and 15\,GHz. Selecting all sources that have an NVSS/9C spectral index less steep than $+0.25$ will pick up all of the rising spectrum sources, as well as a few flat spectrum sources, but very few steep spectrum objects (since the spectral index of a convex-spectrum source increases with frequency). Using the spectral index cut-off of $\alpha_{NVSS}^{9C} < 0.25$, and a 9C flux limit of 25\,mJy creates a sample in which 49\,percent of the sources would have a simultaneous spectral index of $\alpha^{4.8}_{1.4} < -0.1$.

\section{Conclusions}
\label{sec:discussion}

Our follow-up of 176 radio sources taken from the 9C catalogue has shed light on the properties of the high frequency radio population. Multi-frequency simultaneous measurements have provided maps and radio spectra. Optical imaging at the Palomar 60-inch telescope in combination with the DSS2 digitised form POSS-II sky survey plates has led to the identification of counterparts for a total of 155 of the 176 sources studied. 

\setcounter{con}{1}(\arabic{con}) At a flux limit of 25\,mJy, 18\,percent of sources display spectra that peak at 5\,GHz or above. 27\,percent of the 60\,mJy sample peak at 5\,GHz or above. These rising spectrum sources represent a much larger fraction of the radio source population at 15\,GHz than at lower radio frequencies.

\addtocounter{con}{1}(\arabic{con}) Increasing the flux limit raises the fraction of rising spectrum sources: in a sample of sources selected with a 9C flux density at 15\,GHz of more than 150\,mJy, 39\,percent of the sources have rising spectra between 1.4 and 5\,GHz.

\addtocounter{con}{1}(\arabic{con}) At higher frequencies the median spectral index increases, due to turn-over. However, the median value remains constant up to 15\,GHz.

\addtocounter{con}{1}(\arabic{con}) The optical counterparts of the steep spectrum sources are less often blue in colour than those of the flat and rising spectrum sources. The optical colours and sizes suggest that between one third and one half of the peaked spectrum and flat spectrum sources may be quasars.

\addtocounter{con}{1}(\arabic{con}) We have shown that a sample defined as a flux-limited complete sample from 9C and then refined to contain only those sources having a spectral index taken from NVSS at 1.4\,GHz and 9C at 15\,GHz of $\alpha_{NVSS}^{9C} < 0.25$ will still contain all of the peaked-spectrum sources from the complete sample. The efficiency of finding peaked-spectrum objects can thus be raised to 49\,percent -- allowing larger samples of GPS sources to be studied with the same amount of dedicated telescope time, without compromising on completeness. 

\section{Acknowledgements}
\label{sec:ack}
RCB is supported by a PPARC studentship. GC acknowledges support from PPARC observational rolling grants PPA/G/O/2003/00123 and PPA/G/O/2001/00575.

The National Geographic Society - Palomar Observatory Sky Atlas (POSS-I) was made by the California Institute of Technology with grants from the National Geographic Society.

The Second Palomar Observatory Sky Survey (POSS-II) was made by the California Institute of Technology with funds from the National Science Foundation, the National Aeronautics and Space Administration, the National Geographic Society, the Sloan Foundation, the Samuel Oschin Foundation, and the Eastman Kodak Corporation.

The Digitized Sky Survey was produced at the Space Telescope Science Institute under U.S. Government grant NAG W-2166. The images of these surveys are based on photographic data obtained using the Oschin Schmidt Telescope on Palomar Mountain and the UK Schmidt Telescope. The plates were processed into the present compressed digital form with the permission of these institutions.

\end{document}